\documentclass[12pt,a4paper,openright]{book}
\usepackage{fancyhdr}
\pdfoutput=1
\usepackage[utf8]{inputenc}
\usepackage{subfig}
  \usepackage[round]{natbib}
\usepackage{afterpage,multicol}
\usepackage{caption}
\usepackage{amssymb,color,longtable}
\usepackage{hyperref,color,longtable,graphicx,}
\usepackage{makeidx,fancybox}
\usepackage{float, lscape,amsmath}
\usepackage{epstopdf}
\newcommand{\Rs}{R$_{\odot}$\,}
\newcommand{\kms}{km s$^{-1}$\,}
\newcommand{\CD}{$C_{\rm D}$\,}
\newcommand{\htil}{$\widetilde{h}_{0}$\,}

\begin{document}
 
 \chapter*{}
\thispagestyle{empty}
\setlength{\textheight}{650pt}
\setlength{\oddsidemargin}{25pt}
\setlength{\evensidemargin}{0pt}
\setlength{\marginparwidth}{57pt}
\setlength{\footskip}{30pt}

%
%

\pagenumbering{gobble}
\begin{center}
{\bf \huge Dynamics of solar Coronal Mass Ejections: forces that impact their propagation}\\
  \vspace{2cm}

A thesis\\
Submitted in partial fulfillment of the requirements\\
Of the degree of\\
Doctor of Philosophy\\[0.8cm]
By\\[1cm]
{\Large Nishtha Sachdeva }\\
20123214 \\[1.50cm]

   \includegraphics[width=3cm]{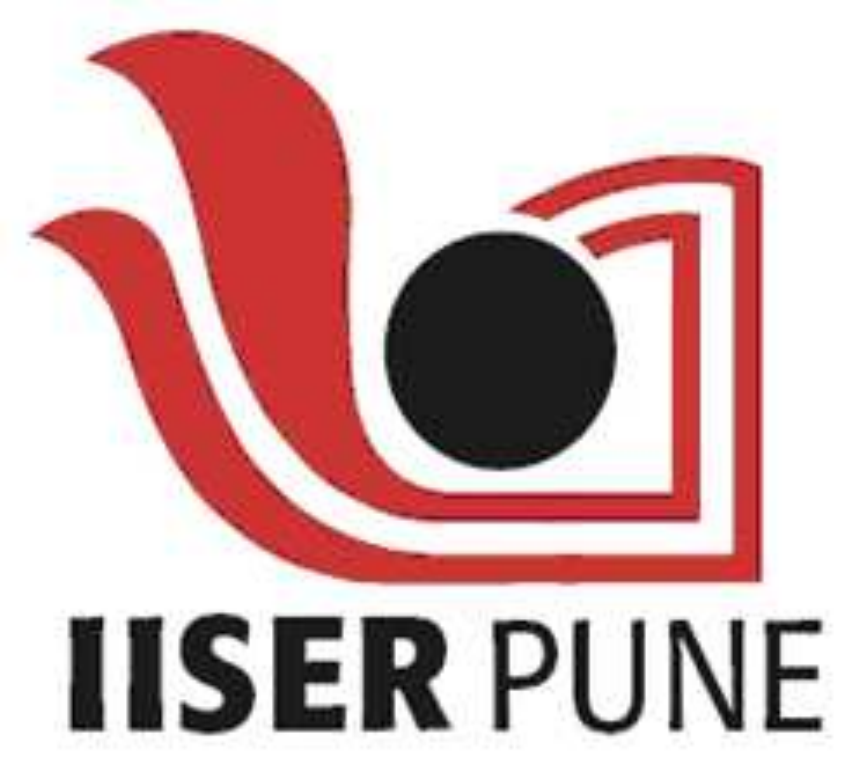}\\  \vspace {0.5cm}  

{INDIAN INSTITUTE OF SCIENCE EDUCATION AND RESEARCH PUNE}\\[1cm]

  \vfill
{January, 2018}

\end{center}

\thispagestyle{empty}
\cleardoublepage

\pagenumbering{roman}
\chapter*{Certificate}\addcontentsline{toc}{section}{\textbf{Certificate}}
Certified that the work incorporated in the thesis entitled 
``\textit{Dynamics of solar Coronal Mass Ejections: forces that impact their propagation}'',
submitted by \textit{Nishtha Sachdeva} was carried out by the candidate, under my supervision. The work presented here or any part of it has not been included in any other thesis submitted previously for the award of any degree or diploma from any other University or institution.\\[3cm]
\textit{Date} \hfill \textbf{Dr. Prasad Subramanian}
\cleardoublepage

\chapter*{Declaration}\addcontentsline{toc}{section}{\textbf{Declaration}}
{
I declare that this written submission represents my ideas in my own words and where others' ideas have been included, I have adequately cited 
and referenced the original sources. I also declare that I have adhered to all principles of academic honesty and integrity and have not misrepresented 
or fabricated or falsified any idea/data/fact/source in my submission. I understand that violation of the above will be cause for disciplinary 
action by the Institute and can also evoke penal action from the sources which have thus not been properly cited or from whom proper permission has 
not been taken when needed.\\[3cm]}
\textit{Date} \hfill \textbf{Nishtha Sachdeva}\\
\begin{flushright}
Roll No.- 20123214 
\end{flushright}
\cleardoublepage
\vspace*{\fill}

  \begingroup
\begin{center}
\vspace{-5cm}
\huge \textbf{\textit{For\,\,\\ \,Ma  \& Papa}}
\vspace*{\fill}
\end{center}

 \endgroup

\pagestyle{fancy}
\fancyhf{}
\fancyhead[LO]{\nouppercase{\rightmark}}
\fancyhead[RE]{\nouppercase{\rightmark}} 
\fancyhead[LE,RO]{\thepage} 

\setlength{\parskip}{10pt}
\pagenumbering{roman}

\chapter*{Abstract} \addcontentsline{toc}{section}{\textbf{Abstract}} \chaptermark{Abstract}

The Sun occasionally ejects parts of its outer atmosphere into the interplanetary medium.
These massive, large-scale eruptions from its corona are called coronal mass ejections (CMEs). As CMEs propagate, they evolve and expand, often driving 
interplanetary shocks and accelerating energetic particles. Earth-directed CMEs can cause extreme geo-magnetic 
storms leading to significant disruptions in satellite operations, space-bound technologies, and near-Earth space weather. It is therefore becoming 
increasingly important to understand the Sun-Earth dynamics of CMEs in order to develop reliable tools for predicting their arrival speed and time at the Earth.

It is generally assumed that Lorentz forces dominate the early stages of CME propagation and solar wind aerodynamic drag takes over later on. However, 
the precise distance range where one force dominates over the other is not well known. 

In this study, we investigate the Sun-Earth dynamics of a set of 38 well-observed CMEs using data from the {\it Solar Terrestrial Relations Observatory}
(STEREO), the {\it Solar and Heliospheric 
Observatory} (SOHO) missions and the \textit{WIND} instrument. We seek to quantify the relative contributions of Lorentz force 
and aerodynamic drag on their 
propagation. The CMEs are 3D reconstructed using a geometrical fitting technique called the Graduated Cylindrical Shell (GCS) 
model to derive observed CME parameters. The fitting procedure provides the height-time profile, radius, and width of the CMEs which are used to derive 
other parameters, such as mass and cross-sectional area. These observed and derived parameters are used in the models for the forces acting on CMEs. 
Using a microphysical prescription of the drag coefficient (and not an empirical value), we find that solar wind aerodynamic drag adequately accounts for the dynamics 
of the fastest CMEs (initial velocity$> 900$ \kms) from as low as 3.5 \Rs \citep{Sac15}. For relatively slower CMEs, however, we find that when the 
drag-based model is initiated below the distances ranging from 12 to 50 \Rs, the observed CME trajectories cannot be accounted for. This 
suggests that aerodynamic drag force dominates the 
dynamics of slower CMEs only above these heights. This is at variance with the general perception that the solar wind drag influence is significant just above a few solar radii.
We also find that for slower CMEs, the drag ``does not do much'', i.e. the CMEs evolve very little above these heights. 

To investigate CME dynamics below the heights where aerodynamic drag dominates, we consider the Torus Instability model for the 
driving Lorentz force \citep{Kli06}. 
Using observational inputs to the model, we find that the Lorentz force increases from equilibrium and peaks between 1.65--2.45 \Rs for all CMEs, following which it decreases 
gradually. We do not find a clear distinction between the peak positions for slow and fast CMEs. We find that for fast CMEs, Lorentz forces become negligible in comparison to aerodynamic 
drag as early as 3.5--4 \Rs. For slow CMEs, however, they become negligible only by 12--50 \Rs. 
This justifies the success of the drag-only model for fast CMEs. In case of slow CMEs, the Lorentz force is only slightly smaller than the drag force even beyond 12--50 \Rs. 
In other words, the difference between the two forces is more pronounced for fast CMEs than for slow ones. For these slow events, our results suggest 
that some of the magnetic flux carried by CMEs might be expended in expansion or heating. These dissipation effects might be important in describing the propagation of slower 
CMEs \citep{Sac17}. To the best of our knowledge, this is the first systematic study in this regard using a diverse CME sample. 

A physical understanding of the forces that affect CME propagation and how they compare with each other at various heliocentric distances is an important ingredient in 
building tools for describing and predicting CME trajectories.

\chapter*{List of Publications}\addcontentsline{toc}{section}{\textbf{List of Publications}}
\label{publications}
\begin{enumerate}

\item \label{Sachdeva15} CME propagation – where does aerodynamic drag ``take over''?\ \\
\textit{\textbf {Sachdeva, N.}}, Subramanian, P., Colaninno, R.,  Vourlidas, A.  2015, The Astropysical Journal, 809, 158

\item \label{Sachdeva17} CME dynamics using STEREO and LASCO observations: The relative importance of Lorentz Forces and Solar wind drag. \\
\textit{ \textbf {Sachdeva, N.}}, 
Subramanian, P., Vourlidas, A., Bothmer, V.  2017, Solar Physics, 292, 118
\end{enumerate}

\tableofcontents

\listoffigures \addcontentsline{toc}{section}{\textbf{List of figures}} 
\listoftables \addcontentsline{toc}{section}{\textbf{List of tables}} 

\clearpage
\fancyhead[LO]{\nouppercase{\leftmark}} 
  \fancyhead[RE]{\nouppercase{\rightmark}}
\fancyhead[LE,RO]{\thepage}
\pagenumbering{arabic}
 .\chapter{Introduction}
\label{chap1}

\noindent\makebox[\linewidth]{\rule{\textwidth}{3pt}} 
{\textit {This chapter introduces some overall properties of the Sun, coronal mass ejections (CMEs) and space weather. We begin with an 
introduction to the properties of the Sun, its structure and atmosphere. Historical and observational details of CMEs are discussed, describing the instruments used to observe 
the Sun and solar transients. We describe the properties of CMEs along with details regarding their onset and propagation. We also discuss the space weather and how the Sun-Earth connection is 
influenced by CMEs. We finally detail the organization of the rest of the thesis.} }\\
\noindent\makebox[\linewidth]{\rule{\textwidth}{3pt}}

\section{The Sun}
\subsection{Introduction}
About 4.6 billion years ago, the most extraordinary ``ordinary'' star of our solar system was born. The Sun is the source of sustenance of life on Earth and plays a prominent role in 
art, religion, and science in human history. While ancient astronomers recorded the movement and features on the Sun by observing with the naked eye, it was only in the 17$^{th}$ century, with 
the invention of the telescope, that systematic records of sunspot observations were made. Since then we have come a long way in understanding, observing and explaining the 
structure, evolution, and properties of the Sun and its related phenomena.

\begin{figure}[h]
\centering
\includegraphics[height=0.35\paperheight,width=0.65\paperwidth]{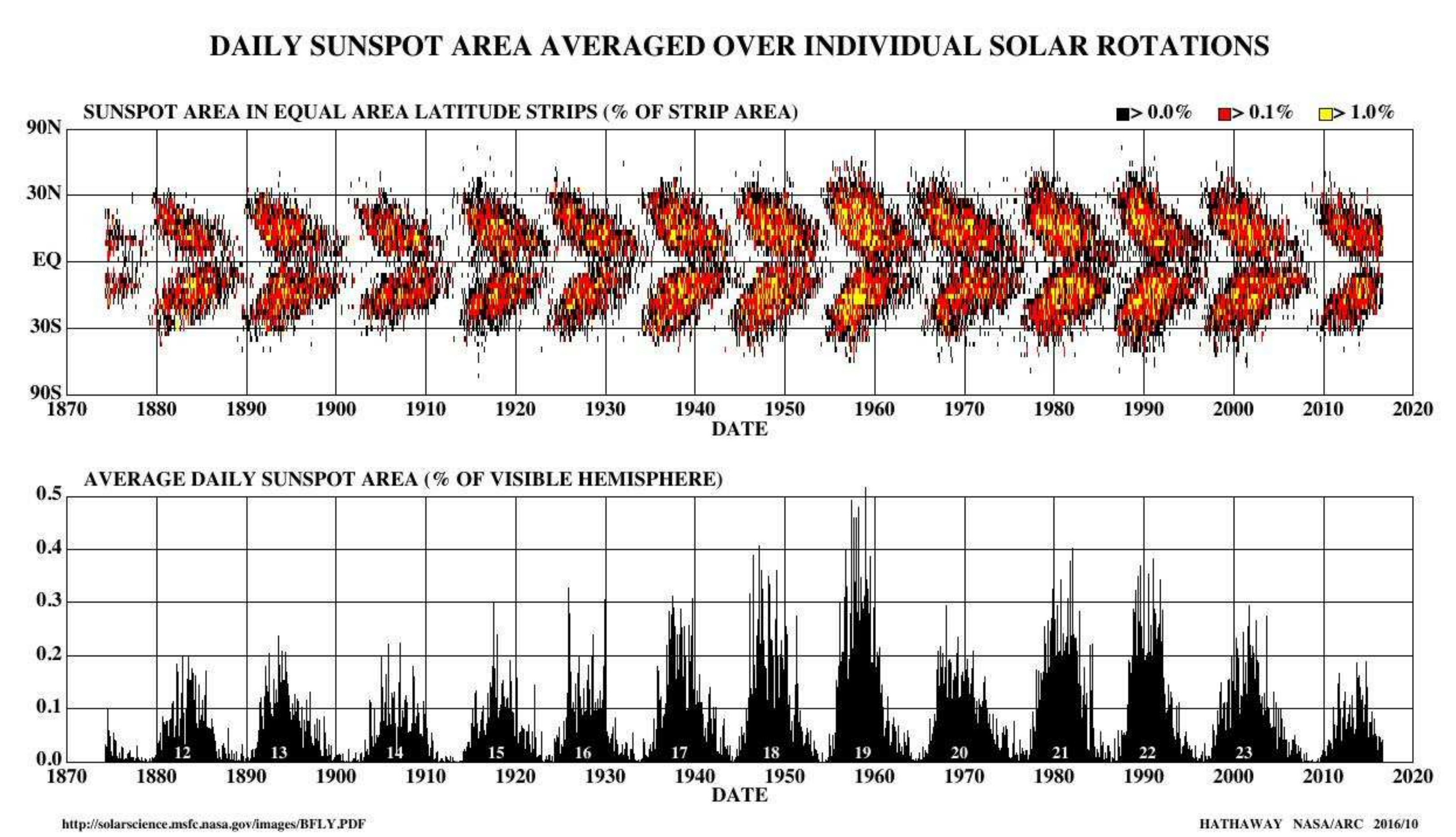}
\caption[Butterfly diagram of sunspots]{The butterfly diagram showing the positions of the sunspots for each rotation of the sun since May 1874.
It can be seen that the spot bands first form at mid-latitudes, widen, and then move towards the equator as each cycle progresses. {\it Image credit- NASA MSFC (https://solarscience.msfc.nasa.gov/).}}
\label{butterfly}
 \end{figure}

\begin{figure}[h]
\centering
\includegraphics[height=0.31\paperheight,width=0.55\paperwidth]{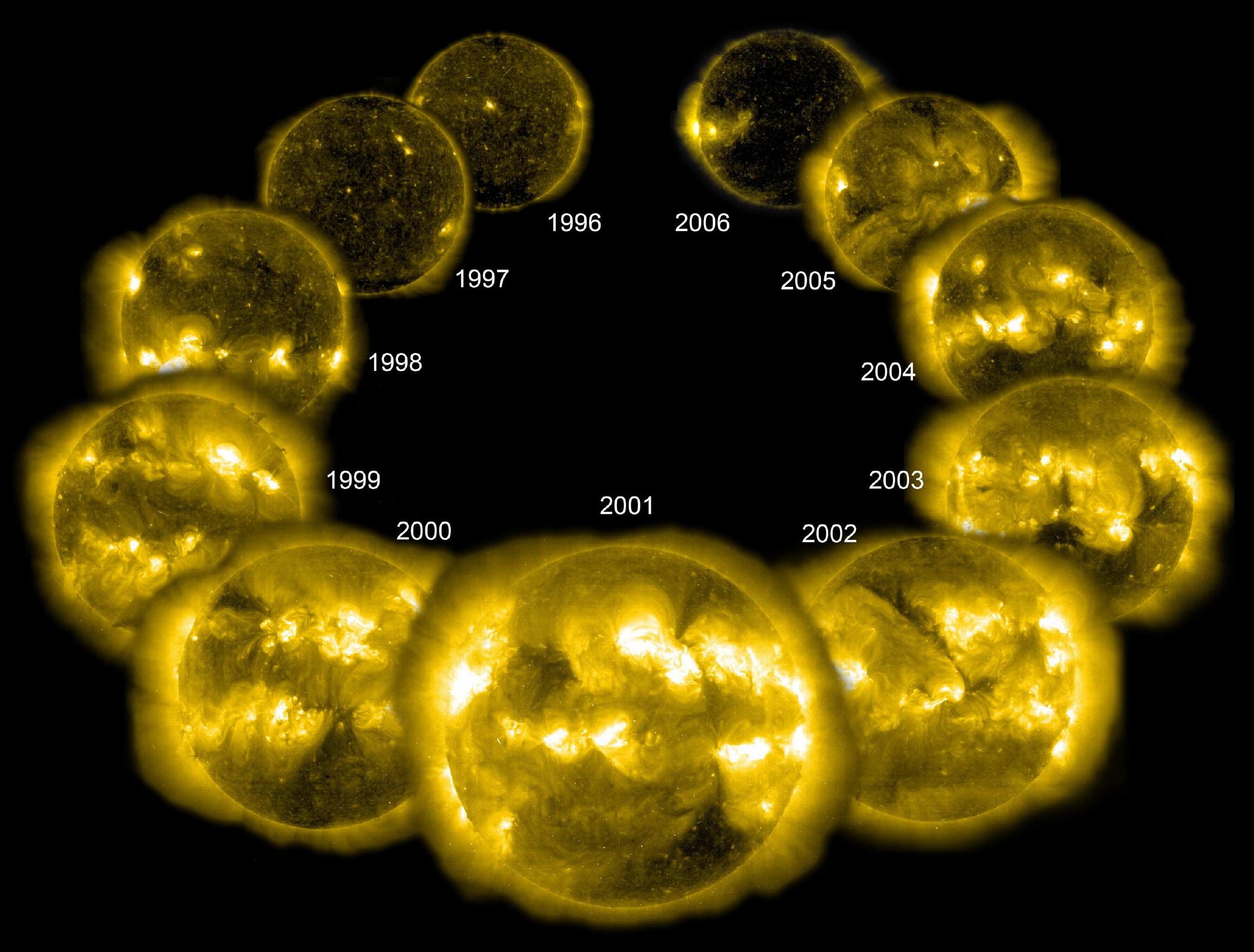}
\caption[Magnetic activity of the Sun over one solar cycle]{{Full disk images of the corona in 284 \AA \, wavelength of extreme UV light of the
the solar cycle 23 (1996 - 2006) showing the variation of magnetic activity on the Sun.{\it Image credit- SOHO EIT}.}}
\label{solarcycle}
 \end{figure}

The sunspot cycle was discovered by Samuel Heinrich Schwabe who found that the number of sunspots varied over a regular period. This was further confirmed by a 
Swiss astronomer Rudolf Wolf (in 1852) who established this period of variation as $\sim$ 11 years. In 1908, George E. Hale measured the magnetic 
field in sunspots for the first time. During a cycle of 11 years, the active, magnetic regions migrate from high latitudes to lower latitudes near the equator which is seen in the famous Butterfly 
diagram of sunspot positions (Figure \ref{butterfly}). The magnetic polarity of the global solar magnetic field reverses over the course 
of a cycle, with the magnetic activity 
varying with the number of sunspots. With observations beginning from 1755, we are currently (in 2017-2018) in the solar cycle 24 which began in December 2008 and reached its maximum in 2014. 
Figure \ref{solarcycle} shows the eleven years of magnetic activity (in extreme UV wavelengths) on the Sun during solar cycle 23, starting with a 
quiet period during the minimum in 1996 to increased activity, followed by a decrease till 2006. 
 
\subsection{Properties}
The nearest star and center of our solar system, the Sun is a main sequence star of spectral type G2V. Below we list a few characteristic properties of 
this star:
\begin{enumerate}
\item {\bf Radius} - The Sun has a equatorial radius of $(6.96\pm 0.001) \times 10^{5}$ km which is about 109 times the radius of the Earth. 
\item {\bf Distance} - The Sun-Earth distance is 1 astronomical unit (AU) which is equal to $1.496 \times 10^{8}$ km or 215 times the solar radii ($R_{\odot}$). 
\item{\bf Mass} - The total mass of the Sun is $(1.98\pm 0.0003) \times 10^{30}$ kg with a volume of $1.4 \times 10^{27}$ cubic meters. It is large enough to fit about 1.3 million Earths.
\item {\bf Luminosity and composition} - Solar luminosity, $L_{\odot}=(3.844 \pm 0.010 \times 10^{26})$ Watt. The Sun is composed primarily of hydrogen and helium and only about 2\% of heavier elements like oxygen, carbon and iron.
\end{enumerate}

\subsection{Solar structure}
Like other stars, the Sun is a massive ball of ionized gas held together by its gravitational pull leading to immense pressure and temperature conditions. The 
structure of the Sun can be conceptually 
divided into six regions: the solar interior which includes the core, the radiative zone, and the convective zone, and the solar atmosphere which includes 
the photosphere, the chromosphere and the corona (Figure \ref{sun}).
\begin{enumerate}
 \item {\bf Solar Interior}
 
The central region of the Sun, called the core has a temperature of about $1.5\times 10^{7}$ K and pressure which exceeds $2.5\times 10^{11}$ atm. 
These conditions sustain the thermonuclear fusion processes which produce large amounts of energy in the core of the Sun. It extends up to 0.25 \Rs.
Beyond the core, up to 0.7 \Rs lies the radiative zone with temperatures of about $2\times 10^{6}$ K. The energy produced in the solar core is transferred via radiation 
through this region. Since the density in this region is high ($2\times 10^{4} - 2\times 10^{2} \, kg\, m^{-2}$) and the mean free path of 
photons is very small ($\sim 9 \times 10^{-2}$ cm), it takes over 170,000 years for the photons to travel from the core to the top of 
the radiative zone.

The core and the radiative zone rotate as a solid body (rigid rotation) whereas the convective zone rotates differentially. 
The layer between the radiative zone and the convective zone is called the tachocline. Due to the difference in the rotation rates, the tachocline is subjected to shear 
flows which are thought to be the source of generating the magnetic fields and powering the solar dynamo. The convective zone extends from 0.7 \Rs up to the solar surface. 
Plasma heated at the tachocline expands and rises up creating convective currents in this zone. This material cools as it reaches the surface, 
which decreases the density and it sinks to the base of the convective zone. As this cycle continues, the solar surface or the photosphere gets a granular appearance.
The magnetic loops generated in the core twist and wind-up due to the differential rotation of the convective zone. This causes the magnetic pressure to increase and the loops
become buoyant, rising up through the solar surface and creating sunspots on the solar disk. These sunspots are visible as dark regions on the 
surface and always appear in pairs of opposite magnetic polarity.
\begin{figure}[h]
\centering
\includegraphics[height=0.28\paperheight,width=0.44\paperwidth]{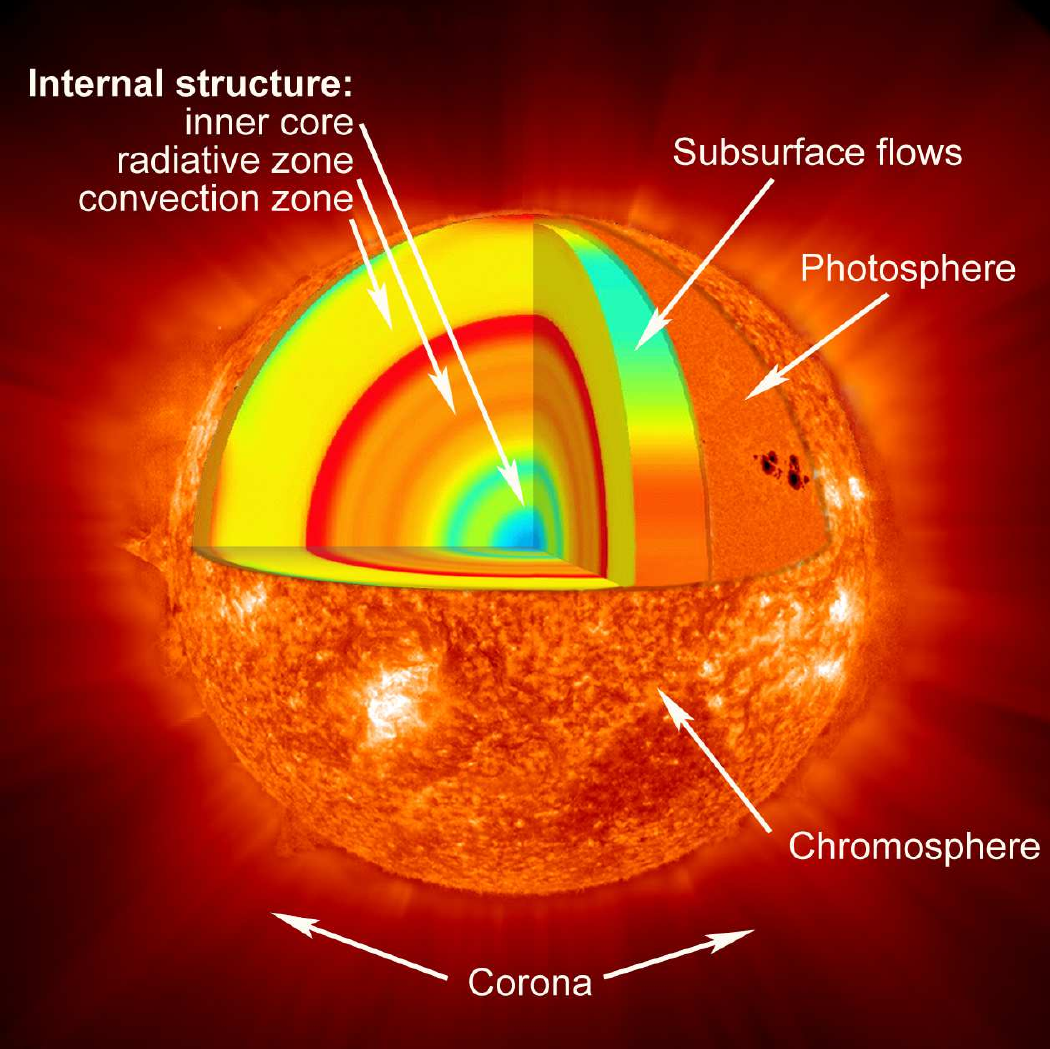}
\caption[Structure of the Sun]{ Image depicting different layers of the Sun. Solar Interior comprising of the hot core, radiative zone and the convective zone. Photosphere forms the solar surface followed by the 
chromosphere and the solar corona. {\it Image credit- NASA (https://www.nasa.gov)}.}
\label{sun}
 \end{figure}

\item{\bf Solar Atmosphere}

The lowest layer of the solar atmosphere is the visible surface of the Sun called the photosphere. It is about 500 km in thickness with a temperature of $\sim$ 5000 K and 
particle density of $10^{23} \,m^{-3}$. The Sun becomes opaque to visible light below the photosphere \textit{i.e.} all light escapes freely above the photosphere. 
The layers of the solar atmosphere above the photosphere are visible only during a solar eclipse. In the chromosphere, the temperature falls up to $\sim$ 4500 K, before it increases to 
about 20000 K. This layer extends to about 2000 km above the photosphere with density $\sim 10^{16}\, m^{-3}$. Above the chromosphere lies a thin layer ($\sim 200$ km) called the transition 
region, where the temperature increases from 20000 K to about $10^{6}$ K. Despite decades of research, it is not yet clear how the corona gets heated to such high temperatures.
The solar corona, at 2500 km from the photosphere has proton densities in the range $10^{14}\, m^{-3}$ to $< 10^{12}\, m^{-3}$ for 
heights $>1$\Rs \citep{Asc05}. The average temperature in the corona ranges from 
$1-2 \times 10^{6}$ to $8-20 \times 10^{6}$ K in the hottest regions (Figure \ref{temp}). The solar corona is visible during eclipses due to Thomson scattering of the photospheric light by the highly 
ionized coronal plasma. 
\begin{figure}[h]
\centering
\includegraphics[height=0.28\paperheight,width=0.5\paperwidth]{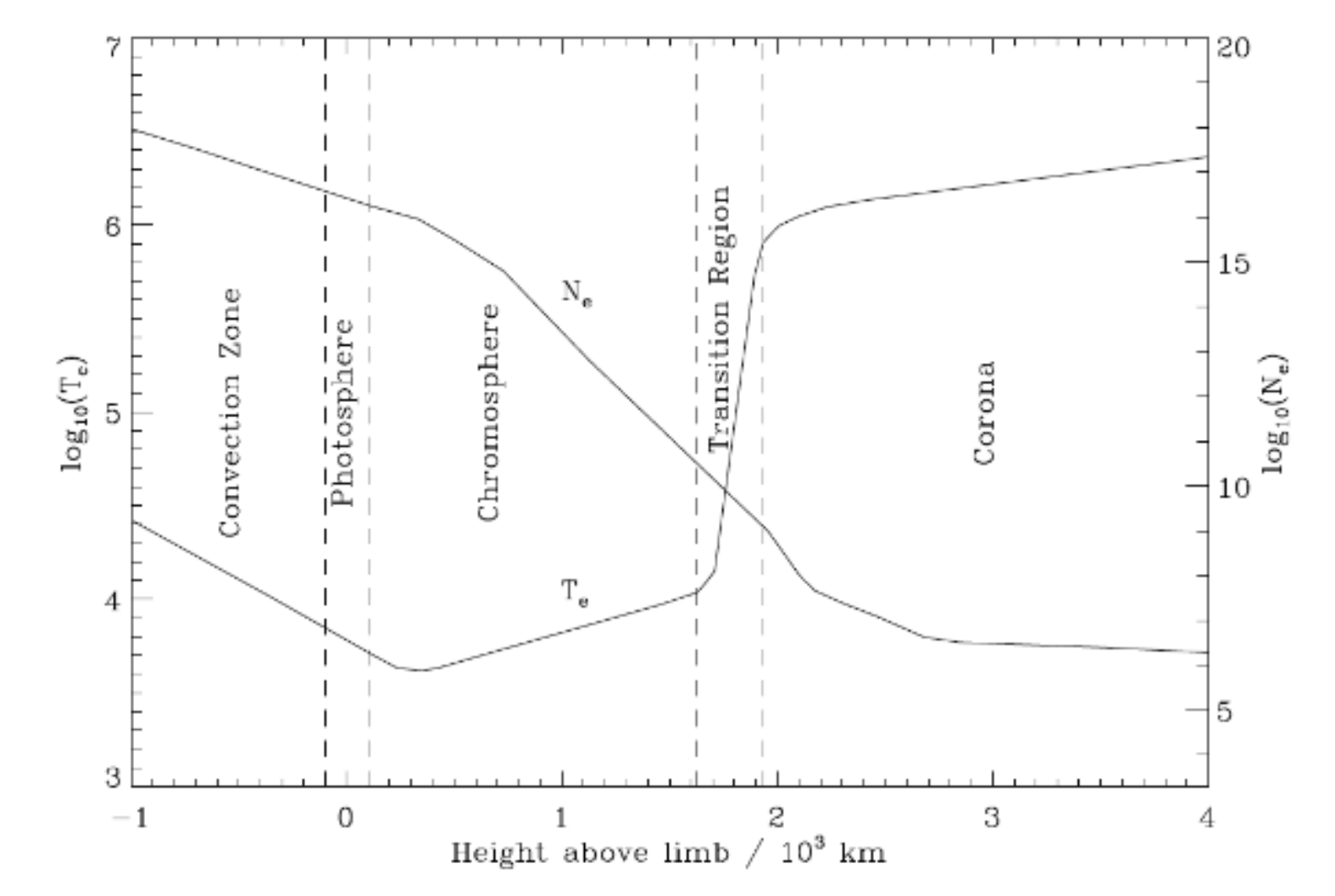}
\caption[Electron density and temperature profile of the Sun]{A 1D model for electron density ($N_{e} (cm^{-3})$ and temperature ($T_{e} (K)$) profile through different layers 
of the Sun) from \citet{Gab82}.}
\label{temp}
 \end{figure}

\subsection{Plasma $\beta$-parameter}
The magnetic fields on the solar surface are not homogeneous. Sunspots can have field strengths of about 2000-3000 G, while active regions on the surface have average photospheric
fields of 100-300 G. On the other hand, the quiet Sun generally has an average field of 0.1-0.5 G.
Ratio of the thermal plasma pressure ($p_{th}$) to the magnetic pressure ($p_{mag}$) is given by the plasma-$\rm \beta$ parameter.
\begin{equation}
 \beta\,=\,\frac{p_{th}}{p_{mag}}\,=\,\frac{n k_{B} T}{B^{2}/8 \pi}
\end{equation}
where, $n$ is the number density, $k_{B}$ is the Boltzmann constant, $T$ is the temperature and $B$ is the magnetic field strength. Plasma-$\beta<1$ indicates that 
the region is magnetically dominated. Most parts of the solar corona have $\beta<1$; however, it increases above $1$ in the outer corona. Gas pressure dominates ($\beta>1$) in the 
photosphere and the chromosphere as well (Figure \ref{beta}).

The solar corona is also the source of a continuous stream of particles in all directions called the solar wind. Due to the high temperature in the corona, particles have enough 
kinetic energy to escape the gravity of the Sun with speeds of about $\sim 300-400$ \kms. The solar wind plasma carries with it the embedded solar magnetic field which forms the 
interplanetary magnetic field (IMF). The magnetic activity of the Sun manifests itself in the form of active regions, flares and solar transients including solar wind and coronal mass 
ejections. These govern the space weather and affect the Sun-Earth climate. 
\end{enumerate}
The next section discusses a specific kind of large-scale solar transients, called coronal mass ejections that primarily drive the space weather and affect the near-Earth environment.

\begin{figure}[h!]
\centering
\includegraphics[height=0.3\paperheight,width=0.55\paperwidth]{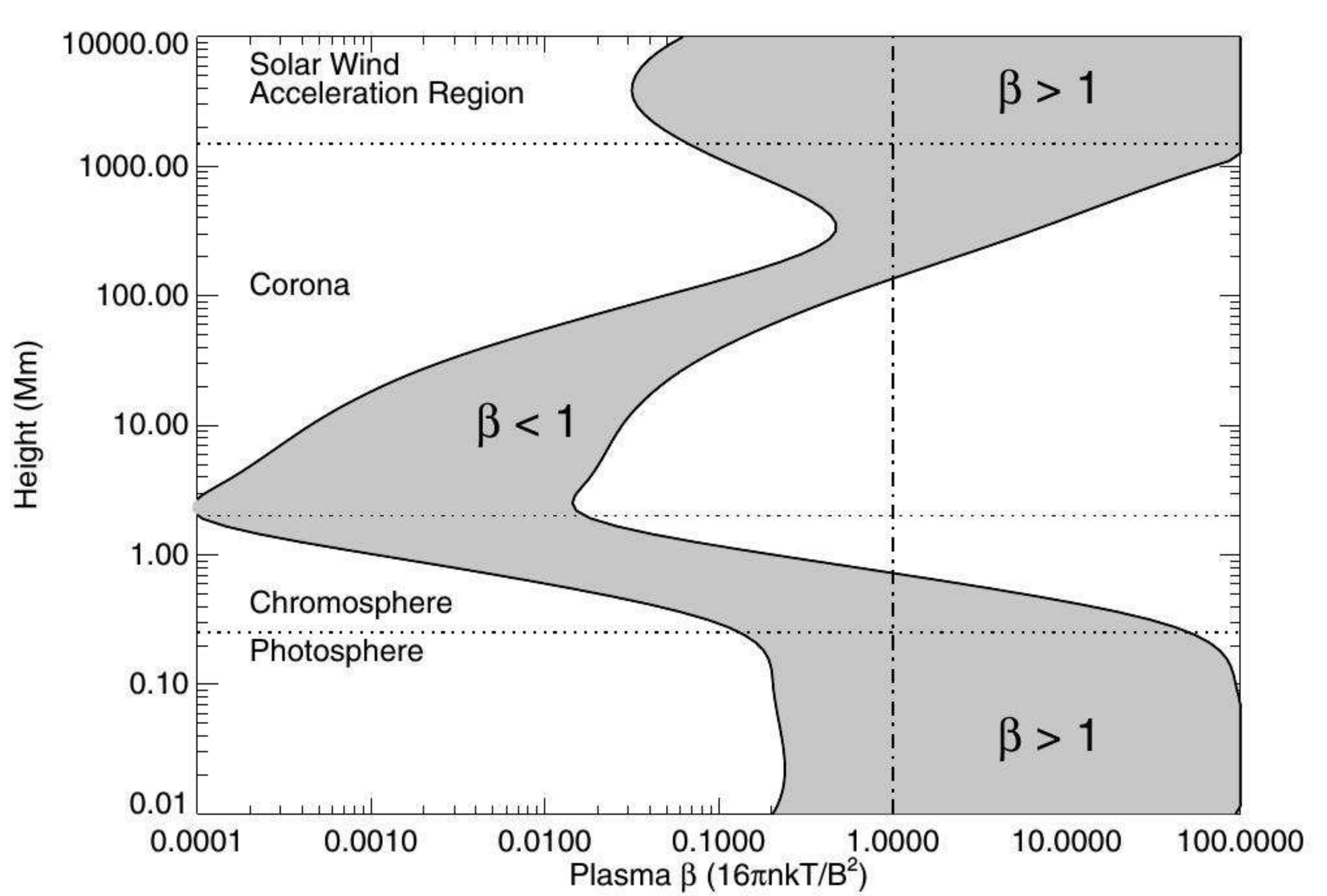}
\caption[Plasma-$\beta$ in the solar atmosphere for 100 G and 2500 G.]{Plasma-$\beta$ in the solar atmosphere for two assumed field strengths-100 G and 2500 G. Magnetic 
pressure dominates over the gas pressure
in the inner corona ($<0.2$ \Rs). {\it Image adapted from \citet{Asc05}}.}
\label{beta}
 \end{figure}
\vfill

\section{Coronal Mass Ejections}
\subsection{Introduction}
\begin{figure}[h]
\centering
\includegraphics[height=0.31\paperheight,width=0.46\paperwidth]{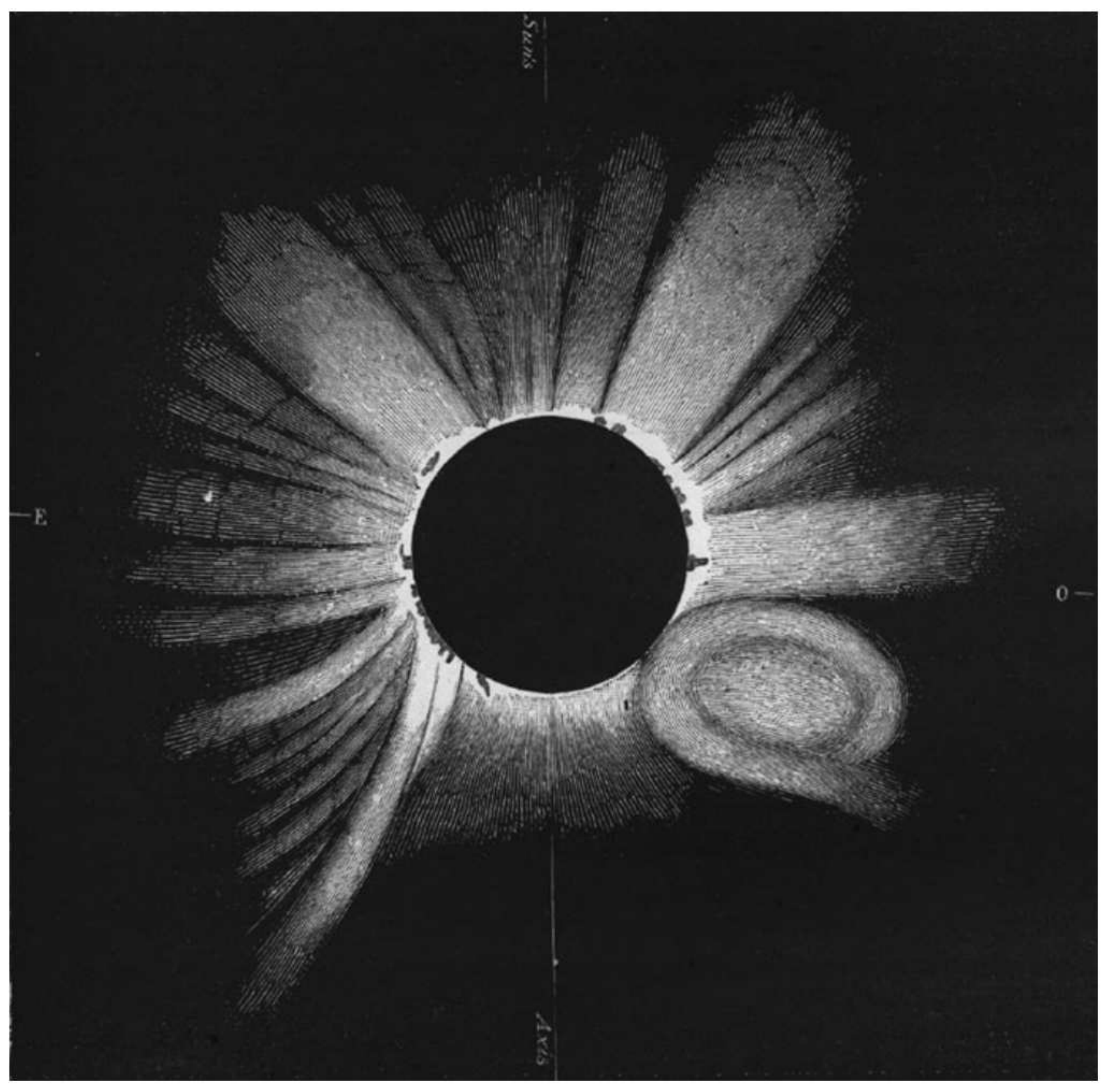}
\caption[First observation of a CME]{First observation of a coronal mass ejection. Drawing of the 1860 solar eclipse. {\it Image taken from \citet{How11}}.}
\label{cmeorig}
 \end{figure}

In 1841, a telegraph system in Exter was affected by strong magnetic fluctuations causing a train delay of about 16 minutes. This is the 
first documented account of manifestation of a solar magnetic activity which was reported by the \citet{Nat71}. Throughout history many such instances have been recorded, most famous being the 
Carrington event of 1859, when a CME was believed to have impacted the Earth's magnetosphere causing one of the biggest solar storms in history. Many such accounts of geo-magnetic storms 
exist. However, coronal mass ejections (CMEs), were discovered, only in 1971 \citep{Han71,Tou73}. Figure \ref{cmeorig} shows one of the earliest drawings of the 
1860 solar eclipse that shows a CME identified later by Jack Eddy.

Coronal mass ejections (CMEs) are massive and energetic expulsions of coronal plasma and magnetic fields into the interplanetary (IP) medium. CMEs may erupt from any region 
of the solar corona, but are often mostly associated with the lower latitudes, especially during the solar minimum. CMEs can have speeds ranging from a few hundred to a 
few thousand kilometers per second with mass of the order of $10^{15}$ gms. CMEs often 
create shocks in the interplanetary medium as they propagate which causes particle acceleration and bursts of radio emission.
They sometimes impact the Earth's magnetosphere; when they do, they can cause geomagnetic storms which disrupt space-based technologies, navigation, telecommunications 
and pose a threat to the safety of airline carriers and astronauts. Due to their direct impact on space weather as well as their indirect affect on humankind, 
CMEs are a widely studied solar phenomena. CME observations using space-based instruments began in the twentieth century and continues with major 
upgradations in observing technology and quality of data.

\subsection{History of CME observations}
\begin{enumerate}
 \item {\bf Remote Sensing}
 
White light emissions from the solar corona arise from photospheric radiation which is Thomson-scattered by free electrons. An enhanced brightness indicates an enhanced
coronal column density along the line of sight. Based on this principle, CMEs are 
imaged in white-light using coronagraphs that observe the Sun by creating an artificial eclipse, blocking the solar disk and imaging 
the solar corona. The first observations of CMEs using space coronagraphs were made by the seventh \textit{Orbiting Solar Observatory} (OSO-7, 1971) in early 
1970's \citep{Tou73}. It observed a total of 20 CMEs before it re-entered the Earth's atmosphere. Improved quality and longer observational periods were achieved by 
the \textit{Apollo Telescope Mount} (ATM) coronagraph 
on board \textit{Skylab} \citep{Mac74} which observed about 77 transients between May 1973 and February 1974, which were all recognized as CMEs.
The term coronal mass ejections first appeared in \citet{Gos76} based on {\it Skylab} observations.
The {\it Solwind} (1979) coronagraph on board the {\it Air Force satellite P78-1} \citep{Mic80} and the {\it Coronagraph/Polarimeter} (CP) on board the {\it Solar 
Maximum Mission} (SMM, 1980) satellite \citep{Mac80} continued observations into the 1980's. Among the most significant discoveries of these coronagraphs was the 
first Earth-directed CME (observed in November 1976) by Russ Howard and co-workers \citep{How82}. The {\it Solwind} and SMM coronagraphs detected over 2000 CMEs providing data for the first statistical 
analysis for studying the CME structure, mass, angular extension and location \citep{How85}.

By 1992, \citet{Kah92} demonstrated through a detailed review of CME and flare observations, metric radio bursts, IP shocks, magnetic fields and solar energetic 
particles and their geomagnetic effects that CMEs (not flares) were the major drivers of heliospheric and geomagnetic phenomena. Regardless, a large portion of the solar community was convinced that solar flares were the primary 
drivers of space weather. This came to be known as the ``Solar Flare Myth'' as described by \citet{Gos93}, who confirmed the source of IP shocks and storm to be CMEs and not solar 
flares. Towards the end of 1995, however, more clarity was achieved following the launch of the {\it Solar and Heliospheric Observatory} (SOHO)
\citep{Bru95} which provided more conclusive CME data using the {\it Large Angle and Spectrometric Coronagraph} (LASCO). LASCO consists of three coronagraphs that observe
the solar corona in white-light from 1.1 -- 30 \Rs. 
Data from LASCO was used to identify ``Halo CMEs'' for the first time and construct larger statistical CME databases ({\it e.g., http://lasco-www.nrl.navy.mil/cmelist.html} and 
{\it http://cdaw.gsfc.nasa.gov/CME\_list}). So far LASCO has detected $10^{4}$ CMEs \citep{Yas04}. The unparalleled quality and resolution of data from SOHO has made it a cornerstone 
for solar observations (Example - Figure \ref{lasco}).

\begin{figure}[h!]
\centering
\includegraphics[height=0.3\textheight,width=1.02\textwidth]{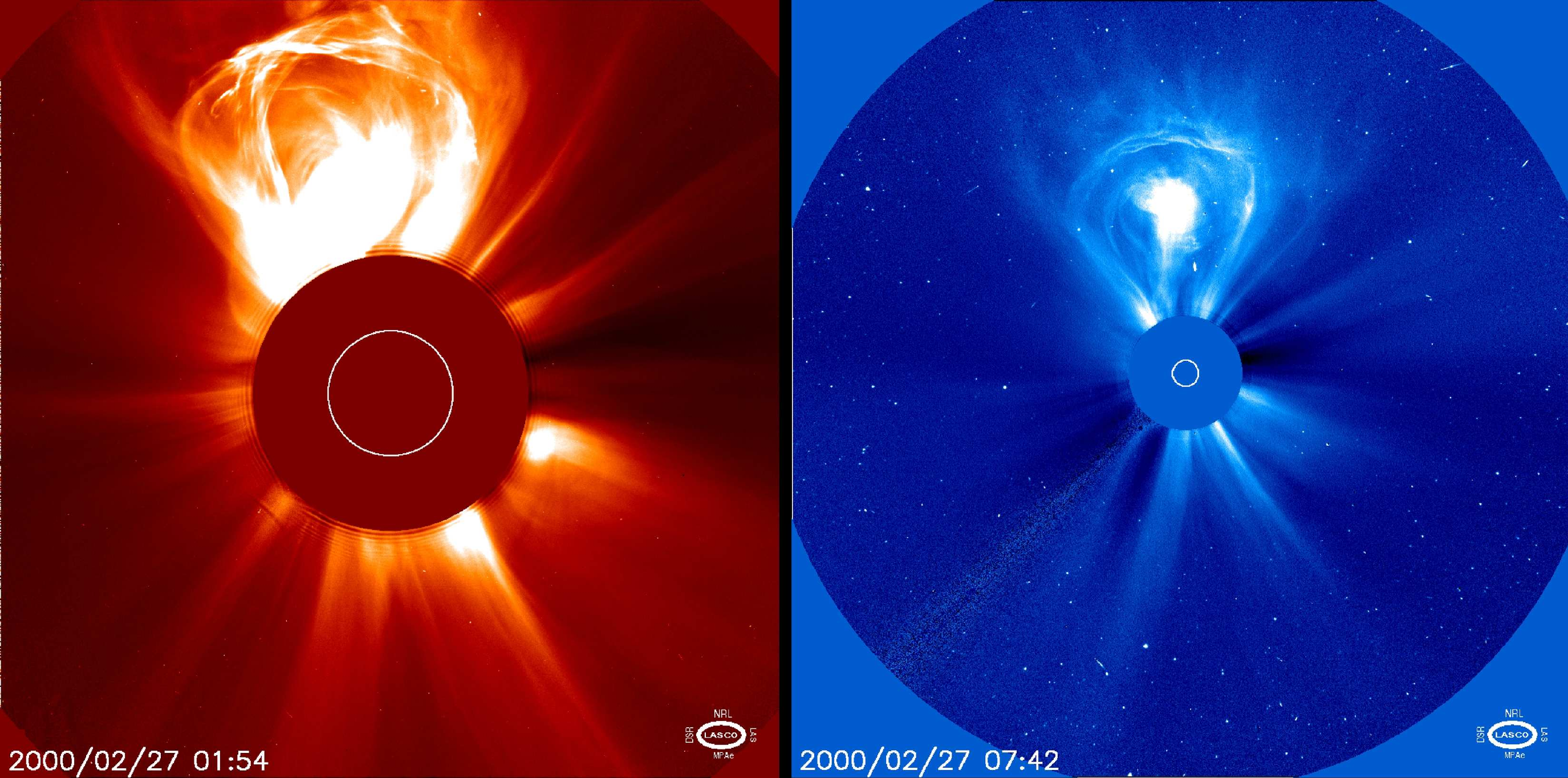}
\caption[SOHO LASCO observation of a CME on February 27, 2000]{A coronal mass ejection on February 27, 2000 taken by SOHO LASCO C2 (left) and C3 (right) coronagraphs. {\it Image credit: https://sohowww.nascom.nasa.gov/}}
\label{lasco}
 \end{figure}

In 2003, the {\it Solar Mass Ejection Imager} (SMEI), a heliospheric imager on board the {\it Coriolis} spacecraft was launched to observe the outer corona in white light
\citep{Eyl03}. SMEI observed about 400 transients during 8.5 years of its life \citep{Web06}. The first white light heliospheric imagers (HI) were launched onboard the
twin {\it Helios} (1974,1976) spacecraft \citep{Ric82,Jac85}. The zodiacal light polarimeters observed 0.3--1 AU with a limited field of view, providing only partial images of 
interplanetary counterparts of CMEs (called ICMEs) in white light. The launch of the twin {\it Solar Terrestrial Relations Observatory} (STEREO) spacecrafts in 2006 \citep{How08} 
heralded a new approach to solar observations. The {\it Sun Earth 
Connection Coronal and Heliospheric Investigation} (SECCHI) mission on board STEREO consists of coronagraphs (COR1, COR2) and heliospheric imagers (HI1, HI2).
The STEREO mission consists of two spacecrafts, STEREO Ahead (A) and STEREO Behind (B), one moving slightly faster than the other in the ecliptic 
plane in opposite directions. The coronagraphs and heliospheric imagers provide white light 
observations of CMEs covering a field of view from 1.4--318 \Rs. Figure \ref{stereoorbit} depicts the location of the STEREO A and B spacecrafts on December 31 in different years 
(2006. 2010, 2014 and 2017). Figure \ref{stereo} shows the COR2 A and B observations of a CME on September 28, 2012. 

\begin{figure}
\centering
  \begin{tabular}{cc}
       \includegraphics[width = 0.34\paperwidth,height=0.23\paperheight]{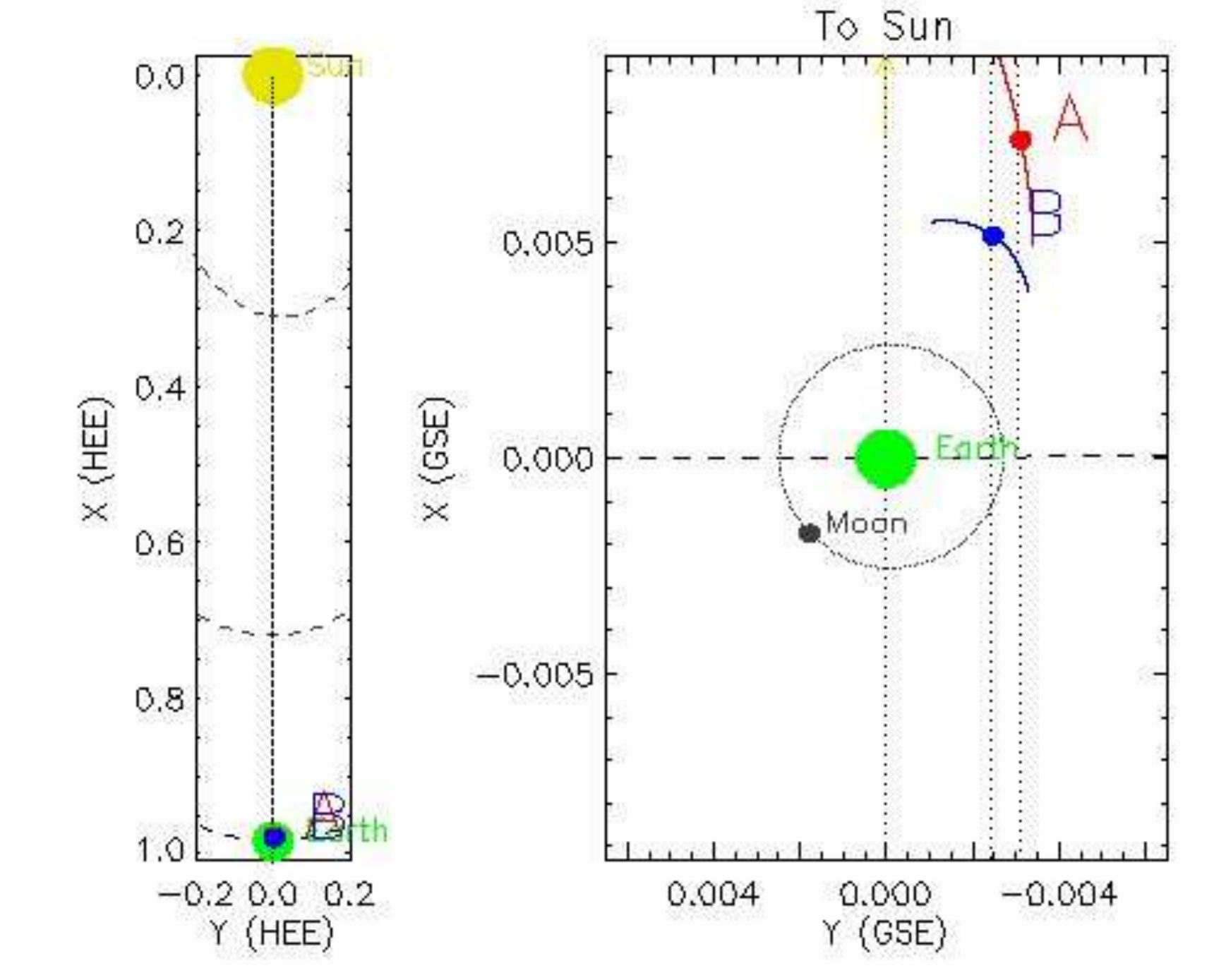} &
              \put(-148.8,195.9){{\rotatebox{0}{{\color{black}\fontsize{11}{11}\fontseries{n}\fontfamily{phv}\selectfont (a) 31 December 2006}}}}
       \includegraphics[width = 0.34\paperwidth,height=0.23\paperheight]{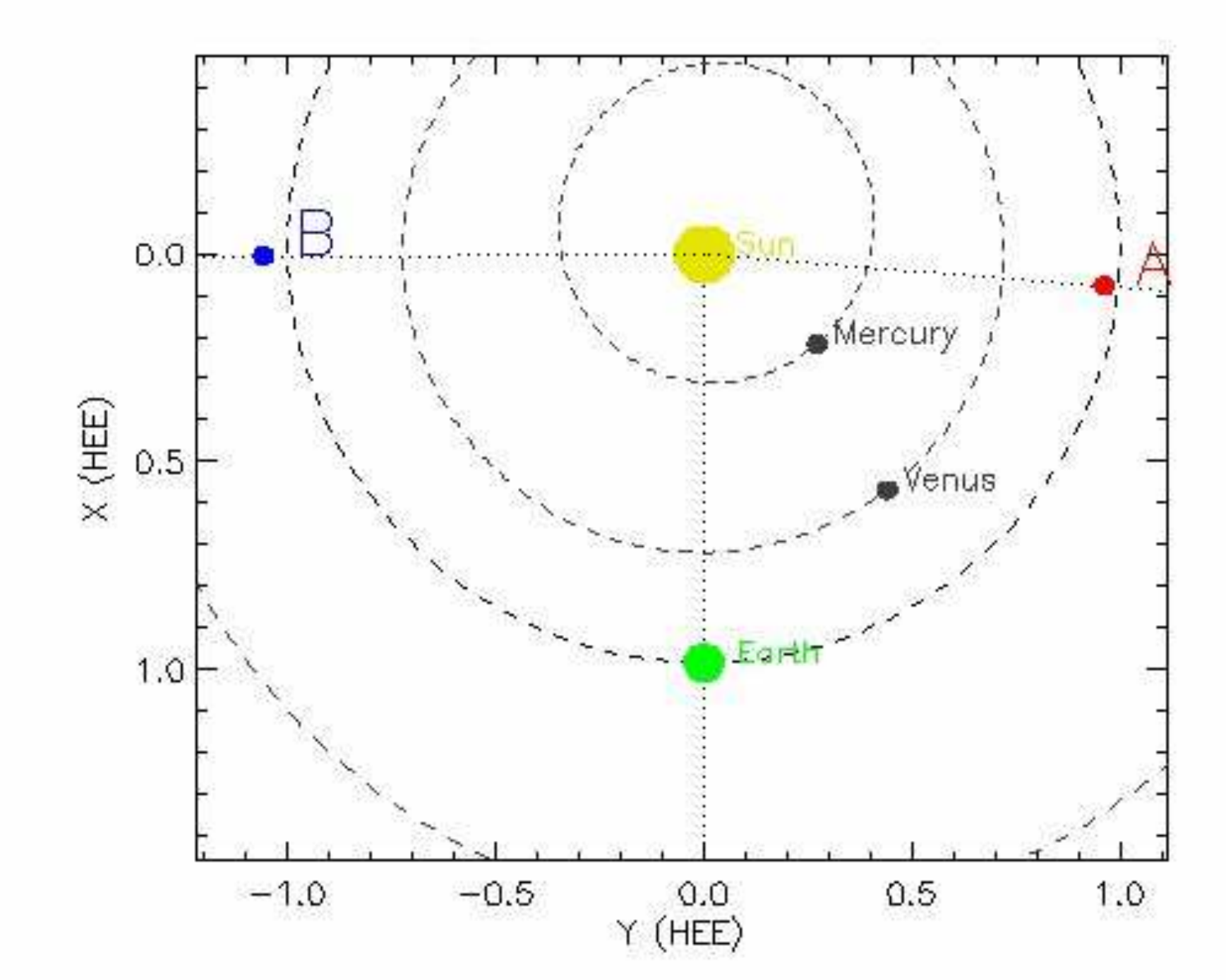}
         \put(-148.8,195.9){{\rotatebox{0}{{\color{black}\fontsize{11}{11}\fontseries{n}\fontfamily{phv}\selectfont (b) 31 December 2010}}}}
\end{tabular}

  \begin{tabular}{cc}

       \includegraphics[width = 0.34\paperwidth,height=0.23\paperheight]{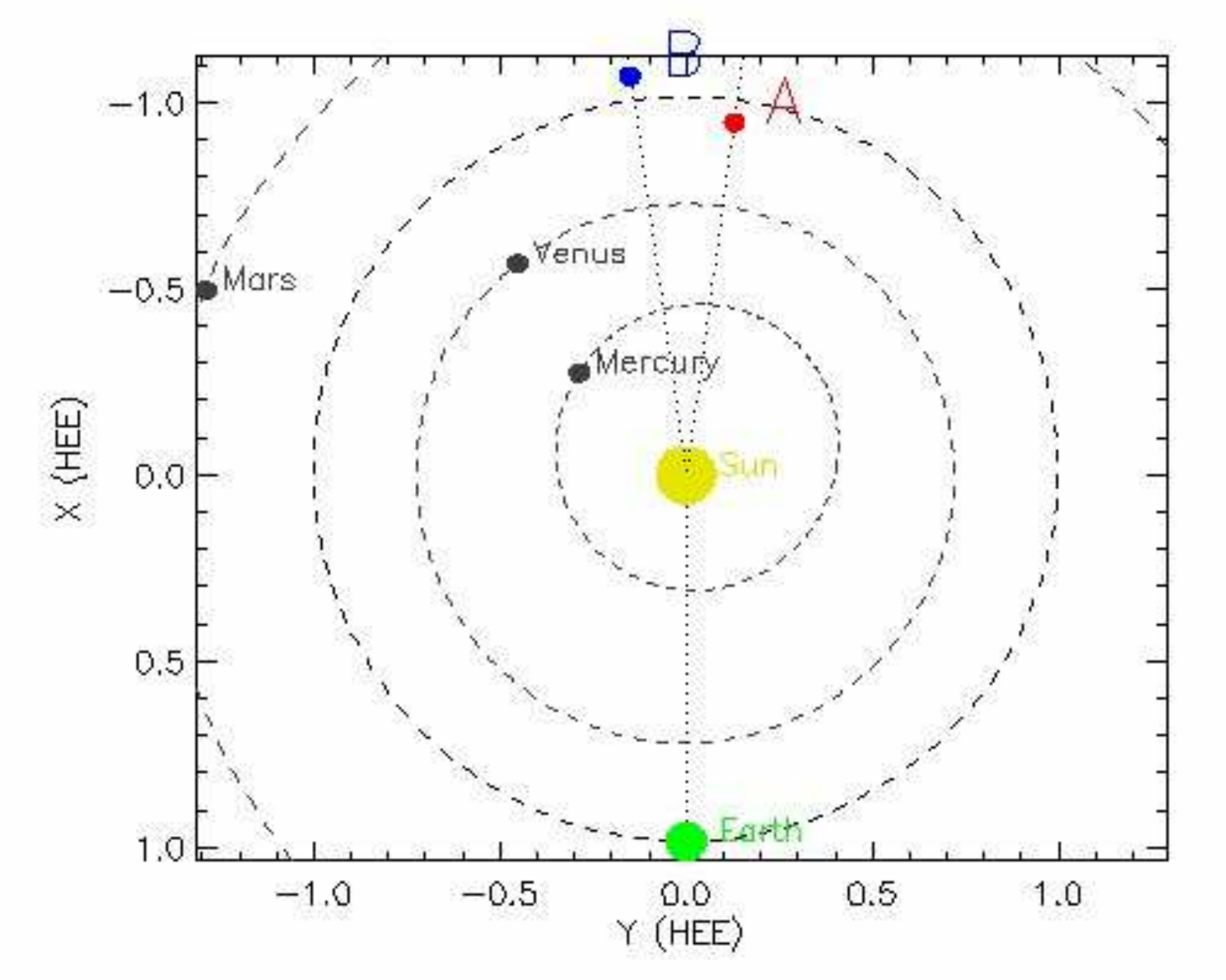} &
               \put(-148.8,195.9){{\rotatebox{0}{{\color{black}\fontsize{11}{11}\fontseries{n}\fontfamily{phv}\selectfont (c) 31 December 2014}}}}
       \includegraphics[width = 0.34\paperwidth,height=0.23\paperheight]{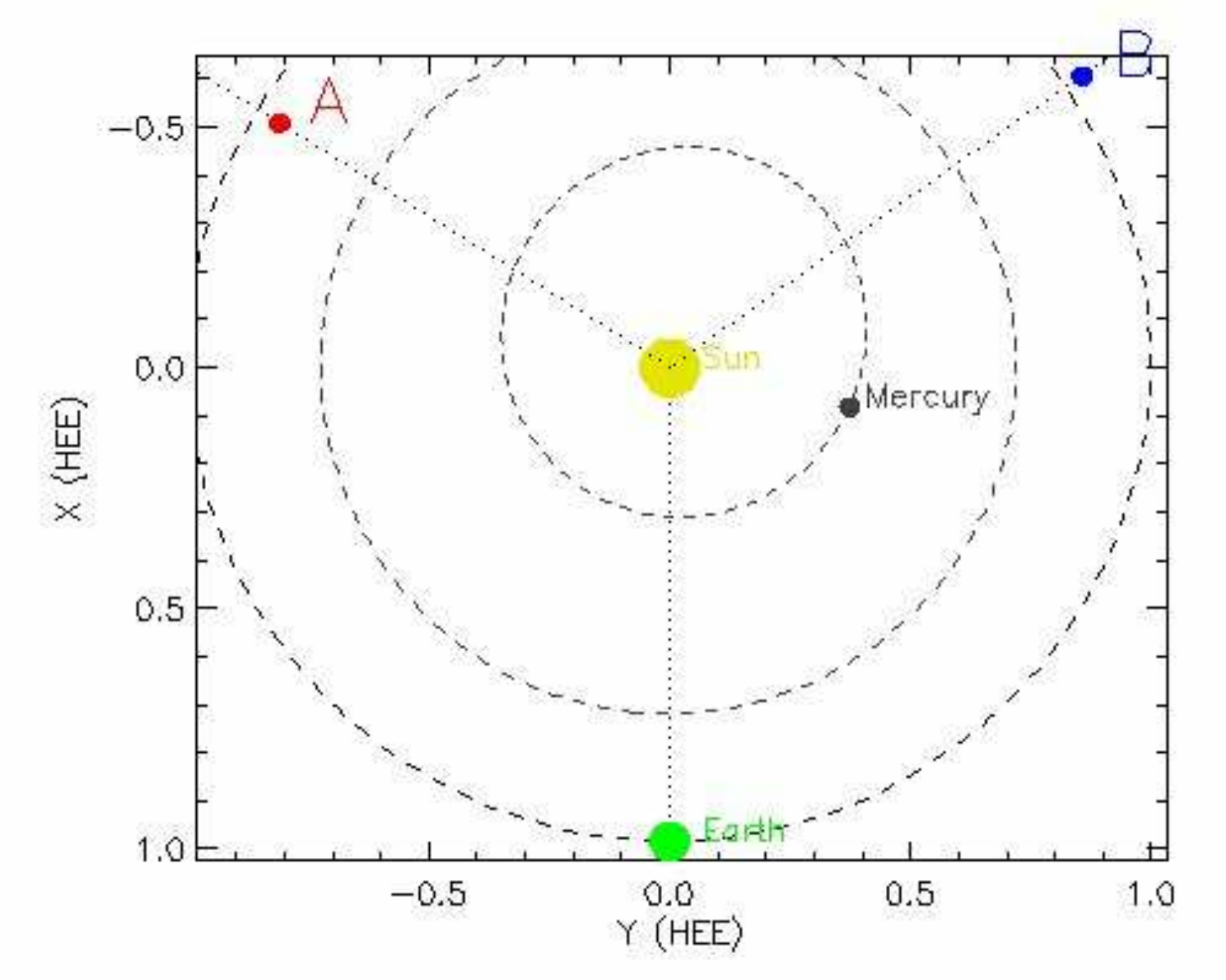}
        \put(-148.8,195.9){{\rotatebox{0}{{\color{black}\fontsize{11}{11}\fontseries{n}\fontfamily{phv}\selectfont (d) 31 December 2017}}}}
\end{tabular}
 \caption[Location of STEREO spacecraft on December 31, 2006, 2010, 2014 and 2017 at 00:00 UT]{Location of STEREO Spacecrafts A and B on December 31, (a) 2006, (b) 
 2010, (c) 2014 and (d) 2017 at 00:00 UT. 
 The color scheme is as follows: Red circle- STEREO-A , Blue circle- STEREO-B, Green circle- Earth, Yellow circle-Sun. The two STEREO spacecrafts separate by around 22.5$^\circ$ from 
 the Earth every year. \textit{Image credit: STEREO orbit tool https://stereo-ssc.nascom.nasa.gov/where.shtml}.}
\label{stereoorbit}
\end{figure}

      \begin{figure}[h!]    
      \centering{
   \centerline{\hspace*{0.001\textwidth}
               \includegraphics[height=0.57\textwidth,width=0.59\textwidth]{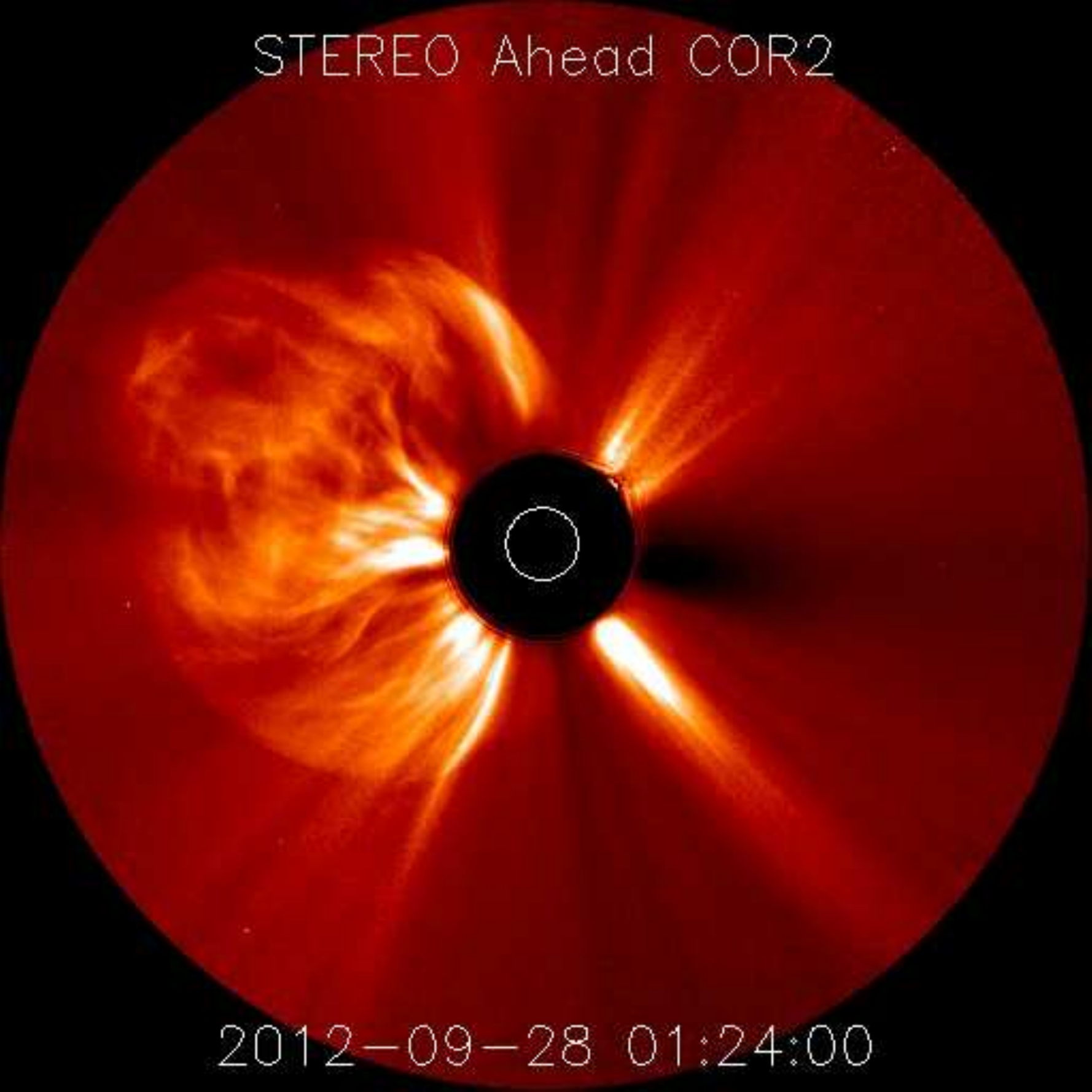}
               \includegraphics[height=0.57\textwidth,width=0.59\textwidth,clip=]{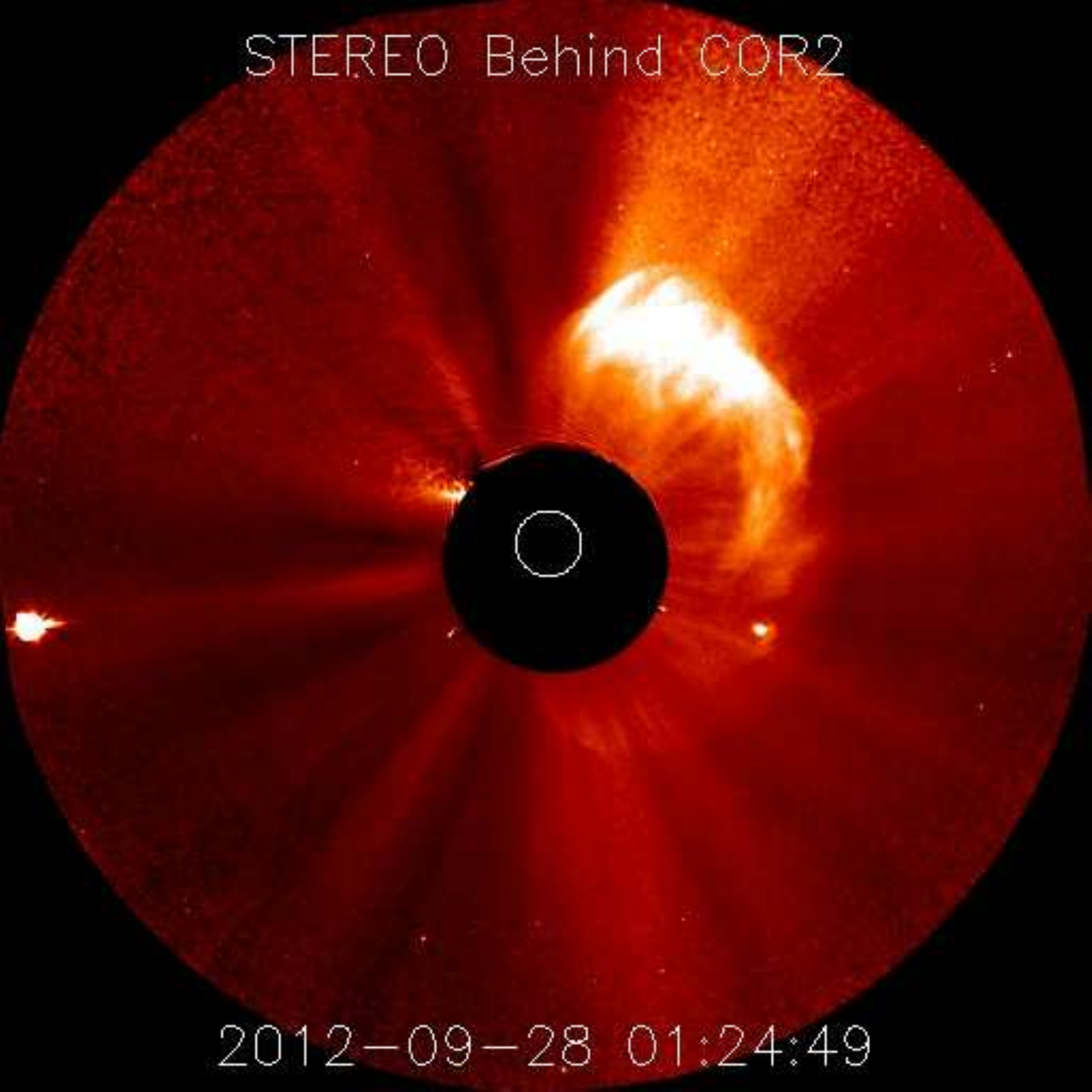}
 }}
\caption[STEREO images of a CME event on September 28, 2012]{Images of a Halo CME on September 28, 2012 from STEREO COR2 A (Ahead) and COR2 B (Behind) 
coronagraphs. {\it Image credit: https://stereo-ssc.nascom.nasa.gov/}}
    \label{stereo}
   \end{figure}

White light CME observations have been accompanied by observations of the solar disk at coronal wavelengths with the SOHO {\it Extreme Ultraviolet Imaging Telescope} 
(EIT), SOHO {\it Coronal Diagnostic Spectrometer} (CDS) imagers, STEREO {\it Extreme-Ultra Violet Imager} (EUVI) and instruments onboard {\it Yohkoh} (1991-2001) and {\it Transition Region And 
Coronal Explorer} (TRACE, 1998) spacecraft \citep{Zha01}. The {\it Reuven Ramaty High Energy Solar Spectroscopic Imager} (RHESSI, 2002) and {\it Hinode} spacecraft (2006) 
include hard and soft X-ray imagers to investigate solar responses to CME launches. In 2010, the {\it Atmospheric Imaging Assembly} (AIA) onboard the 
{\it Solar Dynamics Observatory} (SDO) was launched in a geosynchronous orbit which includes white light, ultra-violet and extreme ultra-violet imagers 
providing a wealth of solar observations over multiple wavelengths. The {\it Solar Orbiter} (2019) and {\it Parker Solar Probe} (2018) are future missions for heliospheric observations.

\item {\bf Interplanetary Scintillation}

Before coronagraphs observed the interplanetary counterparts of coronal mass ejections (ICMEs) at large distances 
from the Sun ($>50$ \Rs), other methods such as interplanetary scintillation (IPS) \citep{Hew64,Hou74,Man06,Man10} were used for this purpose. 
IPS is the variation in the radio signal from distant sources due to density perturbations in the interplanetary medium. These distortions in radio sources at meter
wavelengths are used to monitor the solar wind and density fluctuations in the medium to track ICMEs traveling between the Sun and the Earth. However, IPS 
observations need to be improved to efficiently differentiate between density perturbations due to ICMEs and corotating interaction regions (CIRs). CIRs are mergers 
of fast and slow streams causing density enhancements.

\begin{figure}[h]
\centering
\includegraphics[height=0.32\paperheight,width=0.8\paperwidth]{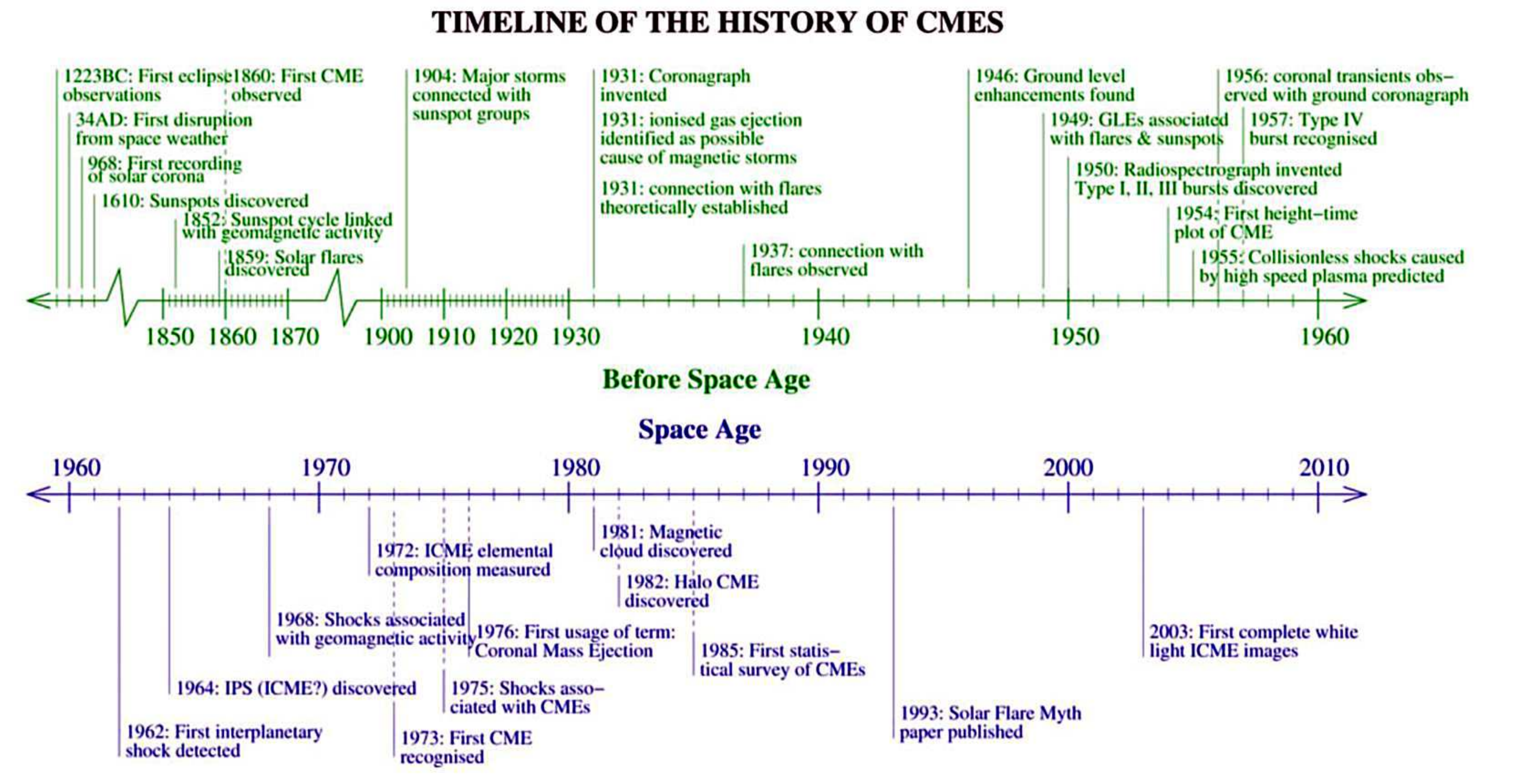}
\caption[Timeline of the history of CMEs]{Timeline of the history of significant events related to CMEs before (green) and during (blue) the space age.
{\it Image credit: \citet{How11}}}
\label{spacecraft}
 \end{figure}
\item {\bf In-situ observations}

Apart from remote sensing observations, \textit{in-situ} measurements of plasma parameters from spacecrafts at 1 AU and beyond 
provide supporting observations. Examples include {\it Ulysses}, launched in 1990, {\it WIND} (1994) and the {\it Advanced Composition Explorer} (ACE, 1997). The {\it WIND} and 
{\it ACE} spacecrafts have sophisticated instrumentation 
for continuous monitoring of solar, interplanetary and magnetospheric activity. 

\end{enumerate}

\subsection{Properties of CMEs}
\begin{enumerate}
 \item {\bf Morphology}
 
 Images from coronagraph observations have shown that CMEs come in a variety of shapes and sizes. However, the classical morphology is the 
 ``three-part'' CME structure - a bright central core, followed  by a dark cavity enveloped by a frontal loop \citep[\textit{e.g.}][]{Low96,Hun99,Cre04} (Figure \ref{fig1} a). 
This is usually interpreted as compressed plasma ahead of a flux rope followed by a cavity surrounded by a bright filament/prominence. It needs to be pointed out that not all
CMEs have the classic three-part structure; many CMEs can have complex or distorted geometries. The line of sight along which CMEs are viewed plays a major role 
in determining the observed CME geometry. 
 \begin{figure}[h!]    
      \centering{
   \centerline{\hspace*{0.001\textwidth}
               \includegraphics[height=0.55\textwidth,width=0.55\textwidth,clip=]{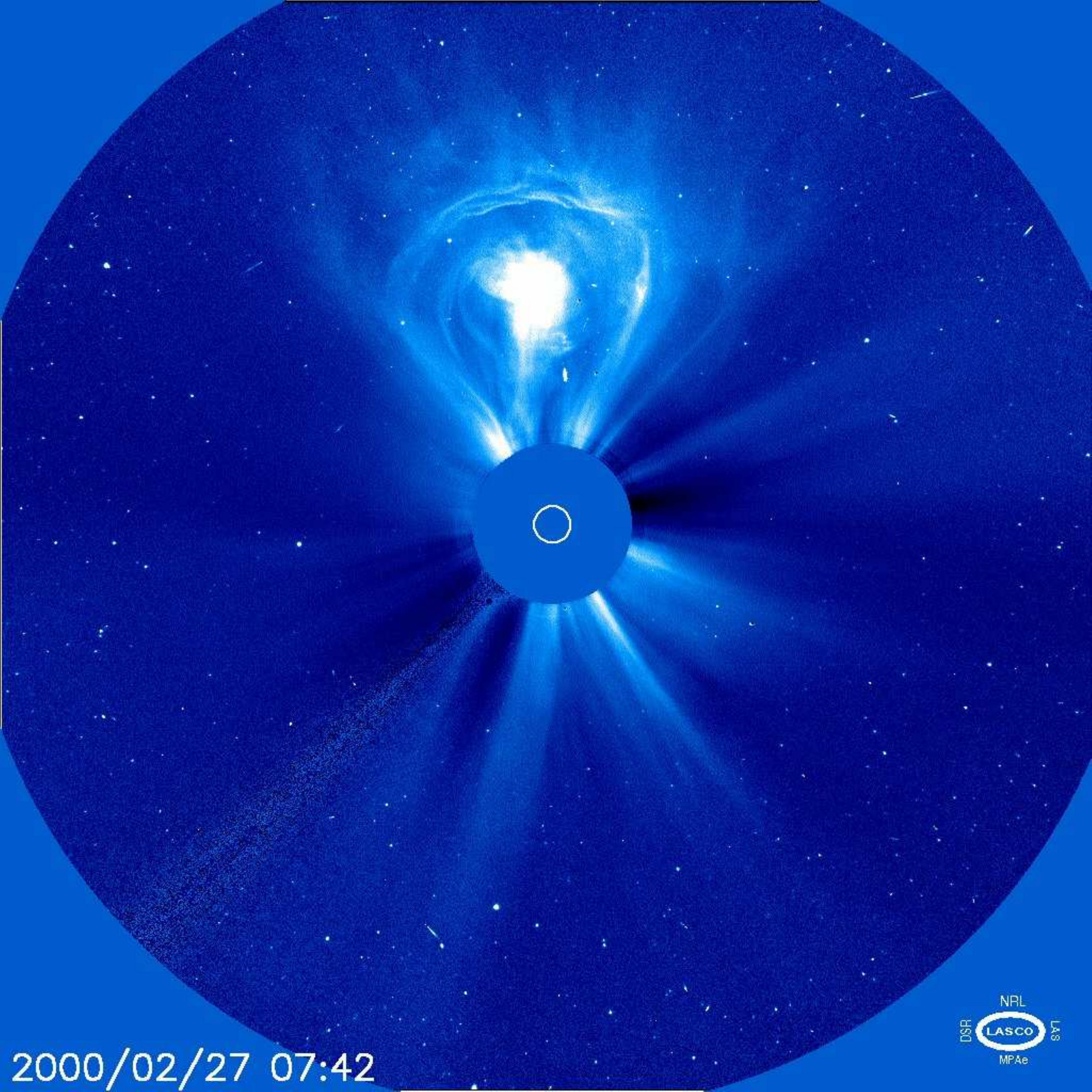}
                \put(-148.8,-15.9){{\rotatebox{0}{{\color{black}\fontsize{9}{9}\fontseries{n}\fontfamily{phv}\selectfont (a) CME morphology}}}}
               \includegraphics[height=0.55\textwidth,width=0.55\textwidth,clip=]{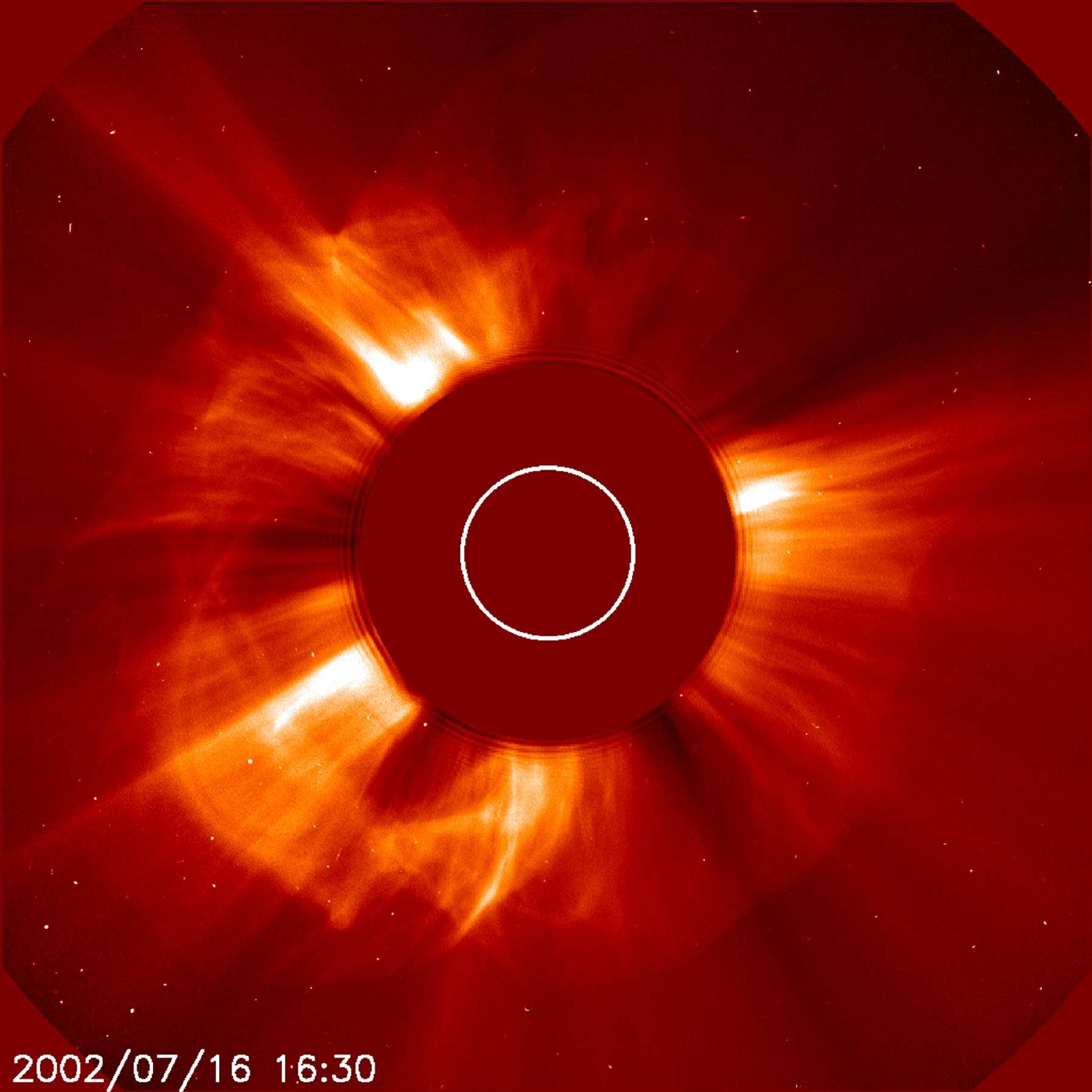}
                               \put(-148.8,-15.9){{\rotatebox{0}{{\color{black}\fontsize{9}{9}\fontseries{n}\fontfamily{phv}\selectfont (b) Halo CME}}}}
 }}
\caption[Classic three-part CME structure and Halo CME.]{Panel (a) shows a CME on February 27, 2012 imaged by LASCO C2 coronagraph showing the three-part morphology-bright frontal loop, 
dark cavity and a central core. Panel (b) shows a Halo CME on July 16, 2002 observed by the SOHO/LASCO C2 coronagraph. {\it Image credit: https://stereo-ssc.nascom.nasa.gov/}.}
    \label{fig1}
   \end{figure}

   \item {\bf Size and Location}
   
CMEs can appear as narrow jets, as well as wide eruptions. Owing to the viewing perspective in the plane of the sky, SOHO LASCO categorizes 
CMEs with angular widths $\approx$360$^\circ$ as `` Halo '' CMEs  and widths $>120^\circ$ as ``Partial Halos'' \citep{Yas04}. Halo CMEs can in fact also be CMEs with only tens of degrees of angular 
width propagating along the Sun-Earth line either towards or away from the Earth. These CMEs usually span the entire occulting disks as they expand 
\citep{Che11} (Figure \ref{fig1} b). During the solar minimum, CMEs typically erupt from regions near the solar equator while they occur over 
a wider range of latitudes near the solar maximum \citep{StC00}.

\item {\bf Occurrence rate}

The number of CMEs erupting has a direct correlation with solar activity. There may be about 4-5 CME events per day during the solar maximum 
and $\approx$ 1 event in a day near the solar minimum \citep{Yas04}. Figure \ref{cmessn} shows the variation in the daily occurrences of CMEs with the sunspot number from 1997 to 2017.

\begin{figure}[h!]
\centering
\hfill \includegraphics[height=0.23\paperheight,width=0.8\paperwidth]{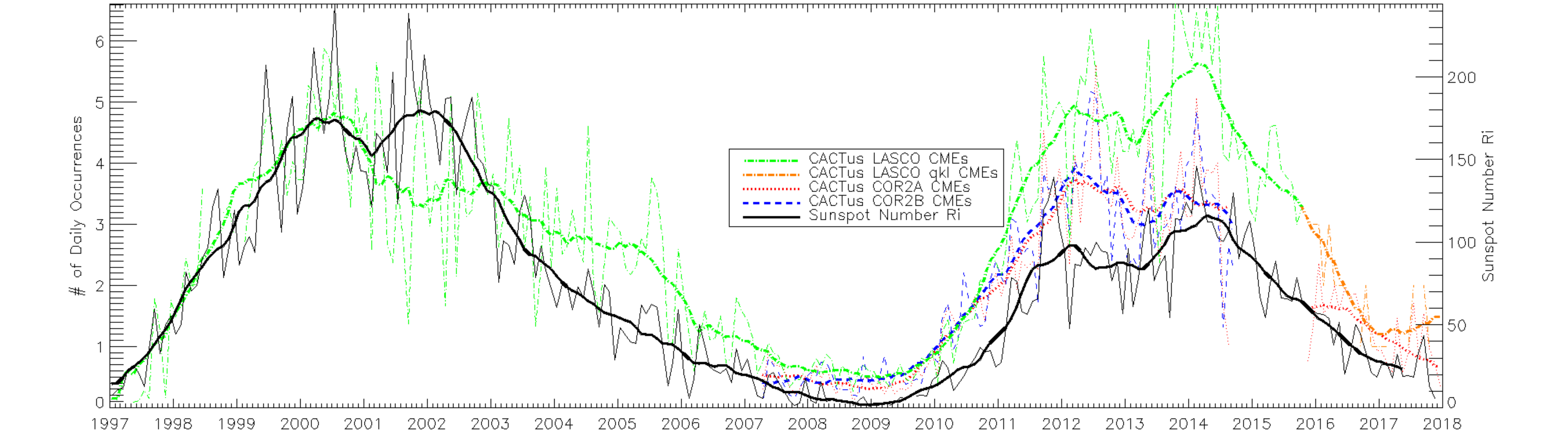}
\caption[CME and Sunspot number overlap]{Overlap of number of daily occurrences of CMEs with the Sunspot number using the CACTUS catalog. 
{\it Image Courtesy: http://sidc.oma.be/cactus/}.}
\label{cmessn}
 \end{figure}

\item {\bf Mass and Energy}

LASCO observations have shown that the average mass and energy of CMEs is of the order of $10^{15}$ gms and $10^{30}$ ergs respectively. \citet{Vou10} suggest that CME 
masses may be underestimated by a factor of two and CME kinetic energies by a factor of 8 due to effects of projection. They note that the CME mass 
increases as a function of height and saturates beyond 10 \Rs.

\item {\bf CME velocity} 

The radial propagation velocity of a CME is the speed of the frontal loop projected in the plane of the sky. It can vary from $\sim$ 20 \kms to $>$ 2500 \kms.
However, efforts are made to correct for projection effects, so as to determine the actual CME speed.

\end{enumerate}


\subsection{CME onset, initiation and propagation}

\begin{enumerate}
 \item {\bf CME onset}
 
It is generally accepted that CMEs are initiated in the corona \textit{i.e.} they are a coronal phenomenon and the energy required to launch these massive structures 
with speeds of hundreds of kilometers per second comes from this region. Since the corona has low plasma-$\beta$, gas pressure alone is not enough to drive these 
eruptions; CMEs are predominantly magnetically dominated and the energy required to accelerate them is provided by the coronal magnetic fields. 

For a CME to erupt from its state of equilibrium in the low corona, the onset mechanisms must include some instability which disturbs this equilibrium, leading to eruption 
of the magnetic structure. Due to lack of observational evidence for such mechanisms of CME onset and acceleration in the low corona, the solar physics community 
relies heavily on physics-based theoretical models. Most of these initiation models assume that the currents that build up in the solar corona continuously evolve until 
they can no longer be stable and lead to an eruption. 
The stored magnetic free energy that drives CMEs can manifest in different forms, like an expanding massive CME (in most cases) as well as erupting filaments. The 
release of energy can also accelerate energetic particles and cause electromagnetic radiation emission in the form of a flare. Multi-wavelength observations of these phenomenon 
associated with CMEs in the early stages of their eruption provide clues for 
understanding the eruption processes \citep[\textit{e.g.}][]{Che11}. These include, flares, prominences, coronal dimming, coronal and shock waves \citep{Web12}.

\item {\bf Initiation models}

Mechanisms that facilitate CME initiation may be broadly divided into two classes: those that rely on magnetic reconnection and those that do not require reconnection.
Magnetic reconnection can be broadly described as the restructuring of magnetic field lines to release the magnetic stresses that build up due to emerging flux and 
differential rotation \citep{Asc05}.

To account for the amount of energy required over a short period of time, models like the tether-cutting (Figure \ref{fig2}) or flux cancellation 
mechanism \citep{Stu89,Moo01} and magnetic breakout model \citep{Ant99,Lyn08} (Figure \ref{fig3}) have been invoked. In the tether-cutting model, 
the overlying field does not reconnect; it only expands and reconnection occurs beneath the overlying field. In the breakout model, the flux or field lines 
that reconnect are a part of the central flux and the overlying magnetic field.

\begin{figure}[h!]    
      \centering
               \includegraphics[width=0.5\textwidth,clip=]{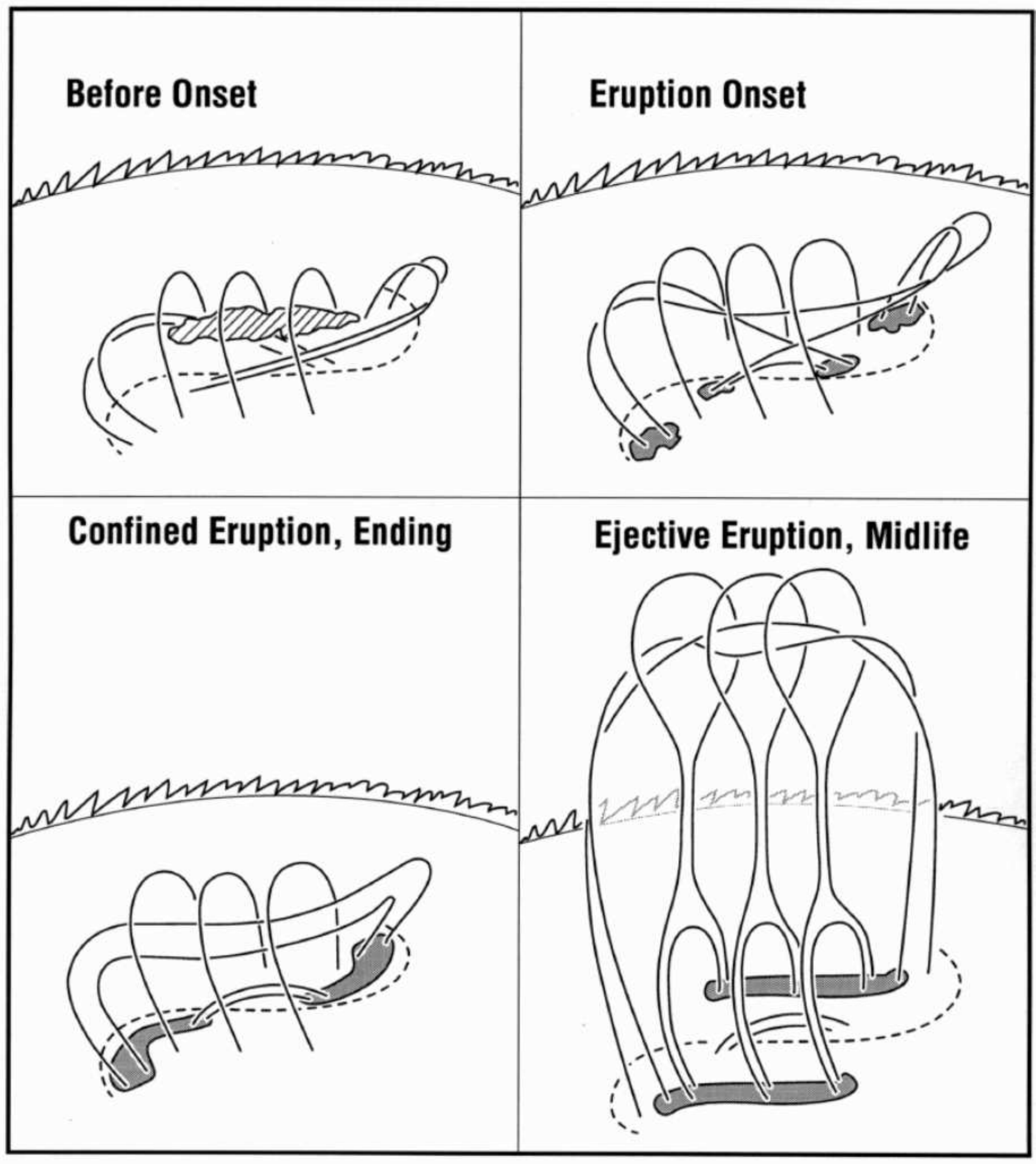} 
 \caption[Tether-cutting mechanism]{Tether-cutting mechanism adapted from \citet{Moo01}. The overlying field restrains the sheared core field followed by reconnection which triggers the 
 rise of the core stretching the overlying field.}
    \label{fig2}
   \end{figure}

   \begin{figure}[h!]    
      \centering
      \includegraphics[width=0.7\textwidth,height=0.23\textheight,clip=]{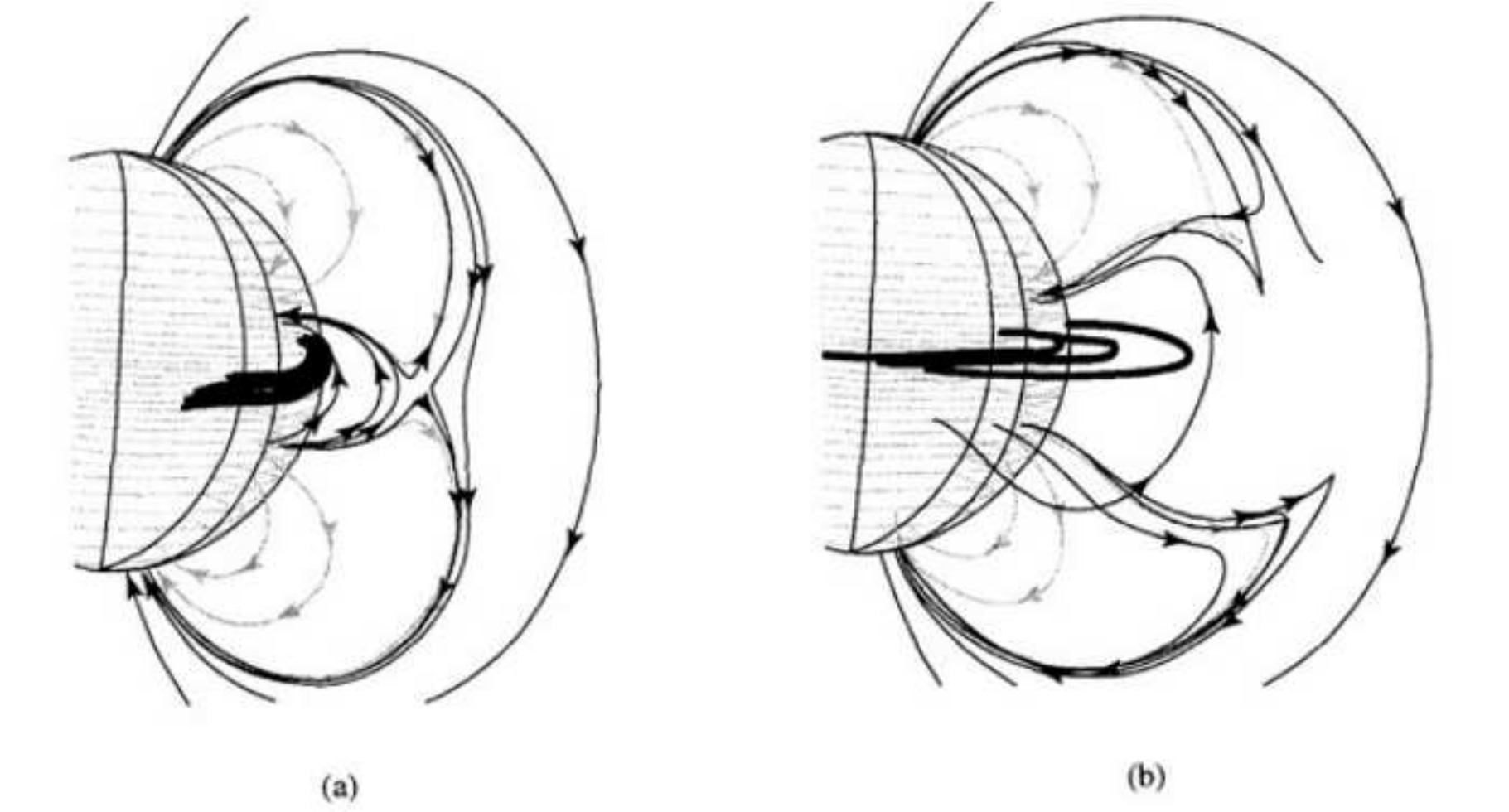}
 \caption[Breakout model]{Breakout model with reconnection above the central flux system. Thick lines indicate the core field that evolve resulting in final 
 eruption. Adapted from \citet{Ant99}.}
    \label{fig3}
   \end{figure}
Models that do not appeal to magnetic reconnection include flux injection \citep{Che96}, kink instability \citep{Tor03} and torus 
instability \citep{Kli06}. Based on the toroidal forces experienced by a curved current-carrying 
loop, \citet{Che89} derived the $J \times B$ Lorentz force acting on a section of a torus. They assume that a flux-rope CME initially at equilibrium, erupts 
as a result of poloidal flux being injected into it. Figure \ref{chen} shows the conditions for the flux-injection model where the \textquoteleft p\textquoteright \,and 
\textquoteleft t\textquoteright\, indicate the poloidal and toroidal components respectively.
These $J\times B$ forces drive the CMEs in the major radial direction \citep{Che89,Che96}. 
Forces acting on a CME due to the toroidal field and the average thermal pressure increase the CME radius, while the forces due to 
the ambient pressure and the poloidal field act in the opposing direction. Action of these competing forces causes the CMEs to radially expand as well. 
The pressure balance between the CME and the surrounding medium causes them CMEs to expand at a rate inversely proportional to the speed of the 
solar wind \citep{Wan90}. This is termed super-radial expansion.

The kink instability model considers continuous shearing of the foot-points of the magnetic flux ropes creating twists, which creates an instability leading to an 
eruption that releases the tension \citep{Hoo81}. \citet{Kli06} describe the torus instability model for CME eruption 
by considering the expansion of a current carrying ring held down by an external overlying field. The ring is unstable to expansion if the external 
field decays sufficiently fast.
\begin{figure}[h!]    
      \centering{
               \includegraphics[height=0.5\textwidth,width=0.55\textwidth,clip=]{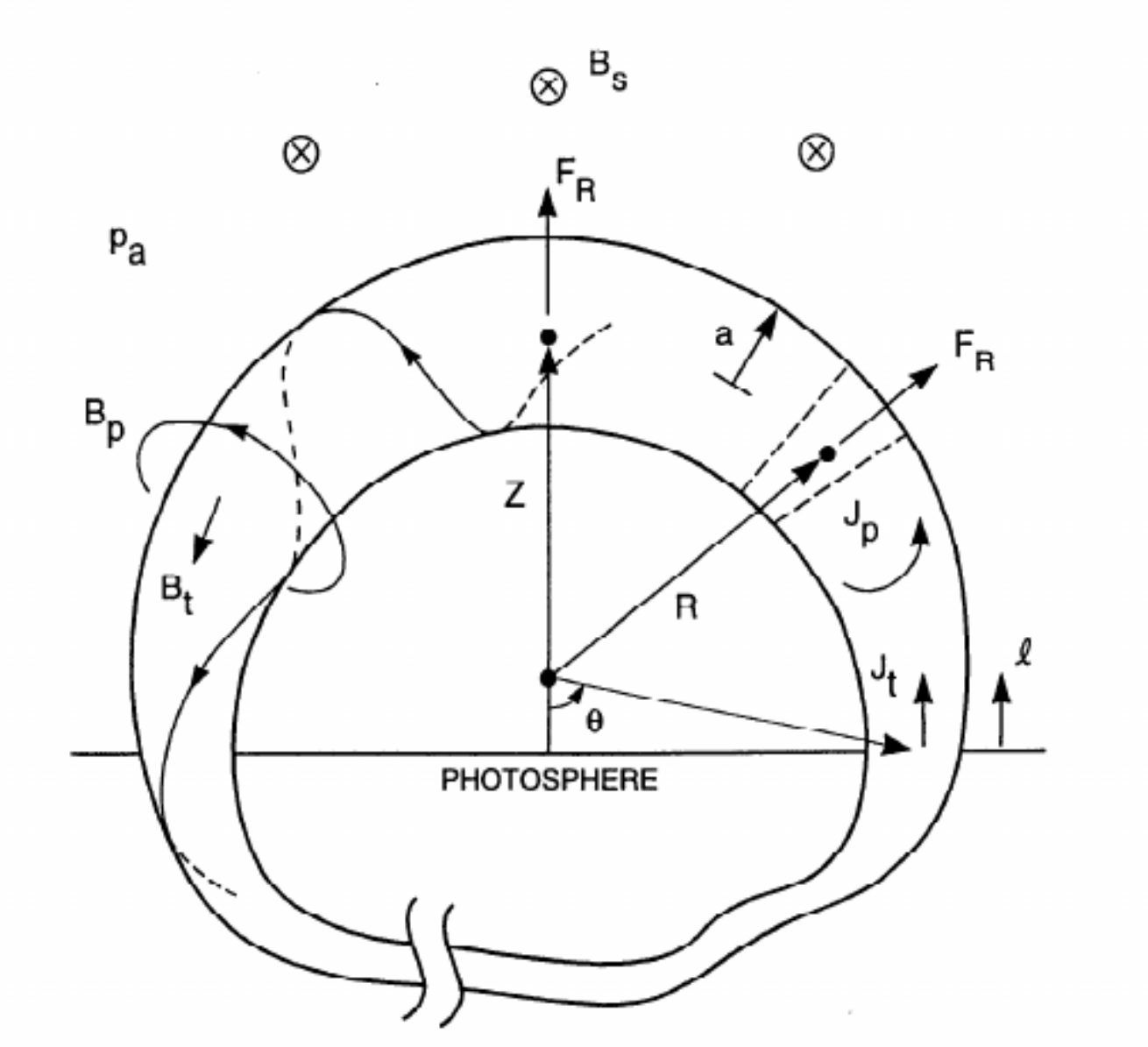} 
}
 \caption[Model current loop depicting toroidal forces]{Conditions for the flux injection model \citet{Che89,Che96}. $F_{R}$ is the Lorentz force in the radial direction, J and B indicate 
 current density and magnetic field respectively. Subscript p corresponds to poloidal and t corresponds to toroidal components.}
    \label{chen}
   \end{figure}   
\item {\bf Propagation models}

CMEs are generally thought to accelerate rapidly in the low corona following which they propagate into the interplanetary medium. 
Figure \ref{CMEacc} shows the speed-time profile of an event on June 11, 1998 along with the soft X-ray flux profile of an associated flare. Three phases of CME acceleration and 
flare intensity evolution are demarcated in the Figure \citep{Zha01}. The evolution of CMEs can be attributed to aerodynamic drag and shocks. 
Interaction of CMEs with the ambient solar wind results in an aerodynamic drag force which can be either accelerating or decelerating. The solar wind is a continuous flow of streaming 
particles from the Sun into the heliosphere. As the solar wind expands through the interplanetary medium, it drags the solar magnetic field with it which 
corotates with the Sun \citep{Par65}. This ambient medium formed by the solar wind interacts via momentum coupling with the propagating CMEs, causing fast CMEs to slow down and slower ones to speed up. 
This represents an 
attempt to equilibrate the CME speeds with the speed of the solar wind. The solar wind drag-force model based on this principle has had considerable success in describing the 
observed propagation of CMEs \citep[\textit{e.g.}][]{Car04,Vrs10,Sub12,Sac15}. 

\begin{figure}[h!]
\centering
\includegraphics[height=0.32\paperheight,width=0.55\paperwidth]{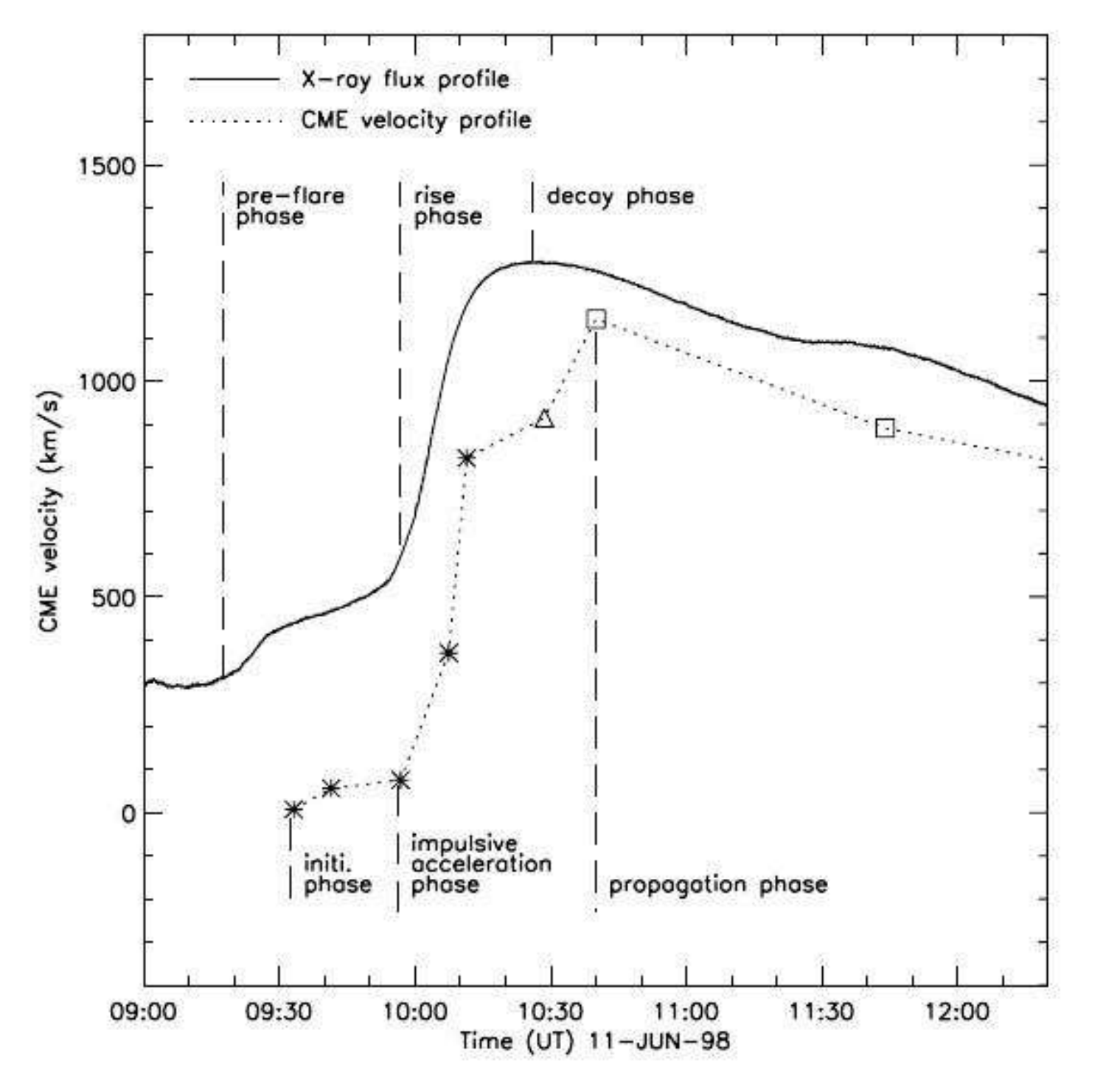}
\caption[Phases of CME acceleration]{Speed-time profile of a CME on June 11, 1998 along with the soft-X ray flux profile of an associated flare. 
Three phases of CME acceleration and flare intensity evolution are demarcated in the figure. {\it Image adapted from \citet{Zha01}}.}
\label{CMEacc}
 \end{figure}

The shock-based models regard ICMEs as a shock wave moving through the ambient solar wind \textit{i.e.} the ICME is considered as a perturbation in the surrounding medium. Some of the 
models describing this approach are outlined in \citet{Dry84} and \citet{Smi90}. Some models like ENLIL \citep{Ods99} treat the CME as a separate ejecta that propagates in 
the background solar wind. CMEs are considered to be dense structures with no intrinsic magnetic field in this model.
These propagation models are used to effectively and accurately predict the observed CME dynamics and their arrival time and speed at the Earth.

The following section describes the cause-effect relationship in terms of how CMEs and other solar phenomena affect space weather.
\end{enumerate} 

\section{Sun-Earth connection - Space Weather}
\begin{figure}[h!]
\centering
\includegraphics[height=0.28\paperheight,width=0.55\paperwidth]{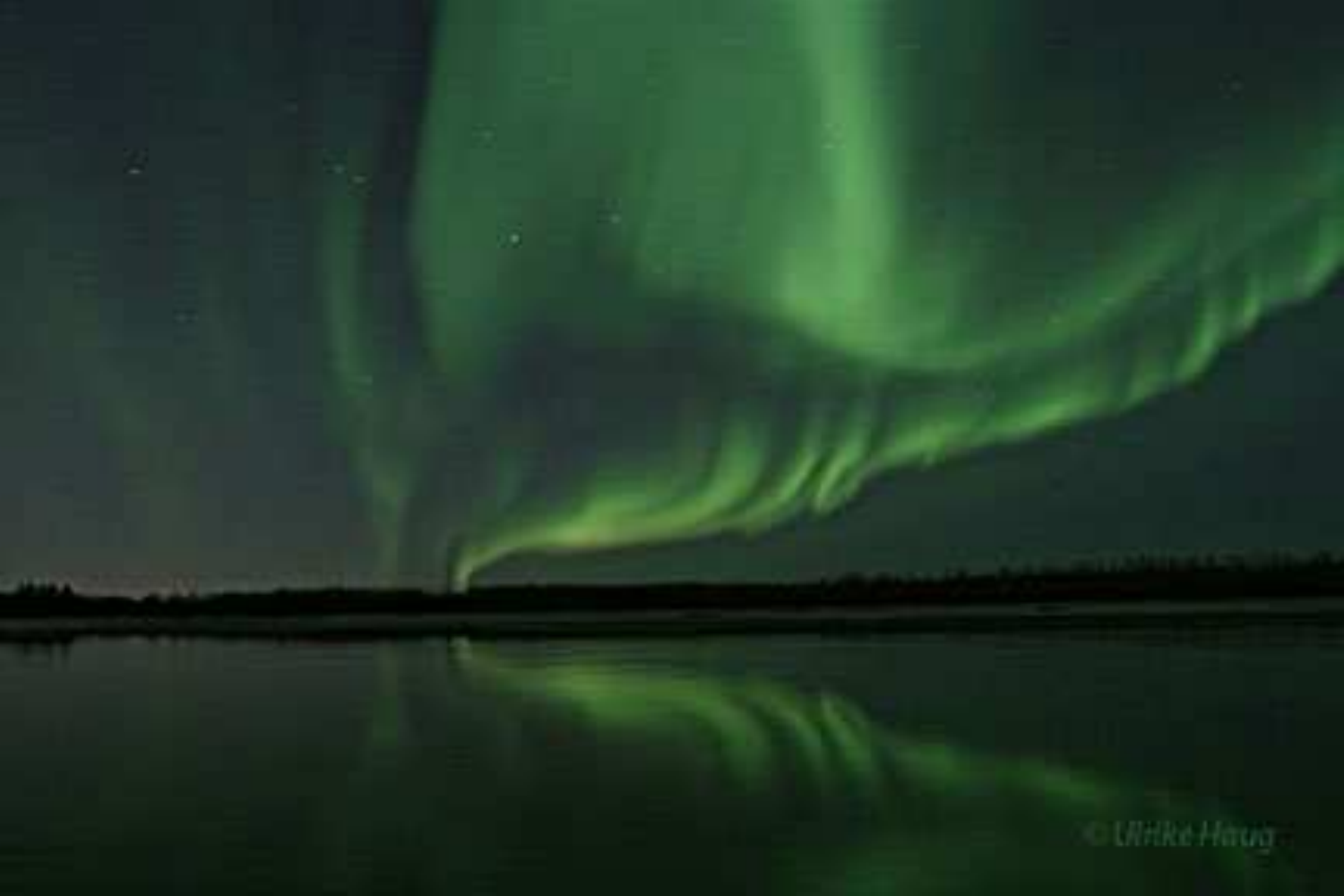}
\caption[Aurora visible in Alaska, USA on October 16-17, 2003.]{Aurora visible in Alaska, USA during October 16-17, 2003 as a result of the famous Halloween solar 
storm. {\it Image courtesy-Ulrike Haug}.}
\label{aurora}
 \end{figure}

Many phenomena related to the Sun can directly and/or indirectly influence space-borne technology and the near-Earth space environment. The first observations of Sun-Earth connection 
were made by Edmund Halley in 1716. He suggested that particles moving along the Earth's magnetic field lines caused auroras (Figure \ref{aurora}). By the mid-nineteenth century, 
geomagnetic disturbances were connected to solar processes and long term observations showed that their occurrences were correlated with the 11-year variability of the solar cycle.

\begin{figure}[h!]
\centering
\includegraphics[height=0.3\paperheight,width=0.58\paperwidth]{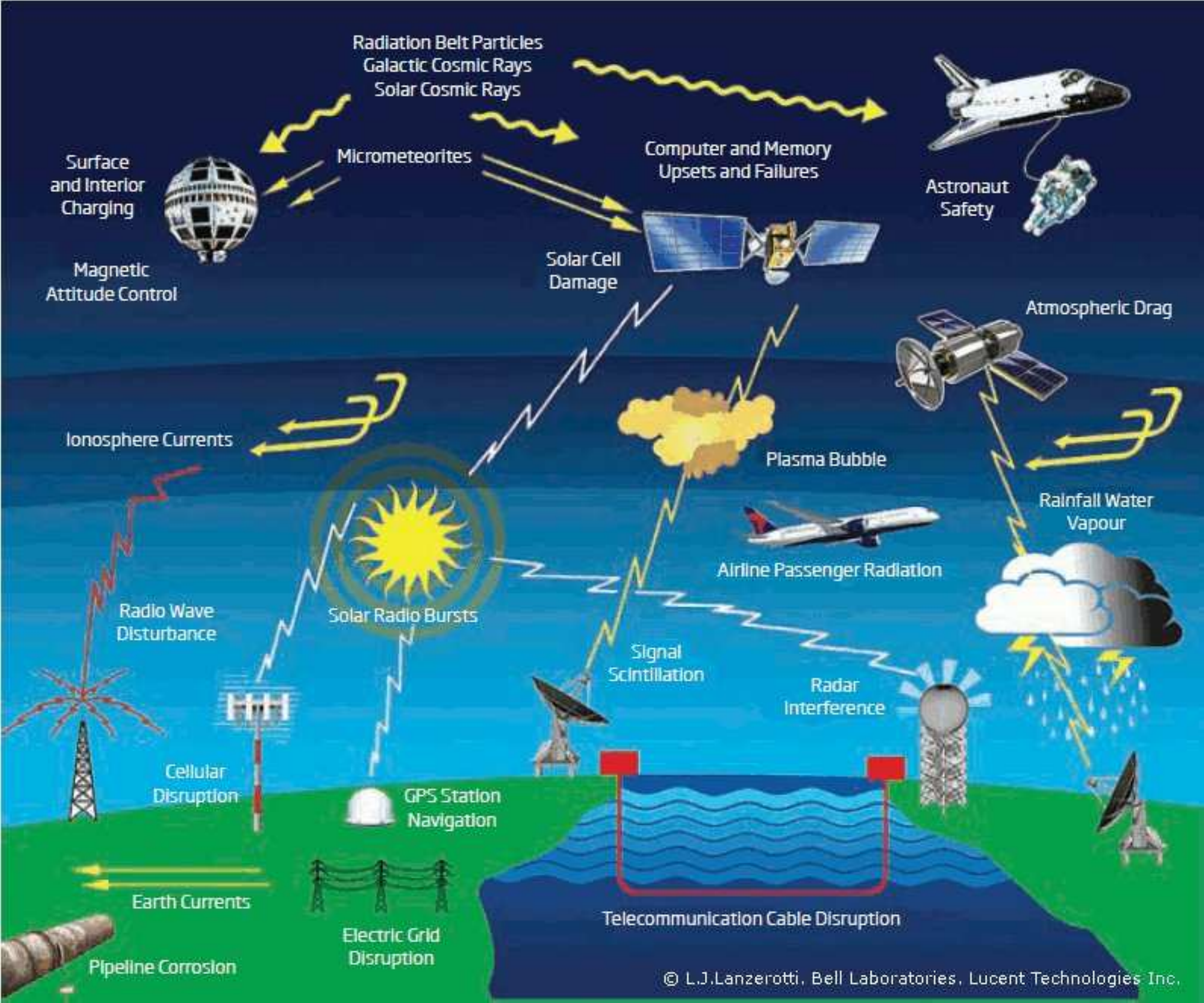}
\caption[Space weather effects]{This image shows the various ways in which solar disturbances and associated phenomena affect the technology and life on Earth. {\it Image courtesy-Bell Laboratories}.}
\label{spaceweather}
 \end{figure}

Space weather refers to conditions on the Sun and the solar wind, magnetosphere, ionosphere and thermosphere that can influence the performance and reliability of space-borne
and ground-based technological systems and can affect human life and health (definition by U.S. National Space Weather Plan). The affects of solar activity include (but are not limited to)
disruption in satellite operations, communications, navigation, radiation hazards to astronauts and airline passengers, failure of power supply grids, leading to societal as well as economic losses. 
Figure \ref{spaceweather} shows the various ways in which solar activity affects the Earth.

The main forms of solar energy output that determine the conditions for space weather include - fast and slow solar wind streams, co-rotating interaction regions (CIRs), flares, 
coronal mass ejections and their interplanetary counterparts and solar energetic particles. Solar phenomena leading to large perturbations in the coupled magnetosphere-ionosphere system of the Earth 
are called geoeffective \citep{Pul07}. Flares release flashes of photons that can heat up the terrestrial atmosphere causing satellites to drop to lower orbits. During major solar storms, particles are accelerated 
to near-relativistic energies, endangering astronauts traveling through the interplanetary space. Coronal mass ejections drive shocks and solar energetic particles causing geomagnetic 
storms. 

Owing to their tremendous impacts, forecasting space weather effects is a major challenge. The accuracy of prediction of CME arrival and its impact on the Earth's magnetosphere is still 
quite poor, although much effort has been made to understand the underlying physics in order to develop better warning tools. 
Space weather storms can be divided into three classes based on their 
size \citep{Bot07}:\\1) M-region storms- comprising of fast solar wind stream and CIRs,\\ 2) CMEs - smaller in size but greater in intensity, and\\ 3) Auroral electrojets - 
smallest in size, greatest in intensity, creating rapid fluctuations of magnetic fields at the ground level.

\subsection{Solar wind and space weather}
\begin{figure}[h!]
\centering
\includegraphics[height=0.3\paperheight,width=0.6\paperwidth]{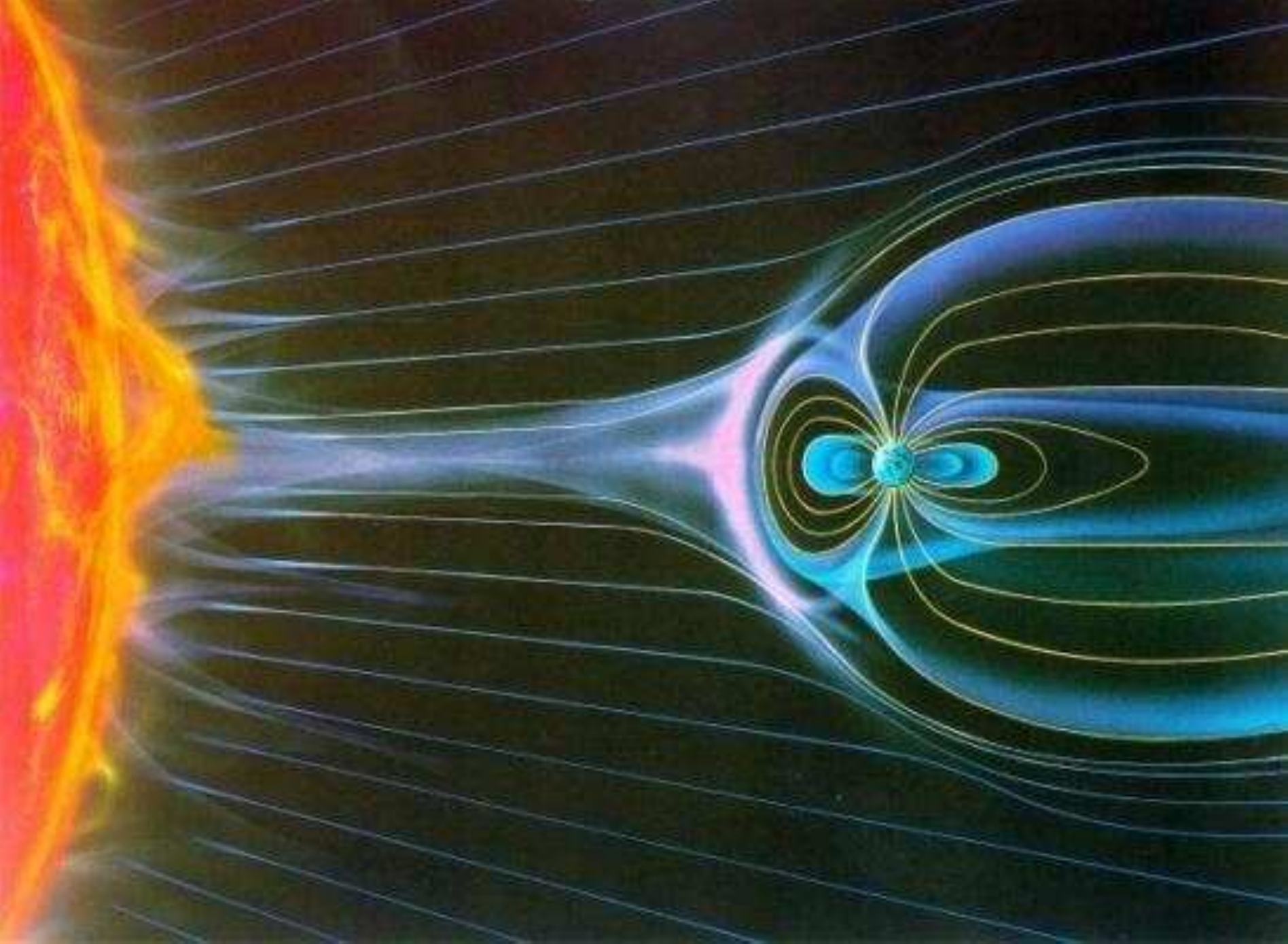}
\caption[Interaction between solar wind and Earth's magnetosphere]{Artist impression of the impact of solar wind on the Earth's magnetosphere. Solar wind compresses the day side and 
stretches the magnetic fields on the night side of the Earth. {\it Image courtesy - NASA}.}
\label{solarwind}
 \end{figure}

Besides electromagnetic radiation, the Sun emits a flow of charged particles and embedded magnetic fields from the solar corona into the interplanetary space. This continuous stream of 
plasma is called the solar wind, which travels with speeds of a few hundred \kms. This supersonic solar wind impinges on the Earth's magnetosphere inducing currents and creating 
fluctuations in the Earth's magnetic fields. Magnetic reconnection between the northward pointing magnetospheric fields and the southward or $B_{z}$ component of the 
interplanetary magnetic field (IMF) carried by the solar wind allows the charged particles to enter the magnetosphere \citep{Sch06}.
As the solar wind flow is diverted around the Earth's magnetosphere, the dynamic pressure of the solar wind and the IMF compresses
the Earth's magnetic fields on the Sun-facing side and stretches the fields to extend as far out as the night side (Figure \ref{solarwind}).
The solar wind interacts with the terrestrial magnetic field by forming a standing bow shock that slows, deflects and heats the plasma. The Earth's magnetosphere acts both 
as a shield against the solar wind and as a net that gathers solar wind momentum flux and stirs the magnetospheric plasma.
Electromagnetic radiation from the Sun that reaches the Earth much faster than the solar wind flow also affects the Earth's environment. It increases solar irradiance which causes 
heating of the upper atmosphere, affecting the drag experienced by low-Earth-orbiting satellites. In the outer magnetosphere, solar energetic particles pose a danger for 
the satellite systems and instrumentation \citep{Bak00}.

\subsection{Coronal mass ejections and space weather}
Coronal mass ejections (CMEs) rapidly develop into large-scale structures expanding to sizes greater than that of the Sun itself. When Earth-directed, 
these eruptions of plasma and magnetic fields can cause intense geomagnetic storms. Fast CMEs can drive interplanetary shocks that interact with the Earth's bow-shock, 
transferring energy to the magnetosphere and compressing the dayside magnetopause. These compression effects travel towards the tailward side at the solar wind speed,
causing strong auroral activity observable almost instantaneously all around the auroral oval \citep{Zho01}.
CMEs can also accelerate solar energetic particles (SEPs) that can penetrate the skins of space-borne probes and damage technical systems. The SEP fluxes from flares and CMEs 
can increase radiation levels endangering the lives of astronauts as well as enhance ionization and excitations in the Earth's middle atmospheric polar caps causing ozone depletion
\citep{Jac05}.

One of the most famous examples of a geo-effective solar event is the Halloween storm of 2003 during solar cycle 23. A series of 
solar flares and CMEs between mid-October to early November 2003 hit the Earth's magnetosphere affecting the satellite systems, causing power outages 
in Sweden and auroras as far south as Texas. There was a temporary failure of the SOHO satellite and damages caused to the 
{\it Advanced Composition Explorer} (ACE) spacecraft as well.

\subsection{Space weather forecasting}
The damaging effects of solar phenomena necessitates an accurate warning and predictive tool to determine the space weather affects, prevent disruptions in 
technical systems and ensure the safety of airlines and astronauts.
Space weather research needs studies to quantitatively predict the magnetospheric and ionospheric conditions based on measurements of the solar wind and interplanetary 
magnetic field. The aim is to understand the physics of solar processes so that primary predictions can be made using early solar observations 
\citep[\textit{e.g.}][]{Che11,Man17}. Advances in observational capacity of heliospheric missions as well as scientific efforts in developing both theoretical and numerical 
models are bringing us closer to building a dependable framework for forecasting the arrival and affects of solar transients on the Earth.

\section{Motivation and Thesis organization}
Based on the discussions in the previous sections highlighting the importance of space weather affects and their prediction, it is crucial to 
develop a thorough understanding of the evolution and propagation dynamics of CMEs. Since CMEs are major drives of the near-Earth space weather, 
physics behind their dynamics is important for improved advance warning tools which are vital in this era of technological dependence. 
With this goal, we select a set of well-observed CMEs using STEREO and SOHO coronagraph observations and derive physical parameters for each event (Chapter \ref{chap2}).
We investigate the forces that affect CME propagation using physics-based models for Lorentz force 
and solar wind aerodynamic drag force (Chapter \ref{chap4}). We quantify the heliocentric distances at which each of these forces 
dominate the CME dynamics and how they compare with each other in magnitude (Chapter \ref{chap5}). 

Below is a brief summary of all the chapters:
\begin{enumerate}
 \item {\bf Chapter 2 : CME Selection and Geometrical fitting}\\
This chapter describes the data sample used for this study. Observations from SOHO and STEREO missions in the rising phase of solar cycle 24 (2010 onwards)
are used to identify CMEs based on detailed selection criteria outlined in this chapter. 
A geometrical fitting technique (Graduated Cylindrical Shell model) is used to fit the selected CMEs to derive physical parameters (positional and structural). 
We provide details of the fitting technique and the CME sample of 38 events selected for this work along with their GCS parameters.

\item {\bf Chapter 3 : CME dynamics and propagation}\\
This chapter discusses the forces that affect CME dynamics as they erupt from the solar corona and propagate into the interplanetary 
space. We focus on Lorentz forces that drive CMEs and aerodynamic drag due to the ambient solar wind. Various approaches 
to these forces along with the specific models that we use in this study are described in this chapter in the form of a force equation. 
Details of the models for individual forces (Lorentz force and solar wind drag), parameters involved in the force prescription and their 
calculations using observational data derived by GCS fitting are also described. It outlines the analysis method for 
the two forces which is applied on all the CMEs in our sample to estimate the height beyond which aerodynamic drag force begins to dominate 
CME propagation, to determine the Lorentz force profile and compare the magnitude of the two forces at different heliocentric distances.

\item {\bf Chapter 4 : Results and Discussions}\\
This chapter describes in detail the results of the force analysis for all the CMEs in the sample. 
Using the solar wind drag analysis, we determine the range of heights beyond which the aerodynamic force 
becomes dominant for both slow and fast CMEs. The Lorentz force profiles for all the events are shown in this chapter and compared to the 
solar wind drag force magnitude. Results for all events are tabulated and discussed in detail in this chapter.

\item {\bf Chapter 5 : Future Work}\\
Based on our findings described in the previous chapter, suggestions for future analysis that can be done using a 
larger data set are listed in this Chapter.

\item {\bf Chapter 6 : Appendix} \\
The Appendix describes some supplementary material including models for virtual mass calculation and a simplified version for the 
Lorentz force model used for calculations. We also show a comparative treatment and equivalence of two different models for Lorentz forces: 
the one used in this work (torus instability model) and another based on poloidal flux injection. This chapter also shows the remote sensing data and the GCS fitting
at a single time stamp for all CMEs.
\end{enumerate}
 \chapter{CME Selection and Geometrical fitting}
\label{chap2}

\noindent\makebox[\linewidth]{\rule{\textwidth}{3pt}} 
{\textit {Using LASCO and STEREO coronagraph data, we identify 38 CMEs based on criteria described in this chapter. The white-light coronagraph images of these 
CMEs are fitted using a geometrical flux-rope model. The Graduated Cylindrical Shell (GCS) model is used to reconstruct the 
3D geometry of each CME to obtain their physical parameters. We describe the detailed criteria for selecting the CME events, the geometrical 
fitting procedure and list the observed parameters for each CME.} }\\
\noindent\makebox[\linewidth]{\rule{\textwidth}{3pt}} 

\section{Introduction}
With the launch of the \textit{Solar and Heliospheric Observatory} \citep[SOHO;][]{Dom95} in 1996 and the \textit{Solar Terrestrial 
Relations Observatory} \citep[STEREO;][]{Kai08} in 2006, it has been possible to observe the solar atmosphere and the Sun\,--\,Earth system continuously. 
Data compiled from these missions has been used to analyze the dynamics of solar coronal mass ejections (CMEs) in this study during the 
rising phase of solar cycle 24 (which began in 2008). The copious amount of data from these instruments requires a systematic and efficient procedure for 
shortlisting events for useful and conclusive analysis. Data for the selected CMEs are acquired from 
Instrument Resource data archives available online for both LASCO and STEREO. In order to observe the complete CME trajectory, we also include 
in our study the near-Earth \textit{in-situ} observations from the \textit{WIND} spacecraft. Changes in observed magnetic and plasma parameters 
provide indications of CME arrival at the Earth \citep{Bur81}.
We use the SOHO LASCO and STEREO coronagraphs and arrival signatures 
from {\it WIND} to trace the CME propagation from the Sun to the Earth. A three-dimensional geometrical fitting technique is then used to reconstruct the 
observed CME structures based on the flux-rope geometry. We use the Graduated Cylindrical Shell (GCS) model for this purpose. 
In this chapter, we first describe the aforementioned instruments used for CME observations (section \ref{sec2}) followed by a detailed 
account of the shortlisting criteria for event selection (section \ref{3}) and the description of GCS fitting technique used for these selected CMEs (section \ref{4}).
\section{Instruments and Observations} \label{sec2}
The SOHO mission carries the \textit{Large Angle and Spectrometric Coronagraph} \citep[LASCO;][]{Bru95} originally consisting of three solar 
coronagraphs which observe the solar corona from L1 Lagrangian point. After 1998 however, only two remain in working condition:
the C2 and C3 coronagraphs with field of view (FOV) 1.5\,--\,6 \Rs and 3.7\,--\,32 \Rs respectively. LASCO was launched to investigate the outer layer 
of the Sun and the solar atmosphere. A solar coronagraph, as the name suggests, looks at the solar corona by creating an artificial eclipse. 
In other words, it uses an occulting disk to cover the Sun, thereby blocking the light from the photosphere so as to image the Thomson-scattered white light from the corona.
The {\it SOHO LASCO CME CATALOG}\footnotemark[1] lists details of all the CMEs recorded by LASCO coronagraphs. It includes the Central 
Position Angle (CPA) and the sky-plane width of the CMEs based on which they are categorized into Halo (H) or Partial Halo (PH) CMEs. It also provides the 
first approximations of their speeds, acceleration and mass. Information from this catalog is the first step in selecting the CMEs for this 
study.
\footnotetext[1]{http://cdaw.gsfc.nasa.gov/CME\_list/}

The \textit{Sun--Earth Connection Coronal and Heliospheric Investigation} \citep[SECCHI;][]{How08} instrument onboard the STEREO mission constitutes a set of 
five telescopes that observe the solar atmosphere and inner heliosphere. In October 2006, two STEREO spacecrafts were launched together into the 
heliocentric orbit. The STEREO Ahead (STA) and STEREO Behind (STB) drift away from the Earth at a rate of 22.5$^\circ$ per year. STA travels slightly faster than the 
Earth around the Sun while STB is slower. SECCHI includes- an Extreme UltraViolet Imager (EUVI), two white
light coronagraphs (COR1 and COR2) and two Heliospheric Imagers (HI1 and HI2).
Taken together they study the evolution of CMEs from the corona, thorough the interplanetary (IP) medium, up to and beyond the Earth. 
COR1 has a FOV of 1.4\,--\,4 \Rs and COR2's FOV is 2\,--\,15 \Rs. HI observes the 
heliosphere from 12\,--\,318 \Rs. STEREO's online resource\footnotemark[2] provides images of the evolving CME, showing the deflection as 
well as expansion of the CME as it propagates through
\footnotetext[2]{https://secchi.nrl.navy.mil/}
the interplanetary medium from two different viewpoints (A and B).

A combination of data from all these instruments, namely : LASCO C2, COR2 A/B and HI1 has been used in our work to provide a continuous, three-perspective observation 
of coronal mass ejections originating in the solar corona and traveling towards the Earth ($\sim$ 215 \Rs).
In addition, we also include the \textit{in-situ} observations from the \textit{WIND}\footnotemark[3] spacecraft.
Launched in 1994, \textit{WIND} orbits the Sun at L1 Lagrangian point with the primary objective of collecting data for magnetospheric and 
ionospheric studies along with plasma processes in the near-Earth solar wind. The \textit{Solar Wind Experiment} (SWE) aboard \textit{WIND} 
includes sensors for recording the density, velocity and temperature of the ions in the solar wind. We use minute-averaged data for the solar wind 
flow speed and proton density about a day in advance of the CME arrival at the Earth to determine the conditions in the heliospheric plasma into which the CME propagates.
Using the arrival signatures, HI1, COR2 and LASCO observations (in that order) the CMEs can be backtracked from the Earth to the Sun. 

\begin{figure}
  \begin{tabular}{cc}
       \includegraphics[width = 0.38\paperwidth,height=0.25\paperheight]{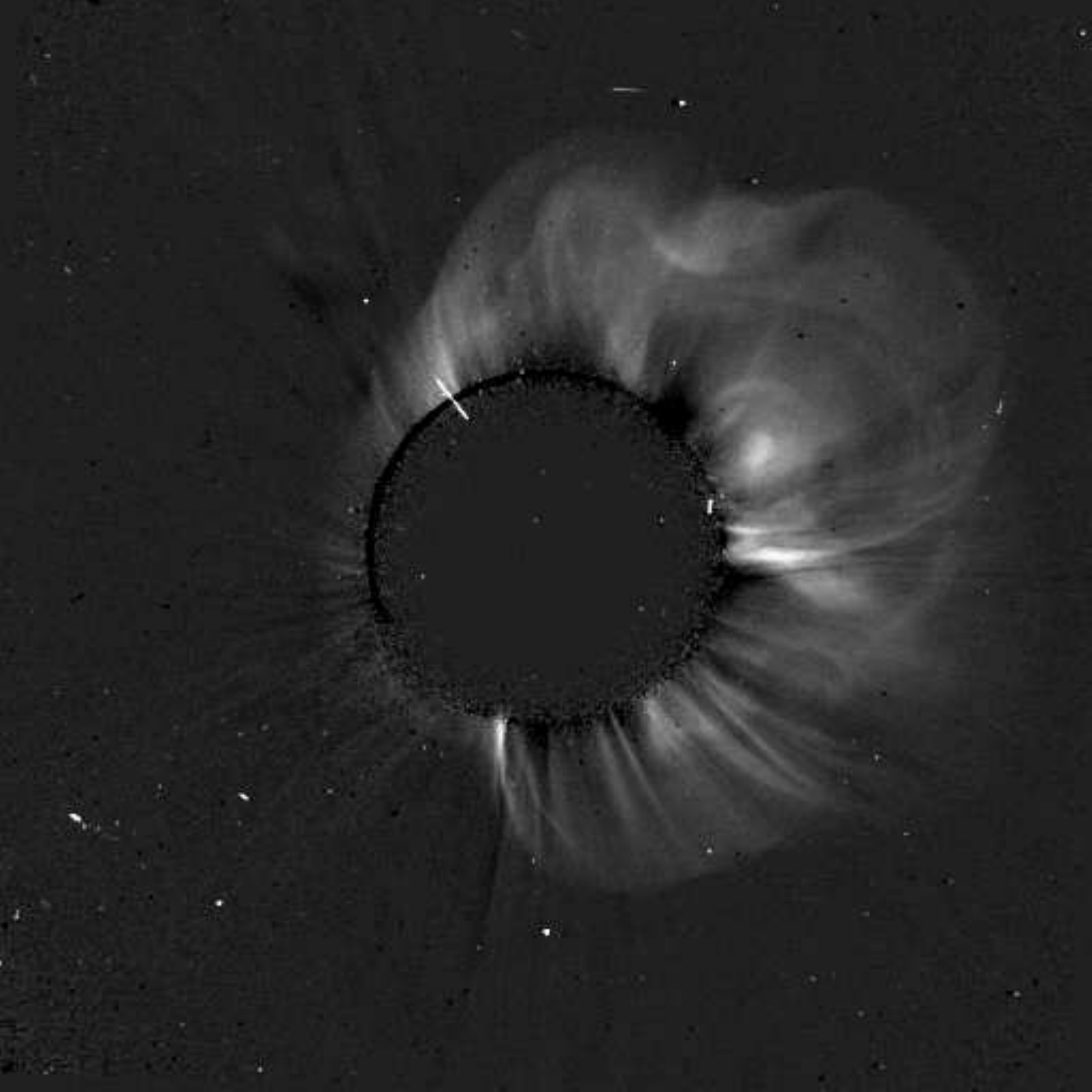} &
              \put(-125.8,-10.9){{\rotatebox{0}{{\color{black}\fontsize{9}{9}\fontseries{n}\fontfamily{phv}\selectfont (a)}}}}
             \put(-215.8,195.7){{\rotatebox{0}{{\color{white}\fontsize{9}{9}\fontseries{n}\fontfamily{phv}\selectfont LASCO C2}}}}
               \put(-215.8,8.){{\rotatebox{0}{{\color{white}\fontsize{9}{9}\fontseries{n}\fontfamily{phv}\selectfont 2012-09-28 00:34 }}}}             
       \includegraphics[width = 0.38\paperwidth,height=0.25\paperheight]{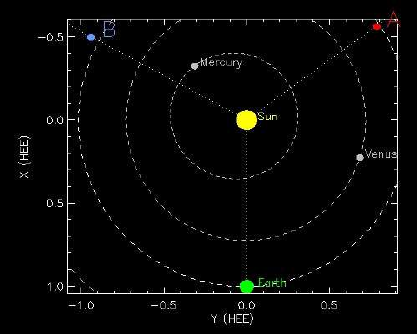}
                   \put(-115.8,-10.9){{\rotatebox{0}{{\color{black}\fontsize{9}{9}\fontseries{n}\fontfamily{phv}\selectfont (b)}}}}
\end{tabular}
  \begin{tabular}{cc}
       \includegraphics[width = 0.38\paperwidth,height=0.25\paperheight]{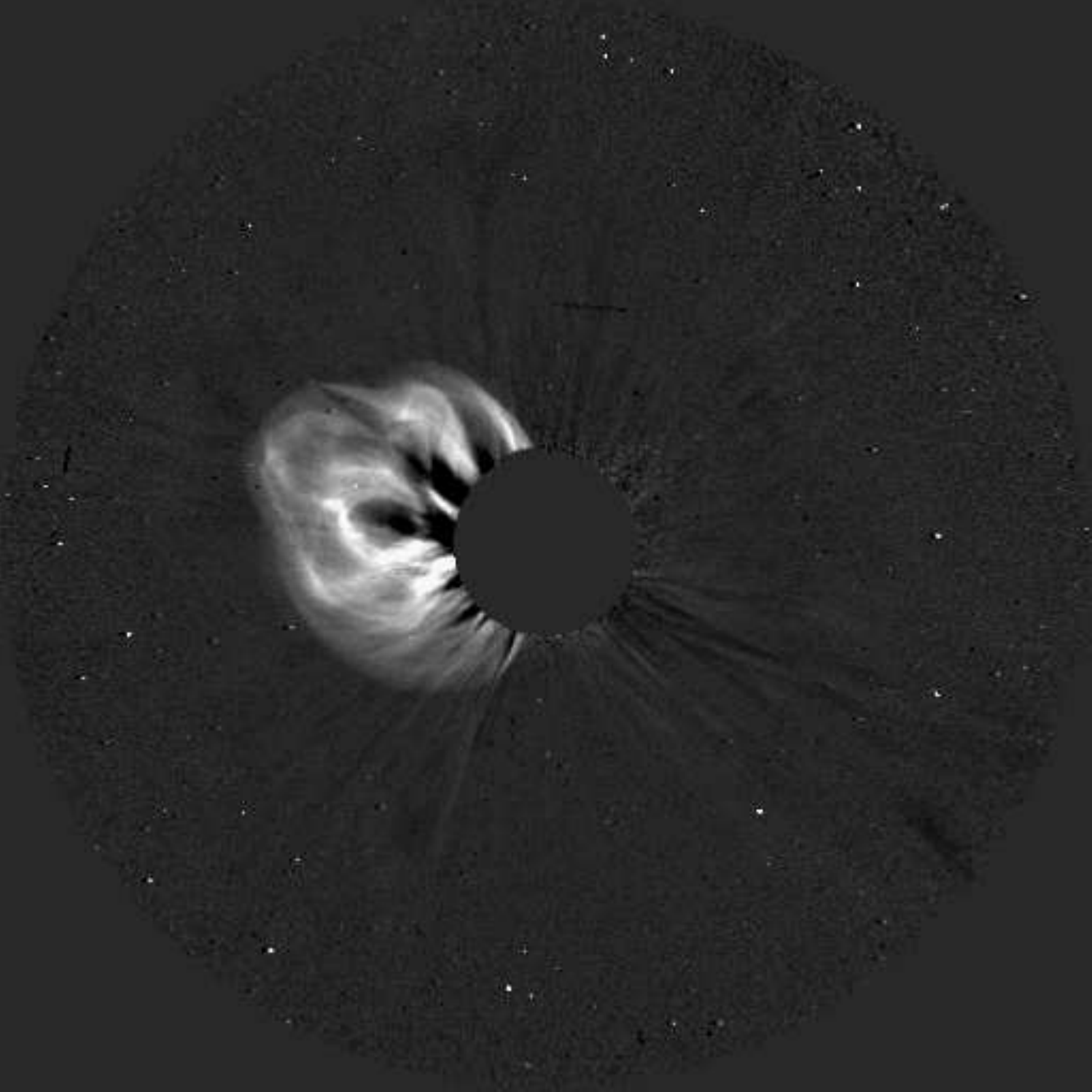} &
              \put(-125.8,-10.9){{\rotatebox{0}{{\color{black}\fontsize{9}{9}\fontseries{n}\fontfamily{phv}\selectfont (c)}}}}
             \put(-215.8,195.7){{\rotatebox{0}{{\color{white}\fontsize{9}{9}\fontseries{n}\fontfamily{phv}\selectfont COR2 A}}}}
               \put(-215.8,8.){{\rotatebox{0}{{\color{white}\fontsize{9}{9}\fontseries{n}\fontfamily{phv}\selectfont 2012-09-28 00:39 }}}}             
       \includegraphics[width = 0.38\paperwidth,height=0.25\paperheight]{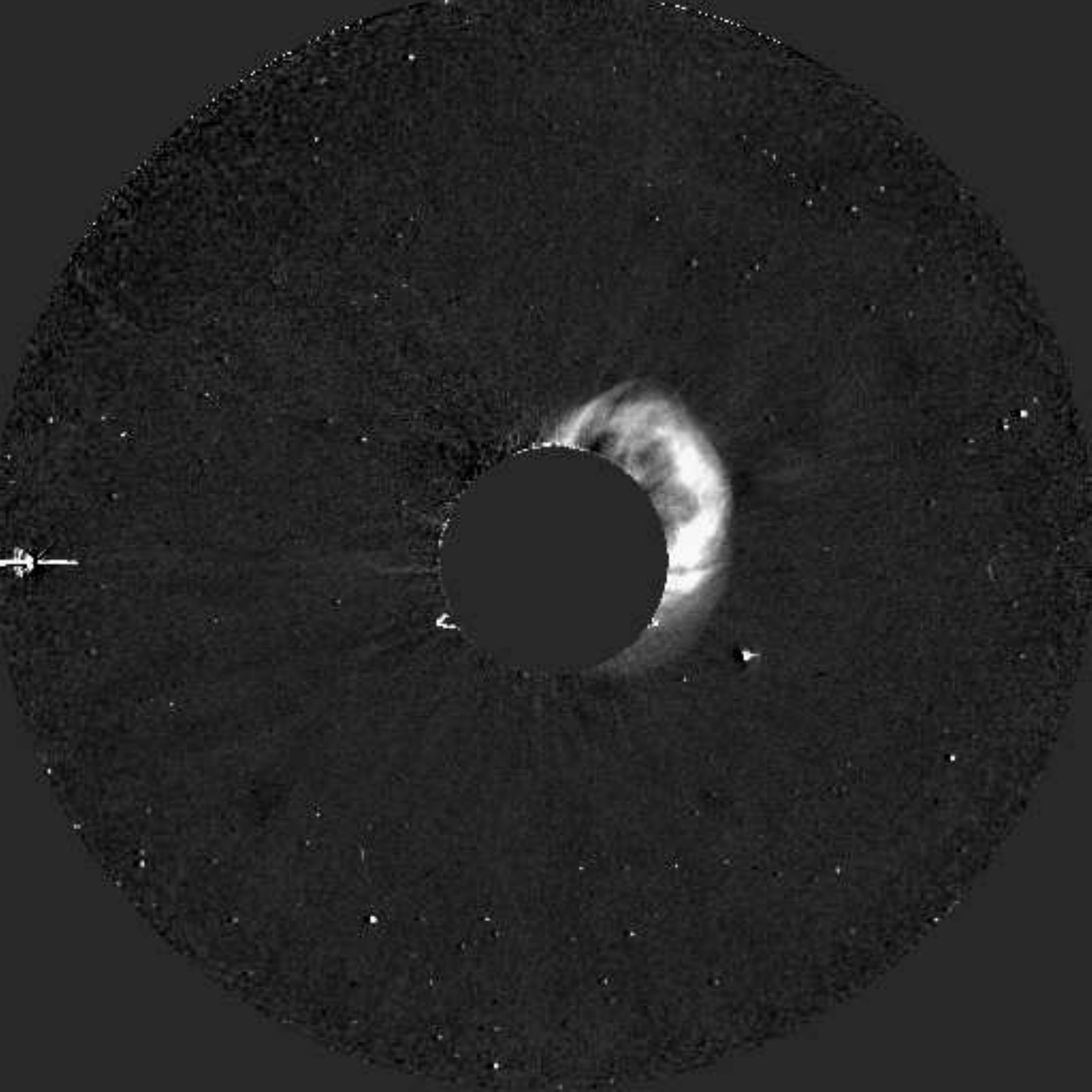}
                   \put(-115.8,-10.9){{\rotatebox{0}{{\color{black}\fontsize{9}{9}\fontseries{n}\fontfamily{phv}\selectfont (d)}}}}
                                 \put(-215.8,195.7){{\rotatebox{0}{{\color{white}\fontsize{9}{9}\fontseries{n}\fontfamily{phv}\selectfont COR2 B}}}}
                                               \put(-215.8,8.4){{\rotatebox{0}{{\color{white}\fontsize{9}{9}\fontseries{n}\fontfamily{phv}\selectfont 2012-09-28 00:39}}}}
\end{tabular}
 \caption[Simultaneous observation of a CME in LASCO C2 and STEREO COR2 A/B and positions of STEREO A/B.]{White-light images of a CME 
 on September 28, 2012 in LASCO C2 (Panel a), COR2 A (Panel c) \& COR2 B (Panel d) coronagraphs. Positions of the two STEREO spacecrafts 
 during this event is shown in Panel b.}
\label{cmefig1}
\end{figure}
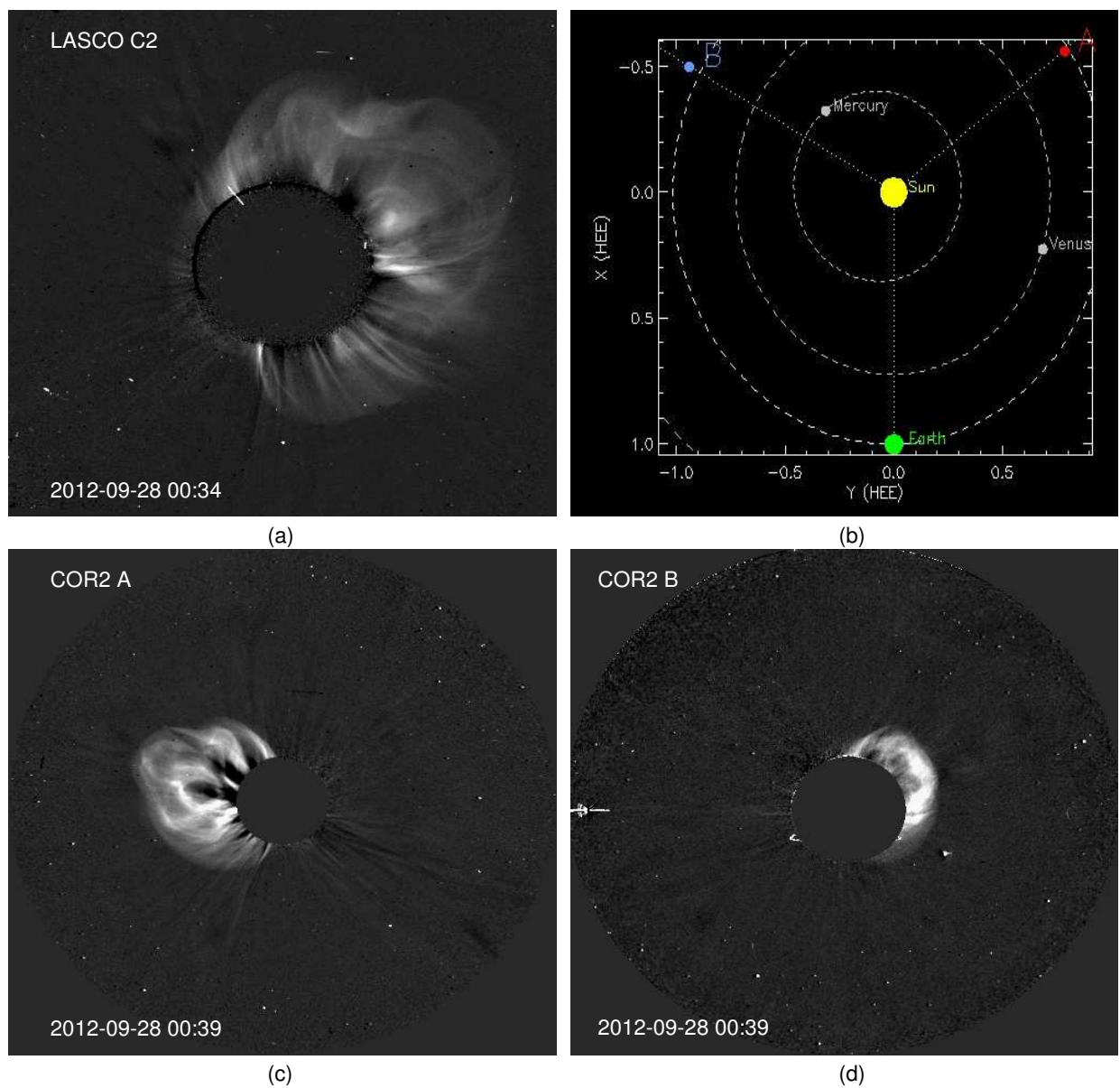

We however, begin with the CME identification in LASCO and COR2, continuing into the HI1 followed by identifying CME arrival using the ICME signatures in \textit{WIND}.
As an example, we show the white-light observation images for a CME on September 28, 2012 at 00:39 in all the three instruments in Figure \ref{cmefig1}.
Panel (a) shows a Halo CME in LASCO C2 FOV. Panel (b) indicates the position of the STEREO spacecraft with respect to the Earth at 00:39 on the date of the event.
Panel (c) shows a limb CME on the left side of the occulted Sun as observed from COR2 A, while Panel (d) shows the same event on the right side as seen from COR2 B.
\footnotetext[3]{http://omniweb.gsfc.nasa.gov/}

For the three-dimensional (3D) reconstruction, we use the LASCO and STEREO A and B viewpoints. The geometry of CMEs as seen from STEREO A and B instruments are similar, therefore to accurately
fit the GCS model, it is important to include LASCO observations. The LASCO view point gives crucial information about the orientation and dimensions 
of the CME, which are ambiguous if only SECHHI data is used. In general, the GCS fitting can be subjective leading to multiple parameter combinations that fit the observations, however, this 
degeneracy is broken by using the LASCO images which constrain the positional CME parameters (particularly the tilt angle). Extreme ultraviolet images (EUVI) have not been
used in this procedure, because the use of LASCO images in conjunction with the STEREO data provides a good optimization of the CME 
parameters. Beyond the LASCO field of view, the tilt, longitude and latitude are taken to be constant, 
unless a change is clearly visible in the CME rotation. The EIT images are not always reliable in indicating the correct position of leading edge if the LASCO viewpoint is not used.

\section{Selection Criteria} \label{3}
We analyze CMEs observed between 2010 and 2013. The criteria for selecting these events are described below. 
\begin{enumerate}
 \item {\bf CME identification in SOHO LASCO CME CATALOG} \\
As a first step, we use the comprehensive SOHO LASCO CME CATALOG to identify Halo (H) or Partial Halo (PH) CMEs. Partial halo CMEs have an angular width $>120^{\circ}$ while 
halo CMEs have an apparent angular span of $\approx360^{\circ}$.
Studies show that in general halo or partial halo CMEs have a component that is Earth-directed \citep{Zha04}. In particular, 
halo-type CMEs are expected to be Earth-directed because they cover the disk center of the Sun which is also the Earth's projection on the Sun's disk.
For each H/PH event identified in the catalog there exists a corresponding movie made from the compilation of LASCO C2 images for the event. 
We choose CMEs that are visible clearly in each image in the LASCO FOV and do not spill outside the image frame. We also include a pre-event image in 
the data. A pre-event image is taken right before the CME appears in the LASCO FOV (\textit{i.e.} no CME is visible) and is used as 
the background image.

\item {\bf CME identification in COR2} \\
The LASCO listed CME event is then identified in the STEREO observations. We select only those events from 
C2 observations which are completely visible in both STEREO A (STA) and STEREO B (STB) simultaneously. Starting with the same initial time as in C2, the CME is tracked in 
both the COR2 instruments till it exits their FOV ($\sim 15$ \Rs). The STEREO orbit Tool\footnotemark[4] indicates the positions of both STA and STB at any time.
Using the tool to identify the COR2 A/B positions and the direction in which the CME evolves, it can be determined if the CME is Earth-directed. 
For example, the event on September 28, 2012 (Figure \ref{cmefig1}) appears as a Halo CME in the LASCO catalog. This CME is visible as a limb CME in COR2. 
It appears on the left limb of the Sun when viewed from STA and on the right side when seen from STB, when the two spacecrafts were separated by 116$^\circ$. 
\footnotetext[4]{https://stereo-ssc.nascom.nasa.gov/where.shtml}

\item {\bf Continuity in CME data} \\
Only CMEs with continuous availability of images in both the coronagraphs were included in the sample \textit{i.e.} we require that there are no gaps in the 
data. It is required for the expanding CME structure to stay within the image frame for all the instruments.
\item {\bf Exclusion of events} \\
CMEs with visible distortions and diffuse structures were excluded because these are difficult to fit into geometrical models and cannot be reconstructed 
properly. During phases of extreme activity of the Sun, we also find multiple events from the same or close by active regions (AR) which leads to a faster CME 
overtaking a preceding slower one, or a fast CME altering the environment into which the following slower CME travels. Therefore,
we exclude events with CME-CME interactions because we are not equipped to handle the CME collision dynamics as of now.

\item {\bf CME identification in HI} \\
Once a CME is identified in both LASCO C2 and COR2 instruments, we track the event in the HI1 FOV. Using SECCHI movie tools, 
a CME can be seen in HI1 (up to $\sim 80$ \Rs) after it leaves the COR2 FOV. In most cases, the images are diffuse and therefore one needs to be careful 
while tracking a CME in HI1.

\item {\bf CME arrival signatures} \\
To establish the complete Sun-Earth timeline of the CME, we investigate the ICME/shock arrival data from the \textit{WIND} spacecraft for each CME. 
ICME signatures in \textit{WIND} data include depressed plasma proton temperature, low plasma-$\beta$ and strong, smoothly rotating magnetic field \citep{Bur81,Kle82}.
\end{enumerate}
\footnotetext[5]{https://wind.nasa.gov/ICMEindex.php}
\footnotetext[6]{http://www.affects-fp7.eu/services/cme-databases/}
\footnotetext[7]{http://www.srl.caltech.edu/ACE/ASC/DATA/level3/icmetable2.htm}
\footnotetext[8]{http://hec.helio-vo.eu/hec/}
Using the above mentioned criteria we shortlist 38 Earth-directed CMEs. We also referred to various compiled lists of CMEs e.g. WIND ICME list\footnotemark[5],
AFFECTS database\footnotemark[6], List of ICMEs by Richardson and Cane\footnotemark[7] and JHUAPL COR-CME Catalog \citep{Vou17} to cross-reference the selected CMEs. 
We also referred to the Heliophysics Event Catalog\footnotemark[8] for information about source region (SR) positions and associated flare activity.
The downloaded images for all events are prepared using the standard SolarSoft $sechhi\_prep.pro$ procedure.
These are then fitted using a geometrical model for all time-stamps from the first appearance of CME in LASCO C2 up to the HI1 FOV. This fitting 
technique is described in the next section \ref{4}. \\
Table \ref{tbl31} lists details of all the CMEs studied in this work. The serial number (column 1) of each event is used throughout as a reference for the corresponding CME.
Columns 2 and 3 indicate the date of the event and the time when it is first fit in C2 FOV using GCS fitting technique respectively. FR (Flux-rope) indicates if a three-part 
flux-rope structure was visible in the COR2 observations of the CME. We also indicate if the observed CME was a Halo (H) or a Partial Halo (PH) CME. 

\section{Fitting Technique} \label{4}
Coronal mass ejections are observed by white-light imaging of the solar corona, i.e. the coronagraphs measure the light from the photosphere which is 
Thomson scattered by free electrons in the coronal plasma \citep{Bil66} along the light of sight. Coronagraphs yield two-dimensional white-light images in the plane
of the sky. It is therefore, difficult to obtain a true three-dimensional geometrical reconstruction of CMEs, the understanding of which is an important tool 
in developing a physical model for CME propagation.\\
A coronal mass ejection typically has a three-part structure- a bright front, followed by a dark cavity and a bright core 
\citep{Ill85}. LASCO observations have revealed detailed structure in the form of circular striations around the cavity. 
This gave rise to the widespread acceptance of CMEs as flux-rope like structures \citep{The06}. Various geometrical models are used for 
fitting and characterizing the CME structure, \textit{e.g.}, the Cone model \citep{How82}, the Elliptical Model \citep{Kah07}, the Graduated Cylindrical Shell (GCS) model 
\citep{The06,The09}. With many theoretical models fitting the idealized flux-ropes and successfully reproducing the observations 
\citep[\textit{e.g.,}][]{Che97}, the term ``flux-rope CME'' has gradually substituted the ``three-part CME''. CME properties are also shown to be consistent with 
these theorized flux-ropes \citep{Vou00,Kra05}. In this study, we use one such theoretical model called the Graduated Cylindrical Shell (GCS) 
model \citep{The09,The11} for the geometrical fitting of the white-light CME images.

\subsection{Graduated Cylindrical Shell (GCS) Model}
\begin{figure}[H]
\centerline{\hspace*{0.0\textwidth}
       \includegraphics[width = 0.45\paperwidth,height=0.28\paperheight]{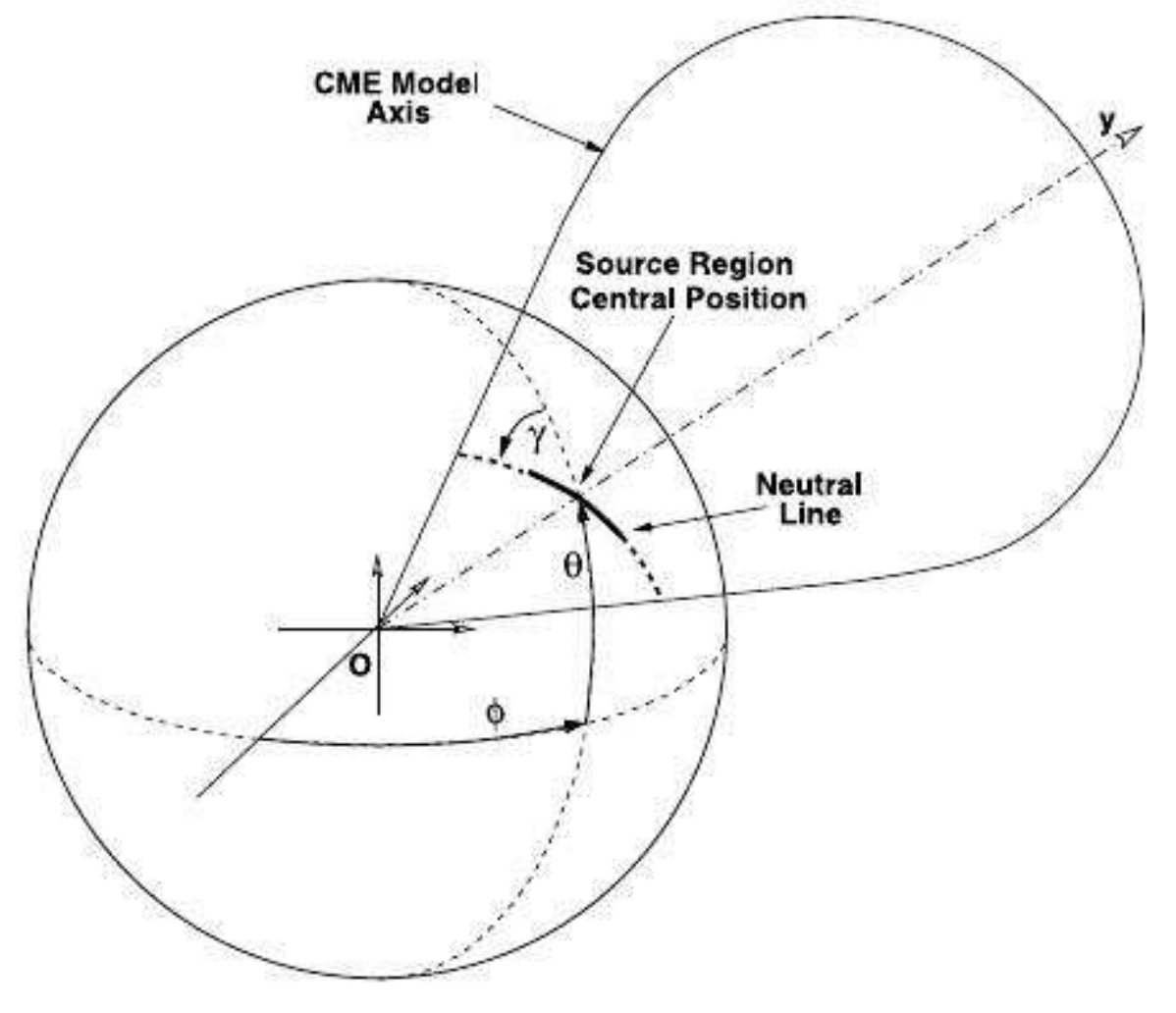} 
              \put(-200.8,-12.9){{\rotatebox{0}{{\color{black}\fontsize{9}{9}\fontseries{n}\fontfamily{phv}\selectfont (a) Location of the GCS model in space}}}}
              }
              \vspace{1.3cm}
              \centerline{\hspace*{0.0\textwidth}
       \includegraphics[width = 0.57\paperwidth,height=0.25\paperheight]{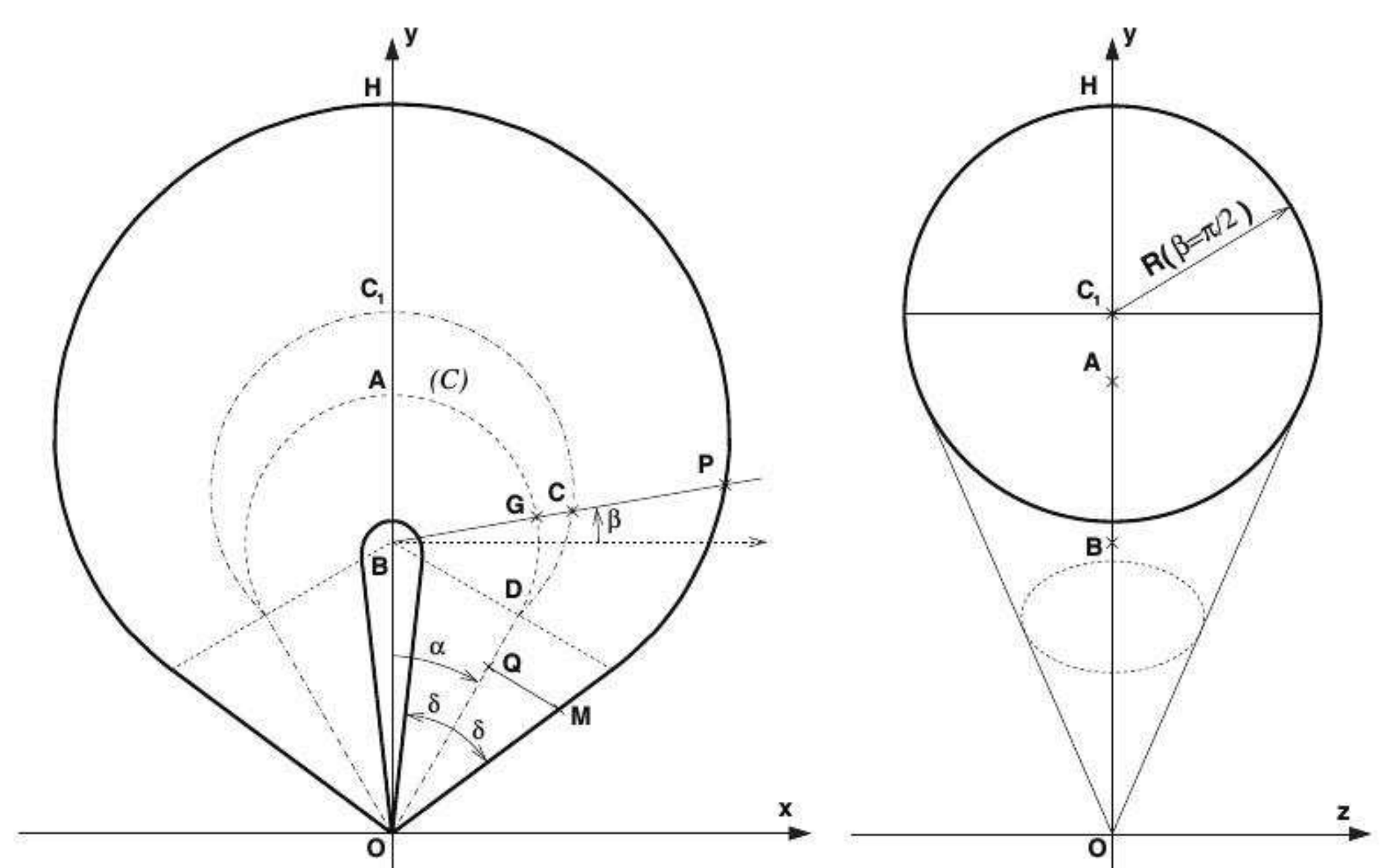}
                   \put(-230.8,-12.9){{\rotatebox{0}{{\color{black}\fontsize{9}{9}\fontseries{n}\fontfamily{phv}\selectfont (b) GCS geometry face-on and 
edge on.}}}}
}
\caption[Schematic for GCS model]{{Geometrical representation of the GCS model adapted from \citet{The11}}}
\label{gcs1}
\end{figure}

\begin{figure}[H]
 \centering
  \begin{minipage}[b]{.45\textwidth}
  \subfloat
    []
    {\label{fig:figA}\includegraphics[width=1.2\textwidth,height=18.4cm]{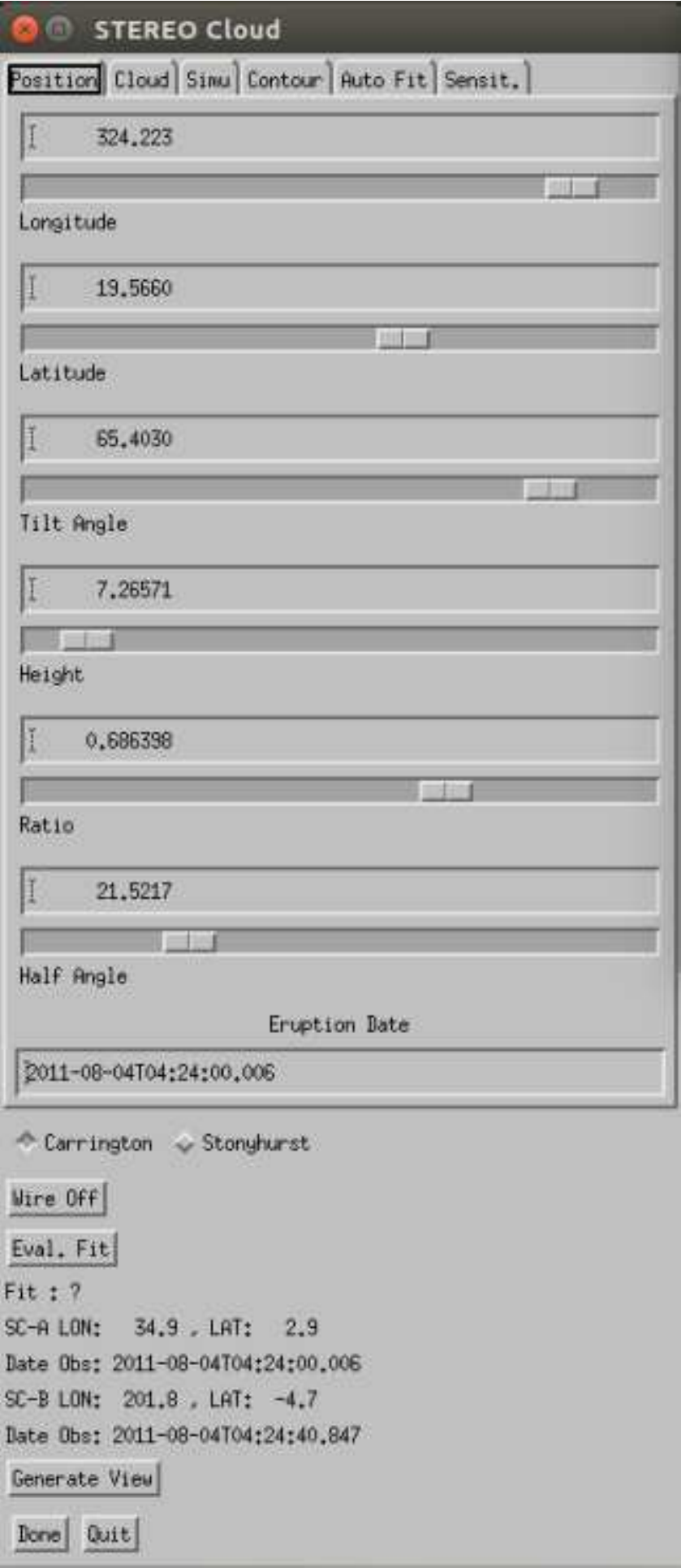}}
  \end{minipage}
\hspace{1.1cm}
\begin{minipage}[b]{.45\textwidth}
\centering

\subfloat
  []
  {\label{fig:figB}\includegraphics[width=1.05\textwidth,height=5.5cm]{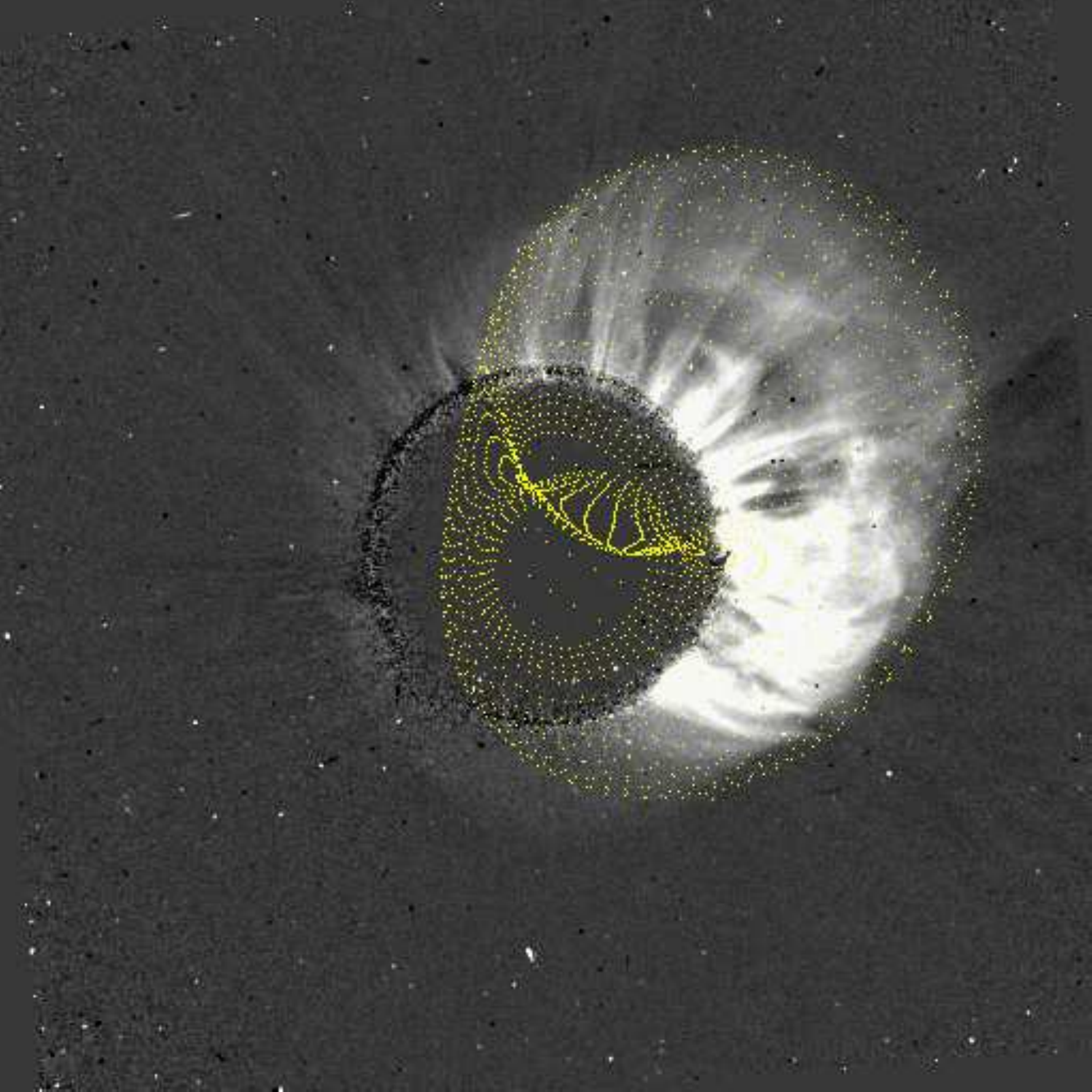}}

\vfill

\subfloat
  []
  {\label{fig:figC}\includegraphics[width=1.05\textwidth,height=5.5cm]{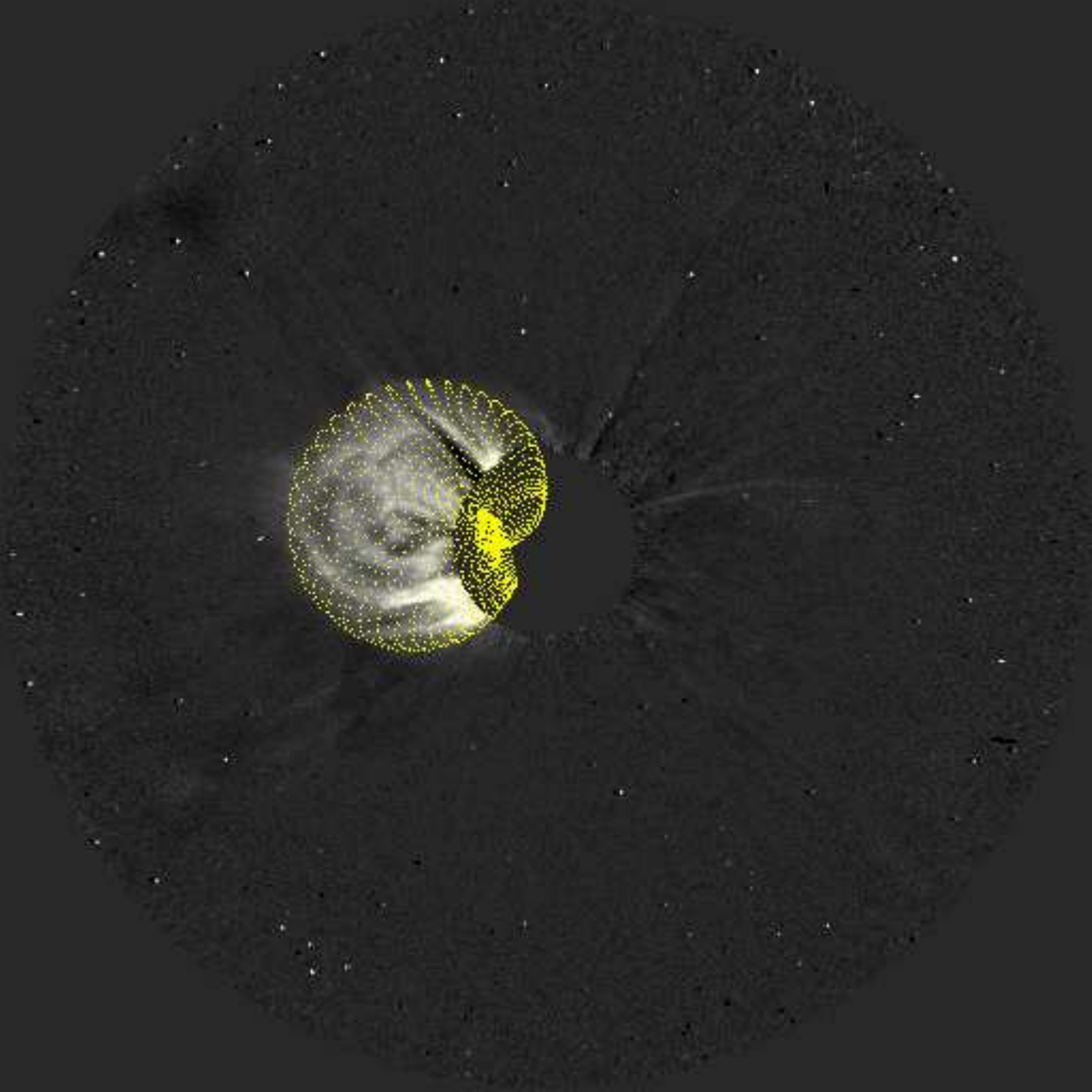}}
  \vfill

\subfloat
  []
  {\label{fig:figD}\includegraphics[width=1.05\textwidth,height=5.5cm]{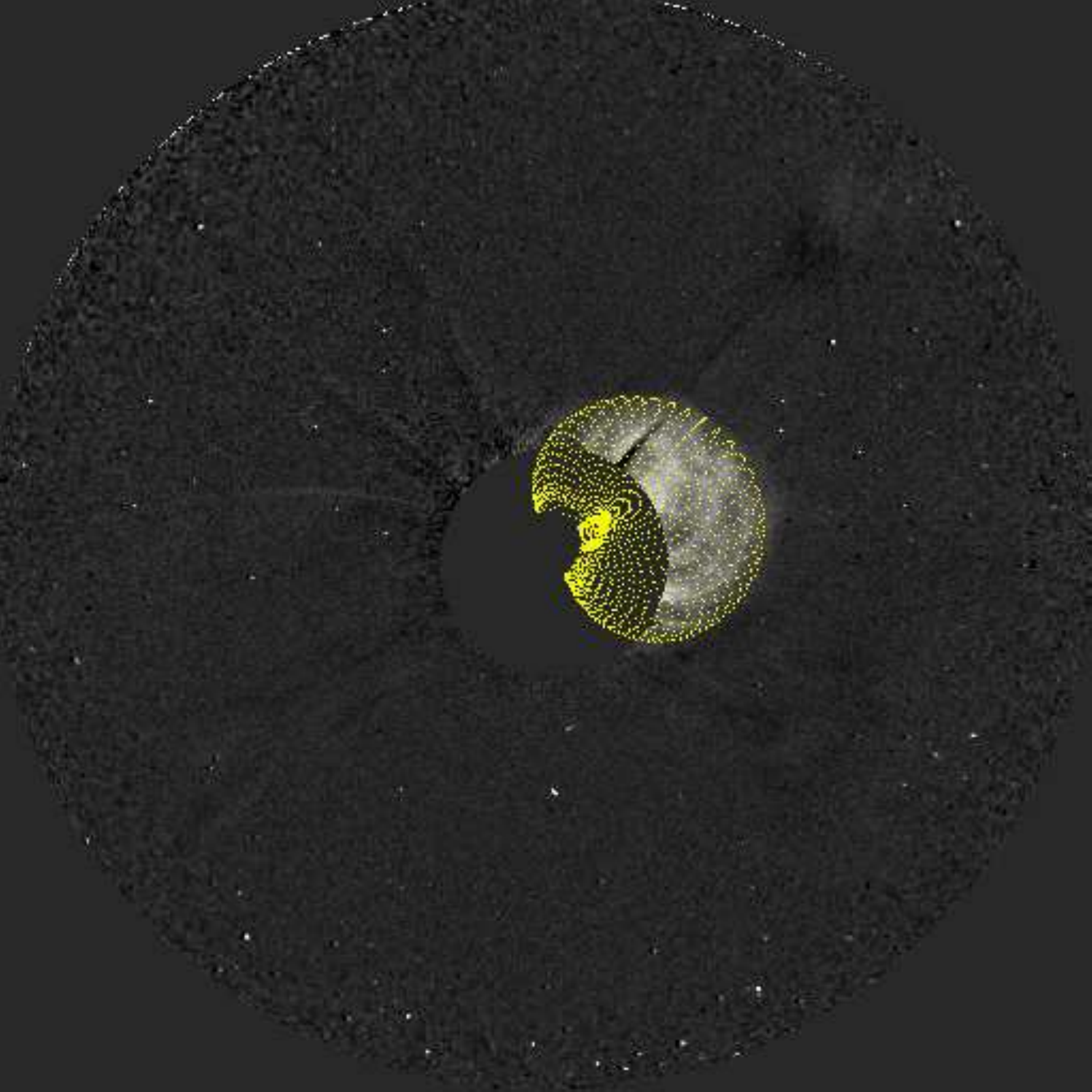}}

\end{minipage}
\caption[GUI interface for GCS fitting technique.]{Panel a is the graphic user interface indicating six GCS parameters that can be varied as shown. Panels b, c and d show the 
corresponding fit of the wire-mesh (yellow) structure to the CMEs in C2, COR2 A and COR2 B images respectively for an event on August 4, 2011 at 4:24 UT.}
\label{gui}
\end{figure}

The Graduated Cylindrical Shell (GCS) model envisages a CME as two conical legs attached to the Sun with a tubular section in between, 
forming the main body of the CME \citep{The06,The09}. It fits a helical flux-rope like structure to the visible CME by varying a set of 
six parameters. Panel (a) of Figure \ref{gcs1} shows the location of the GCS model with respect to the solar surface. Three parameters
define the CME position: Carrington Longitude ($\phi$), Latitude ($\theta$) and Tilt angle ($\gamma$). These quantities represent the Euler angles that 
relate the Heliocentric Earth Ecliptic (HEE) co-ordinate system to the model axes.
Panel (b) of Figure \ref{gcs1} shows the schematic of the GCS model in both face-on and edge on views. The structural parameters that define 
the 3D model are: Height of the leading edge ($R$), Aspect ratio ($\kappa$) and Half angle ($\alpha$). The quantity $R$ is measured from the center of the Sun 
while $\kappa=sin(\delta)$, where $\delta$ is the half angle of the cone. The half angle $\alpha$ is the angle between the axis of the leg and the y-axis.
Using $R, \kappa$ \& $\alpha$, all physical parameters that represent the CME structure can be derived using the detailed geometrical description in \citet{The11}.
Each image containing the CME structure is fitted using these six parameters. For a 3D geometrical reconstruction of the flux-rope structure, 
CME images from LASCO C2, COR2 A \& B are fitted simultaneously. These parameters are varied to obtain the best fit to the evolving CME. 
The graphic user interface (GUI) using the SSWIDL \textit{scraytrace} routine provides an interactive way of varying the GCS parameters and displays these 
changes simultaneously in the wire-frame which is overlaid on top of the images in each instrument frame. 
An example of the fitting technique is shown in Figure \ref{gui}. It shows the GUI wherein the GCS parameters can be varied to get the optimal fitting of the 
yellow wire-mesh like flux-rope structure to overlap the visible CME in the images.
It is important to not lose sight of the basic structure of the CME while fitting it at each time-stamp. Therefore, it is recommended to focus on a strikingly 
visible feature in the CME image and use the parameter variation to fit it in all images. In particular, we try to fit the leading edge of the CME 
with the outer front of the wire-mesh structure. In most cases, depending on how the CME deflects the longitude and latitude vary accordingly. 
Since our CMEs are Earth-directed, we use the longitudinal position of the Earth (Carrington Longitude) as a first estimate of the CME longitude. If there is not much 
deflection, these estimates provide a good approximation. We also consider the location of the associated flare (if any)
(\textit{$http://hec.helio-vo.eu/hec/hec\_gui.php$}) and/or the position of the source region of the CME
(\textit{$https://www.solarmonitor.org$}). The actual position can still be different from these values, however, they give a good initial approximation 
for the longitude and latitude. We first vary the longitude $\phi$, latitude $\theta$ and the height of the leading 
edge ($R$) to fit the CME front so that the outer edge of the wire-frame matches with the CME leading edge in all three instrument frames simultaneously. The tilt angle 
($\gamma$) is also adjusted according to the visible CME structure. Since the CME expands as it propagates, the spatial extent is covered by varying the aspect 
ratio ($\kappa$) parameter and the angular width is adjusted by changing the half angle ($\alpha$). All the variations in parameters should be such that the 
wire-frame overlaps the CME structure in all three instrument images simultaneously. With an optimized set of these six parameters we get a well-fitting match with 
the observed CME. This procedure is repeated for all images sequentially and the model is fit at all time-stamps. In cases where deflection is not so significant 
the positional parameters vary only slightly.

When the CME transitions from the COR2 to the HI1 FOV, it is important to maintain the chronological continuity. There should not be a large data gap (more than 1-2 hours)
between the data from these two instruments. In HI1, the CME images appear somewhat diffuse; therefore, it is important to properly track the leading edge
from COR2 to HI1. Since the CME structure is visibly less distinct, we keep the positional parameters constant unless a significant deflection is seen in the 
images and vary the other parameters accordingly. In HI1, images are fitted using the GCS procedure at successive time-stamps up to 80 \Rs. Finally we get a 3D fit to all 
CME images beginning from LASCO C2, COR2 A \& B, followed by HI1 at each time-stamp. This method is repeated carefully for all the 
CMEs in our sample set to get the 3D reconstructed CME evolution from $\approx$ 3 \Rs up to 80 \Rs. With the observational parameters derived from the GCS fitting routine, other physical parameters can be estimated as well. Using the 
geometrical description in \citet{The11}, we can calculate (at each time-stamp), the cross-sectional width of the CME and the circular cross-section
radius (minor radius). 

Once we have the complete height-time evolution of the CME (height of the leading edge), we make estimates of the initial velocity using a third degree 
polynomial. The GCS height-time profile is fit to determine the local velocity which is later fed into the force model to predict the complete CME trajectory.
Details of the GCS parameters of each CME at the first observed time are given in Table \ref{tbl32}. The serial number corresponds to event described in 
Table \ref{tbl31}. The 8 events from \citet{Sac15} (marked with an asterisk $\ast$) 
have observations up to the HI2 FOV, while the remaining events have been fitted up to HI1 FOV. 
$h_{0}$ is the first observed height at which the GCS fit is done at the time indicated in Table \ref{tbl31}. $v_{0}$ is the CME initial velocity at 
$h_{0}$. GCS parameters - Carrington Longitude $\phi$, Latitude $\theta$, Tilt angle $\gamma$, Aspect ratio $\kappa$ and Half angle $\alpha$ are given at the 
first observed height $h_{0}$ for each CME. The height-time evolution data as well as other physical parameters for each reconstructed CME can now be used 
for analyzing the CME dynamics.

\begin{table}
\caption[Observational details of CMEs in the sample]{Details of observed characteristics of the CMEs in the sample set.
First column is the serial number of each event with which it is referenced throughout the thesis.
For each event the observation date (Date) and time when it is first fitted in C2 FOV is mentioned.
FR indicates if a distinct flux-rope structure was observed in COR2 FOV. Cross ($\times$) indicates that a FR structure was seen. 
H/PH indicate if it is a Halo or Partial Halo CME as seen in LASCO.}
\label{tbl31}
\centering
 \begin{tabular}{ccccc} 
  \hline
 No.& Date &Time & FR & H/PH\\
    & &(U.T.)& & \\
  \hline
  1   & 2010 Mar. 19 &11:39 & $\times$ & H \\
  2 & 2010 Apr. 03 &10:24 & $\times$ & H \\
  3   & 2010 Apr. 08& 03:24 & -- & PH \\
  4   & 2010 Jun. 16 &15:24 & $\times$ & PH \\
  5   & 2010 Sep. 11 &02:24 & $\times$ & PH  \\
  6   & 2010 Oct. 26 &07:39 & $\times$ & --    \\
  7         & 2010 Dec. 23& 05:54 & $\times$ & PH   \\
  8         & 2011 Jan. 24 &03:54 & -- & --  \\
  9   & 2011 Feb. 15 &02:24 & -- & H   \\
  10        & 2011 Mar. 03& 05:54 & $\times$ & PH  \\
  11  & 2011 Mar. 25 &07:00 & -- & H  \\
  12        & 2011 Apr. 08& 23:39 & -- & H  \\
  13        & 2011 Jun. 14 &07:24 & $\times$ & PH   \\
  14  & 2011 Jun. 21 &03:54 & $\times$ & H  \\
  15  & 2011 Jul. 09 &00:54 & $\times$ & PH  \\
  16  & 2011 Aug. 08 &04:24 & -- & H  \\
  17        & 2011 Sep. 13 &23:39 & $\times$ & PH   \\
  18  & 2011 Oct. 22 &10:54 & -- & H   \\
  19        & 2011 Oct. 26& 12:39 &-- & PH  \\
  20        & 2011 Oct. 27 &12:39 & $\times$ & H   \\
  21  & 2012 Jan. 19 &15:24 & $\times$ & H  \\
  22  & 2012 Jan. 23& 03:24 & $\times$ & H   \\
  23  & 2012 Jan. 27 &17:54 & $\times$ & H   \\
  24  & 2012 Mar. 13 &17:39 & $\times$ & H    \\
  25        & 2012 Apr. 19& 15:39 & -- & PH    \\
  26  & 2012 Jun. 14 &14:24 & $\times$ & H   \\
  27  & 2012 Jul. 12 &16:54 & $\times$ & H   \\
  28  & 2012 Sep. 28& 00:24 & $\times$ & H   \\
  29        & 2012 Oct. 05& 03:39 & $\times$ & PH  \\
  30        & 2012 Oct. 27 &17:24 & $\times$ & H  \\
  31        & 2012 Nov. 09 &14:54 & $\times$ & --  \\
  32        & 2012 Nov. 23 &14:39 & $\times$ & H   \\
  33  & 2013 Mar. 15 &06:54 & -- & H \\
  34  & 2013 Apr. 11 &07:39 & -- & H   \\
  35  & 2013 Jun. 28 &02:24 & -- & H   \\
  36  & 2013 Sep. 29 &22:24 & $\times$ & H   \\
  37  & 2013 Nov. 07 &00:24 & -- & H   \\
  38  & 2013 Dec. 07 &08:24 & -- & H  \\
\hline
\end{tabular}
\end{table}

\begin{table}
\caption[GCS fitted parameters for all CMEs at first observed height]{CME no. indicates the serial number of the CME as referenced in Table 
\ref{tbl31}. $h_{0}$ is the observed GCS height at the first observation and $v_{0}$ is the derived velocity at 
$h_{0}$. GCS parameters at $h_{0}$ are given by Carrington longitude ($\phi$), heliographic latitude ($\theta$), 
tilt ($\gamma$), aspect ratio ($\kappa$) and half angle ($\alpha$). Fast CMEs are indicated by a 
superscript (f) in their serial number. Events from \citet{Sac15} are indicated by a superscript($\ast$) 
in their corresponding serial number.}
\label{tbl32}
\centering
 \begin{tabular}{lccccccc} 
  \hline
     &     & & \multicolumn{5}{c}{GCS Parameters at $h_{0}$} \\
 No.& $h_{0}$ & $v_{0}$&$\phi$&$\theta$& $\gamma$ & $\kappa$ &$\alpha$ \\
    & (\Rs) & ($km\, s^{-1}$) &($^{\circ}$)&($^{\circ}$)&($^{\circ}$)& &($^{\circ}$) \\
  \hline
  1$^{*}$   & 3.5 & 162 & 119 & -10 & -35 & 0.28 & 10  \\
  2$^{* f}$ & 5.5 & 916 & 267 & -25 &  33 & 0.34 & 25 \\
  3$^{*}$   & 2.9 & 468 & 180 &  17 & -18 & 0.20 & 22  \\
  4$^{*}$   & 5.7 & 193 & 336 & 0.5 & -15 & 0.23 & 9.5  \\
  5$^{*}$   & 4.0 & 444 & 260 & 23  & -49 & 0.41 & 18  \\
  6$^{*}$   & 5.3 & 215 & 74  & -31 & -55 & 0.25 & 22 \\
  7         & 3.7 & 147 & 29  & -28 & -15 & 0.40 & 18  \\
  8         & 4.4 & 276 & 336 & -15 & -15 & 0.30 & 22 \\
  9$^{*}$   & 4.4 & 832 & 30  & -6  &  30 & 0.47 & 27 \\
  10        & 4.9 & 349 & 175 & -22 &  8  & 0.35 & 21 \\
  11$^{*}$  & 4.8 & 47  & 207 &   1 &  9  & 0.21 & 37  \\
  12        & 4.7 & 300 & 41 &   6 & -6  & 0.30 & 35\\
  13        & 3.6 & 562 & 202 & 1   & 36  & 0.26 & 57 \\
  14$^{f}$  & 8.4 & 1168& 129 & 5   & -8  & 0.45 & 14 \\
  15$^{f}$  & 4.1 & 903 & 264 & 17  & 15  & 0.35 & 18 \\
  16$^{f}$  & 7.3 & 1638& 324 & 19  & 65  & 0.69 & 29 \\
  17        & 3.8 & 493 & 134 & 19  & -38 & 0.43 & 41 \\
  18$^{f}$  & 4.0 & 1276& 54  & 44  & 16  & 0.60 & 45 \\
  19        & 7.8 & 889 & 302 & 7   & -1  & 0.46 & 9\\
  20        & 5.3 & 882 & 223 & 29  & 16  & 0.36 & 16  \\
  21$^{f}$  & 4.6 & 1823& 212 & 44  & 90  & 0.47 & 58 \\
  22$^{f}$  & 4.0 & 1910& 206 & 28  & 58  & 0.48 & 41 \\
  23$^{f}$  & 3.5 & 2397& 193 & 30  & 69  & 0.38 & 41 \\
  24$^{f}$  & 3.9 & 1837& 302 & 21  & -40 & 0.74 & 73  \\
  25        & 4.1 & 648 & 82  & -28 & 0.0 & 0.27 & 30  \\
  26$^{f}$  & 6.2 & 1152& 92  & -22 & -87 & 0.38 & 20 \\
  27$^{f}$  & 4.4 & 1248& 88  & -10 & 78  & 0.45 & 35 \\
  28$^{f}$  & 6.7 & 1305& 165 & 17  & 86  & 0.42 & 42 \\
  29        & 4.4 & 461 & 56  & -24 & 37  & 0.30 & 31 \\
  30        & 7.3 & 380 & 118 & 8   & -36 & 0.20 & 40 \\
  31        & 3.8 & 602 & 285 & -18 &  7  & 0.48 & 35 \\
  32        & 6.3 & 492 & 91  & -21 & -66 & 0.52 & 10 \\
  33$^{f}$  & 4.7 & 1504& 76  & -7  & -86 & 0.31 & 40 \\
  34$^{f}$  & 5.9 & 1115& 77  & -1  & 90  & 0.14 & 47 \\
  35$^{f}$  & 6.6 & 1637& 177 & -35 & -20 & 0.41 &  5 \\
  36$^{f}$  & 4.9 & 1217& 360 & 21  & 90  & 0.38 & 47 \\
  37$^{f}$  & 5.9 & 975 & 304 & -30 & -75 & 0.34 & 12 \\
   38$^{f}$ & 6.8 & 1039& 221 & 32  & 51  & 0.36 & 47  \\
\hline
\end{tabular}
\end{table}
 \chapter{CME dynamics and propagation}
\label{chap4}

\noindent\makebox[\linewidth]{\rule{\textwidth}{3pt}}
{\textit {We discuss the major forces that govern CME propagation. These are : Lorentz driving force, solar wind aerodynamic drag and gravity. 
We motivate and investigate the phenomenological expressions used to describe these forces. We consider the torus instability model for Lorentz forces and a 1D 
drag-based model to describe the CME-solar wind interaction. Each of the parameters that appear in the force equation is discussed in this chapter and details of the models used for 
their calculation are provided. The method for force analysis is also outlined for both the forces.}}\\
\noindent\makebox[\linewidth]{\rule{\textwidth}{3pt}} 

\section{Introduction}
CMEs have been widely observed and studied to understand the mechanisms that trigger their eruption and cause their propagation in the interplanetary 
(IP) space. The complete CME evolution is typically divided into two phases; the initial eruption and acceleration phase followed by the propagation phase.
\citet{Gop13} describes the CME acceleration as the sum of accelerations due to each of the three forces that affect CME propagation. These are, 
Lorentz force, aerodynamic drag and gravity. Similarly, \citet{Zha06} describe three distinct phases of CME propagation: 1) Initiation phase of slow rise of CMEs, 2) phase of fast acceleration of CMEs
and 3) propagation phase in which the CMEs undergo relatively minor evolution. They club the latter two phases and term it ``residual acceleration'' \citep[\textit{e.g.}][]{Sub07,Bei11}.
We describe the Lorentz driving force that is responsible for initial CME acceleration in section \ref{1} and the aerodynamic drag due to interaction between 
the CME and ambient solar wind in section \ref{2}. The solar gravitational pull causes CMEs to decelerate very close to the Sun \citep{She12}; however, its effects are much smaller compared to the accelerating Lorentz forces 
\textit{i.e.} gravity does not affect CME dynamics \citep{Che03}. Therefore, for simplicity we neglect gravity in this study.
We outline the models that are used to calculate the two governing forces (Lorentz force and aerodynamic drag) along with the models/definitions 
used for evaluating the physical parameters required in the force equations. We also describe the methods for the aerodynamic drag and Lorentz force 
analysis in this chapter.

\section{Forces acting on CMEs}
\subsection{Lorentz driving force} \label{1}
Initial CME triggering can be attributed to various mechanisms that lead to magnetic rearrangement, loss of equilibrium or some instability that 
causes the CME to erupt \citep{Che11}. In other words, CME eruption can be described as the disruption in the equilibrium of the magnetic field 
configuration \citep[\textit{e.g.}][]{Che89,Che93}.  
Comparisons between the data and theory related to the dynamics of expanding CME structure has shown that CMEs are driven by 
Lorentz forces ($J\times B$) from very early on in their evolution. In order to drive these magnetized structures, the Lorentz forces must involve misaligned currents and 
magnetic fields. This means that the flux-rope CMEs are required to be non force-free to be accelerated by Lorentz forces \citep{Sub14}. \citet{Man04,Man06,Sub07} have 
shown, using observational evidence that the magnetic energy carried by flux-rope CMEs in sufficient to drive them from the Sun to the Earth. 
However, the actual nature of these Lorentz forces is still unclear. While some authors include it in the initial eruption process 
\citep[\textit{e.g.}][]{Ise07,Kli14}, others assume that the Lorentz force time profile follows the temporal evolution of the soft X-ray 
flux \citep[\textit{e.g.}][]{Che10}. \\
Most Lorentz force prescriptions model the CME morphology like an expanding current carrying loop-like structure with toroidal and poloidal magnetic fields 
and corresponding current components \citep[\textit{e.g.,}][]{Che89,Gar94}). \citet{Olm13} describe the action of Lorentz force by considering the forces acting on elliptical current carrying loops.
In addition, \citet{Kum96} use the arguments of magnetic helicity conservation in their model which considers a current-core helical flux-rope CME and apply it to a 
torus-shaped cloud. Results of numerical simulations also support the driving action of Lorentz forces. These include MHD simulations of CME 
initiation which involves reconnection (magnetic breakout model)  \citep[\textit{e.g.}][]{Ant99,Moo01,Kar12}. Other MHD models that 
involve loss of equilibrium are given by \citet{For91, Ise93} and \citet{She12}.

CME eruption has also been explained by the kink instability \citep{Tor04} and the torus instability (TI) \citep{Kli06}. The torus instability models CMEs as 
current carrying rings that expand as they evolve. \citet{Kli06} consider the expansion instability of the ring in low-beta magnetic plasma in the presence of 
an overlying magnetic field and compare the properties with those of solar CMEs. For TI it is required for the overlying poloidal magnetic field to decrease
rapidly enough so that the CME can erupt. 
When this field falls sufficiently fast, an instability occurs and the CME launches outwards with a whiplash-like action.
In this study, we follow this prescription of Lorentz force as the major driving force acting on the CMEs. \\ 
The TI model does not include the effects of reconnection on the CME dynamics. \citet{Wel17} state that in order for the CME to erupt through the field overlying 
the magnetized structure, the \citet{Kli06} model must include the dynamic effects that occur due to the reconfiguration of field lines. We consider the original TI model as described in \citet{Kli06} for modeling the Lorentz forces acting on solar coronal mass ejections and include it in the 
force equation described in  Equation\ref{feq}.

\subsection{Aerodynamic drag force} \label{2}
As the CME propagates into the interplanetary space, it couples with the ambient solar wind via momentum transfer. The action of viscous forces on the CME boundary 
causes the CMEs to slow down or accelerate depending on their speed relative to the speed of the surrounding medium (solar wind) \citep{Bor09}. \citet{Man06,Mal09,Gop13} and 
others found that CMEs erupting with speeds lesser than the ambient solar wind speed were dragged up while those with speeds exceeding the solar wind speed were decelerated. \citet{Lew02} arrived at 
similar conclusions based on their study of an average ``representative'' CME using coronagraph observations. Various drag-based models that take into 
account aerodynamic drag force only due to interaction with the solar wind include \citet{Car04, Vrs10, Vrs13, Mis13} and \citet{Tem15}. 

Other dynamical CME models used for modeling the CME propagation include ENLIL, which is a three-dimensional (3D) magneto-hydrodynamic (MHD) model \citep[e.g.,][]{Ods99,Ods04,Tak09,Lee13,Vrs14,May15} and, 
the Global MHD model using data-driven Eruptive Event Generator Gibson-Low (EEGGL) \citep[e.g.][]{Jin17}. CME and Shock propagation Models like the Shock Time of Arrival Model (STOA),
Interplanetary Shock Propagation model (ISPM) and Hakamada-Akasofu-Fry version 2 model \citep[e.g.,][]{Fry03,McK06}, other hybrid models \citep{Wu07} and the Space Weather 
Modeling Framework \citep[SWMF;][]{Lug07,Tot07} are also often used to model CME propagation.

Most models that describe the momentum coupling between the CME and the solar wind use empirical drag parameters. \citet{Car04} uses 2.5D MHD simulation results 
to determine the drag coefficient, $C_{\rm D}$. \citet{Sub12, Sac15} describe a microphysical prescription for the collisionless solar wind viscosity in order to determine the 
drag coefficient ($C_{\rm D}$). The dimensionless drag coefficient is representative of the strength of the coupling interaction between the CME and the solar wind.
The aerodynamic drag force on CMEs is typically assumed to be proportional to the square of the CME velocity relative to the solar wind speed \citep[\textit{e.g.}][]{Car04,Sub12,Vrs14,Sac15}.
This formulation is used for high Reynolds number flows past a solid body.

\citet{Lan59} describe a scenario for a solid body moving through a fluid at high Reynolds number. The bulk flow (considered to be a potential flow) and 
the solid body are separated by a turbulent boundary layer. This description is relevant for CMEs, as they are over-pressured structures in a high Reynolds number 
flow. The ``solid body'' law requires that the tangential and normal velocities vanish at the boundary (which is thin, owing to large Reynolds number).
The turbulence in the boundary layer arises because the velocity transitions from its bulk value to zero on the boundary over a very short distance and  
nonlinear terms in the Navier-Stokes equation assume importance giving rise to turbulence. 

While some believe that the Lorentz forces play an important role in governing the CME dynamics only up to a few solar radii \citep[\textit{e.g.}][]{Vrs06}, others state that the solar wind drag 
is the dominant force affecting CMEs from very early on and model the CME trajectory using drag-based models \citep{Lew02,Car04}. \citet{Zha06} suggest 
that the CME initiation and initial acceleration takes place $<2$ \Rs. On the other hand, \citet{Byr10} find that drag dominates the CME dynamics beyond $7$ \Rs.
\citet{Tem11,Vrs13,Iju14,Rol16} mention that ICMEs equilibrate with the solar wind speeds within $\sim 0.5$ AU. In spite of significant progress in modeling the Sun-Earth 
propagation of CMEs, predictions of travel time and speeds of Earth-directed CMEs are still limited in their accuracy \citep{Zha14} even for relatively simple events that do not involve CME-CME interactions \citep[\textit{e.g.}][]{Tem12}.
Our understanding of how the Lorentz forces operate, at what heights and when they cease to be important is not very robust as well. \\
We now describe the equation for the net force acting on CMEs (section \ref{feq}) based on particular prescriptions of the two contributing forces.

\section{Force Equation}\label{feq}
The equation of motion for CMEs is given by:
\begin{equation}
F\,= m_{cme} \frac{d^{2}R}{dt^{2}} \nonumber
 \end{equation}
where $F$ is the total force, $m_{cme}$ is the CME mass, $R$ is the heliocentric distance of the leading edge of the CME and $t$ represents time. $F$ 
includes the Lorentz and aerodynamic drag forces (gravity is neglected),
 \begin{eqnarray}
 F\, &=&F_{Lorentz} + F_{drag} \nonumber \\
 &=& \biggl\{\biggl[\frac{\pi I^{2}}{c^{2}} \biggl(ln \biggl(\frac{8 R}{R_{cme}}\biggr)-\frac{3}{2}+ \frac{l_{i}}{2}\biggr)\biggr]\,-\, \frac{(\pi R)I B_{ext}(R)}{c} \biggr\} \nonumber \\
    &&-\,\,\frac{1}{2}\,\, C_{\rm D}\,\, A_{cme} \,\,n_{sw}\,\, m_{p}\, 
  \bigl( V_{cme}-V_{sw} \bigr)\,\,\bigl|V_{cme}-V_{sw}\bigr| \label{eqforce}
 \end{eqnarray}
$F_{Lorentz}$ and $F_{drag}$ denote the Lorentz and aerodynamic drag forces respectively (in cgs units).
$F_{Lorentz}$ is the net Lorentz force ($(1/c) J\times B$) that acts on a CME in the direction of the major radius. It includes the two terms 
in the curly brackets \citep[see, \textit{e.g.}][]{Sha66,Kli06}. The first term denotes the Lorentz self-forces acting on the expanding CME in the radially outward direction 
while the second term is the inward force due to an external poloidal magnetic field ($B_{ext}$) that holds down the expanding CME. CME current is represented by $I$, 
$R_{cme}$ is the minor radius of the CME, c is the speed of light and $l_{i}$ is the internal inductance.\\
The term involving $C_{\rm D}$ in the force equation represents the drag force acting on the CME ($F_{drag}$). The drag coefficient $C_{\rm D}$ denotes the strength of the 
interaction between the CME and the ambient solar wind \citep{Sub12,Sac15}. The negative sign ensures that the solar wind can ``pull up'' a CME 
(if $V_{cme}<V_{sw}$) and also ``drag it down'' (if $V_{cme}>V_{sw}$). $A_{cme}$ 
is the cross-sectional area of the CME, $n_{sw}$ is the solar wind density, $m_{p}$ is the proton mass, $V_{cme}$ and $V_{sw}$ denote the CME and 
solar wind speeds respectively. Non-radial components of forces have not been considered in this model \citep{Byr10}.

Equation \ref{eqforce} thus describes a full force equation in terms of various CME parameters. 
Models used for the calculation of each of these parameters are described in detail in the sections \ref{secdrag} and \ref{seclsf}.
We note that most physical quantities used in this analysis are observationally derived using the Graduated 
Cylindrical Shell (GCS) fitting procedure described in Chapter \ref{chap2}. 

\section{Solar wind Drag Model} \label{secdrag}
The interaction between ambient solar wind and CMEs is given by the following equation (see, Equation \ref{eqforce}):
\begin{equation}
 F_{drag}\,=\,m_{cme}\,\frac{dV_{cme}}{dt}\,=-\,\,\frac{1}{2}\,\, C_{\rm D}\,\, A_{cme} \,\,n_{sw}\,\, m_{p}\, 
 \biggl( V_{cme}-V_{sw} \biggr)\,\,\biggl|V_{cme}-V_{sw}\biggr| \, ,  \label{eq1}
\end{equation}
$F_{drag}$ is the aerodynamic drag force due to momentum coupling between the CME and solar wind, $m_{cme}$ is CME mass,
$V_{cme}$ is CME speed, $C_{\rm D}$ is the dimensionless drag coefficient, $A_{cme}$ is CME cross-sectional area, $n_{sw}$ is the solar wind proton density, 
$m_{p}$ denotes the proton mass and $V_{sw}$ is the ambient solar wind speed.
The negative sign ensures that the solar wind can be both driving as well as a decelerating force. A fast CME propagating into a relatively 
slower solar wind is ``dragged down'' by it while a slow CME is ``picked up'' by a faster solar wind. The drag acceleration is considered 
to have a quadratic dependence on the CME-solar wind relative speed in accordance with the law for solid bodies moving at high Reynolds numbers \citep{Lan59}.

\citet{Car04,Vrs10,Vrs13} prescribe the drag-based model in terms of the $\gamma$ parameter described below:
\begin{equation}
 \frac{F_{drag}}{m_{cme}} \equiv  a_{d} =\,-\,\gamma \,\bigr(V_{cme}-V_{sw}\bigr) \bigl|V_{cme}-V_{sw}\bigr| \, , \label{eq5}
\end{equation}
where, $a_{d}$ is the acceleration and $\gamma \, ({\rm cm}^{-1})$ is given by
\begin{equation}
\gamma\,\,=\,\,C_{\rm D} \frac{n_{sw} m_{p} A_{cme} }{m_{cme}}  \, ,
\label{eq6}
\end{equation}
which depends on the solar wind density ($n_{sw}$), cross-sectional area ($A_{sme}$) and the CME mass ($m_{cme}$). \citet{Vrs13} use this analytical equation to explicitly solve for the Sun-Earth transit 
time and speed of CMEs with the assumption that $\gamma(r)=$ constant.
In the following sections, we describe the models used for each of the parameters that appear in the drag equation (Equation \ref{eq1}) and their calculation.

\subsection{Drag Coefficient $C_{\rm D}$}
The strength of the momentum coupling between a CME and the ambient solar wind is represented by the drag parameter $C_{\rm D}$. Most drag-based models consider an 
empirical drag coefficient. However, we consider a microphysical prescription for $C_{\rm D}$ using collisionless solar wind viscosity as described in 
\citet{Sub12} and \citet{Sac15}.
Let $\nu_{sw}$ be the viscosity in the ambient collisionless solar wind, which arises from resonant scattering of the solar wind protons 
with the turbulent Alfv\'en wave spectrum. It can be described with an expression for turbulent viscosity \citep[\textit{e.g.,}][]{Ver96} given by :
\begin{equation}
\nu_{sw} = \sqrt{6}\,\frac{2}{15}\,v_{\rm rms}\,\lambda\, ,
\label{eqvisc}
\end{equation}
where $v_{\rm rms}\equiv (3kT_{i}/m_{p})^{1/2}$ is the rms speed of solar wind protons and $\lambda$ is the mean free path. The ion inertial length is taken to be the mean free path here 
(\citealp{Lea00}; \citealp{Smi01}; \citealp{Bru14}) and is given by :
\begin{equation}
\lambda\,\,=\,\,\frac{v_{a}}{\Omega_{i}} \,\,=\,\,\frac{c}{\omega_{p}} \,\,\sim\,\, 228\,\,n_{sw}^{-1/2}\,\,k\,m \, ,
\label{eqioninertial}
\end{equation}
where, $v_{a}\,$ is the Alfv\'{e}n speed, $\,\Omega_{i}\,$ is  the ion cyclotron frequency, c is the light speed, $\,\omega_{p}\,$ is the ion 
plasma frequency and $n_{sw}$ is the ambient solar wind density.
\citet{Col89} use a similar prescription for the mean free path which is also used in \citet{Sub12}. However, their prescription is a factor of 3 larger 
than the ion inertial length given by Equation \ref{eqioninertial}.
%
The Reynolds number ({\it Re}) is a dimensionless ratio of the inertial and viscous forces in a fluid. For CMEs, it is defined by:
\begin{equation}
Re \equiv \frac{(V_{cme} - V_{sw})\,R_{cme}}{\nu_{sw}}\,=\, \frac{(V_{cme} - V_{sw})\,R_{cme}\, n_{sw} \,m_{p}}{\eta_{hyb}} ,
\label{eqrey}
\end{equation}
where the quantity $\nu_{sw}$ ($\rm cm^{2}\, s^{-1}$) is the solar wind viscosity and $\eta_{hyb}\equiv n_{sw}\,m_{p}\,\nu_{sw}$. 
The dimensionless coefficient of drag, $C_{\rm D}$ is expressed as a function of the Reynolds number using a fit to the data from \citet{Ach72} for the drag 
on a sphere at high Reynolds numbers in the super-critical regime:
\begin{equation}
C_{\rm D} = 0.148 - 4.3 \times 10^4 \, Re^{-1} + 9.8 \times 10^{-9} \, Re \, 
\label{eqcdfit}
\end{equation}

The above definition for $C_{\rm D}$ was experimentally determined for subsonic, high Reynolds number flow past a solid metal 
sphere (\citealp{Ach72}). The drag force acting on CMEs described by Equation \ref{eq1} appeals to a high Reynolds number, solid body law and the 
prescription for drag coefficient in Equation \ref{eqcdfit} is consistent with these assumptions.
\citet{Rus05} and \citet{Jia06} show that CMEs are overpressured structures that do not deform in response to tangential stresses. 
That is, the total (magnetic + particle) pressure exhibits a substantial jump across a typical magnetic cloud boundary, suggestive of a solid-body like behavior.
The drag formula given by \citet{Ach72} has also been verified by modern detached-eddy simulations of high Reynolds number flows in the super-critical 
regime (e.g., \citealp{Con04}), which is the regime we are interested in.
Although Equation \ref{eqcdfit} is derived experimentally for subsonic flows, we note that the CME motion through the solar wind is supersonic. 
In order to justify the applicability of this definition in the subsonic regime, we appeal to the Morokovin hypothesis, which has been verified extensively via 
numerical simulations \citep[\textit{e.g.,}][]{Dua11}. According to this hypothesis, a subsonic turbulent drag law is valid even 
for supersonic flows as long as the fluctuations in the turbulent boundary layer are incompressible, or subsonic.
To a very good approximation, the turbulent fluctuations in the ambient solar wind are certainly incompressible \citep[\textit{e.g.,}][]{Sha10}, 
therefore, the drag parameter prescription holds for CMEs as well. 
This is also evident from the small values of the density fluctuations $\Delta n/n$ in the ambient solar wind 
\citep[\textit{e.g.,}][]{Bis14,Sas16}. \citet{Aru13} show that the fluctuations in the magnetic field ($\Delta B/B$) in the turbulent sheath region, 
between the shock and the CME are also as small as 10\%. Since magnetic field fluctuations can be considered to be a proxy for the 
$\Delta n/n$ values \citep[\textit{e.g.,}][]{Spa02}, it follows that the turbulent density fluctuations in the sheath region are also fairly incompressible. 
Therefore, the drag coefficient model as described here is quite relevant to describe the momentum coupling between CMEs and the solar wind.
For completeness, in addition to the $C_{\rm D}$ prescription described above, we also use (empirical) constant $C_{\rm D}$ values lying within 
and outside the range of values predicted by this model.
\subsection{CME Cross-sectional Area}
The CME area ($A_{cme}$) in Equation \ref{eq1} is the area of the elliptical cross-section in the \textit{x-z} plane 
of the flux-rope CME and is calculated using,
\begin{equation}
A_{cme}= \pi \, R_{cme}\,W_{cme} 
\label{eqarea}
\end{equation}
The quantity $R_{cme}$ is the radius of the circular cross-section of the CME as seen edge-on (also called the minor radius). $W_{cme}$ is the maximum value of the 
face-on half-width of the CME, or the major radius of the elliptical 
cross-section (Figure \ref{gcs1}). Both these quantities are derived using the height of the leading edge $R$, half angle $\alpha$ and aspect 
ratio $\kappa$ from the GCS fitting procedure \citep{The11}. The CME cross-sectional area is thus calculated using observationally fitted parameters for each observed height/time.

\subsection{Models for Solar wind parameters}
The solar wind proton number density is a function of the observed GCS height of the 
leading edge, $R$. It is calculated using a modified version of the LeBlanc electron density model. The original model of \citet{Leb98} 
(Equation \ref{eq2}) considers the number density at 1 AU to be 7.2 $cm^{-3}$; and is given by:
\begin{equation}
 n(R)=\, \biggl [ 3.3 \times 10^{5} R^{-2}+4.1 \times 10^{6} R^{-4}+8 \times 10^{7} R^{-6} \, \biggr ]
 \,\,\,\, {\rm cm}^{-3} \, ,\label{eq2}
\end{equation}
However, we take the proton density at 1 AU ($n_{wind}$) to be that observed \textit{in-situ} about one-two days before the 
arrival of the CME and its associated shock (if there is a shock). The density model we use is thus given by,
\begin{equation}
 n_{sw}(R)=\biggl ( \frac{n_{wind}}{7.2} \biggr ) \, \biggl [ 3.3 \times 10^{5} R^{-2}+4.1 \times 10^{6} R^{-4}+8 \times 10^{7} R^{-6} \, \biggr ]
 \,\,\,\, {\rm cm}^{-3} \, \label{eqn}
\end{equation}
The quantity $n_{sw}(R)$ denotes the density of the ambient solar wind into which a CME propagates.
The correction factor, $n_{wind}/7.2$ ensures that the modeled density at 1 AU from Equation \ref{eqn} is same as the proton number density 
observed {\it in-situ} by the {\it WIND} spacecraft near the Earth. The modified LeBlanc model is used to extrapolate the measured 1 AU solar wind 
proton number density Sunwards. The quantity $n_{wind}$ for all CME events in the sample is given in Table \ref{tbl41}.

Another important parameter that affects the CME propagation significantly is the solar wind speed \citep[\textit{e.g.,}][]{Tem11}.
Denoted by $V_{sw}$ in the drag model (Equation \ref{eq1}), the solar wind speed is modeled as a function of the heliocentric distance ($R$) using the 
prescription given by \citet{She97,She99}:
\begin{equation}
 V^{2}_{sw}(R)\, = \,v^{2}_{wind}\,\, \bigl[1\,-\,e^{-\frac{(R\,-\,r_{0})}{r_{a}}}\bigr]\, \, ,\label{eq3}
\end{equation}
where $v_{wind}$ is the near-Earth solar wind speed observed {\it in-situ} by the {\it WIND} spacecraft (Table \ref{tbl41}). Like $n_{wind}$, the 
quantity $v_{wind}$ is also observed approximately one-two days in advance of the CME arrival, which essentially describes the solar wind environment that the 
CME will propagate into. In the solar wind speed model, $r_{0}$ denotes the heliocentric distance where the solar wind speed is zero and
is taken to be 1.5 R$_{\odot}$ in this work. The quantity $r_{a}$, is the e-folding distance over which the asymptotic speed $v_{wind}$ is reached.
$r_{a}$ is taken to be 50 $R_{\odot}$ for all CMEs.
Both the quantities, $n_{sw}$ and $V_{sw}$ are therefore, observationally derived using the GCS parameters as well as {\it in-situ} observations.

\subsection{Calculation of CME Mass}
The CME structure is visible in the white-light images observed by coronagraphs as a projection on the Plane of the Sky (POS), which is the plane perpendicular to the 
Sun-spacecraft line. Quantities like height, width, brightness, speed, mass etc. are derived from these projected CME structures on the POS. This leads to a slight 
discrepancy between the actual and observed quantities due to projection effects.

With multi-spacecraft observations from different positions, methods to determine the actual CME parameters (correcting for projection effects) have been developed.
The STEREO COR2 A \& B coronagraphs provide two projected images of an event erupting from the Sun along the Line of Sight (LOS). 
Using STEREO A/B and LASCO observations it is possible to obtain the actual CME parameters and minimize projection effects.
The observed CME intensities, once calibrated to units of mean-solar brightness give two values of the projected CME mass, one from each STEREO COR2 
spacecraft (A \& B). The mass values of the same observed CME structure from both instruments are generally slightly different. The ``true'' CME mass is the mass which 
is corrected for the projection effects due to the difference in the LOS of each instrument. 
We use the method outlined in \cite{Col09} to calculate the ``true'' CME mass for all the events in our sample. For each CME, we have the projected mass values 
from COR2 A \& B coronagraphs. In the method briefly described here, it is required that the total CME mass must be equal for both these coronagraphs.
\citet{Col09} follow the Thomson scattering calculation outlined in \citet{Bil66} and given by:
\begin{equation}
 m\,=\,\frac{B_{obs}}{B_{e}(\theta)}\times 1.97 \times 10^{-24}\,\,\,\,\,\, gm, 
 \label{eq_bil}
\end{equation} 
where, m is the mass at each pixel, $B_{obs}$ is the observed brightness and $B_{e} (\theta)$ is the brightness of a single electron at an angular distance $\theta$. 
$\theta$ is the angle along the LOS away from the POS. This method of mass estimation has been used by \citet{Vou00,Sub07,Col09}.

From the GCS fitting procedure, and using the $make\_cme\_mass\_fits.pro$ IDL routine we obtain the derived CME mass simultaneously for each view-point (COR2 A/B). In 
principle, since the two instruments observe the same volume of material from different angles, the mass observed in each should be the same. If this is not the case, 
the error must lie in the usage of an incorrect angle in the Thomson Scattering calculation (Equation \ref{eq_bil}).
From \citet{Col09},
\begin{equation}
 \frac{M_{A}}{M_{B}}\,=\, \frac{f_{m}(\theta + \Delta_{sc}/2) M_{T}}{f_{m}(\theta - \Delta_{sc}/2) M_{T}} \,\, , \label{eq_mass}
\end{equation}
where, $M_{A}$ and $M_{B}$ are observed masses from COR2 A and B respectively and,
\begin{equation}
f_{m}(\theta)\,=\, \frac{B_{e}(\theta)}{B_{e}(\theta=0^{\circ})} \,\, ,
\label{eqfm}
\end{equation}
is the ratio of brightness of an electron at $\theta$ and on the POS ($\theta=0^{\circ}$). $\Delta_{sc}$  is the angular separation of the two spacecrafts. 
We estimate the $\theta$ for which Equation \ref{eq_mass} holds true. For this value of $\theta$ we calculate,
\begin{equation}
M_{A}-M_{B}\,=\,M_{T}\,[f_{m}(\theta + \Delta_{sc}/2)\,-\,f_{m}(\theta - \Delta_{sc}/2)],
\label{eqmassdiff}
\end{equation}
at each height for each CME in the COR2 FOV. The quantity $M_{T}$ is the true CME mass. It increases with height \citep{Vou00}, and beyond the COR2 FOV ($\sim 15$ \Rs) it 
is taken to be constant and equal to the last observed value in the COR2 FOV.
The true mass ($M_{T}$) for each CME is calculated at each GCS observed height using this technique. 
We also include the Virtual Mass ($M_{v}$) \citep[\textit{e.g.,}][]{Car96,Car04} in our calculations (Refer Appendix \ref{Mv}).

Total CME mass, $m_{cme} = M_{T}+\, M_{v}$, is written as,
\begin{equation}
 m_{cme}\,=\, M_{T} \biggl[1\,+\, \frac{1}{2}\frac{n_{sw}(R) \,m_{p}}{M_{T}} A_{cme} R \biggl]
\end{equation}
For all CMEs, $m_{cme}$ at the last observed height ($h_{COR}$) in the COR2 FOV, is given in Table \ref{tbl41}.

The above section describes models used for evaluating each parameter in the drag force equation using various observed parameters (GCS and {\it in-situ}). In 
order to calculate the speed of the CME, we solve Equation \ref{eq1} using these prescriptions. The details are described in the sections below.
 
\subsection{Solving for $V_{cme}$}
The drag force Equation \ref{eq1} is a simple 1D model which describes the momentum coupling between CMEs and solar wind and is prescribed in terms of parameters derived using 
observations (wherever applicable). To calculate the CME velocity profile ($V_{cme}$ as a function of time) we require an estimate for the CME 
initial speed ($v_{0}$) at the first observed height ($h_{0}$). This initial velocity ($v_{0}$ at $h_{0}$) is calculated by fitting a 3$^{rd}$ degree polynomial to 
the GCS fitted height-time profile for each CME in the event sample and is used as the initial condition when the differential equation for 
$V_{cme}$ is solved from $h_{0}$ onwards (Table \ref{tbl32}). For each event, the drag-only model (Equation \ref{eq1}) is initiated (a) from the first data point ($h_{0}$) and (b) at progressively larger heights. At each height, the corresponding parameters at that height are used in the model. The solutions for $V_{cme}$ from Equation 
\ref{eq1} are integrated to obtain a height-time profile predicted by this model, that includes only the solar wind drag force. We then compare this modeled height-time 
solution with the observed CME profile obtained from the GCS fitting technique described earlier in Chapter \ref{chap2}. 
The initiation height at which the model and observed height-time profiles show a good agreement is denoted by $\widetilde{h}_{0}$. 
The magnitude of the aerodynamic drag force above $\widetilde{h}_{0}$ is computed from the force Equation \ref{eq1} using the corresponding parameters and model solutions. 
Results from this analysis are described in detail in Chapter \ref{chap5} for all the events. 

\section{The TI Lorentz Force Model}\label{seclsf}
The Lorentz force model based on the torus instability (TI) (section\ref{1}) is described in terms of two competing ($J\times B$) factors. One is 
due to the Lorentz self-forces acting radially outwards on the flux-rope and the other is due to the overlying external poloidal field that tends to hold down the CME. The 
total Lorentz force acting on an expanding CME can be expressed in terms of the CME current $I$, the height of the leading edge $R$ and the external poloidal 
magnetic field $B_{ext}$($\propto R^{-n}$) which is required to fall sufficiently fast for the CME to escape \citep[see, \textit{e.g.}][]{Sha66,Kli06}:
\begin{equation}
F_{Lorentz} \,=\,\frac{\pi I^{2}}{c^{2}} \biggl(ln \biggl(\frac{8 R}{R_{cme}}\biggr)-\frac{3}{2}+ \frac{l_{i}}{2}\biggr)\,-\, \frac{(\pi R)I B_{ext}(R)}{c} \, ,
\label{eqlsf}
 \end{equation}
where c is the speed of light, $R_{cme}$ is CME minor radius and $l_{i}$ is the internal inductance which is taken to be $l_{i}=1/2$. The ratio $R/R_{cme}$ is the 
aspect ratio determined by GCS derived parameters. The decay index $n$ needs to be above a critical value $n_{cr}$, for the torus instability to be operative, 
ensuring CME eruption \citep{Kli06}. For each CME, we choose a value of $n$ that is $> n_{cr}$. Both these quantities are listed in Table \ref{tbl41} for each event.
In the sections below, we describe how the CME current (and hence, the Lorentz force) is calculated as described in 
\citet{Sac17}. We note that the CME minor radius ($R_{cme}$) is referenced as ``$b$'' in \citet{Sac17}.

\subsection{CME current} \label{subI}
The CME axial current, $I$ is determined by the conservation of total (i.e. flux-rope + external) magnetic flux 
enclosed by a current carrying ring which is given by \citep{Kli06}:
\begin{equation}
\Psi_{total}\,=\,\Psi_{int}+\Psi_{ext} \,\, ,\nonumber 
\end{equation}
where, $\Psi_{int}$ and $\Psi_{ext}$ are internal (flux-rope) and external magnetic flux given by (in cgs units):
\begin{eqnarray}
\Psi_{int}&=&c L \,I \,\,  \nonumber  \\
\Psi_{ext}\,&=&\, -\int B_{ext}.da = \,- 2 \pi \int_{0}^{R}\, B_{ext}(r)\, r \,dr  \, \, ,\nonumber
\label{eqflux1}
\end{eqnarray}
where, L is the inductance.

Conservation of total magnetic flux requires, 
 \begin{eqnarray}
\Psi_{total}\,(h_{eq})\,&=&\,\Psi_{total}\,(R)\,\nonumber \\
\Rightarrow \Psi_{int}(h_{eq})- 2 \pi \int_{0}^{h_{eq}} B_{ext}(r) r dr &=&\Psi_{int}(R) -  2 \pi \int_{0}^{R} B_{ext}(r) r dr  \nonumber\\
\Rightarrow L_{eq}I_{eq}-\frac{2 \pi}{c} \int_{0}^{h_{eq}} B_{ext}(r) r dr&=& L(R)\,I(R)- \frac{2 \pi}{c} \int_{0}^{R} B_{ext}(r) r dr \nonumber\\
\label{eqflux}
 \end{eqnarray}
The quantities with subscript ``eq'' represent values at the pre-eruption equilibrium position ($h_{eq}$), 
where the total Lorentz force acting on a CME is zero (\textit{i.e.} from Equation \ref{eqlsf}, $F_{Lorentz}=0$). As described by \citet{Che89}, the inductance L is given by:
\begin{equation}
L\,=\, \frac{4 \pi R}{c^{2}}\bigl[ln \bigl(\frac{8R}{R_{cme}}\bigr)-2+l_{i}/2\bigr]
\end{equation}
When substituted in Equation \ref{eqflux}, it gives the axial CME current carried by the CME, 
\begin{equation}
  I\,=\,\frac{c^{'}_{eq} I_{eq} h_{eq}}{c^{'} R}\biggl(1+\frac{(c^{'}_{eq}+\frac{1}{2})}{2 c^{'}_{eq}(2-n)}\biggl
  [\biggl(\frac{R}{h_{eq}}\biggr)^{2-n}-1\,\biggr]\biggr) \, ,
\label{eqI}
\end{equation}

where,
\begin{eqnarray}
c^{'}(R)\,&=&\,\bigl[ln (8R/R_{cme})-2+l_{i}/2\bigr]   \nonumber\\
c^{'}_{eq}\,&=&\,c^{'}(R=h_{eq})=\bigl[ln (8h_{eq}/R_{cme}(h_{eq}))-2+l_{i}/2\bigr] \nonumber
\end{eqnarray}
The equilibrium current $I_{eq}$ and external field at equilibrium ($B_{ext}(h_{eq})$) are related via,
\begin{equation}
  I_{eq}\,=\,\frac{B_{\rm ext}(h_{eq})h_{eq} c}{c^{'}_{eq}+\frac{1}{2}}.
  \label{eqIe}
\end{equation}
In this study, the equilibrium position, $h_{eq}$ is taken to be equal to $1.05$ \Rs. $n> n_{cr}=3/2-1/(4c^{'}_{eq})$ is chosen to be such that, $|F_{drag}|>F_{Lorentz}$ for
$R > \widetilde{h}_{0}$ (see, Appendix \ref{decayn} for details). The CME current is evaluated using observationally derived CME parameters determined from the GCS fit (R and $R_{cme}$). The details are described in Appendix \ref{LF}.  
 
\subsection{Calculating the CME current $I$}
The external poloidal field, $B_{ext}\, (\propto\, R^{-n})$ is related to the equilibrium current by Equation \ref{eqIe}. In order to determine the CME current $I$, either $I_{eq}$ or 
$B_{ext}(h_{eq})$ is required. For a given value of $n$, the external field at equilibrium 
($B_{ext}(h_{eq})$) is determined by equating the solar wind drag and Lorentz forces at the height $\widetilde{h}_{0}$. 
Basically, it is required that $|F_{drag}(\widetilde{h}_{0})|=F_{Lorentz}(\widetilde{h}_{0})$. 
This constrains the equilibrium current $I_{eq}$.

Using Equations \ref{eqI}, \ref{eqIe} and $B_{ext}(R)=B_{ext}(h_{eq})(R/h_{eq})^{-n}$ in Equation \ref{eqlsf}, the Lorentz force acting on a CME 
can be calculated (in dynes). With the external field, $B_{ext}(h_{eq})$ at equilibrium (and hence, the current $I_{eq}$ at $h_{eq}$), we can 
calculate the CME current ($I$) values at each 
observed height. The Lorentz force magnitude can then be calculated using the current ($I$) estimates and the ratio $R/R_{cme}$ 
from the GCS fittings for all heights starting from the first observed height $h_{0}$. Between $h_{eq}$ and $h_{0}$ due to FOV constraints of 
the coronagraphs, we do not have the GCS observations of the CME. In this range, we assume that the CME expands self-similarly. For $R<h_{0}$, we assume 
$R/R_{cme}$ to be the same as its value at $h_{0}$ and calculate the Lorentz force magnitude. The expression used to determine the Lorentz force magnitude is described 
in detail in the Appendix \ref{LF}. 

\subsection{Lorentz force profile}
\begin{figure}[h]
\centering
\includegraphics[height=0.23\paperheight,width=0.5\paperwidth]{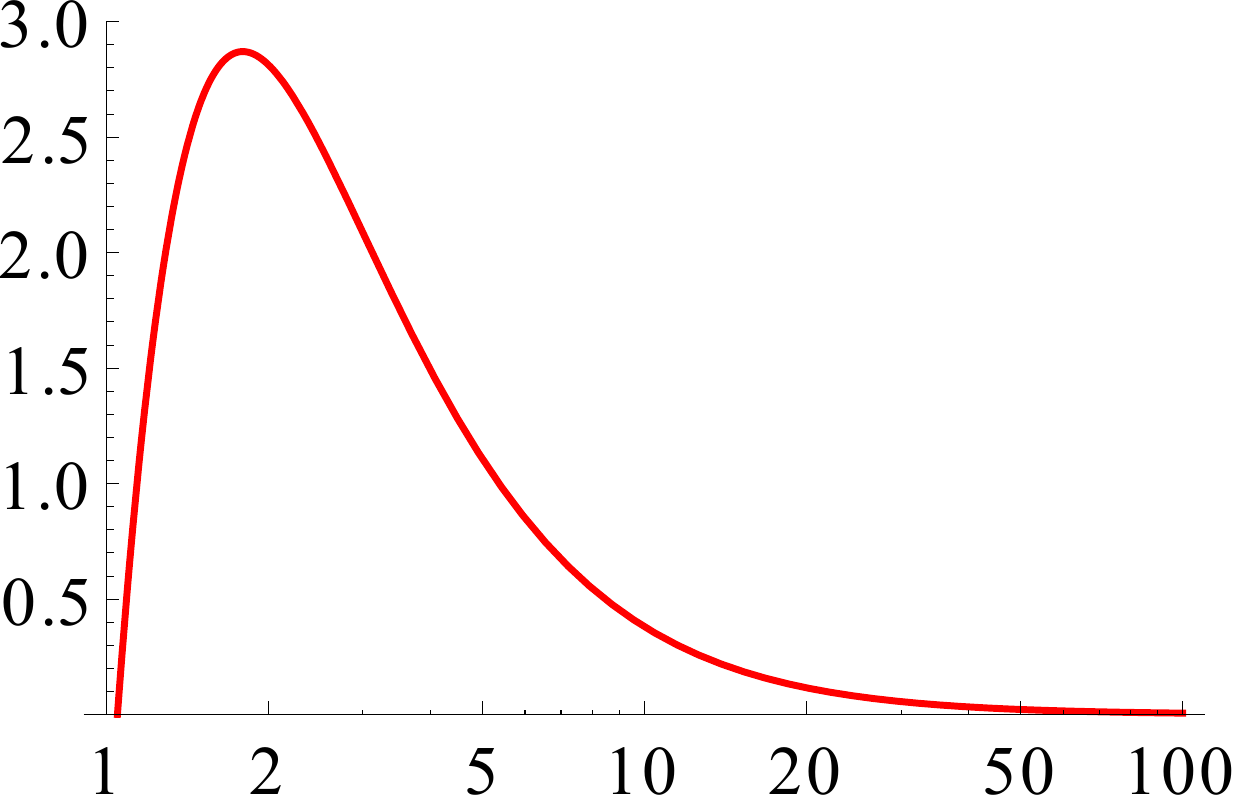}
  \put(-325,30.7){{\rotatebox{90}{{\color{black}\fontsize{11}{11}\fontseries{n}\fontfamily{phv}\selectfont   Lorentz Force (units of $10^{17}$ dynes)}}}}
               \put(-205.8,-8.9){{\rotatebox{0}{{\color{black}\fontsize{11}{11}\fontseries{n}\fontfamily{phv}\selectfont  Heliocentric distance ($R$) (\Rs)}}}}
\caption[Lorentz force profile]{Lorentz force profile with respect to heliocentric distance for a set of parameters given by: $n=2.5$, $B_{ext}(h_{eq})=0.013 \,G$ and 
$R/R_{cme}=4.56$. The force increases from $h_{eq}=1.05$ \Rs to peak at $1.75$ \Rs and then decreases. X-axis is plotted on a logarithmic scale for clarity.}
\label{lsfprofile} 
\end{figure}

The Lorentz force profile shows a steep increase from the equilibrium position ($h_{eq}$), and peaks at $h_{peak}$, beyond which 
it decreases gradually. An example is shown in Figure \ref{lsfprofile} for CME 1 (Table \ref{tbl31}). The two terms in Equation \ref{eqlsf} add up to give this 
resultant observed Lorentz force profile. 

\begin{figure}[H]
\centering
\includegraphics[height=0.23\paperheight,width=0.5\paperwidth]{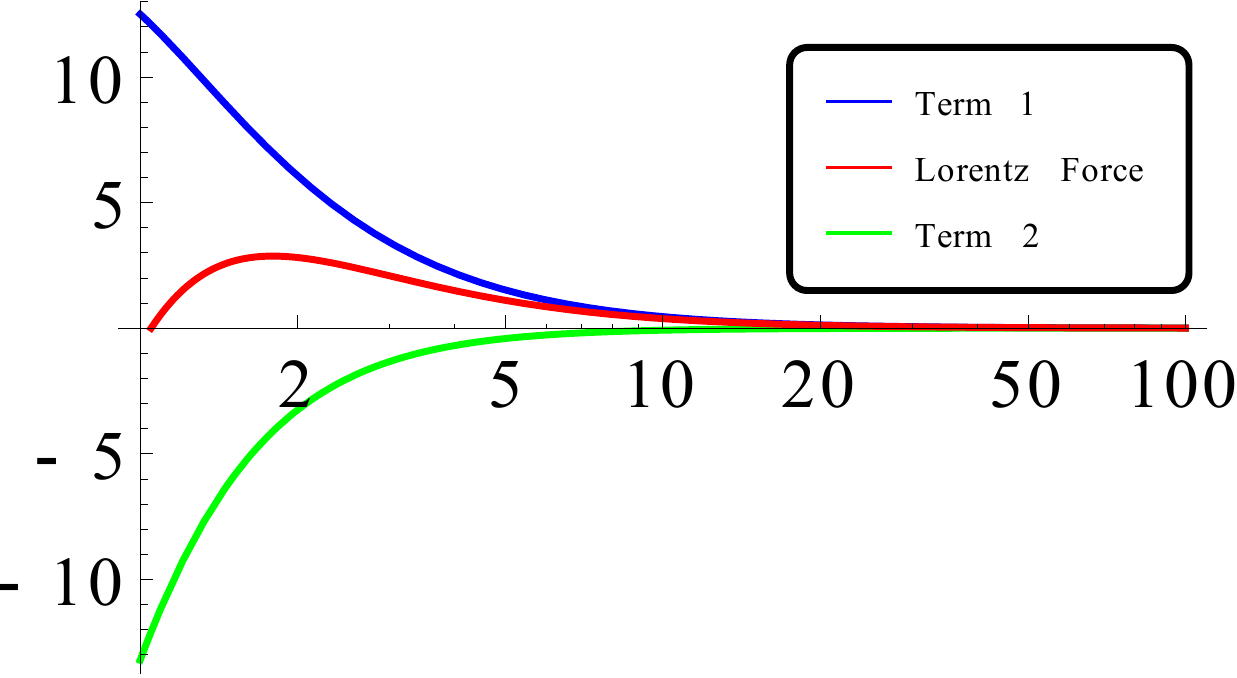}
  \put(-320,42.7){{\rotatebox{90}{{\color{black}\fontsize{12}{12}\fontseries{n}\fontfamily{phv}\selectfont  Force (units of $10^{17}$ dynes)}}}}
               \put(-205.8,-8.9){{\rotatebox{0}{{\color{black}\fontsize{12}{12}\fontseries{n}\fontfamily{phv}\selectfont  Heliocentric distance ($R$) (\Rs)}}}}
\caption[Individual $J \times B$ terms in Lorentz force equation]{Individual terms in the Lorentz force equation (Equation \ref{eqlsf}) and the 
resultant force profile with respect to heliocentric distance (R) for a 
set of parameters given by: $n=2.5$, $B_{ext}(h_{eq})=0.013 \,G$ and 
$R/R_{cme}=4.56$ are shown here. First term represents the Lorentz self-force (blue) and the second term is the force due to the 
external poloidal field (green). The resultant Lorentz force profile is shown in red color.
X-axis is plotted on a logarithmic scale for clarity.}
\label{all} 
\end{figure}

We rewrite the terms in Lorentz force equation (Equation \ref{eqlsf}) individually:
\begin{eqnarray}
Term\,1 \,&=&\,\frac{\pi I^{2}}{c^{2}} \biggl(ln \biggl(\frac{8 R}{R_{cme}}\biggr)-\frac{3}{2}+ \frac{l_{i}}{2}\biggr)\, \\
Term\, 2 &=&\,-\, \frac{(\pi R)I B_{ext}(R)}{c} \, \\
F_{Lorentz}&=& Term\,1\, +\,Term\,2 \nonumber
\label{lsfterms}
\end{eqnarray}
Figure \ref{all} plots Term 1 (Lorentz self-force) (blue) and Term 2 (force due to external field) (green) \textit{versus} the heliocentric distance ($R$), 
for a set of parameter values. When added together they result in the Lorentz force profile (red). The first term (Term 1) dominates over Term 2.
The peak denotes the height where,
\begin{equation}
\frac{d\, F_{Lorentz}}{dR} \,=\,0 \,\,\,\, at \,\,\,\,\,\,R=h_{peak} \nonumber
\end{equation}	
and, 
\begin{eqnarray}
\frac{d\, F_{Lorentz}}{dR} \,&>&\,0 \,\,\,\, for \,\,\,\,\,\,R<h_{peak} \nonumber \\
\frac{d\, F_{Lorentz}}{dR} \,&<&\,0 \,\,\,\, for \,\,\,\,\,\,R>h_{peak} \nonumber
\end{eqnarray}
$h_{peak}$ is therefore, the height where the Lorentz force has the maximum magnitude, beyond which it decreases.

The model framework for the two major forces governing CME propagation described in the preceding sections is used to determine the heliocentric distances 
at which they dominate CME dynamics. Results of solar wind drag and Lorentz force analysis are described in Chapter \ref{chap5}.

\begin{table}
\caption[Solar wind parameters, CME mass and decay index]{Column 1 represents the CME no. with reference to Table \ref{tbl31}. 
Columns 2 and 3 denote the \textit {in-situ} solar wind proton density and speed observed at 1 AU, one-two days in advance of the CME arrival.
Column 4 is the final observed height ($h_{COR}$) in the COR2 FOV and column 5 indicates the mass at this height in units of $10^{15}$ gms.
Columns 6 and 7 list the critical decay index ($n_{cr}\,=\, 3/2 -1/4c^{'}_{eq}$) and the decay index ($n$) chosen for each CME, respectively.}
\label{tbl41}
\centering
\begin{tabular}{lcccccc} 
  \hline
 No.& $n_{wind}$ & $v_{wind}$& COR2 final height($h_{COR}$) & Mass at $h_{COR}$ &$n_{cr}$ & $n$  \\
    & ($cm^{-3}$)&(km$s^{-1}$)& (\Rs) & (10$^{15}$gm)& & \\
  \hline
  1$^{*}$   & 3.60 & 380 & 15.43 & 2.90  & 1.36  & 2.5 \\
  2$^{* f}$ & 7.10 & 470 & 14.28 & 5.85 & 1.35  & 1.6 \\
  3$^{*}$   & 3.60 & 440 & 12.57 & 7.30 & 1.38 & 1.9\\
  4$^{*}$   & 3.50 & 500 & 15.86 & 2.30 & 1.37  & 2.5 \\
  5$^{*}$   & 4.00 & 320 & 15.93 & 7.65 & 1.34  & 1.6 \\
  6$^{*}$   & 3.80 & 350 & 15.43 & 7.14 & 1.37  & 1.7 \\
  7         & 6.10 & 321 & 17.60 & 4.15 & 1.34  & 1.6 \\
  8         & 9.00 & 320 & 15.29 & 7.86 & 1.36  & 1.6 \\
  9$^{*}$   & 2.50 & 440 & 16.36 & 5.35 & 1.33  & 2.1\\
  10        & 2.25 & 550 & 14.07 & 3.13 & 1.35  & 2.5 \\
  11$^{*}$  & 3.00 & 360 & 16.71 & 4.72 & 1.38  & 1.9\\
  12        & 5.00 & 375 & 14.52 & 5.61 & 1.36  & 2.5 \\
  13        & 3.70 & 455 & 18.52 & 10.30 & 1.37  & 1.6\\
  14$^{f}$  & 8.00 & 470 & 16.86 & 4.98 & 1.33 & 1.6 \\
  15$^{f}$  & 7.50 & 445 & 13.94 & 3.17 & 1.35 & 1.9 \\
  16$^{f}$  & 2.00 & 355 & 16.01 & 6.17 & 1.29 & 1.6 \\
  17        & 2.13 & 468 & 14.89 & 4.87 & 1.33 & 1.7 \\
  18$^{f}$  & 8.00 & 300 & 18.22 & 1.31 & 1.31 & 2.1 \\
  19        & 3.00 & 260 & 16.37 & 3.61 & 1.33 & 2.1\\
  20        & 8.42 & 411 & 14.22 & 3.01 & 1.35 & 2.2 \\
  21$^{f}$  & 7.00 & 310 & 18.22 & 9.17 & 1.33 & 3.0\\
  22$^{f}$  & 6.00 & 416 & 13.92 & 14.76 & 1.32 & 3.0\\
  23$^{f}$  & 4.00 & 420 & 17.60 & 12.45 & 1.34 & 3.0\\
  24$^{f}$  & 1.00 & 533 & 15.79 & 10.04 & 1.29 & 1.9 \\
  25        & 10.00& 325 & 16.37 & 5.93 & 1.37 & 1.6\\
  26$^{f}$  & 3.23 & 324 & 14.83 & 5.65 & 1.35 & 1.6\\
  27$^{f}$  & 3.20 & 355 & 13.91 & 14.80 & 1.33 & 1.6 \\
  28$^{f}$  & 7.00 & 320 & 15.78 & 8.96 & 1.34 & 1.6\\
  29        & 6.00 & 320 & 15.46 & 6.72 & 1.36 & 1.6\\
  30        & 5.00 & 280 & 12.37 & 3.70 & 1.38 & 1.6 \\
  31        & 13.00& 290 & 15.14 & 5.19 & 1.33 & 2.9 \\
  32        & 7.00 & 370 & 16.68 & 3.45 & 1.32 & 1.7\\
  33$^{f}$  & 4.50 & 470 & 19.14 & 1.74 & 1.36 & 1.8\\
  34$^{f}$  & 3.30 & 445 & 16.37 & 15.10 & 1.39 & 1.6\\
  35$^{f}$  & 10.00& 420 & 12.68 & 3.33 & 1.34 & 2.5\\
  36$^{f}$  & 11.00& 260 & 16.37 & 13.73 & 1.34 & 2.1\\
  37$^{f}$  & 5.50 & 381 & 18.83 & 5.25 & 1.35 & 1.7\\
  38$^{f}$  & 15.00& 367 & 16.68 & 4.87 & 1.35 & 1.9 \\
\hline
\end{tabular}
\end{table}

     \chapter{Results and Discussion}
  \label{chap5}

  \noindent\makebox[\linewidth]{\rule{\textwidth}{3pt}} 
  {\textit {Using coronagraph data we shortlist a set of 38 CMEs for this study. For all CMEs in our sample, we solve the 1D solar wind drag model to first determine 
  the heliocentric distances beyond which the solar wind drag dominates CME dynamics. We then compute the Lorentz force acting on the CMEs based on the 
  torus instability model. Comparative analysis of the two forces at different heights and the individual force profiles are discussed in this chapter for all the events. 
  These results have been published in \citet{Sac15} and \citet{Sac17}.} }\\
  \noindent\makebox[\linewidth]{\rule{\textwidth}{3pt}} 

  \section{Introduction}
  In order to understand CME dynamics, it is essential to investigate the forces that act on it as it propagates through the interplanetary medium.
  The two major forces in this regard are the solar wind aerodynamic drag and Lorentz force. We seek to quantify the relative contributions of each of these 
  forces as a function of the heliocentric distance. First, we use the solar wind drag-only model to determine the height beyond which 
  the drag is dominant. Then, we evaluate the 
  magnitude of the Lorentz force that accelerates the CMEs, using the torus instability model prescription. We begin with a brief discussion 
  of the CME sample set, followed by the force analysis method used to determine the heights where these forces govern the CME propagation 
  by comparing the model predictions with observations. Results for all the CMEs in our sample are described in this chapter.

  \section{Sample Description}
  Using STEREO, SOHO LASCO and {\it WIND} observations we build a sample of 38 Earth-directed CMEs based on the selection criteria described in Chapter \ref{chap2}. These instruments 
  provide continuous tracking of the CMEs from three view-points which makes this a well-observed sample with detailed observations of 3D evolution of these CMEs.
  Using the geometrical fitting technique of Graduated Cylindrical Shell (GCS) model, these CMEs are 3D reconstructed. The GCS fitting provides information regarding various geometrical parameters like the observed height-time profile, CME minor 
  radius and derived parameters like CME area and width. Our sample includes CMEs with initial 
  velocities ranging between $47-2400$ \kms. 18 of these CMEs are labeled as ``fast'' CMEs having initial speeds, $v_{0}>900$ \kms
  while the remaining are slow CMEs. 
  
\begin{figure}[H]    
  \centering
  \includegraphics[height=0.45\textwidth,width=0.65\textwidth,clip=]{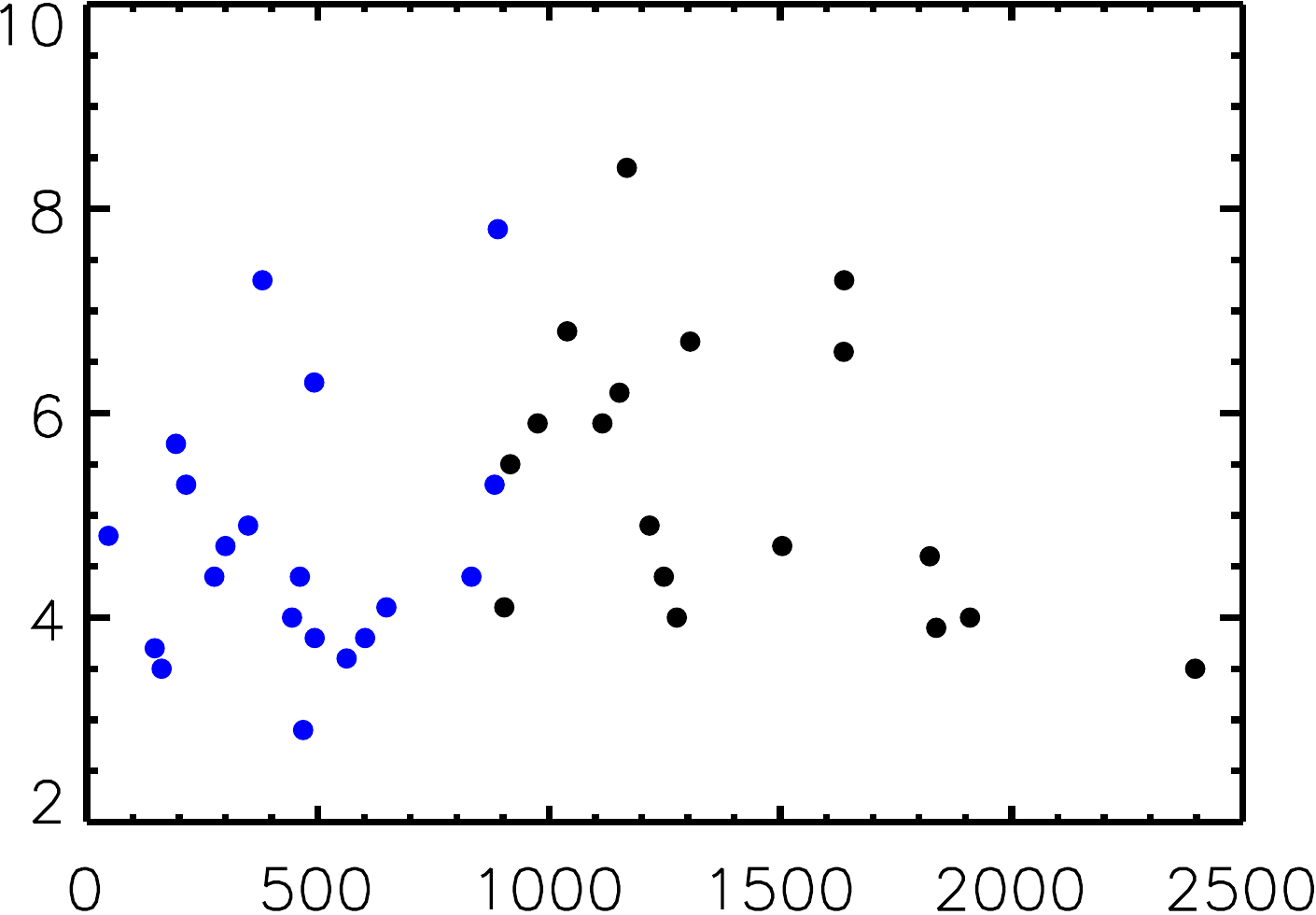}
  \put(-268,72.7){{\rotatebox{90}{{\color{black}\fontsize{14}{14}\fontseries{n}\fontfamily{phv}\selectfont   $h_{0}$ (R$_{\odot}$)}}}}
  \put(-145.8,-13.9){{\rotatebox{0}{{\color{black}\fontsize{14}{14}\fontseries{n}\fontfamily{phv}\selectfont  $v_{0}$ (km s$^{-1})$}}}}
  \caption[Plot of observed initial height ($h_{0}$) for all CMEs.]{Initial observed GCS fitted height ($h_{0}$) at the first time-stamp is shown in the figure for each CME in our sample referenced by the 
  CME initial velocity ($v_{0}$). Blue spheres indicate slow CMEs and black spheres represent fast CMEs. Details in Table \ref{tbl32}}
  \label{fig611}
  \end{figure}

  Figure \ref{fig611} shows the initial observed height ($h_{0}$), fitted using the GCS 
  technique plotted as a function of the initial velocity of each CME. Blue spheres indicate slow CMEs and black spheres represent fast CMEs. 
  We do not find any significant distinction in this regard, between the slow and fast CMEs. This color scheme (blue spheres - slow CMEs and black spheres - fast CMEs) 
  is followed throughout this work. About 60 \% CMEs in our sample have the first observed height 
  as low as 3\,--\,5 \Rs. The limited time cadence of the LASCO C2 coronagraph might affect the $h_{0}$ values for fast CMEs. Owing to this, faster CMEs have fewer data points 
  in the range 3\,--\,10 \Rs.

\section{Part 1 - Aerodynamic Drag} 
  The first part of this work focuses on studying the affects of the solar wind aerodynamic drag on CMEs 
  using the drag-only model described in detail in Chapter \ref{chap4}. Assuming that only drag force is acting, we consider the 
  1D drag-only model (Equation \ref{eq1}). In other words, Lorentz forces are not included, 
  and we consider only the $F_{drag}$ term in Equation \ref{eqforce}. This drag model incorporates a physical prescription for the collisionless solar wind viscosity which determines the non-constant drag 
  coefficient ($C_{\rm D}$) definition (Equation \ref{eqcdfit}). Using observationally derived parameters (from GCS fitting) and \textit{in-situ} observations, we determine the quantities, $A_{cme}$, $V_{sw}$, $n_{sw}$ and $C_{\rm D}$.
  Then we solve the drag equation and compare the model solutions with the height-time data.

  The 1D differential equation for drag force (Equation \ref{eq1}) is solved for all events, (a) from the 
  first observed height and (b) at progressively larger heights. At each height the parameters corresponding to that particular heliocentric distance are used.
  The initial velocity, determined via polynomial fitting to the data is used as an initial condition to the equation. It corresponds to the height 
  at which the model is initiated. The solution of this 1D aerodynamic drag equation is a model-predicted velocity profile of each CME
  (subjected exclusively to a drag force). The output of Equation \ref{eq1} is then integrated to obtain a height-time trajectory predicted by the model which is compared 
  to the observed height-time profile. 
  
  These observed height-time profiles (from GCS fitting) and the model predictions are 
  shown in Panel (a) of Figures \ref{fig51} \,--\, \ref{fig69} for all CMEs in the sample. Each plot is marked by the CME number with which it is 
  referenced in Table \ref{tbl31}. Plots for fast CMEs (initial velocity $v_{0}> 900\, km\,s^{-1}$)
  are indicated by a ($f$) in the CME serial number. Observed height-time data points from the GCS fitting (see Chapter \ref{chap2}) 
  are denoted by black diamond symbols in all the plots. The dash-dotted line (red) shows the model predicted height-time profiles when the drag-only model 
  is initiated from the first observed height ($h_{0}$). On the other hand, the solid (blue) line indicates the model solutions when it is initiated at a later 
  height ($\widetilde{h}_{0}$). 
  
  A jump in the height-points is seen in some plots. This is due to a change in the observing instrument. In particular, while transitioning from COR2 to HI1 instrument, there is 
 a time-gap in the observations which leads to a jump in fitted height. Sometimes, the HI1 observations are too faint initially for a clear GCS fit and the later images when 
 fitted can lead to a few missing height points. Error bars ($\pm 0.2$ \Rs for COR2, $\pm 1$ \Rs for HI1 and $\pm 10$ \Rs for HI2) are shown in all the plots.
  \subsection{Fast CMEs}
  CMEs with initial velocity $v_{0}>\,900$ \kms are referred to as ``fast'' CMEs in this study. These are indicated by a superscript ($f$) in Table 
  \ref{tbl32} in Chapter \ref{chap2} and all subsequent tables. Plots titled ``$f$'' in Figures \ref{fig51} \,--\, \ref{fig69}
  show the observed and predicted height-time profiles for all the 18 fast CMEs in our sample.
  Each plot indicates the modeled (red dash-dotted line) and observed height-time (black diamond symbols) profile for fast CMEs. 
  When the 1D drag-only force equation is solved for all the events, we find that the model solutions agree reasonably well with the observed CME trajectory 
  from the first data point ($h_{0}$) for the fast CMEs. In other words, for fast CMEs, the drag force 
  is dominant from as early as the first observed height (in the range 3.5-8.4 \Rs). How well the predicted model solutions fit the data is determined using the coefficient of determination (also called $R$ squared). 
  Model solutions with $R^{2}>98 \%$ are considered acceptable. 

  For some events we note that the height-time plots look roughly like a straight line. This might suggest that the CME speed is nearly 
  constant ({\it e.g.}, CMEs 15, 16, 27, 34); however, this is not the case. The profile looks like a straight line when the complete height-range is scaled down 
  to fit the frame. In fact, in most cases the CME speeds decrease by 20-60 \% of the initial speeds. We also see curved profiles in events CME 21, 22, 35, 36 etc. 
  In some cases, the modeled CMEs are slightly faster than the observed CMEs ({\it e.g.,} CMEs 2, 9, 19, 20). 
  That is, the model predicts that the CME reaches the final observed height ($h_{f}$; Table \ref{tbl51}) 
  earlier than it is actually observed. Some predicted profiles agree quite well with the data ({\it e.g.}, CMEs 16, 26, 28, 34, 36 and 37).
  About 5 of the 18 fast CMEs, decelerate by more than $\sim$ 500 \kms from their initial height up to the last predicted height $\widetilde{h}_{f}$ (see, Table \ref{tbl51}).
  The first observed height ($h_{0}$) for fast CMEs in our sample lies between $3.5$ and $8.5$ \Rs.
  Since the drag-only model considers only the effects of aerodynamic drag force, it can be claimed 
  that solar wind aerodynamic drag is the dominant force for these CMEs, from as low as 3.5 \Rs. 
  This result is in general consensus with most existing reports on fast CMEs being drag dominated above only a few solar radii.

  \subsection{Slow CMEs}
  Next, we consider the remaining 20 (slower) CMEs in our sample. These CMEs have initial velocities ranging between $47-900$ \kms. 
  In general, it is believed that Lorentz self-forces are dominant for the initial part of CME trajectory, while aerodynamic drag takes over at larger heights 
  \citep[\textit{e.g.},][]{Mic15}. Some authors \citep[\textit{e.g.},][]{Lew02} claim that slow CMEs are exclusively dragged {\em up} by the solar wind. 
  \citet{Mis13} apply the drag-based model to explain observations of slow CMEs over their entire trajectory while others, \citep[\textit{e.g.},][]{Byr10,Car12,Gop13} 
  think that solar wind aerodynamic drag becomes dominant for slow CMEs beyond a few solar radii. 

\begin{figure}[h]
  \centering
  \includegraphics[height=0.33\paperheight,width=0.62\paperwidth]{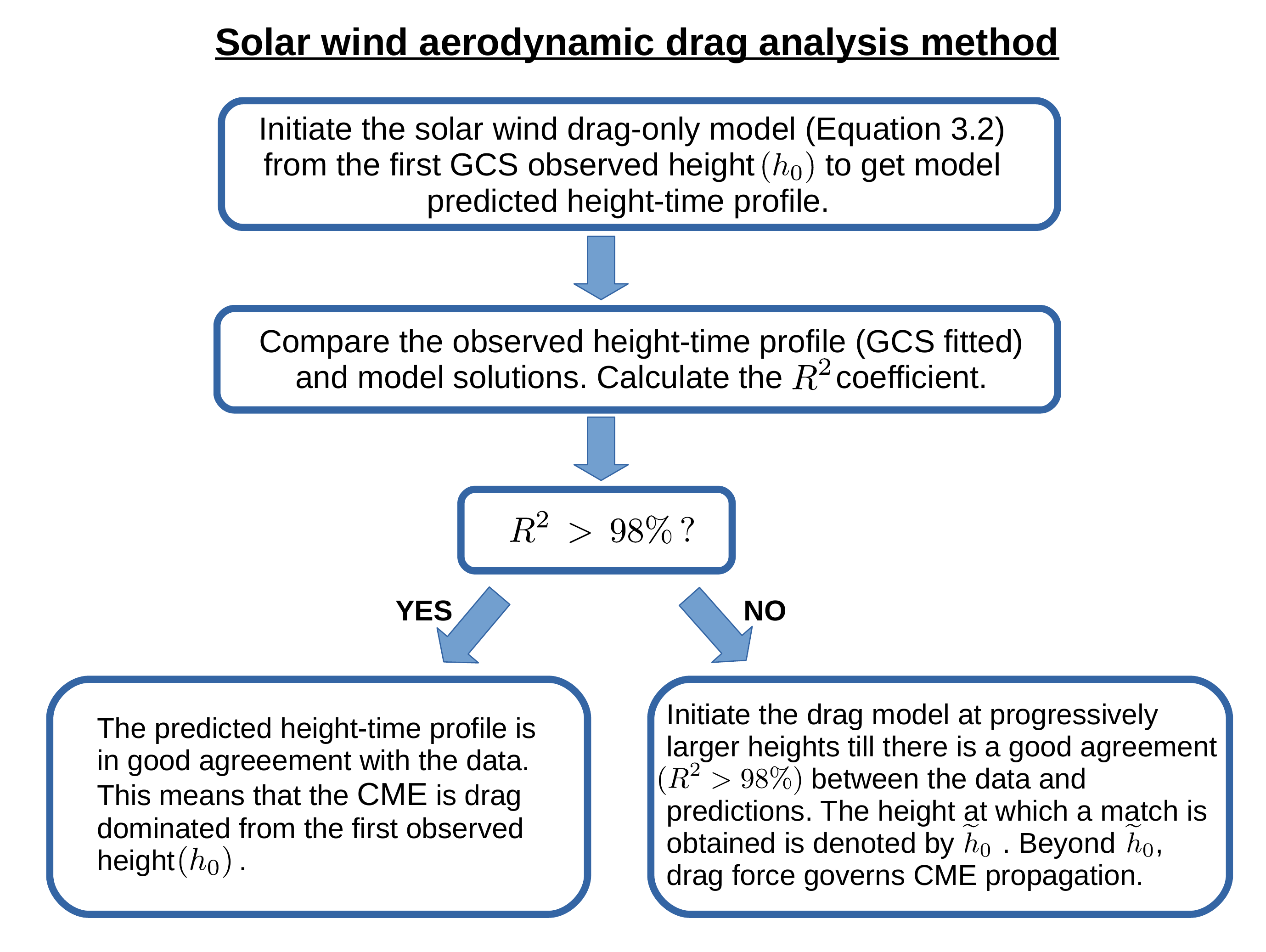}
  \caption[Flowchart for drag force analysis method]{Flowchart describing the solar wind drag force analysis method for all CMEs.}
  \label{flowchart}
  \end{figure}

  When the drag-only model analysis (as described for fast CMEs) is 
  repeated for the slower events in the sample, we find a considerable discrepancy between the modeled and observed CME profiles.
  This is evident from plots for slower CMEs in Figures \ref{fig51} \,--\, \ref{fig69}.
  The height-time data and the model predicted solutions (red dash-dotted line) disagree substantially for these CMEs when the model is initiated from the first observed height
  ($h_{0}$) which lies in the range $2.9-7.8$ \Rs for slow CMEs. This means that the momentum coupling between the CME and solar wind alone is not sufficient to describe 
  the observed CME dynamics for these CMEs in our sample. Therefore, these events cannot be considered drag-dominated from the start. In order to find the 
  height beyond which the solar wind aerodynamic drag ``takes over'' their propagation, we initiate the drag-only model at progressively later heights. In other words, we begin solving the 
  drag equation at larger heights to find the height at which the predicted and observed height-time profiles agree reasonably well with each other. Our method is depicted as a 
  flowchart in Figure \ref{flowchart}. The height at which this agreement is achieved is denoted by $\widetilde{h}_{0}$ and the corresponding CME velocity at this height which is used as an initial 
  condition is denoted by $\widetilde{v}_{0}$. 

  In the plots for slower CMEs in Figures \ref{fig51} \,--\, \ref{fig69}, the predicted solutions when the model is initiated at $\widetilde{h}_{0}$, are indicated by a blue solid line. The quantities $\widetilde{h}_{0}$ and corresponding $\widetilde{v}_{0}$ are listed 
  in Table \ref{tbl51} for all the CMEs. For fast CMEs (described in the previous section), the initiation height $\widetilde{h}_{0}$ is same as 
  the first observed height $h_{0}$ and so is the initial velocity i.e. $\widetilde{v}_{0}$ is same as $v_{0}$.
  Since the model we use (Equation \ref{eq1}) does not include any other force except the solar wind drag, it can be definitively 
  stated that the slower CMEs in our sample ($v_{0}<900$ \kms) are not drag-dominated below the height $\widetilde{h}_{0}$. 
  We find that the height $\widetilde{h}_{0}$ lies between $12-50$ \Rs for these slow CMEs (Table \ref{tbl51}). In other words, the ambient solar wind drag force can be 
  considered to be the dominant force influencing the CME trajectory beyond 12\,--\,50 \Rs for the slow CMEs in our sample.

\begin{figure}[h]    
  \centering
  \includegraphics[height=0.45\textwidth,width=0.65\textwidth,clip=]{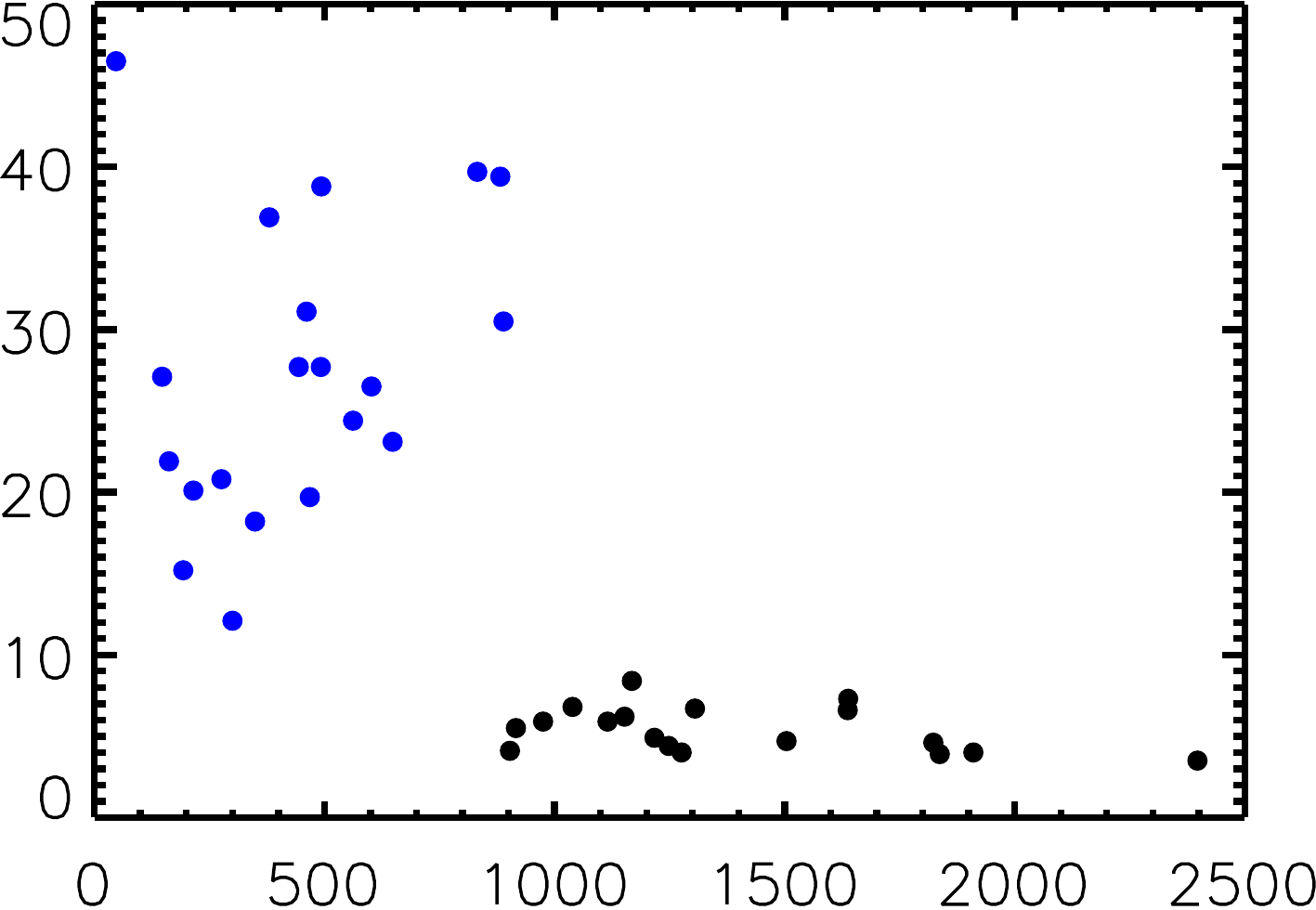}
  \put(-278,72.7){{\rotatebox{90}{{\color{black}\fontsize{14}{14}\fontseries{n}\fontfamily{phv}\selectfont   $\widetilde{h}_{0}$ (R$_{\odot}$)}}}}
  \put(-145.8,-13.9){{\rotatebox{0}{{\color{black}\fontsize{14}{14}\fontseries{n}\fontfamily{phv}\selectfont  $v_{0}$ (km s$^{-1})$}}}}
  \caption[Drag initiation height (\htil) for all CMEs.]{Height beyond which the solar wind drag model solutions match the observed CME propagation ($\widetilde{h}_{0}$) is plotted 
  as a function of CME initial velocity. Blue spheres indicate slow CMEs and black spheres represent fast CMEs. Details in Table \ref{tbl51}.}
  \label{fig61b}
  \end{figure}

  Figure \ref{fig61b} shows this heliocentric distance ($\widetilde{h}_{0}$) corresponding to the initial
  velocity ($v_{0}$) of each CME. Independent 
  studies using empirical fitting parameters have also confirmed that slower CMEs are drag-dominated only beyond $\approx 20$ \Rs (\textit{e.g.}, \citealp{Tem15}, 
  \citealp{Zic15}). \citet{Bor09} discuss 
  the CME dynamics using drag-only models only beyond $\sim 30$ \Rs. It can be concluded therefore, that below these heights, other forces, such as Lorentz forces 
  must be taken into account. The solar wind interaction with CMEs tends to equilibrate their speeds, \textit{i.e.} a CME will accelerate or decelerate to match the 
  speed of the solar wind. An important point to note here is that most of the slower CMEs show little or no evolution in their speeds beyond $\widetilde{h}_{0}$. 
  CME 31 decelerates the most above $\widetilde{h}_{0}$, from $\widetilde{v}_{0} \sim 597$ \kms \, to $\sim 470$ \kms at $\widetilde{h}_{f}$. 
  For several slow CMEs, the speeds stay roughly constant for heliocentric distances $> \widetilde{h}_{0}$. Above this height, 
  the drag-based model tends to perform quite well, therefore, it can be said that the ambient solar wind aerodynamic drag does not do much for these CMEs. 
  Since the deceleration is so small, using models with constant \CD $\sim$ 0 yield good results. Therefore, initiating drag-only models above \htil does not constrain 
  these models to any significant extent, and only reinforces the fact that most slow CMEs do not accelerate or decelerate much beyond \htil.

  \subsubsection{Final height: observations versus model predictions}
  Column 4 in Table \ref{tbl51} indicates the final observed height ($h_{f}$) derived from the GCS fitting for all CMEs. For CMEs that are indicated with a 
  superscript $\ast$, the observations extend up to HI2 \citep{Sac15} while the remaining events are observed till the HI1 FOV ($\sim 80$ \Rs). 
  $\widetilde{h}_{f}$ (Column 5 of Table \ref{tbl51}) denotes the final predicted height when the drag-only model is initiated at $\widetilde{h}_{0}$. 
  The difference between them is represented by the quantity $\Delta \widetilde{h}_{f}\,=\, h_{f}\,-\,\widetilde{h}_{f}$. For all CMEs except CME 2, $0.2<\Delta \widetilde{h}_{f}<10$ \Rs.
  For CME 2, $\Delta \widetilde{h}_{f}$ is -31.7 \Rs. A positive $\Delta \widetilde{h}_{f}$ indicates that the model underpredicts the final height (when compared to 
  the observed height $h_{f}$) and vice-versa. We find that for about 79 \% of the CMEs in the sample, $\Delta \widetilde{h}_{f}$ is less than 5 \Rs. This is another indication of 
  how well the drag-only model predicts the observed trajectory.

  \subsubsection{$C_{\rm D}$ prescription (Equation \ref{eqcdfit}) versus constant $C_{\rm D}$}
  The last column of Table \ref{tbl51} lists the range of values of the drag coefficient $C_{\rm D}$ derived using 
  Equation \ref{eqcdfit} (\citet{Sub12} prescription for $C_{\rm D}$), for all the CMEs. We also use constant $C_{\rm D}$ values and solve 
  the Equation \ref{eq1}. Constant $C_{\rm D}$ models with values outside the range mentioned in Table \ref{tbl51} perform poorly. As an example, 
  we show this result for CME 2 in Figure \ref{fig51}. 
  The model solutions with constant $C_{D}s$ of 0.1 (green dash-dotted line) and 5 (brown dash-dotted line) disagree considerably with the observed 
  height-time data for CME 2. The \citet{Sub12} prescription for drag coefficient (Equation \ref{eqcdfit}) is therefore a good guide for constant $C_{\rm D}$ models. 
  In other words, only those constant $C_{D}$ drag-only models with $C_{D}$ values that are close to those predicted by Equation \ref{eqcdfit} agree reasonably well 
  with the data.

  \subsubsection{How robust are our conclusions?}
  We consider as an example, CME 2 which has a starting speed of 916 $\rm km\, s^{-1}$ and decelerates to $\sim 715\,\, km\,s^{-1}$.
  The Sun-Earth travel time for CME 2 was observed to be $\approx$ 60 hours and the model predicts that the CME arrives at the Earth around 8 hours earlier. 
  This means that modeled CME is somewhat faster than the observed one representing an error of 14$\%$.
  Our time of arrival (ToA) errors are comparable to the values of $\approx$ 10-12 hours obtained from MHD models \citep{May15} and drag-based models 
  \citep{Shi15}. Although we have used observational data (obtained by GCS fitting) as much as possible for the model,
  there is some room for uncertainty in the quantity $r_{a}$ (Equation \ref{eq3}). We find that a 50 $\%$ increase(decrease) in $r_{a}$ results in a 3.7$\%$ increase(decrease)
  in the predicted CME travel time. Thus, our results are not very sensitive to the precise value of $r_{a}$. Errors in the GCS flux-rope 
  fitting procedure can also lead to errors in the measured cross-sectional area of the CME.
  When $A_{cme}$ is decreased by 50 \%, the CME travel time decreases by 6 \%.

  To summarize, we find that CMEs that travel with initial velocity $v_{0}>900$ \kms, are drag-dominated from very early on.
  The solar wind decelerates fast CMEs from heights as low as 3.5\,--\,4 \Rs. However, in case of slower CMEs, 
  momentum coupling with the ambient solar wind does not satisfactorily describe the observed dynamics from very low heights. Instead, we 
  find that these CMEs are drag dominated only above 12\,--\,50 \Rs. This conclusion is independent of the specific $C_{D}$ model used (\citet{Sub12,Sac15} or 
  a constant $C_{D}$). We now investigate the effect of forces at heights $<\widetilde{h}_{0}$. We also use Equation \ref{eq1} to calculate the 
  magnitude of the drag force for heights $>\widetilde{h}_{0}$ for all CMEs.

  \section{Part 2 - Lorentz Force}
  We now consider the $F_{\rm Lorentz}$ term in Equation \ref{eqforce} which describes the driving Lorentz force 
  acting on CMEs. Most Lorentz force models concerned with CME initiation predict that the total Lorentz force profile increases up to a peak at a certain heliocentric
  distance and decreases thereafter. Various models described in literature cater to such a profile by either tailoring the 
  injected poloidal flux (or equivalently, the driving current) \citep{Che10} or rely on the external Lorentz forces to decrease rapidly enough 
  (with heliocentric distance) for the CME to ``launch'' (\citealp{Kli06}; torus instability (TI) model). 
  \citet{Kli14} have also shown the equivalence of TI and the catastrophe mechanism for CME eruption \citep{For91}.

  To evaluate the magnitude of the Lorentz force acting on CMEs, we require the CME current ($I$) and equilibrium current 
  ($I_{eq}$) at $h_{eq}$ (which 
  relates to the equilibrium value of external field $B_{ext}(h_{eq})$), calculated using Equations \ref{eqI} and \ref{eqIe} (section \ref{subI}). The final expression used for the Lorentz force acting on a CME is given by Equation \ref{eqfl} in Appendix \ref{LF}.
  For each CME, we use the quantities derived from the GCS fitting technique to calculate the Lorentz force. 
  The decay index $n$ (Table \ref{tbl41}) is such that the drag force is larger in magnitude compared to the Lorentz force above 
  $\widetilde{h}_{0}$; it also ensures, $n>n_{cr}$. Thus, $n$ is the minimum value that requires $|F_{drag}|>F_{Lorentz}$ for $R>\widetilde{h}_{0}$. 
  For a given value of $n$, the quantity $I_{eq}$ (and equivalently $B_{ext}(h_{eq})$) is determined by the condition 
  $|F_{drag}(\widetilde{h}_{0})|=F_{Lorentz}(\widetilde{h}_{0})$. Thus, by equating the two forces at $\widetilde{h}_{0}$, the equilibrium current ($I_{eq}$) and hence, the 
  external poloidal field ($B_{ext}$) at equilibrium ($h_{eq}$) can be constrained. For the sake of concreteness the equilibrium position of the flux rope (where 
  the Lorentz self-force and external forces balance each other) is taken to be $h_{eq}=1.05$ \Rs for all CMEs in our sample.
  
  The GCS fittings to the CME events begin at the time of first observation in the COR2 FOV. This observed GCS height denoted by 
  $h_{0}$ (Table \ref{tbl31}) lies between 2.9\,--\,8.4 \Rs for our sample. Other derived GCS parameters like the flux-rope aspect ratio ($R/R_{cme}$), 
  ($R_{cme}$ is the minor radius), are also available at each observed height (R) beginning from $h_{0}$ (first observed height). We assume that the 
  aspect ratio at heliocentric distances between $h_{eq}$ and $h_{0}$ is the same as the ratio at $h_{0}$. 
  In other words, we assume that the CME expands self-similarly from $h_{eq}$ to $h_{0}$. For heights beyond $h_{0}$, the aspect ratio is obtained using the observed quantities $R$ and minor radius $R_{cme}$ derived using the GCS technique.

  For each CME, we calculate the Lorentz force between $h_{eq}$ and the final observed height ($h_{f}$) with observed values of 
  $R$, minor radius ($R_{cme}$) and $l_{i}=1/2$. In Figures \ref{fig51}\,--\,\ref{fig69}, each plot in Panel (b) shows the Lorentz force profile versus the heliocentric 
  distance of the CME leading edge ($R$) for all CMEs. The red solid line indicates the Lorentz force evaluated at 
  heights between $h_{eq}$ and $h_{0}$ (using a constant $R/R_{cme}$). Beyond $h_{0}$, the open diamond symbols (connected by a red dotted line) 
  represent the Lorentz force derived using observations. The blue vertical dashed line denotes the position of $\widetilde{h}_{0}$ which is the height 
  at which solar wind drag takes over. In case of fast CMEs (indicated by $f$, alongside the CME number), this line indicates the position of the first observed height ($h_{0}$).

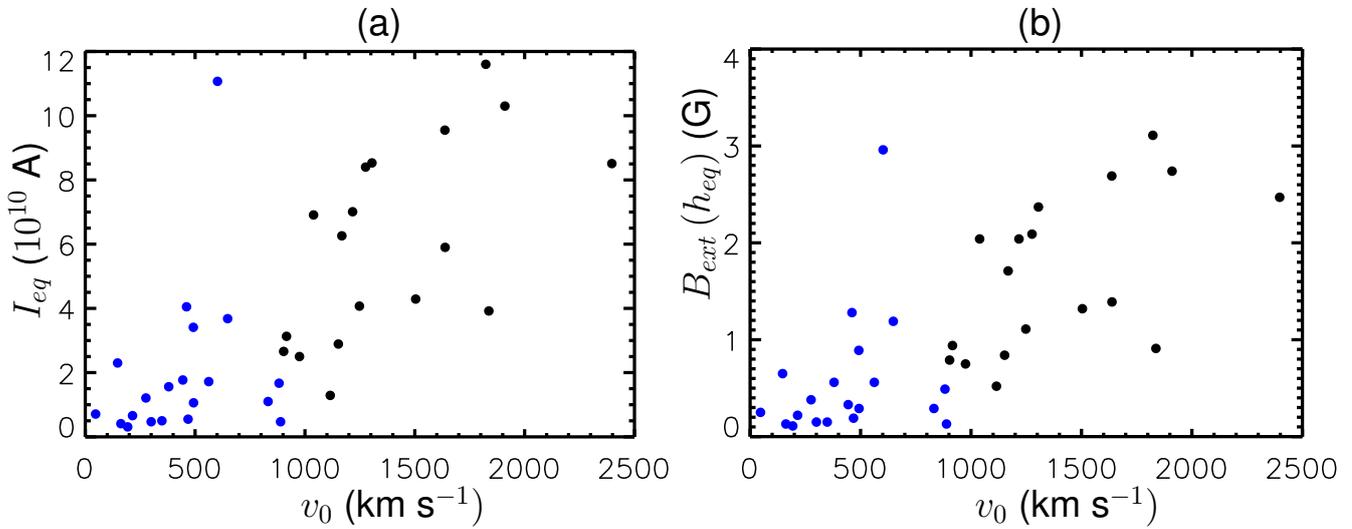
\begin{figure}[h]    
    \centerline{\hspace*{0.065\textwidth}
		\includegraphics[width=0.6\textwidth,clip=]{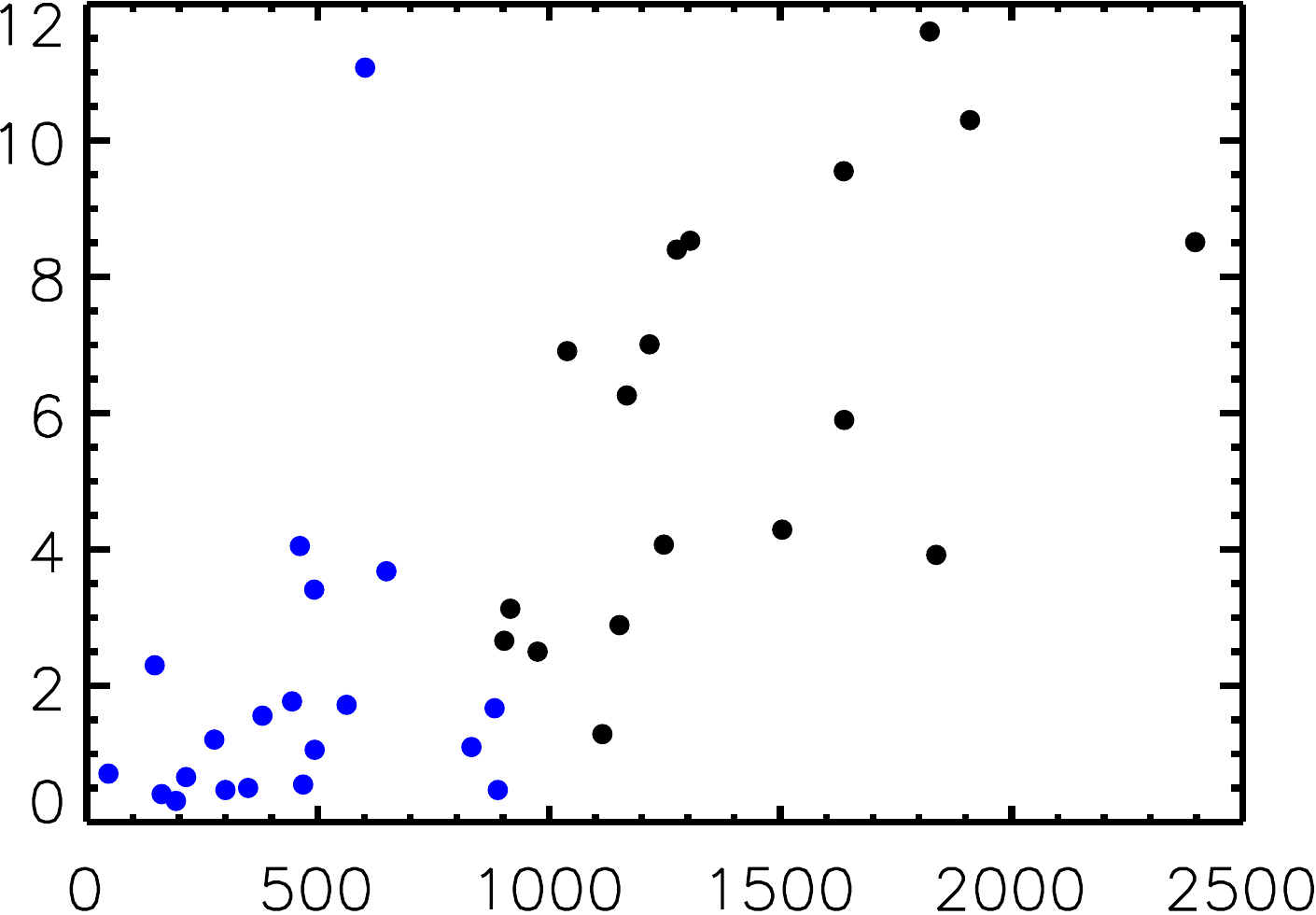}
		\hspace*{0.01\textwidth}
		\put(-255,60.7){{\rotatebox{90}{{\color{black}\fontsize{14}{14}\fontseries{n}\fontfamily{phv}\selectfont   $I_{eq}$ ($10^{10}$ A)}}}}
		\put(-145.8,-13.9){{\rotatebox{0}{{\color{black}\fontsize{14}{14}\fontseries{n}\fontfamily{phv}\selectfont  $v_{0}$ (km s$^{-1})$}}}}
		\put(-125,168.){{\rotatebox{0}{{\color{black}\fontsize{14}{14}\fontseries{n}\fontfamily{phv}\selectfont  (a)}}}}
		\hspace{0.3cm}
		\includegraphics[width=0.59\textwidth,clip=]{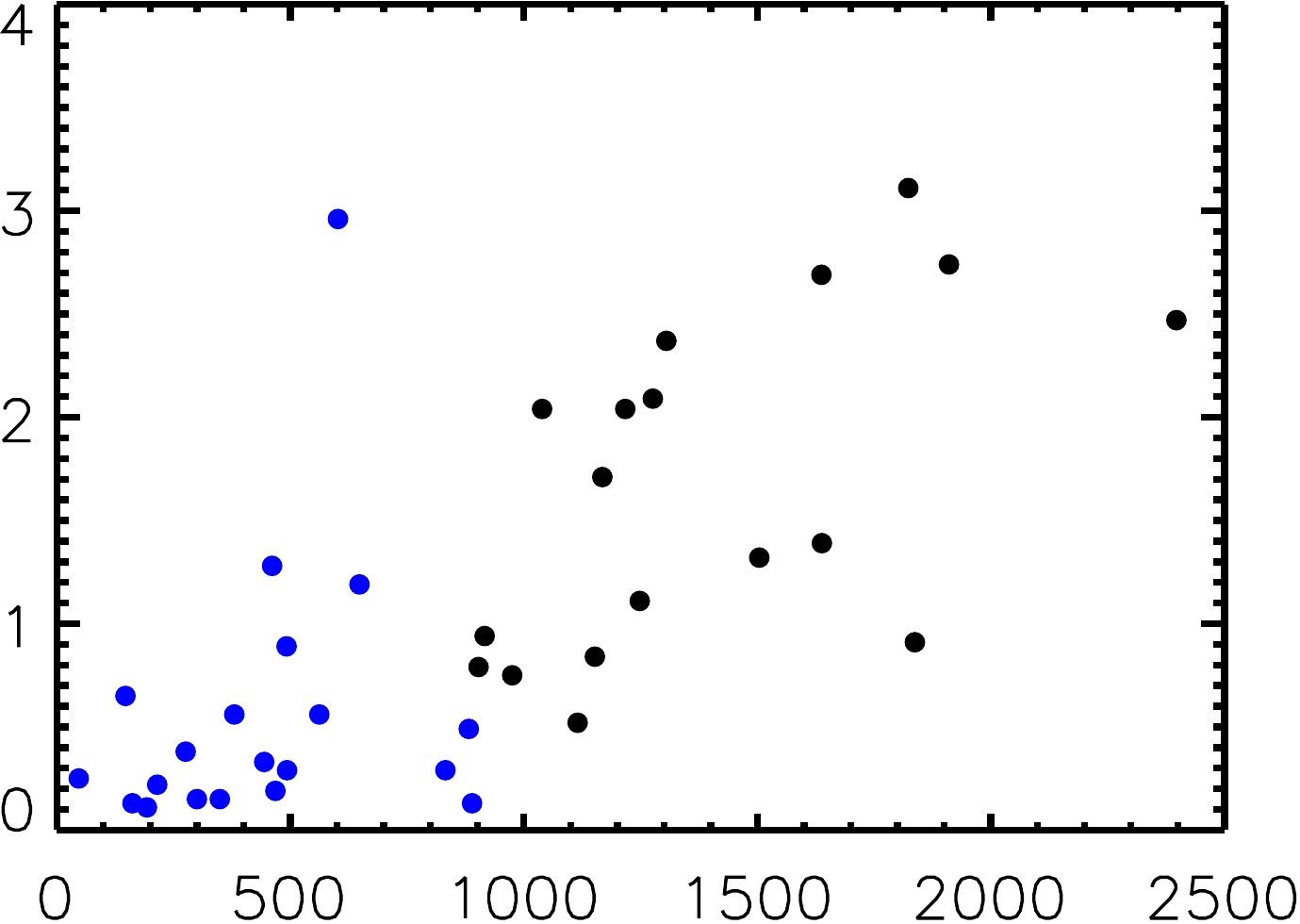}
		\put(-246,67.7){{\rotatebox{90}{{\color{black}\fontsize{14}{14}\fontseries{n}\fontfamily{phv}\selectfont  $B_{ext}\, (h_{eq})$ (G)}}}}
		\put(-135.8,-13.9){{\rotatebox{0}{{\color{black}\fontsize{14}{14}\fontseries{n}\fontfamily{phv}\selectfont  $v_{0}$ (km s$^{-1})$}}}}
		\put(-120,168.){{\rotatebox{0}{{\color{black}\fontsize{14}{14}\fontseries{n}\fontfamily{phv}\selectfont  (b)}}}}
  }
	    \vspace{-0.465\textwidth}   
		\centerline{\small      
	    \hfill}
  \vspace{0.46\textwidth}  
  \caption[Plot of Lorentz force parameters ($I$ and $B_{ext}$) at equilibrium]{Plots of Lorentz force parameters, CME current and external magnetic 
  field at equilibrium, as a function of CME initial velocity ($v_{0}$). Panel (a) shows the CME current ($I_{eq}$) at 
  equilibrium position ($h_{eq}$) \textit{versus} the CME initial velocity ($v_{0}$). Panel (b) is a plot of the external 
  poloidal magnetic field ($B_{ext}(h_{eq})$) at the equilibrium height ($h_{eq}$) \textit{versus}
  the initial velocity ($v_{0}$). Details in Table \ref{tbl52}.}
    \label{fig621}
    \end{figure}

  For each CME, $I_{eq}$ and $B_{ext}(h_{eq})$ are listed in Table \ref{tbl52} . The equilibrium current ($I_{eq}$) is in units of $10^{10}$ Amperes. These 
  estimates are in general agreement with the average axial current calculated by \citet{Sub07}. $B_{ext}$ at $h_{eq}(=1.05$ \Rs) is in units of $10^{-1}$
  Gauss. Figure \ref{fig621} shows the equilibrium CME current and magnetic field for all the CMEs as a function of the initial velocity.
  In general, for slower CMEs (blue spheres), both these quantities are smaller in magnitude when compared to their values for faster CMEs. 
  The Lorentz force magnitude (in units of $10^{17}$ dynes) at $\widetilde{h}_{0}$ is also indicated in Table \ref{tbl52} for all CMEs. 
  It is equal to the absolute drag force at this height. Panel (a) of Figure \ref{fig511} depicts the magnitude of the Lorentz 
  force at $\widetilde{h}_{0}$ versus the initial velocity ($v_{0}$) for each CME in our sample. 
  It is equal to the absolute magnitude of the drag force at this height.

  \subsubsection{Lorentz force peak at $h_{peak}$}
\begin{figure}[h]    
    \centerline{\hspace*{0.065\textwidth}
		    \includegraphics[width=0.61\textwidth,clip=]{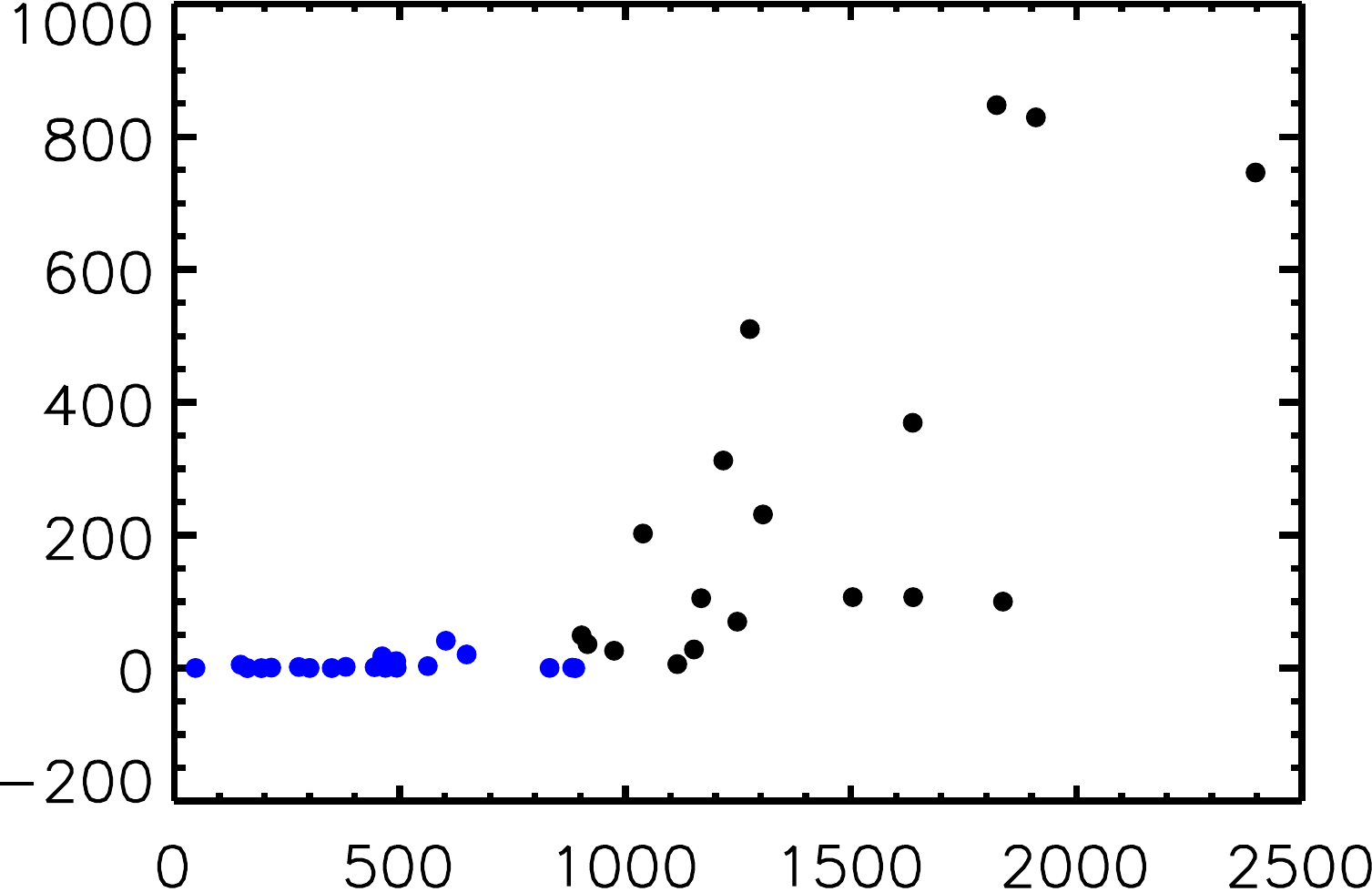}
				  \hspace*{0.01\textwidth}
		\put(-260,23.7){{\rotatebox{90}{{\color{black}\fontsize{14}{14}\fontseries{n}\fontfamily{phv}\selectfont   $F_{\rm Lorentz}$ ($10^{17}$ dyn)}}}}
		\put(-135.8,-13.9){{\rotatebox{0}{{\color{black}\fontsize{14}{14}\fontseries{n}\fontfamily{phv}\selectfont  $v_{0}$ (km s$^{-1})$}}}}
		\put(-125,165.){{\rotatebox{0}{{\color{black}\fontsize{14}{14}\fontseries{n}\fontfamily{phv}\selectfont  (a)}}}}
		\hspace{0.3cm}
		\includegraphics[width=0.59\textwidth,clip=]{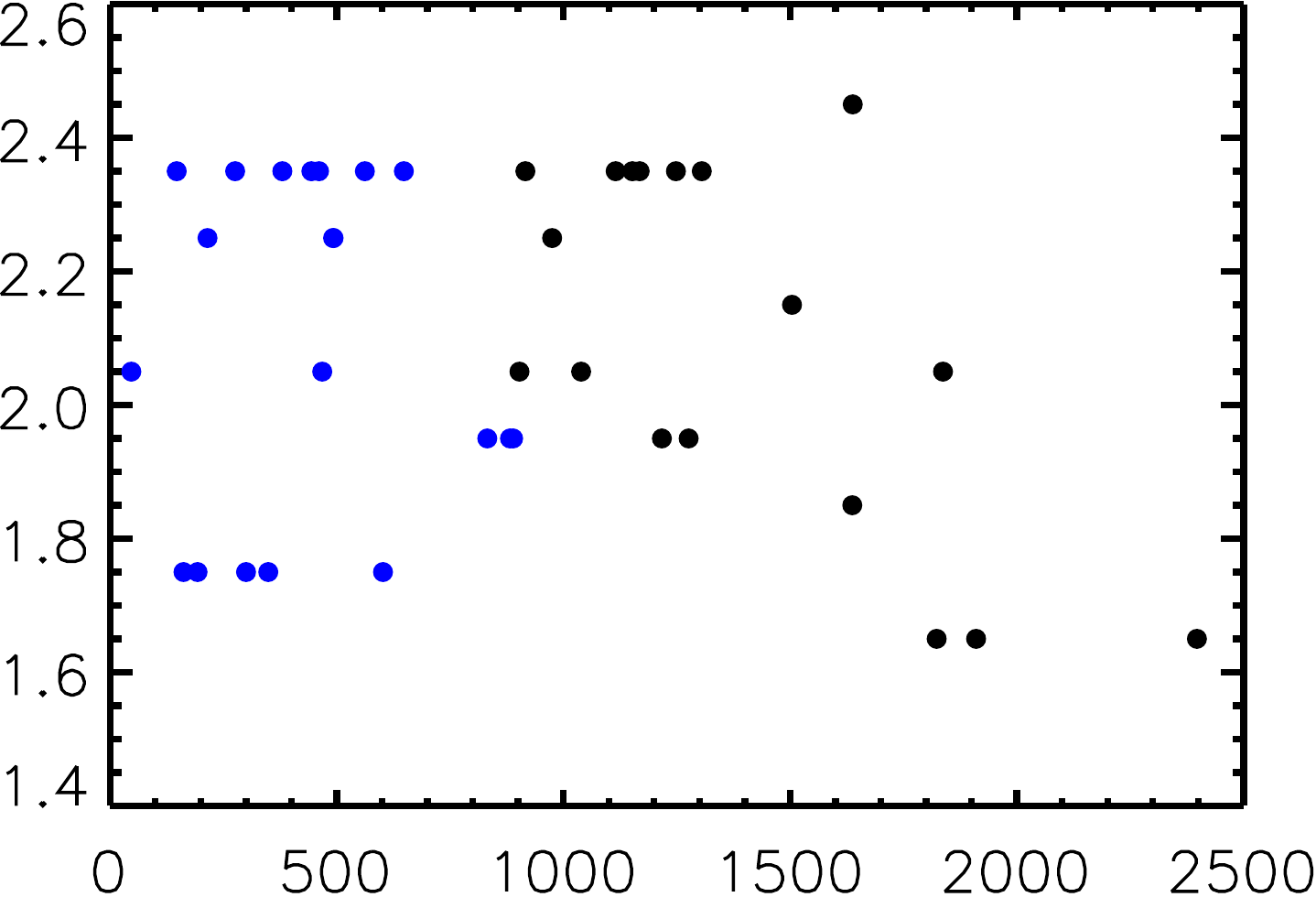}
		\put(-250,57.7){{\rotatebox{90}{{\color{black}\fontsize{14}{14}\fontseries{n}\fontfamily{phv}\selectfont  $h_{peak}$ (R$_{\odot}$)}}}}
		\put(-145.8,-13.9){{\rotatebox{0}{{\color{black}\fontsize{14}{14}\fontseries{n}\fontfamily{phv}\selectfont  $v_{0}$ (km s$^{-1})$}}}}
		\put(-125,165.){{\rotatebox{0}{{\color{black}\fontsize{14}{14}\fontseries{n}\fontfamily{phv}\selectfont  (b)}}}}
  }
	    \vspace{-0.465\textwidth}   
		\centerline{\small      
	    \hfill}

  \vspace{0.46\textwidth}  
  \caption[Plot of Lorentz force magnitude and peak position.]{Panel (a) depicts the magnitude of the Lorentz force ($F_{\rm Lorentz}$) at $\widetilde{h}_{0}$ in units of $10^{17}$ dyn as a function of CME initial velocity, $v_{0}$. 
  Panel (b) shows the heliocentric distance, $h_{peak}$ where the Lorentz force achieves its maximum value as a function of CME initial velocity, $v_{0}$ (see Table \ref{tbl52}.  
  Symbols in blue represent slow CMEs (\textit{i.e.} $v_{0}<900$ km s$^{-1}$) and symbols in black represent fast CMEs ($v_{0}>900$ km s$^{-1}$).}
    \label{fig511}
      \end{figure}
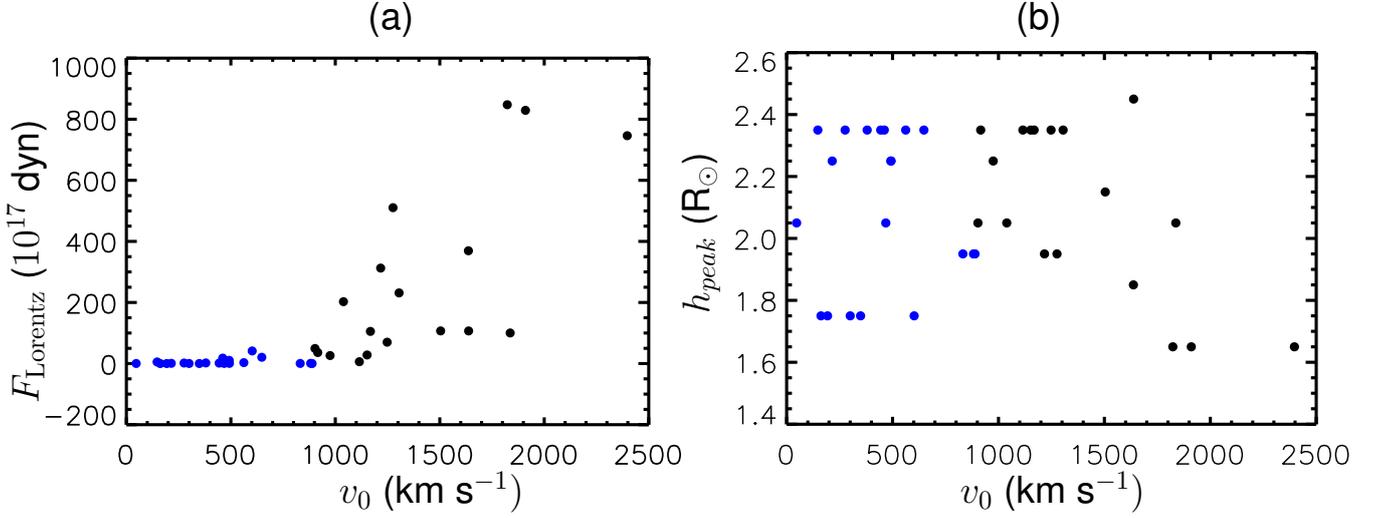

  For both fast and slow CMEs, we note that the Lorentz force profile increases steeply from $h_{eq}$ to reach its maximum at a 
  height denoted by ``$h_{peak}$'' (in Table \ref{tbl52}) and then decreases gradually. This quantity $h_{peak}$ lies in the 
  range, 1.65\,--\,2.45 \Rs. Panel (b) of Figure \ref{fig511} plots the quantity $h_{peak}$ for each CME as 
  a function of the CME initial velocity ($v_{0}$; Table \ref{tbl32}). The blue symbols represent slow CMEs ($v_{0}<900$ \kms) 
  while symbols in black represent fast CMEs ($v_{0}>900$ \kms). There are no noticeable trends distinguishing slow and fast CMEs.

  \subsubsection{The Fall \%}
\begin{figure}[h]    
    \centerline{\hspace*{0.065\textwidth}
		\includegraphics[width=0.59\textwidth,clip=]{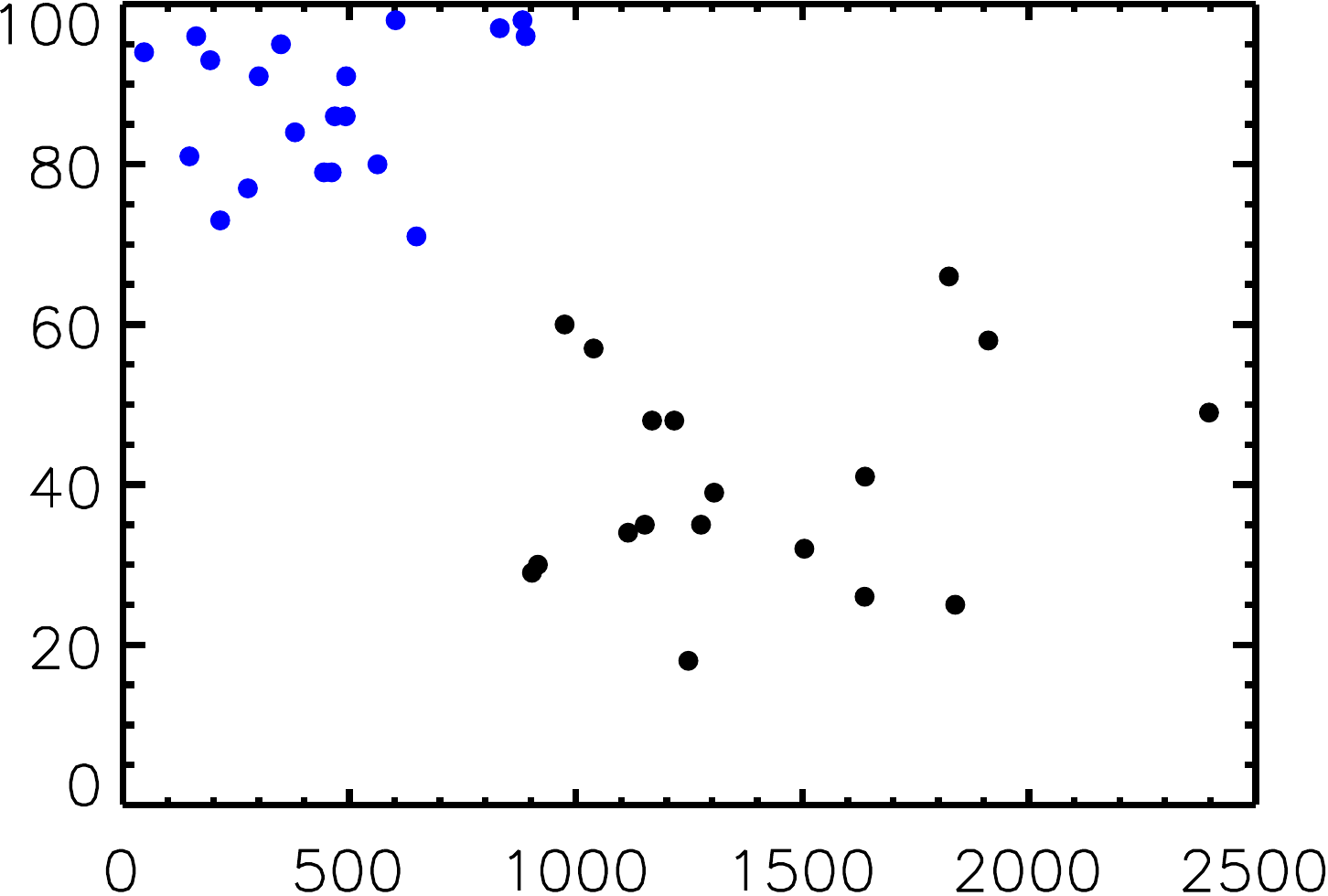}
		\hspace*{0.01\textwidth}
		\put(-250,60.7){{\rotatebox{90}{{\color{black}\fontsize{14}{14}\fontseries{n}\fontfamily{phv}\selectfont $Fall\, \%$}}}}
		\put(-135.8,-14.9){{\rotatebox{0}{{\color{black}\fontsize{14}{14}\fontseries{n}\fontfamily{phv}\selectfont  $v_{0}$ (km s$^{-1})$}}}}
		\put(-125,165.){{\rotatebox{0}{{\color{black}\fontsize{14}{14}\fontseries{n}\fontfamily{phv}\selectfont  (a)}}}}
		\hspace{0.3cm}
		\includegraphics[width=0.59\textwidth,clip=]{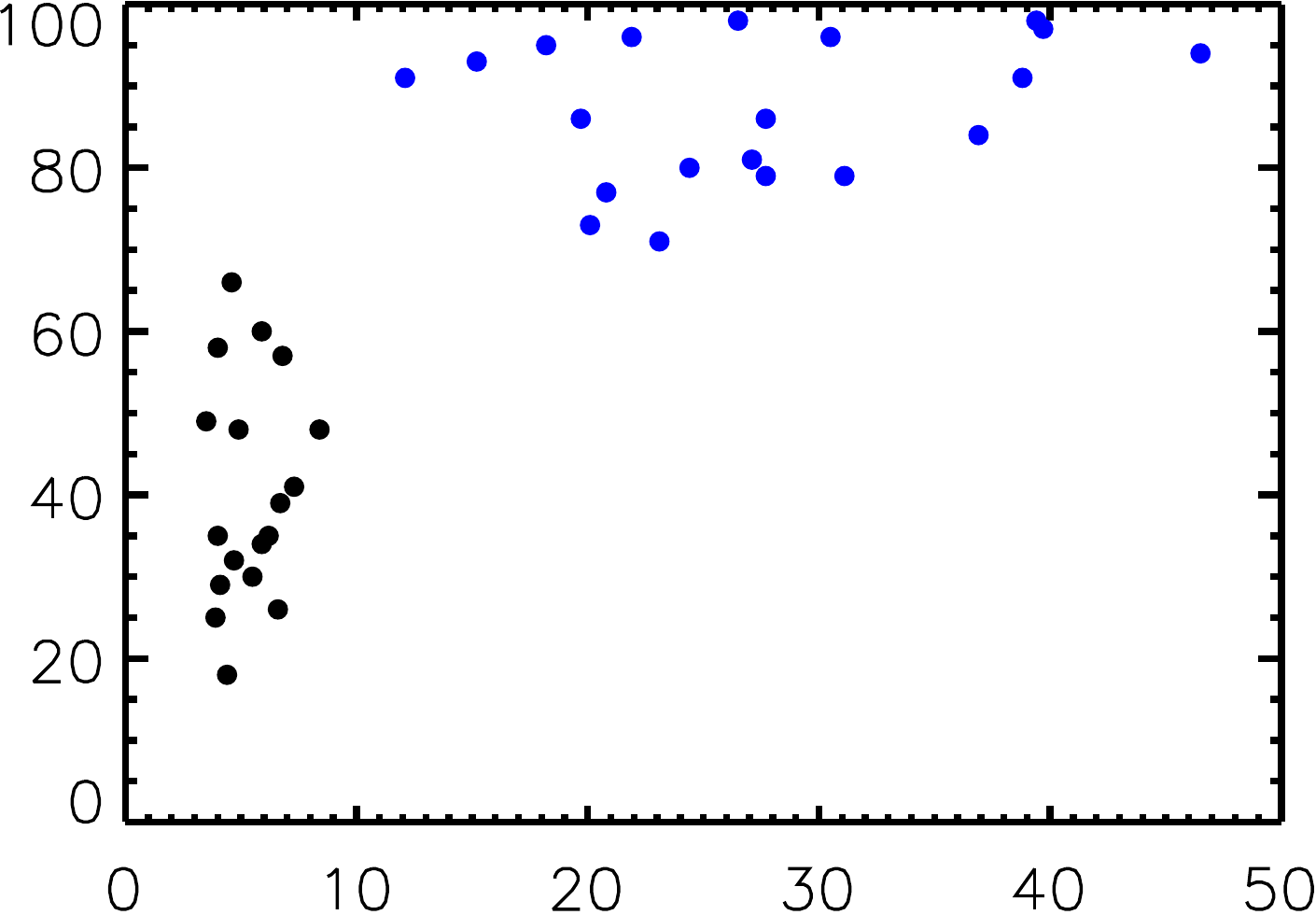}
		\put(-240,60.7){{\rotatebox{90}{{\color{black}\fontsize{14}{14}\fontseries{n}\fontfamily{phv}\selectfont  $Fall\, \%$}}}}
		\put(-125.8,-14.9){{\rotatebox{0}{{\color{black}\fontsize{14}{14}\fontseries{n}\fontfamily{phv}\selectfont  $\widetilde{h}_{0}$ (R$_{\odot}$)}}}}
		\put(-120,166.){{\rotatebox{0}{{\color{black}\fontsize{14}{14}\fontseries{n}\fontfamily{phv}\selectfont  (b)}}}}
			      }
	    \vspace{-0.465\textwidth}   
		\centerline{\small      
	    \hfill}

  \vspace{0.46\textwidth}  
  \caption[Plot of the Fall \% \textit{versus} $v_{0}$ and \htil]{Plot of the Lorentz force $Fall\, \%$ \textit{versus} CME initial velocity ($v_{0}$) in Panel (a) 
  and $\widetilde{h}_{0}$ in Panel (b) (see, Table \ref{tbl52}).}
    \label{fig512}
    \end{figure}

  A measure of how much the Lorentz force falls from its maximum value (at $h_{peak}$) till the height $\widetilde{h}_{0}$ is 
  indicated by the quantity $Fall \,\%$ in Table \ref{tbl52}. Figure \ref{fig512} plots the $Fall\, \%$ 
  for all events in our sample as a function of the CME initial speed ($v_{0}$) in Panel (a), and $\widetilde{h}_{0}=h_{0}$ in Panel (b). 
  It can be seen clearly that the $Fall\, \%$ 
  is larger for slower CMEs (larger \htil) in comparison to faster ones. Since fast CMEs are drag dominated from relatively early on ($\widetilde{h}_{0}=h_{0}$), the Lorentz force decrease from its maximum value at $h_{peak}$ is smaller 
  compared to the decrease in case of slower CMEs ($\widetilde{h}_{0} \sim$  12\,--\,50 \Rs).
  For slower CMEs, the $Fall \,\%$ at $\widetilde{h}_{0}$ (12\,--\,50 \Rs) lies between 70--98 \% while for faster CMEs, it is between 20--60 \% .

\begin{figure}[h]    
    \centering                              
		\includegraphics[width=0.6\textwidth,clip=]{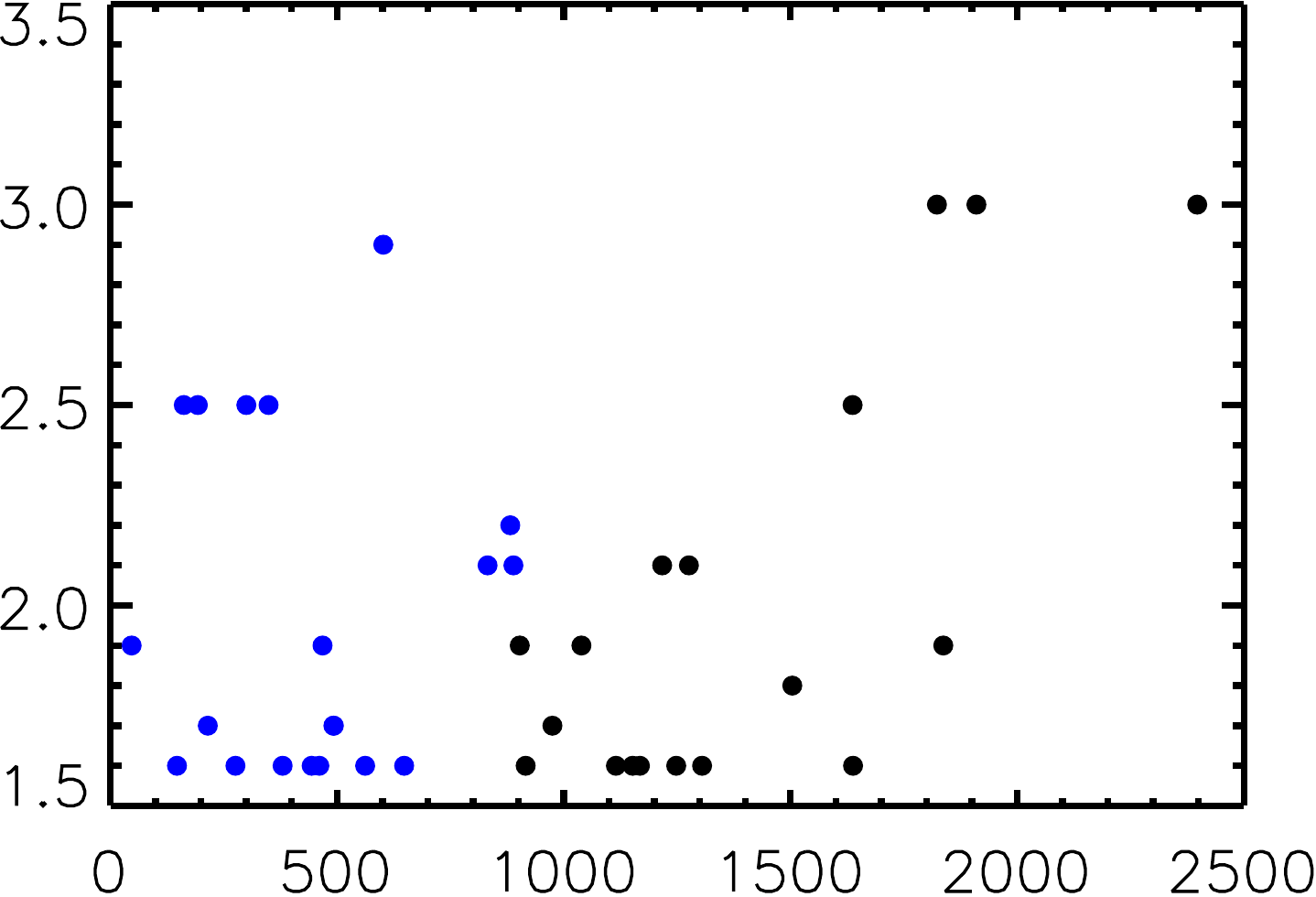}
		\hspace*{0.01\textwidth}
		\put(-257,78.7){{\rotatebox{90}{{\color{black}\fontsize{14}{14}\fontseries{n}\fontfamily{phv}\selectfont   $n$ }}}}
		\put(-145.8,-14.9){{\rotatebox{0}{{\color{black}\fontsize{14}{14}\fontseries{n}\fontfamily{phv}\selectfont  $v_{0}$ (km s$^{-1})$}}}}
  \caption[Plot of decay index $n$ for all CMEs.]{Plot of decay index ($n$) corresponding to CME initial velocity ($v_{0}$) (see, Table \ref{tbl41}).} 
    \label{fig631}
    \end{figure}

  The decay index ($n$; Table \ref{tbl41}) for the external poloidal field is represented in a scatterplot in Figure \ref{fig631}. We see no 
  clear trend for $n$ with regard to slow and fast CMEs. However, the highest $n$ values are typically associated with the faster CMEs in the sample. 
  $n$ lies in the range $1.6-3$.

  For a fixed value of $n$, we note that an increase in $h_{eq}$ by 14 $\%$ 
  increases the peak force position value by $\sim 15 \%$. It decreases the $Fall\, \%$ of the Lorentz force 
  at $\widetilde{h}_{0}$ (relative to its peak value) by 5$\%$.
  For a fixed value of $h_{eq} (=1.05 R_{\odot})$, an increase in $n$ by $31 \%$ decreases the peak position by 17$\%$. 
  It also increases the $Fall\, \%$ of the Lorentz force at $\widetilde{h}_{0}$ (relative to its peak value) by 19.5$\%$.

    
  \subsection{Lorentz force vs aerodynamic drag} \label{lsfvsdrag}
  In Panel (b) of Figures \ref{fig51}\,--\,\ref{fig69}, the filled green circles represent absolute value of the aerodynamic drag force ($F_{drag}$) 
  beginning at $\widetilde{h}_{0}$. We note that for all CMEs, the Lorentz force peaks between 1.65--2.45 \Rs but becomes negligible only by 12--50 \Rs 
  for the slower CMEs in the sample. This means that the Lorentz force dominates the CME propagation at heights up to 12--50 \Rs for 
  slower CMEs, beyond which the solar wind aerodynamic drag takes over. The effects of Lorentz force for slower CMEs is therefore quite pronounced 
  well beyond a few solar radii. However, in case of faster CMEs in our sample, the Lorentz force becomes 
  negligible (compared to the solar wind drag) from as early as 3.5\,--\,4 \Rs.  In fact, the solar wind drag is significantly larger in magnitude 
  in comparison to the Lorentz force beyond $h_{0}$ for fast CMEs.
  
\citet{Vrs16} investigate the CME-flare relationship by illustrating how the CME eruption, driven by MHD (magneto-hydrodynamic) instabilities (i.e., Kink or Torus) as 
well as the CME acceleration and propagation are affected by magnetic reconnection. They also discuss a feedback relationship 
between the flare and ejection through reconnection. They find that in the absence of magnetic reconnection, the Lorentz force (or acceleration) 
that drives the CMEs decreases rapidly. This does not explain the observed acceleration which can be as high as several $km s^{-2}$. To achieve 
these strong accelerations, they introduce reconnection which increasingly adds poloidal flux strengthening the forces acting on the flux-rope \citep{Che89}. 

\citet{Vrs16} also suggest that the eruption acceleration is closely related to the impulsive phase of the flare (e.g., Hard X-ray, microwave bursts, Soft X-ray flux) \citep[\textit{e.g.,}][]{Zha01,Vrs04,Zhan04}. 
Based on their study, CME acceleration (\textit{i.e.}, Lorentz force) peaks at larger heights for more gradual events. 
Due to the addition of the poloidal flux, the Lorentz force driving is prolonged for slower events, which is also what we observe. 
We find that in the case of slower CMEs, the Lorentz forces are dominant upto 12-50 $R_{\odot}$, beyond which they become negligible and the aerodynamic drag takes over.

We conclude that the drag-only model accounts well for the observed CME trajectory when initiated at (or beyond) $\widetilde{h}_{0}$. This means 
  that the Lorentz force is not important beyond this height. However, below \htil, the Lorentz force is 
  appreciable and governs the CME propagation. We find this to be true for 36 out of 38 CMEs in our sample.
  Two slow CMEs (CME 4 and CME 10) are exceptions. 

  \subsubsection{Relative difference between the two forces ($F_{\rm diff} \%$)} \label{secdiff}
\begin{figure}[h]    
    \centering                              
		    \includegraphics[width=0.6\textwidth,clip=]{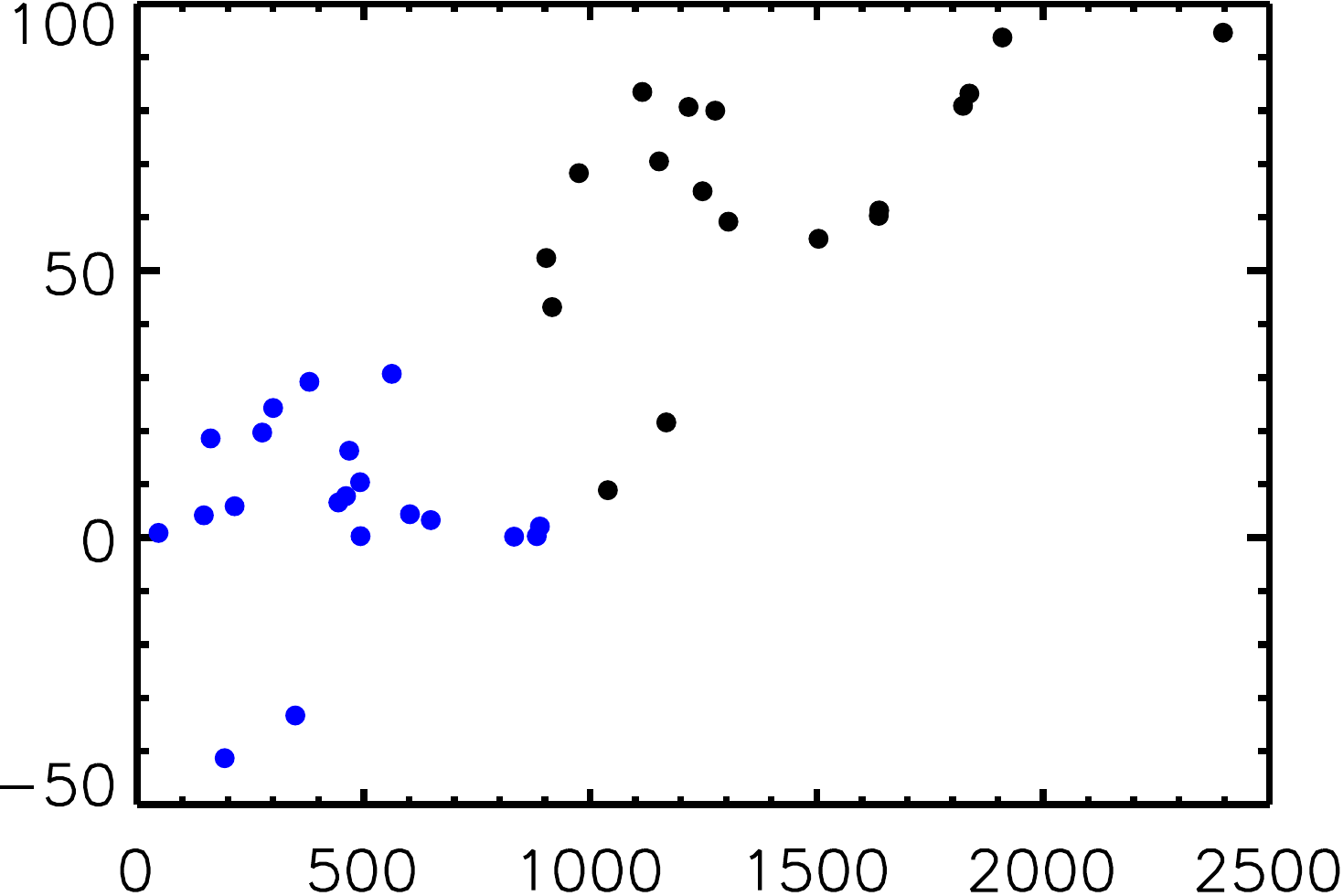}
		\put(-243,65.7){{\rotatebox{90}{{\color{black}\fontsize{14}{14}\fontseries{n}\fontfamily{phv}\selectfont   $F_{\rm diff}$ (\%)}}}}
		\put(-135.8,-14.9){{\rotatebox{0}{{\color{black}\fontsize{14}{14}\fontseries{n}\fontfamily{phv}\selectfont  $v_{0}$ (km s$^{-1})$}}}}
  
  \caption[Plot of $F_{\rm diff}$\% for all CMEs.]{Plot of relative difference between drag and Lorentz forces as a function of CME initial 
  velocity, $v_{0}$. $F_{\rm diff} \, \%$ is calculated at 40 R$_{\odot}$ for all CMEs except CME 11, for which it is evaluated at 50 R$_{\odot}$ (see, Table \ref{tbl52}). 
  The blue circles represent slow CMEs and the black ones represent fast CMEs.} 
    \label{fig513}
    \end{figure}

\begin{figure}[h]    
    \centering                              
    \centerline{\hspace*{0.06\textwidth}
		\includegraphics[width=0.55\textwidth,clip=]{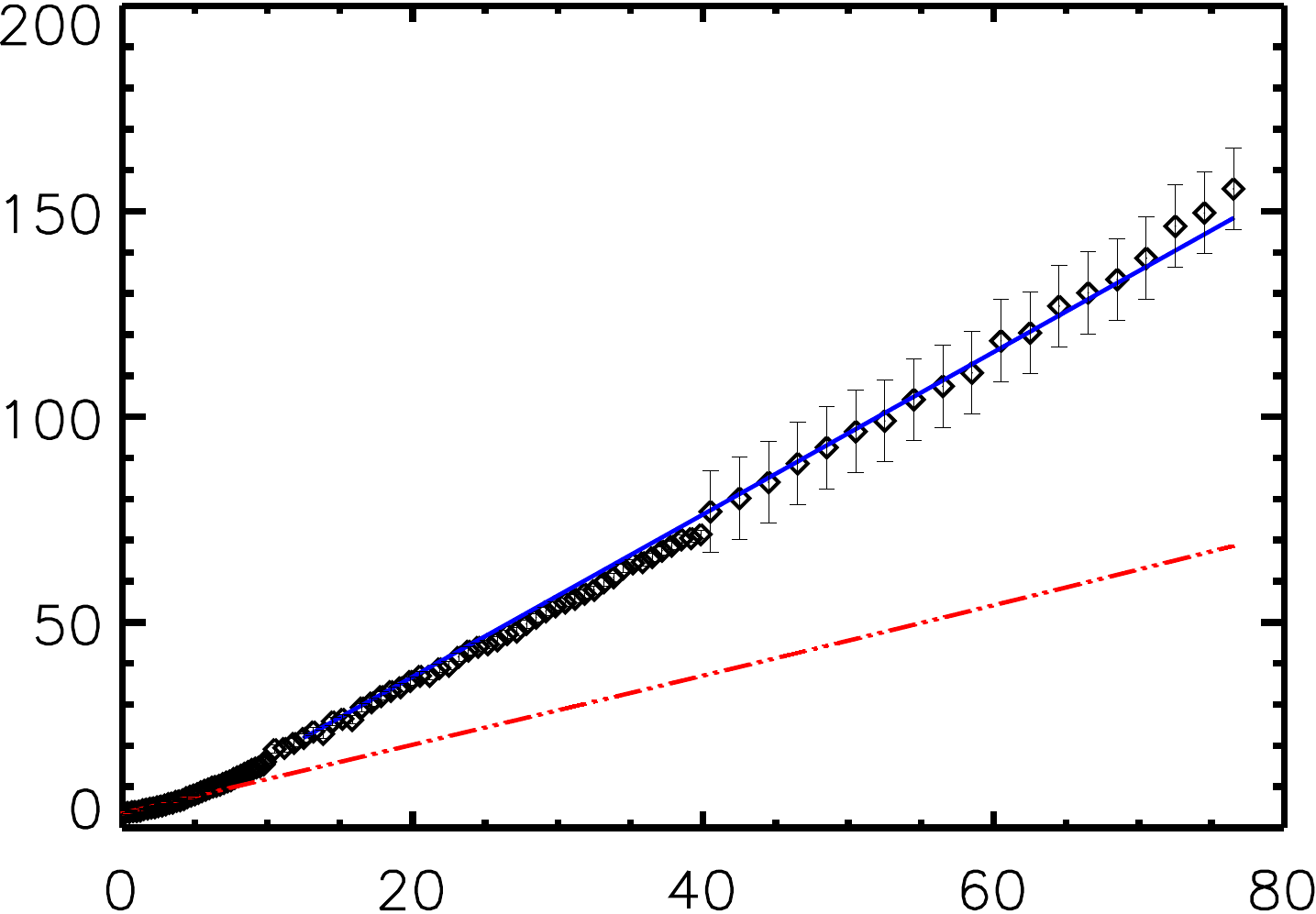}
		\put(-230,62.7){{\rotatebox{90}{{\color{black}\fontsize{12}{12}\fontseries{n}\fontfamily{phv}\selectfont  Height (R$_{\odot}$)}}}}
		\put(-160.8,-14.9){{\rotatebox{0}{{\color{black}\fontsize{12}{12}\fontseries{n}\fontfamily{phv}\selectfont  Elapsed Time (hrs)}}}}
		\put(-130,130.7){{\rotatebox{0}{{\color{black}\fontsize{13}{13}\fontseries{n}\fontfamily{phv}\selectfont \underbar{CME 1}}}}}
		\put(-115,158.7){{\rotatebox{0}{{\color{black}\fontsize{13}{13}\fontseries{n}\fontfamily{phv}\selectfont (a)}}}}
		\hspace*{0.069\textwidth}
		\includegraphics[width=0.53\textwidth,clip=]{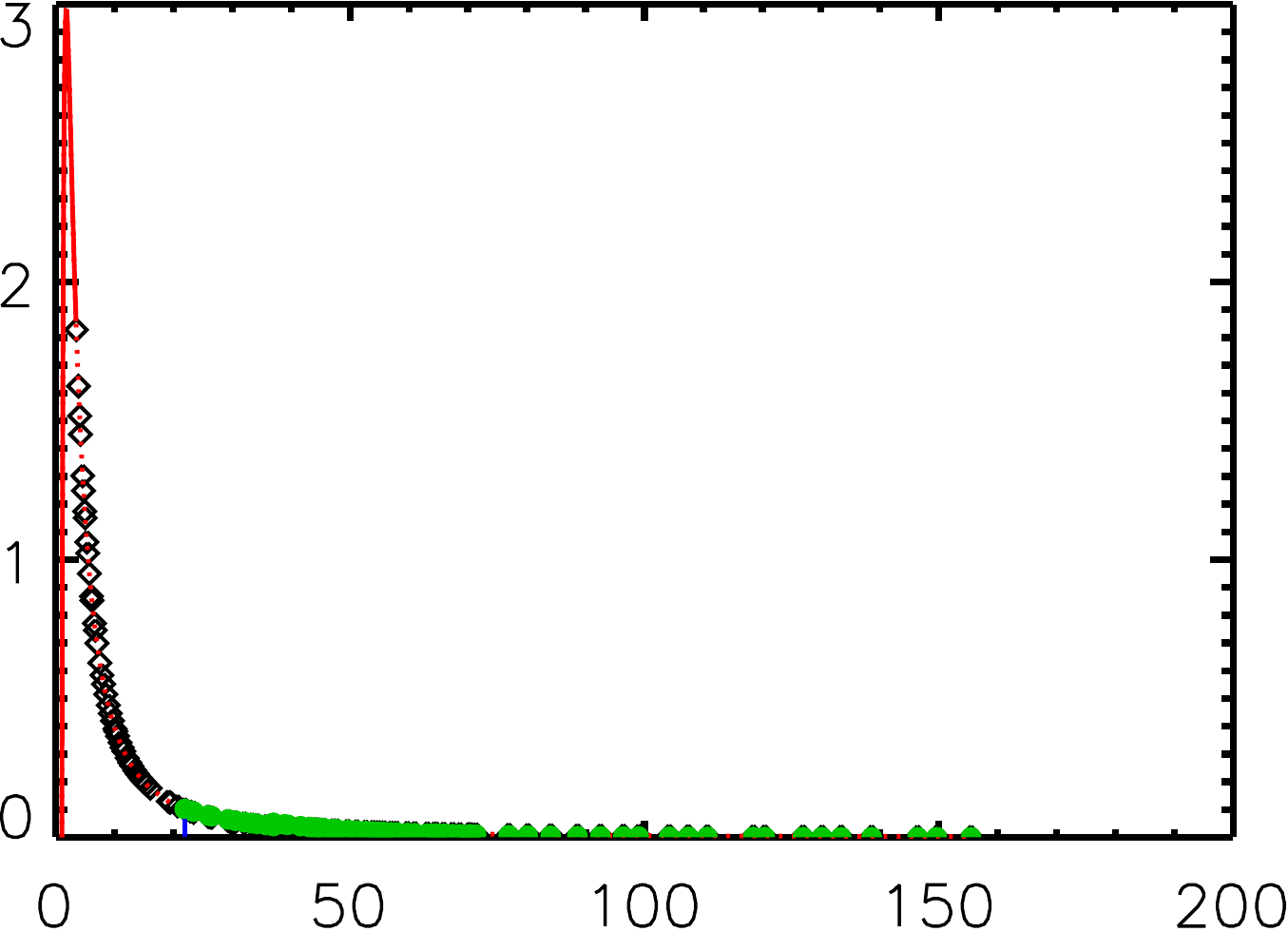}
		\put(-231,43.7){{\rotatebox{90}{{\color{black}\fontsize{12}{12}\fontseries{n}\fontfamily{phv}\selectfont  Force ($10^{17}$ dyn)}}}}
		\put(-140.8,-12.9){{\rotatebox{0}{{\color{black}\fontsize{12}{12}\fontseries{n}\fontfamily{phv}\selectfont Height (R$_{\odot}$)}}}}
		\put(-130,130.7){{\rotatebox{0}{{\color{black}\fontsize{13}{13}\fontseries{n}\fontfamily{phv}\selectfont  \underbar{CME 1}}}}}
		\put(-117,158.7){{\rotatebox{0}{{\color{black}\fontsize{13}{13}\fontseries{n}\fontfamily{phv}\selectfont (b)}}}}          
		  }
		\vspace{1.3cm}
		  \centerline{\hspace*{0.06\textwidth}
		\includegraphics[width=0.55\textwidth,clip=]{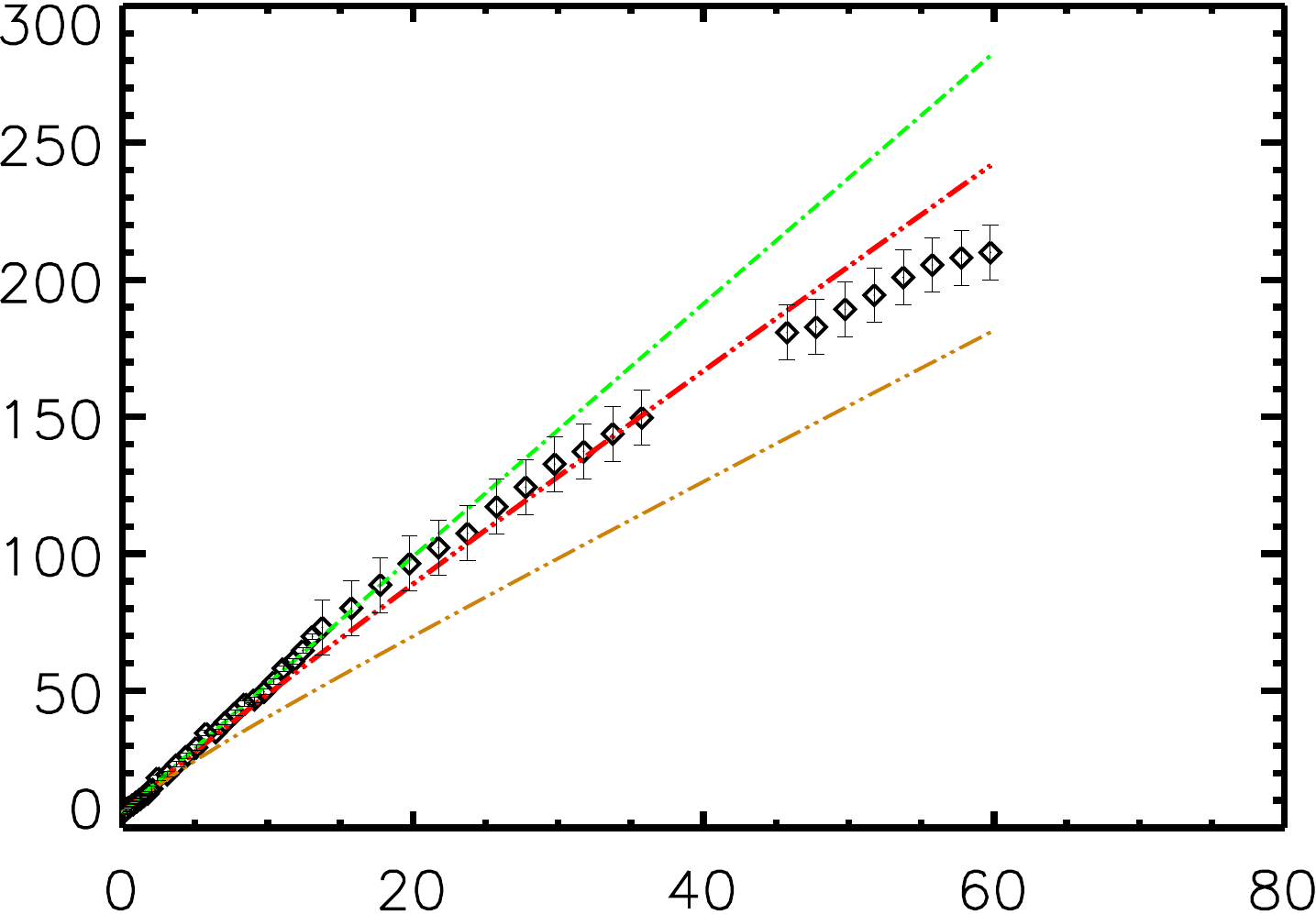}
		\put(-230,62.7){{\rotatebox{90}{{\color{black}\fontsize{12}{12}\fontseries{n}\fontfamily{phv}\selectfont  Height (R$_{\odot}$)}}}}
		\put(-160.8,-14.9){{\rotatebox{0}{{\color{black}\fontsize{12}{12}\fontseries{n}\fontfamily{phv}\selectfont  Elapsed Time (hrs)}}}}
		\put(-130,130.7){{\rotatebox{0}{{\color{black}\fontsize{13}{13}\fontseries{n}\fontfamily{phv}\selectfont \underbar{CME 2($f$)}}}}}
		\hspace*{0.069\textwidth}
		\includegraphics[width=0.545\textwidth,clip=]{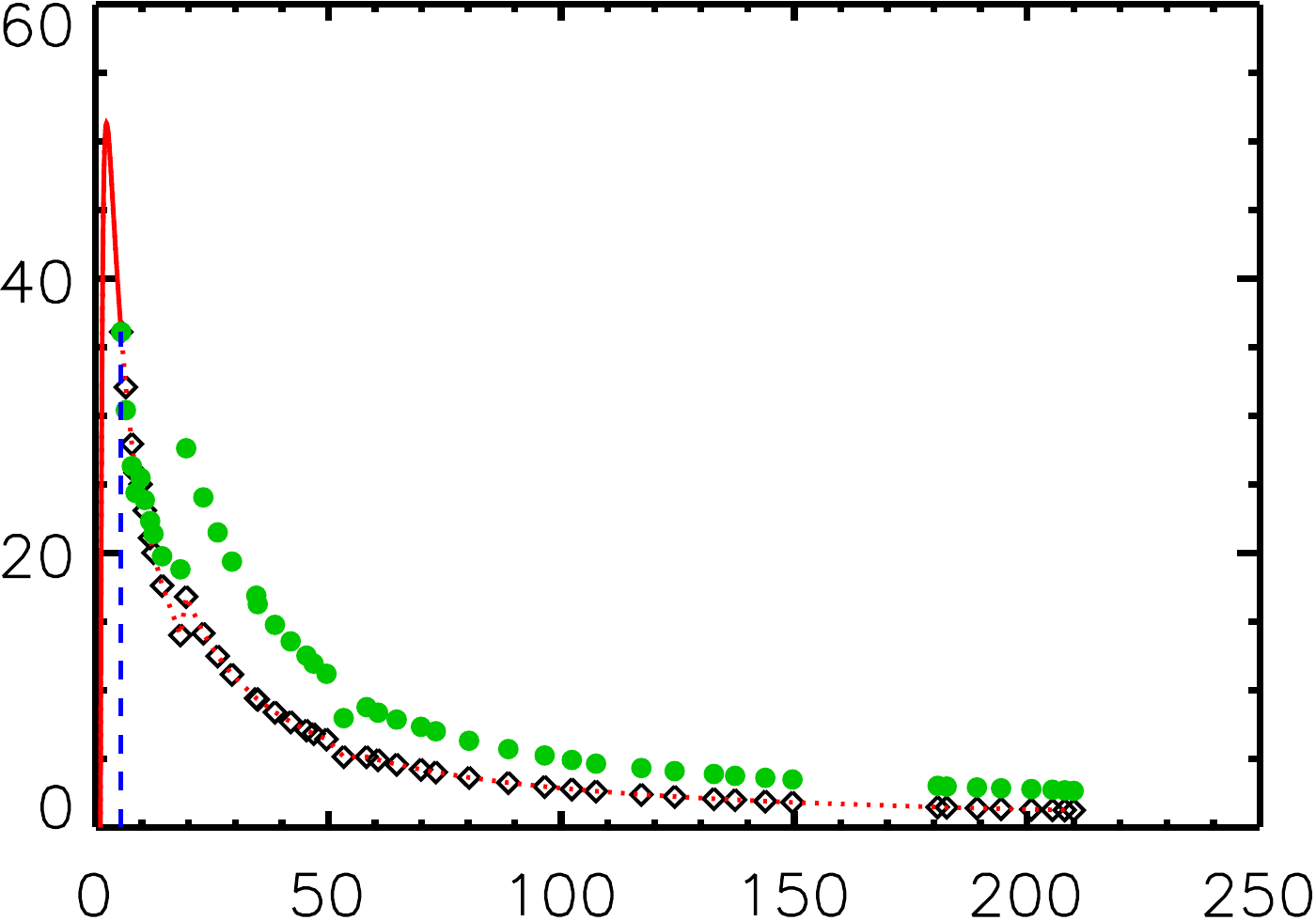}
		\put(-231,43.7){{\rotatebox{90}{{\color{black}\fontsize{12}{12}\fontseries{n}\fontfamily{phv}\selectfont  Force ($10^{17}$ dyn)}}}}
		\put(-140.8,-12.9){{\rotatebox{0}{{\color{black}\fontsize{12}{12}\fontseries{n}\fontfamily{phv}\selectfont Height (R$_{\odot}$)}}}}
		\put(-130,130.7){{\rotatebox{0}{{\color{black}\fontsize{13}{13}\fontseries{n}\fontfamily{phv}\selectfont  \underbar{CME 2 ($f$)}}}}}
		  }
  \vspace{0.0261\textwidth}  
  \caption[Height-time and Force profiles for CMEs 1 and 2]{Panel (a) (first column) depicts the observed and predicted height-time profiles and Panel (b) (second column) 
  shows the force profiles for CME 1 and CME 2. Each plot title indicates CME number referenced by Table \ref{tbl51}.
  In Panel (a), the observed height-time data is shown with diamonds. The red dash-dotted line is the drag model solution when it is initiated from the first 
  observed height, $h_{0}$. The blue solid line shows the predicted height-time trajectory when the drag model is initiated from height $\widetilde{h}_{0}$. 
  The fast CMEs are indicated by a ($f$) along with the CME number. For CME 2, the height-time solutions using constant $C_{\rm D}$ of 0.1 (green dash-dotted line) and 
  5 (brown dash-dotted line) are also shown. In Panel (b), the open diamond symbols (connected by a red dotted line) represent the Lorentz force values derived 
  observationally starting from $h_{0}$. The red solid line indicates the Lorentz force values for heights between $h_{eq}$ and $h_{0}$. 
  The filled green circles represent the absolute value of the solar wind drag force (beyond \htil). The blue dashed vertical line indicates the height $\widetilde{h}_{0}$ 
  ($=h_{0}$ for fast CMEs) at which the solar wind drag force takes over.}
  \label{fig51}
  \end{figure}

  The difference between the solar wind drag force and Lorentz driving force beyond 
  $\widetilde{h}_{0}$ is significantly pronounced in case of fast CMEs as can be seen from Figures \ref{fig51}\,--\,\ref{fig69}.
  We define a quantity, 
  \begin{equation}
  F_{\rm diff} \% \,=\,\frac{(F_{drag}-F_{Lorentz})}{F_{drag}} \times 100\, \% \,\, , \nonumber 
  \end{equation}
  which measures the relative difference between the two forces. $F_{\rm diff}$ (last column in Table \ref{tbl52}) is evaluated at 40 \Rs for  
  all events except for CME 11. Since for CME 11, $\widetilde{h}_{0}\sim 46$ \Rs, 
  $F_{\rm diff}$ is evaluated at 50 \Rs for this event and indicated by a superscript $^{\circledast}$ in Table \ref{tbl52}.
  Figure \ref{fig513}, plots $F_{\rm diff} \%$ \textit{versus} the CME initial velocity ($v_{0}$). As before, the blue circles indicate slow CMEs and 
  black ones indicate fast CMEs. The solar wind aerodynamic drag is about 50-90 \% larger than the Lorentz force at 40 \Rs for most of the 
  fast CMEs. For slower events, however, this range lies between 0.2\% and 30 \%. 

  For some slow CMEs, the Lorentz force is only slightly smaller that the drag force even 
  much beyond $\widetilde{h}_{0}$. For CMEs 4 and 10, we find that the Lorentz force magnitude is in fact larger than the solar wind drag force. 
  This could be because the computed Lorentz force is an overestimate. 
  In determining the Lorentz force, the torus instability model assumes the total magnetic flux to be frozen in. However, this assumption may not be entirely true. 
  The magnetic energy is dissipated in heating the CME plasma and/or CME expansion and the total magnetic flux may not be conserved. For slower CMEs, 
  our results suggest that it is important to take into account the energy expended in CME expansion and internal heating \citep[\textit{e.g.},][]{Kum96,Wan09,Ems12}.

  Reiterating the results described in Tables \ref{tbl51} and \ref{tbl52}: 
  \begin{enumerate}
  \item We determine the height at which solar wind drag dominates the CME dynamics.
  For each CME, this height ($\widetilde{h}_{0}$) is given in Table \ref{tbl51} together with the initiation velocity ($\widetilde{v}_{0}$) at this height. 
  Fast CMEs ($v_{0} >  900$ \kms) are drag dominated from very early on (\htil $=h_{0} \,=$ 3.5--8 
  \Rs). For slower CMEs ($v_{0}<900$ \kms), $\widetilde{h}_{0}$ ranges between 12--50 \Rs.

  \item The difference between observed final height ($h_{f}$) and the model predicted final height $\widetilde{h}_{f}$ is 
  given by $\Delta \widetilde{h}_{f}$.
  This difference lies between 0.2-10 for most of the CMEs, indicating that the drag model fits the observations well in this respect. 
  %
  %

  \item We note that the decay index $n$ ranges from 1.6--3. The fastest CMEs typically have the highest values of n. 

  \item The height at which the Lorentz force peaks ($h_{peak}$) lies between 1.65--2.45 \Rs.

  \item The $Fall\, \%$ is the amount by which the Lorentz force decreases from its maximum value at $h_{peak}$ up to $\widetilde{h}_{0}$.
  This quantity is larger for slower CMEs, $50< Fall \, \%<90 $ as compared to faster CMEs, $20<Fall\, \%<60$.

  \item Finally, the quantity $F_{\rm diff} \%$ evaluated at 40 \Rs (except for CME 11, where it is evaluated at 50 \Rs) denotes the relative 
  percentage difference between the two forces. For fast CMEs, $50 <F_{\rm diff}\, \% <90 $ whereas $0.2 <F_{\rm diff}\, \% <30$ for slower ones.
  \end{enumerate}

  A thorough understanding of ``how'' and ``where'' of the forces that affect CME propagation is needed to build reliable models for 
  accurately predicting the CME arrival 
  time and speed at the Earth. 
  This work is the first systematic study in this regard using a diverse CME sample set (to the best of our knowledge). 
  The force analysis described in this work will 
  provide a physical basis for inputs to space weather forecast models.

  \begin{table}
  \caption[Results of aerodynamic drag analysis]{Parameters and results of the drag analysis for all CMEs in the sample.
  The first column is the serial number of each event with which it is referenced.
  $\widetilde{h}_{0}$ is the height at which the drag force takes over the CME dynamics and $\widetilde{v}_{0}$ is the corresponding speed at this height.
  $h_{f}$ is the final observed GCS height.  $\widetilde{h}_{f}$ is the model predicted final height at the same time instant as $h_{f}$. 
  $\Delta \widetilde{h}_{f}\,=\, h_{f}\,=\, \widetilde{h}_{f}$. The last column indicates the predicted range of $C_{D}$ values 
  using the \citet{Sub12} prescription.}
  \label{tbl51}
  \centering
  \begin{tabular}{lcccccc} 
    \hline
	  &     &  &    &  & &  \\
  No.& $\widetilde{h}_{0}$ & $\widetilde{v}_{0}$& ${h}_{f}$ &  $\widetilde{h}_{f}$ & $\Delta \widetilde{h}_{f}$ &  $C_{D}$ range \\
      & (\Rs) &(km$s^{-1}$)& (\Rs) & (\Rs)& (\Rs) &  \\
    \hline
    1$^{*}$   & 21.9 & 383 & 155.4 & 148.3 & 7.1   & 0.1-0.3 \\
    2$^{* f}$ & 5.5  & 916 & 210.0 & 241.7 & -31.7 & 0.6-1.3 \\
    3$^{*}$   & 19.7 & 506 & 195.1 & 191.8 & 3.3  & 0.21-0.37\\
    4$^{*}$   & 15.2 & 437 & 163.3 & 159.0 & 4.3  & 0.19-0.29 \\
    5$^{*}$   & 27.7 & 490 & 150.3 & 149.1 & 1.2  & 0.38-0.55 \\
    6$^{*}$   & 20.1 & 445 & 131.5 & 141.8 & -10.3  & 0.27-0.45 \\
    7         & 27.1 & 583 & 59.7  & 60.2 &  -0.5 & 0.63-0.76 \\
    8         & 20.8 & 454 & 55.9  & 56.4 & -0.5  & 0.46-0.53 \\
    9$^{*}$   & 39.7 & 530 & 188.6 & 190.7 & -2.1  & 0.25-0.38\\
    10        & 18.2 & 511 & 42.2 & 42.9 & -0.7  & 0.22-0.31 \\
    11$^{*}$  & 46.5 & 456 & 211.0 & 212.0 & -1.0  & 0.27-0.34\\
    12        & 12.1 & 373 & 48.9 & 52.3 & -3.4  & 0.21-0.33 \\
    13        & 24.4 & 767 & 77.8 & 76.8 & 1.0  & 0.55-0.64\\
    14$^{f}$  & 8.4 & 1168 & 80.9 & 84.5 & -3.6 & 1.0-1.9 \\
    15$^{f}$  & 4.1 & 903  & 67.7 & 70.4 & -2.7 & 0.67-1.44 \\
    16$^{f}$  & 7.3 & 1638 & 67.1 & 61.3 & 5.8 & 1.41-1.93 \\
    17        & 38.8 & 636 & 73.6 & 69.4 & 4.2 & 0.38-0.42 \\
    18$^{f}$  & 4.0 & 1276 & 55.1 & 58.6 & -3.5 & 1.63-3.05 \\
    19        & 30.5 & 313 & 52.6 & 53.1 & -0.5  & 0.25-0.3\\
    20        & 39.4 & 491 & 75.4 & 74.7 & 0.7 & 0.34-0.45 \\
    21$^{f}$  & 4.6 & 1823 & 100.0 & 91.6 & 8.4  & 1.4-3.4\\
    22$^{f}$  & 4.0 & 1910 & 86.8 & 90.7 & -3.9  & 1.87-3.5\\
    23$^{f}$  & 3.5 & 2397 & 92.0 & 99.7 & -7.7 & 1.7-3.23\\
    24$^{f}$  & 3.9 & 1837 & 71.4 & 81.4 & -10.0 & 1.24-1.83 \\
    25        & 23.1 & 684 & 77.9 & 77.5 & 0.4 & 0.76-1.1\\
    26$^{f}$  & 6.2 & 1152 & 85.5 & 85.2 & 0.3 & 0.93-1.22\\
    27$^{f}$  & 4.4 & 1248 & 72.3 & 82.2 & -9.9 & 1.12-1.58 \\
    28$^{f}$  & 6.7 & 1305 & 87.4 & 86.2 & 1.2  & 1.44-2.3\\
    29        & 31.1 & 790 & 70.8 & 70.3 & 0.5 & 0.85-1.03\\
    30        & 36.9 & 570 & 88.6 & 89.9 & -1.3 & 0.51-0.66 \\
    31        & 26.5 & 597 & 88.9 & 89.2 & -0.3 & 0.77-1.38 \\
    32        & 27.7 & 668 & 58.5 & 58.3 & 0.2 & 0.8-0.98\\
    33$^{f}$  & 4.7 & 1504 & 90.5 & 84.4 & 6.1 & 0.76-1.68\\
    34$^{f}$  & 5.9 & 1115 & 86.2 & 84.1 & 2.1 & 0.59-0.76\\
    35$^{f}$  & 6.6 & 1637 & 60.9 & 65.0 & -4.1 & 1.21-2.99\\
    36$^{f}$  & 4.9 & 1217 & 77.6 & 77.9 & -0.3 & 1.35-2.42\\
    37$^{f}$  & 5.9 & 975  & 78.2 & 79.1 & -0.9 & 0.84-1.22\\
    38$^{f}$  & 6.8 & 1039 & 48.7 & 51.4 & -2.7 & 0.96-2.1 \\
    \hline
  \end{tabular}
  \end{table}

  \begin{table}
  \caption[Results of Lorentz force analysis]{Column 1 is the CME serial number. $I_{eq}$ is the axial CME current at equilibrium ($h_{eq}$).
  $B_{ext}(h_{eq})$ is the external magnetic field at equilibrium. $F_{\rm Lorentz}$ is the Lorentz force magnitude at $\widetilde{h}_{0}$ (equal to $|F_{drag}(\widetilde{h}_{0})|$). The
  position of maximum Lorentz force is indicated by $h_{peak}$. The $Fall\%$ is the amount by which the Lorentz force decreases from 
  its peak value ($h_{peak}$) to $\widetilde{h}_{0}$. The last column indicates $F_{\rm diff} = (\frac{F_{drag}-F_{Lorentz}}{F_{drag}})\times 100 \%$ 
  at 40 $\rm R_{\odot}$. For CME 11 ($^\circledast$), $F_{\rm diff} \, \%$ is evaluated at 50 \Rs.}
  \label{tbl52}
  \centering
  \begin{tabular}{lcccccc} 
    \hline
  No.& $I_{eq}$ & $B_{ext}(h_{eq})$ & $F_{Lorentz}(\widetilde{h}_{0})$ & $h_{peak}$ & $Fall \%$ &  $F_{\rm diff} \%$ \\
      & ($10^{10}$ A) &($10^{-1}$ G)& ($10^{17}$ dyn) & (\Rs)& (\%) & (\%) \\
    \hline
    1$^{*}$   & 0.41 & 0.13 & 0.10  & 1.75 &  96  & 18.6\\
    2$^{* f}$ & 3.13 & 0.94 & 36.12 & 2.35 & 30 &  43.2\\
    3$^{*}$   & 0.55 & 0.19 & 0.49 & 2.05 & 86  & 16.3\\
    4$^{*}$   & 0.31 & 0.11 & 0.12 & 1.75 & 93  & -41.3 \\
    5$^{*}$   & 1.77 & 0.33 & 1.52 & 2.35 & 79  & 6.6 \\
    6$^{*}$   & 0.66 & 0.22 & 0.85 & 2.25 & 73  & 5.9 \\
    7         & 2.30 & 0.65 & 5.31  & 2.35 & 81 & 4.2 \\
    8         & 1.21 & 0.38 & 1.78  & 2.35 & 77 & 19.7 \\
    9$^{*}$   & 1.10 & 0.29 & 0.42 & 1.95 & 97 & 0.2\\
    10        & 0.50 & 0.15 & 0.19 & 1.75 & 95  & -33.3 \\
    11$^{*}$  & 0.71 & 0.25 & 0.29 & 2.05 & 94  & 0.9 ${^\circledast}$\\
    12        & 0.47 & 0.15 & 0.33 & 1.75 & 91  & 24.3 \\
    13        & 1.72 & 0.56 & 3.11 & 2.35 & 80  & 30.7\\
    14$^{f}$  & 6.26 & 1.71 & 105.37 & 2.35 & 48 & 21.6 \\
    15$^{f}$  & 2.66 & 0.79 & 49.41 & 2.05 & 29 & 52.4 \\
    16$^{f}$  & 5.90 & 1.39 & 106.87 & 2.45 & 41 & 61.3 \\
    17        & 1.06 & 0.29 & 0.67 & 2.25 & 91 & 0.3 \\
    18$^{f}$  & 8.40 & 2.09 & 510.45 & 1.95 & 35 & 80.0 \\
    19        & 0.47 & 0.13 & 0.11 & 1.95 & 96 & 2.1\\
    20        & 1.67 & 0.49 & 0.79 & 1.95 & 98 & 0.3 \\
    21$^{f}$  & 11.60& 3.11 & 847.54 & 1.65 & 66 & 80.9\\
    22$^{f}$  & 10.30& 2.74 & 829.15 & 1.65 & 58  & 93.7\\
    23$^{f}$  & 8.51 & 2.47 & 746.12 & 1.65 & 49 & 94.6\\
    24$^{f}$  & 3.92 & 0.91 & 100.22& 2.05 & 25 & 83.2 \\
    25        & 3.68 & 1.19 & 20.68 & 2.35 & 71 & 3.3\\
    26$^{f}$  & 2.89 & 0.84 & 28.15 & 2.35 & 35 & 70.5\\
    27$^{f}$  & 4.07 & 1.11 & 70.02 & 2.35 & 18 & 64.9 \\
    28$^{f}$  & 8.53 & 2.37 & 231.39 & 2.35 & 39  & 59.2\\
    29        & 4.05 & 1.28 & 17.89 & 2.35 & 79 & 7.8\\
    30        & 1.56 & 0.56 & 2.05 & 2.35 & 84 & 29.2 \\
    31        & 11.07& 2.96 & 41.31 & 1.75 & 98 & 4.4 \\
    32        & 3.41 & 0.89 & 10.64 & 2.25 & 86 & 10.4\\
    33$^{f}$  & 4.29 & 1.32 & 106.99 & 2.15 & 32 & 56.0\\
    34$^{f}$  & 1.29 & 0.52 & 6.13 & 2.35 & 34 & 83.5\\
    35$^{f}$  & 9.55 & 2.69 & 369.41 & 1.85 & 26 & 60.3\\
    36$^{f}$  & 7.06 & 2.04 & 312.65 & 1.95 & 48 & 80.7\\
    37$^{f}$  & 2.50 & 0.75 & 26.29 & 2.25 & 60 & 68.3\\
    38$^{f}$  & 6.91 & 2.04 & 202.65 & 2.05 & 57 & 8.9 \\
  \hline
  \end{tabular}
  \end{table}

  \clearpage

  \clearpage
  \begin{figure}[h]    
    \centering                              
    \centerline{\hspace*{0.06\textwidth}
		\includegraphics[width=0.55\textwidth,clip=]{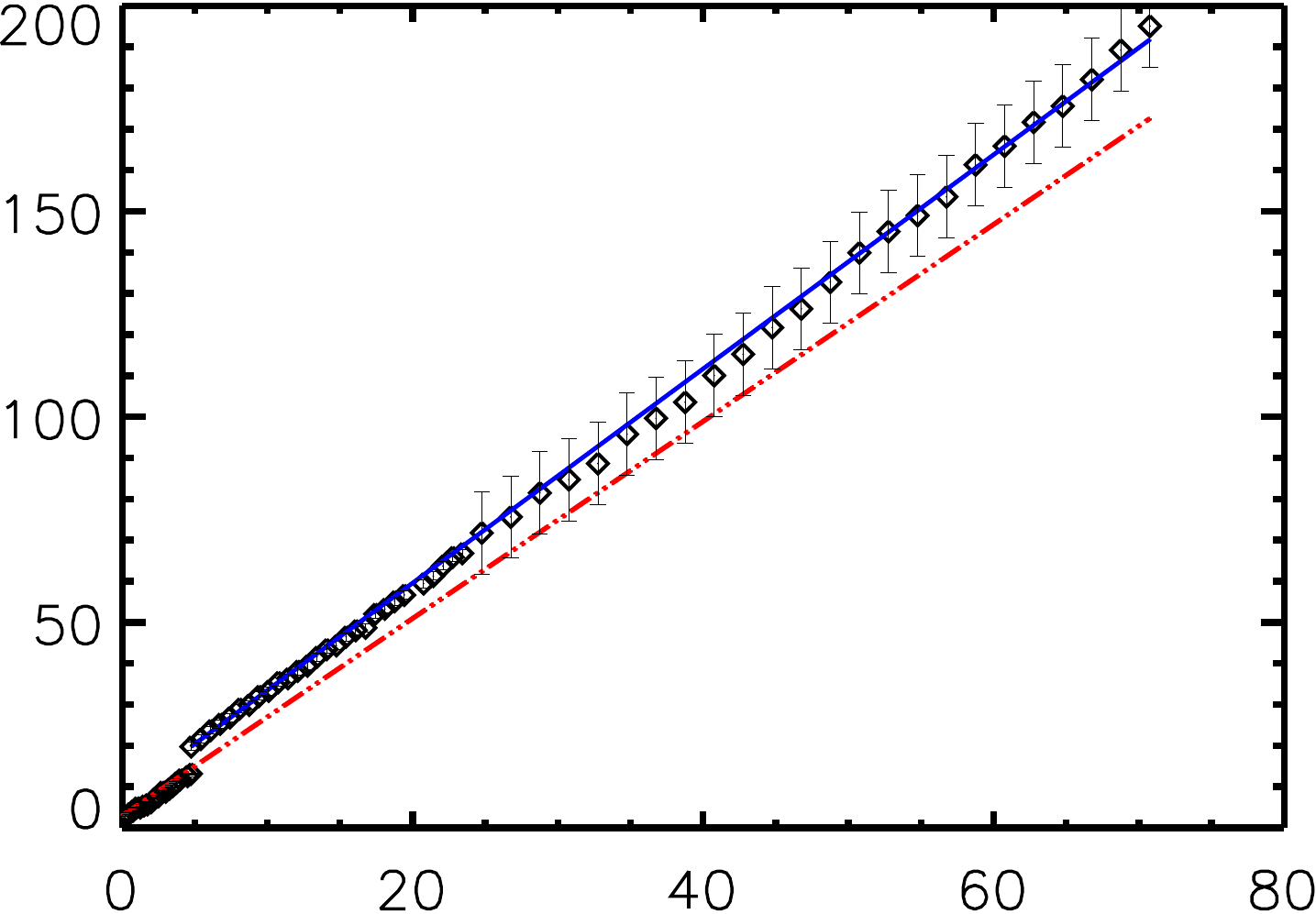}
		\put(-230,62.7){{\rotatebox{90}{{\color{black}\fontsize{12}{12}\fontseries{n}\fontfamily{phv}\selectfont  Height (R$_{\odot}$)}}}}
		\put(-160.8,-14.9){{\rotatebox{0}{{\color{black}\fontsize{12}{12}\fontseries{n}\fontfamily{phv}\selectfont  Elapsed Time (hrs)}}}}
		\put(-130,130.7){{\rotatebox{0}{{\color{black}\fontsize{13}{13}\fontseries{n}\fontfamily{phv}\selectfont \underbar{CME 3}}}}}
		\put(-115,158.7){{\rotatebox{0}{{\color{black}\fontsize{13}{13}\fontseries{n}\fontfamily{phv}\selectfont (a)}}}}
		\hspace*{0.069\textwidth}
		\includegraphics[width=0.53\textwidth,clip=]{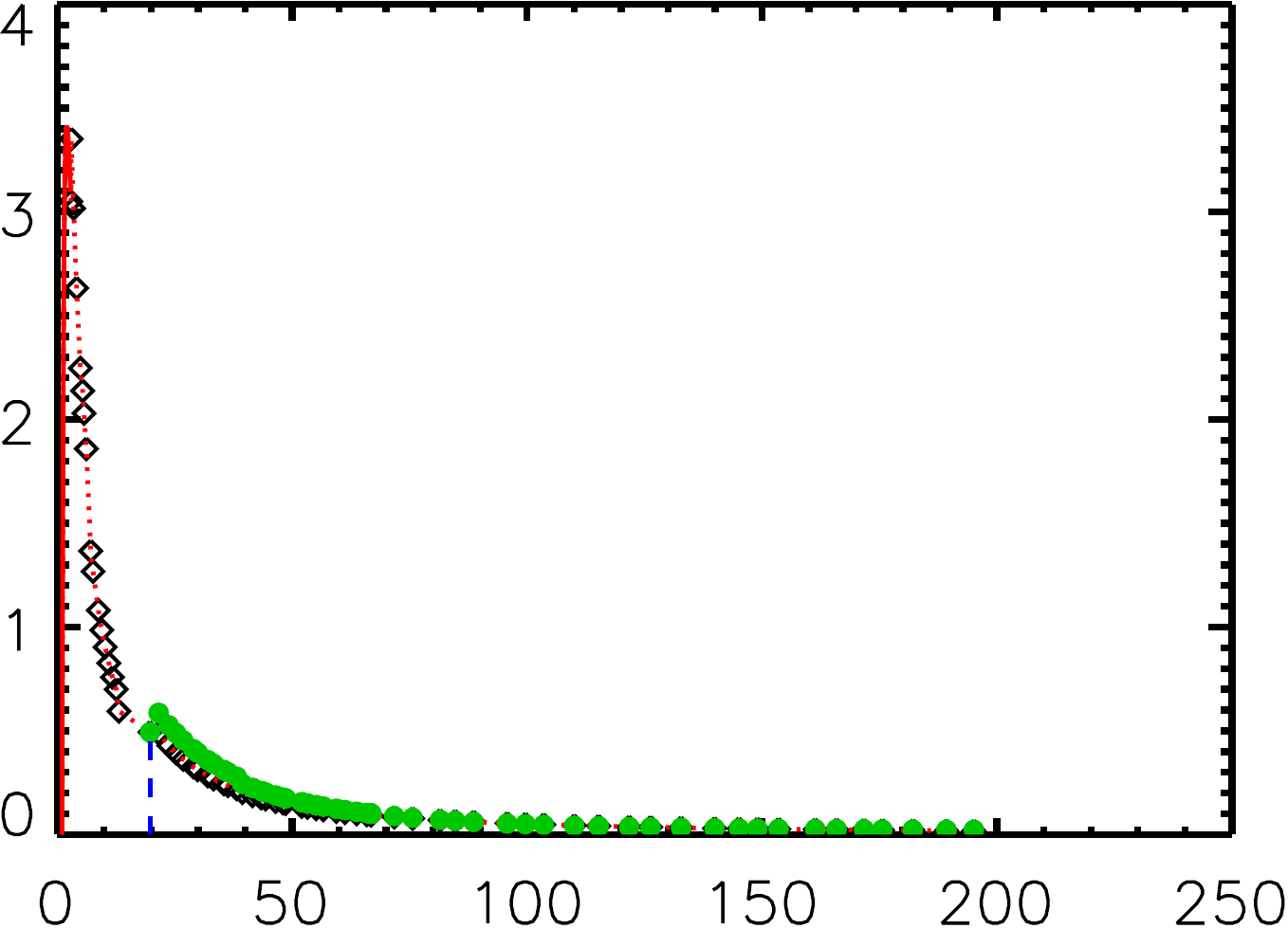}
		\put(-231,43.7){{\rotatebox{90}{{\color{black}\fontsize{12}{12}\fontseries{n}\fontfamily{phv}\selectfont  Force ($10^{17}$ dyn)}}}}
		\put(-140.8,-12.9){{\rotatebox{0}{{\color{black}\fontsize{12}{12}\fontseries{n}\fontfamily{phv}\selectfont Height (R$_{\odot}$)}}}}
		\put(-130,130.7){{\rotatebox{0}{{\color{black}\fontsize{13}{13}\fontseries{n}\fontfamily{phv}\selectfont  \underbar{CME 3}}}}}
		\put(-117,158.7){{\rotatebox{0}{{\color{black}\fontsize{13}{13}\fontseries{n}\fontfamily{phv}\selectfont (b)}}}}          
		  }
		\vspace{1.3cm}
		  \centerline{\hspace*{0.065\textwidth}
		\includegraphics[width=0.55\textwidth,clip=]{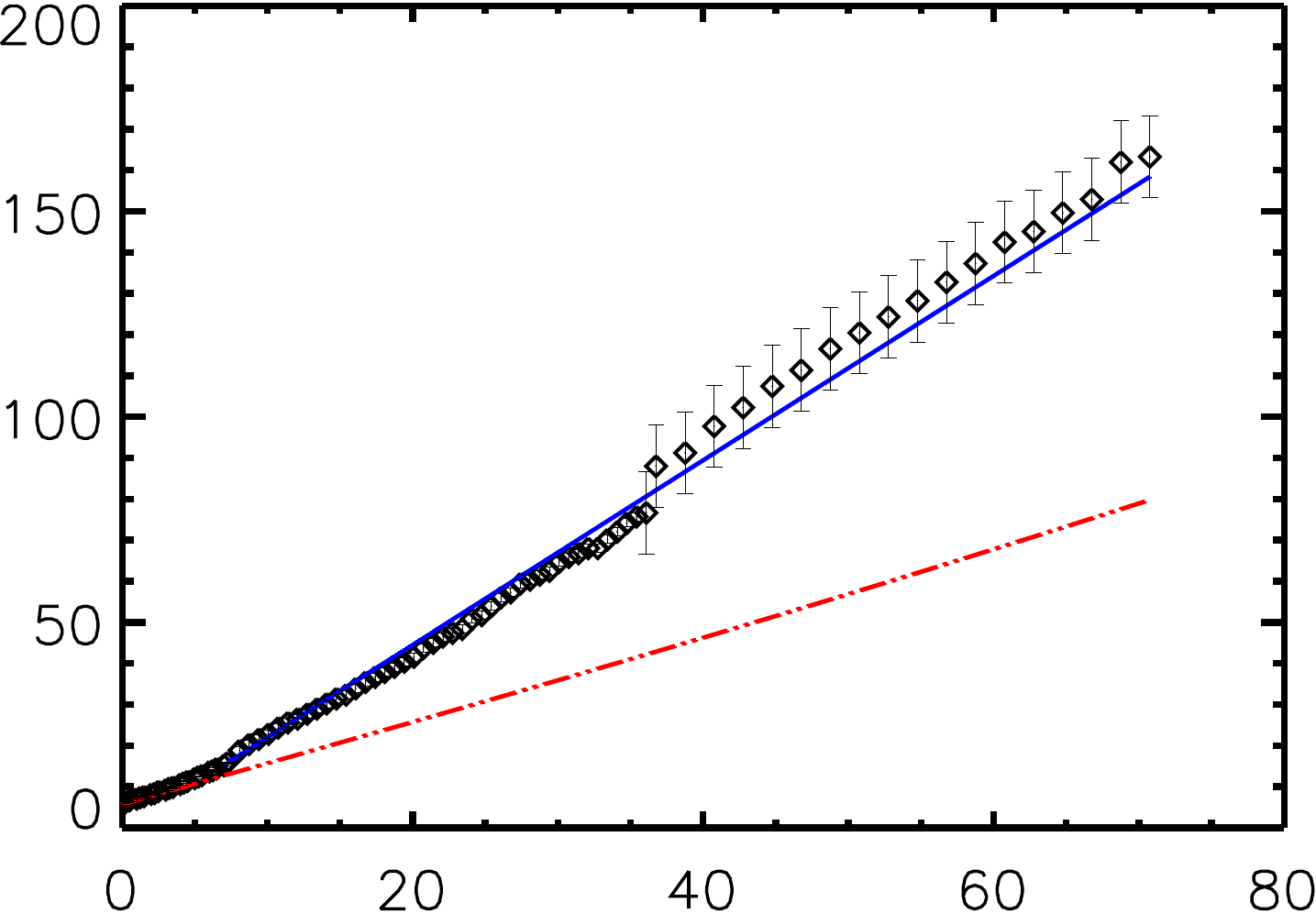}
		\put(-230,62.7){{\rotatebox{90}{{\color{black}\fontsize{12}{12}\fontseries{n}\fontfamily{phv}\selectfont  Height (R$_{\odot}$)}}}}
		\put(-160.8,-14.9){{\rotatebox{0}{{\color{black}\fontsize{12}{12}\fontseries{n}\fontfamily{phv}\selectfont  Elapsed Time (hrs)}}}}
		\put(-130,130.7){{\rotatebox{0}{{\color{black}\fontsize{13}{13}\fontseries{n}\fontfamily{phv}\selectfont \underbar{CME 4}}}}}
		\hspace*{0.049\textwidth}
		\includegraphics[width=0.55\textwidth,clip=]{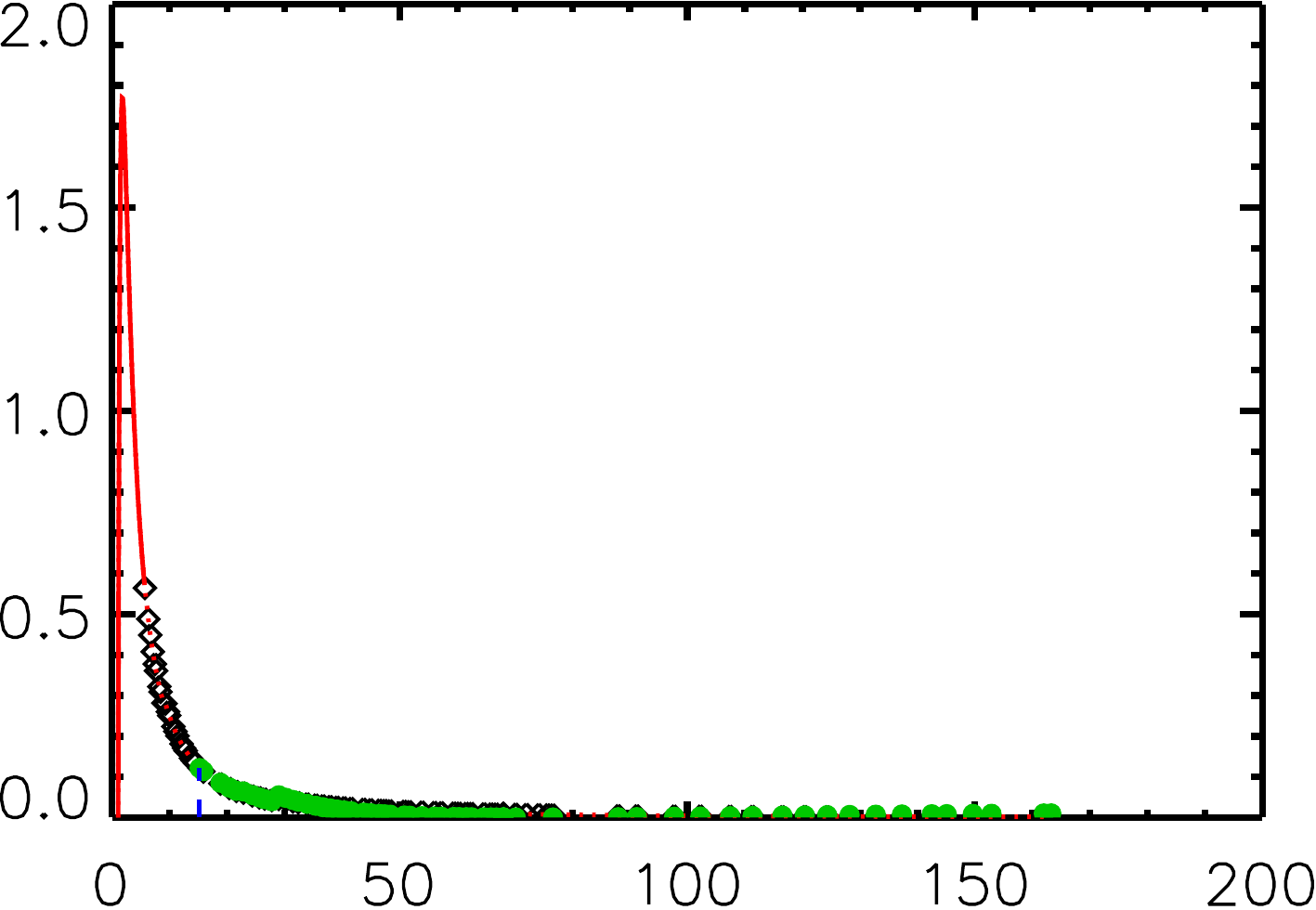}
		\put(-231,43.7){{\rotatebox{90}{{\color{black}\fontsize{12}{12}\fontseries{n}\fontfamily{phv}\selectfont  Force ($10^{17}$ dyn)}}}}
		\put(-140.8,-12.9){{\rotatebox{0}{{\color{black}\fontsize{12}{12}\fontseries{n}\fontfamily{phv}\selectfont Height (R$_{\odot}$)}}}}
		\put(-130,130.7){{\rotatebox{0}{{\color{black}\fontsize{13}{13}\fontseries{n}\fontfamily{phv}\selectfont  \underbar{CME 4}}}}}
		  }
  \vspace{0.0261\textwidth}  
  \caption[Height-time and Force profiles for CMEs 3 and 4]{Panel (a) depicts the observed and predicted height-time profiles and Panel (b) shows the 
  force profiles for CME 3 and CME 4. Each plot title indicates CME number referenced by Table \ref{tbl51}.
  In Panel (a), the observed height-time data is shown with diamonds. The red dash-dotted line is the drag model solution when it is initiated from the first 
  observed height, $h_{0}$. The blue solid line shows the predicted height-time trajectory when the drag model is initiated from height $\widetilde{h}_{0}$. 
  The fast CMEs are indicated by a ($f$) along with the CME number. In Panel (b), the open diamond symbols (connected by red dotted line) represent the Lorentz force values derived 
  observationally starting from $h_{0}$. The red solid line indicates the Lorentz force values for heights between $h_{eq}$ and $h_{0}$. 
  The filled green circles represent the absolute value of the solar wind drag force. The blue dashed vertical line indicates the height $\widetilde{h}_{0}$ 
  ($=h_{0}$ for fast CMEs) at which the solar wind drag force takes over.}
  \label{fig52}
  \end{figure}

  \clearpage
  \begin{figure}[h]    
    \centering                              
    \centerline{\hspace*{0.06\textwidth}
		\includegraphics[width=0.55\textwidth,clip=]{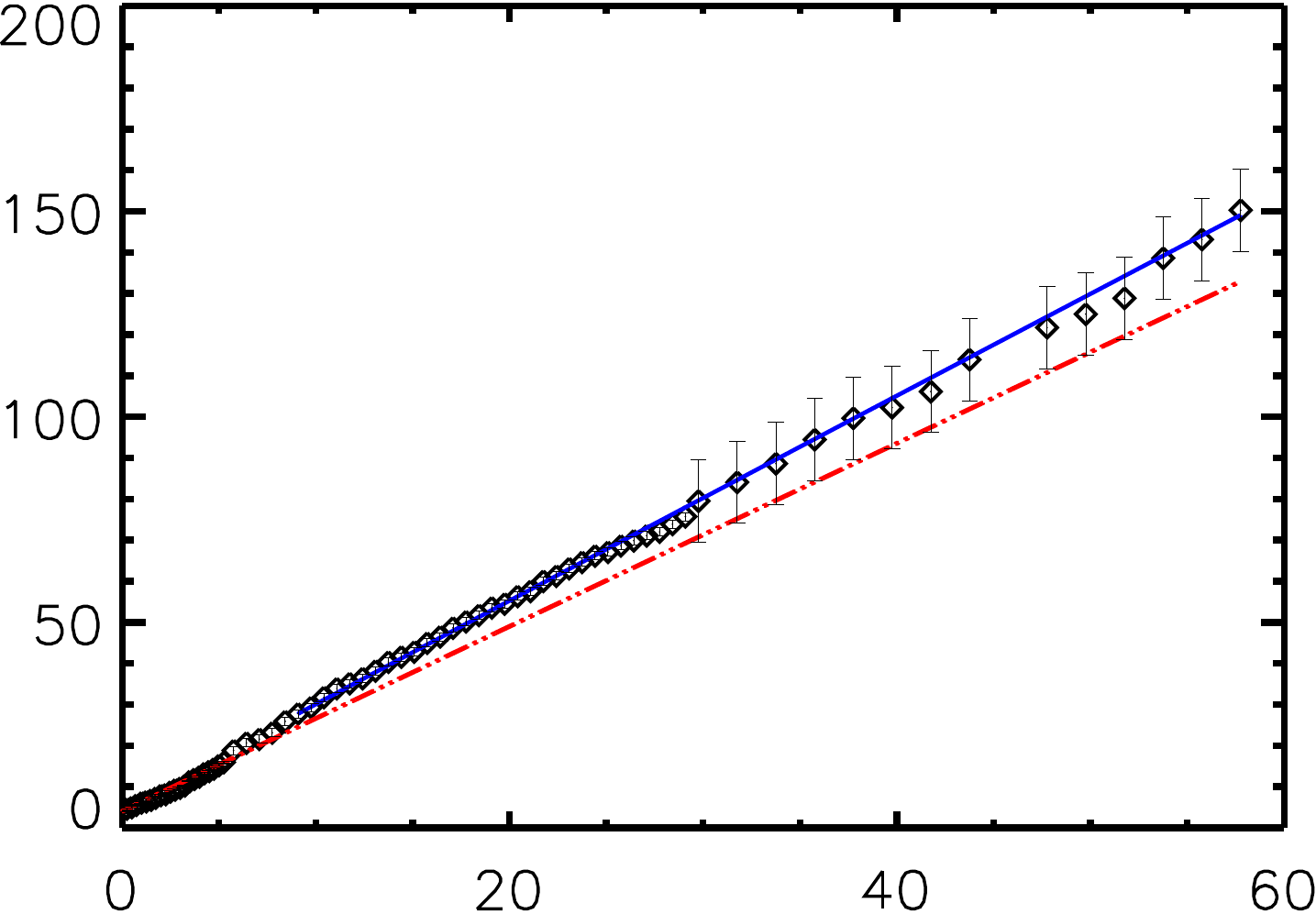}
		\put(-230,62.7){{\rotatebox{90}{{\color{black}\fontsize{12}{12}\fontseries{n}\fontfamily{phv}\selectfont  Height (R$_{\odot}$)}}}}
		\put(-160.8,-14.9){{\rotatebox{0}{{\color{black}\fontsize{12}{12}\fontseries{n}\fontfamily{phv}\selectfont  Elapsed Time (hrs)}}}}
		\put(-130,130.7){{\rotatebox{0}{{\color{black}\fontsize{13}{13}\fontseries{n}\fontfamily{phv}\selectfont \underbar{CME 5}}}}}
		\put(-115,158.7){{\rotatebox{0}{{\color{black}\fontsize{13}{13}\fontseries{n}\fontfamily{phv}\selectfont (a)}}}}
		\hspace*{0.069\textwidth}
		\includegraphics[width=0.53\textwidth,clip=]{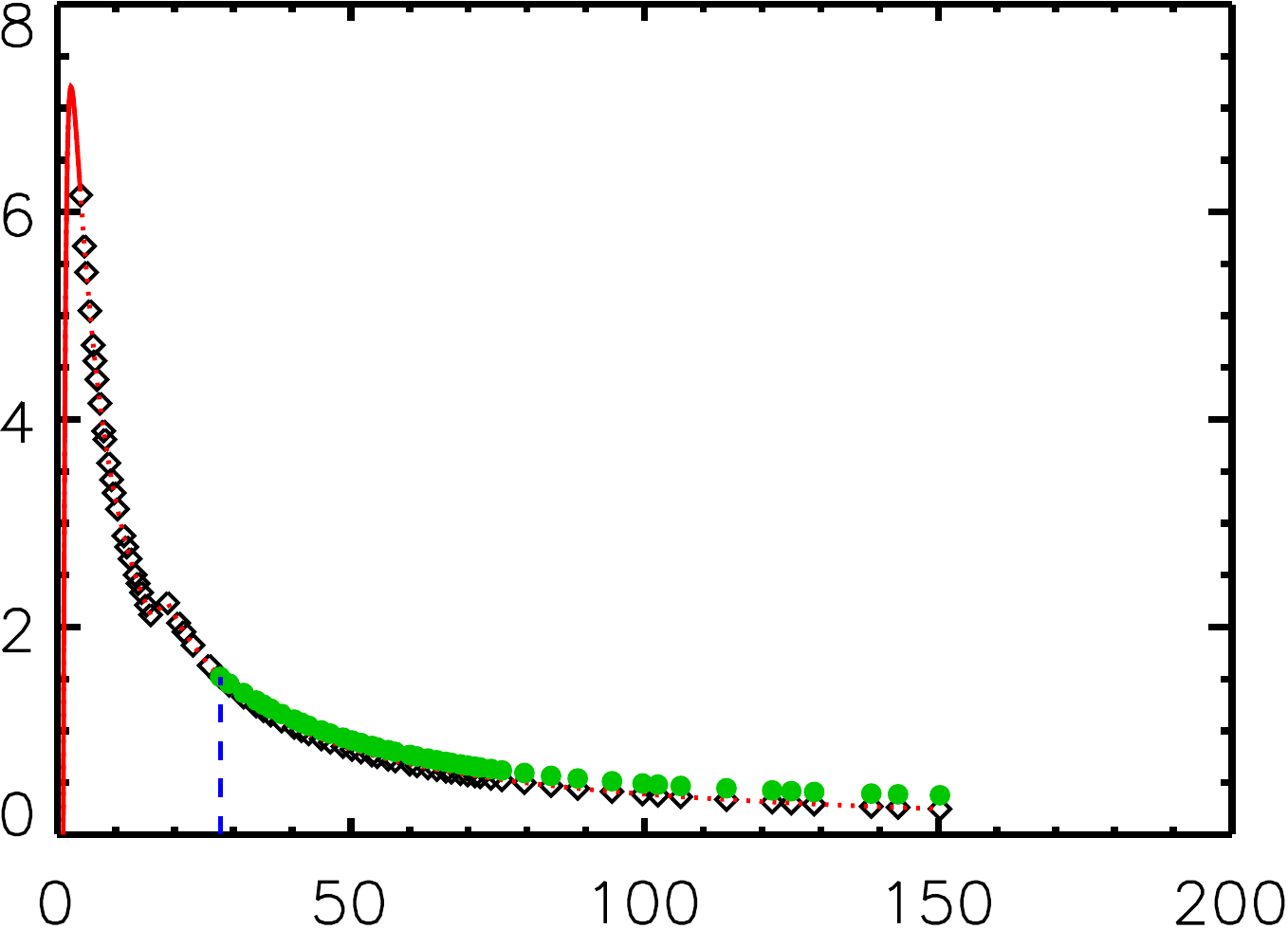}
		\put(-225,43.7){{\rotatebox{90}{{\color{black}\fontsize{12}{12}\fontseries{n}\fontfamily{phv}\selectfont  Force ($10^{17}$ dyn)}}}}
		\put(-140.8,-12.9){{\rotatebox{0}{{\color{black}\fontsize{12}{12}\fontseries{n}\fontfamily{phv}\selectfont Height (R$_{\odot}$)}}}}
		\put(-130,130.7){{\rotatebox{0}{{\color{black}\fontsize{13}{13}\fontseries{n}\fontfamily{phv}\selectfont  \underbar{CME 5}}}}}
		\put(-117,158.7){{\rotatebox{0}{{\color{black}\fontsize{13}{13}\fontseries{n}\fontfamily{phv}\selectfont (b)}}}}          
		  }
		\vspace{1.3cm}
		  \centerline{\hspace*{0.06\textwidth}
		\includegraphics[width=0.56\textwidth,clip=]{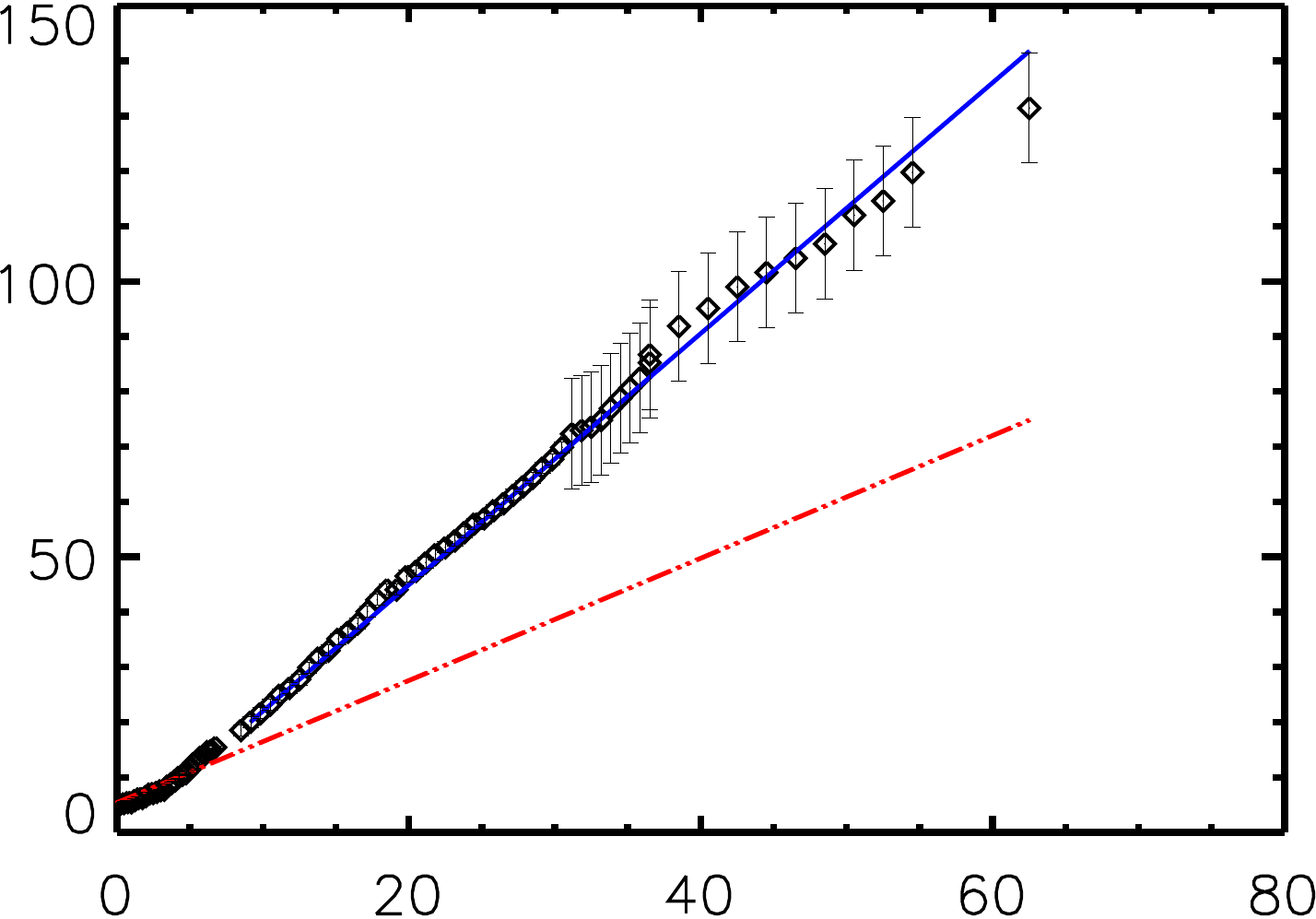}
		\put(-230,62.7){{\rotatebox{90}{{\color{black}\fontsize{12}{12}\fontseries{n}\fontfamily{phv}\selectfont  Height (R$_{\odot}$)}}}}
		\put(-160.8,-14.9){{\rotatebox{0}{{\color{black}\fontsize{12}{12}\fontseries{n}\fontfamily{phv}\selectfont  Elapsed Time (hrs)}}}}
		\put(-130,130.7){{\rotatebox{0}{{\color{black}\fontsize{13}{13}\fontseries{n}\fontfamily{phv}\selectfont \underbar{CME 6}}}}}
		\hspace*{0.069\textwidth}
		\includegraphics[width=0.545\textwidth,clip=]{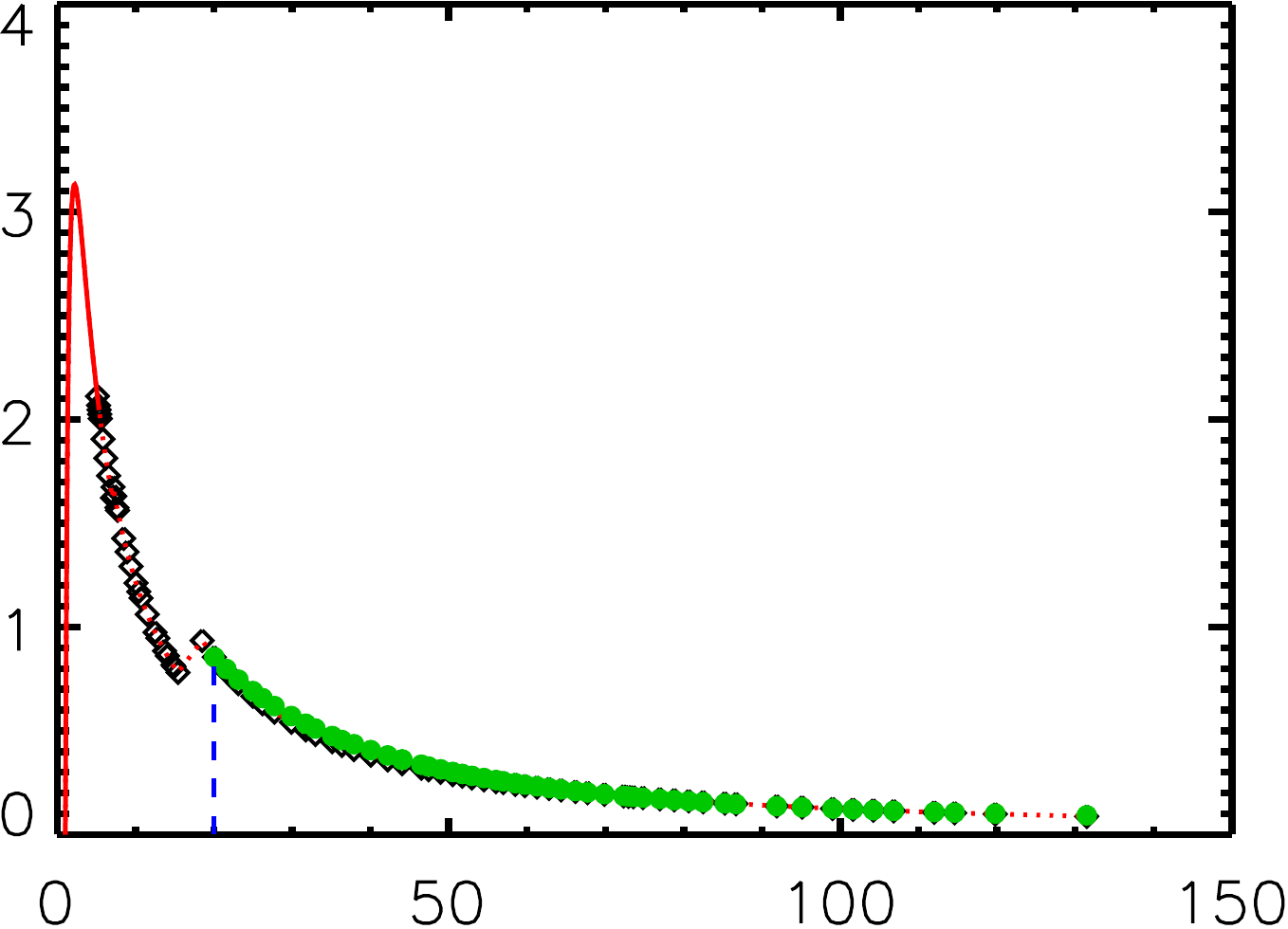}
		\put(-231,43.7){{\rotatebox{90}{{\color{black}\fontsize{12}{12}\fontseries{n}\fontfamily{phv}\selectfont  Force ($10^{17}$ dyn)}}}}
		\put(-140.8,-12.9){{\rotatebox{0}{{\color{black}\fontsize{12}{12}\fontseries{n}\fontfamily{phv}\selectfont Height (R$_{\odot}$)}}}}
		\put(-130,130.7){{\rotatebox{0}{{\color{black}\fontsize{13}{13}\fontseries{n}\fontfamily{phv}\selectfont  \underbar{CME 6}}}}}
		  }
  \vspace{0.0261\textwidth}  
  \caption[Height-time and Force profiles for CMEs 5 and 6]{Height-time and Force profiles for CMEs 5 and 6. Caption same as Figure \ref{fig52}.}
  \label{fig53}
  \end{figure}

  \clearpage
  \begin{figure}[h]    
    \centering                              
    \centerline{\hspace*{0.06\textwidth}
		\includegraphics[width=0.55\textwidth,clip=]{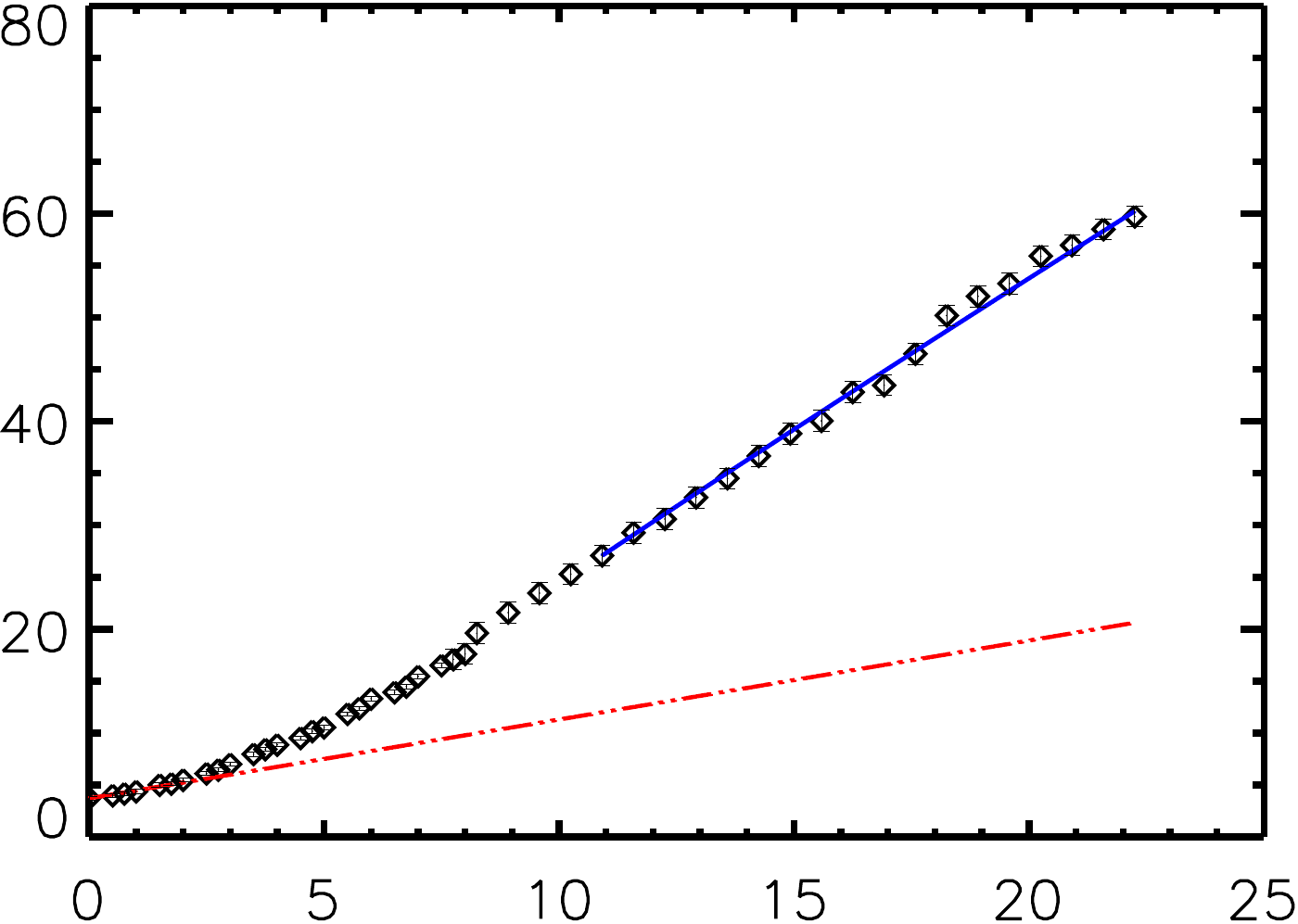}
		\put(-230,62.7){{\rotatebox{90}{{\color{black}\fontsize{12}{12}\fontseries{n}\fontfamily{phv}\selectfont  Height (R$_{\odot}$)}}}}
		\put(-160.8,-14.9){{\rotatebox{0}{{\color{black}\fontsize{12}{12}\fontseries{n}\fontfamily{phv}\selectfont  Elapsed Time (hrs)}}}}
		\put(-130,130.7){{\rotatebox{0}{{\color{black}\fontsize{13}{13}\fontseries{n}\fontfamily{phv}\selectfont \underbar{CME 7}}}}}
		\put(-115,158.7){{\rotatebox{0}{{\color{black}\fontsize{13}{13}\fontseries{n}\fontfamily{phv}\selectfont (a)}}}}
		\hspace*{0.069\textwidth}
		\includegraphics[width=0.55\textwidth,clip=]{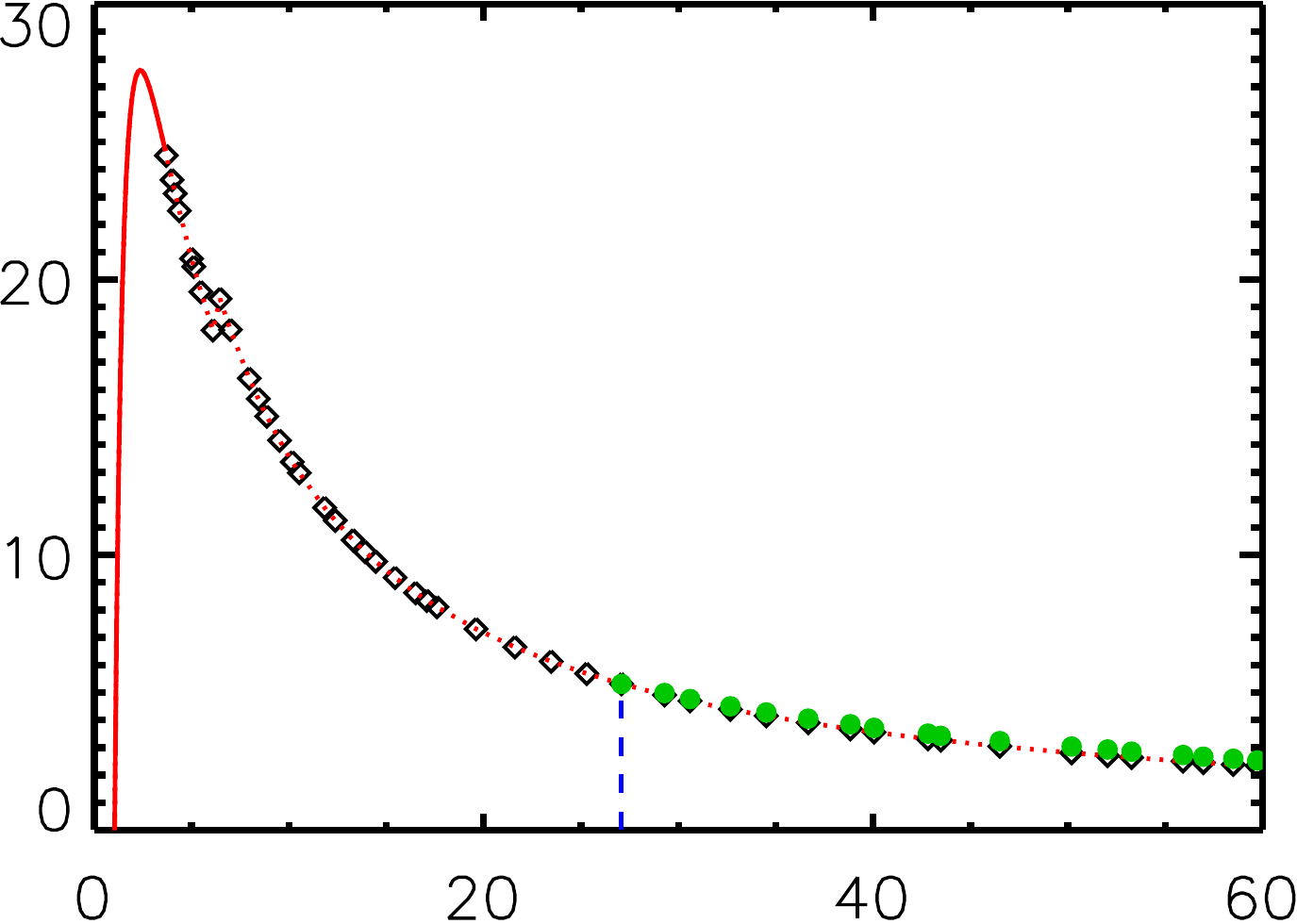}
		\put(-231,43.7){{\rotatebox{90}{{\color{black}\fontsize{12}{12}\fontseries{n}\fontfamily{phv}\selectfont  Force ($10^{17}$ dyn)}}}}
		\put(-140.8,-12.9){{\rotatebox{0}{{\color{black}\fontsize{12}{12}\fontseries{n}\fontfamily{phv}\selectfont Height (R$_{\odot}$)}}}}
		\put(-130,130.7){{\rotatebox{0}{{\color{black}\fontsize{13}{13}\fontseries{n}\fontfamily{phv}\selectfont  \underbar{CME 7}}}}}
		\put(-117,158.7){{\rotatebox{0}{{\color{black}\fontsize{13}{13}\fontseries{n}\fontfamily{phv}\selectfont (b)}}}}          
		  }
		\vspace{1.3cm}
		  \centerline{\hspace*{0.06\textwidth}
		\includegraphics[width=0.565\textwidth,clip=]{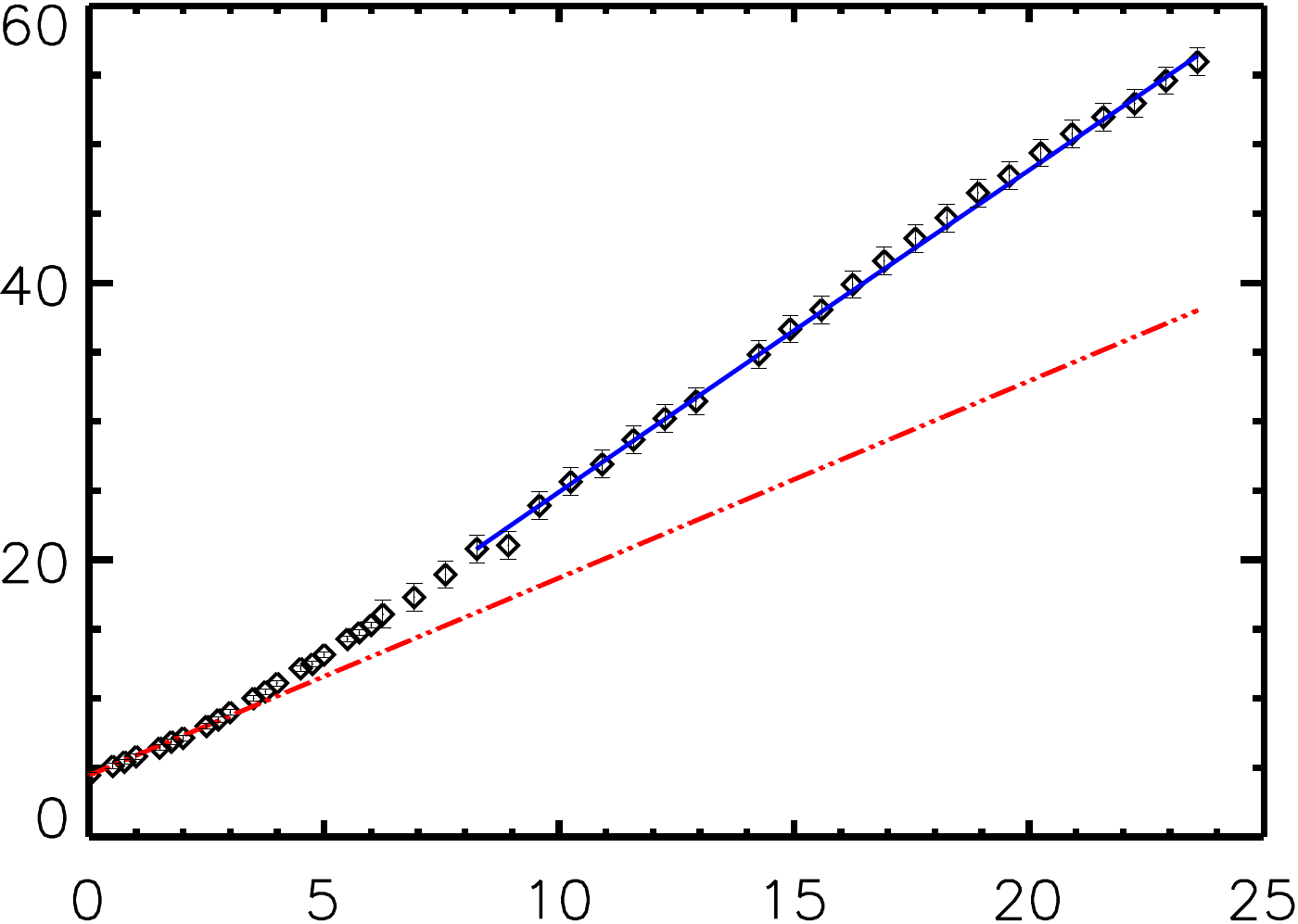}
		\put(-230,62.7){{\rotatebox{90}{{\color{black}\fontsize{12}{12}\fontseries{n}\fontfamily{phv}\selectfont  Height (R$_{\odot}$)}}}}
		\put(-160.8,-14.9){{\rotatebox{0}{{\color{black}\fontsize{12}{12}\fontseries{n}\fontfamily{phv}\selectfont  Elapsed Time (hrs)}}}}
		\put(-130,130.7){{\rotatebox{0}{{\color{black}\fontsize{13}{13}\fontseries{n}\fontfamily{phv}\selectfont \underbar{CME 8}}}}}
		\hspace*{0.069\textwidth}
		\includegraphics[width=0.55\textwidth,clip=]{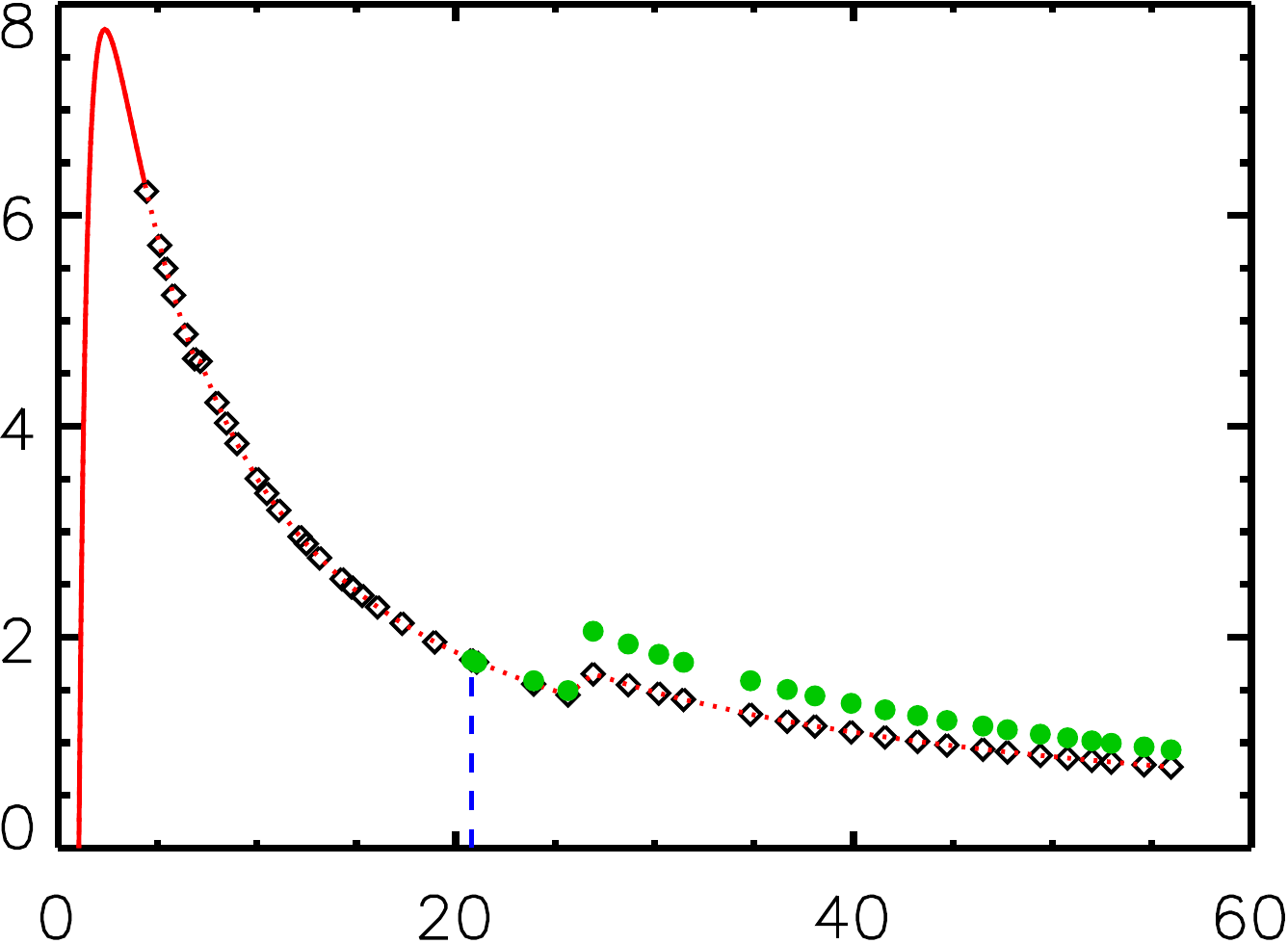}
		\put(-231,43.7){{\rotatebox{90}{{\color{black}\fontsize{12}{12}\fontseries{n}\fontfamily{phv}\selectfont  Force ($10^{17}$ dyn)}}}}
		\put(-140.8,-12.9){{\rotatebox{0}{{\color{black}\fontsize{12}{12}\fontseries{n}\fontfamily{phv}\selectfont Height (R$_{\odot}$)}}}}
		\put(-130,130.7){{\rotatebox{0}{{\color{black}\fontsize{13}{13}\fontseries{n}\fontfamily{phv}\selectfont  \underbar{CME 8}}}}}
		  }
  \vspace{0.0261\textwidth}  
  \caption[Height-time and Force profiles for CMEs 7 and 8]{Height-time and Force profiles for CMEs 7 and 8. Caption same as Figure \ref{fig52}.}
  \label{fig54}
  \end{figure}

  \clearpage
  \begin{figure}[h]    
    \centering                              
    \centerline{\hspace*{0.06\textwidth}
		\includegraphics[width=0.56\textwidth,clip=]{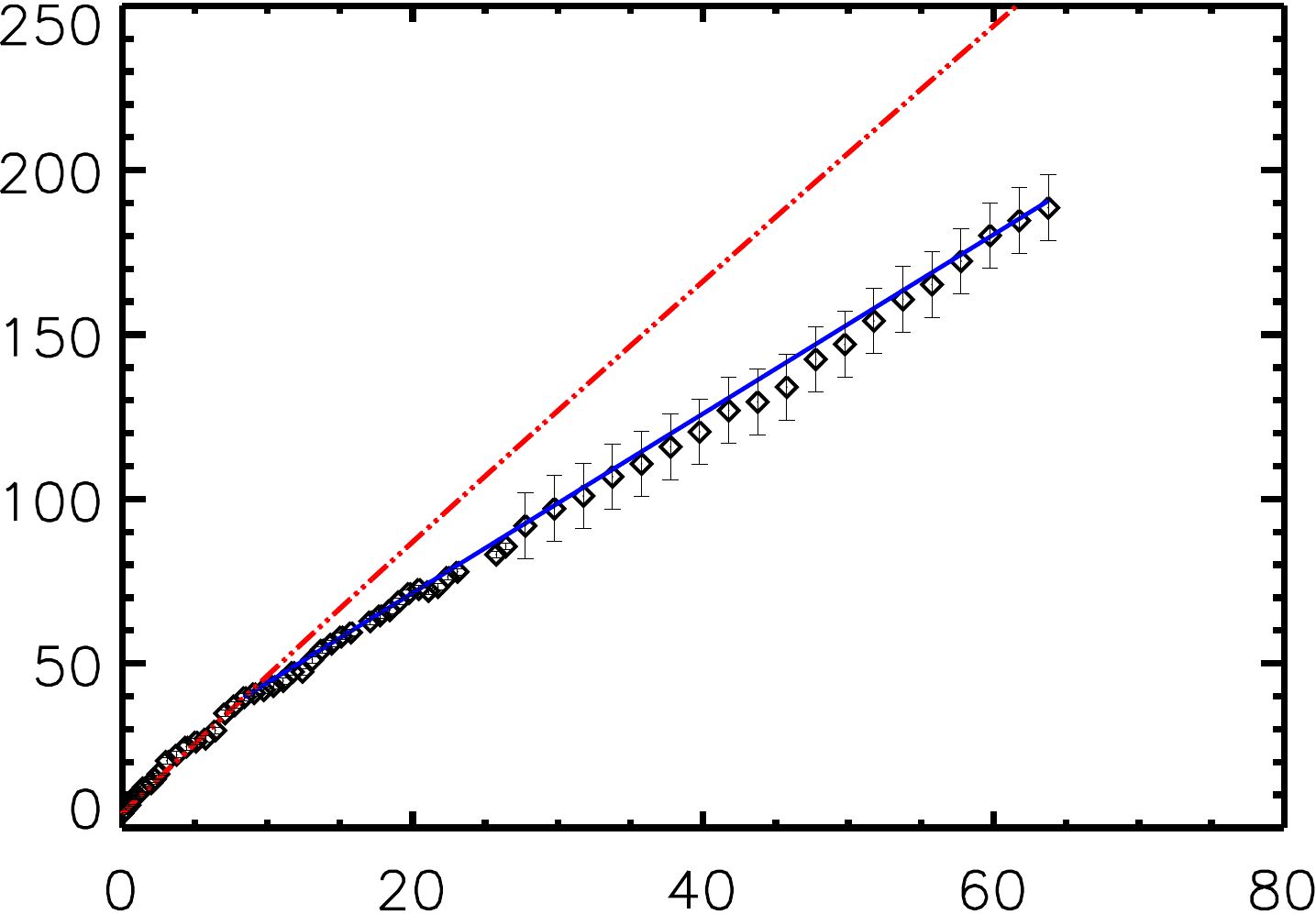}
		\put(-230,62.7){{\rotatebox{90}{{\color{black}\fontsize{12}{12}\fontseries{n}\fontfamily{phv}\selectfont  Height (R$_{\odot}$)}}}}
		\put(-160.8,-14.9){{\rotatebox{0}{{\color{black}\fontsize{12}{12}\fontseries{n}\fontfamily{phv}\selectfont  Elapsed Time (hrs)}}}}
		\put(-130,130.7){{\rotatebox{0}{{\color{black}\fontsize{13}{13}\fontseries{n}\fontfamily{phv}\selectfont \underbar{CME 9}}}}}
		\put(-115,158.7){{\rotatebox{0}{{\color{black}\fontsize{13}{13}\fontseries{n}\fontfamily{phv}\selectfont (a)}}}}
		\hspace*{0.069\textwidth}
		\includegraphics[width=0.553\textwidth,clip=]{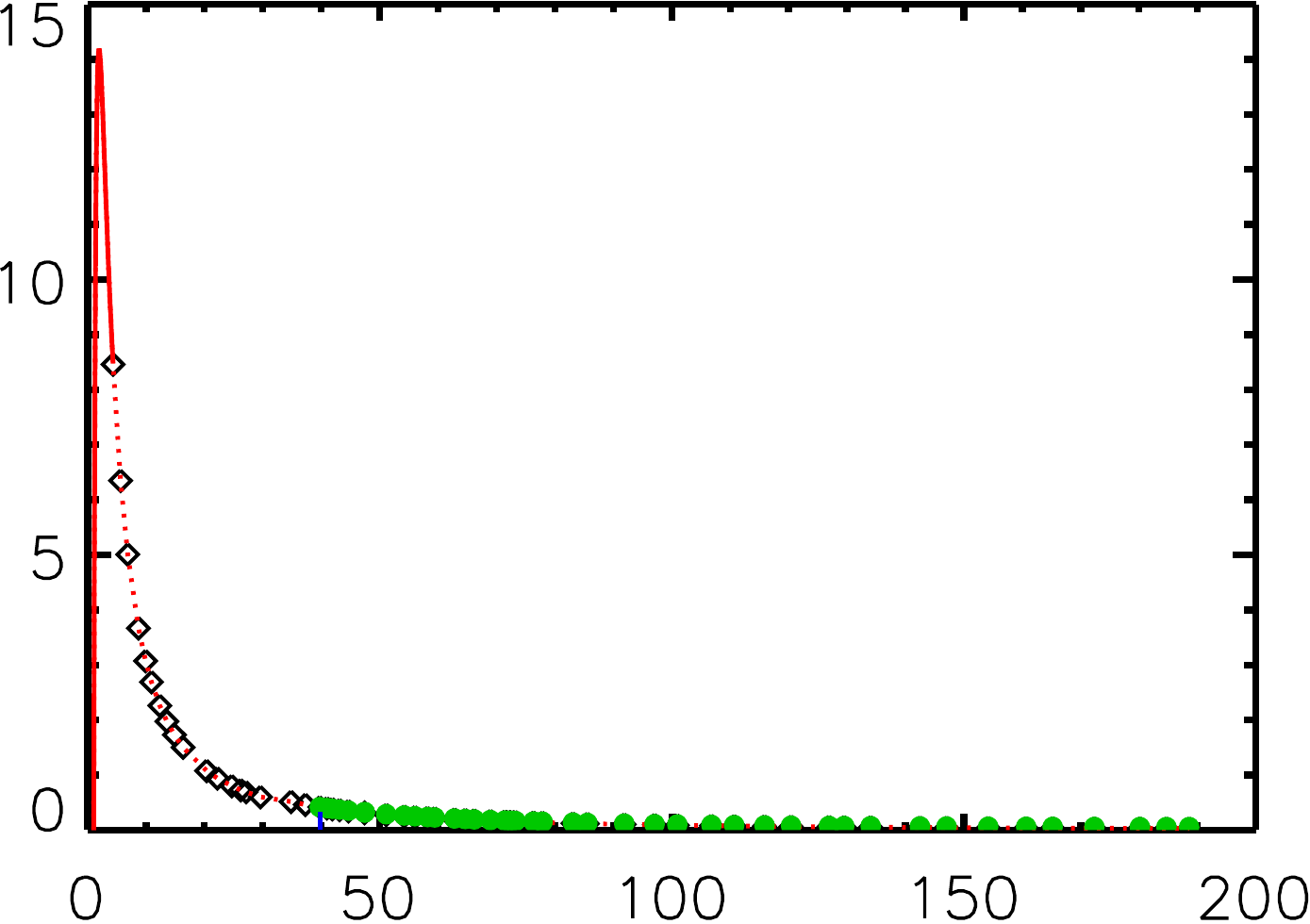}
		\put(-231,43.7){{\rotatebox{90}{{\color{black}\fontsize{12}{12}\fontseries{n}\fontfamily{phv}\selectfont  Force ($10^{17}$ dyn)}}}}
		\put(-140.8,-12.9){{\rotatebox{0}{{\color{black}\fontsize{12}{12}\fontseries{n}\fontfamily{phv}\selectfont Height (R$_{\odot}$)}}}}
		\put(-130,130.7){{\rotatebox{0}{{\color{black}\fontsize{13}{13}\fontseries{n}\fontfamily{phv}\selectfont  \underbar{CME 9}}}}}
		\put(-117,158.7){{\rotatebox{0}{{\color{black}\fontsize{13}{13}\fontseries{n}\fontfamily{phv}\selectfont (b)}}}}          
		  }
		\vspace{1.3cm}
		  \centerline{\hspace*{0.06\textwidth}
		\includegraphics[width=0.56\textwidth,clip=]{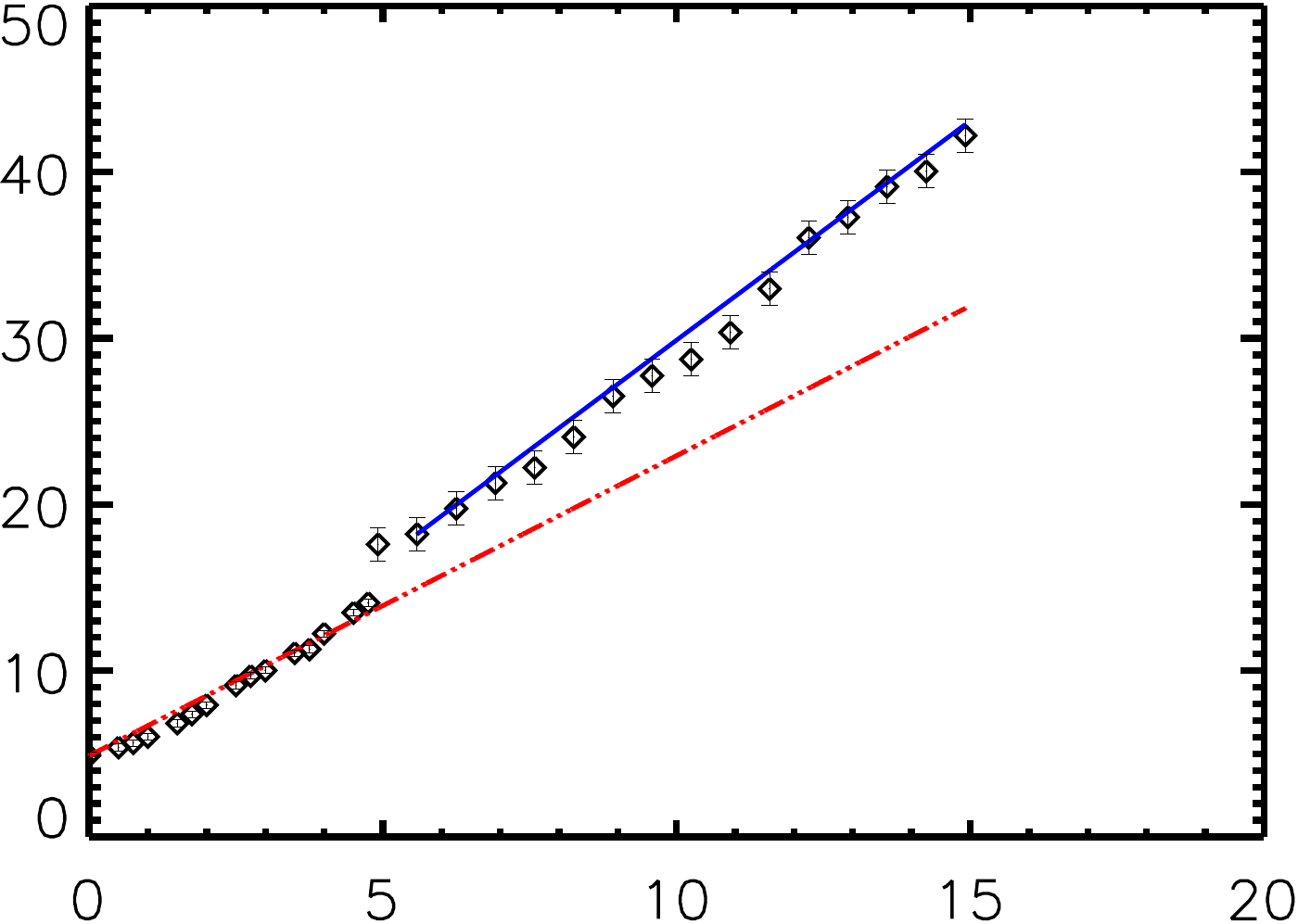}
		\put(-230,62.7){{\rotatebox{90}{{\color{black}\fontsize{12}{12}\fontseries{n}\fontfamily{phv}\selectfont  Height (R$_{\odot}$)}}}}
		\put(-160.8,-14.9){{\rotatebox{0}{{\color{black}\fontsize{12}{12}\fontseries{n}\fontfamily{phv}\selectfont  Elapsed Time (hrs)}}}}
		\put(-130,130.7){{\rotatebox{0}{{\color{black}\fontsize{13}{13}\fontseries{n}\fontfamily{phv}\selectfont \underbar{CME 10}}}}}
		\hspace*{0.069\textwidth}
		\includegraphics[width=0.545\textwidth,clip=]{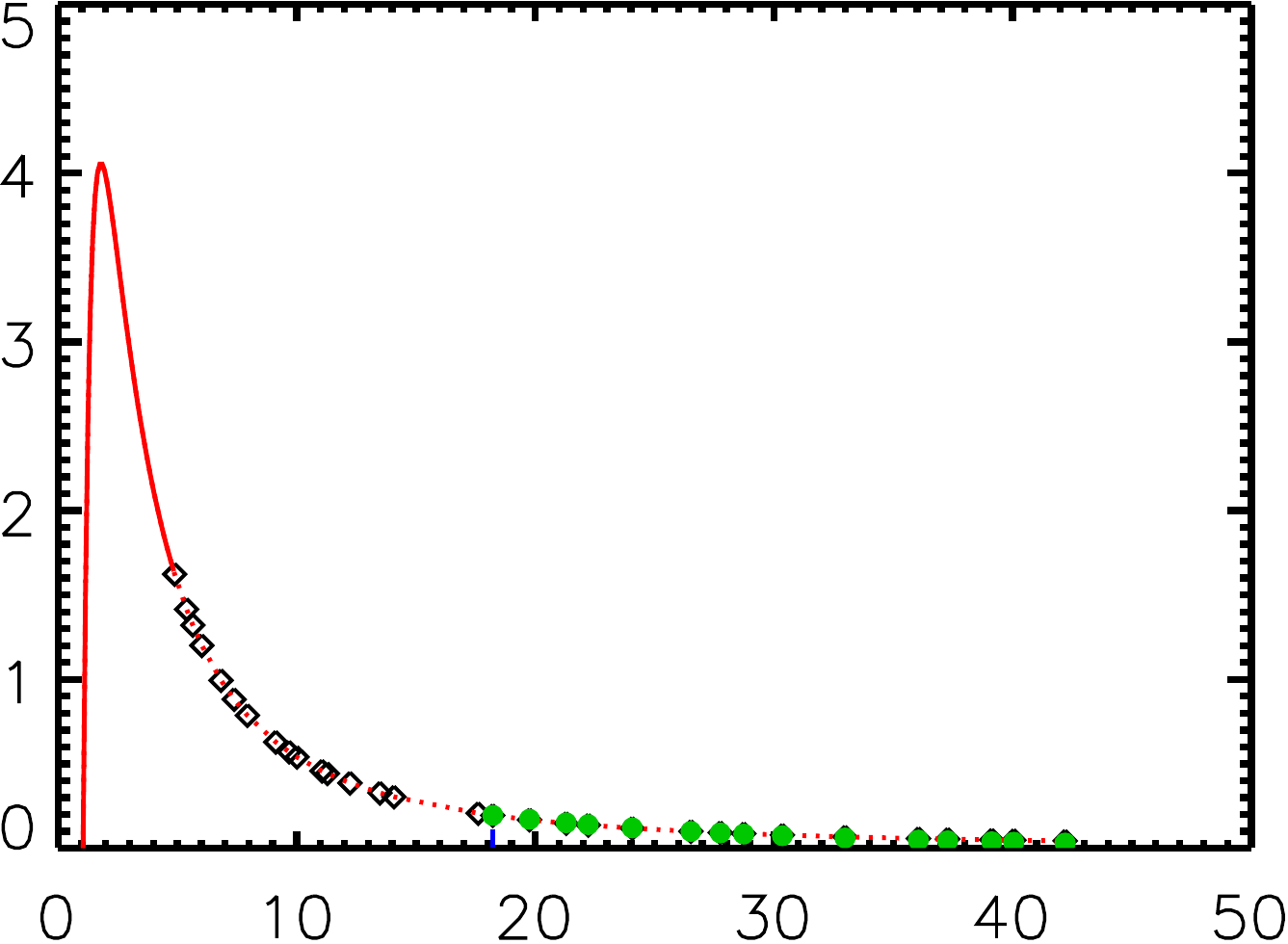}
		\put(-231,43.7){{\rotatebox{90}{{\color{black}\fontsize{12}{12}\fontseries{n}\fontfamily{phv}\selectfont  Force ($10^{17}$ dyn)}}}}
		\put(-140.8,-12.9){{\rotatebox{0}{{\color{black}\fontsize{12}{12}\fontseries{n}\fontfamily{phv}\selectfont Height (R$_{\odot}$)}}}}
		\put(-130,130.7){{\rotatebox{0}{{\color{black}\fontsize{13}{13}\fontseries{n}\fontfamily{phv}\selectfont  \underbar{CME 10}}}}}
		  }
  \vspace{0.0261\textwidth}  
  \caption[Height-time and Force profiles for CMEs 9 and 10]{Height-time and Force profiles for CMEs 9 and 10. Caption same as Figure \ref{fig52}}
  \label{fig55}
  \end{figure}

  \clearpage
  \begin{figure}[h]    
    \centering                              
    \centerline{\hspace*{0.06\textwidth}
		\includegraphics[width=0.56\textwidth,clip=]{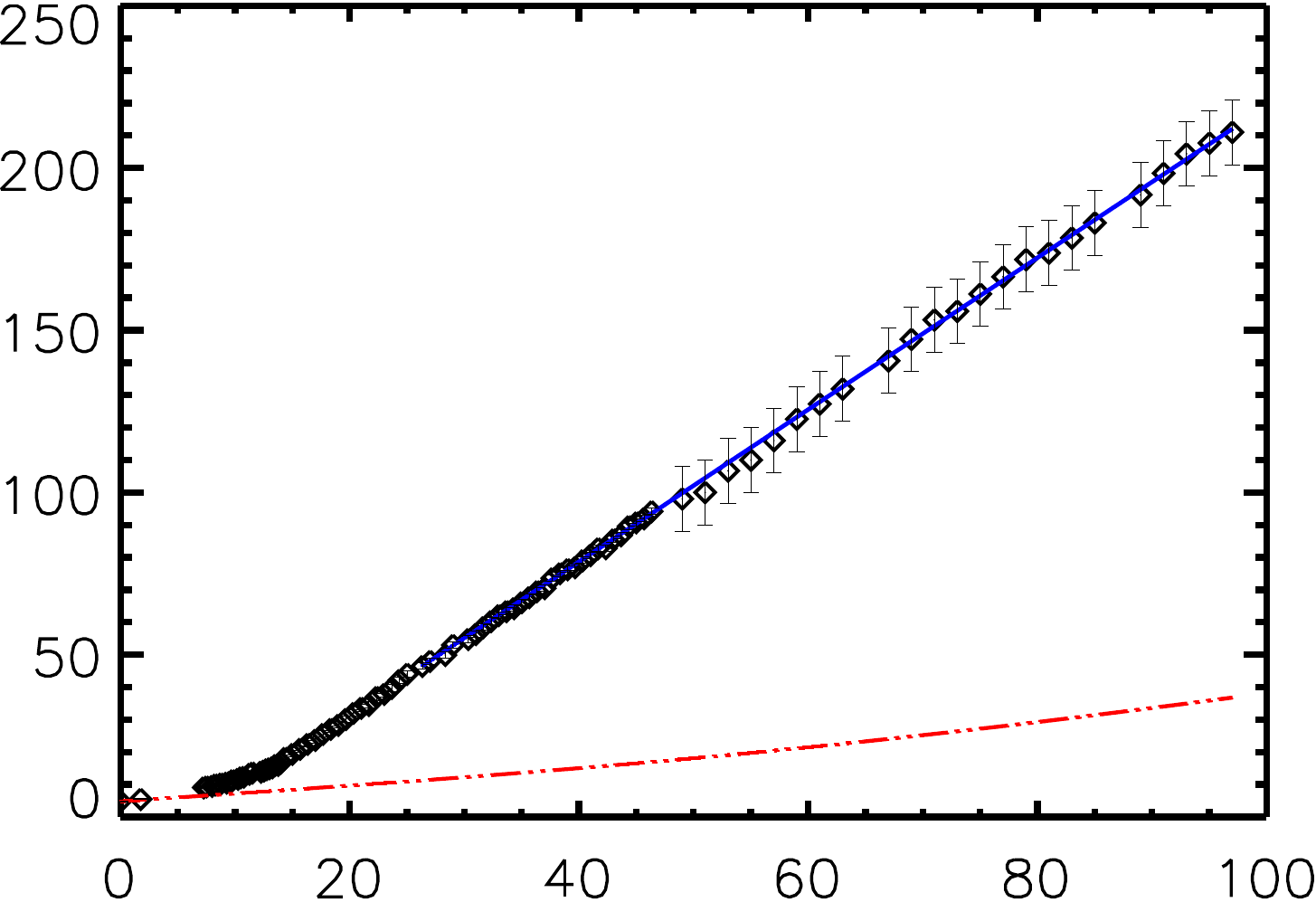}
		\put(-230,62.7){{\rotatebox{90}{{\color{black}\fontsize{12}{12}\fontseries{n}\fontfamily{phv}\selectfont  Height (R$_{\odot}$)}}}}
		\put(-160.8,-14.9){{\rotatebox{0}{{\color{black}\fontsize{12}{12}\fontseries{n}\fontfamily{phv}\selectfont  Elapsed Time (hrs)}}}}
		\put(-130,130.7){{\rotatebox{0}{{\color{black}\fontsize{13}{13}\fontseries{n}\fontfamily{phv}\selectfont \underbar{CME 11}}}}}
		\put(-115,158.7){{\rotatebox{0}{{\color{black}\fontsize{13}{13}\fontseries{n}\fontfamily{phv}\selectfont (a)}}}}
		\hspace*{0.069\textwidth}
		\includegraphics[width=0.53\textwidth,clip=]{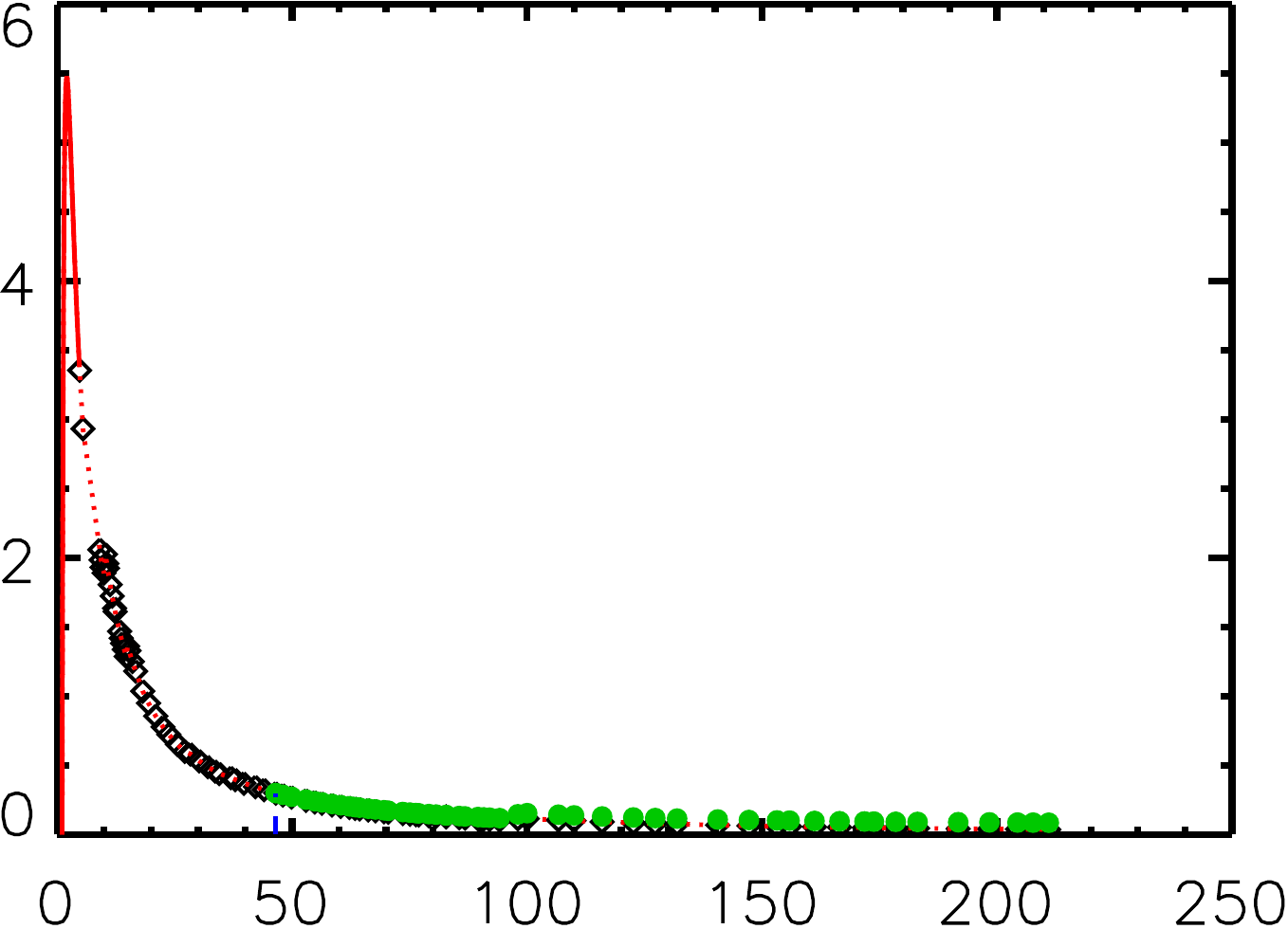}
		\put(-231,43.7){{\rotatebox{90}{{\color{black}\fontsize{12}{12}\fontseries{n}\fontfamily{phv}\selectfont  Force ($10^{17}$ dyn)}}}}
		\put(-140.8,-12.9){{\rotatebox{0}{{\color{black}\fontsize{12}{12}\fontseries{n}\fontfamily{phv}\selectfont Height (R$_{\odot}$)}}}}
		\put(-130,130.7){{\rotatebox{0}{{\color{black}\fontsize{13}{13}\fontseries{n}\fontfamily{phv}\selectfont  \underbar{CME 11}}}}}
		\put(-117,158.7){{\rotatebox{0}{{\color{black}\fontsize{13}{13}\fontseries{n}\fontfamily{phv}\selectfont (b)}}}}          
		  }
		\vspace{1.3cm}
		  \centerline{\hspace*{0.06\textwidth}
		\includegraphics[width=0.565\textwidth,clip=]{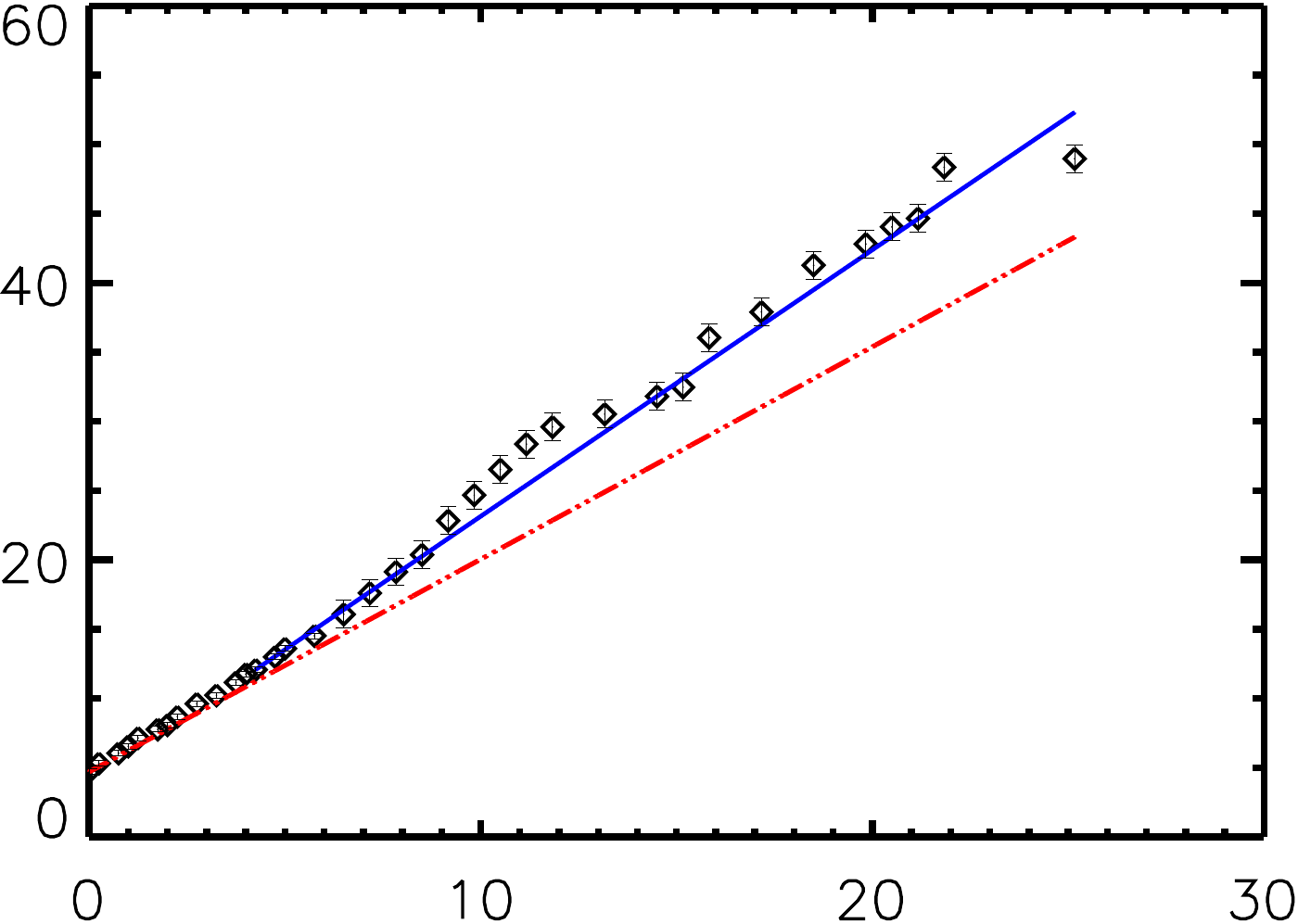}
		\put(-230,62.7){{\rotatebox{90}{{\color{black}\fontsize{12}{12}\fontseries{n}\fontfamily{phv}\selectfont  Height (R$_{\odot}$)}}}}
		\put(-160.8,-14.9){{\rotatebox{0}{{\color{black}\fontsize{12}{12}\fontseries{n}\fontfamily{phv}\selectfont  Elapsed Time (hrs)}}}}
		\put(-130,130.7){{\rotatebox{0}{{\color{black}\fontsize{13}{13}\fontseries{n}\fontfamily{phv}\selectfont \underbar{CME 12}}}}}
		\hspace*{0.069\textwidth}
		\includegraphics[width=0.55\textwidth,clip=]{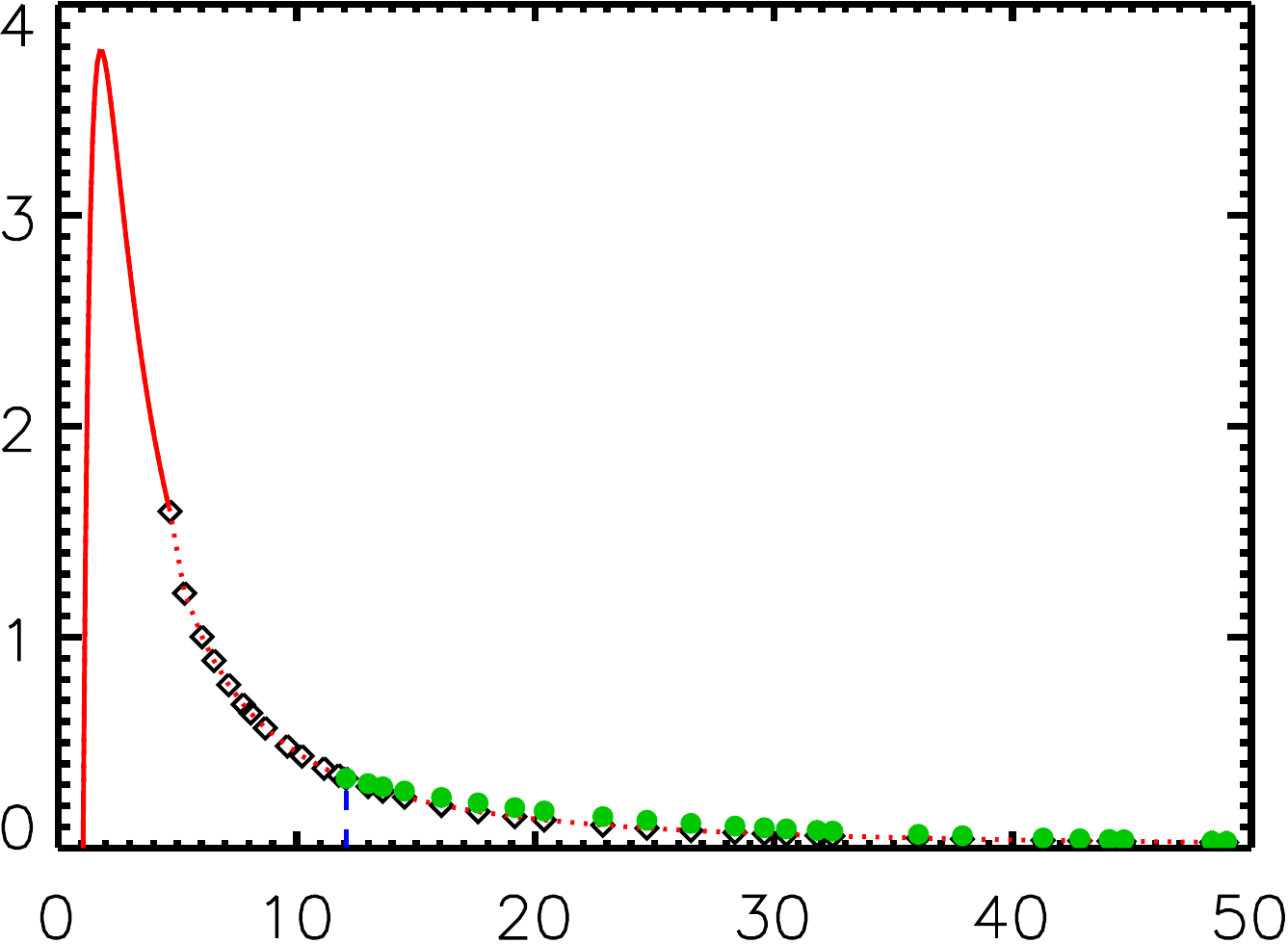}
		\put(-231,43.7){{\rotatebox{90}{{\color{black}\fontsize{12}{12}\fontseries{n}\fontfamily{phv}\selectfont  Force ($10^{17}$ dyn)}}}}
		\put(-140.8,-12.9){{\rotatebox{0}{{\color{black}\fontsize{12}{12}\fontseries{n}\fontfamily{phv}\selectfont Height (R$_{\odot}$)}}}}
		\put(-130,130.7){{\rotatebox{0}{{\color{black}\fontsize{13}{13}\fontseries{n}\fontfamily{phv}\selectfont  \underbar{CME 12}}}}}
		  }
  \vspace{0.0261\textwidth}  
  \caption[Height-time and Force profiles for CMEs 11 and 12]{Height-time and Force profiles for CMEs 11 and 12. Caption same as Figure \ref{fig52}}
  \label{fig56}
  \end{figure}

  \clearpage
  \begin{figure}[h]    
    \centering                              
    \centerline{\hspace*{0.06\textwidth}
		\includegraphics[width=0.55\textwidth,clip=]{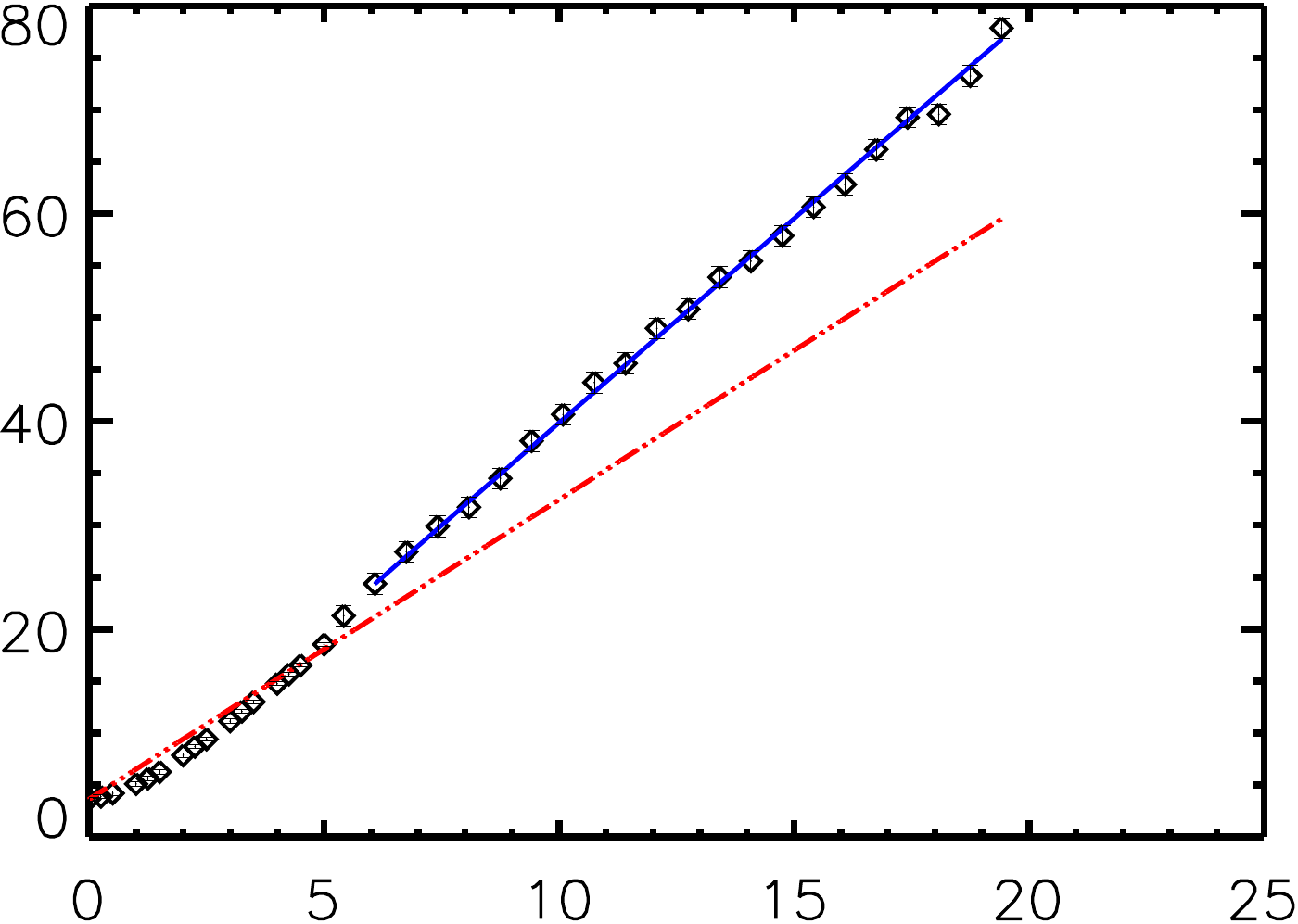}
		\put(-230,62.7){{\rotatebox{90}{{\color{black}\fontsize{12}{12}\fontseries{n}\fontfamily{phv}\selectfont  Height (R$_{\odot}$)}}}}
		\put(-160.8,-14.9){{\rotatebox{0}{{\color{black}\fontsize{12}{12}\fontseries{n}\fontfamily{phv}\selectfont  Elapsed Time (hrs)}}}}
		\put(-130,130.7){{\rotatebox{0}{{\color{black}\fontsize{13}{13}\fontseries{n}\fontfamily{phv}\selectfont \underbar{CME 13}}}}}
		\put(-115,158.7){{\rotatebox{0}{{\color{black}\fontsize{13}{13}\fontseries{n}\fontfamily{phv}\selectfont (a)}}}}
		\hspace*{0.069\textwidth}
		\includegraphics[width=0.55\textwidth,clip=]{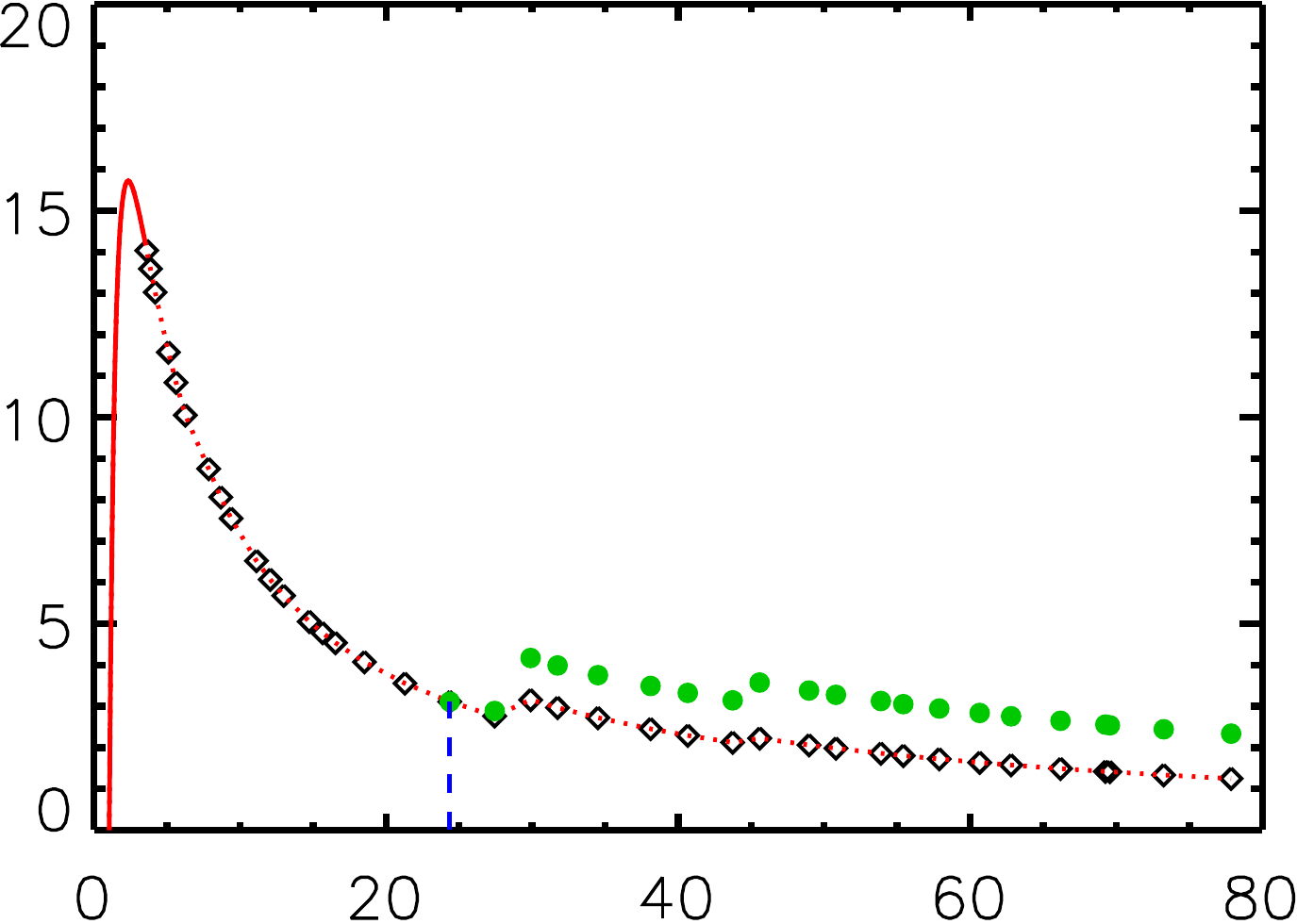}
		\put(-231,43.7){{\rotatebox{90}{{\color{black}\fontsize{12}{12}\fontseries{n}\fontfamily{phv}\selectfont  Force ($10^{17}$ dyn)}}}}
		\put(-140.8,-12.9){{\rotatebox{0}{{\color{black}\fontsize{12}{12}\fontseries{n}\fontfamily{phv}\selectfont Height (R$_{\odot}$)}}}}
		\put(-130,130.7){{\rotatebox{0}{{\color{black}\fontsize{13}{13}\fontseries{n}\fontfamily{phv}\selectfont  \underbar{CME 13}}}}}
		\put(-117,158.7){{\rotatebox{0}{{\color{black}\fontsize{13}{13}\fontseries{n}\fontfamily{phv}\selectfont (b)}}}}          
		  }
		\vspace{1.3cm}
		  \centerline{\hspace*{0.06\textwidth}
		\includegraphics[width=0.55\textwidth,clip=]{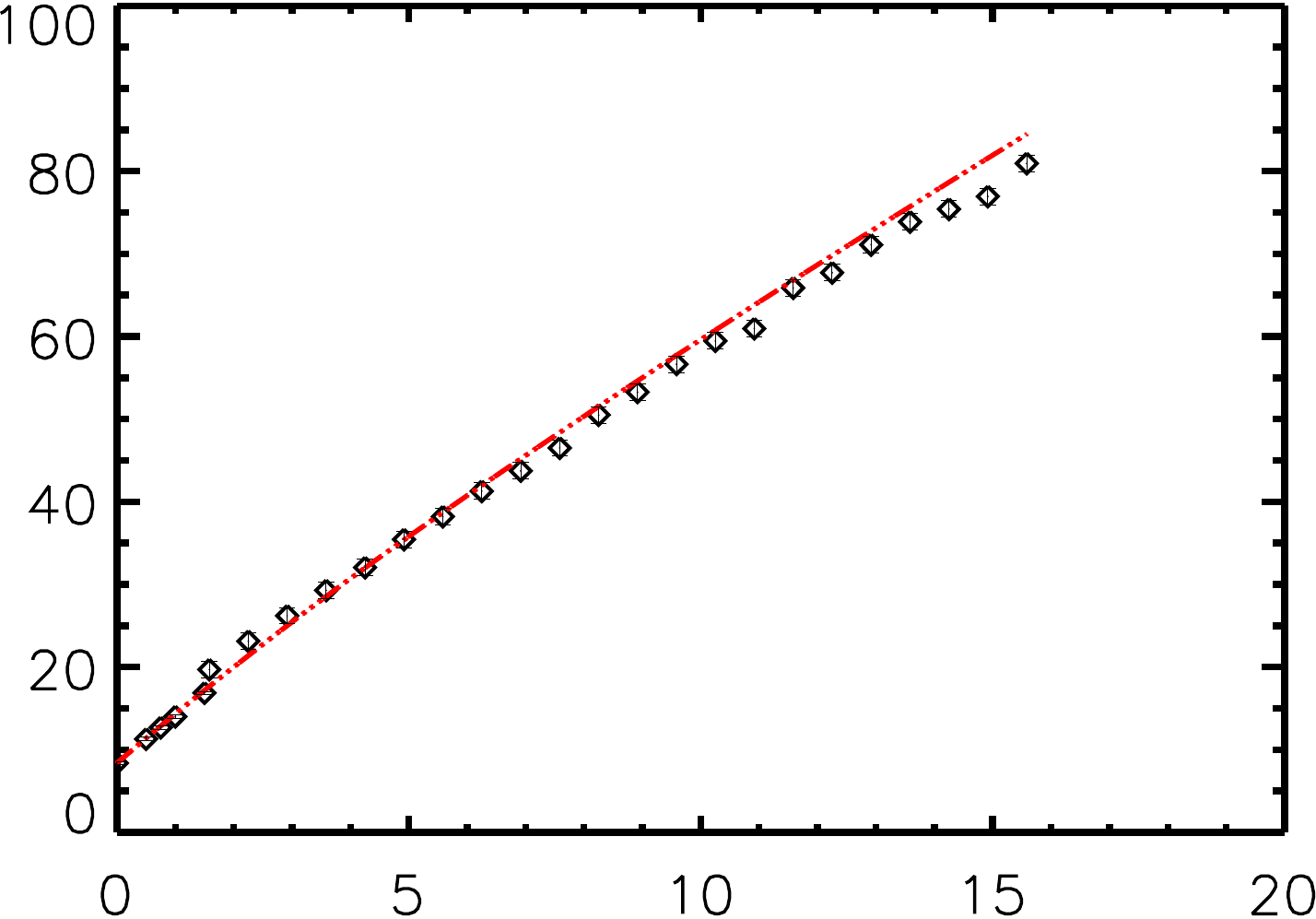}
		\put(-225,62.7){{\rotatebox{90}{{\color{black}\fontsize{12}{12}\fontseries{n}\fontfamily{phv}\selectfont  Height (R$_{\odot}$)}}}}
		\put(-160.8,-14.9){{\rotatebox{0}{{\color{black}\fontsize{12}{12}\fontseries{n}\fontfamily{phv}\selectfont  Elapsed Time (hrs)}}}}
		\put(-130,130.7){{\rotatebox{0}{{\color{black}\fontsize{13}{13}\fontseries{n}\fontfamily{phv}\selectfont \underbar{CME 14 ($f$)}}}}}
		\hspace*{0.069\textwidth}
		\includegraphics[width=0.563\textwidth,clip=]{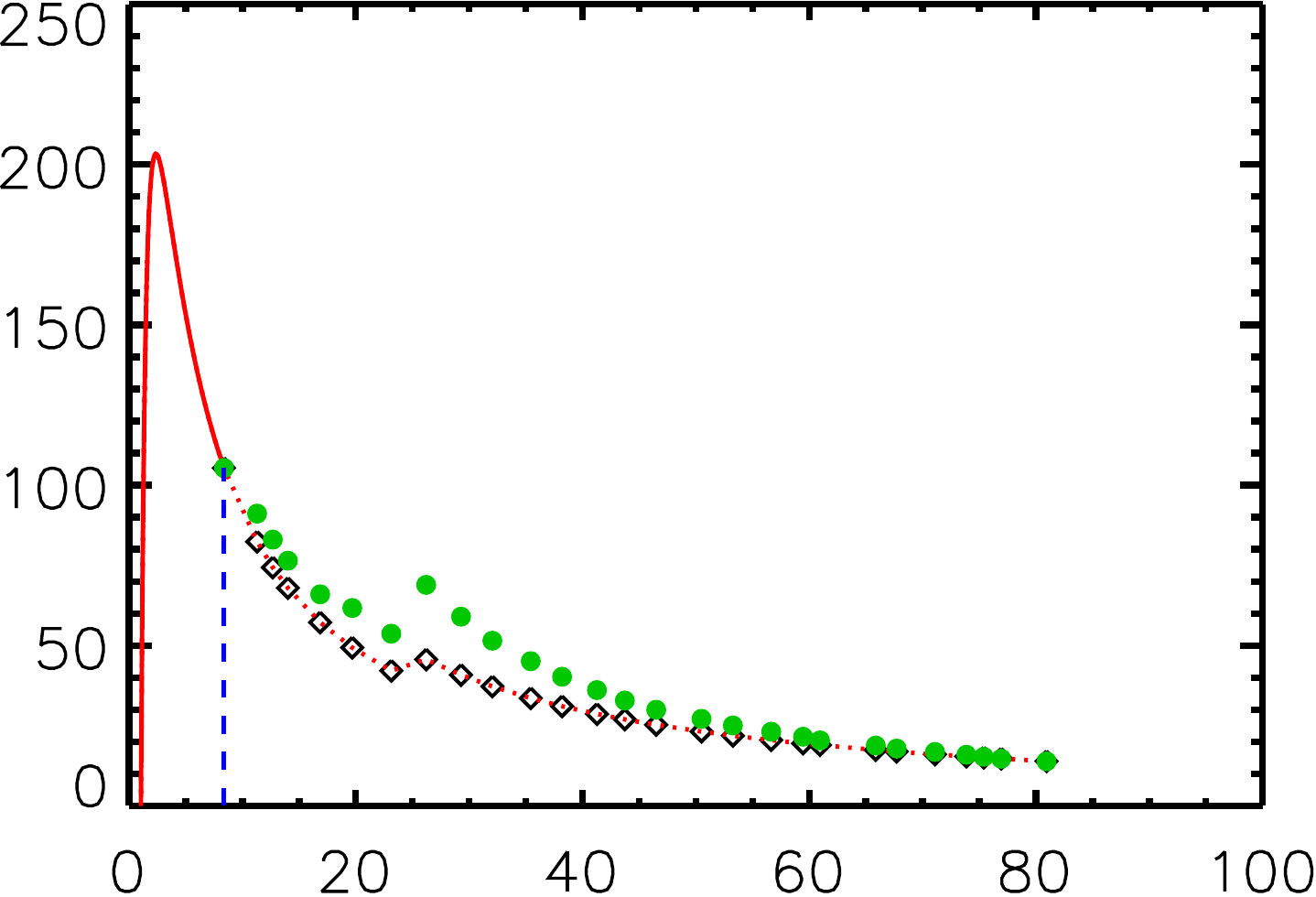}
		\put(-231,43.7){{\rotatebox{90}{{\color{black}\fontsize{12}{12}\fontseries{n}\fontfamily{phv}\selectfont  Force ($10^{17}$ dyn)}}}}
		\put(-140.8,-12.9){{\rotatebox{0}{{\color{black}\fontsize{12}{12}\fontseries{n}\fontfamily{phv}\selectfont Height (R$_{\odot}$)}}}}
		\put(-130,130.7){{\rotatebox{0}{{\color{black}\fontsize{13}{13}\fontseries{n}\fontfamily{phv}\selectfont  \underbar{CME 14 ($f$)}}}}}
		  }
  \vspace{0.0261\textwidth}  
  \caption[Height-time and Force profiles for CMEs 13 and 14]{Height-time and Force profiles for CMEs 13 and 14. Caption same as Figure \ref{fig52}}
  \label{fig57}
  \end{figure}

  \clearpage
  \begin{figure}[h]    
    \centering                              
    \centerline{\hspace*{0.06\textwidth}
		\includegraphics[width=0.55\textwidth,clip=]{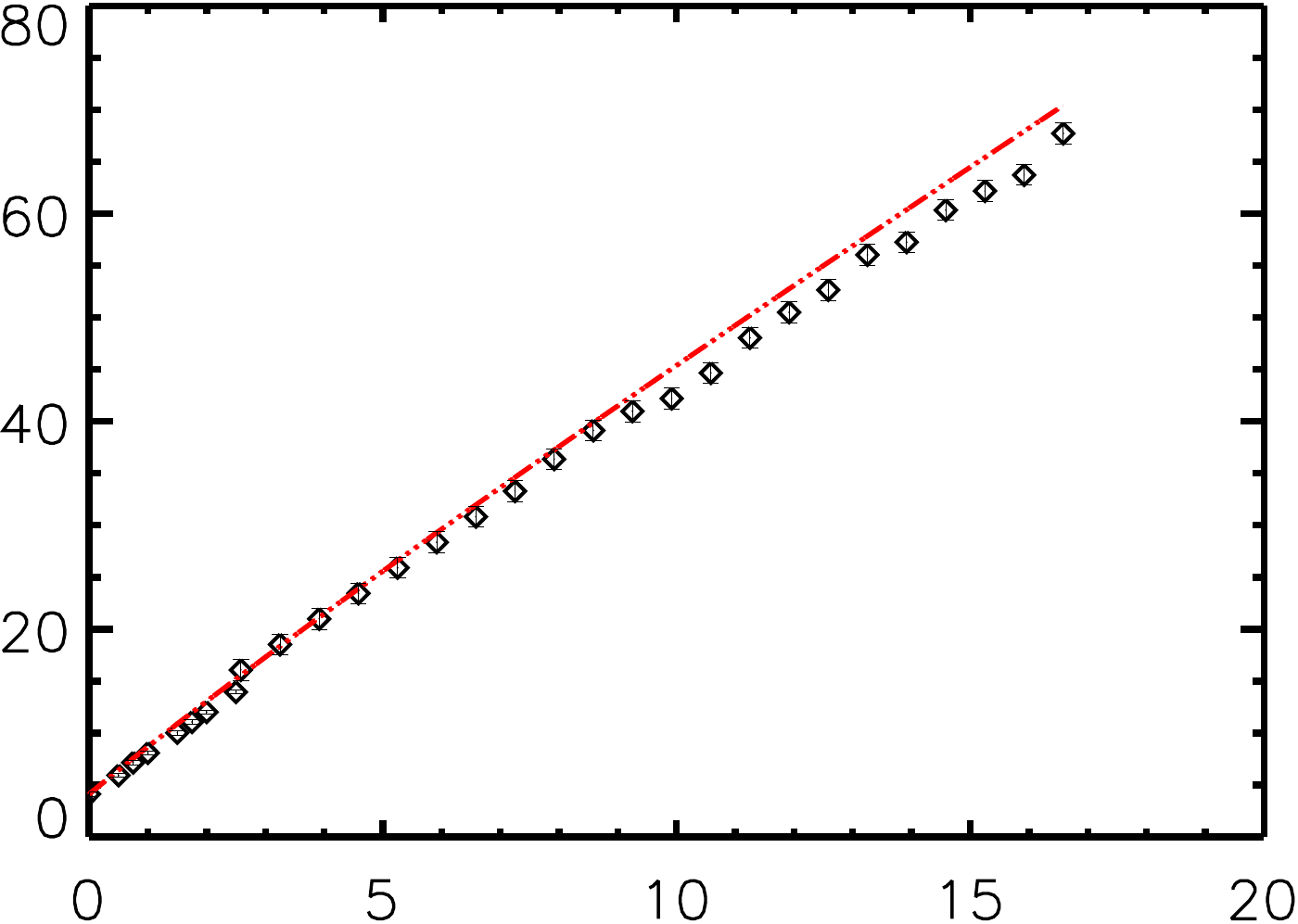}
		\put(-230,62.7){{\rotatebox{90}{{\color{black}\fontsize{12}{12}\fontseries{n}\fontfamily{phv}\selectfont  Height (R$_{\odot}$)}}}}
		\put(-160.8,-14.9){{\rotatebox{0}{{\color{black}\fontsize{12}{12}\fontseries{n}\fontfamily{phv}\selectfont  Elapsed Time (hrs)}}}}
		\put(-130,130.7){{\rotatebox{0}{{\color{black}\fontsize{13}{13}\fontseries{n}\fontfamily{phv}\selectfont \underbar{CME 15 ($f$)}}}}}
		\put(-115,158.7){{\rotatebox{0}{{\color{black}\fontsize{13}{13}\fontseries{n}\fontfamily{phv}\selectfont (a)}}}}
		\hspace*{0.069\textwidth}
		\includegraphics[width=0.55\textwidth,clip=]{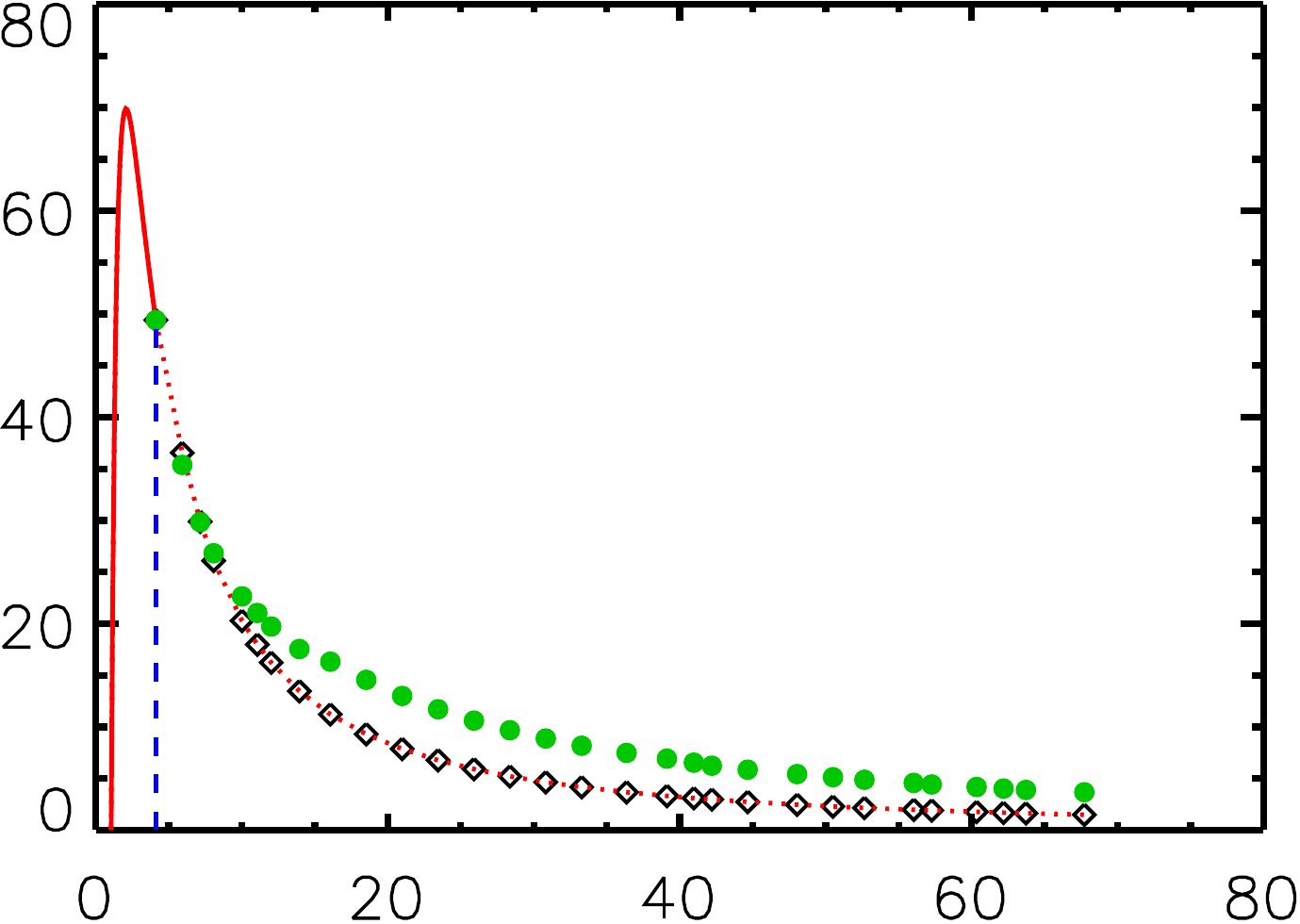}
		\put(-231,43.7){{\rotatebox{90}{{\color{black}\fontsize{12}{12}\fontseries{n}\fontfamily{phv}\selectfont  Force ($10^{17}$ dyn)}}}}
		\put(-140.8,-12.9){{\rotatebox{0}{{\color{black}\fontsize{12}{12}\fontseries{n}\fontfamily{phv}\selectfont Height (R$_{\odot}$)}}}}
		\put(-130,130.7){{\rotatebox{0}{{\color{black}\fontsize{13}{13}\fontseries{n}\fontfamily{phv}\selectfont  \underbar{CME 15 ($f$)}}}}}
		\put(-117,158.7){{\rotatebox{0}{{\color{black}\fontsize{13}{13}\fontseries{n}\fontfamily{phv}\selectfont (b)}}}}          
		  }
		\vspace{1.3cm}
		  \centerline{\hspace*{0.06\textwidth}
		\includegraphics[width=0.54\textwidth,clip=]{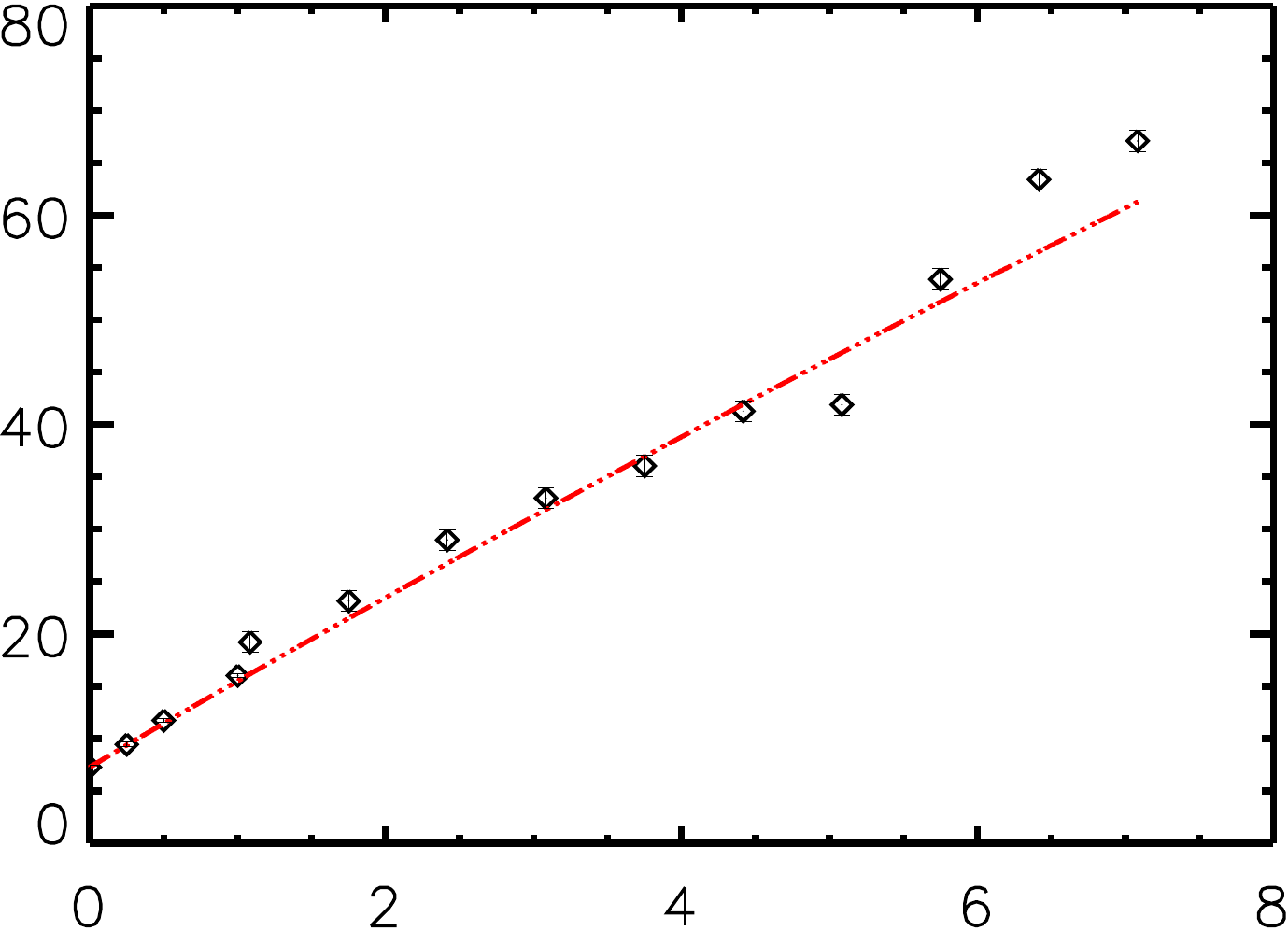}
		\put(-225,62.7){{\rotatebox{90}{{\color{black}\fontsize{12}{12}\fontseries{n}\fontfamily{phv}\selectfont  Height (R$_{\odot}$)}}}}
		\put(-160.8,-14.9){{\rotatebox{0}{{\color{black}\fontsize{12}{12}\fontseries{n}\fontfamily{phv}\selectfont  Elapsed Time (hrs)}}}}
		\put(-130,130.7){{\rotatebox{0}{{\color{black}\fontsize{13}{13}\fontseries{n}\fontfamily{phv}\selectfont \underbar{CME 16 ($f$)}}}}}
		\hspace*{0.069\textwidth}
		\includegraphics[width=0.565\textwidth,clip=]{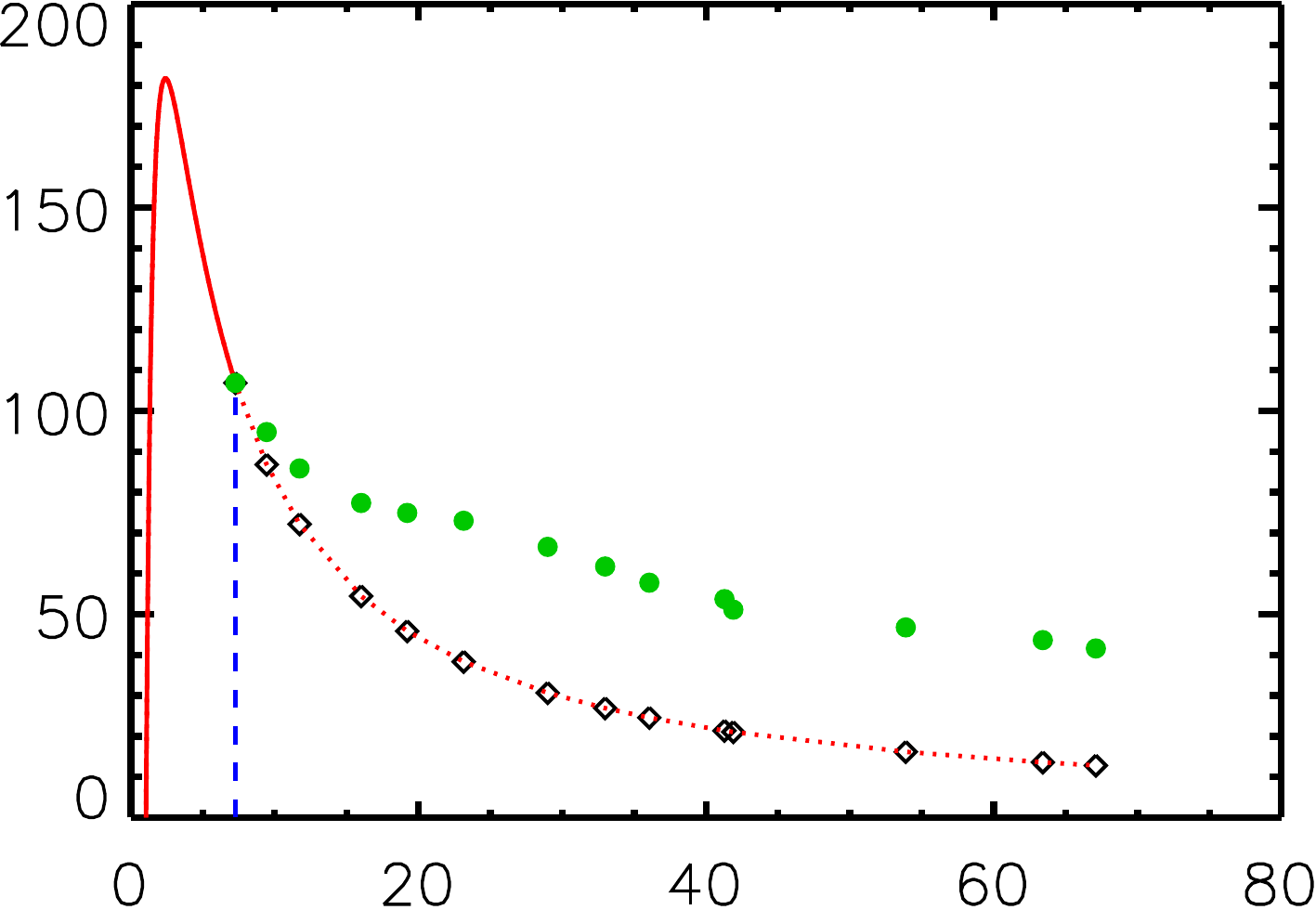}
		\put(-231,43.7){{\rotatebox{90}{{\color{black}\fontsize{12}{12}\fontseries{n}\fontfamily{phv}\selectfont  Force ($10^{17}$ dyn)}}}}
		\put(-140.8,-12.9){{\rotatebox{0}{{\color{black}\fontsize{12}{12}\fontseries{n}\fontfamily{phv}\selectfont Height (R$_{\odot}$)}}}}
		\put(-130,130.7){{\rotatebox{0}{{\color{black}\fontsize{13}{13}\fontseries{n}\fontfamily{phv}\selectfont  \underbar{CME 16 ($f$)}}}}}
		  }
  \vspace{0.0261\textwidth}  
  \caption[Height-time and Force profiles for CMEs 15 and 16]{Height-time and Force profiles for CMEs 15 and 16. Caption same as Figure \ref{fig52}}
  \label{fig58}
  \end{figure}

  \clearpage
  \begin{figure}[h]    
    \centering                              
    \centerline{\hspace*{0.06\textwidth}
		\includegraphics[width=0.55\textwidth,clip=]{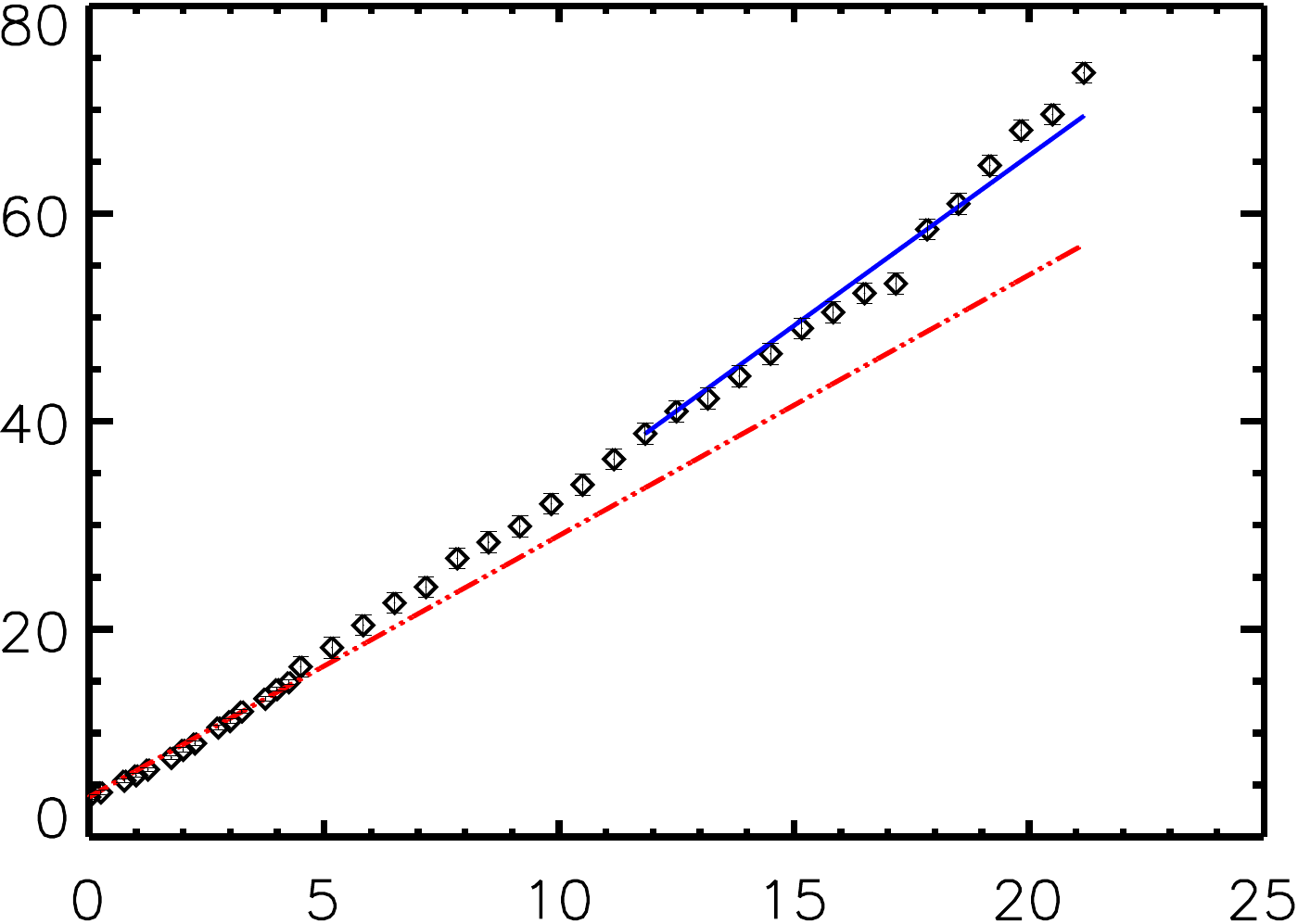}
		\put(-230,62.7){{\rotatebox{90}{{\color{black}\fontsize{12}{12}\fontseries{n}\fontfamily{phv}\selectfont  Height (R$_{\odot}$)}}}}
		\put(-160.8,-14.9){{\rotatebox{0}{{\color{black}\fontsize{12}{12}\fontseries{n}\fontfamily{phv}\selectfont  Elapsed Time (hrs)}}}}
		\put(-130,130.7){{\rotatebox{0}{{\color{black}\fontsize{13}{13}\fontseries{n}\fontfamily{phv}\selectfont \underbar{CME 17}}}}}
		\put(-115,158.7){{\rotatebox{0}{{\color{black}\fontsize{13}{13}\fontseries{n}\fontfamily{phv}\selectfont (a)}}}}
		\hspace*{0.069\textwidth}
		\includegraphics[width=0.535\textwidth,clip=]{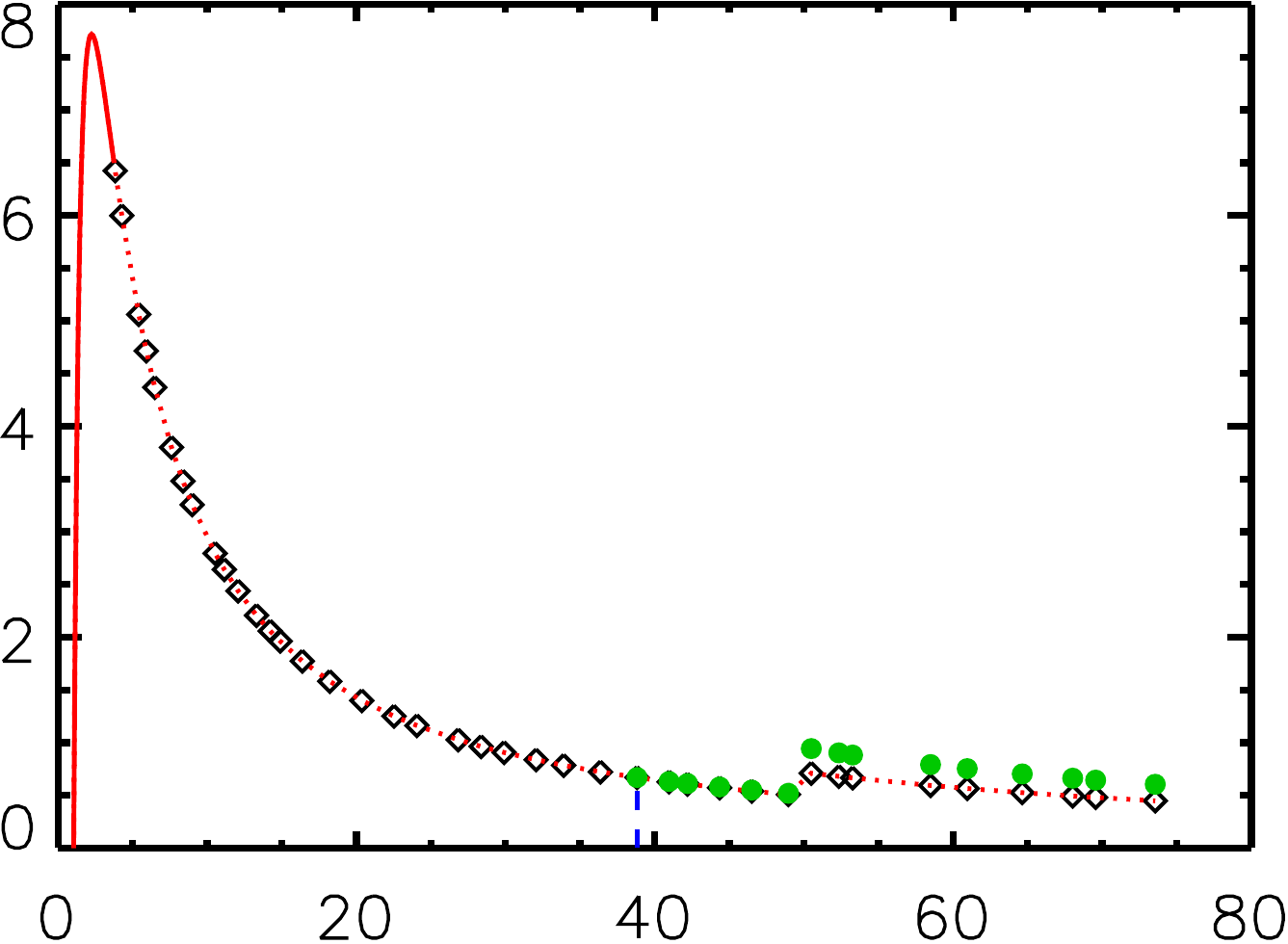}
		\put(-231,43.7){{\rotatebox{90}{{\color{black}\fontsize{12}{12}\fontseries{n}\fontfamily{phv}\selectfont  Force ($10^{17}$ dyn)}}}}
		\put(-140.8,-12.9){{\rotatebox{0}{{\color{black}\fontsize{12}{12}\fontseries{n}\fontfamily{phv}\selectfont Height (R$_{\odot}$)}}}}
		\put(-130,130.7){{\rotatebox{0}{{\color{black}\fontsize{13}{13}\fontseries{n}\fontfamily{phv}\selectfont  \underbar{CME 17}}}}}
		\put(-117,158.7){{\rotatebox{0}{{\color{black}\fontsize{13}{13}\fontseries{n}\fontfamily{phv}\selectfont (b)}}}}          
		  }
		\vspace{1.3cm}
		  \centerline{\hspace*{0.06\textwidth}
		\includegraphics[width=0.54\textwidth,clip=]{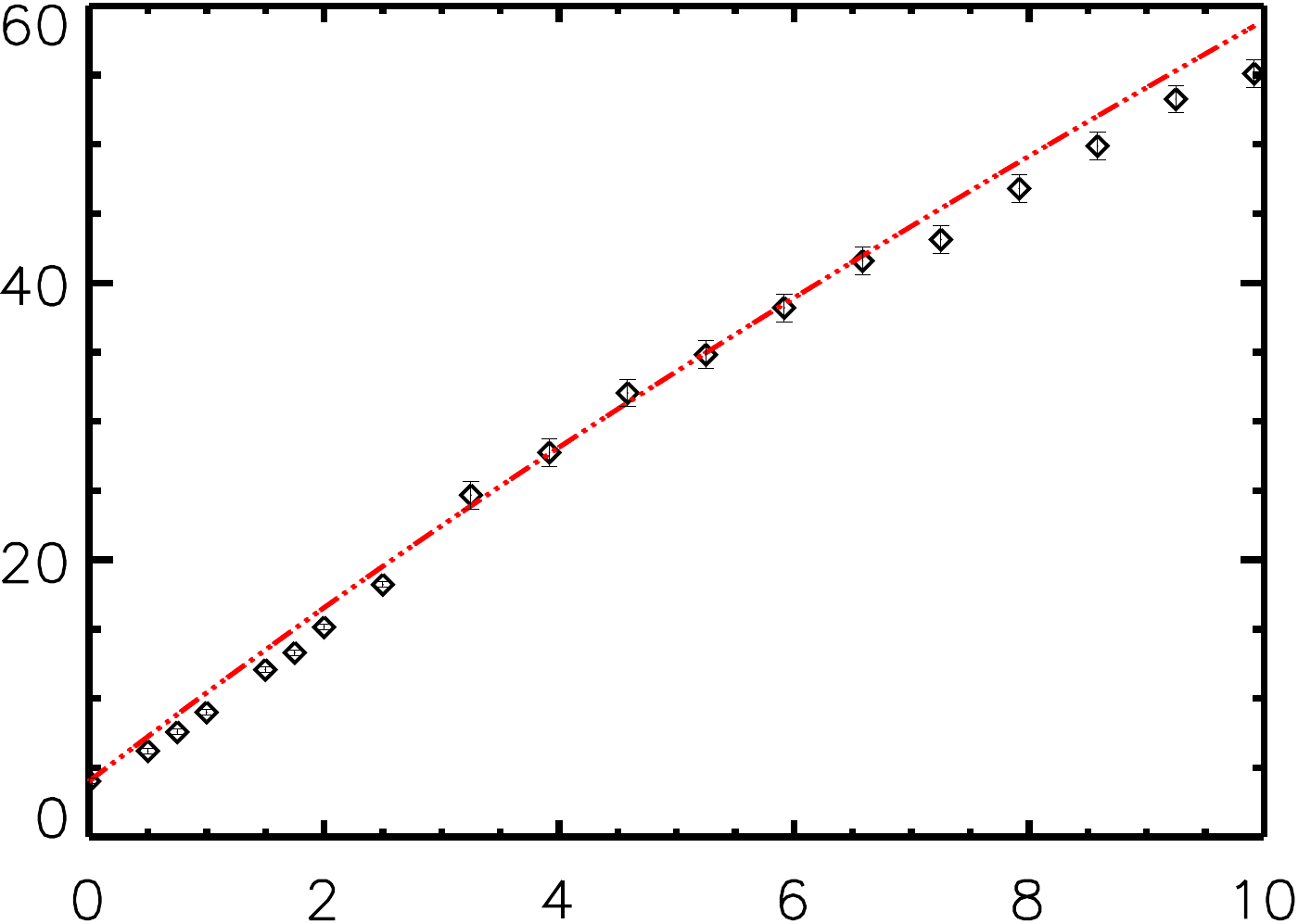}
		\put(-225,62.7){{\rotatebox{90}{{\color{black}\fontsize{12}{12}\fontseries{n}\fontfamily{phv}\selectfont  Height (R$_{\odot}$)}}}}
		\put(-160.8,-14.9){{\rotatebox{0}{{\color{black}\fontsize{12}{12}\fontseries{n}\fontfamily{phv}\selectfont  Elapsed Time (hrs)}}}}
		\put(-130,130.7){{\rotatebox{0}{{\color{black}\fontsize{13}{13}\fontseries{n}\fontfamily{phv}\selectfont \underbar{CME 18 ($f$)}}}}}
		\hspace*{0.069\textwidth}
		\includegraphics[width=0.563\textwidth,clip=]{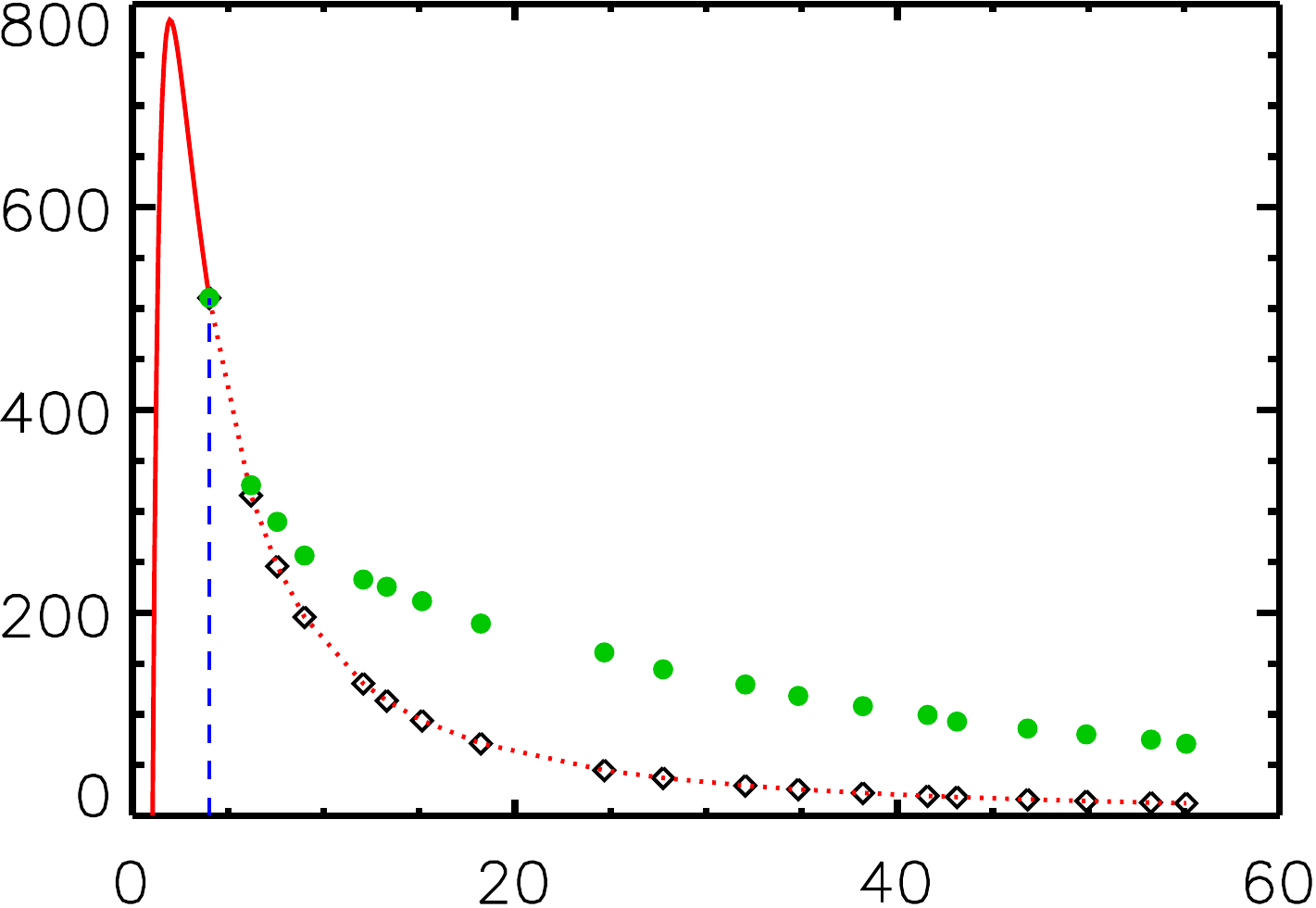}
		\put(-234,43.7){{\rotatebox{90}{{\color{black}\fontsize{12}{12}\fontseries{n}\fontfamily{phv}\selectfont  Force ($10^{17}$ dyn)}}}}
		\put(-140.8,-12.9){{\rotatebox{0}{{\color{black}\fontsize{12}{12}\fontseries{n}\fontfamily{phv}\selectfont Height (R$_{\odot}$)}}}}
		\put(-130,130.7){{\rotatebox{0}{{\color{black}\fontsize{13}{13}\fontseries{n}\fontfamily{phv}\selectfont  \underbar{CME 18 ($f$)}}}}}
		  }
  \vspace{0.0261\textwidth}  
  \caption[Height-time and Force profiles for CMEs 17 and 18]{Height-time and Force profiles for CMEs 17 and 18. Caption same as Figure \ref{fig52}}
  \label{fig59}
  \end{figure}

  \clearpage
  \begin{figure}[h]    
    \centering                              
    \centerline{\hspace*{0.06\textwidth}
		\includegraphics[width=0.55\textwidth,clip=]{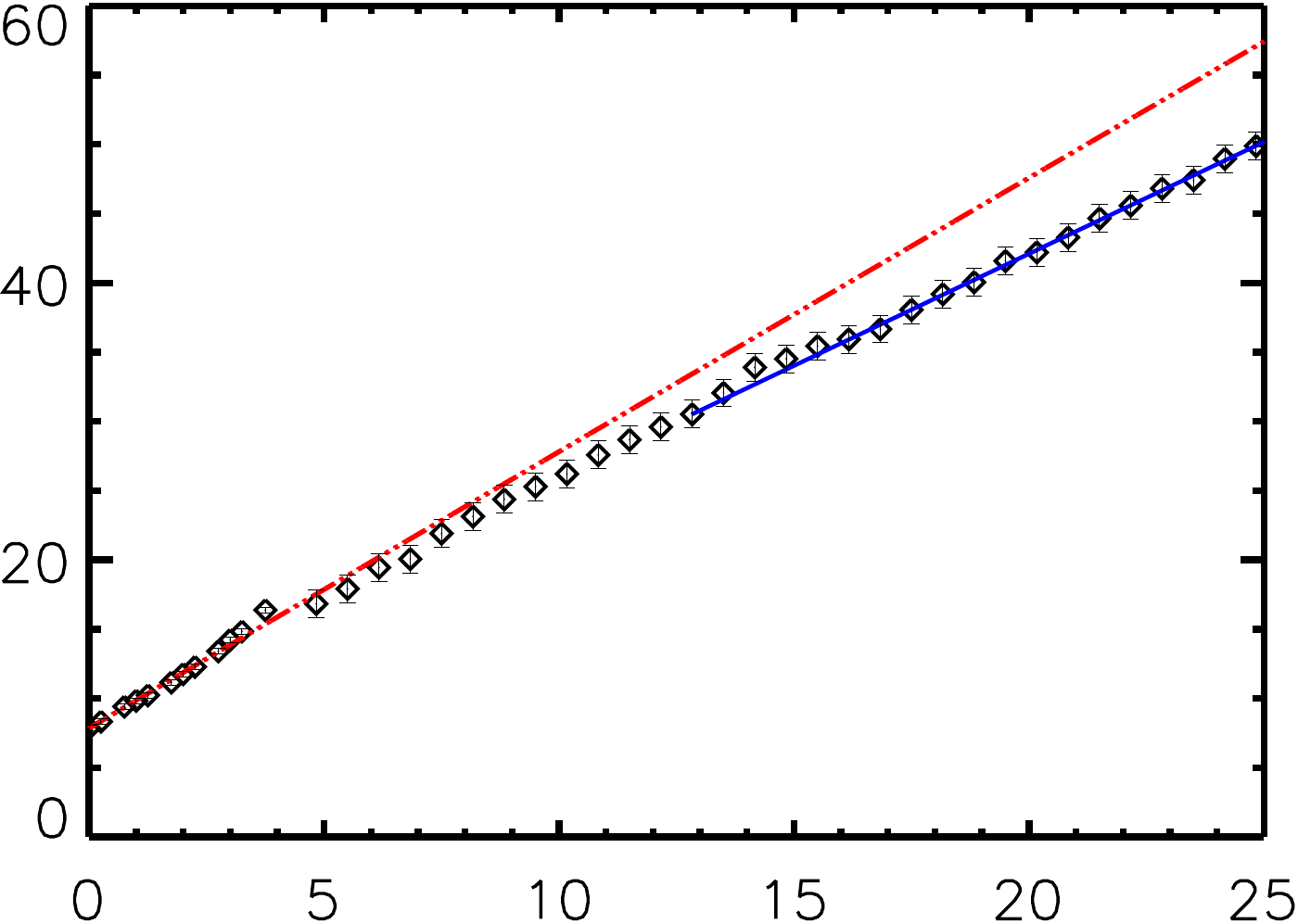}
		\put(-230,62.7){{\rotatebox{90}{{\color{black}\fontsize{12}{12}\fontseries{n}\fontfamily{phv}\selectfont  Height (R$_{\odot}$)}}}}
		\put(-160.8,-14.9){{\rotatebox{0}{{\color{black}\fontsize{12}{12}\fontseries{n}\fontfamily{phv}\selectfont  Elapsed Time (hrs)}}}}
		\put(-130,130.7){{\rotatebox{0}{{\color{black}\fontsize{13}{13}\fontseries{n}\fontfamily{phv}\selectfont \underbar{CME 19}}}}}
		\put(-115,158.7){{\rotatebox{0}{{\color{black}\fontsize{13}{13}\fontseries{n}\fontfamily{phv}\selectfont (a)}}}}
		\hspace*{0.069\textwidth}
		\includegraphics[width=0.535\textwidth,clip=]{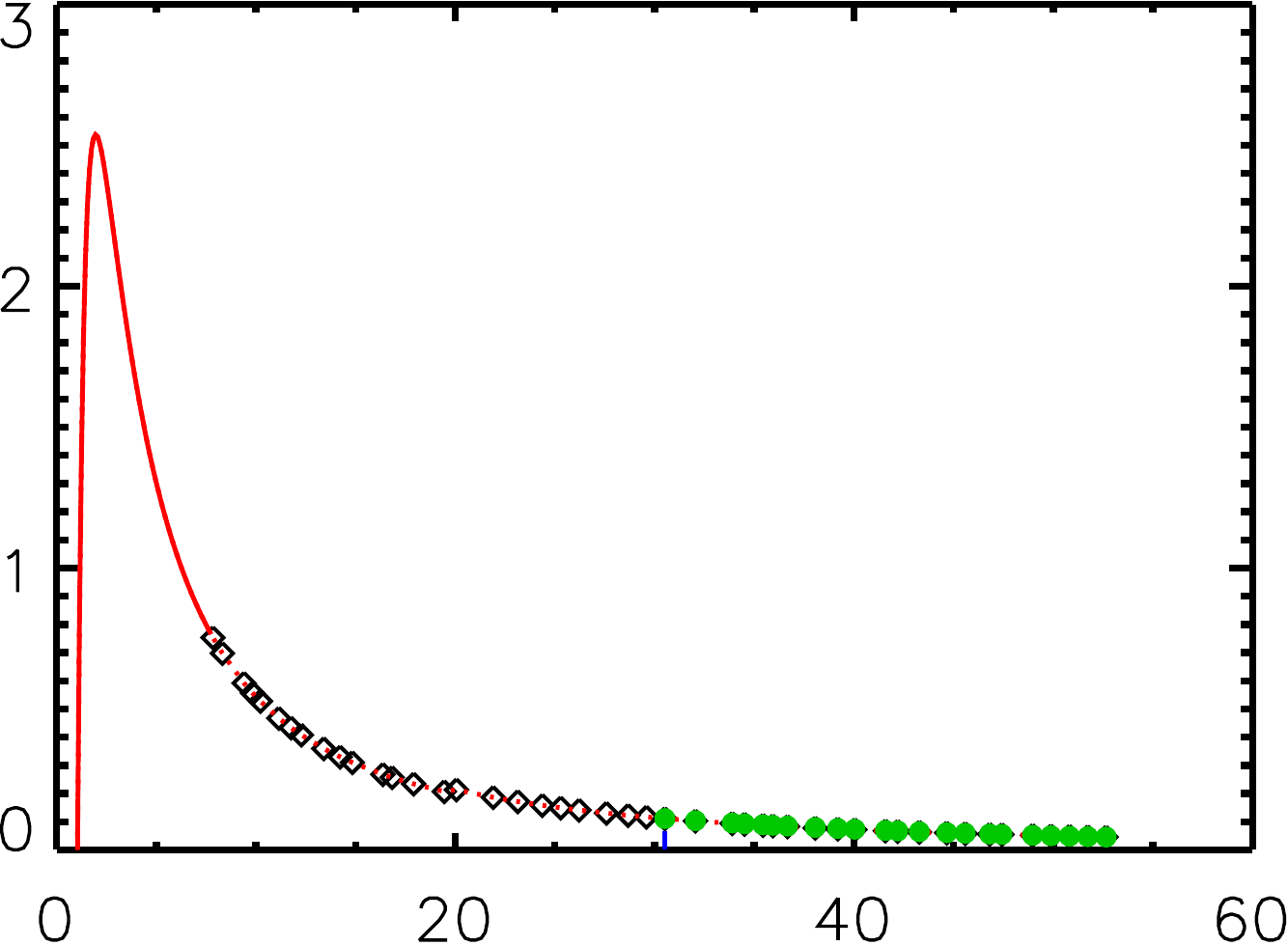}
		\put(-231,43.7){{\rotatebox{90}{{\color{black}\fontsize{12}{12}\fontseries{n}\fontfamily{phv}\selectfont  Force ($10^{17}$ dyn)}}}}
		\put(-140.8,-12.9){{\rotatebox{0}{{\color{black}\fontsize{12}{12}\fontseries{n}\fontfamily{phv}\selectfont Height (R$_{\odot}$)}}}}
		\put(-130,130.7){{\rotatebox{0}{{\color{black}\fontsize{13}{13}\fontseries{n}\fontfamily{phv}\selectfont  \underbar{CME 19}}}}}
		\put(-117,158.7){{\rotatebox{0}{{\color{black}\fontsize{13}{13}\fontseries{n}\fontfamily{phv}\selectfont (b)}}}}          
		  }
		\vspace{1.3cm}
		  \centerline{\hspace*{0.06\textwidth}
		\includegraphics[width=0.555\textwidth,clip=]{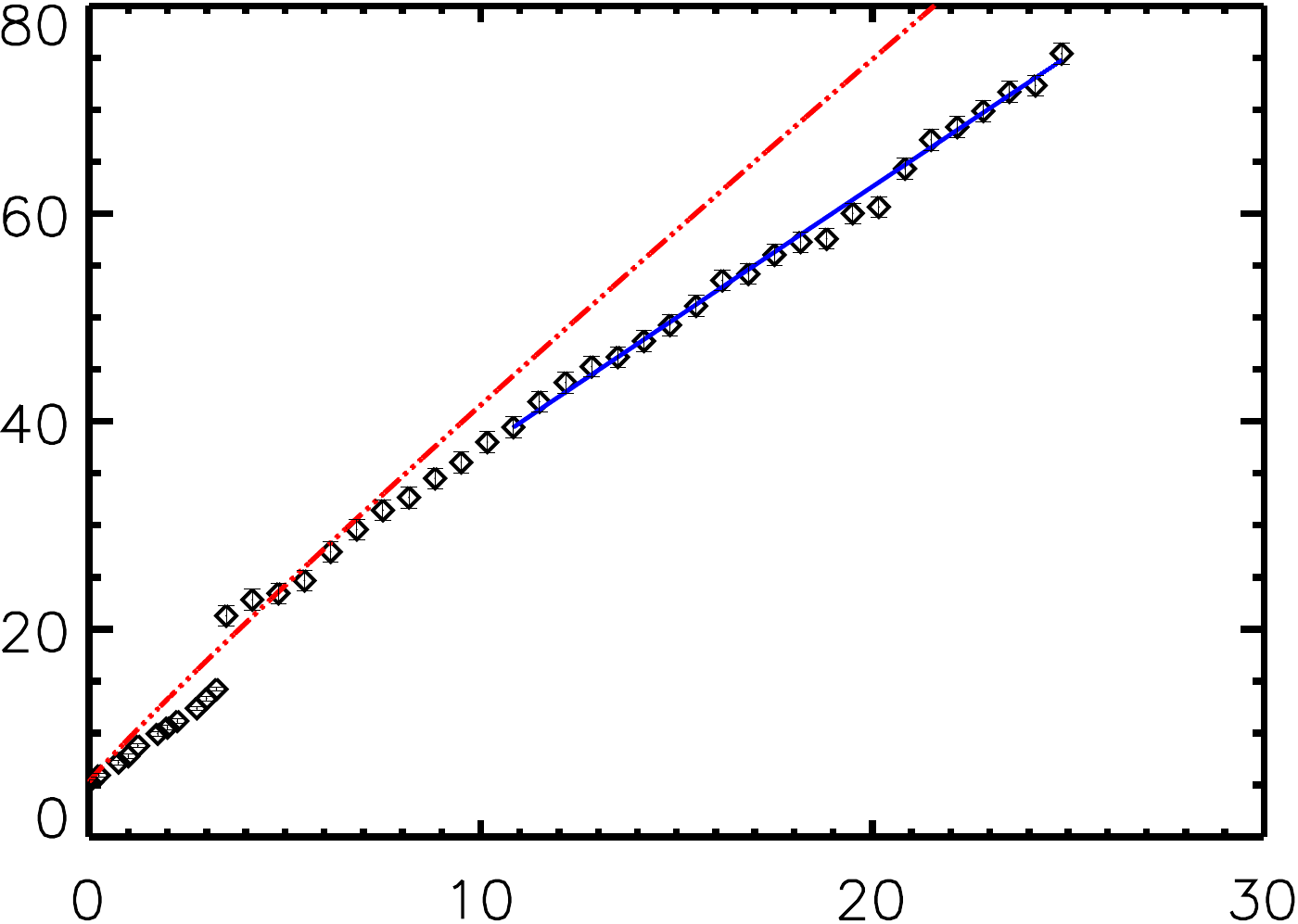}
		\put(-230,62.7){{\rotatebox{90}{{\color{black}\fontsize{12}{12}\fontseries{n}\fontfamily{phv}\selectfont  Height (R$_{\odot}$)}}}}
		\put(-160.8,-14.9){{\rotatebox{0}{{\color{black}\fontsize{12}{12}\fontseries{n}\fontfamily{phv}\selectfont  Elapsed Time (hrs)}}}}
		\put(-130,130.7){{\rotatebox{0}{{\color{black}\fontsize{13}{13}\fontseries{n}\fontfamily{phv}\selectfont \underbar{CME 20}}}}}
		\hspace*{0.06\textwidth}
		\includegraphics[width=0.555\textwidth,clip=]{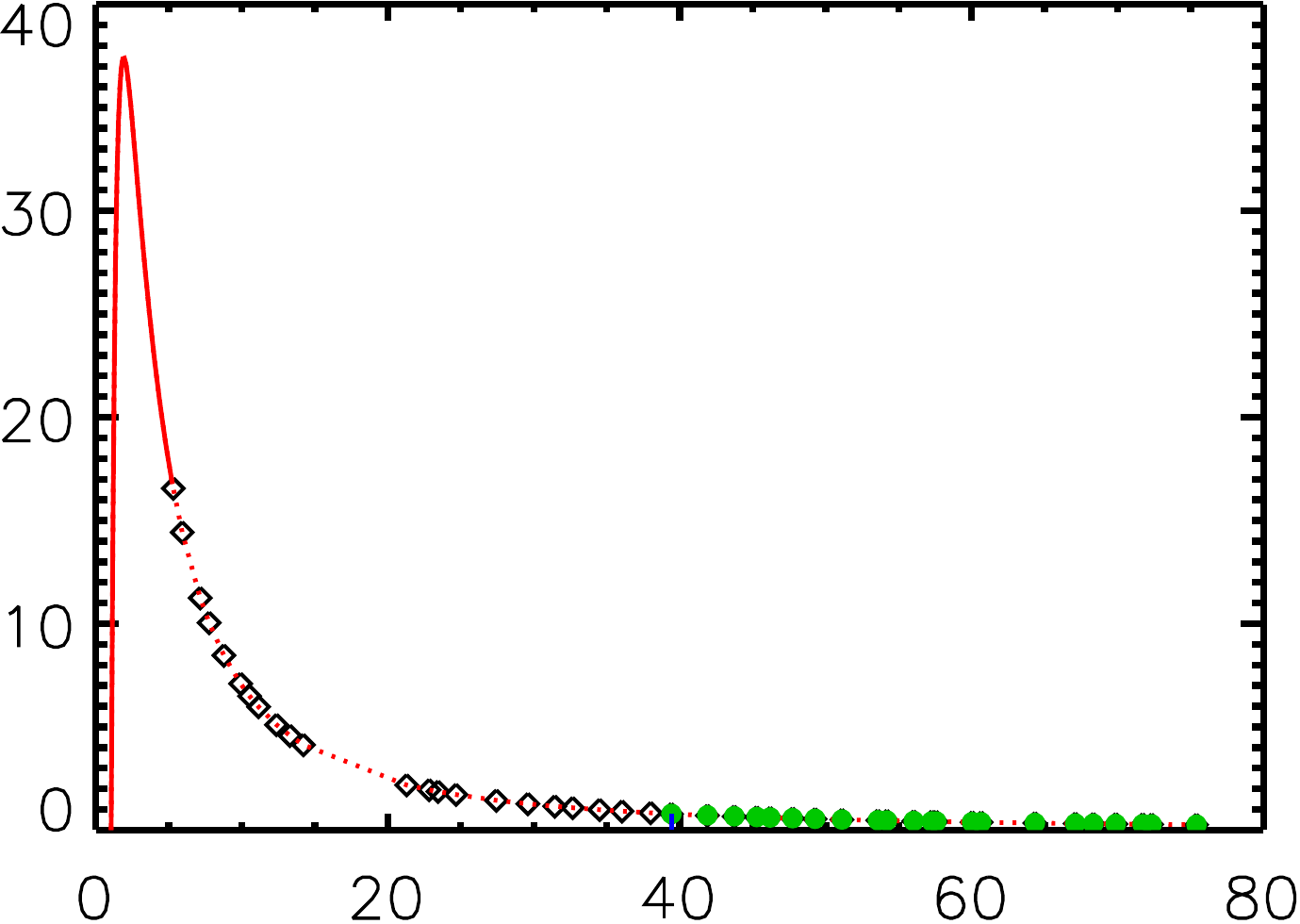}
		\put(-235,43.7){{\rotatebox{90}{{\color{black}\fontsize{12}{12}\fontseries{n}\fontfamily{phv}\selectfont  Force ($10^{17}$ dyn)}}}}
		\put(-140.8,-12.9){{\rotatebox{0}{{\color{black}\fontsize{12}{12}\fontseries{n}\fontfamily{phv}\selectfont Height (R$_{\odot}$)}}}}
		\put(-130,130.7){{\rotatebox{0}{{\color{black}\fontsize{13}{13}\fontseries{n}\fontfamily{phv}\selectfont  \underbar{CME 20}}}}}
		  }
  \vspace{0.0261\textwidth}  
  \caption[Height-time and Force profiles for CMEs 19 and 20]{Height-time and Force profiles for CMEs 19 and 20. Caption same as Figure \ref{fig52}}
  \label{fig60}
  \end{figure}

  \clearpage
  \begin{figure}[h]    
    \centering                              
    \centerline{\hspace*{0.06\textwidth}
		\includegraphics[width=0.56\textwidth,clip=]{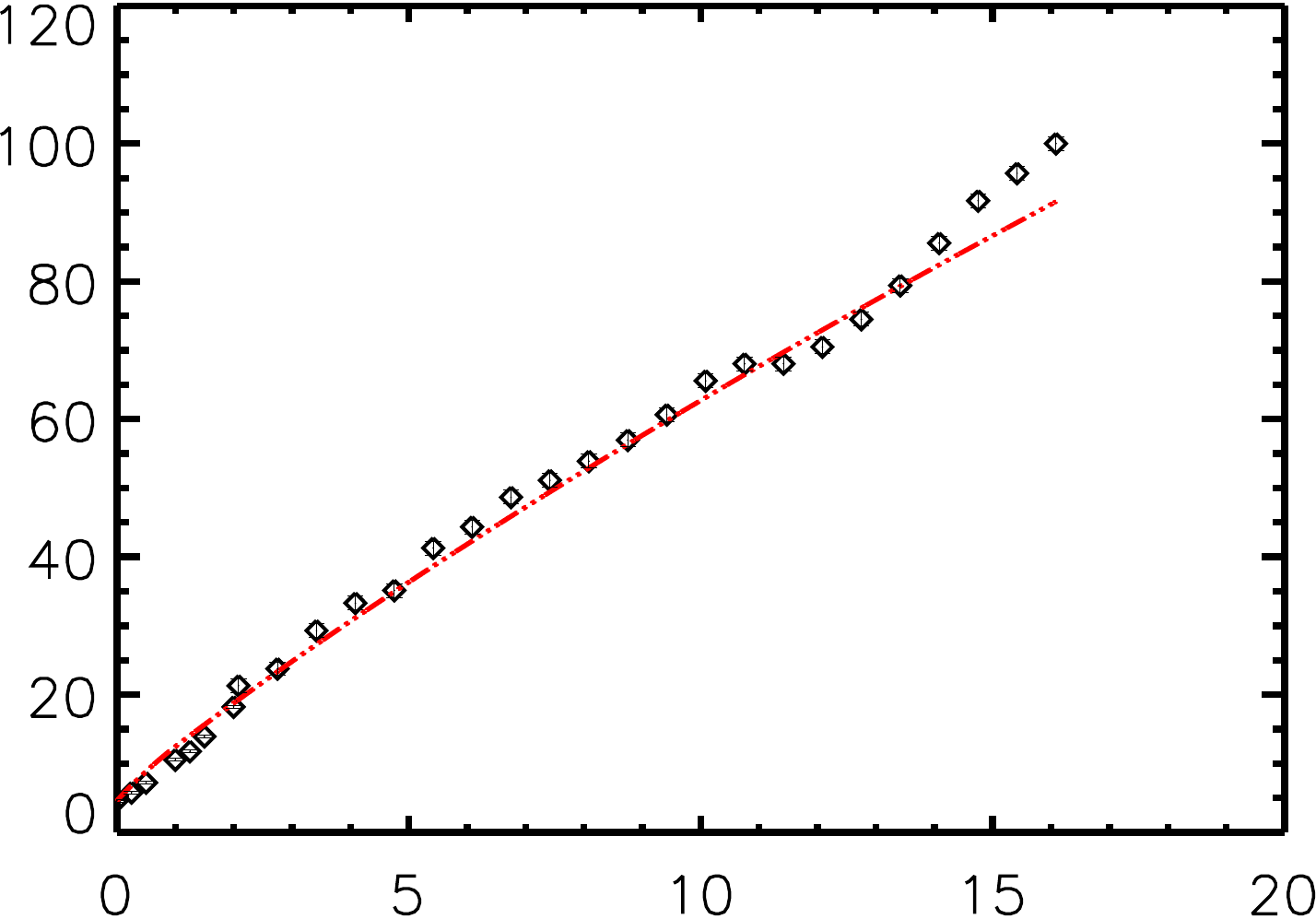}
		\put(-230,62.7){{\rotatebox{90}{{\color{black}\fontsize{12}{12}\fontseries{n}\fontfamily{phv}\selectfont  Height (R$_{\odot}$)}}}}
		\put(-160.8,-14.9){{\rotatebox{0}{{\color{black}\fontsize{12}{12}\fontseries{n}\fontfamily{phv}\selectfont  Elapsed Time (hrs)}}}}
		\put(-130,130.7){{\rotatebox{0}{{\color{black}\fontsize{13}{13}\fontseries{n}\fontfamily{phv}\selectfont \underbar{CME 21 ($f$)}}}}}
		\put(-115,158.7){{\rotatebox{0}{{\color{black}\fontsize{13}{13}\fontseries{n}\fontfamily{phv}\selectfont (a)}}}}
		\hspace*{0.069\textwidth}
		\includegraphics[width=0.59\textwidth,clip=]{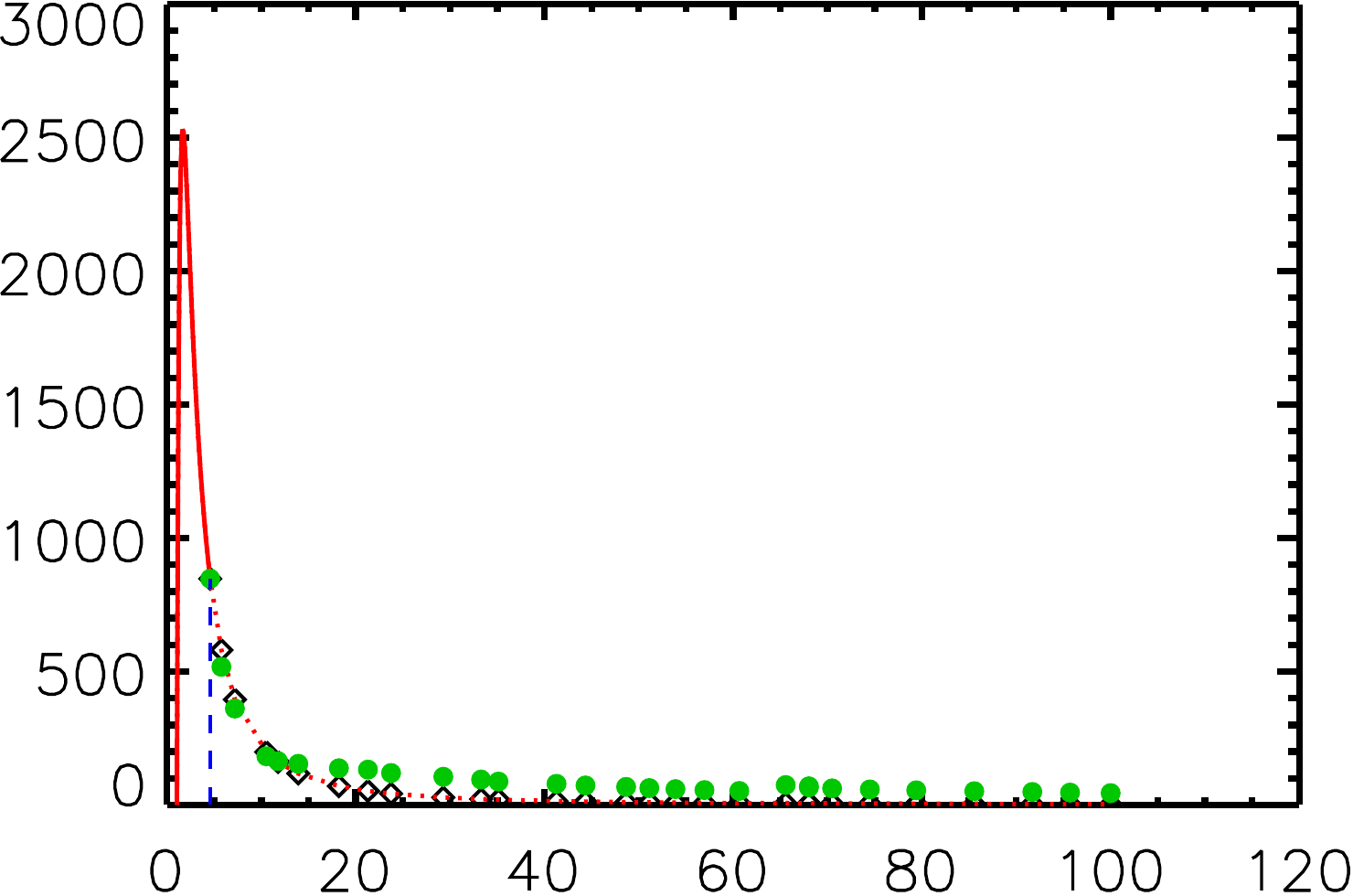}
		\put(-246,43.7){{\rotatebox{90}{{\color{black}\fontsize{12}{12}\fontseries{n}\fontfamily{phv}\selectfont  Force ($10^{17}$ dyn)}}}}
		\put(-140.8,-12.9){{\rotatebox{0}{{\color{black}\fontsize{12}{12}\fontseries{n}\fontfamily{phv}\selectfont Height (R$_{\odot}$)}}}}
		\put(-130,130.7){{\rotatebox{0}{{\color{black}\fontsize{13}{13}\fontseries{n}\fontfamily{phv}\selectfont  \underbar{CME 21 ($f$)}}}}}
		\put(-117,158.7){{\rotatebox{0}{{\color{black}\fontsize{13}{13}\fontseries{n}\fontfamily{phv}\selectfont (b)}}}}          
		  }
		\vspace{1.3cm}
		  \centerline{\hspace*{0.06\textwidth}
		\includegraphics[width=0.56\textwidth,clip=]{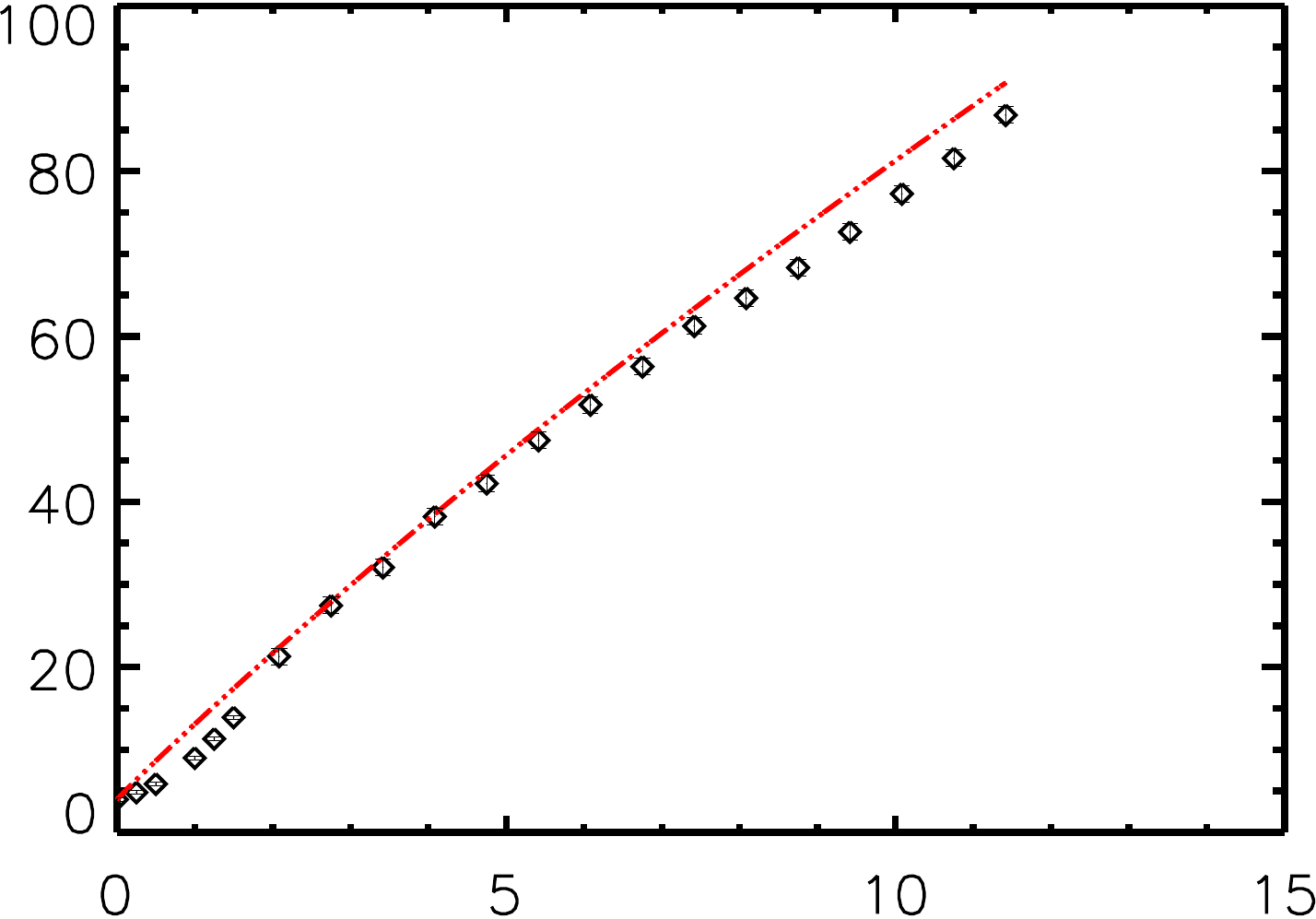}
		\put(-230,62.7){{\rotatebox{90}{{\color{black}\fontsize{12}{12}\fontseries{n}\fontfamily{phv}\selectfont  Height (R$_{\odot}$)}}}}
		\put(-160.8,-14.9){{\rotatebox{0}{{\color{black}\fontsize{12}{12}\fontseries{n}\fontfamily{phv}\selectfont  Elapsed Time (hrs)}}}}
		\put(-130,130.7){{\rotatebox{0}{{\color{black}\fontsize{13}{13}\fontseries{n}\fontfamily{phv}\selectfont \underbar{CME 22 ($f$)}}}}}
		\hspace*{0.069\textwidth}
		\includegraphics[width=0.59\textwidth,clip=]{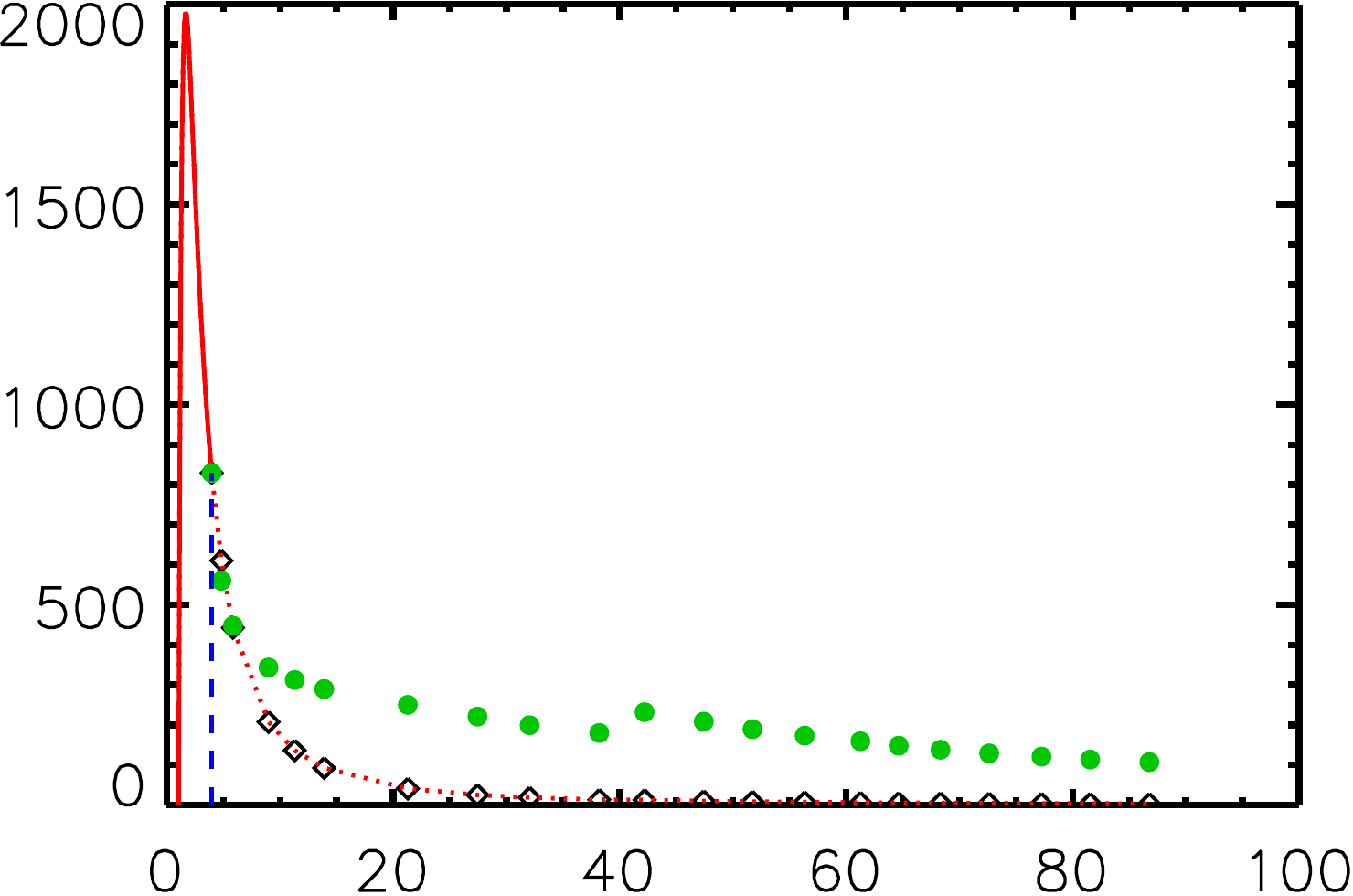}
		\put(-242,43.7){{\rotatebox{90}{{\color{black}\fontsize{12}{12}\fontseries{n}\fontfamily{phv}\selectfont  Force ($10^{17}$ dyn)}}}}
		\put(-140.8,-12.9){{\rotatebox{0}{{\color{black}\fontsize{12}{12}\fontseries{n}\fontfamily{phv}\selectfont Height (R$_{\odot}$)}}}}
		\put(-130,130.7){{\rotatebox{0}{{\color{black}\fontsize{13}{13}\fontseries{n}\fontfamily{phv}\selectfont  \underbar{CME 22 ($f$)}}}}}
		  }
  \vspace{0.0261\textwidth}  
  \caption[Height-time and Force profiles for CMEs 21 and 22]{Height-time and Force profiles for CMEs 21 and 22. Caption same as Figure \ref{fig52}}
  \label{fig61}
  \end{figure}

  \clearpage
  \begin{figure}[h]    
    \centering                              
    \centerline{\hspace*{0.06\textwidth}
		\includegraphics[width=0.557\textwidth,clip=]{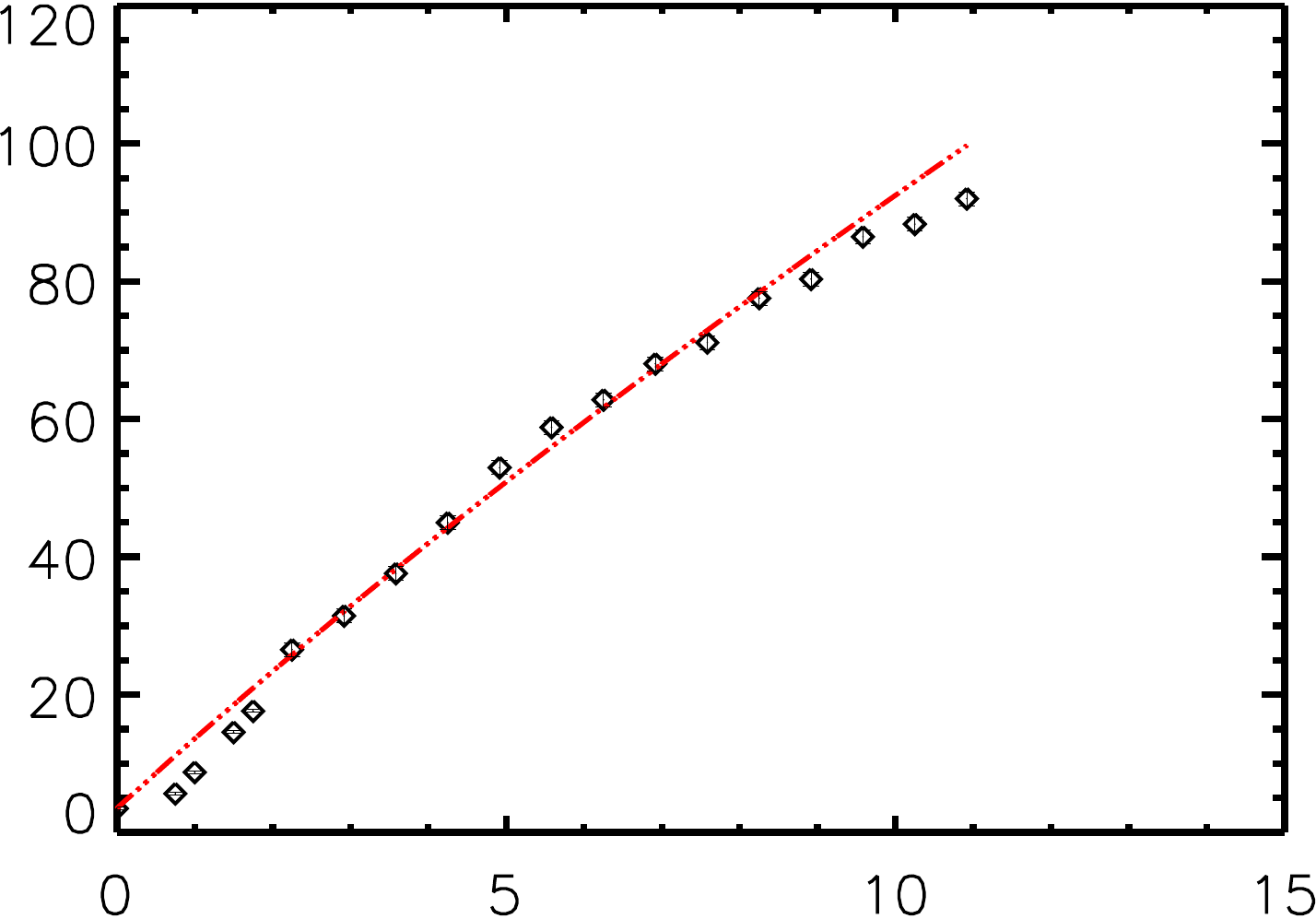}
		\put(-230,62.7){{\rotatebox{90}{{\color{black}\fontsize{12}{12}\fontseries{n}\fontfamily{phv}\selectfont  Height (R$_{\odot}$)}}}}
		\put(-160.8,-14.9){{\rotatebox{0}{{\color{black}\fontsize{12}{12}\fontseries{n}\fontfamily{phv}\selectfont  Elapsed Time (hrs)}}}}
		\put(-130,130.7){{\rotatebox{0}{{\color{black}\fontsize{13}{13}\fontseries{n}\fontfamily{phv}\selectfont \underbar{CME 23 ($f$)}}}}}
		\put(-115,158.7){{\rotatebox{0}{{\color{black}\fontsize{13}{13}\fontseries{n}\fontfamily{phv}\selectfont (a)}}}}
		\hspace*{0.069\textwidth}
		\includegraphics[width=0.58\textwidth,clip=]{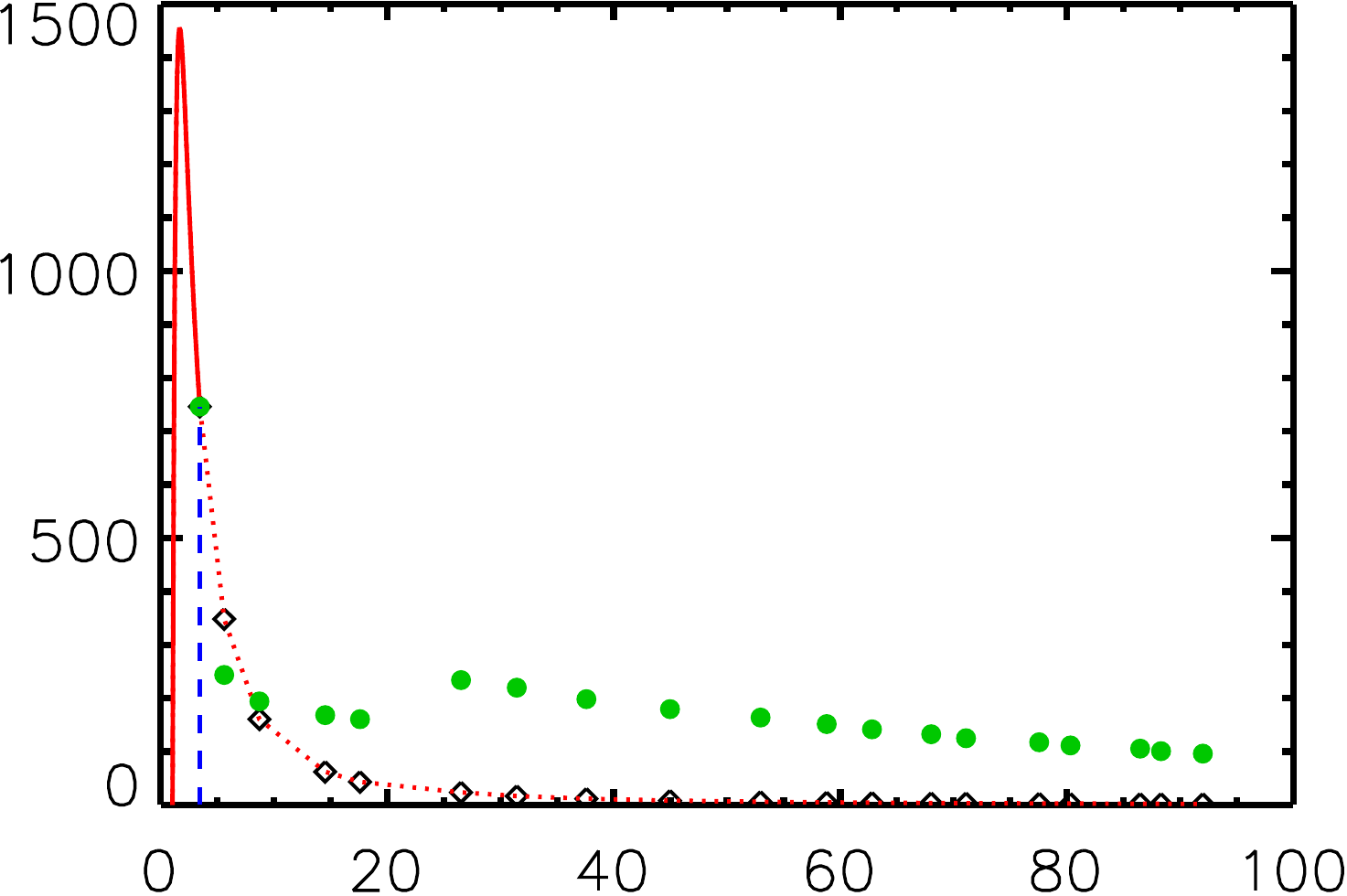}
		\put(-238,43.7){{\rotatebox{90}{{\color{black}\fontsize{12}{12}\fontseries{n}\fontfamily{phv}\selectfont  Force ($10^{17}$ dyn)}}}}
		\put(-140.8,-12.9){{\rotatebox{0}{{\color{black}\fontsize{12}{12}\fontseries{n}\fontfamily{phv}\selectfont Height (R$_{\odot}$)}}}}
		\put(-130,130.7){{\rotatebox{0}{{\color{black}\fontsize{13}{13}\fontseries{n}\fontfamily{phv}\selectfont  \underbar{CME 23 ($f$)}}}}}
		\put(-117,158.7){{\rotatebox{0}{{\color{black}\fontsize{13}{13}\fontseries{n}\fontfamily{phv}\selectfont (b)}}}}          
		  }
		\vspace{1.3cm}
		  \centerline{\hspace*{0.06\textwidth}
		\includegraphics[width=0.56\textwidth,clip=]{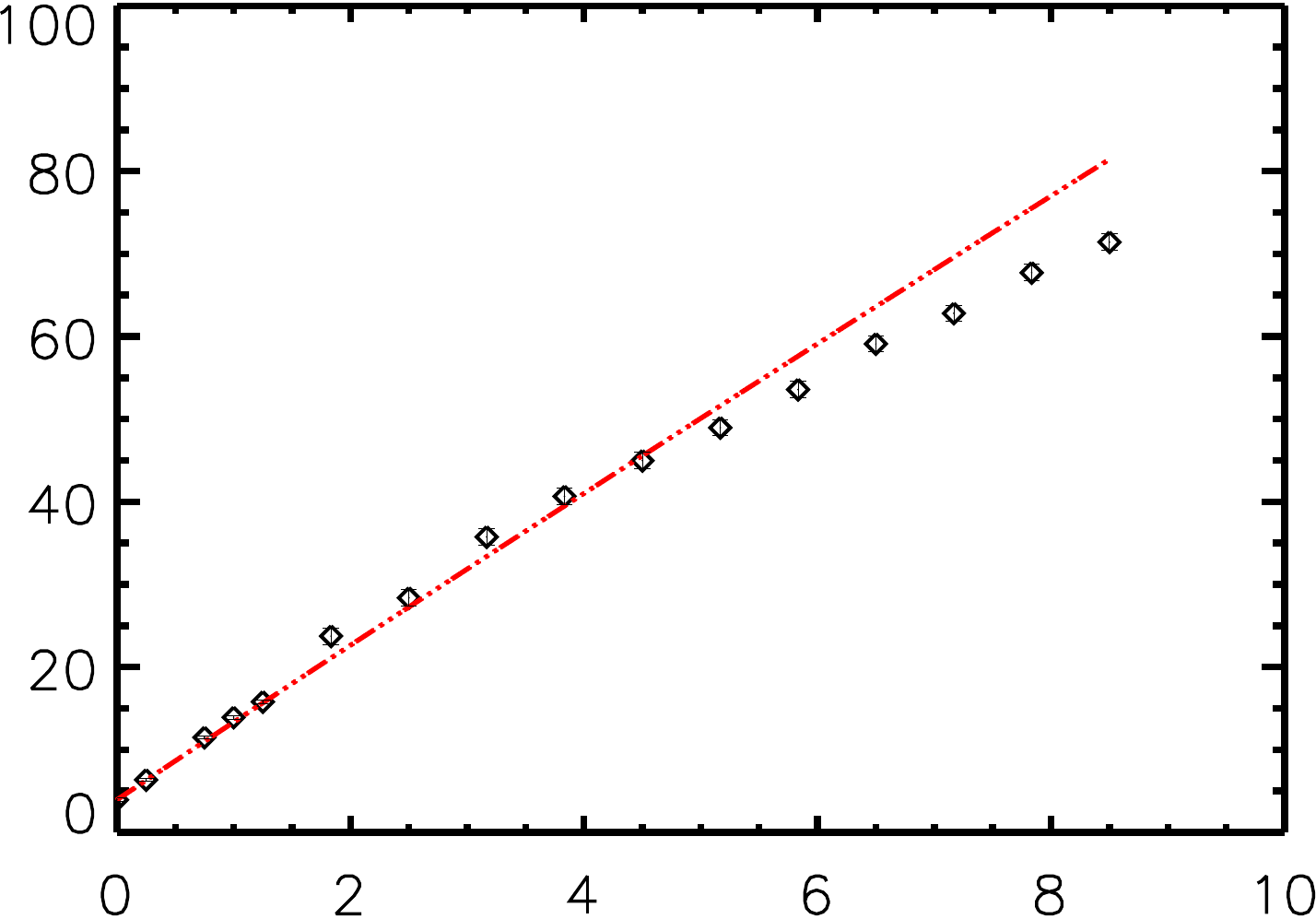}
		\put(-230,62.7){{\rotatebox{90}{{\color{black}\fontsize{12}{12}\fontseries{n}\fontfamily{phv}\selectfont  Height (R$_{\odot}$)}}}}
		\put(-160.8,-14.9){{\rotatebox{0}{{\color{black}\fontsize{12}{12}\fontseries{n}\fontfamily{phv}\selectfont  Elapsed Time (hrs)}}}}
		\put(-130,130.7){{\rotatebox{0}{{\color{black}\fontsize{13}{13}\fontseries{n}\fontfamily{phv}\selectfont \underbar{CME 24 ($f$)}}}}}
		\hspace*{0.069\textwidth}
		\includegraphics[width=0.564\textwidth,clip=]{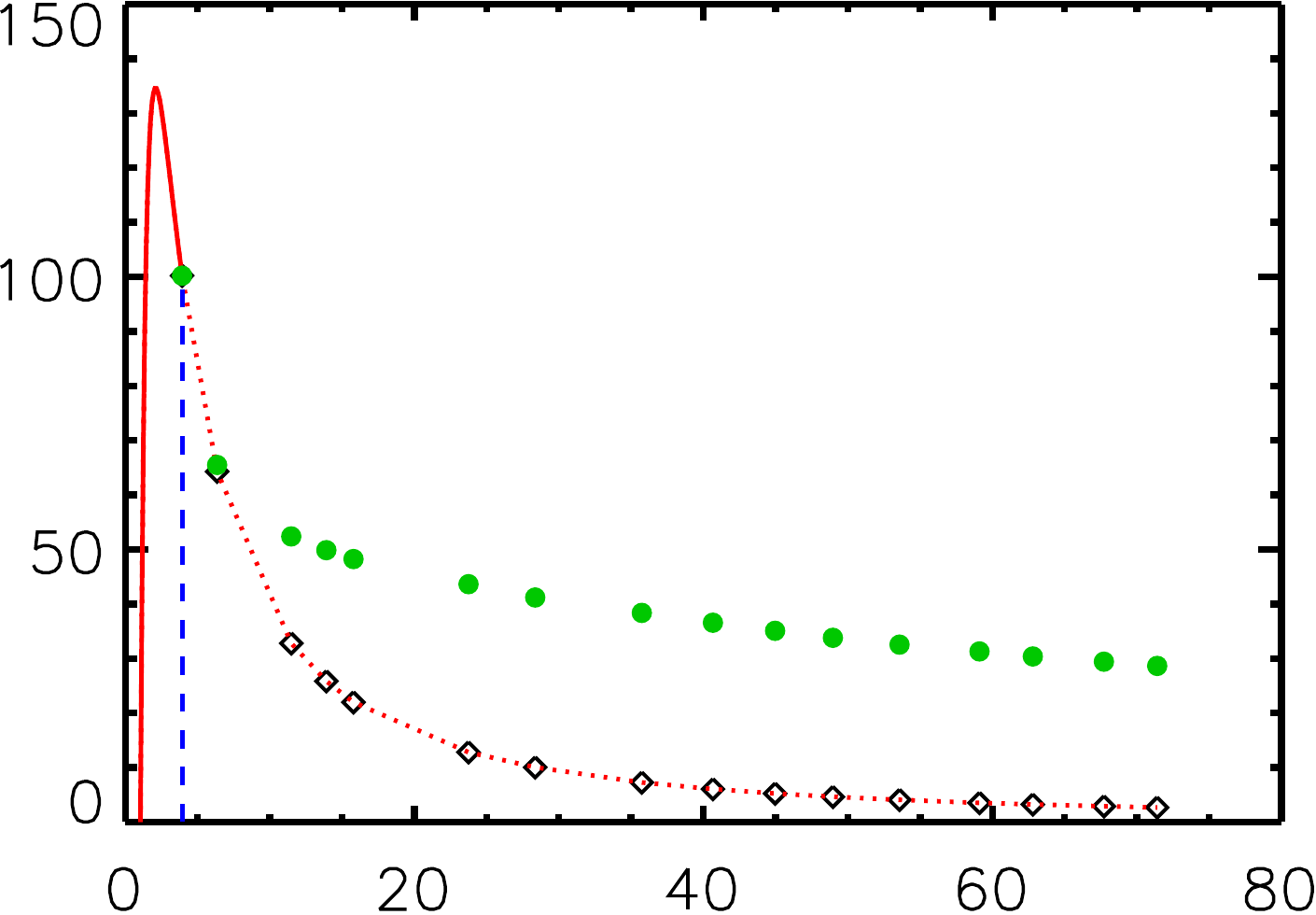}
		\put(-231,43.7){{\rotatebox{90}{{\color{black}\fontsize{12}{12}\fontseries{n}\fontfamily{phv}\selectfont  Force ($10^{17}$ dyn)}}}}
		\put(-140.8,-12.9){{\rotatebox{0}{{\color{black}\fontsize{12}{12}\fontseries{n}\fontfamily{phv}\selectfont Height (R$_{\odot}$)}}}}
		\put(-130,130.7){{\rotatebox{0}{{\color{black}\fontsize{13}{13}\fontseries{n}\fontfamily{phv}\selectfont  \underbar{CME 24 ($f$)}}}}}
		  }
  \vspace{0.0261\textwidth}  
  \caption[Height-time and Force profiles for CMEs 23 and 24]{Height-time and Force profiles for CMEs 23 and 24. Caption same as Figure \ref{fig52}}
  \label{fig62}
  \end{figure}

  \clearpage
  \begin{figure}[h]    
    \centering                              
    \centerline{\hspace*{0.06\textwidth}
		\includegraphics[width=0.55\textwidth,clip=]{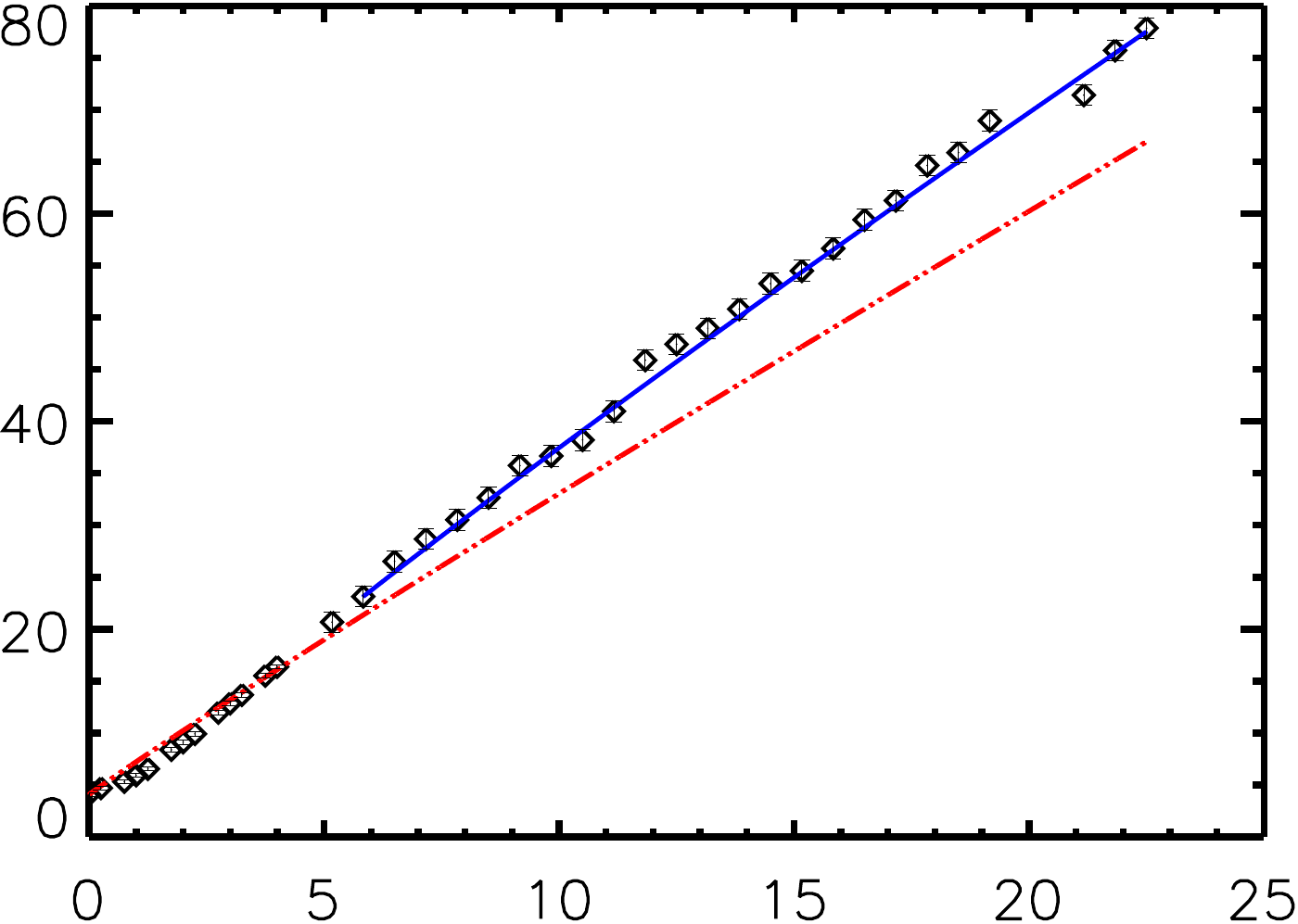}
		\put(-230,62.7){{\rotatebox{90}{{\color{black}\fontsize{12}{12}\fontseries{n}\fontfamily{phv}\selectfont  Height (R$_{\odot}$)}}}}
		\put(-160.8,-14.9){{\rotatebox{0}{{\color{black}\fontsize{12}{12}\fontseries{n}\fontfamily{phv}\selectfont  Elapsed Time (hrs)}}}}
		\put(-130,130.7){{\rotatebox{0}{{\color{black}\fontsize{13}{13}\fontseries{n}\fontfamily{phv}\selectfont \underbar{CME 25}}}}}
		\put(-115,158.7){{\rotatebox{0}{{\color{black}\fontsize{13}{13}\fontseries{n}\fontfamily{phv}\selectfont (a)}}}}
		\hspace*{0.069\textwidth}
		\includegraphics[width=0.555\textwidth,clip=]{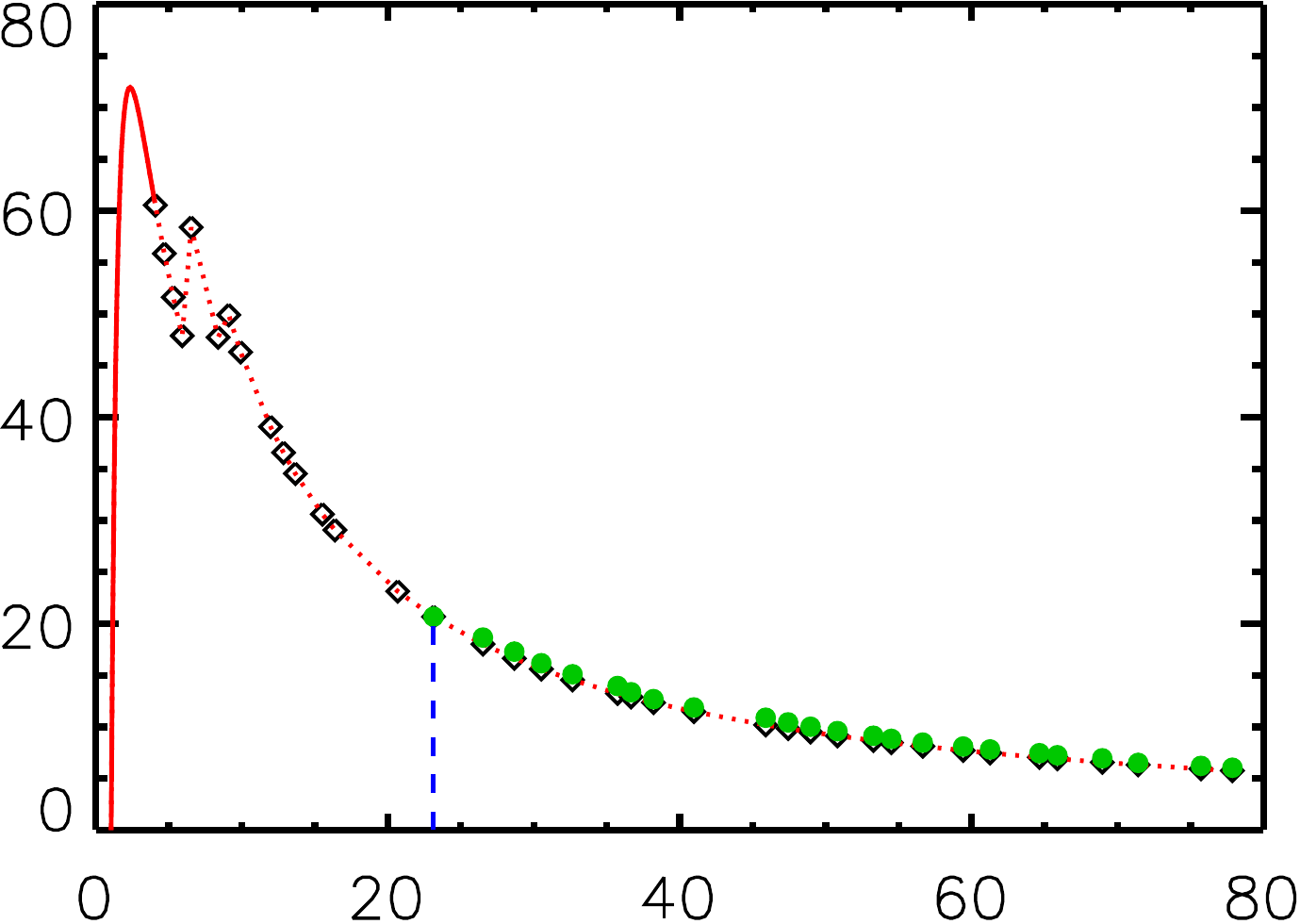}
		\put(-231,43.7){{\rotatebox{90}{{\color{black}\fontsize{12}{12}\fontseries{n}\fontfamily{phv}\selectfont  Force ($10^{17}$ dyn)}}}}
		\put(-140.8,-12.9){{\rotatebox{0}{{\color{black}\fontsize{12}{12}\fontseries{n}\fontfamily{phv}\selectfont Height (R$_{\odot}$)}}}}
		\put(-130,130.7){{\rotatebox{0}{{\color{black}\fontsize{13}{13}\fontseries{n}\fontfamily{phv}\selectfont  \underbar{CME 25}}}}}
		\put(-117,158.7){{\rotatebox{0}{{\color{black}\fontsize{13}{13}\fontseries{n}\fontfamily{phv}\selectfont (b)}}}}          
		  }
		\vspace{1.3cm}
		  \centerline{\hspace*{0.06\textwidth}
		\includegraphics[width=0.56\textwidth,clip=]{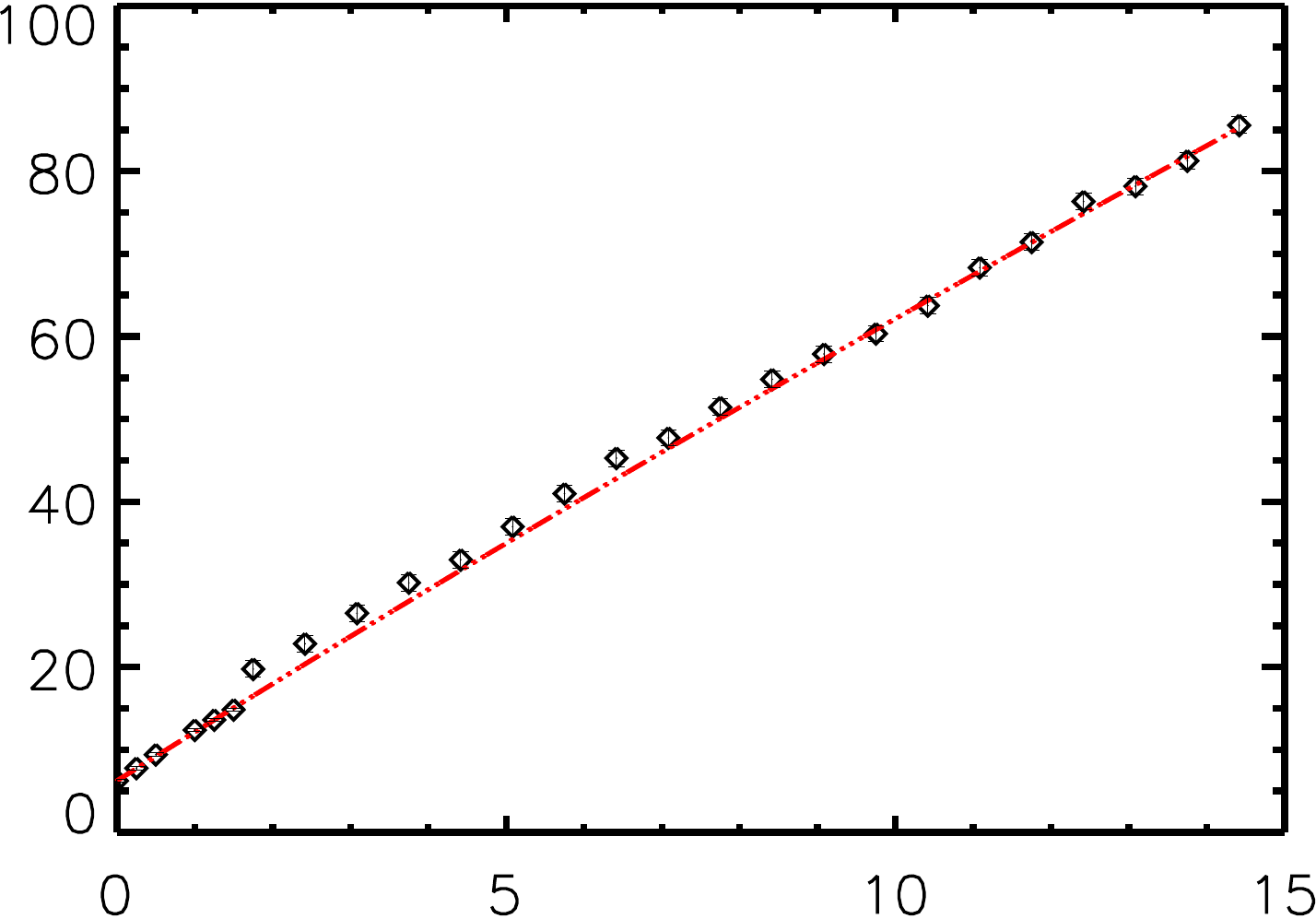}
		\put(-230,62.7){{\rotatebox{90}{{\color{black}\fontsize{12}{12}\fontseries{n}\fontfamily{phv}\selectfont  Height (R$_{\odot}$)}}}}
		\put(-160.8,-14.9){{\rotatebox{0}{{\color{black}\fontsize{12}{12}\fontseries{n}\fontfamily{phv}\selectfont  Elapsed Time (hrs)}}}}
		\put(-130,130.7){{\rotatebox{0}{{\color{black}\fontsize{13}{13}\fontseries{n}\fontfamily{phv}\selectfont \underbar{CME 26 ($f$)}}}}}
		\hspace*{0.069\textwidth}
		\includegraphics[width=0.56\textwidth,clip=]{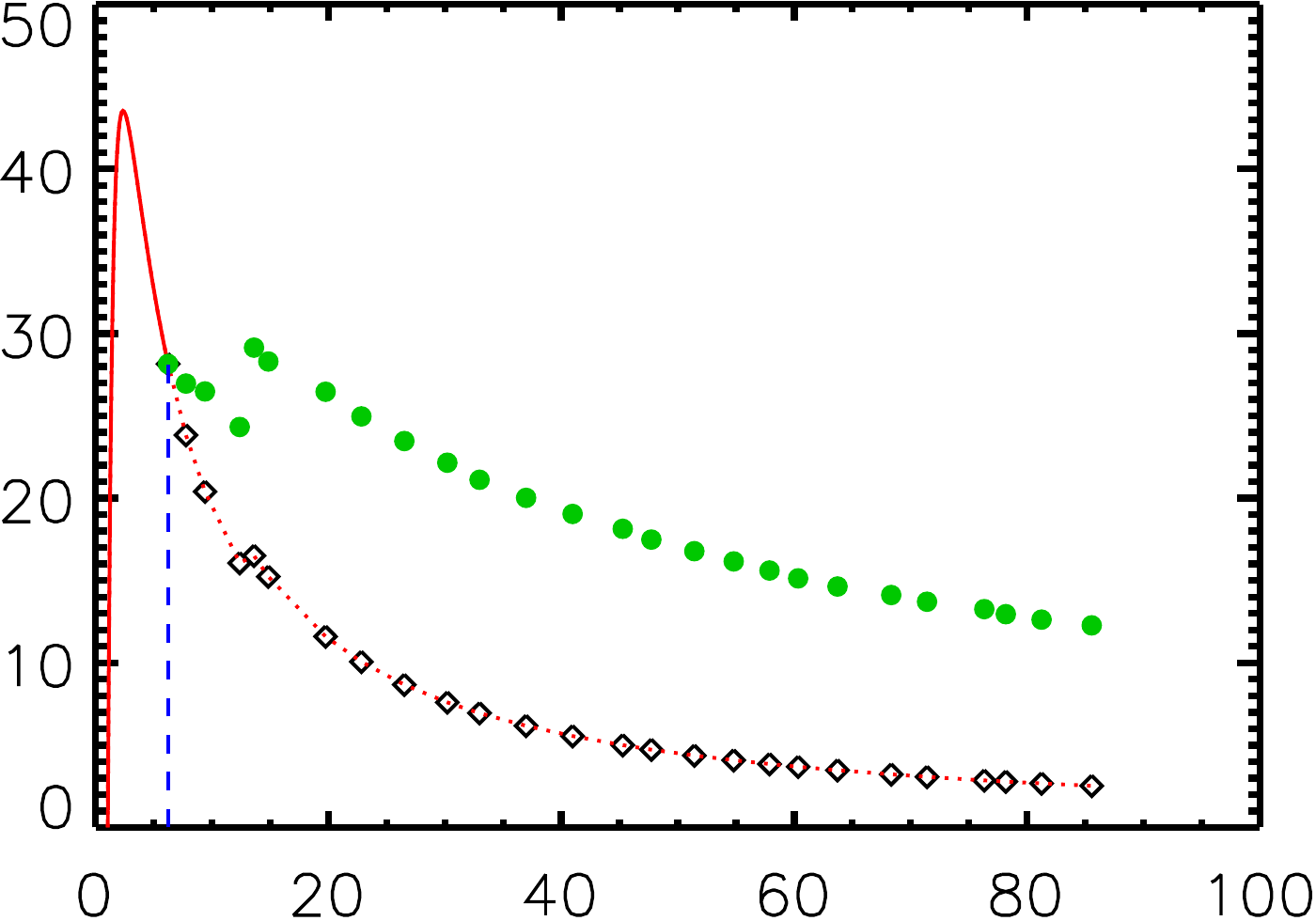}
		\put(-234,43.7){{\rotatebox{90}{{\color{black}\fontsize{12}{12}\fontseries{n}\fontfamily{phv}\selectfont  Force ($10^{17}$ dyn)}}}}
		\put(-140.8,-12.9){{\rotatebox{0}{{\color{black}\fontsize{12}{12}\fontseries{n}\fontfamily{phv}\selectfont Height (R$_{\odot}$)}}}}
		\put(-130,130.7){{\rotatebox{0}{{\color{black}\fontsize{13}{13}\fontseries{n}\fontfamily{phv}\selectfont  \underbar{CME 26 ($f$)}}}}}
		  }
  \vspace{0.0261\textwidth}  
  \caption[Height-time and Force profiles for CMEs 25 and 26]{Height-time and Force profiles for CMEs 25 and 26. Caption same as Figure \ref{fig52}}
  \label{fig63}
  \end{figure}

  \clearpage
  \begin{figure}[h]    
    \centering                              
    \centerline{\hspace*{0.06\textwidth}
		\includegraphics[width=0.557\textwidth,clip=]{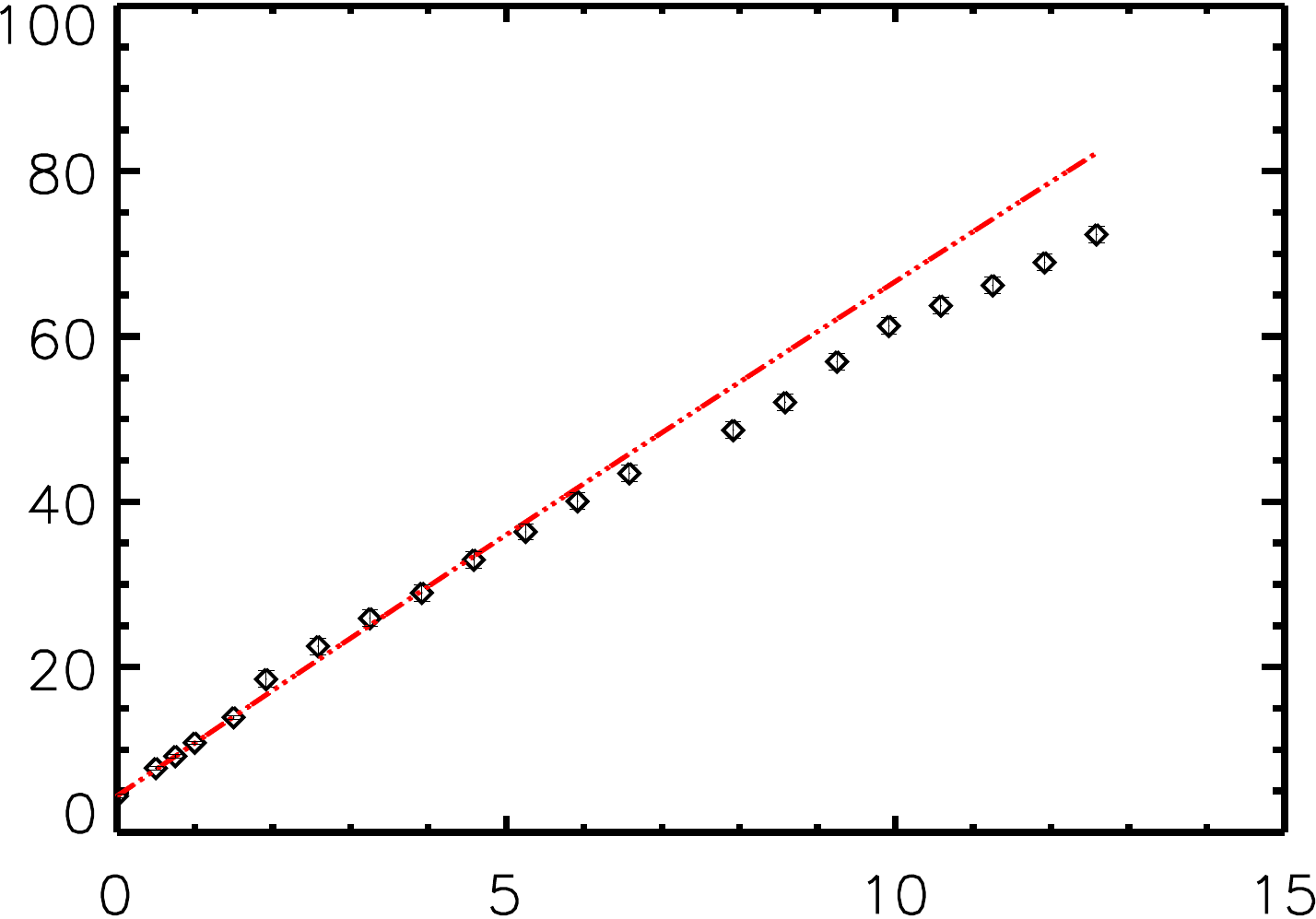}
		\put(-230,62.7){{\rotatebox{90}{{\color{black}\fontsize{12}{12}\fontseries{n}\fontfamily{phv}\selectfont  Height (R$_{\odot}$)}}}}
		\put(-160.8,-14.9){{\rotatebox{0}{{\color{black}\fontsize{12}{12}\fontseries{n}\fontfamily{phv}\selectfont  Elapsed Time (hrs)}}}}
		\put(-130,130.7){{\rotatebox{0}{{\color{black}\fontsize{13}{13}\fontseries{n}\fontfamily{phv}\selectfont \underbar{CME 27 ($f$)}}}}}
		\put(-115,158.7){{\rotatebox{0}{{\color{black}\fontsize{13}{13}\fontseries{n}\fontfamily{phv}\selectfont (a)}}}}
		\hspace*{0.069\textwidth}
		\includegraphics[width=0.56\textwidth,clip=]{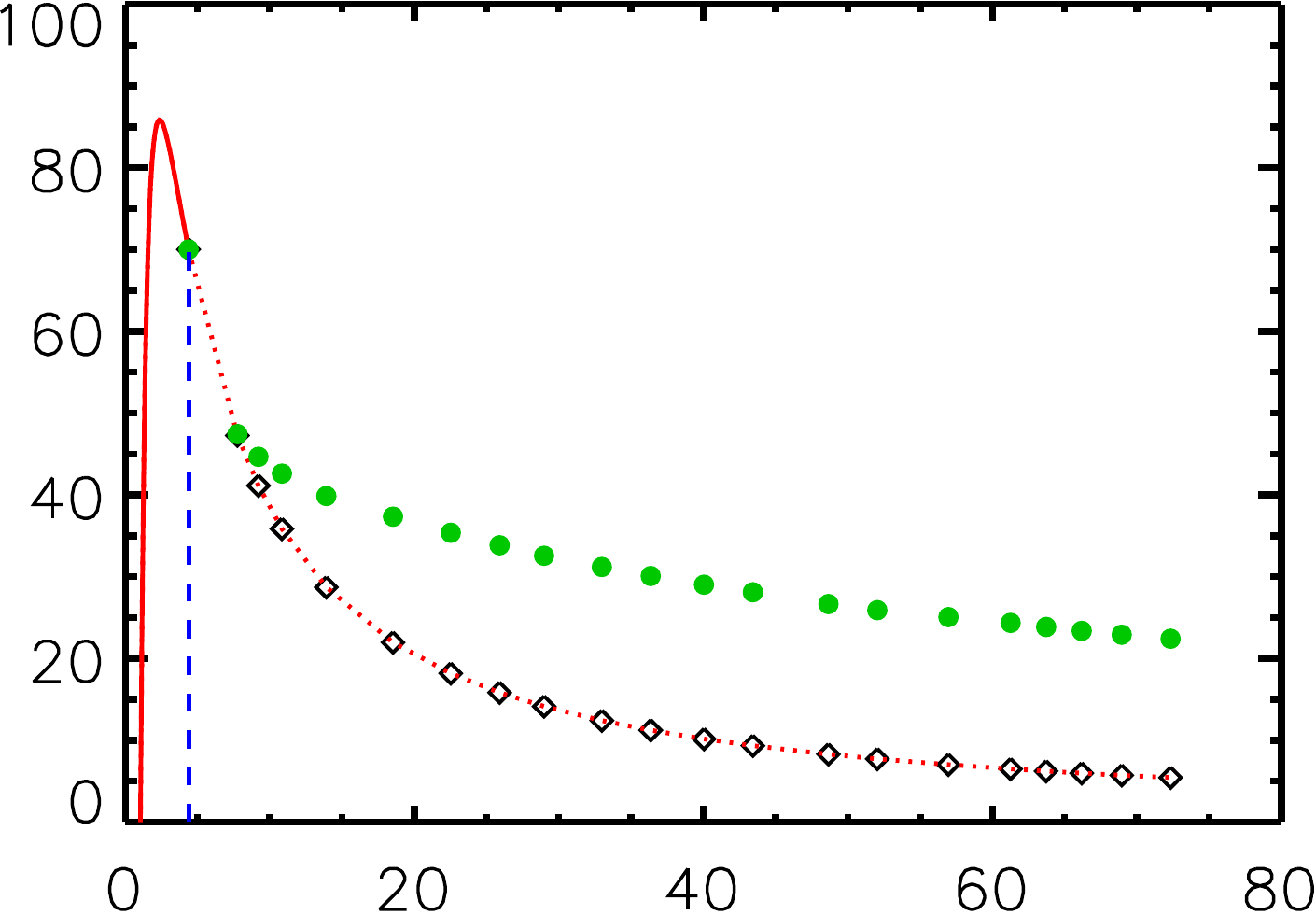}
		\put(-231,43.7){{\rotatebox{90}{{\color{black}\fontsize{12}{12}\fontseries{n}\fontfamily{phv}\selectfont  Force ($10^{17}$ dyn)}}}}
		\put(-140.8,-12.9){{\rotatebox{0}{{\color{black}\fontsize{12}{12}\fontseries{n}\fontfamily{phv}\selectfont Height (R$_{\odot}$)}}}}
		\put(-130,130.7){{\rotatebox{0}{{\color{black}\fontsize{13}{13}\fontseries{n}\fontfamily{phv}\selectfont  \underbar{CME 27 ($f$)}}}}}
		\put(-117,158.7){{\rotatebox{0}{{\color{black}\fontsize{13}{13}\fontseries{n}\fontfamily{phv}\selectfont (b)}}}}          
		  }
		\vspace{1.3cm}
		  \centerline{\hspace*{0.06\textwidth}
		\includegraphics[width=0.55\textwidth,clip=]{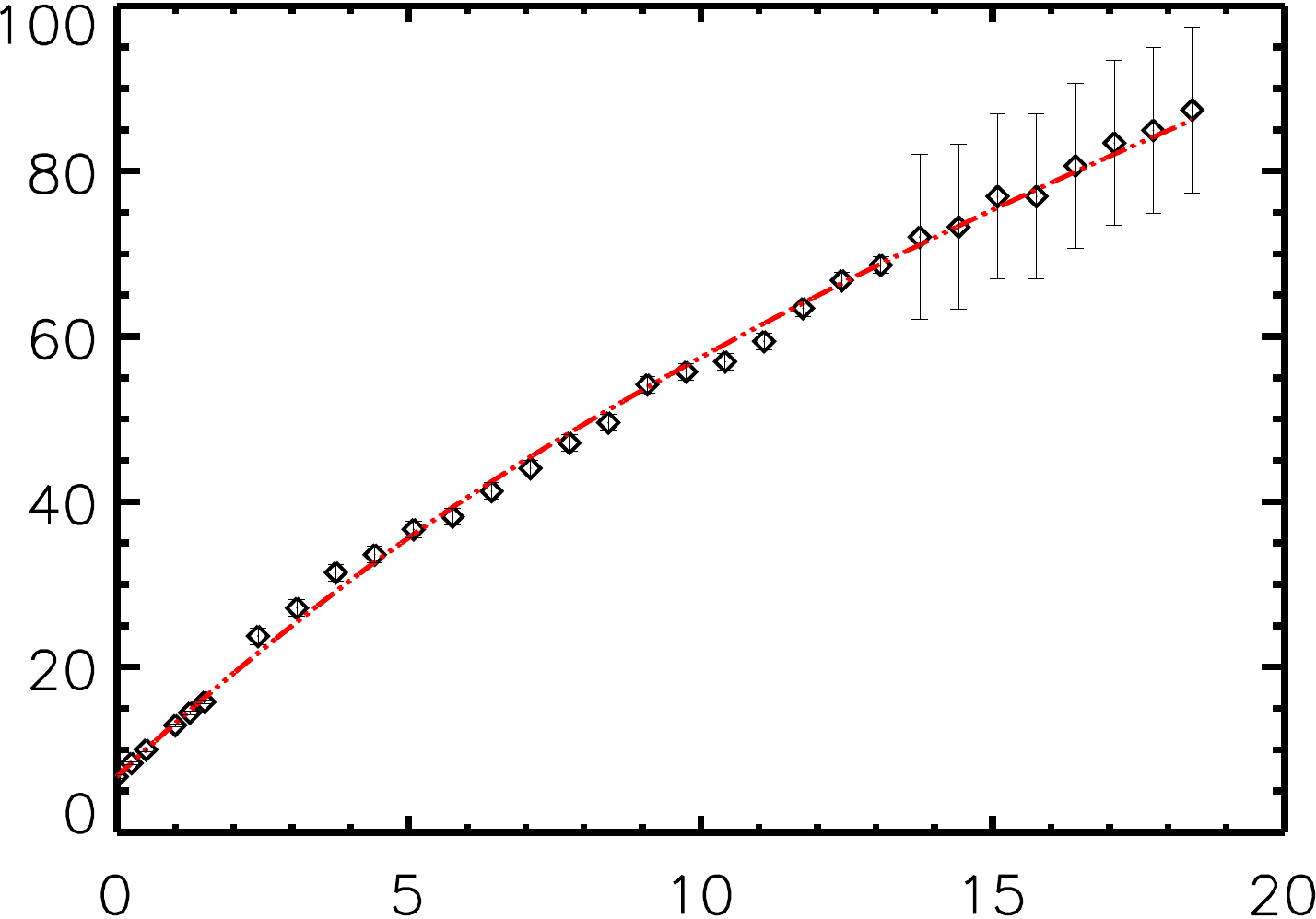}
		\put(-230,62.7){{\rotatebox{90}{{\color{black}\fontsize{12}{12}\fontseries{n}\fontfamily{phv}\selectfont  Height (R$_{\odot}$)}}}}
		\put(-160.8,-14.9){{\rotatebox{0}{{\color{black}\fontsize{12}{12}\fontseries{n}\fontfamily{phv}\selectfont  Elapsed Time (hrs)}}}}
		\put(-130,130.7){{\rotatebox{0}{{\color{black}\fontsize{13}{13}\fontseries{n}\fontfamily{phv}\selectfont \underbar{CME 28 ($f$)}}}}}
		\hspace*{0.069\textwidth}
		\includegraphics[width=0.565\textwidth,clip=]{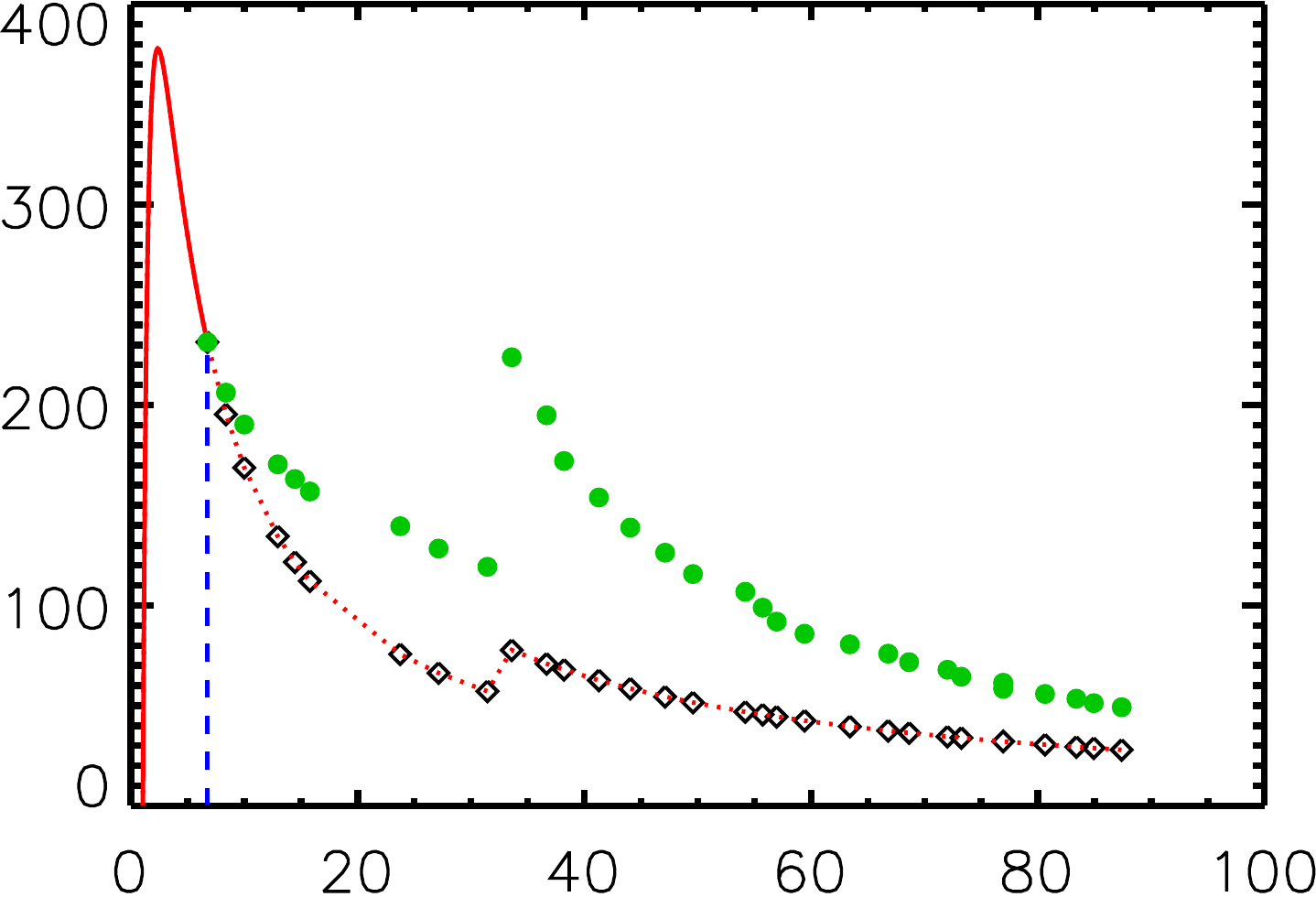}
		\put(-231,43.7){{\rotatebox{90}{{\color{black}\fontsize{12}{12}\fontseries{n}\fontfamily{phv}\selectfont  Force ($10^{17}$ dyn)}}}}
		\put(-140.8,-12.9){{\rotatebox{0}{{\color{black}\fontsize{12}{12}\fontseries{n}\fontfamily{phv}\selectfont Height (R$_{\odot}$)}}}}
		\put(-130,130.7){{\rotatebox{0}{{\color{black}\fontsize{13}{13}\fontseries{n}\fontfamily{phv}\selectfont  \underbar{CME 28 ($f$)}}}}}
		  }
  \vspace{0.0261\textwidth}  
  \caption[Height-time and Force profiles for CMEs 27 and 28]{Height-time and Force profiles for CMEs 27 and 28. Caption same as Figure \ref{fig52}}
  \label{fig64}
  \end{figure}

  \clearpage
  \begin{figure}[h]    
    \centering                              
    \centerline{\hspace*{0.06\textwidth}
		\includegraphics[width=0.55\textwidth,clip=]{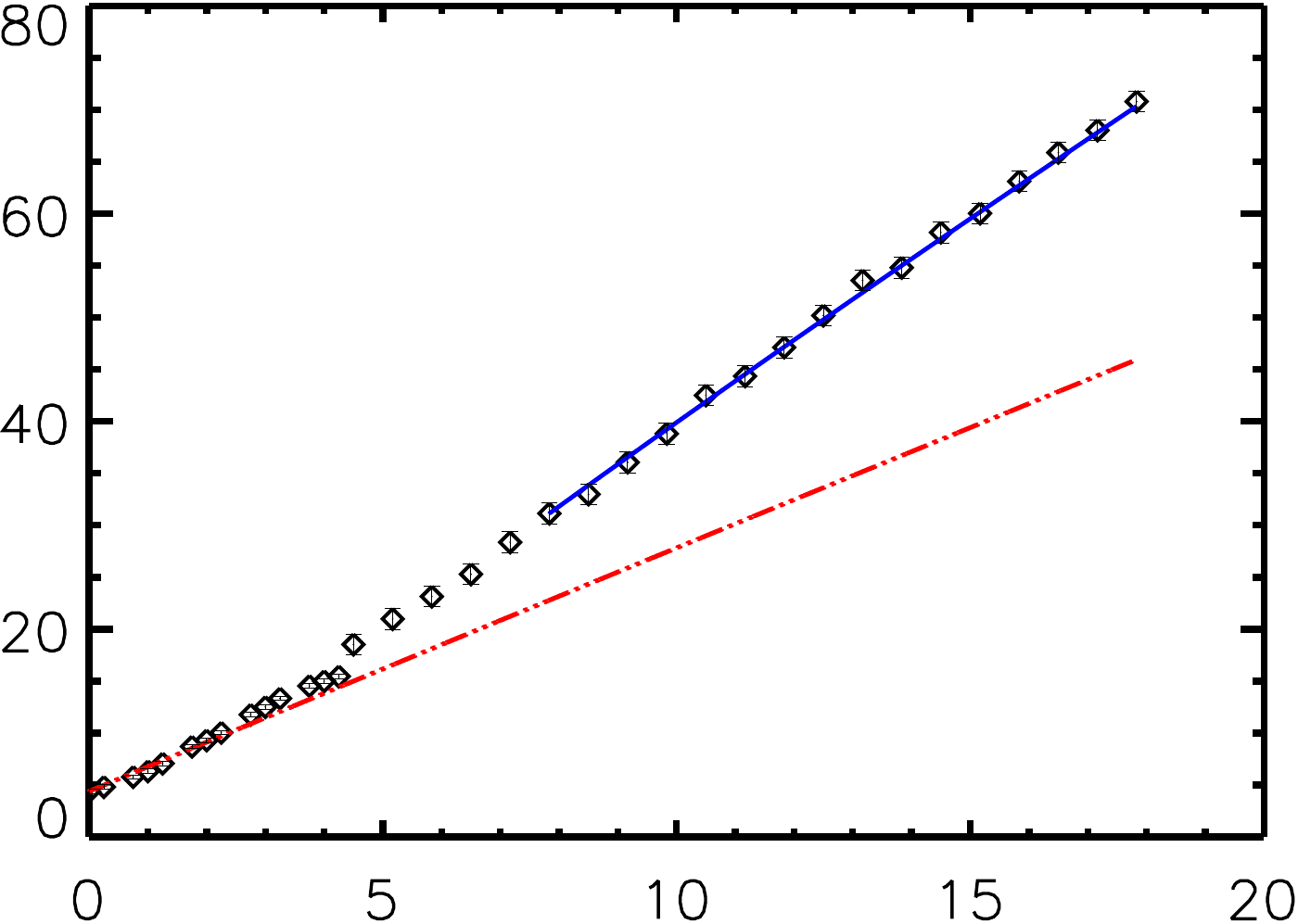}
		\put(-230,62.7){{\rotatebox{90}{{\color{black}\fontsize{12}{12}\fontseries{n}\fontfamily{phv}\selectfont  Height (R$_{\odot}$)}}}}
		\put(-160.8,-14.9){{\rotatebox{0}{{\color{black}\fontsize{12}{12}\fontseries{n}\fontfamily{phv}\selectfont  Elapsed Time (hrs)}}}}
		\put(-130,130.7){{\rotatebox{0}{{\color{black}\fontsize{13}{13}\fontseries{n}\fontfamily{phv}\selectfont \underbar{CME 29}}}}}
		\put(-115,158.7){{\rotatebox{0}{{\color{black}\fontsize{13}{13}\fontseries{n}\fontfamily{phv}\selectfont (a)}}}}
		\hspace*{0.069\textwidth}
		\includegraphics[width=0.565\textwidth,clip=]{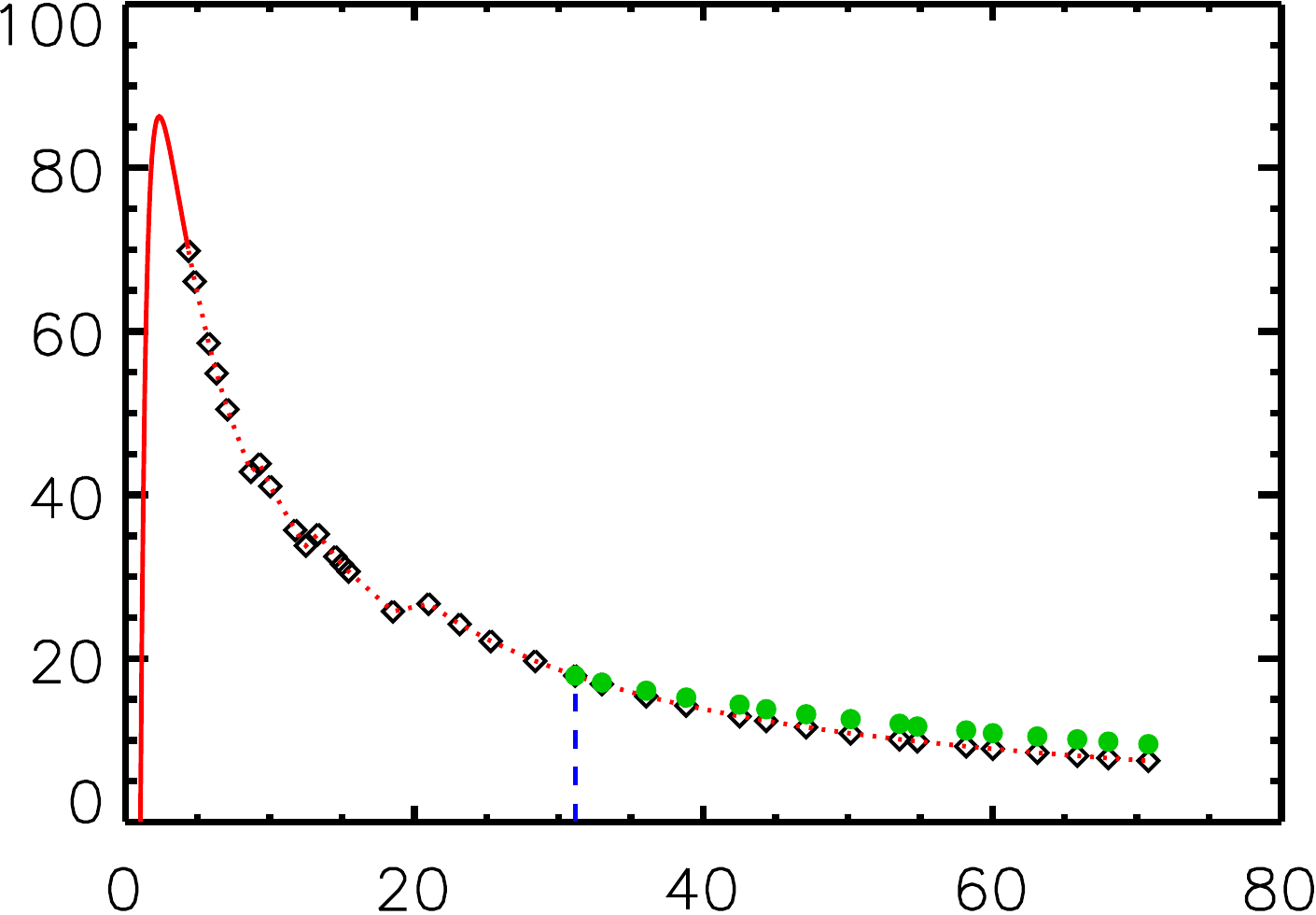}
		\put(-231,43.7){{\rotatebox{90}{{\color{black}\fontsize{12}{12}\fontseries{n}\fontfamily{phv}\selectfont  Force ($10^{17}$ dyn)}}}}
		\put(-140.8,-12.9){{\rotatebox{0}{{\color{black}\fontsize{12}{12}\fontseries{n}\fontfamily{phv}\selectfont Height (R$_{\odot}$)}}}}
		\put(-130,130.7){{\rotatebox{0}{{\color{black}\fontsize{13}{13}\fontseries{n}\fontfamily{phv}\selectfont  \underbar{CME 29}}}}}
		\put(-117,158.7){{\rotatebox{0}{{\color{black}\fontsize{13}{13}\fontseries{n}\fontfamily{phv}\selectfont (b)}}}}          
		  }
		\vspace{1.3cm}
		  \centerline{\hspace*{0.06\textwidth}
		\includegraphics[width=0.57\textwidth,clip=]{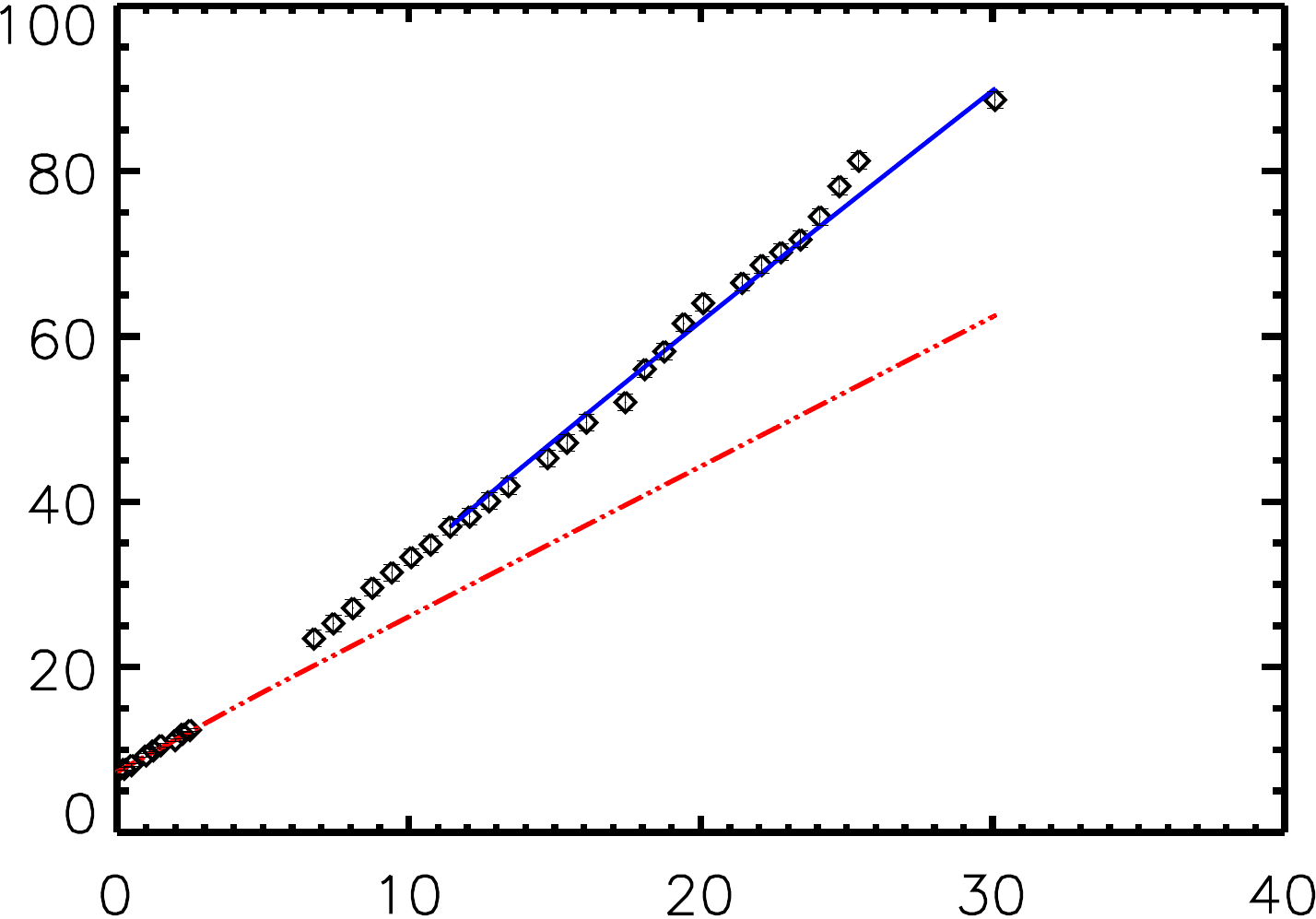}
		\put(-230,62.7){{\rotatebox{90}{{\color{black}\fontsize{12}{12}\fontseries{n}\fontfamily{phv}\selectfont  Height (R$_{\odot}$)}}}}
		\put(-160.8,-14.9){{\rotatebox{0}{{\color{black}\fontsize{12}{12}\fontseries{n}\fontfamily{phv}\selectfont  Elapsed Time (hrs)}}}}
		\put(-130,130.7){{\rotatebox{0}{{\color{black}\fontsize{13}{13}\fontseries{n}\fontfamily{phv}\selectfont \underbar{CME 30}}}}}
		\hspace*{0.069\textwidth}
		\includegraphics[width=0.565\textwidth,clip=]{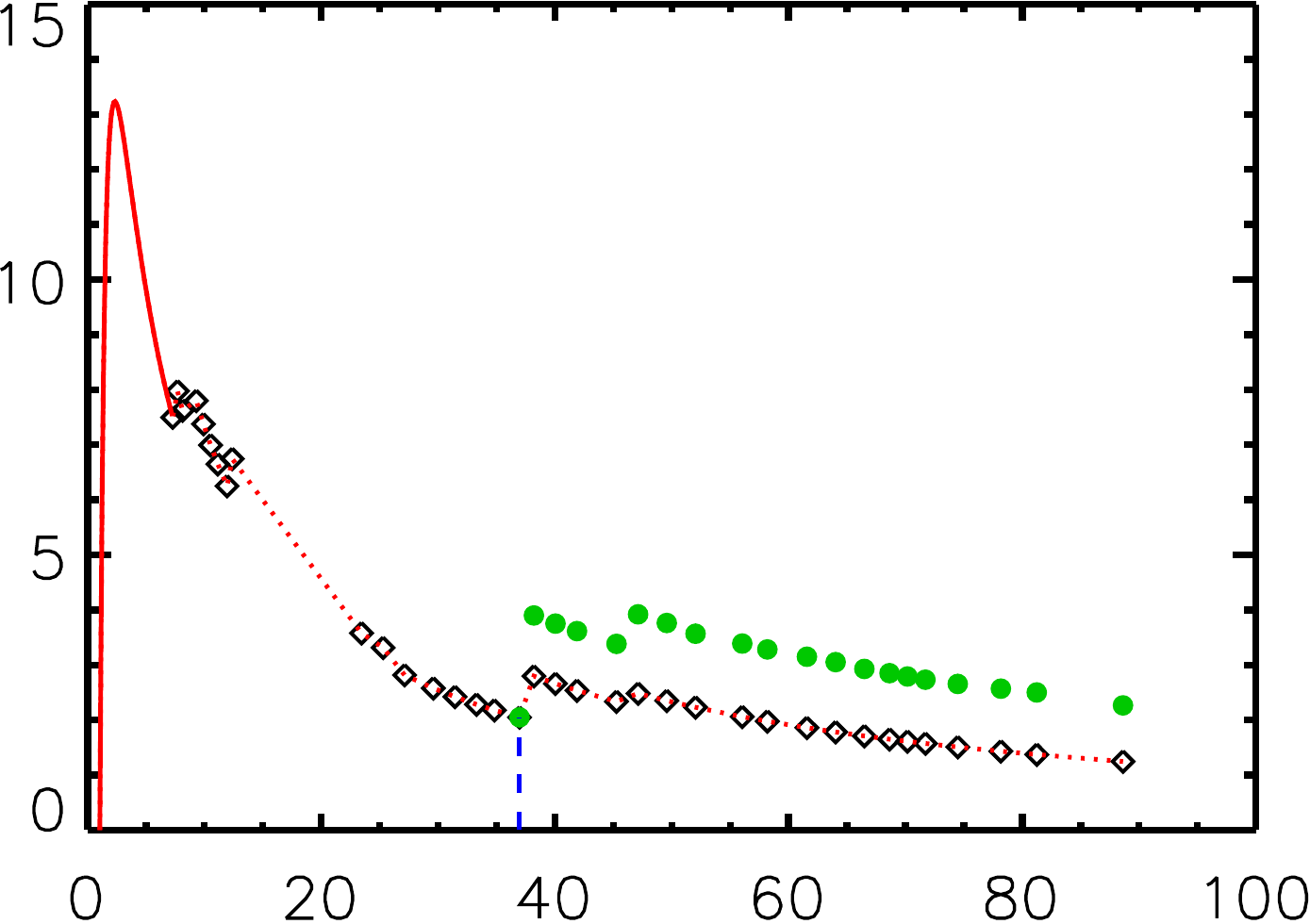}
		\put(-234,43.7){{\rotatebox{90}{{\color{black}\fontsize{12}{12}\fontseries{n}\fontfamily{phv}\selectfont  Force ($10^{17}$ dyn)}}}}
		\put(-140.8,-12.9){{\rotatebox{0}{{\color{black}\fontsize{12}{12}\fontseries{n}\fontfamily{phv}\selectfont Height (R$_{\odot}$)}}}}
		\put(-130,130.7){{\rotatebox{0}{{\color{black}\fontsize{13}{13}\fontseries{n}\fontfamily{phv}\selectfont  \underbar{CME 30}}}}}
		  }
  \vspace{0.0261\textwidth}  
  \caption[Height-time and Force profiles for CMEs 29 and 30]{Height-time and Force profiles for CMEs 29 and 30. Caption same as Figure \ref{fig52}}
  \label{fig65}
  \end{figure}
  \clearpage

  \clearpage
  \begin{figure}[h]    
    \centering                              
    \centerline{\hspace*{0.06\textwidth}
		\includegraphics[width=0.557\textwidth,clip=]{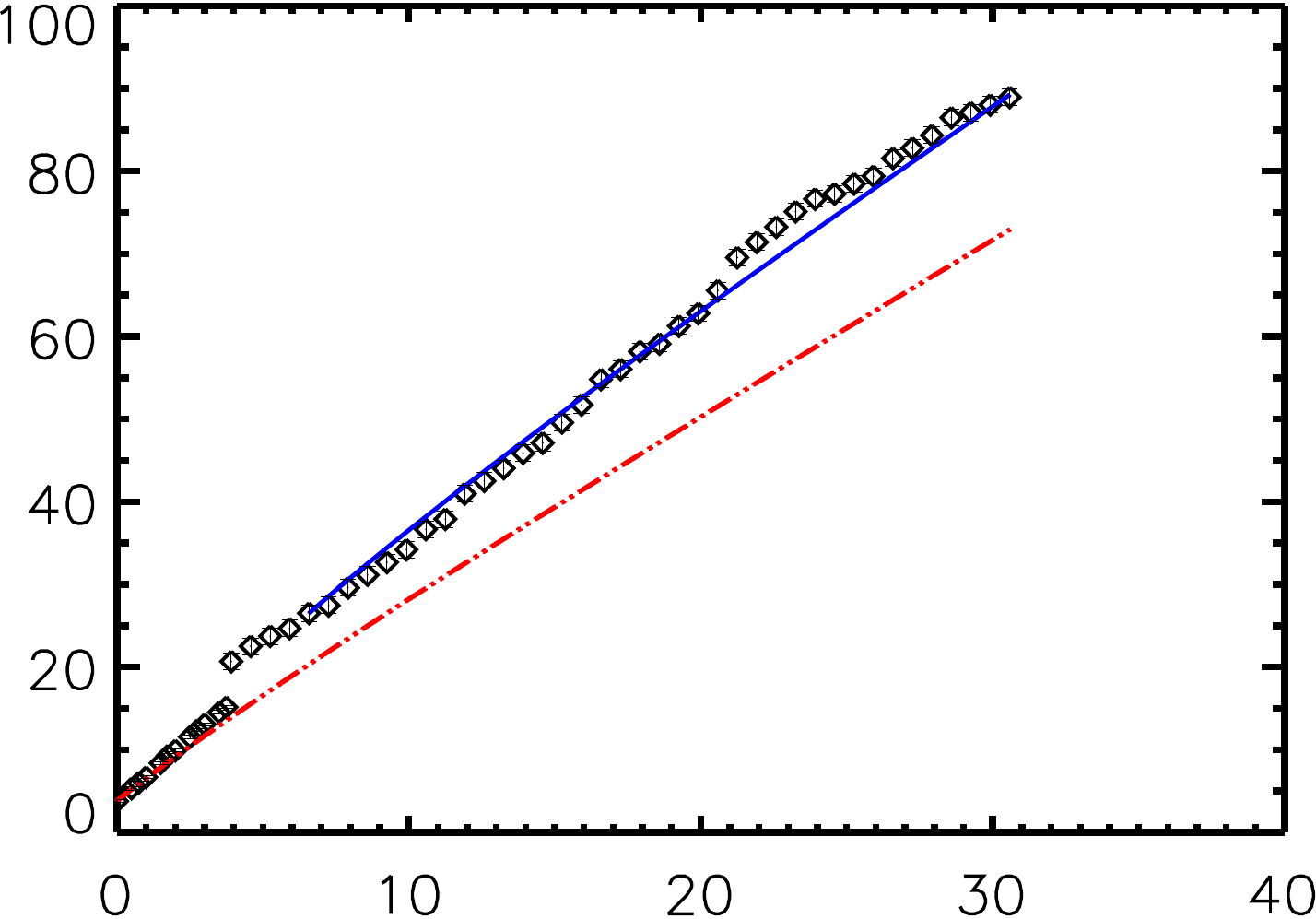}
		\put(-225,62.7){{\rotatebox{90}{{\color{black}\fontsize{12}{12}\fontseries{n}\fontfamily{phv}\selectfont  Height (R$_{\odot}$)}}}}
		\put(-160.8,-14.9){{\rotatebox{0}{{\color{black}\fontsize{12}{12}\fontseries{n}\fontfamily{phv}\selectfont  Elapsed Time (hrs)}}}}
		\put(-130,130.7){{\rotatebox{0}{{\color{black}\fontsize{13}{13}\fontseries{n}\fontfamily{phv}\selectfont \underbar{CME 31}}}}}
		\put(-115,158.7){{\rotatebox{0}{{\color{black}\fontsize{13}{13}\fontseries{n}\fontfamily{phv}\selectfont (a)}}}}
		\hspace*{0.069\textwidth}
		\includegraphics[width=0.589\textwidth,clip=]{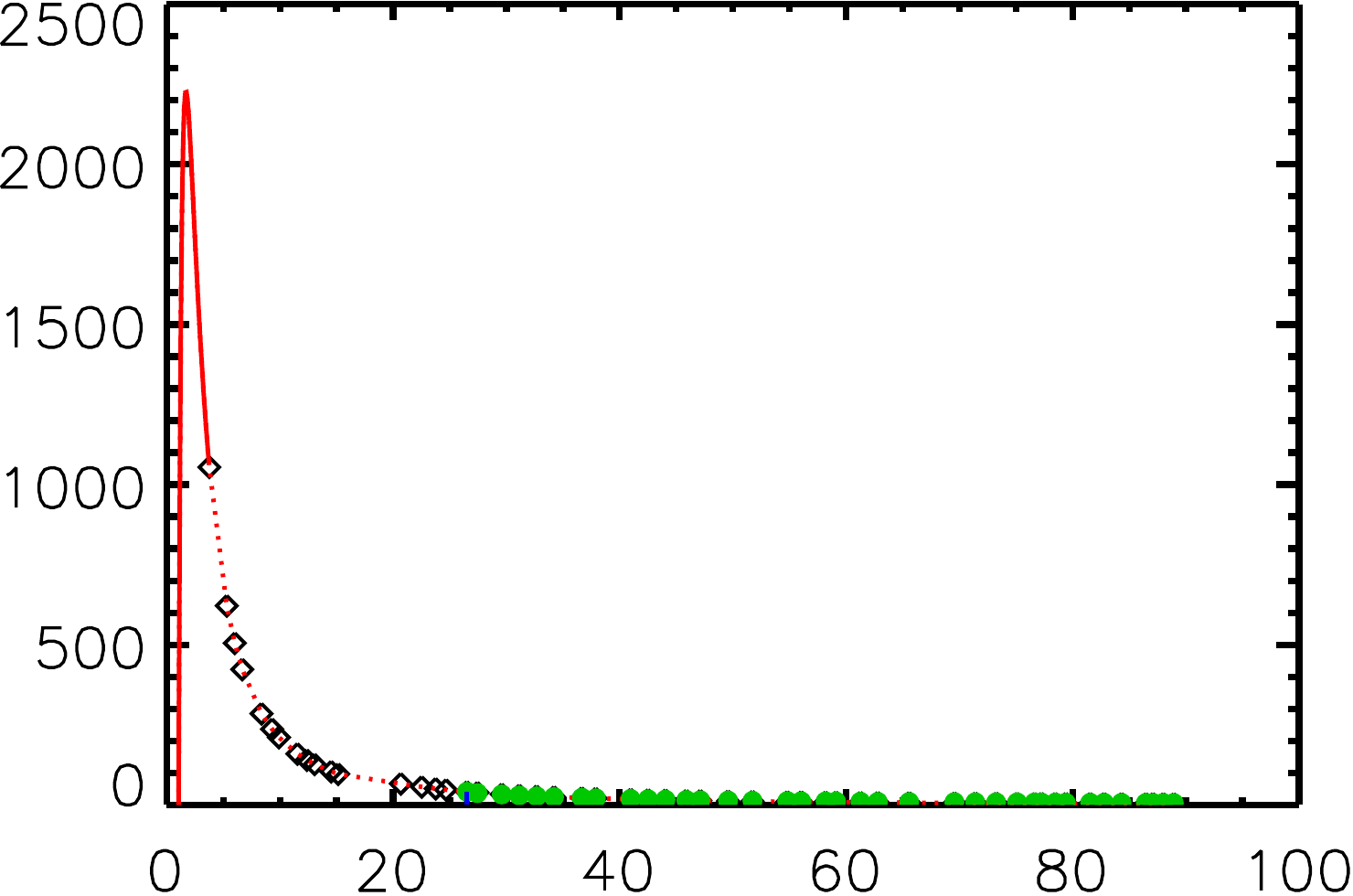}
		\put(-246,43.7){{\rotatebox{90}{{\color{black}\fontsize{12}{12}\fontseries{n}\fontfamily{phv}\selectfont  Force ($10^{17}$ dyn)}}}}
		\put(-140.8,-12.9){{\rotatebox{0}{{\color{black}\fontsize{12}{12}\fontseries{n}\fontfamily{phv}\selectfont Height (R$_{\odot}$)}}}}
		\put(-130,130.7){{\rotatebox{0}{{\color{black}\fontsize{13}{13}\fontseries{n}\fontfamily{phv}\selectfont  \underbar{CME 31}}}}}
		\put(-117,158.7){{\rotatebox{0}{{\color{black}\fontsize{13}{13}\fontseries{n}\fontfamily{phv}\selectfont (b)}}}}          
		  }
		\vspace{1.3cm}
		  \centerline{\hspace*{0.06\textwidth}
		\includegraphics[width=0.565\textwidth,clip=]{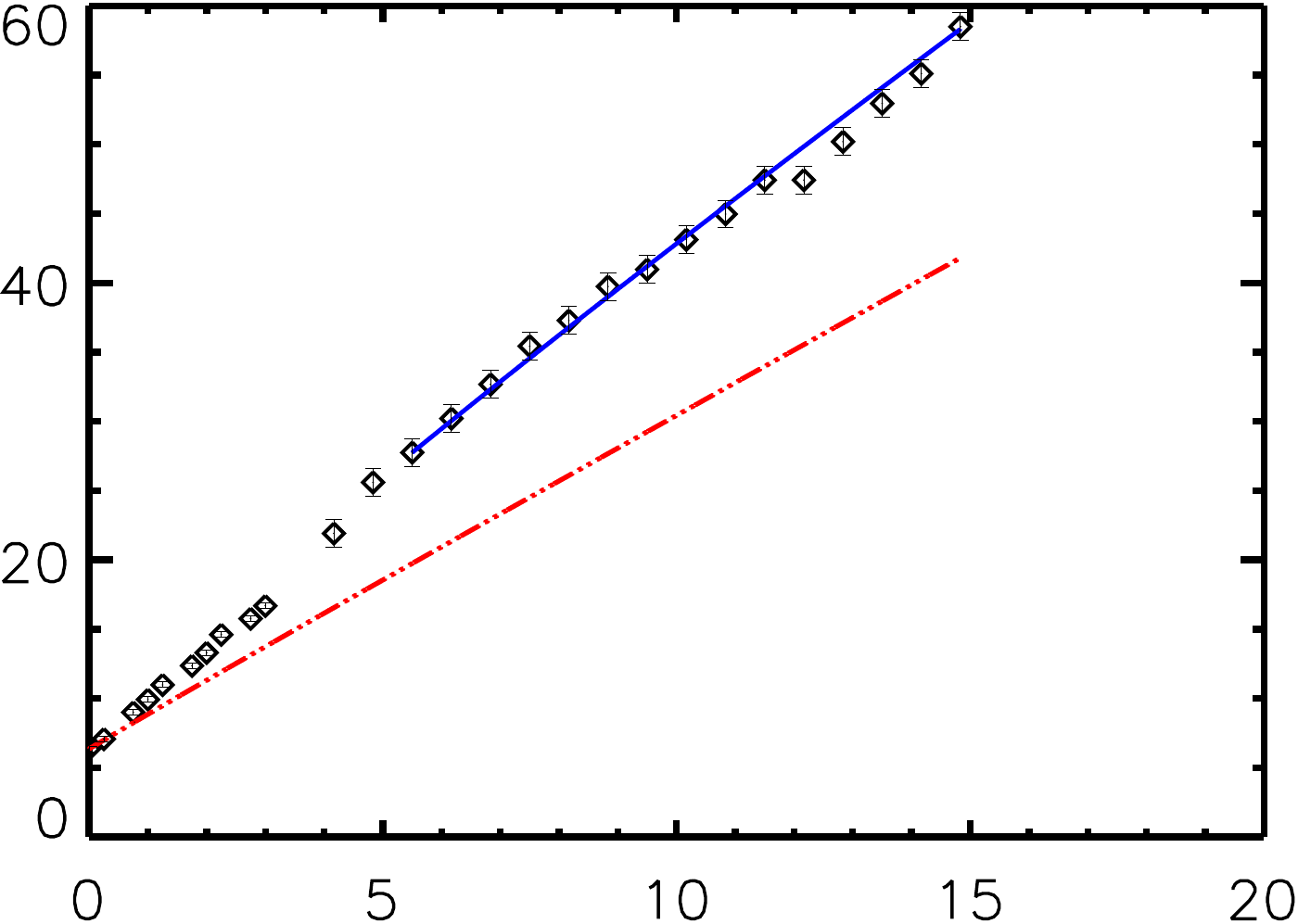}
		\put(-230,62.7){{\rotatebox{90}{{\color{black}\fontsize{12}{12}\fontseries{n}\fontfamily{phv}\selectfont  Height (R$_{\odot}$)}}}}
		\put(-160.8,-14.9){{\rotatebox{0}{{\color{black}\fontsize{12}{12}\fontseries{n}\fontfamily{phv}\selectfont  Elapsed Time (hrs)}}}}
		\put(-138	,135.7){{\rotatebox{0}{{\color{black}\fontsize{13}{13}\fontseries{n}\fontfamily{phv}\selectfont \underbar{CME 32}}}}}
		\hspace*{0.069\textwidth}
		\includegraphics[width=0.565\textwidth,clip=]{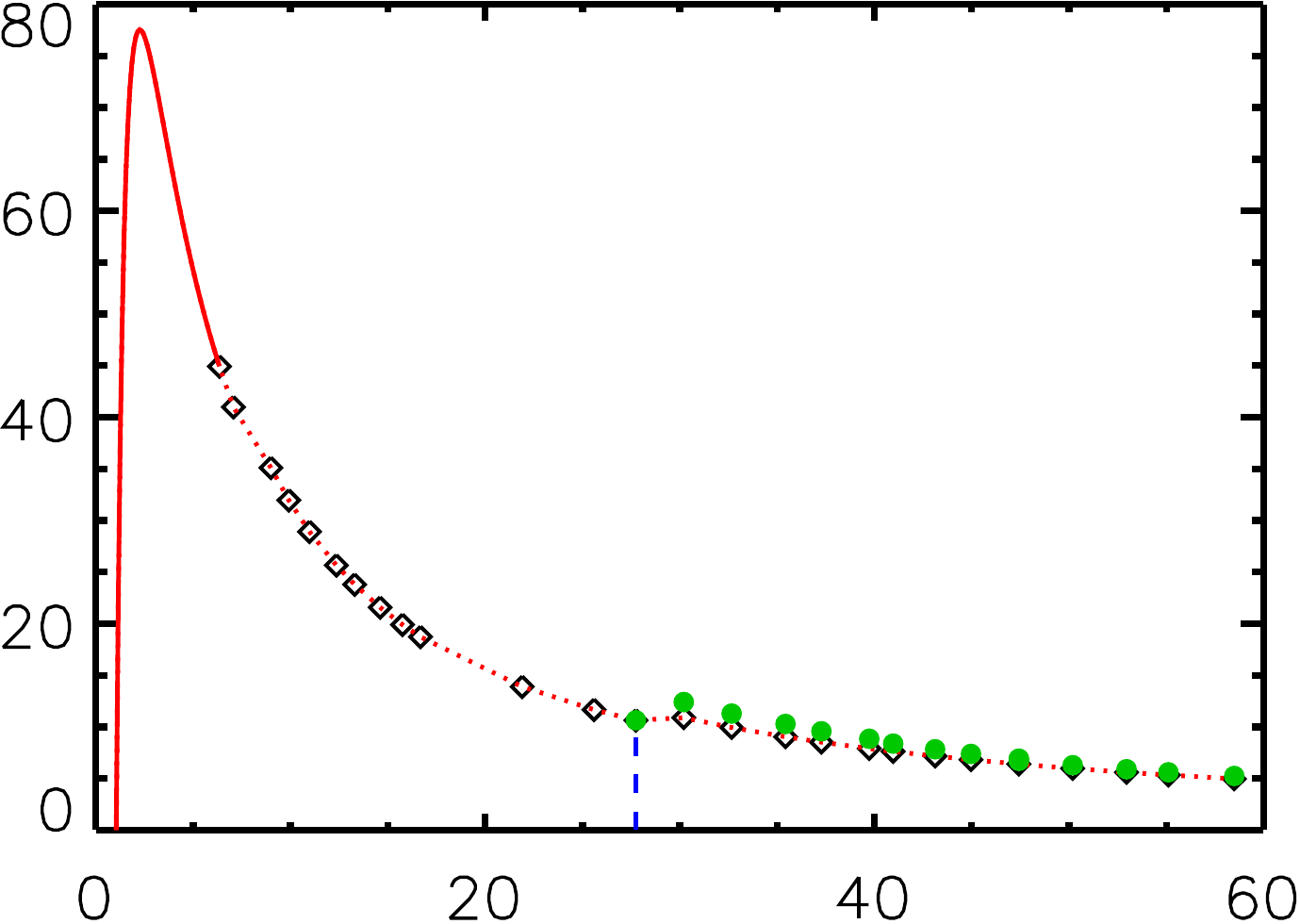}
		\put(-237,43.7){{\rotatebox{90}{{\color{black}\fontsize{12}{12}\fontseries{n}\fontfamily{phv}\selectfont  Force ($10^{17}$ dyn)}}}}
		\put(-140.8,-12.9){{\rotatebox{0}{{\color{black}\fontsize{12}{12}\fontseries{n}\fontfamily{phv}\selectfont Height (R$_{\odot}$)}}}}
		\put(-130,135.7){{\rotatebox{0}{{\color{black}\fontsize{13}{13}\fontseries{n}\fontfamily{phv}\selectfont  \underbar{CME 32}}}}}
		  }
  \vspace{0.0261\textwidth}  
  \caption[Height-time and Force profiles for CMEs 31 and 32]{Height-time and Force profiles for CMEs 31 and 32. Caption same as Figure \ref{fig52}}
  \label{fig66}
  \end{figure}

  \clearpage
  \begin{figure}[h]    
    \centering                              
    \centerline{\hspace*{0.06\textwidth}
		\includegraphics[width=0.55\textwidth,clip=]{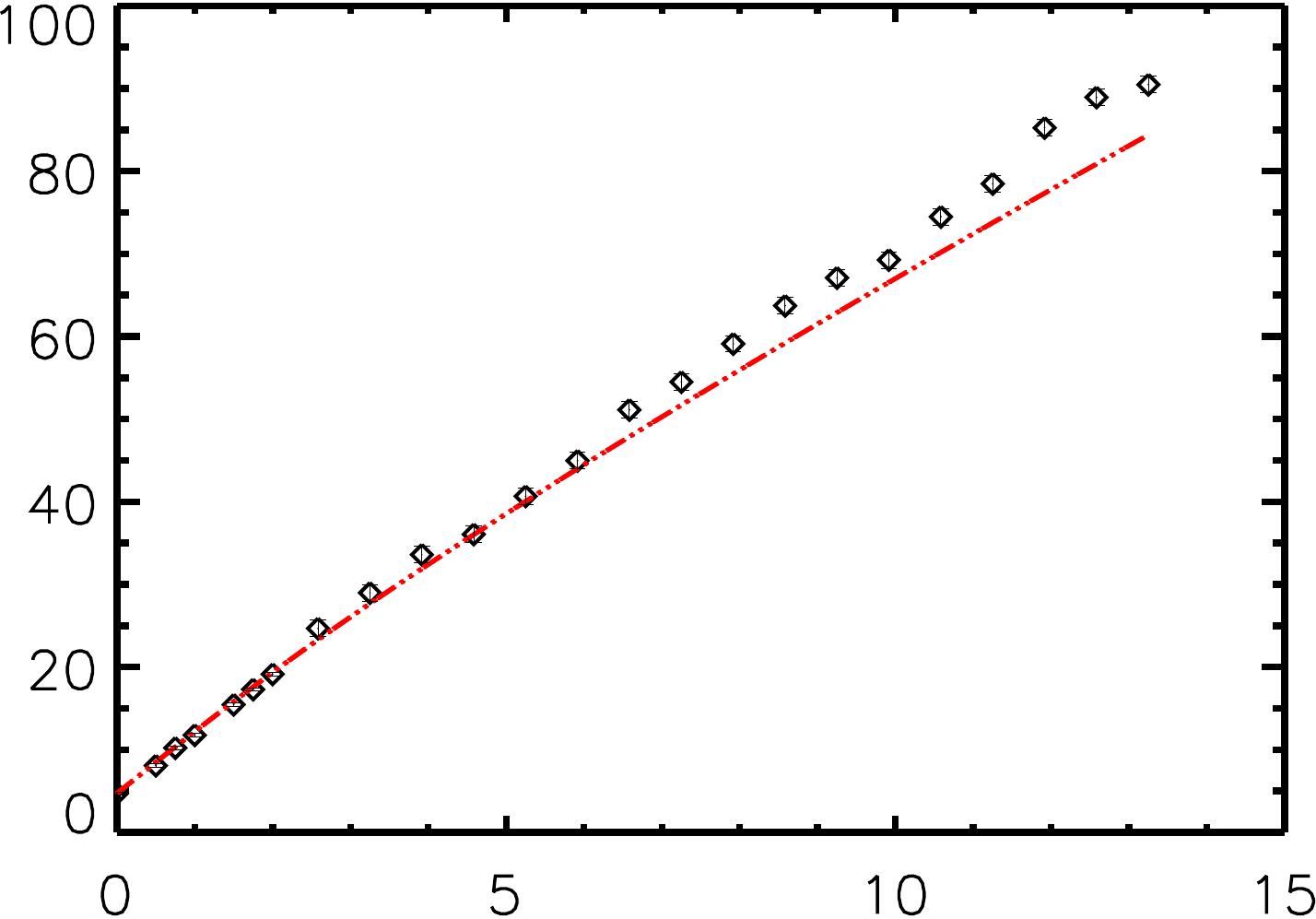}
		\put(-230,62.7){{\rotatebox{90}{{\color{black}\fontsize{12}{12}\fontseries{n}\fontfamily{phv}\selectfont  Height (R$_{\odot}$)}}}}
		\put(-160.8,-14.9){{\rotatebox{0}{{\color{black}\fontsize{12}{12}\fontseries{n}\fontfamily{phv}\selectfont  Elapsed Time (hrs)}}}}
		\put(-130,130.7){{\rotatebox{0}{{\color{black}\fontsize{13}{13}\fontseries{n}\fontfamily{phv}\selectfont \underbar{CME 33 ($f$)}}}}}
		\put(-115,158.7){{\rotatebox{0}{{\color{black}\fontsize{13}{13}\fontseries{n}\fontfamily{phv}\selectfont (a)}}}}
		\hspace*{0.069\textwidth}
		\includegraphics[width=0.565\textwidth,clip=]{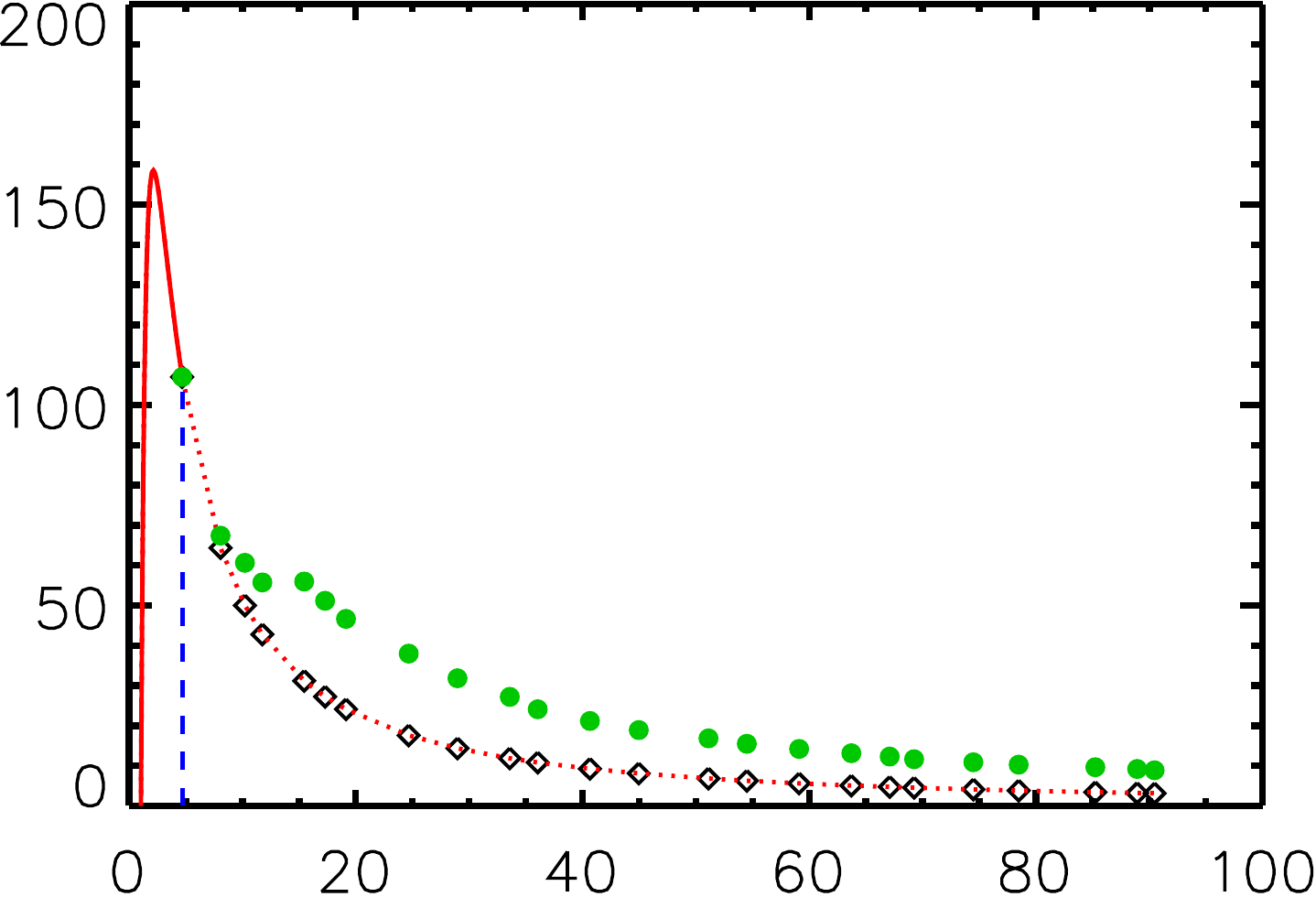}
		\put(-231,43.7){{\rotatebox{90}{{\color{black}\fontsize{12}{12}\fontseries{n}\fontfamily{phv}\selectfont  Force ($10^{17}$ dyn)}}}}
		\put(-140.8,-12.9){{\rotatebox{0}{{\color{black}\fontsize{12}{12}\fontseries{n}\fontfamily{phv}\selectfont Height (R$_{\odot}$)}}}}
		\put(-130,130.7){{\rotatebox{0}{{\color{black}\fontsize{13}{13}\fontseries{n}\fontfamily{phv}\selectfont  \underbar{CME 33 ($f$)}}}}}
		\put(-117,158.7){{\rotatebox{0}{{\color{black}\fontsize{13}{13}\fontseries{n}\fontfamily{phv}\selectfont (b)}}}}          
		  }
		\vspace{1.3cm}
		  \centerline{\hspace*{0.06\textwidth}
		\includegraphics[width=0.565\textwidth,clip=]{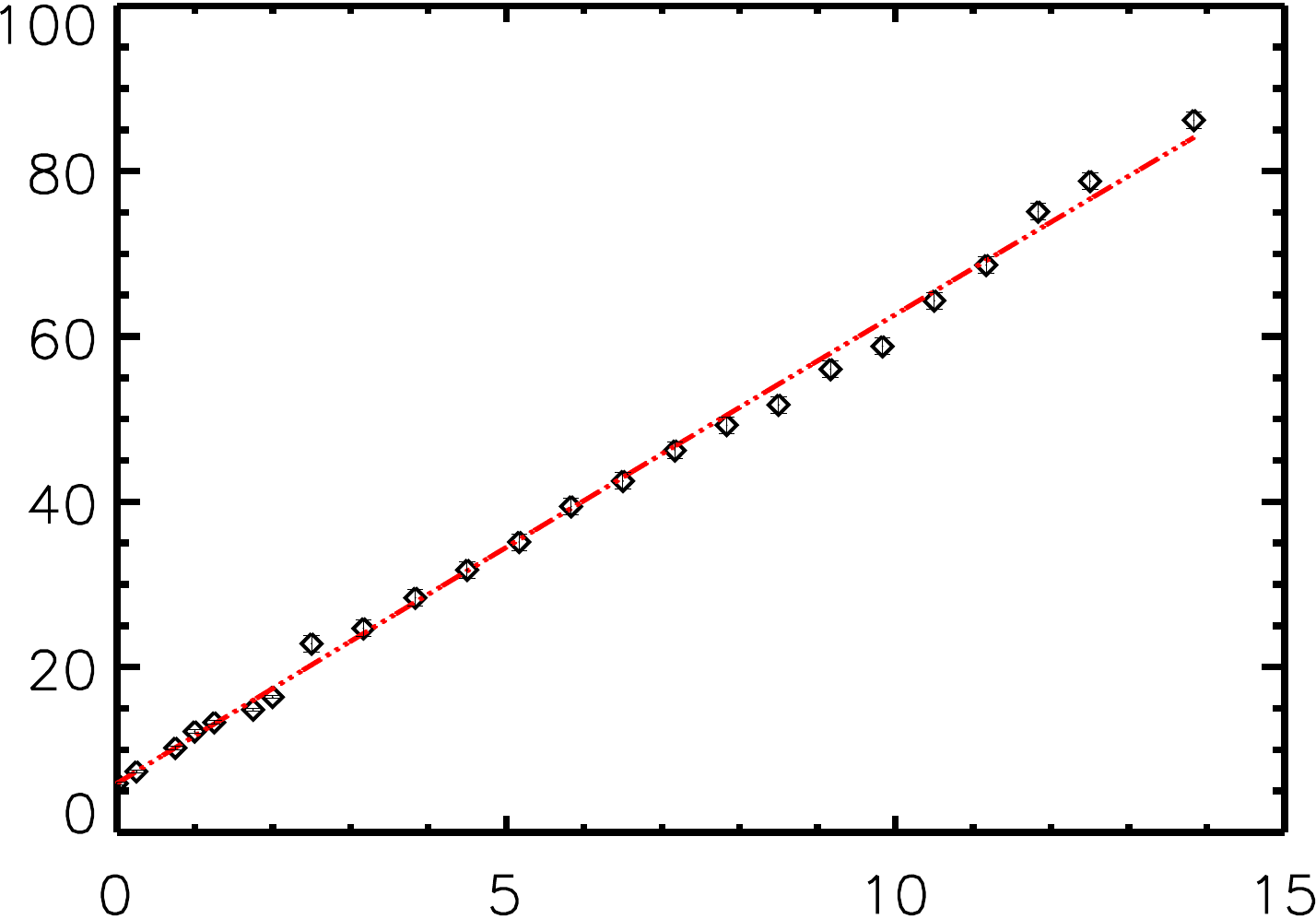}
		\put(-230,62.7){{\rotatebox{90}{{\color{black}\fontsize{12}{12}\fontseries{n}\fontfamily{phv}\selectfont  Height (R$_{\odot}$)}}}}
		\put(-160.8,-14.9){{\rotatebox{0}{{\color{black}\fontsize{12}{12}\fontseries{n}\fontfamily{phv}\selectfont  Elapsed Time (hrs)}}}}
		\put(-130,130.7){{\rotatebox{0}{{\color{black}\fontsize{13}{13}\fontseries{n}\fontfamily{phv}\selectfont \underbar{CME 34 ($f$)}}}}}
		\hspace*{0.069\textwidth}
		\includegraphics[width=0.563\textwidth,clip=]{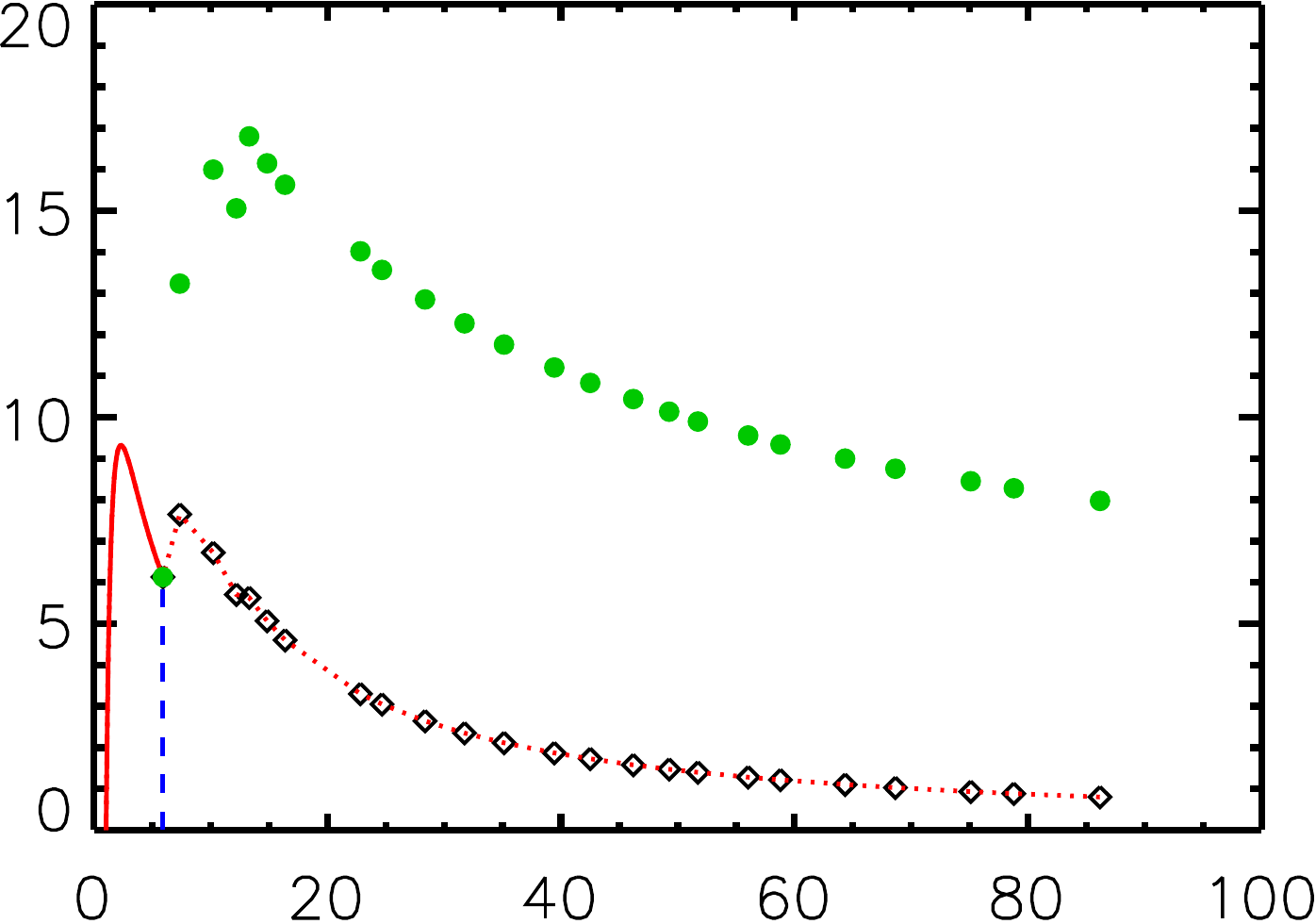}
		\put(-234,43.7){{\rotatebox{90}{{\color{black}\fontsize{12}{12}\fontseries{n}\fontfamily{phv}\selectfont  Force ($10^{17}$ dyn)}}}}
		\put(-140.8,-12.9){{\rotatebox{0}{{\color{black}\fontsize{12}{12}\fontseries{n}\fontfamily{phv}\selectfont Height (R$_{\odot}$)}}}}
		\put(-130,130.7){{\rotatebox{0}{{\color{black}\fontsize{13}{13}\fontseries{n}\fontfamily{phv}\selectfont  \underbar{CME 34 ($f$)}}}}}
		  }
  \vspace{0.0261\textwidth}  
  \caption[Height-time and Force profiles for CMEs 33 and 34]{Height-time and Force profiles for CMEs 33 and 34. Caption same as Figure \ref{fig52}}
  \label{fig67}
  \end{figure}

  \clearpage
  \begin{figure}[h]    
    \centering                              
    \centerline{\hspace*{0.06\textwidth}
		\includegraphics[width=0.55\textwidth,clip=]{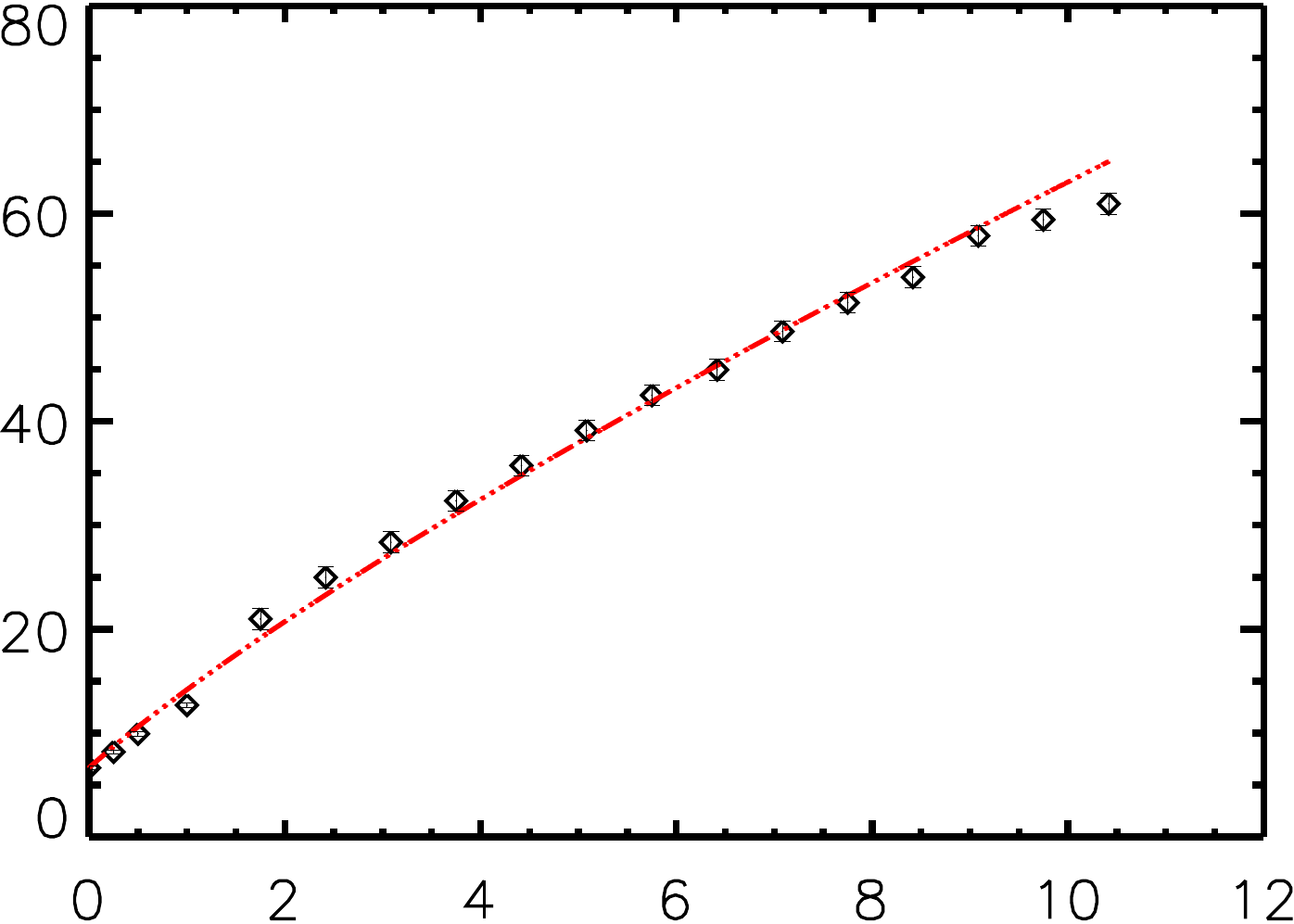}
		\put(-230,62.7){{\rotatebox{90}{{\color{black}\fontsize{12}{12}\fontseries{n}\fontfamily{phv}\selectfont  Height (R$_{\odot}$)}}}}
		\put(-160.8,-14.9){{\rotatebox{0}{{\color{black}\fontsize{12}{12}\fontseries{n}\fontfamily{phv}\selectfont  Elapsed Time (hrs)}}}}
		\put(-130,130.7){{\rotatebox{0}{{\color{black}\fontsize{13}{13}\fontseries{n}\fontfamily{phv}\selectfont \underbar{CME 35 ($f$)}}}}}
		\put(-115,158.7){{\rotatebox{0}{{\color{black}\fontsize{13}{13}\fontseries{n}\fontfamily{phv}\selectfont (a)}}}}
		\hspace*{0.069\textwidth}
		\includegraphics[width=0.584\textwidth,clip=]{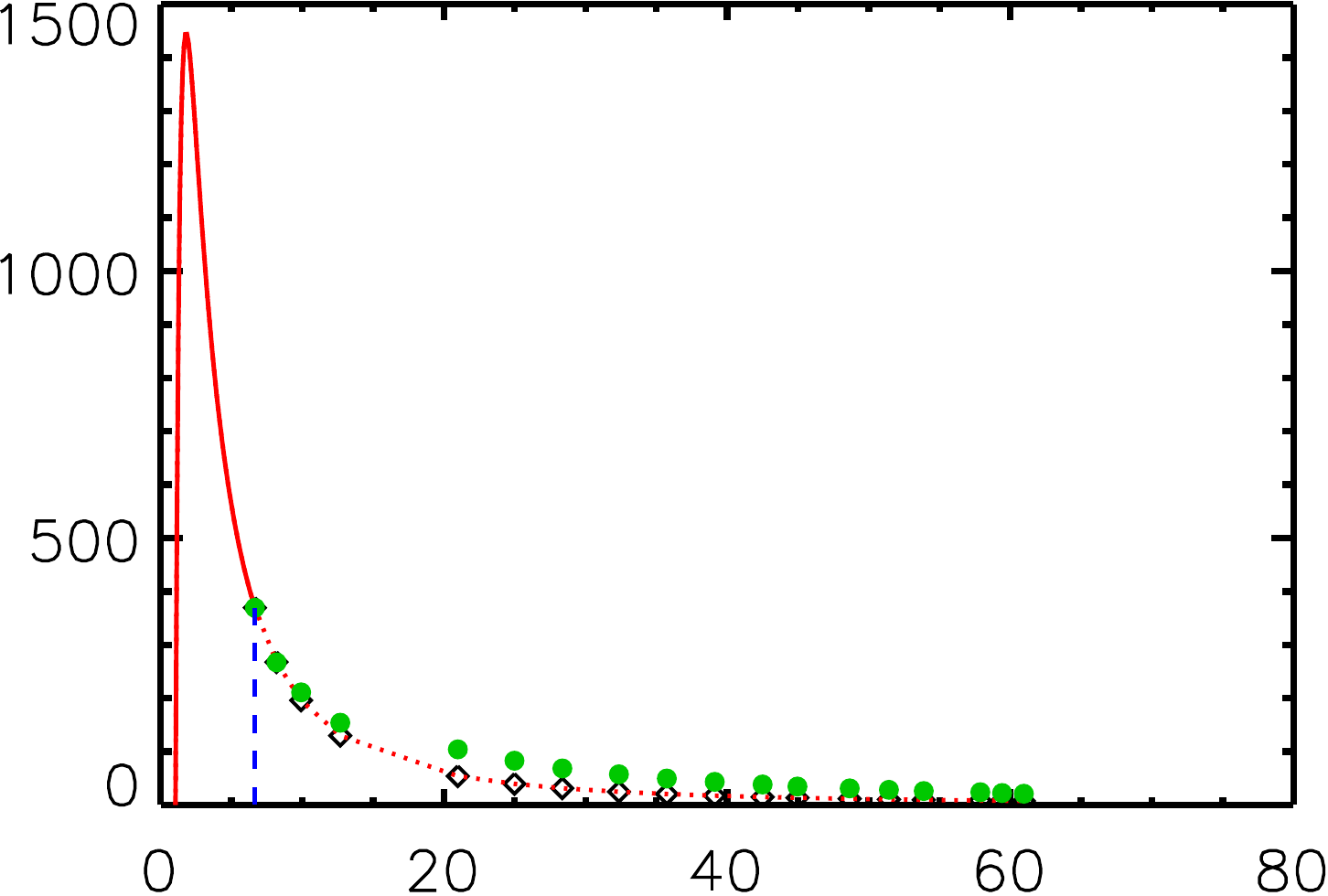}
		\put(-241,43.7){{\rotatebox{90}{{\color{black}\fontsize{12}{12}\fontseries{n}\fontfamily{phv}\selectfont  Force ($10^{17}$ dyn)}}}}
		\put(-140.8,-12.9){{\rotatebox{0}{{\color{black}\fontsize{12}{12}\fontseries{n}\fontfamily{phv}\selectfont Height (R$_{\odot}$)}}}}
		\put(-130,130.7){{\rotatebox{0}{{\color{black}\fontsize{13}{13}\fontseries{n}\fontfamily{phv}\selectfont  \underbar{CME 35 ($f$)}}}}}
		\put(-117,158.7){{\rotatebox{0}{{\color{black}\fontsize{13}{13}\fontseries{n}\fontfamily{phv}\selectfont (b)}}}}          
		  }
		\vspace{1.3cm}
		  \centerline{\hspace*{0.06\textwidth}
		\includegraphics[width=0.56\textwidth,clip=]{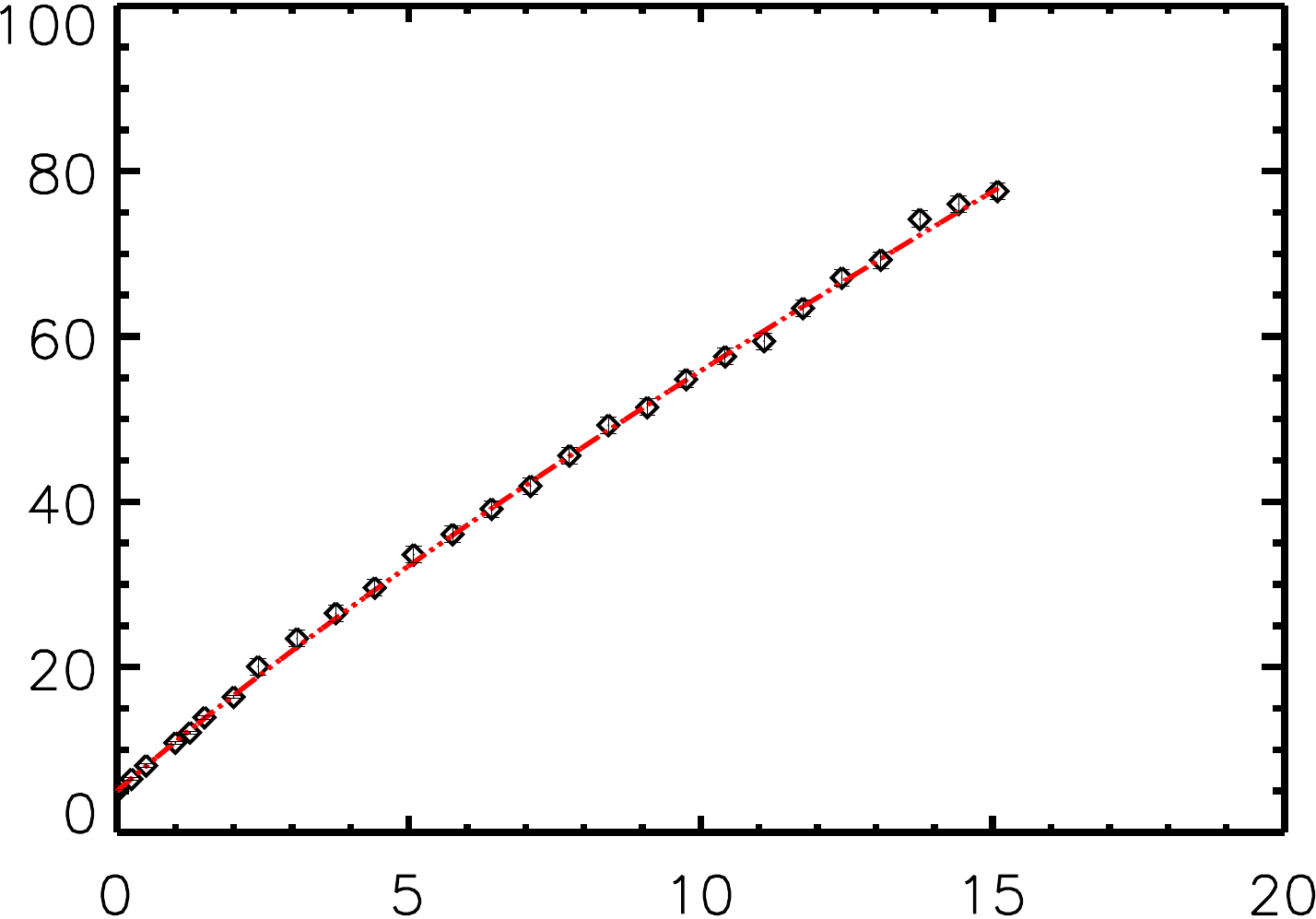}
		\put(-230,62.7){{\rotatebox{90}{{\color{black}\fontsize{12}{12}\fontseries{n}\fontfamily{phv}\selectfont  Height (R$_{\odot}$)}}}}
		\put(-160.8,-14.9){{\rotatebox{0}{{\color{black}\fontsize{12}{12}\fontseries{n}\fontfamily{phv}\selectfont  Elapsed Time (hrs)}}}}
		\put(-138,130.7){{\rotatebox{0}{{\color{black}\fontsize{13}{13}\fontseries{n}\fontfamily{phv}\selectfont \underbar{CME 36 ($f$)}}}}}
		\hspace*{0.069\textwidth}
		\includegraphics[width=0.565\textwidth,clip=]{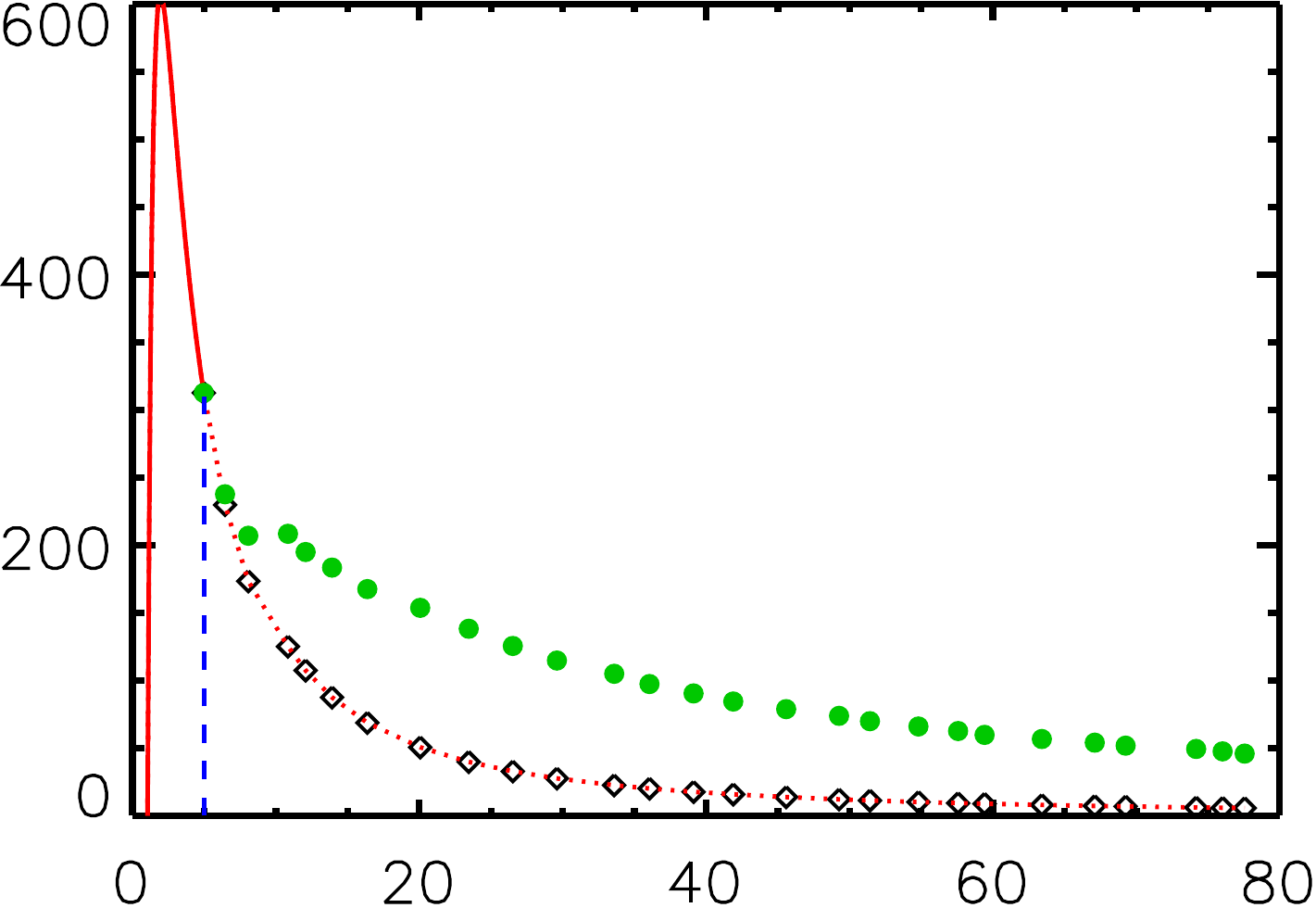}
		\put(-231,43.7){{\rotatebox{90}{{\color{black}\fontsize{12}{12}\fontseries{n}\fontfamily{phv}\selectfont  Force ($10^{17}$ dyn)}}}}
		\put(-140.8,-12.9){{\rotatebox{0}{{\color{black}\fontsize{12}{12}\fontseries{n}\fontfamily{phv}\selectfont Height (R$_{\odot}$)}}}}
		\put(-130,130.7){{\rotatebox{0}{{\color{black}\fontsize{13}{13}\fontseries{n}\fontfamily{phv}\selectfont  \underbar{CME 36 ($f$)}}}}}
		  }
  \vspace{0.0261\textwidth}  
  \caption[Height-time and Force profiles for CMEs 35 and 36]{Height-time and Force profiles for CMEs 35 and 36. Caption same as Figure \ref{fig52}}
  \label{fig68}
  \end{figure}

  \clearpage
  \begin{figure}[h]    
    \centering                              
    \centerline{\hspace*{0.06\textwidth}
		\includegraphics[width=0.56\textwidth,clip=]{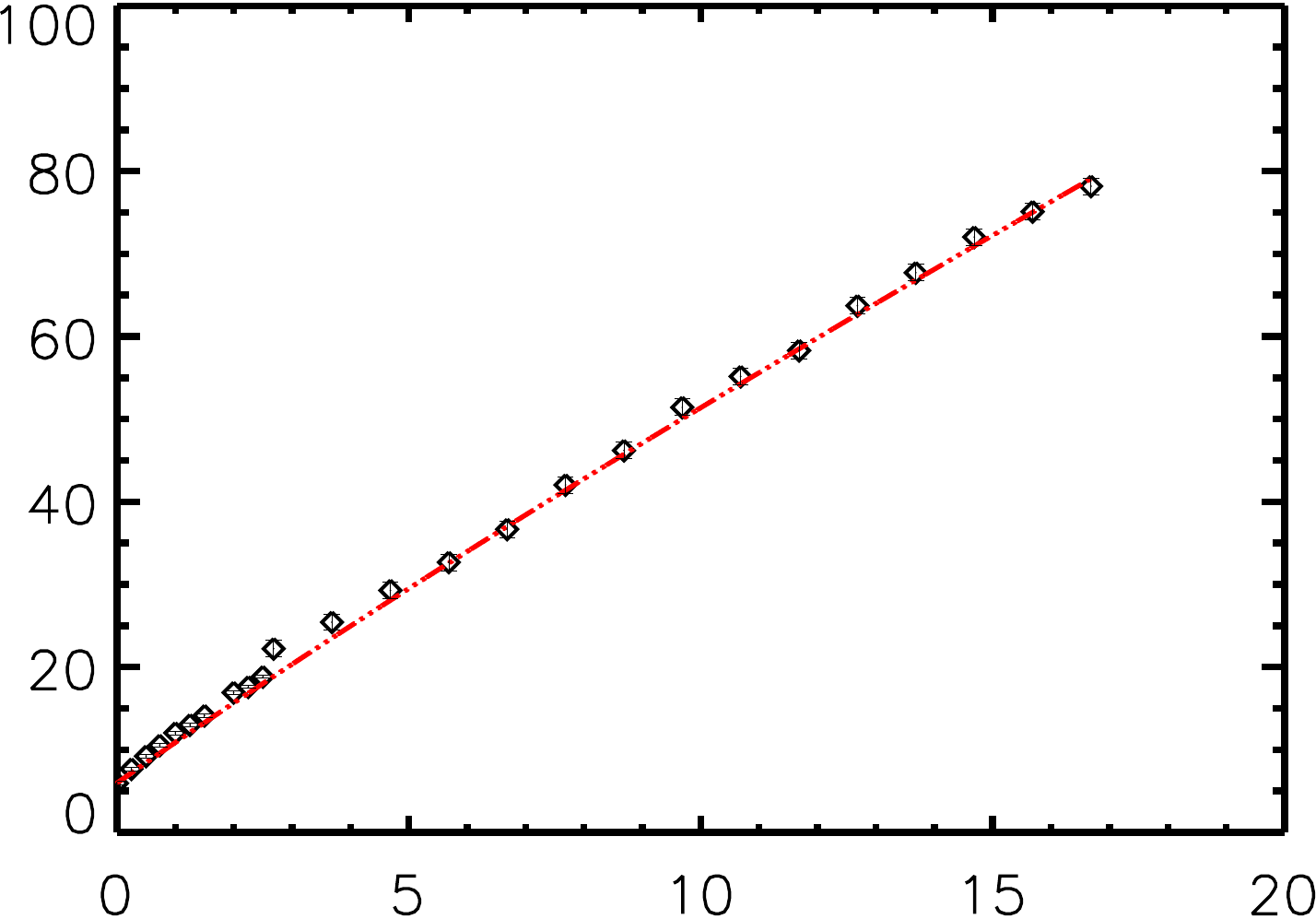}
		\put(-230,62.7){{\rotatebox{90}{{\color{black}\fontsize{12}{12}\fontseries{n}\fontfamily{phv}\selectfont  Height (R$_{\odot}$)}}}}
		\put(-160.8,-14.9){{\rotatebox{0}{{\color{black}\fontsize{12}{12}\fontseries{n}\fontfamily{phv}\selectfont  Elapsed Time (hrs)}}}}
		\put(-130,130.7){{\rotatebox{0}{{\color{black}\fontsize{13}{13}\fontseries{n}\fontfamily{phv}\selectfont \underbar{CME 37 ($f$)}}}}}
		\put(-115,158.7){{\rotatebox{0}{{\color{black}\fontsize{13}{13}\fontseries{n}\fontfamily{phv}\selectfont (a)}}}}
		\hspace*{0.069\textwidth}
		\includegraphics[width=0.55\textwidth,clip=]{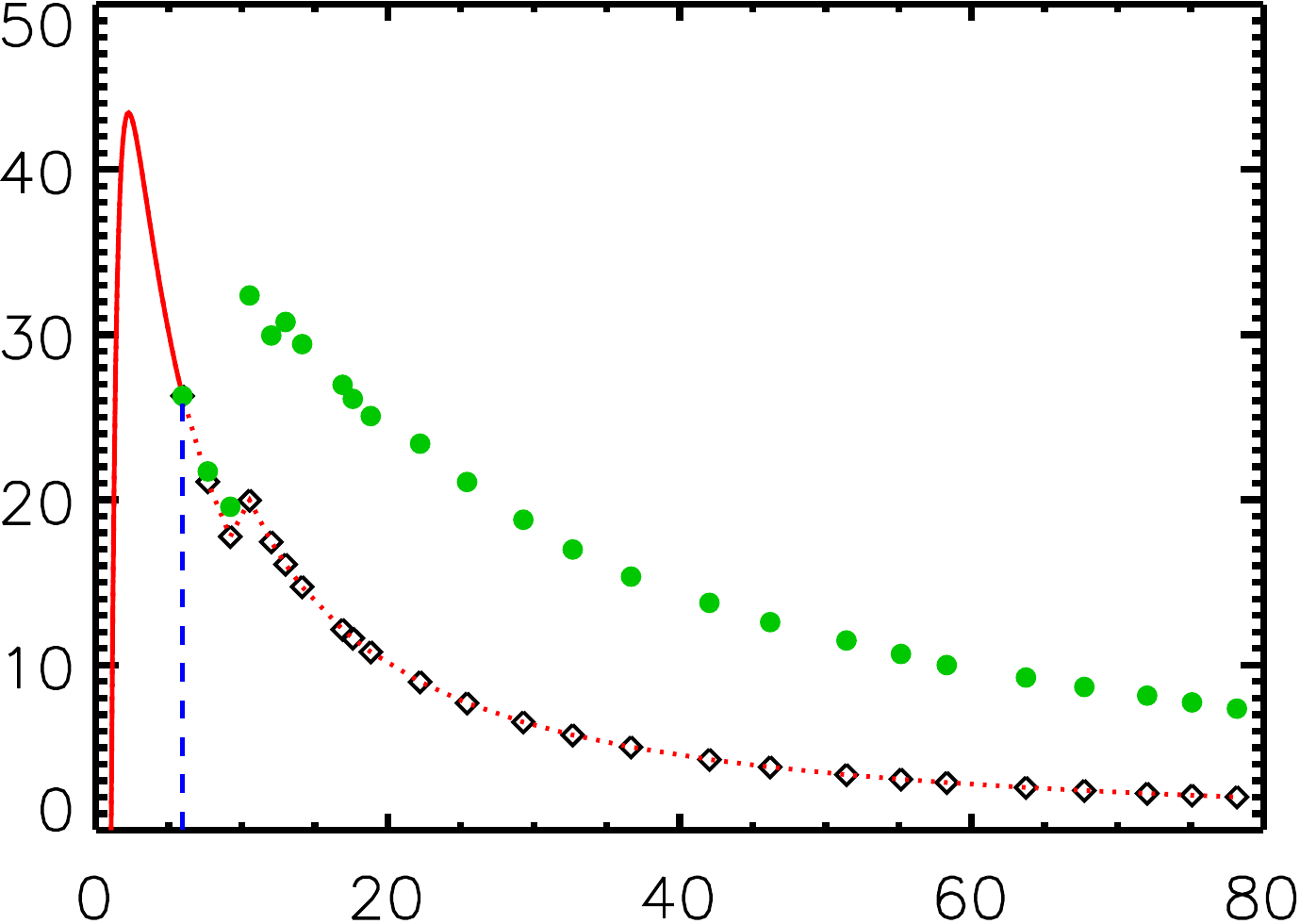}
		\put(-231,43.7){{\rotatebox{90}{{\color{black}\fontsize{12}{12}\fontseries{n}\fontfamily{phv}\selectfont  Force ($10^{17}$ dyn)}}}}
		\put(-140.8,-12.9){{\rotatebox{0}{{\color{black}\fontsize{12}{12}\fontseries{n}\fontfamily{phv}\selectfont Height (R$_{\odot}$)}}}}
		\put(-130,130.7){{\rotatebox{0}{{\color{black}\fontsize{13}{13}\fontseries{n}\fontfamily{phv}\selectfont  \underbar{CME 37 ($f$)}}}}}
		\put(-117,158.7){{\rotatebox{0}{{\color{black}\fontsize{13}{13}\fontseries{n}\fontfamily{phv}\selectfont (b)}}}}          
		  }
		\vspace{1.3cm}
		  \centerline{\hspace*{0.06\textwidth}
		\includegraphics[width=0.545\textwidth,clip=]{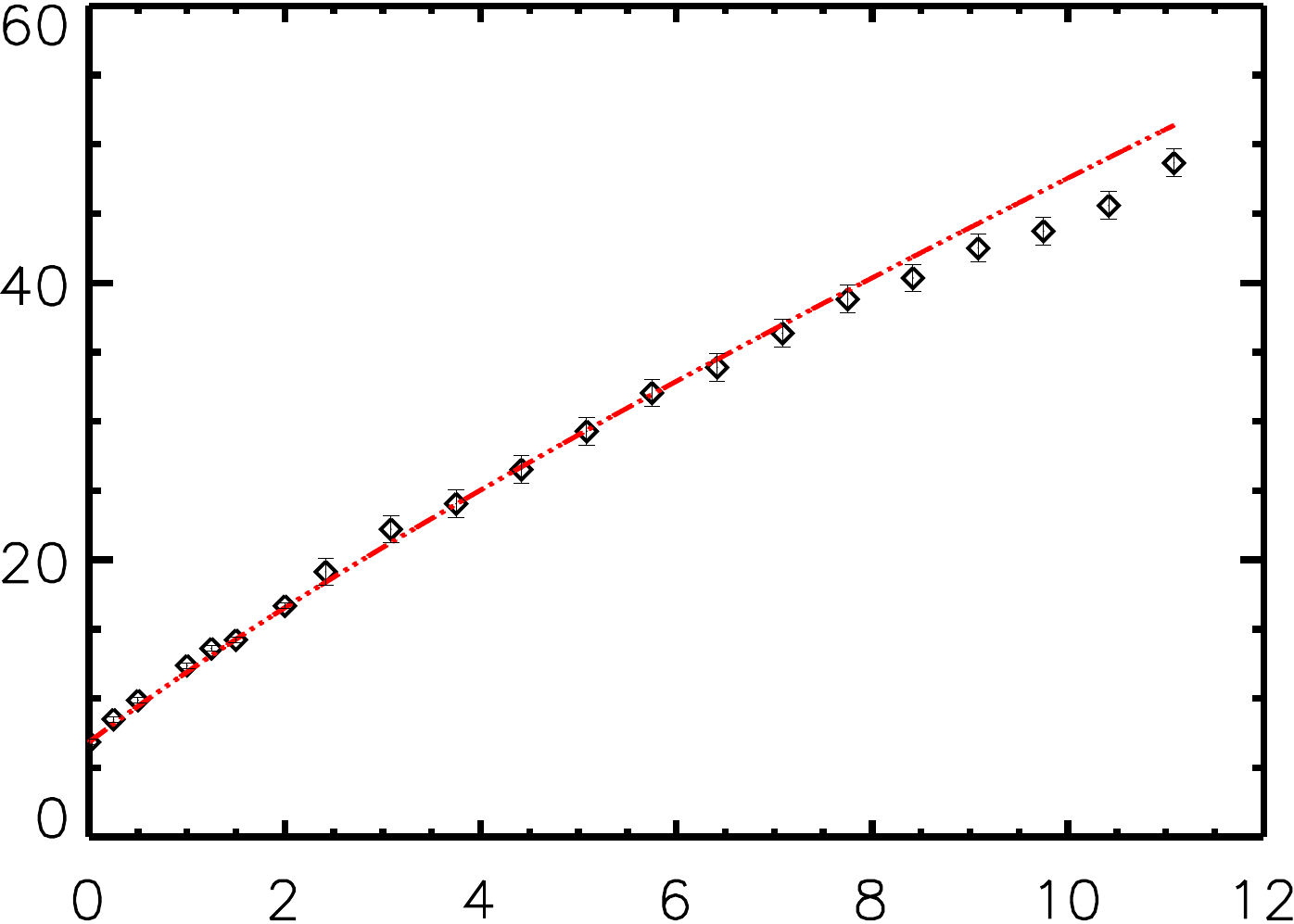}
		\put(-230,62.7){{\rotatebox{90}{{\color{black}\fontsize{12}{12}\fontseries{n}\fontfamily{phv}\selectfont  Height (R$_{\odot}$)}}}}
		\put(-160.8,-14.9){{\rotatebox{0}{{\color{black}\fontsize{12}{12}\fontseries{n}\fontfamily{phv}\selectfont  Elapsed Time (hrs)}}}}
		\put(-130,130.7){{\rotatebox{0}{{\color{black}\fontsize{13}{13}\fontseries{n}\fontfamily{phv}\selectfont \underbar{CME 38 ($f$)}}}}}
		\hspace*{0.069\textwidth}
		\includegraphics[width=0.565\textwidth,clip=]{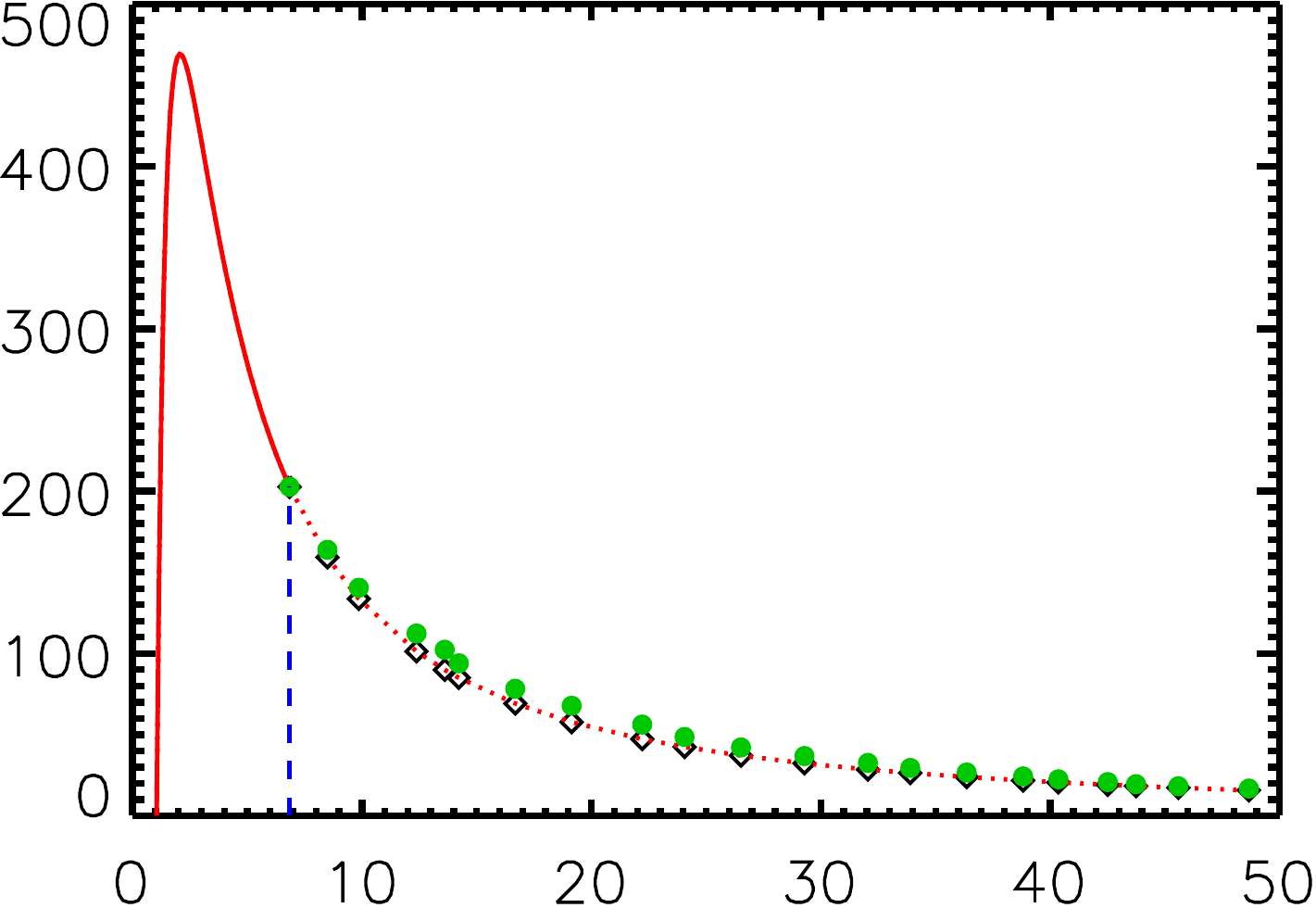}
		\put(-234,43.7){{\rotatebox{90}{{\color{black}\fontsize{12}{12}\fontseries{n}\fontfamily{phv}\selectfont  Force ($10^{17}$ dyn)}}}}
		\put(-140.8,-12.9){{\rotatebox{0}{{\color{black}\fontsize{12}{12}\fontseries{n}\fontfamily{phv}\selectfont Height (R$_{\odot}$)}}}}
		\put(-130,130.7){{\rotatebox{0}{{\color{black}\fontsize{13}{13}\fontseries{n}\fontfamily{phv}\selectfont  \underbar{CME 38 ($f$)}}}}}
		  }
  \vspace{0.0261\textwidth}  
  \caption[Height-time and Force profiles for CMEs 37 and 38]{Height-time and Force profiles for CMEs 37 and 38. Caption same as Figure \ref{fig52}}
  \label{fig69}
  \end{figure}
 \chapter{Future Work}
\label{chap7}

\noindent\makebox[\linewidth]{\rule{\textwidth}{3pt}}
{\it In this chapter, we outline a plan for further analysis of the CME data sample described in this work. 
The first step is to expand the CME data set for improved statistical analysis to derive generalized conclusions for CMEs of all speed ranges. 
We can investigate the dissipation of magnetic energy and plasma heating using the expansion and CME evolution data. 
We can also explore the possibility of finding proxies, (\textit{e.g.} Microwave radio emission, X-ray profiles) for the initiation phase of 
CMEs.}\\
\noindent\makebox[\linewidth]{\rule{\textwidth}{3pt}}
\section{Expansion of CME sample}
In the force analysis of a set of well-studied CMEs described in this thesis, we find an abrupt change in the value of the 
drag-model ``initiation'' height (\htil) at 900 \kms. This speed is also used for differentiating between fast ($v_{0}>900$ \kms) and slow ($v_{0}<900$ \kms) CMEs. 
Figure \ref{fig61b}, which is a plot of the drag initiation height (\htil) {\it versus} CME initial velocity ($v_{0}$), shows this abrupt change 
in the value of $\widetilde{h}_{0}$ at 900 \kms. This is most likely a data selection artefact.
In our sample of 38 CMEs, we have 9 events with initial velocity, $v_{0}<400$ \kms, 14 CMEs with $v_{0}<500$ \kms and 6 CMEs with $500<v_{0}<900$ \kms. 
We thus have few medium speed CMEs ($500<v_{0}<900$ \kms) - these are the ones that typically have \htil between $\sim$ 4 and 20 \Rs.
A larger sample having more medium speed CMEs ($500<v_{0}<900$ \kms) will likely resolve this issue. 
Another factor that needs to be taken into consideration is the limited time cadence of the COR2 coronagraphs. This restriction leads to fewer data points in the 
initial stages of CME propagation, especially for faster CMEs. We find that \htil  for the fast CMEs is the first observed height, and the missing data points 
could in fact lead to even lower values of \htil for some of these events. Data from future coronagraphs with a higher time cadence (\textit{e.g.} the {\it Visible Emission Line Coronagraph}
(VELC) aboard ADITYA L1; \citet{Adi17}) could be useful in estimating \htil more accurately for fast CMEs.
In summary, we plan to expand the CME sample to include CMEs that have initial velocities spanning a wider range with 
more events that have medium range initial speeds. This will be the key to generalizing our results.
\vspace{-0.1cm}
\section{Proxies for the initial acceleration phase of CMEs}
CMEs in white-light can be well observed via coronagraphs only above 2-3 \Rs; this is typically beyond the heights at which the Lorentz force
peaks. This restriction leads to an uncertainty in calculating the CME launch speed.
Due to unavailability of the CME height-time and expansion data in the low corona, we assume a self-similar expansion below the first observed height 
for all the events in our sample. An alternate approach could consist of estimating the CME initial speeds using X-ray or radio emission \citep{Zha06}.
\citet{Che96} use the soft X-ray profile as a proxy for CME current and \citet{Mat17} investigate the feasibility of using the fluence of 
nonthermal microwave bursts to estimate the CME initial speeds. However, these studies need to be done with a larger CME sample. 

We plan to investigate proxies in different wavelengths (particularly low-frequency radio lightcurves) for the CME initiation profile. 
CMEs with observations of the microwave lightcurve or X-ray profiles can be compared to the model derived Lorentz force profiles. 
A good match might indicate these lightcurves to be appropriate proxies for the initial CME evolution profile and can be used to determine CME 
parameters (like initial speeds) very early on. These proxies can provide important observational constraints on 
the Lorentz forces acting on CMEs. This is specially important for slow to medium speed CMEs, for which it has been 
shown that drag forces dominate only beyond 12--50 \Rs.

The results from drag and Lorentz force analysis will be helpful in developing better physics-based models for the CME arrival predictions. 
However, this has been done so far only for past events which have been carefully studied. The purpose of this work 
is to ultimately promote models that forecast CME arrival based on the initial observations and estimates at the first detection of the CMEs in the LASCO FOV or from X-ray/radio
signatures.

\section{Plasma Heating}
As described in Chapter \ref{chap5} (section \ref{secdiff}), we find that for some CMEs, the drag force 
is only slightly larger than the Lorentz force. In fact for two CMEs (CME 4 and CME10) in our sample, the drag force is slightly smaller than the Lorentz force. We suggest that one 
reason for this is that the Lorentz force is overestimated in our analysis. This could be due 
to the inherent assumption that the total magnetic flux is conserved, which might not be true. Therefore, we need to take into account any flux dissipation that 
may be present. We might also need to consider the energy expended, in CME expansion and in heating of the CME plasma \citep[\textit{e.g.}][]{Kum96,Ems12}.
CME event analysis of \citet{Wan09}, suggests that heat is continuously injected into the CME plasma.
With the availability of GCS measurements for the expansion of the CME along with translation, it will be interesting to see how the model predictions 
for describing the CME internal state match these observations. We would also like to explore if something can be concluded about the microphysics of the 
turbulent dissipation.


\section{1D $\rightarrow$ 3D force Equation}
We have discussed in detail a one-dimensional (1D) solar wind drag equation which entails the momentum coupling between CME and the 
ambient solar wind. Factors like front-flattening, rotation etc. of the CME are not taken into account. While this 
simple prescription provides reasonably accurate results when compared to observations, it is important to also account for the 
three-dimensional (3D) effects of propagation on the CMEs and how they alter the predicted trajectory. It is thus essential to 
also understand and modify the 1D force equation to a 3D equation \citep{Ise07}.

\section{Solving a full force equation}
The action of the two major forces governing CME propagation have been solved for, independently in this study as described in Chapter 
\ref{chap4}. Beginning with a drag-only model we determine the heights beyond which solar wind drag force dominates the CME dynamics. We then 
study the nature of Lorentz forces below and above this heliocentric distance. The next logical step is to combine observationally derived models 
for each of the two forces and solve the full-force equation. In other words, Equation \ref{eqforce} needs to be solved in its entirety, with each 
term described by physics-based prescriptions and constrained by 
observations. This endeavor will require good approximations for the CME current or external magnetic field in order to calculate 
the Lorentz forces \citep{Ise07, Sav15}. Various methods to estimate these include the calculation of CME flux content from 
flare ribbon brightenings \citep{Lon07} or via poloidal flux injection method to estimate the field strength \citet{Kun10}.
 
 \chapter{Appendix}
\label{App}

\noindent\makebox[\linewidth]{\rule{\textwidth}{3pt}} 
{\textit {Detailed calculations for CME virtual mass and Lorentz force model as used in the estimation of drag and Lorentz force respectively is described in this Appendix. We show the equivalence of two 
Lorentz force models: torus instability (TI) model \citep{Kli06} and flux-injection model \citep{Che96}). GCS fitting at one timestamp is shown for each CME and the corresponding parameters at that timestamp are given in Table \ref{tblapp}.}}\\
\noindent\makebox[\linewidth]{\rule{\textwidth}{3pt}} 

\section{CME Virtual Mass} \label{Mv}
The total CME mass ($m_{cme}$) is a summation of the ``true'' mass ($M_{T}$) which is corrected for projection effects and the virtual mass $M_{v}$, which is described 
by \citet{Lan59} (p 31), as half the mass of the fluid displaced by a sphere moving in that fluid. The CME traveling in an ambient solar wind is also envisioned
as a solid sphere propagating in a fluid, (see, Chapter \ref{chap4}), therefore we include the virtual mass correction to the true CME mass ($M_{T}$).
\begin{eqnarray}
m_{cme}\,&=&\, M_{T}\,+\, M_{v}  \nonumber \\
&=&\, {\rm Vol}_{cme}\bigl[\rho_{i}\,+\,\frac{\rho_{e}}{2}] \label{eqvm1}
\end{eqnarray}
where, $Vol_{cme}$ is the CME volume, $\rho_{i}$ is density inside the CME and $\rho_{e}$ is the external (solar wind) density.
\begin{gather*}
\rho_{i}=\rho_{cme}=\frac{M_{T}}{{\rm Vol}_{cme}} \\
\rho_{e}=\rho_{sw}=n_{sw}(R)\, m_{p}
\end{gather*}
where, $n_{sw}(R)$ is the solar wind proton number density and $m_{p}
$ is the proton mass. Substituting in Equation \ref{eqvm1},
\begin{eqnarray}
 m_{cme}(R)\,&=&\, {\rm Vol}_{cme}\biggl[\frac{M_{T}}{{\rm Vol}_{cme}}\,+\, \frac{n_{sw}(R) m_{p}}{2}\biggr] \nonumber\\
  &=&\, M_{T} \biggl[ 1\,+\, \frac{n_{sw}(R)\,m_{p}}{2}\frac{{\rm Vol}_{cme}}{M_{T}}\biggr] \nonumber\\
 &=&\, M_{T} \biggl[ 1\,+\, \frac{n_{sw}(R)\,m_{p}}{2}\frac{A_{cme} R}{M_{T}}\biggr] \label{eqvm2}
\end{eqnarray}
using, ${\rm Vol}_{cme}=A_{cme}\, R$, where $A_{cme}$ is the CME cross-sectional area and R is the heliocentric distance of the CME leading edge.
Equation \ref{eqvm2} defines the total CME mass as a function of the heliocentric distance (R). We find in our analysis however, that the addition of virtual 
mass does not change the results significantly.

\section{The TI Lorentz Force Model} \label{LF}
\subsection{Net Lorentz force}
The toroidal force acting on a current carrying loop is given by (see e.g., \citealp{Sha66}, \citealp{Che89}):
\begin{equation}
 F_{int}\,=\, \frac{I^{2}}{c^{2}R}\biggl( ln \frac{8 R}{R_{cme}} + \frac{l_{i}}{2} - \frac{3}{2} \biggr)   \,\,\,\, (in\,\,cgs)
\label{lsf1}
 \end{equation}
where, $F_{int}$ is the Lorentz force ($J\times B$) acting radially outwards per unit length of the flux rope, $I$ is the toroidal current, $R$ is the height 
and $l_{i}$ is the internal inductance. In the presence of an external poloidal magnetic field ($B_{ext}$), the force with which the CME is held down is given by:
\begin{eqnarray}
 F_{ext}&=& \frac{J\times B_{ext}(R)}{c} \nonumber\\
 &=& \int I \frac{dl\times B_{ext}(R)}{c} \nonumber \\
 \frac{F_{ext}}{unit \,\,length}&=& -\frac{I B_{ext}(R)}{c} \label{lsf2}
 \end{eqnarray}
The complete force balance equation is given by \citep{Kli06}:
\begin{equation}
 F_{R}\,=\,\frac{I^{2}}{c^{2}R}\biggl( ln \biggl(\frac{8 R}{R_{cme}}\biggr) + \frac{l_{i}}{2} - \frac{3}{2} \biggr)-\frac{I B_{ext}(R)}{c}
 \label{lsf3}
\end{equation}
where, $F_{R}$ is the Lorentz force per unit length acting on a current carrying loop (in cgs units). The total 
Lorentz force acting on the flux-rope is given by (Equation \ref{eqlsf}):
 \begin{equation}
F_{Lorentz} \,=\,\frac{\pi I^{2}}{c^{2}} \biggl(ln \biggl(\frac{8 R}{R_{cme}}\biggr)-\frac{3}{2}+ \frac{l_{i}}{2}\biggr)\,-\, \frac{(\pi R)I B_{ext}(R)}{c} 
\label{lsf4}
\end{equation}
\subsection{CME current at equilibrium ($I_{eq}$)}
The equilibrium current ($I_{eq}$) is determined by equating the total force $F_{Lorentz}$ (Equation \ref{lsf4}) to zero. 
At equilibrium ($R=h_{eq}$), 
\begin{equation}
\frac{\pi I_{eq}^{2}}{c^{2}} \biggl(ln \biggl(\frac{8 h_{eq}}{R_{cme}(h_{eq})}\biggr)-\frac{3}{2}+ \frac{l_{i}}{2}\biggr)\,=\, \frac{(\pi h_{eq})I_{eq} B_{ext}(h_{eq})}{c}
\label{eqeq}
\end{equation}
Using,
\begin{gather*}
c^{'}(R)\,=\,\bigl[ln (8R/R_{cme})-2+l_{i}/2\bigr] \\
c^{'}_{eq}\,=\,c^{'}(R=h_{eq})=\bigl[ln (8h_{eq}/R_{cme}(h_{eq})-2+l_{i}/2\bigr],
\end{gather*}
in Equation \ref{eqeq} 	we get,
\begin{eqnarray}
 \frac{I_{eq}}{c \,\,h_{eq}}(c^{'}_{eq}+ \frac{1}{2})&=& B_{ext}(h_{eq}) \nonumber \\
 \end{eqnarray}
 The current at equilibrium is given by,
\begin{equation}
I_{eq}\,=\,\frac{B_{ext}(h_{eq})h_{eq} c}{c^{'}_{eq}+\frac{1}{2}} \label{lsf5}
\end{equation}

\subsection{CME current $I$} \label{cmecurr}
The current carried by a flux-rope is determined by conserving the total magnetic flux (internal + external). Using 
$B_{ext}(R)= \hat{B} R^{-n}$ and equations of flux conservation (Equation \ref{eqflux}), we get:
\begin{equation}
L_{eq}I_{eq}-\frac{2 \pi}{c} \int_{0}^{h_{eq}} B_{ext}(r) r dr= L(R)\,I(R)- \frac{2 \pi}{c} \int_{0}^{R} B_{ext}(r) r dr
\label{lsf6}
\end{equation}
where, 

\begin{gather}
L\,=\, \frac{4 \pi R}{c^{2}}\bigl[ln (8R/R_{cme})-2+l_{i}/2\bigr]  \nonumber \\
L(R)= \frac{4 \pi R}{c^{2}} c^{'}  \nonumber \\ 
L_{eq}= \frac{4 \pi h_{eq}}{c^{2}} c^{'}_{eq}
\label{eqL}
\end{gather}
From Equation \ref{lsf6},
\begin{eqnarray}
 L(R) \,I(R)&=& L_{eq} \,I_{eq}+\frac{2 \pi}{c} \hat{B} \biggl[\frac{R^{(2-n)}}{(2-n)} -\frac{h_{eq}^{(2-n)}}{(2-n)}\biggr] \nonumber \\
&=& L_{eq} \,I_{eq}+\frac{2 \pi}{(2-n)\,c} \hat{B} h_{eq}^{(2-n)} \biggl[\biggl(\frac{R}{h_{eq}}\biggr)^{(2-n)}-1\biggr] \nonumber \\
I(R)&=& \frac{L_{eq}\,I_{eq}}{L(R)} + \frac{2 \pi}{(2-n)\,c}  \frac{\hat{B}\,h_{eq}^{(2-n)}}{L(R)} \biggl[\biggl(\frac{R}{h_{eq}}\biggr)^{(2-n)}-1\biggr] \label{lsf7}
\end{eqnarray}

Using, Equations \ref{eqL} and \ref{lsf5} in Equation \ref{lsf7},
\begin{eqnarray}
I(R)&=& \frac{c^{'}_{eq} \,h_{eq} I_{eq}}{c^{'}\, R} +\frac{2 \pi}{(2-n)}\frac{(\hat{B}\, h_{eq}^{-n})\, h_{eq}^{2}}{L(R)\,c}\biggl[\biggl(\frac{R}{h_{eq}}\biggr)^{(2-n)}-1\biggr] \nonumber \\
&=& \frac{c^{'}_{eq} \,h_{eq} I_{eq}}{c^{'}\, R}\,\biggl(1+\frac{2 \pi}{(2-n)} \frac{B_{ext}(h_{eq})\,h_{eq} c^{'} R}{L(R)\, c\, c^{'}_{eq}\, I_{eq}}\biggl[\biggl(\frac{R}{h_{eq}}\biggr)^{(2-n)}-1\biggr]\biggr) \nonumber
\end{eqnarray}
The CME current is given by,
\begin{equation}
I(R)= \frac{c^{'}_{eq} \,h_{eq} I_{eq}}{c^{'}\, R}\,\biggl(1+ \frac{(c^{'}_{eq}+1/2)}{2\, c^{'}_{eq}\,(2-n)} \biggl[\biggl(\frac{R}{h_{eq}}\biggr)^{(2-n)}-1\biggr]\biggr) \label{lsf8}
\end{equation}

\subsection{Final Lorentz force expression}
We rewrite the Lorentz force (Equation \ref{lsf4}) by substituting $I(R)$ (Equation \ref{lsf8}) and $I_{eq}$ (Equation \ref{lsf5}),

\begin{eqnarray}
 F_{Lorentz}&=& \frac{\pi\, I}{c} \biggl[ \frac{I\, (c^{'}+1/2)}{c} - R\,B_{ext}(R) \biggr] \nonumber \\
 &=& \frac{\pi \,c^{'}_{eq} \,h_{eq} \,I_{eq} \, Q}{c^{'}\, R\,c} \biggl[\frac{c^{'}_{eq} \,h_{eq} \,I_{eq} \, Q\, (c^{'}+1/2)}{c^{'}\, R\, c}- R\, B_{ext}\,(h_{eq})\biggl(\frac{R}{h_{eq}}\biggr)^{-n}\biggr] \nonumber
\end{eqnarray}

where, 
\begin{gather*}
 B_{ext}(R)=B_{ext}(h_{eq})[R/h_{eq}]^{-n}  \\
 Q=\biggl(1+ \frac{(c^{'}_{eq}+1/2)}{2\, c^{'}_{eq}\,(2-n)} \biggl[\biggl(\frac{R}{h_{eq}}\biggr)^{(2-n)}-1\biggr]\biggr)
\end{gather*}
Finally, we get the net Lorentz force acting on a CME in terms of the observed and derived GCS parameters 
($R$, $R_{cme}$, $h_{eq}=1.05$ \Rs) and using $B_{ext}(h_{eq})$ from the force analysis,

\begin{equation}
 F_{Lorentz}= \frac{\pi\, c^{'}_{eq}\, (B_{ext}(h_{eq}))^{2}\, (h_{eq})^{2}\, Q}{c^{'}\, (c^{'}_{eq}+1/2)} \biggl[\frac{c^{'}_{eq}\, Q }{(c^{'}_{eq}+1/2)} \frac{(c^{'}+1/2)}{c^{'}}\,\biggl(\frac{h_{eq}}{R}\biggr)^{2}- \biggl(\frac{R}{h_{eq}}\biggr)^{-n}\biggr]
\label{eqfl}
 \end{equation}

 \subsection{Decay Index $n$} \label{decayn}

\citet{Kli06} describe the evolution of the major radius of an expanding current ring by :
\begin{equation}
 \begin{split}
 \frac{d^{2} \rho}{d \tau^{2}}&=\frac{c^{'2}_{eq} \rho^{-2}} {c^{'}(c^{'}_{eq}+1/2)} \biggl[1+\frac{(c^{'}_{eq}+1/2)}{2 c^{'}_{eq} (2-n)} (\rho^{2-n}-1)\biggr]  \nonumber \\ \nonumber
& \times \biggl[ \frac{c^{'}+1/2}{c^{'}} \biggl(1+\frac{(c^{'}_{eq}+1/2)}{2 c^{'}_{eq} (2-n)} (\rho^{2-n}-1)\biggr)- \frac{c^{'}_{eq}+1/2}{c^{'}_{eq}} \rho^{2-n}\biggr]    \,\,\,\,\, (n\neq2)
 \end{split}
\end{equation}
where, $\rho=R/h_{eq}$, $\tau=t/T$ and,\\

\begin{equation}
 T^{2}=\frac{c^{'}_{eq}+1/2}{4} \frac{R_{cme}^{2}(h_{eq})}{B_{eq}^{2}/(4 \pi \rho_{m0})} \nonumber
\end{equation}
where, $\rho_{m0}$ is the mass density at equilibrium. Using Alf\'{v}en speed defined as, $V_{A}=B_{eq}^{2}/(4 \pi \rho_{m0})$ at equilibrium, 
\begin{equation}
T=\frac{(c^{'}_{eq}+1/2)^{1/2}}{2} \frac {R_{cme}(h_{eq})}{V_{A}} \nonumber
\end{equation}

Assuming self-similar expansion, that is, $c^{'}(R) = constt.$, they derive the condition for instability using,\\
 \begin{equation}
\frac{d}{d\rho}\bigl(\frac{d^{2}\rho}{d\tau^{2}}\bigr) > 0 \,\,\,\,\,\, at \,\, \rho=1  \nonumber
 \end{equation}
which gives the critical decay index, 
\begin{equation}
n > n_{cr}=\frac{3}{2}-\frac{1}{4 c^{'}_{eq}}   \nonumber
\end{equation}

Since, $c^{'}$ depends logarithmically on $R/R_{cme}$, it varies very slowly, therefore, the assumption $c^{'}(R)=c^{'}_{eq}$ is valid. It is also seen that 
CMEs expand self-similarly (using GCS observations).

The decay index, $n$ is chosen based on the following two conditions:\\
\begin{enumerate}
 \item $n>n_{cr}$. The decay index for the external magnetic field should be greater than the critical decay index derived using the instability condition, which 
 ensures the CME launch.
 \item Since the solar wind aerodynamic drag is found to be dominant above the heights $R>\widetilde{h}_{0}$, the Lorentz force $F_{Lorentz}$ should be smaller in 
 magnitude above these heights. We use the condition, $|F_{drag}|>F_{Lorentz}$ for $R>\widetilde{h}_{0}$.
\end{enumerate}

We choose different values of $n>n_{cr}$, and check if $|F_{drag}|>F_{Lorentz}$ for $R>\widetilde{h}_{0}$ for that $n$.
The smallest value of $n>n_{cr}$ for which this condition holds true is taken to be the value of the decay index 
which takes care of the instability condition as well as constrains the Lorentz force.
A larger value of $n$ indicates that the external field decays rapidly, typical to active regions from where fast CMEs erupt. It is seen that for our sample, the 
largest values of $n (\sim 3)$ correspond to fast events. Smaller $n$ values indicate slower CMEs originating from erupting prominences.

\section{Equivalence of two Lorentz force models}
\begin{figure}[h!]
\centering
\includegraphics[height=0.23\paperheight,width=0.5\paperwidth]{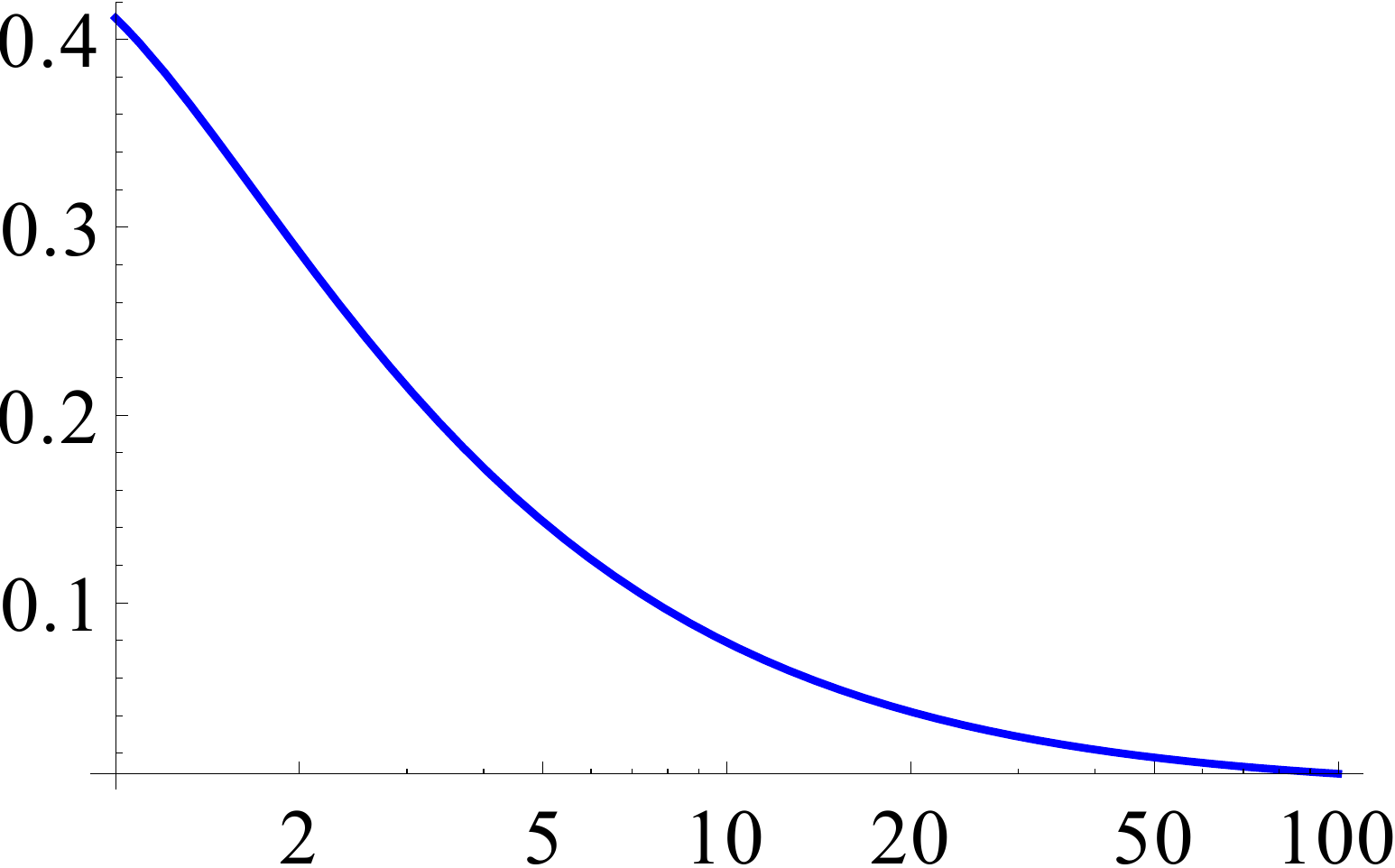}
  \put(-320,30.7){{\rotatebox{90}{{\color{black}\fontsize{12}{12}\fontseries{n}\fontfamily{phv}\selectfont   Current (units of $10^{10}$ Amperes)}}}}
               \put(-205.8,-8.9){{\rotatebox{0}{{\color{black}\fontsize{12}{12}\fontseries{n}\fontfamily{phv}\selectfont  Heliocentric distance ($R$) (\Rs)}}}}
\caption[Current profile using \citet{Kli06} prescription]{CME current profile (Equation \ref{lsf8}) for parameters : $n=2.5$, $I_{eq}=0.41\times10^{10}$ 
A and $R/R_{cme}=4.56$. 
The current decreases as a function of the heliocentric distance ($R$). X-axis scale is logarithmic.}
\label{currentti}
\end{figure}

The Lorentz force prescription we use is based on torus instability model (TI) described in \citet{Kli06}. It appeals to an overlying external magnetic 
field which decreases rapidly enough for the CME to launch (like a whiplash action). The external poloidal field is required to decay as $\propto R^{-n}$, 
where $n$ is the decay index. Equation \ref{lsf4} represents the net Lorentz force acting on a current carrying loop.
The two competing $J\times B$ terms in Equation \ref{lsf4} give a Lorentz force profile which increases to a peak and then 
decreases (Figure \ref{all}).
The TI model assumes conservation of magnetic flux to derive the CME current ($I$) (see section \ref{cmecurr}, Equation \ref{lsf8}).
An example of CME current profile is shown in Figure \ref{currentti} for CME 1. It decreases as a function of the heliocentric distance.

Another approach to Lorentz forces relies on a tailored injection of poloidal flux at the base of the flux-rope \citep{Che96,Che10}. 
The temporal profile of the accompanying soft X-ray flare is often used as a guide for the time profile of injected flux 
(or equivalently, the poloidal CME current). The force acting on a toroidal section of current carrying 
loop as described by \citet{Che96} 
(based on \citet{Sha66}) is given by :
\begin{equation}
F_{R}\,=\, \frac{I^{2}_{t}}{c^{2}R}\biggl[ln \bigl(\frac{8 R}{R_{cme}}\bigr) + \frac{1}{2} \beta_{p} -\frac{1}{2}\frac{\bar{B}^{2}_{t}}{B^{2}_{pa}} + 2\bigl(\frac{R}{R_{cme}}\bigr) \frac{B_{s}}{B_{pa}} -1 +\frac{li}{2}\biggr]
\label{chenforce}
\end{equation}
where, $F_{R}$ is the Lorentz force per unit length in the major radial direction, $I_{t}$ is the toroidal current, $\beta_{p}=8\pi (\bar{p}-p_{a})/B^{2}_{pa}$, $\bar{p}$ is the average pressure inside the loop, 
$p_{a}$ is the ambient coronal pressure, $\bar{B}_{t}$ is the average toroidal field and $B_{pa}$ is the 
poloidal magnetic field inside the loop. $B_{s}$ is the ambient field. 
The poloidal flux enclosed by the torus section is given by:
\begin{equation}
\Phi_{p}(t)\,=\,c L(t)I_{t}(t) \label{phip}
\end{equation}
Using Equation \ref{phip} in \ref{chenforce}, we get,
\begin{equation}
F_{R}\,=\, \frac{\Phi^{2}_{p}}{c^{4}R L^{2}}\biggl[ln \bigl(\frac{8 R}{b}\bigr) + \frac{1}{2} \beta_{p} -\frac{1}{2}\frac{\bar{B}^{2}_{t}}{B^{2}_{pa}} + 2\bigl(\frac{R}{a}\bigr) \frac{B_{s}}{B_{pa}} -1 +\frac{li}{2}\biggr]
\label{chenforce1}
\end{equation}

The injection of flux (by increasing the poloidal magnetic flux $\Phi_{p}(t)$) drives the initial flux rope out of equilibrium. For each CME, 
\citet{Che10} adjust the function, $d\Phi_{p}(t)/dt$ to obtain best-fit solutions to the observed height-time profiles.
The Lorentz force profile from this model also increases to a peak and then decreases. However, the toroidal current 
$I_{t}$ from this prescription emulates the injected flux profile (Equation \ref{phip}).

On comparing the TI and flux-injection models for Lorentz forces, we find that the ambient field is required 
to decay sufficiently in the TI model. However, no such restriction exists in the second approach. In fact, \citet{Che96} describe a 
simple functional form for the external ambient field chosen to increase before decreasing as a function of height (Equation 16 in \citet{Che96}). This provides 
stability to the initial flux rope against expansion in the major radial direction.

We assume a simplified version of Equation \ref{chenforce} specialized to small perturbations from equilibrium, given by :
\begin{equation}
F_{R}\,=\, \frac{I^{2}_{t}}{c^{2}R}\biggl[ln \bigl(\frac{8 R}{R_{cme}}\bigr) + \beta_{p} -\frac{3}{2} +\frac{li}{2}\biggr]
\label{eq71}
\end{equation}
where, $\beta_{p}\simeq 1- \frac{B^{2}_{t}}{B^{2}_{p}}$. 

We compare the Lorentz forces from these two equations \ref{eq71} and \ref{lsf3}, to determine the functional form of current $I_{t}$
(from flux-injection model) in terms of the current prescription from the TI model. For simplicity, we consider $\beta_{p}$ is small 
($\approx 0$).

\begin{eqnarray}
\frac{I^{2}_{t}}{c^{2}R}\biggl[ln \bigl(\frac{8 R}{R_{cme}}\bigr) + \beta_{p} -\frac{3}{2} +\frac{li}{2}\biggr]&=&\frac{I^{2}}{c^{2}R}\biggl( ln \biggl(\frac{8 R}{R_{cme}}\biggr) + \frac{l_{i}}{2} - \frac{3}{2} \biggr)-\frac{I B_{ext}(R)}{c} \nonumber \\
\nonumber\\
I^{2}_{t}& =&I^{2}\,-\, \frac{c R I B_{ext}(R)}{\biggl( ln \frac{8 R}{R_{cme}} + \frac{l_{i}}{2} - \frac{3}{2} \biggr)}
\label{ichen}
\end{eqnarray}
where, $I$ is the CME current determined from the TI model (Equation \ref{lsf8}) and $B_{ext}(R)=B_{ext}(h_{eq})\bigl(R/h_{eq} \bigr)^{-n}$. 

\begin{figure}[h!]
\centering
\includegraphics[height=0.23\paperheight,width=0.5\paperwidth]{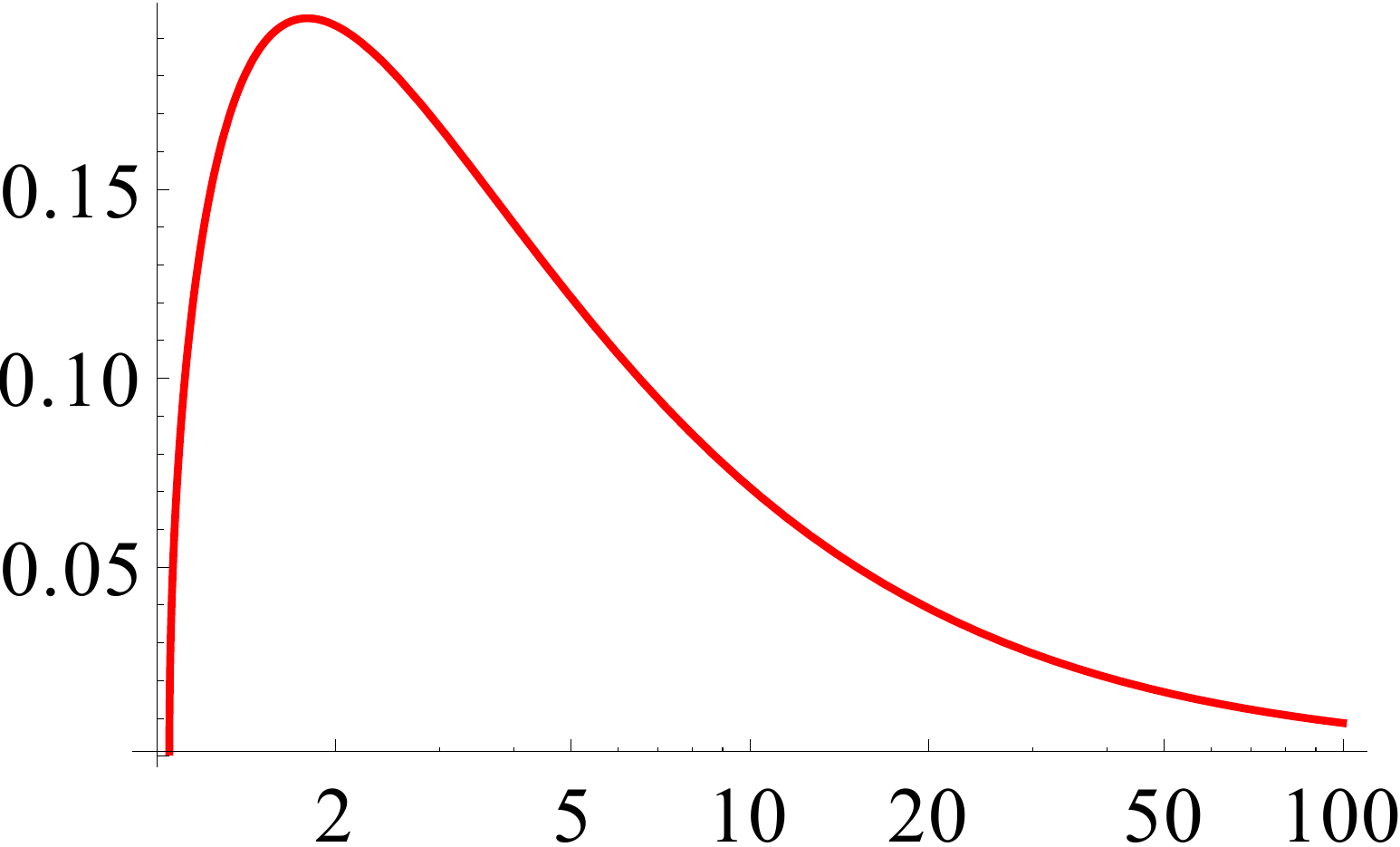}
  \put(-320,30.7){{\rotatebox{90}{{\color{black}\fontsize{12}{12}\fontseries{n}\fontfamily{phv}\selectfont   Current (units of $10^{10}$ Amperes)}}}}
               \put(-205.8,-8.9){{\rotatebox{0}{{\color{black}\fontsize{12}{12}\fontseries{n}\fontfamily{phv}\selectfont  Heliocentric distance ($R$) (\Rs)}}}}
\caption[Current profile using \citet{Che96} prescription]{CME current using injection of poloidal flux \citep{Che96} follows the Lorentz force profile. Current increases to a peak and then decreases
as a function of the heliocentric distance ($R$). The x-axis scale is logarithmic.}
\label{currentchen} 
\end{figure}

As an example, we show the current profile ($I_{t}$) from the flux injection model for CME 1 determined using Equation \ref{ichen} and CME parameters: $n=2.5$, $I_{eq}=0.41\times10^{10}$ A and 
$R/R_{cme}=4.56$ in Figure \ref{currentchen}.

The functional form of CME current $I_{t}$ is given by (for $n=2.5$):
\begin{equation}
 I_{t} =\sqrt{ \frac{C_{1}}{R^{3}} - \frac{C_{2}}{R^{2.5}}+\frac{C_{3}}{R^{2}}}
\end{equation}
The constants $C_{1}$, $C_{2}$ and $C_{3}$ are determined for corresponding values of parameters, $n$, aspect ratio ($R/R_{cme}$) and 
$B_{ext}(h_{eq})$. For this event (with $n=2.5$),
\begin{eqnarray}
 C_{1}&=&4.9\times10^{38} \nonumber \\
 C_{2}&=&1.3\times10^{39} \nonumber \\
 C_{3}&=&8.4\times10^{38} \nonumber
\end{eqnarray}

Both the approaches for Lorentz forces acting on CMEs yield a solution that initially increases with time/height, 
reaches a maximum and subsequently decreases. The mathematical equivalence of their current prescriptions has been demonstrated above.

\section{GCS fittings for all CMEs}
\begin{figure}[h]    
  \centering                              
  \centerline{\hspace*{0.0\textwidth}
               \includegraphics[width=0.4\textwidth,clip=]{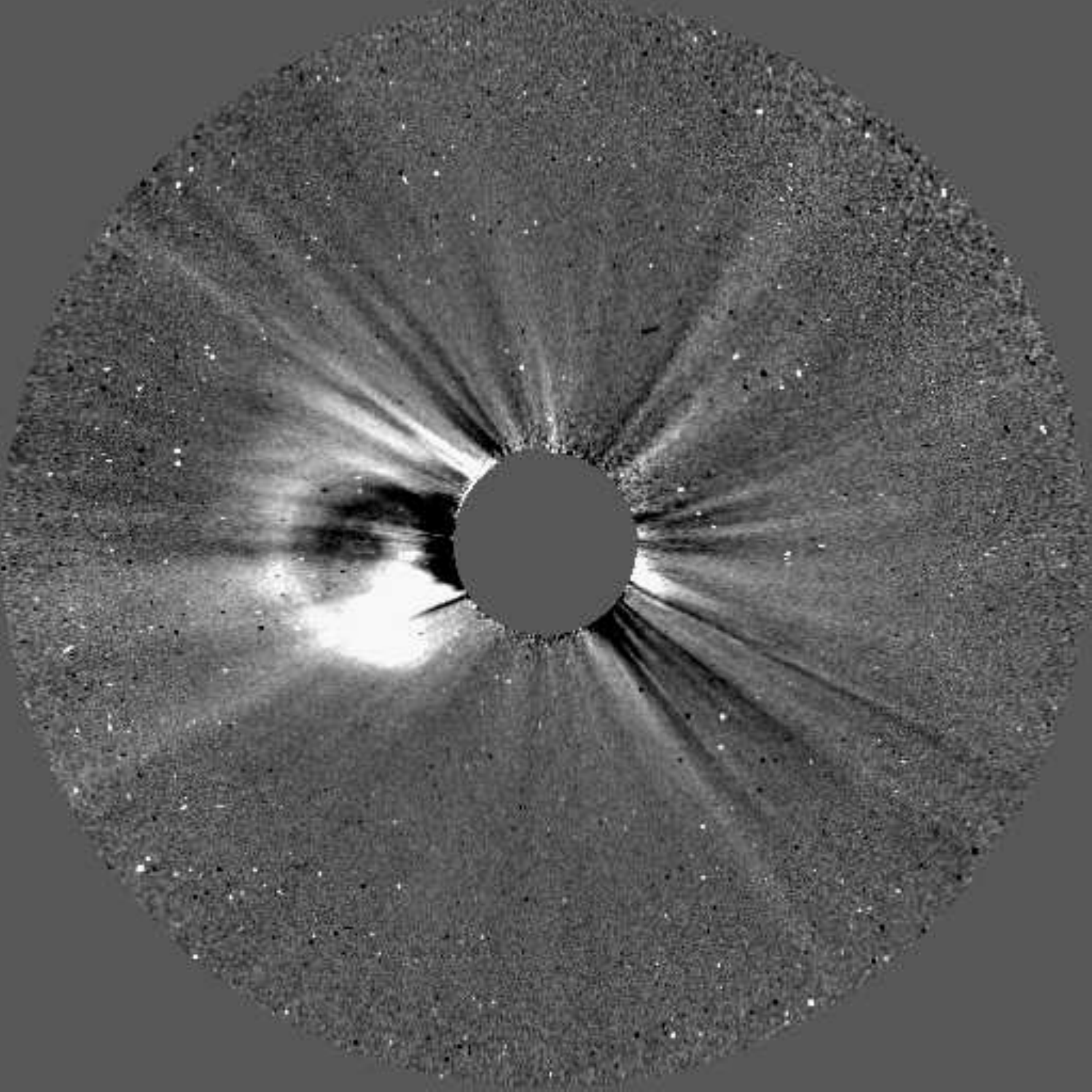}
                \hspace*{-0.02\textwidth}
               \includegraphics[width=0.4\textwidth,clip=]{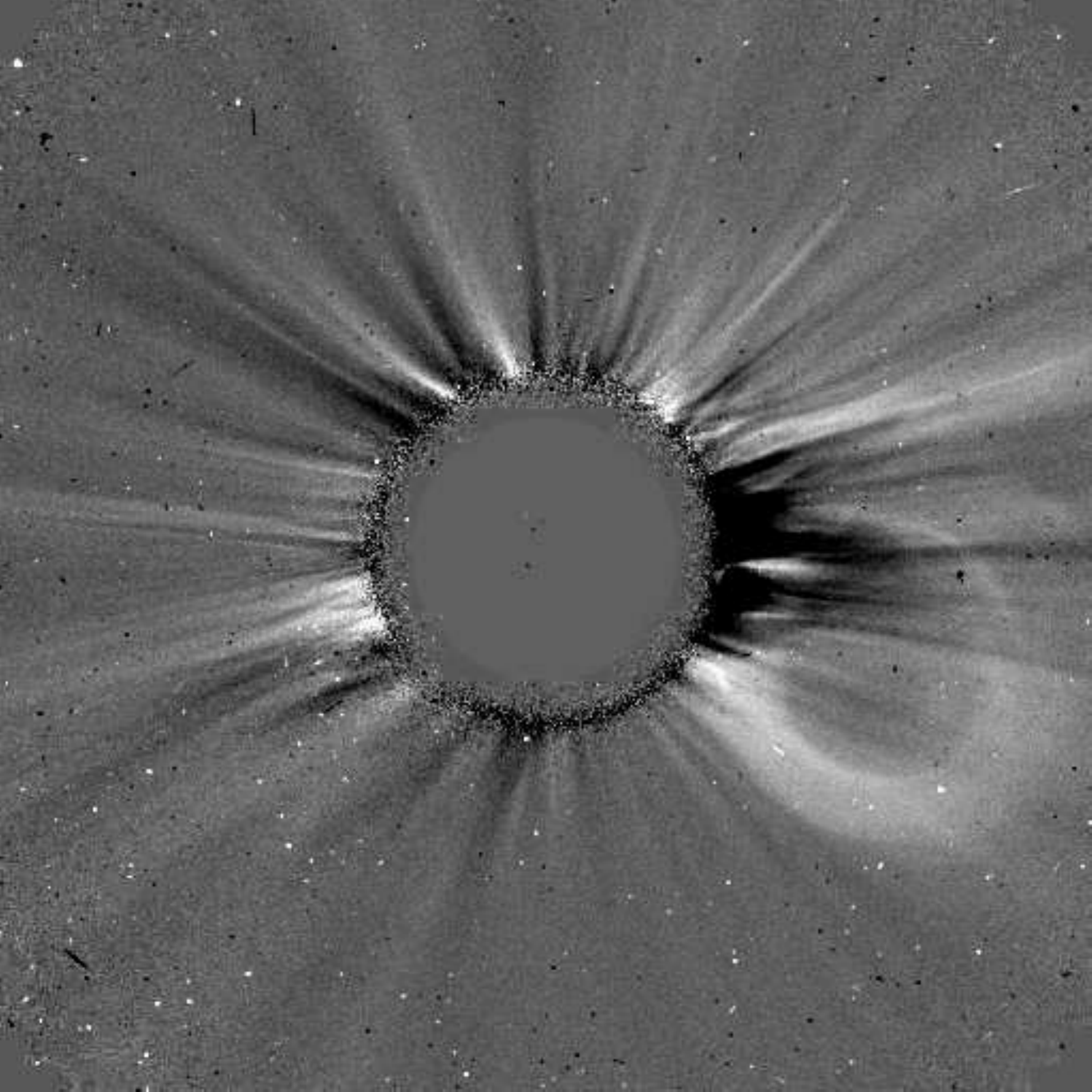}
             \hspace*{-0.02\textwidth}
               \includegraphics[width=0.4\textwidth,clip=]{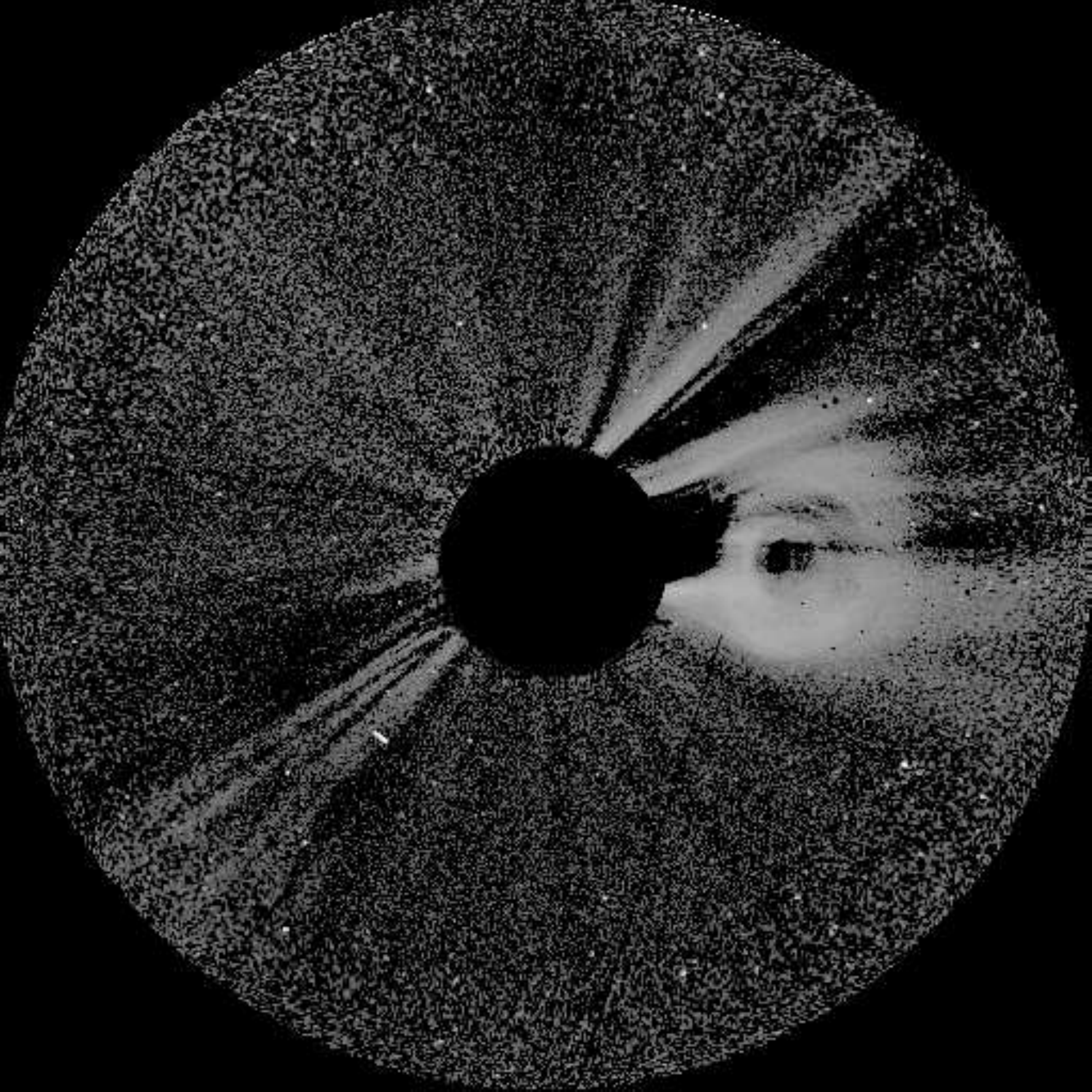}
                }
                 \centerline{\hspace*{0.0\textwidth}
              \includegraphics[width=0.4\textwidth,clip=]{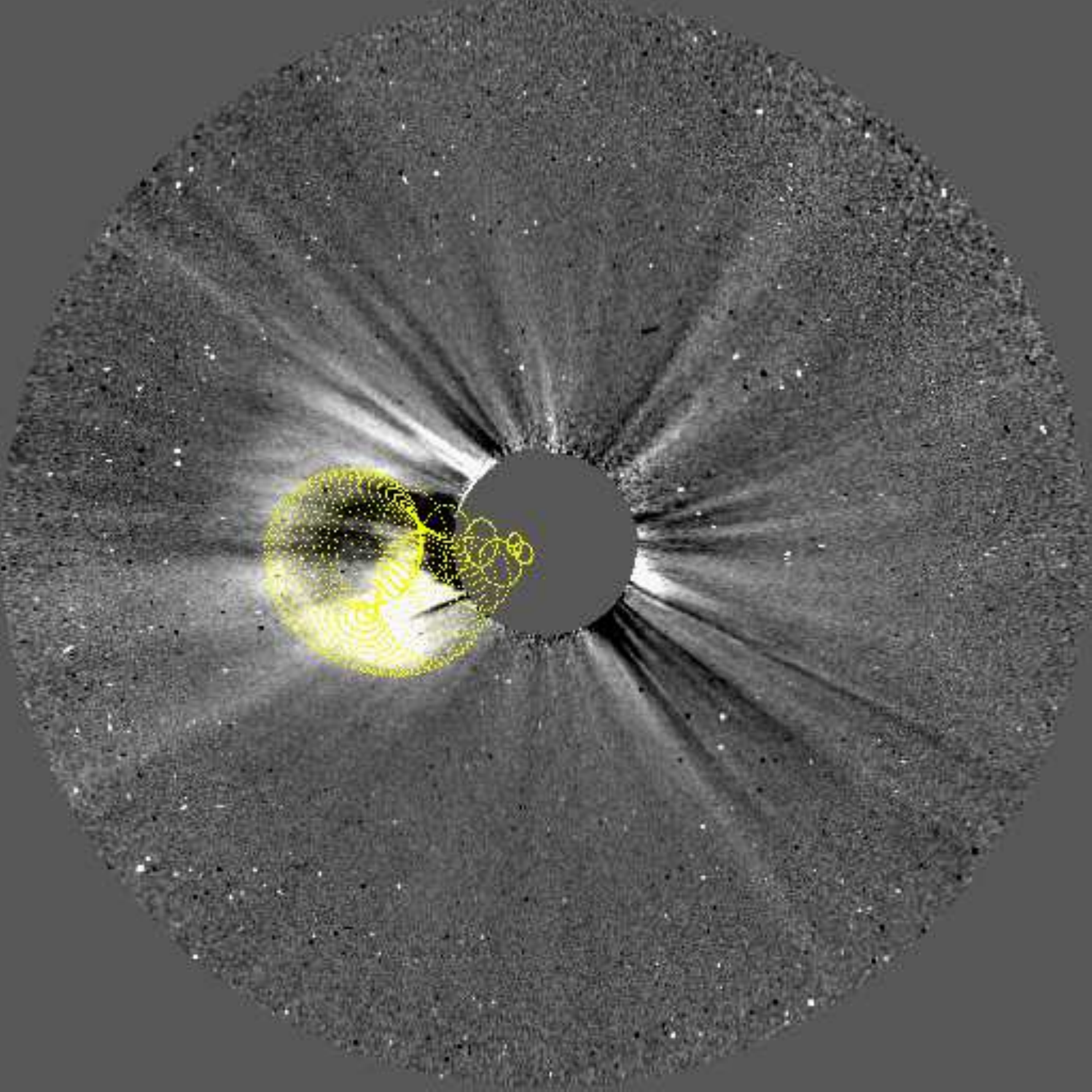}
               \hspace*{-0.02\textwidth}
               \includegraphics[width=0.4\textwidth,clip=]{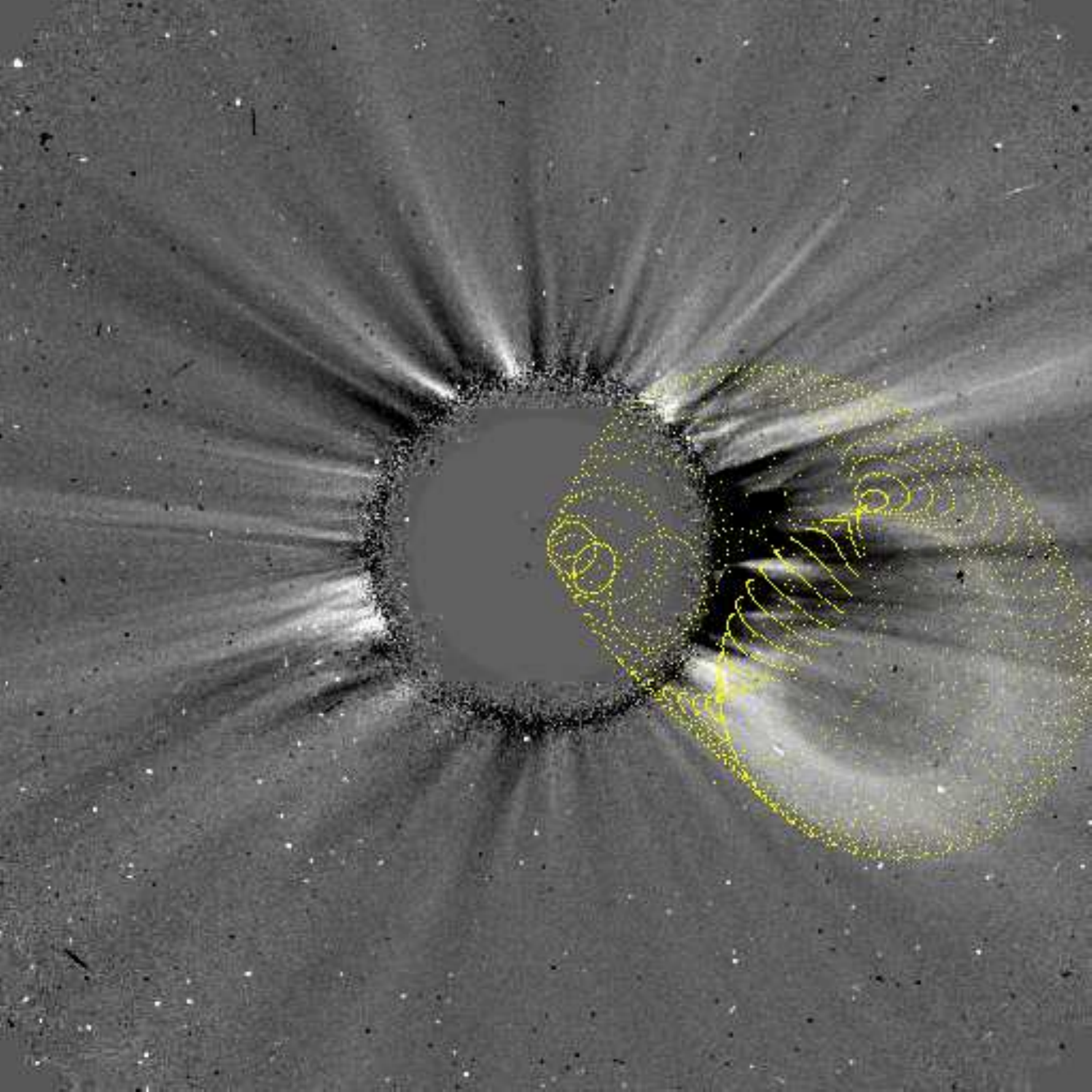}
              \hspace*{-0.02\textwidth}
               \includegraphics[width=0.4\textwidth,clip=]{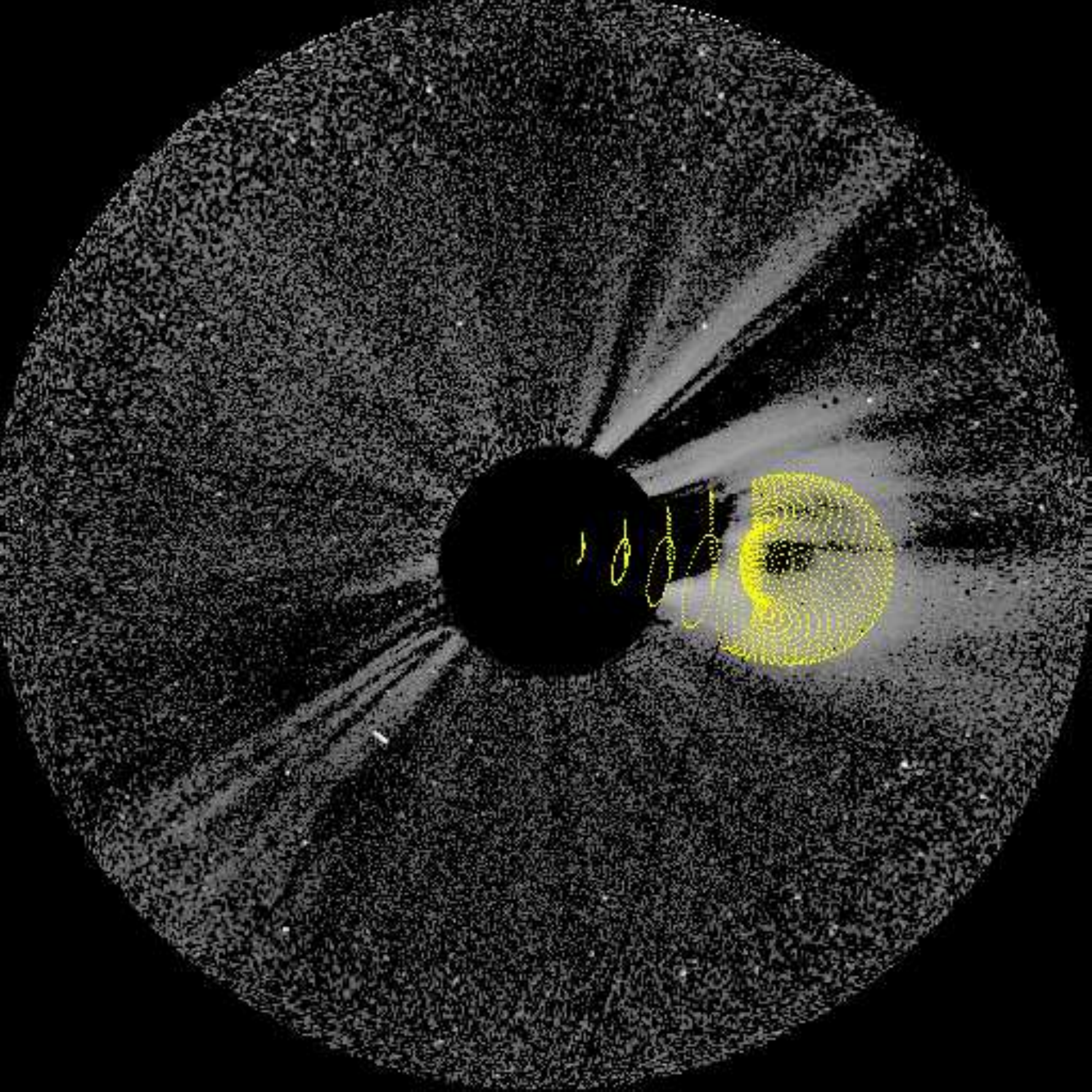}
                }
\vspace{0.0261\textwidth}  
\caption[GCS fit for CME 1 at 17:54]{GCS fit for CME 1 on March 19, 2010 at 17:54 UT at height $H= 10.07$ \Rs. Table \ref{tblapp} 
lists the GCS parameters for this event.}
\label{figa1}
\end{figure}

\begin{figure}[h]    
  \centering                              
   \centerline{\hspace*{0.04\textwidth}
               \includegraphics[width=0.4\textwidth,clip=]{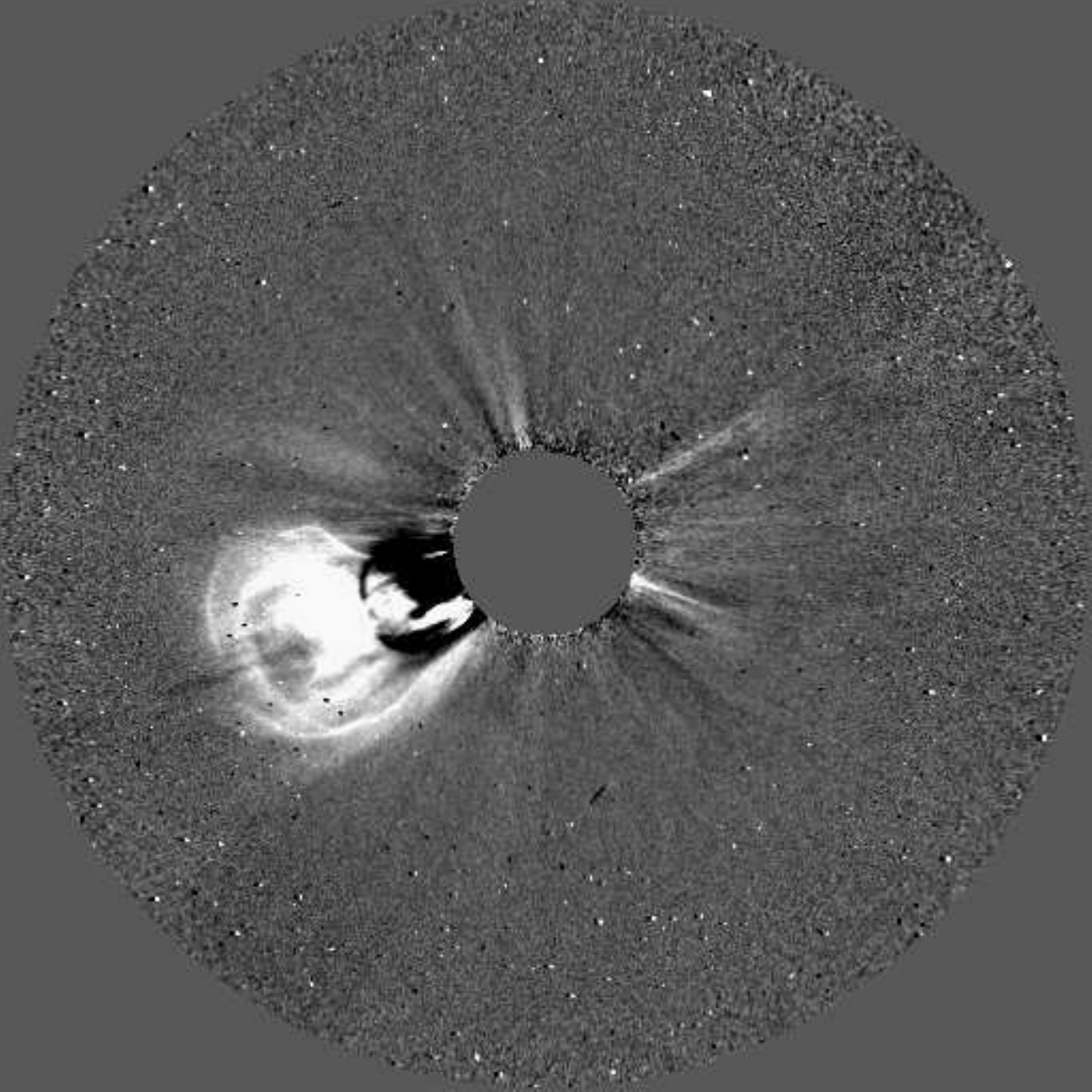}
                \hspace*{-0.02\textwidth}
               \includegraphics[width=0.4\textwidth,clip=]{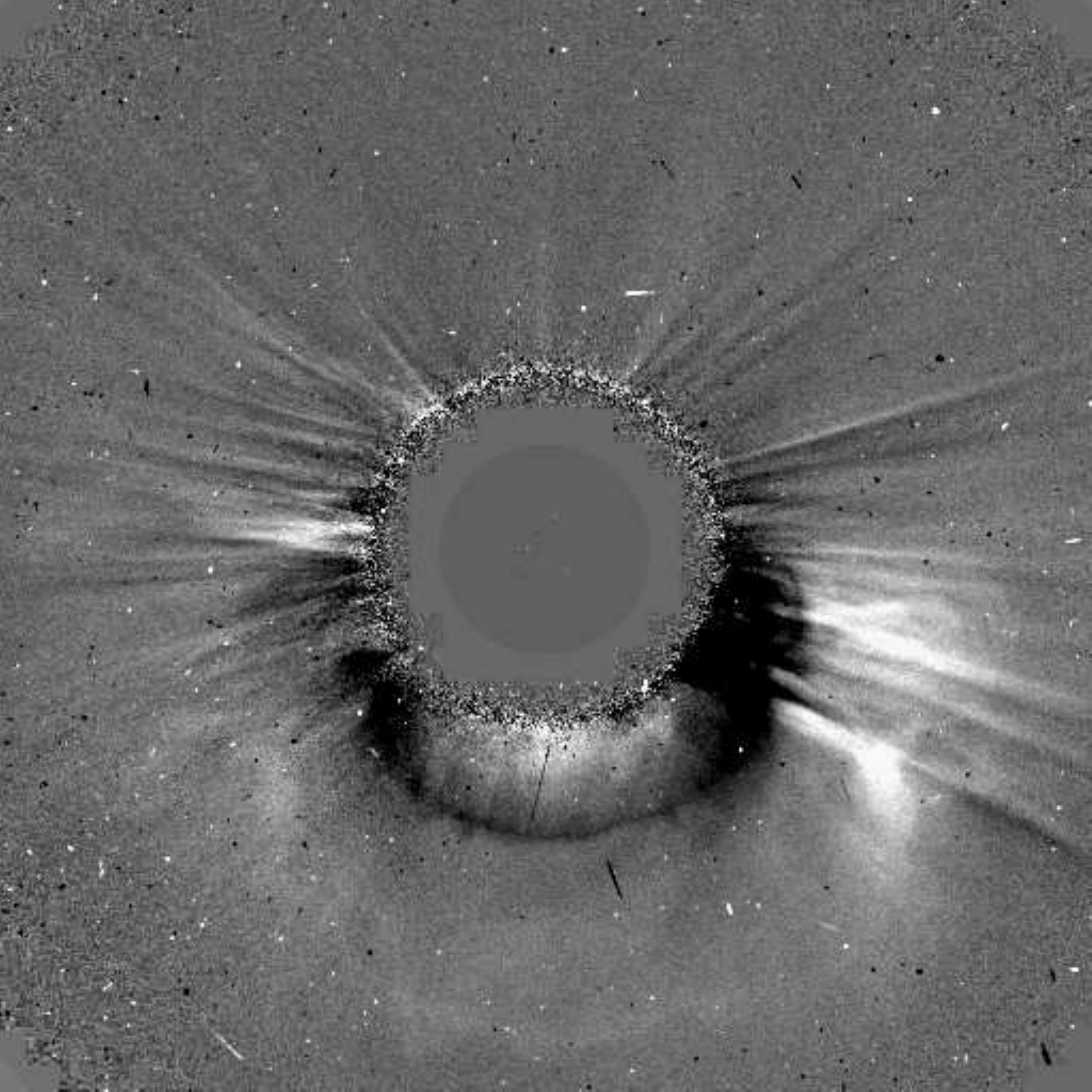}
             \hspace*{-0.02\textwidth}
               \includegraphics[width=0.4\textwidth,clip=]{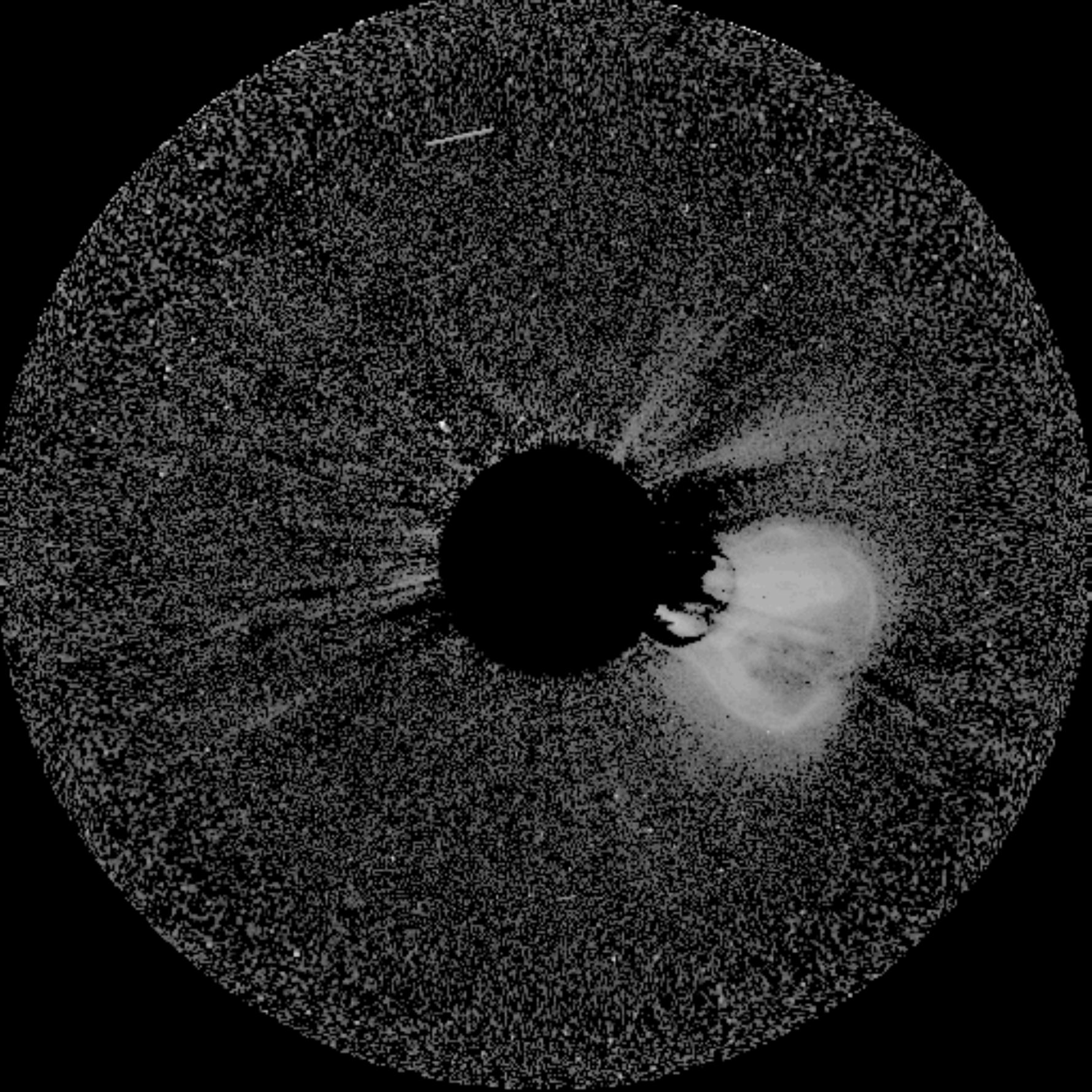}
               }
                 \centerline{\hspace*{0.04\textwidth}
              \includegraphics[width=0.4\textwidth,clip=]{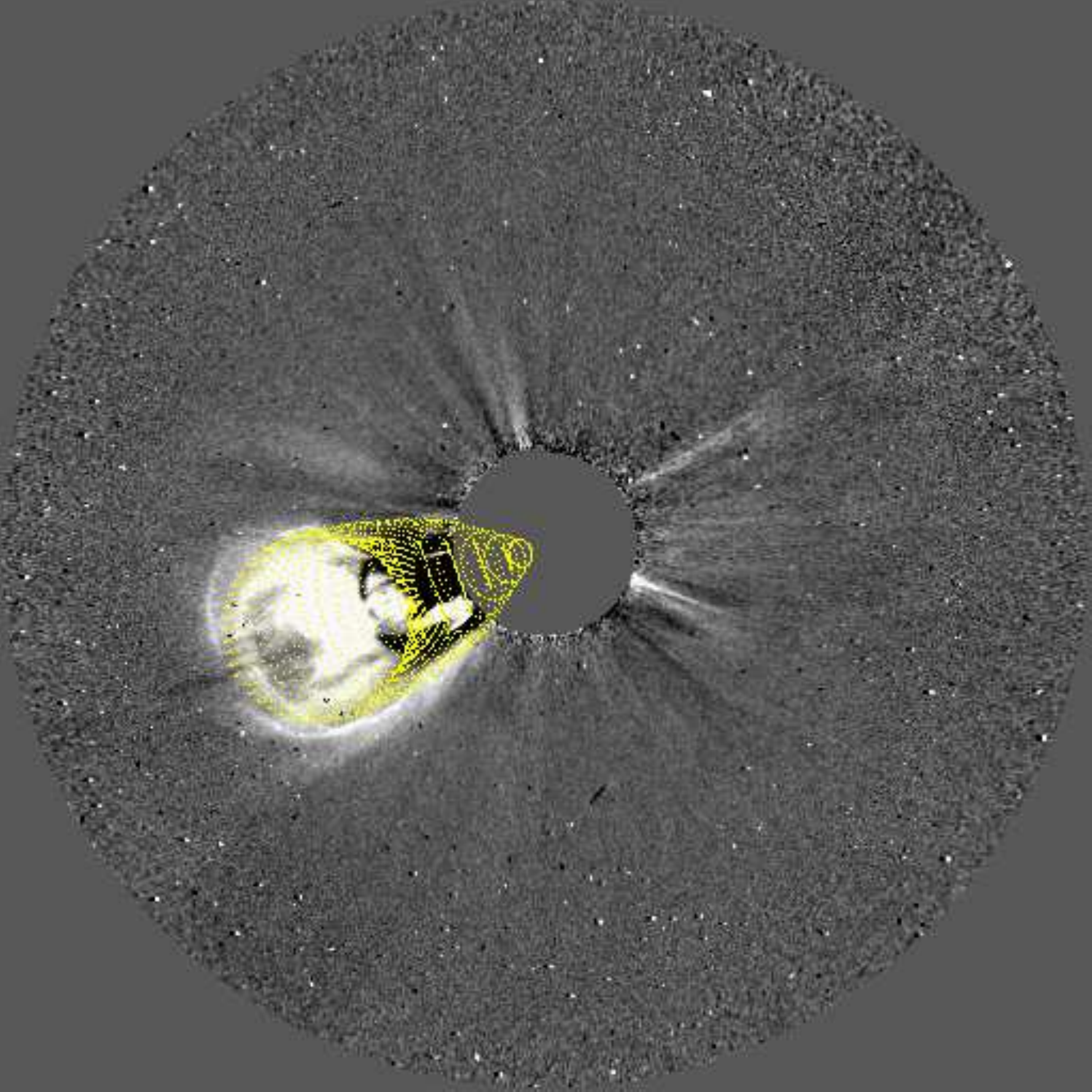}
               \hspace*{-0.02\textwidth}
               \includegraphics[width=0.4\textwidth,clip=]{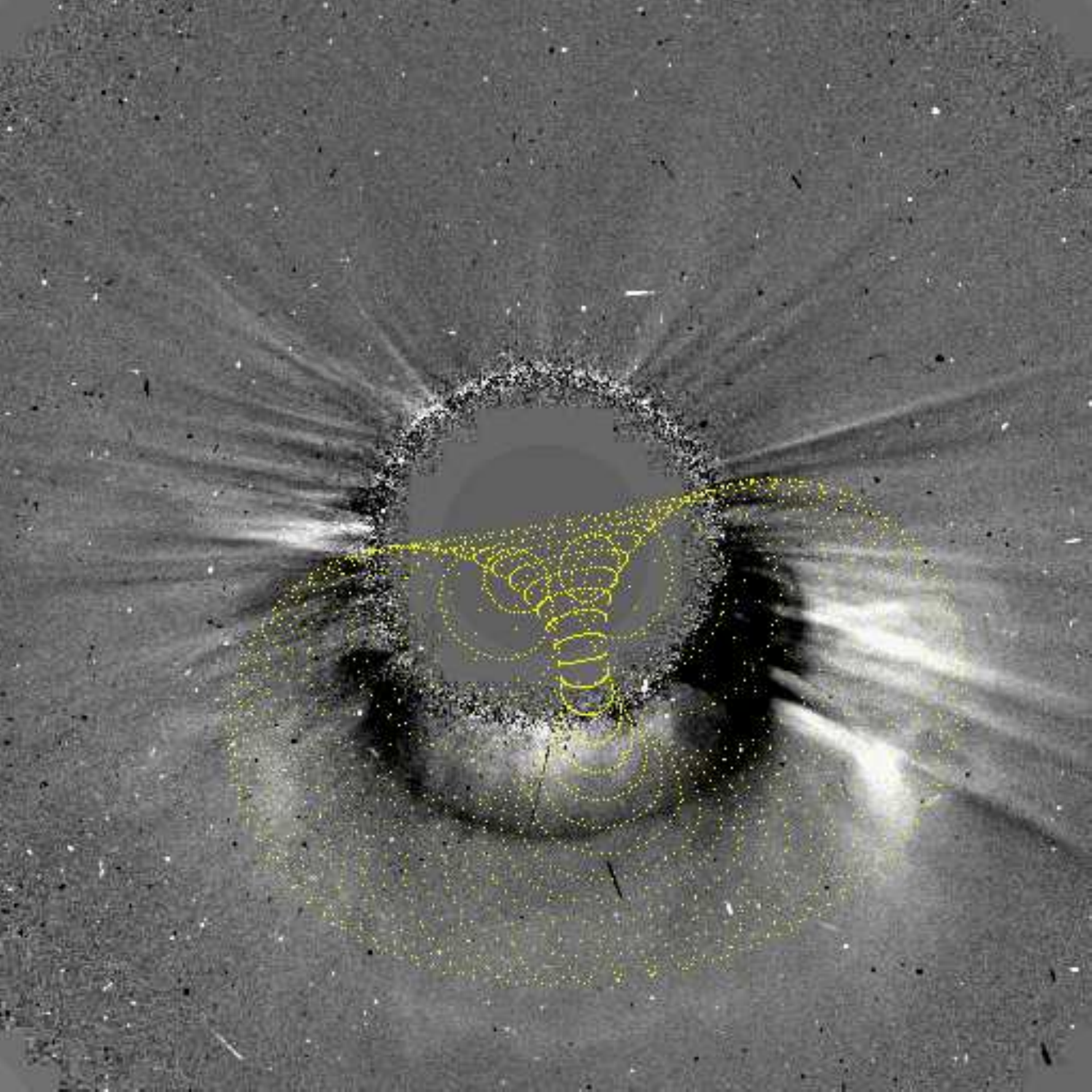}
              \hspace*{-0.02\textwidth}
               \includegraphics[width=0.4\textwidth,clip=]{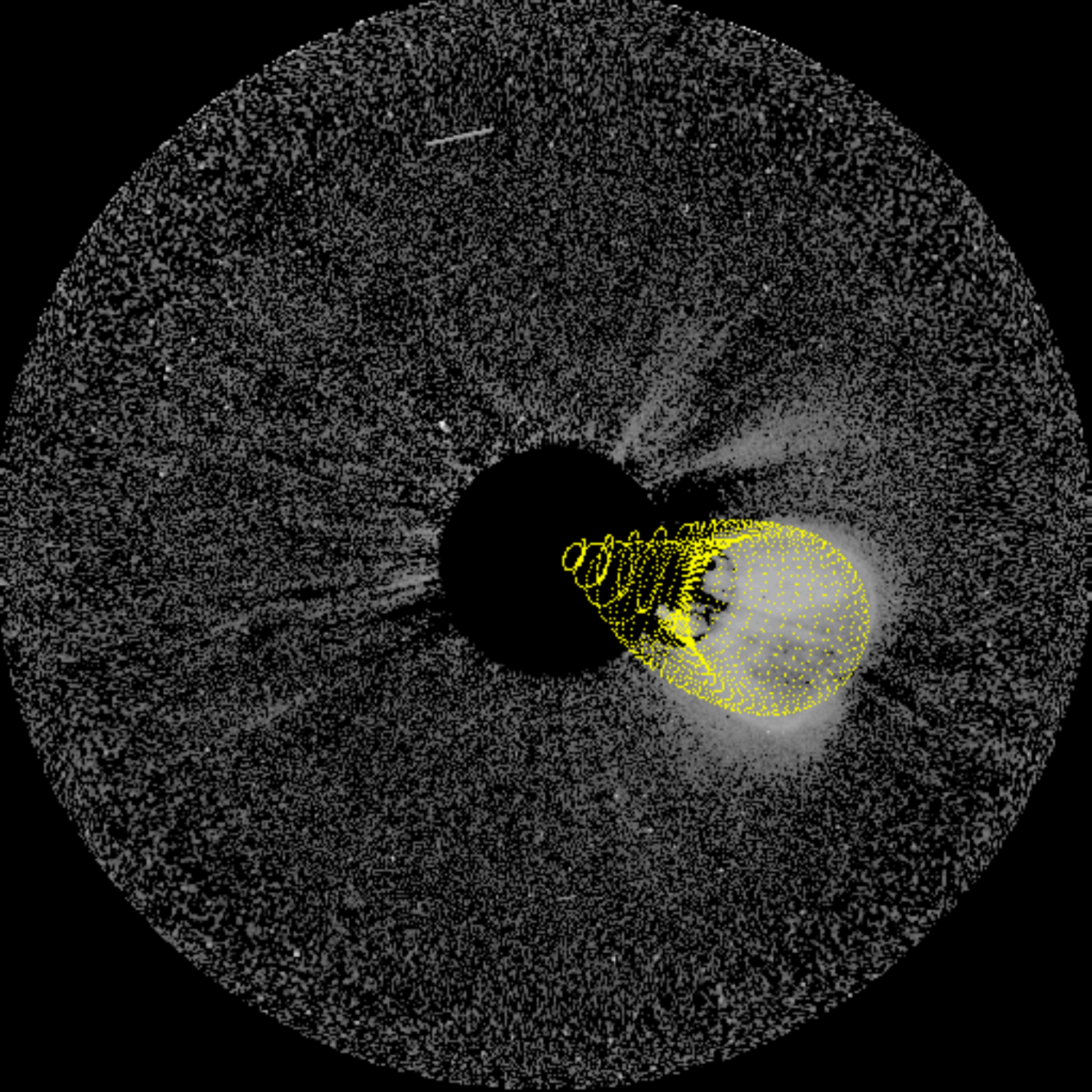}
                }
\vspace{0.0261\textwidth}  
\caption[GCS fit for CME 2 at 11:24]{GCS fit for CME 2 on April 03, 2010 at 11:24 UT at height $H=9.6$ \Rs. Table \ref{tblapp} 
lists the GCS parameters for this event.}
\label{figa2}
\end{figure}

\begin{figure}[h]    
  \centering                              
   \centerline{\hspace*{0.00\textwidth}
               \includegraphics[width=0.4\textwidth,clip=]{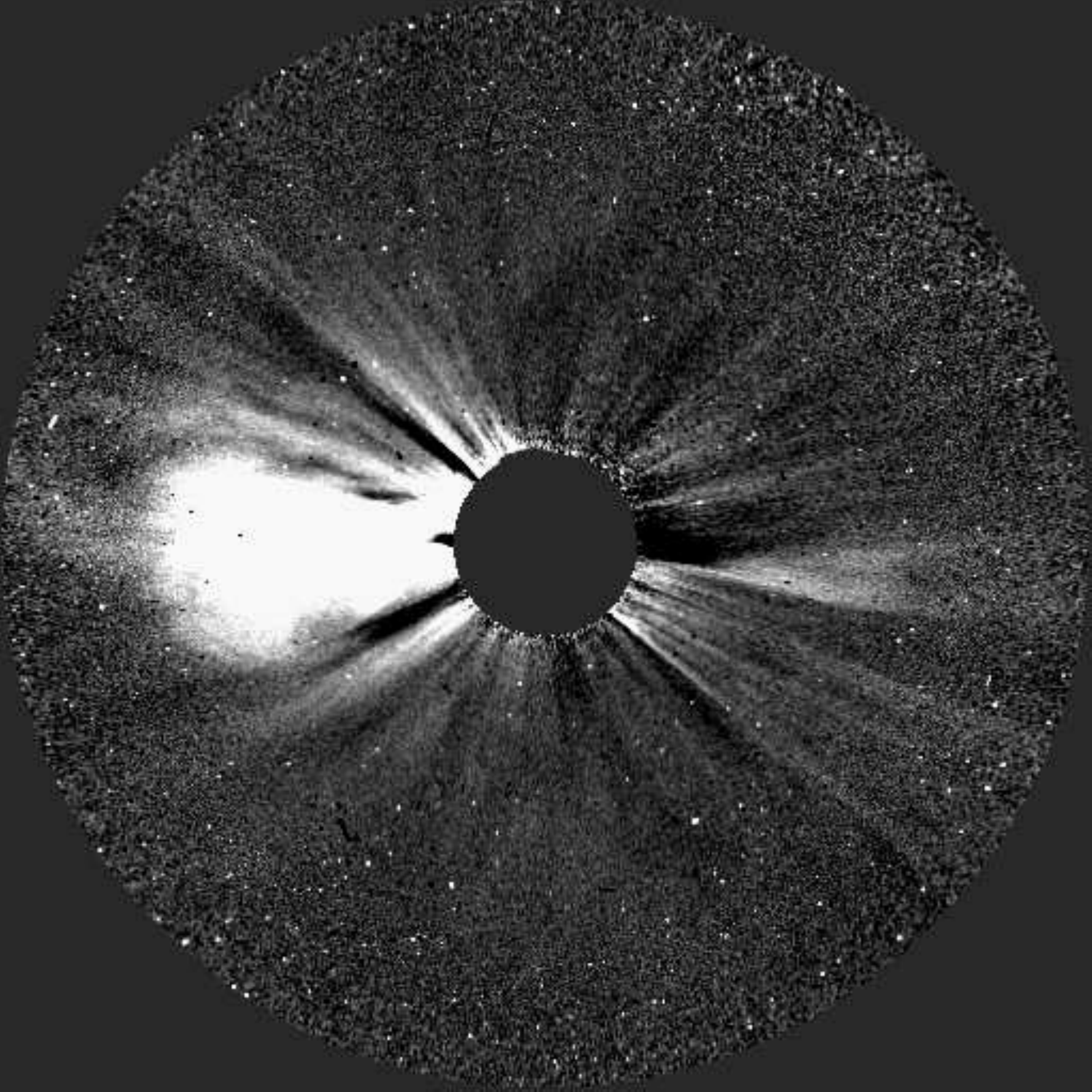}
                \hspace*{-0.02\textwidth}
               \includegraphics[width=0.4\textwidth,clip=]{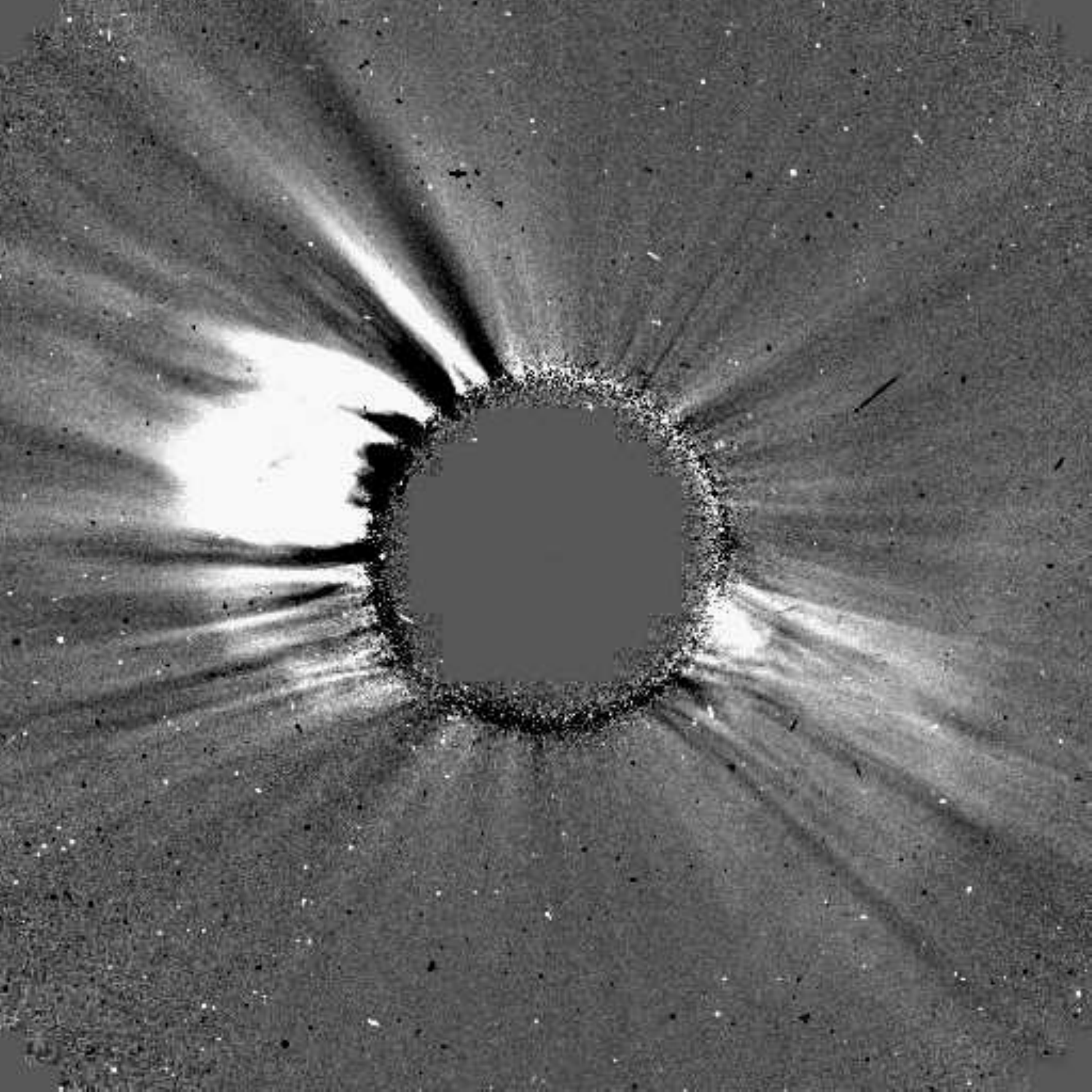}
             \hspace*{-0.02\textwidth}
               \includegraphics[width=0.4\textwidth,clip=]{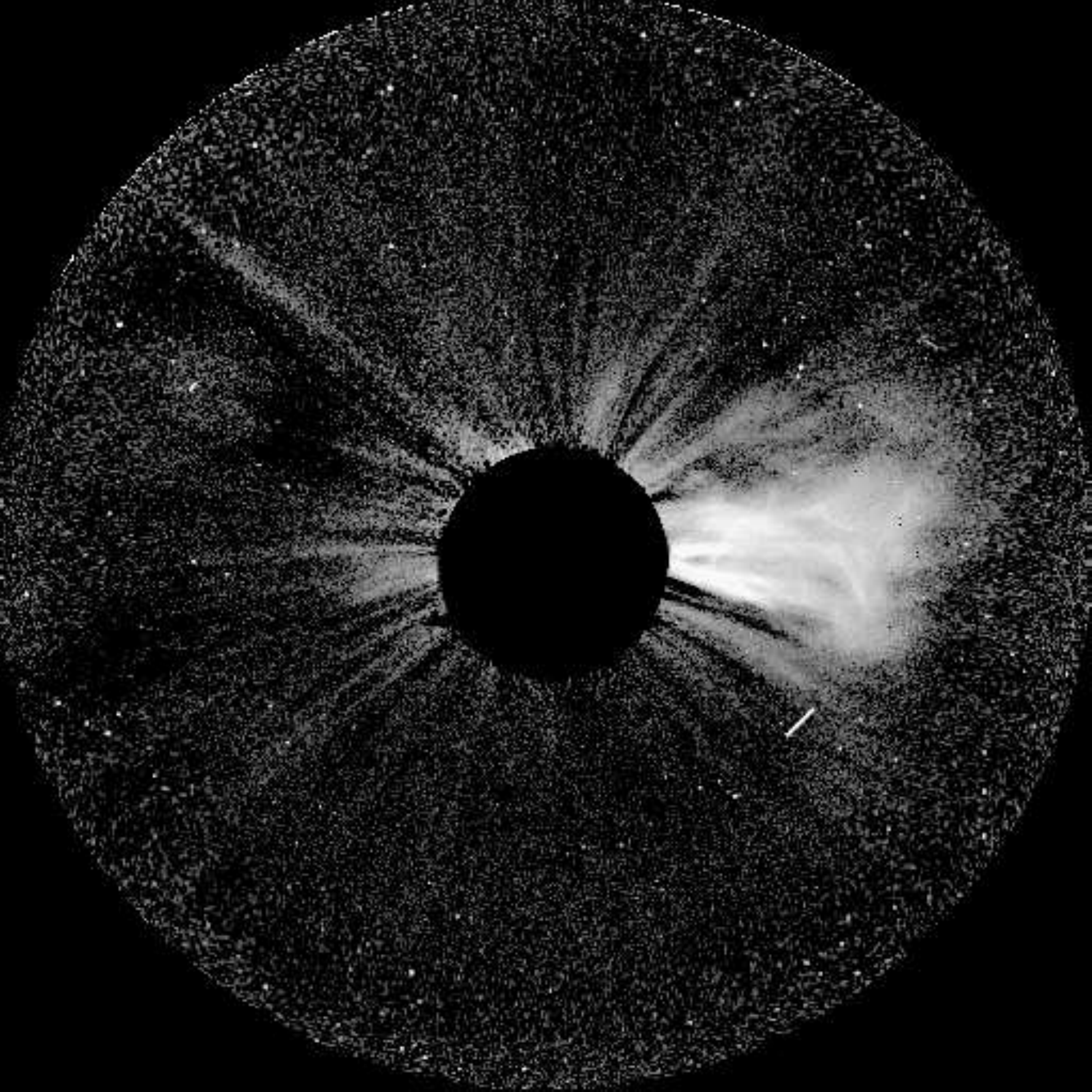}
               }
                 \centerline{\hspace*{0.0\textwidth}
              \includegraphics[width=0.4\textwidth,clip=]{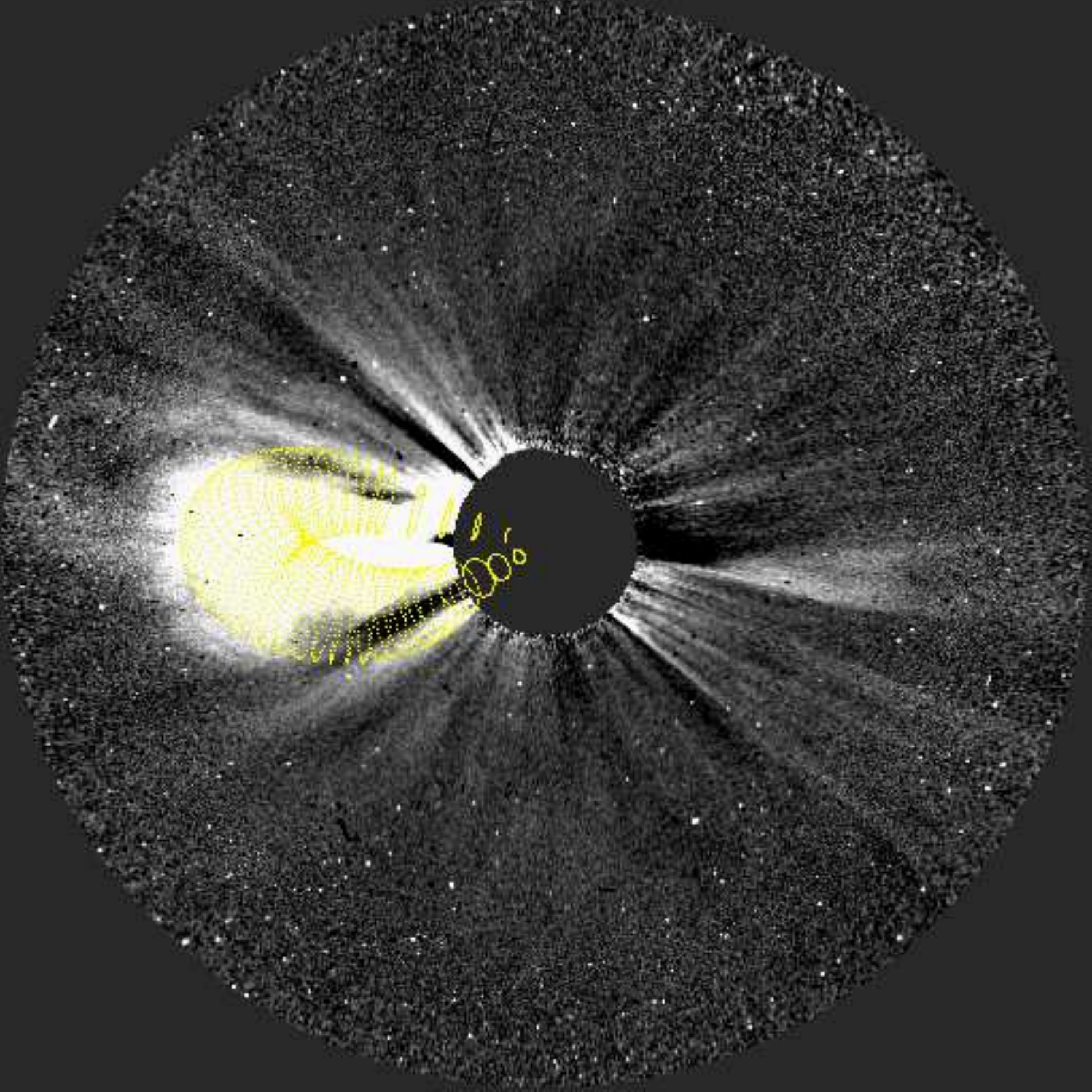}
               \hspace*{-0.02\textwidth}
               \includegraphics[width=0.4\textwidth,clip=]{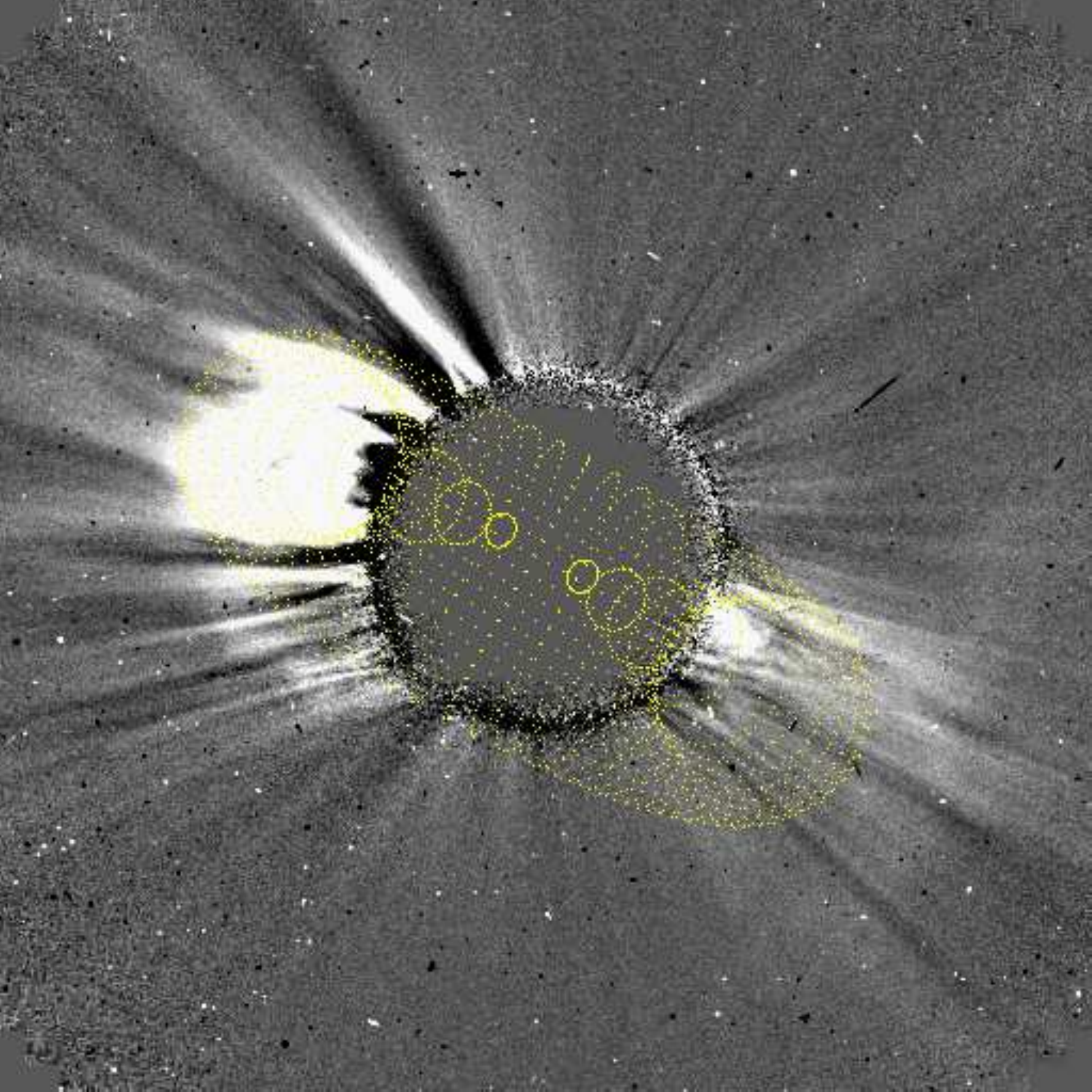}
              \hspace*{-0.02\textwidth}
               \includegraphics[width=0.4\textwidth,clip=]{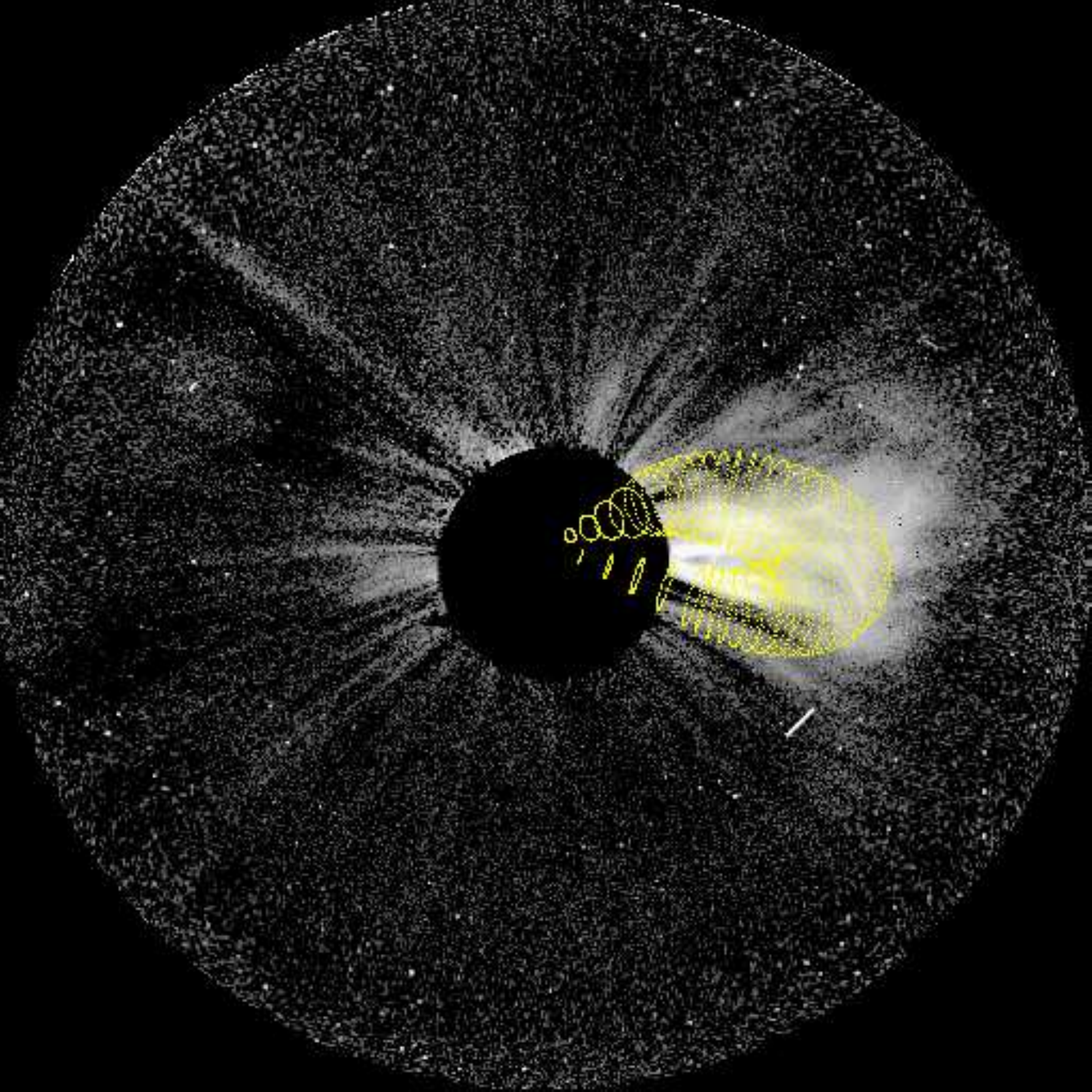}
                }
\vspace{0.0261\textwidth}  
\caption[GCS fit for CME 3 at 06:54]{GCS fit for CME 3 on April 08, 2010 at 06:54 UT at height $H=10.2$ \Rs. Table \ref{tblapp} 
lists the GCS parameters for this event.}
\label{figa3}
\end{figure}

\begin{figure}[h]    
  \centering                              
   \centerline{\hspace*{0.04\textwidth}
               \includegraphics[width=0.4\textwidth,clip=]{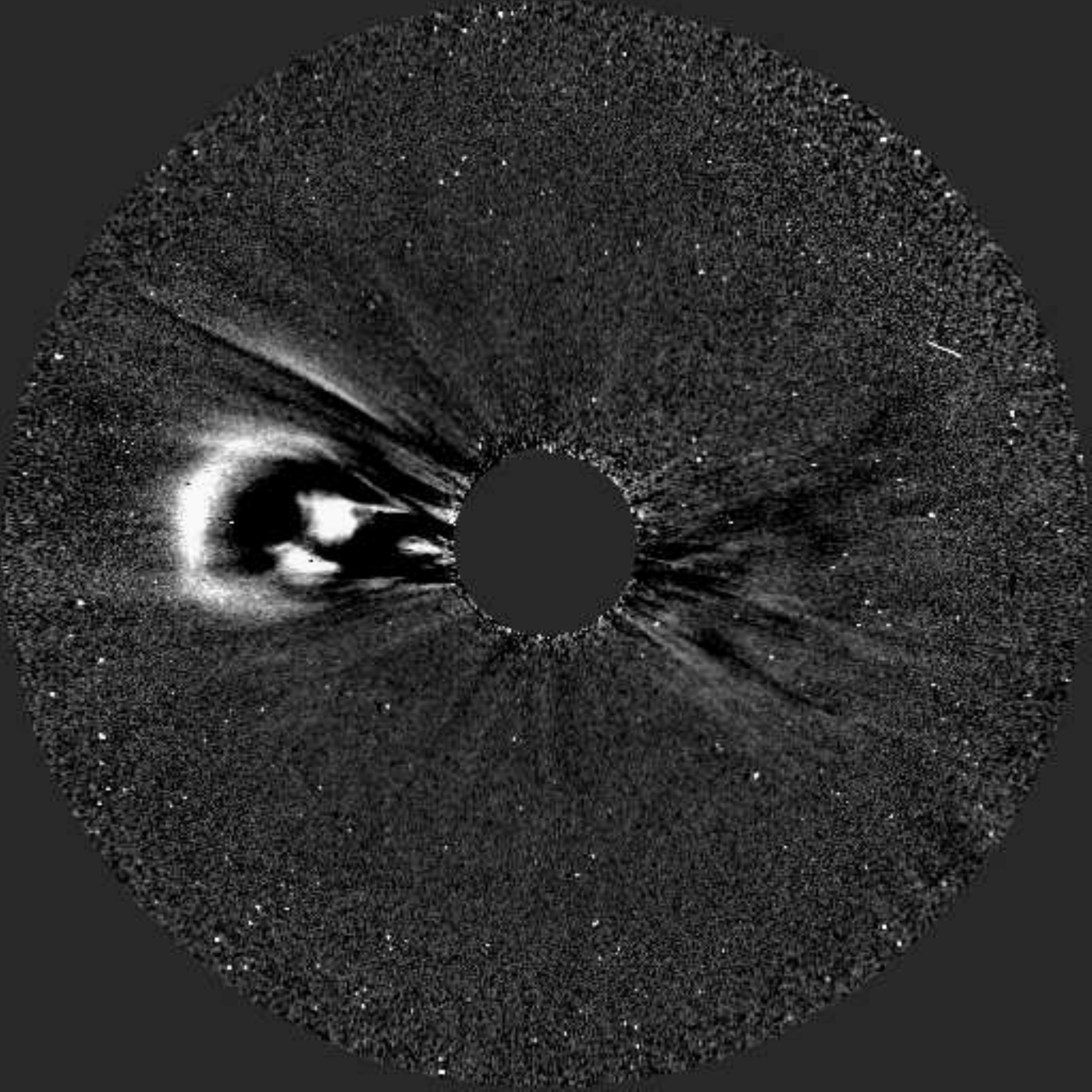}
                \hspace*{-0.02\textwidth}
               \includegraphics[width=0.4\textwidth,clip=]{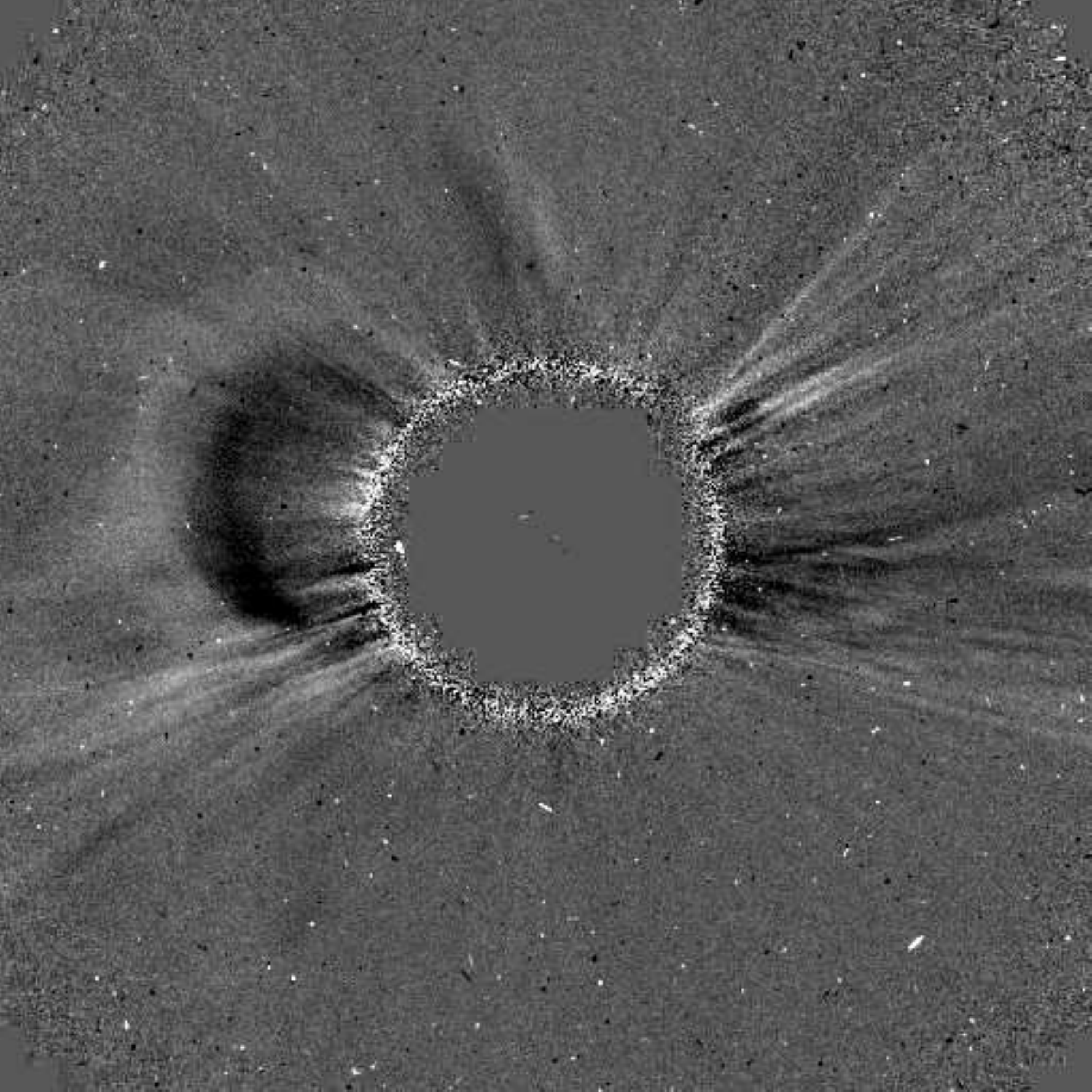}
             \hspace*{-0.02\textwidth}
               \includegraphics[width=0.4\textwidth,clip=]{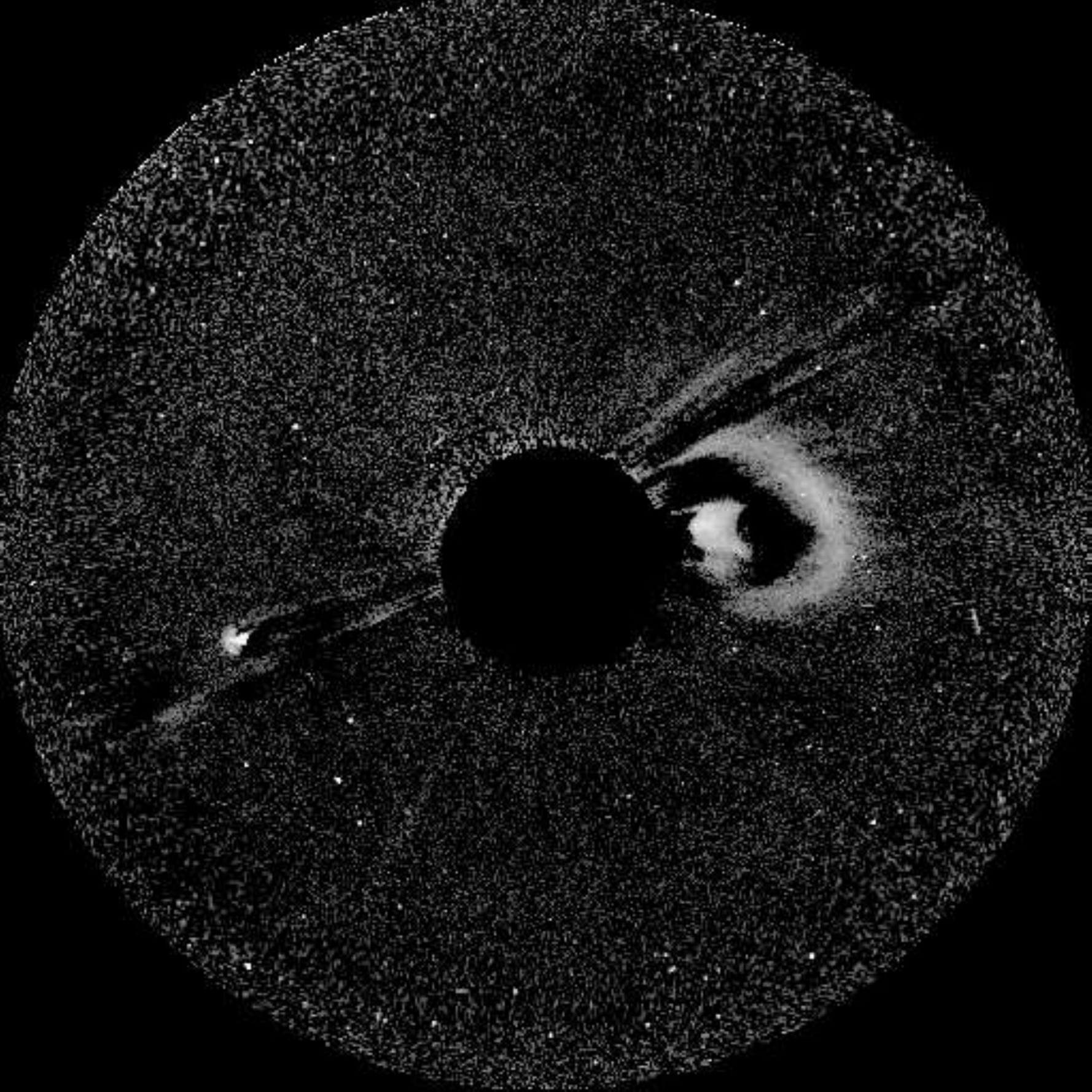}
               }
                 \centerline{\hspace*{0.04\textwidth}
              \includegraphics[width=0.4\textwidth,clip=]{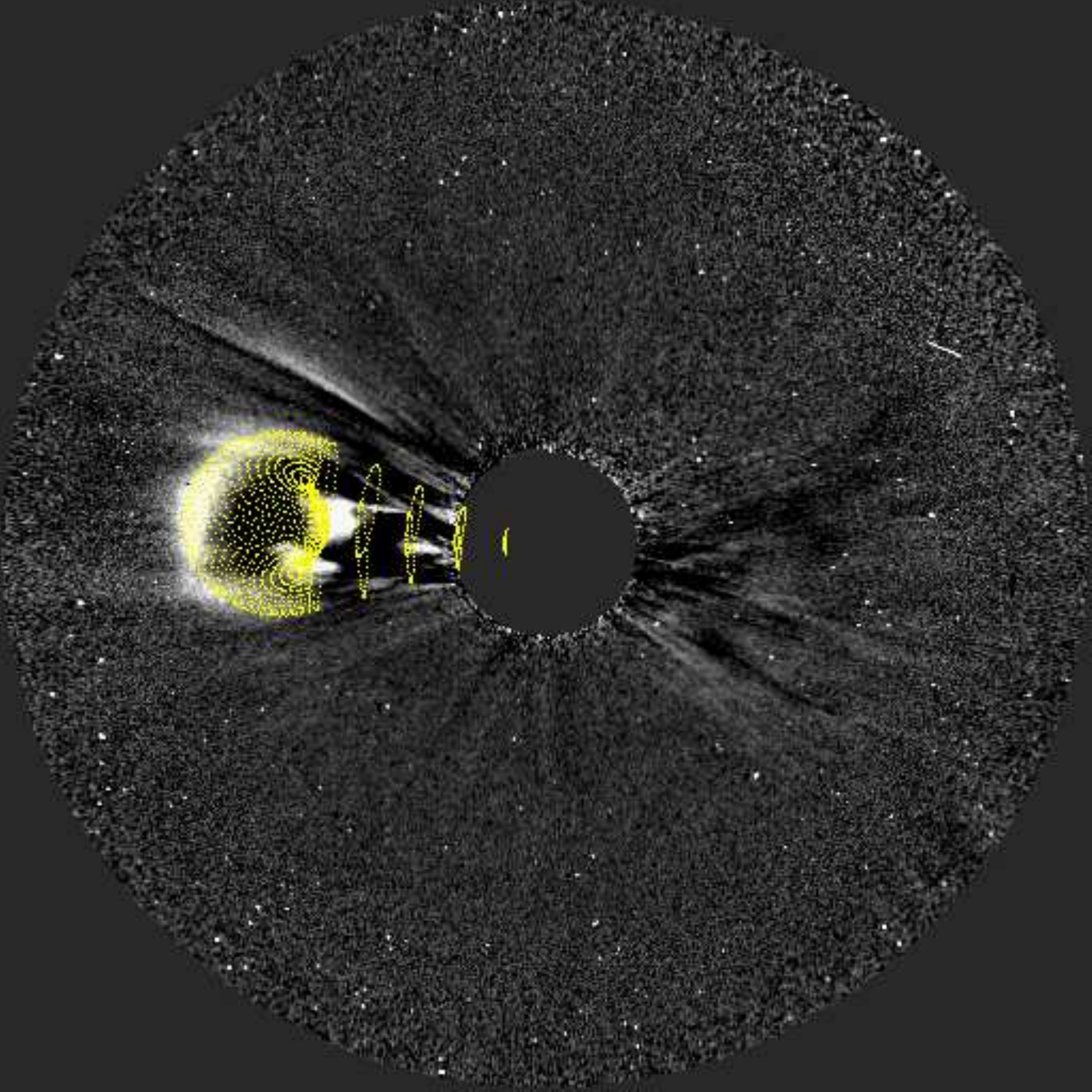}
               \hspace*{-0.02\textwidth}
               \includegraphics[width=0.4\textwidth,clip=]{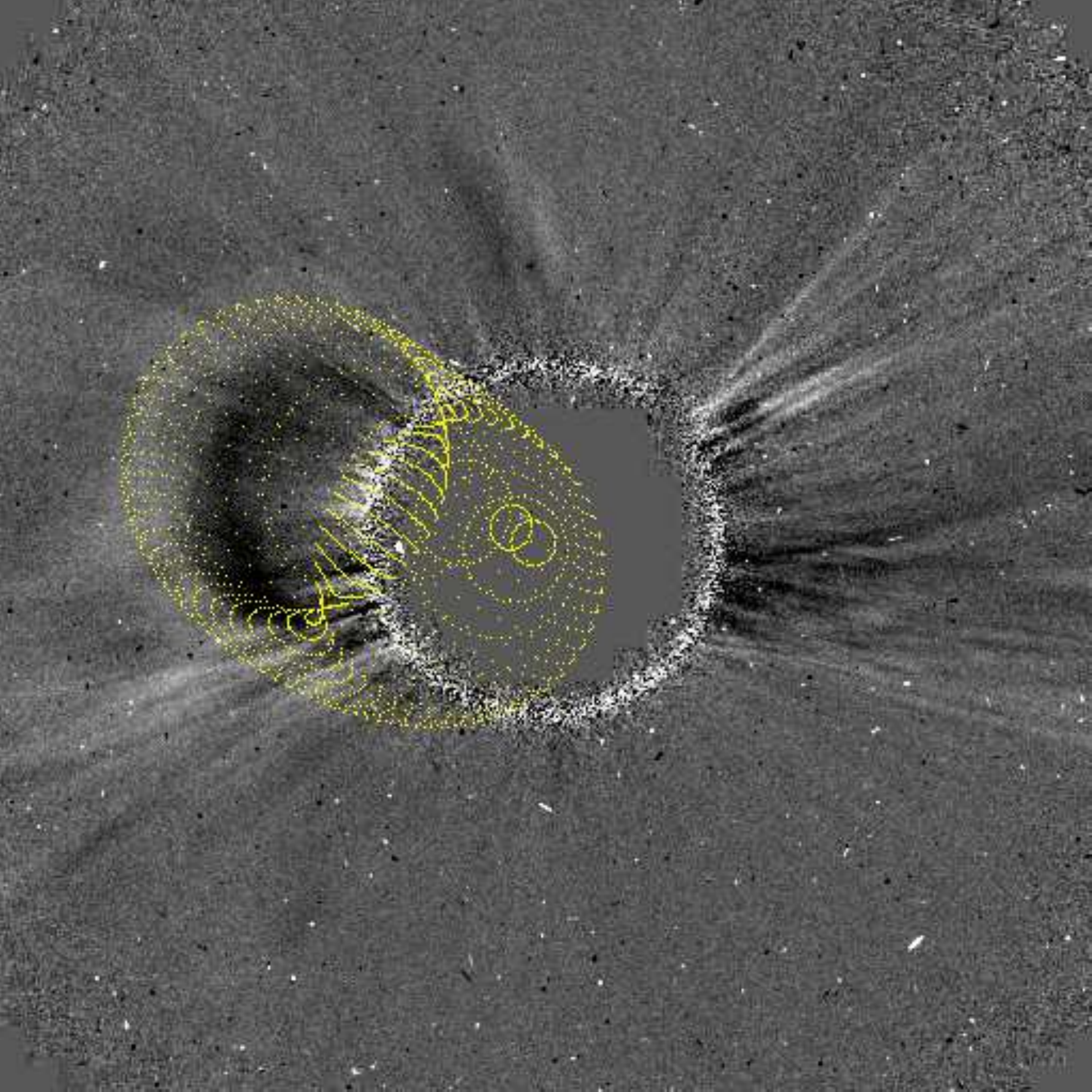}
              \hspace*{-0.02\textwidth}
               \includegraphics[width=0.4\textwidth,clip=]{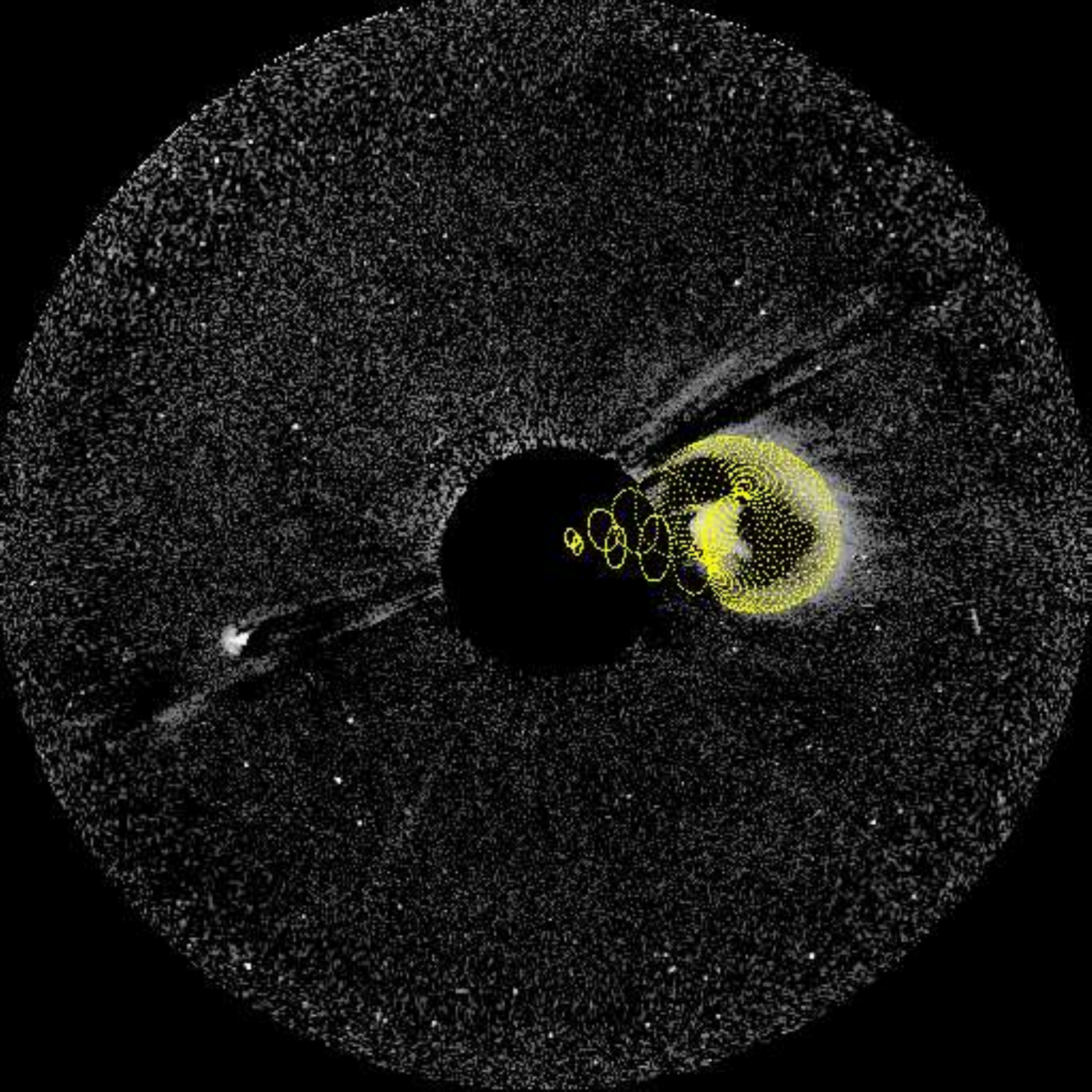}
                }
\vspace{0.0261\textwidth}  
\caption[GCS fit for CME 4 at 18:54]{GCS fit for CME 4 on June 16, 2010 at 18:54 UT at height $H=9.9$ \Rs. Table \ref{tblapp} 
lists the GCS parameters for this event.}
\label{figa4}
\end{figure}

\begin{figure}[h]    
  \centering                              
   \centerline{\hspace*{0.00\textwidth}
               \includegraphics[width=0.4\textwidth,clip=]{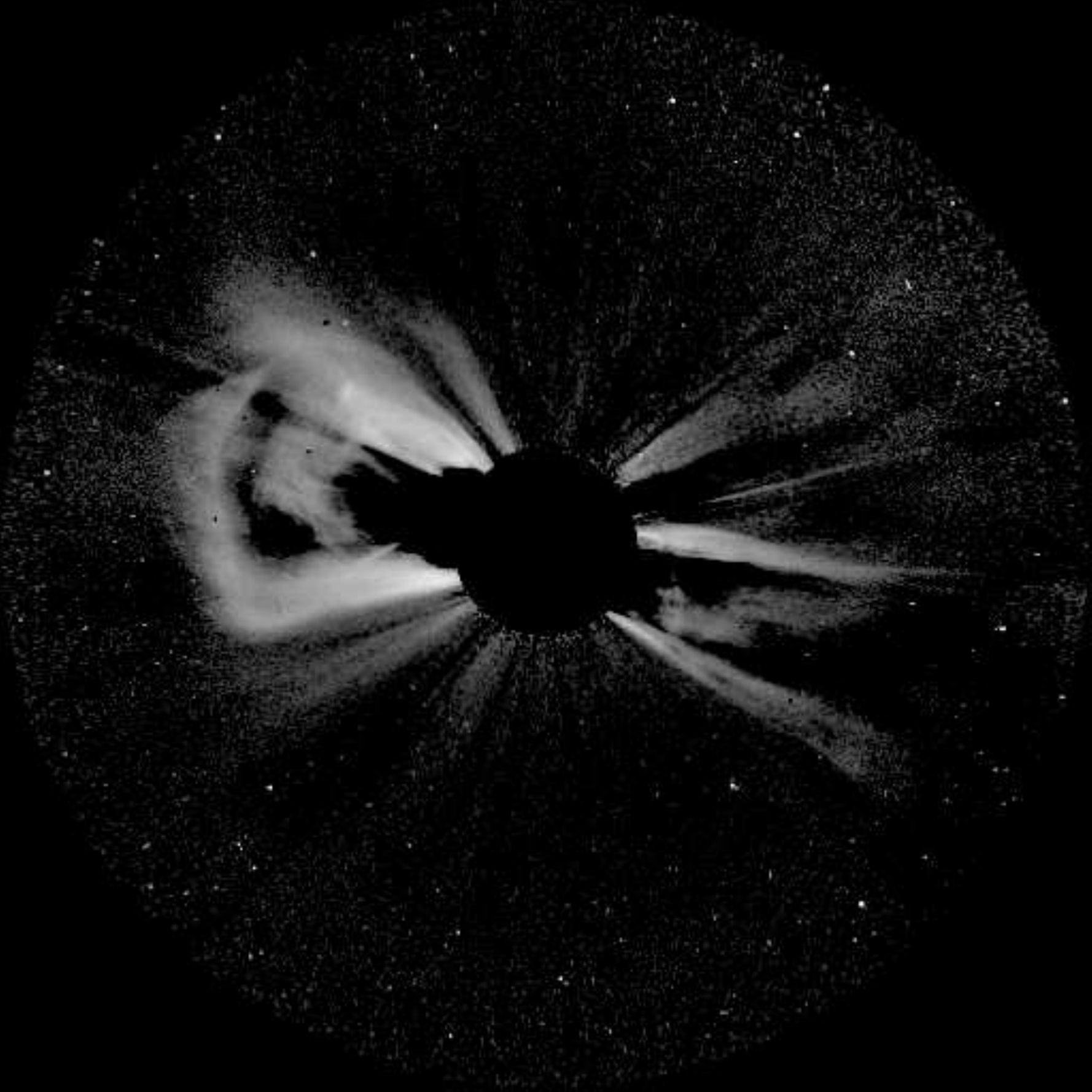}
                \hspace*{-0.02\textwidth}
               \includegraphics[width=0.4\textwidth,clip=]{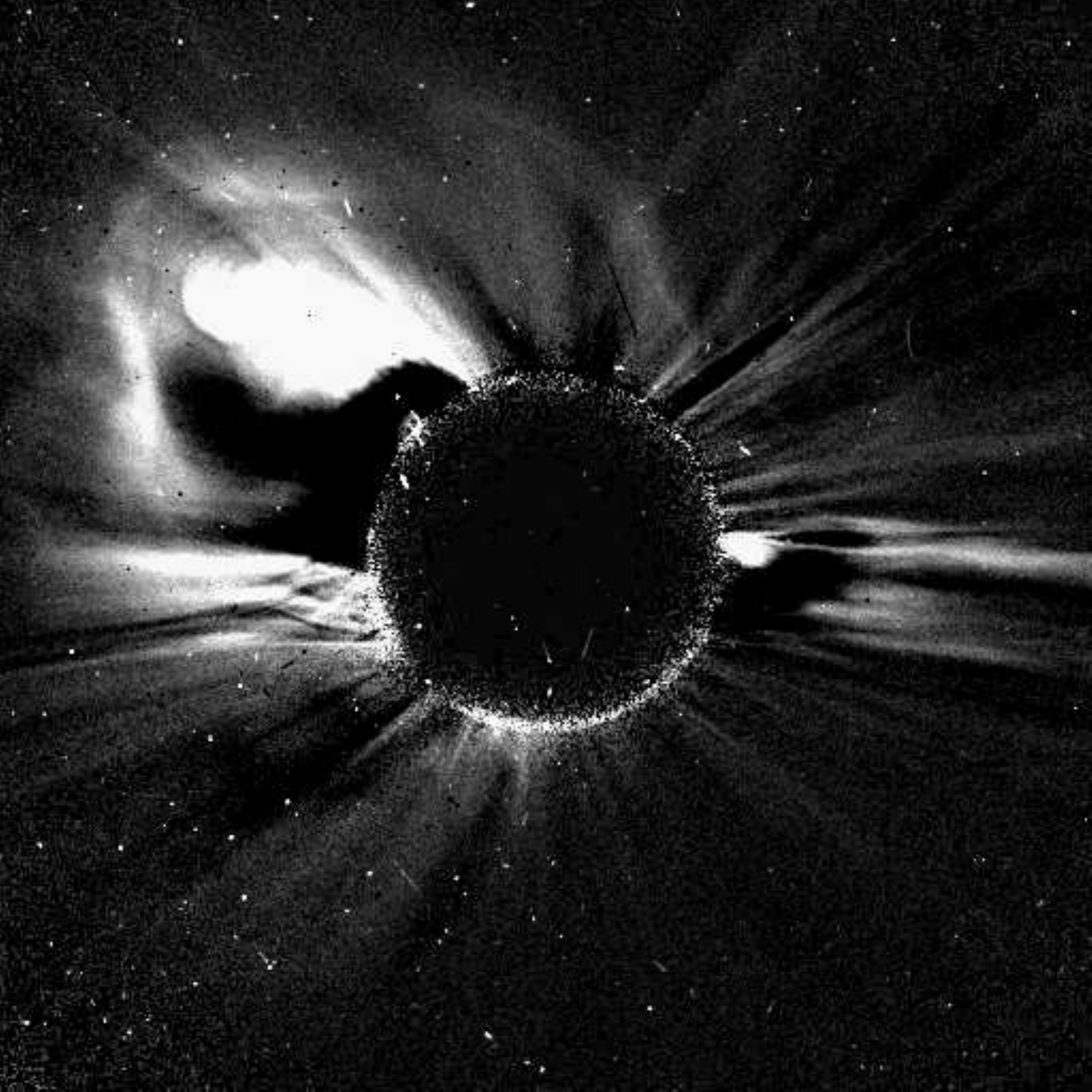}
             \hspace*{-0.02\textwidth}
               \includegraphics[width=0.4\textwidth,clip=]{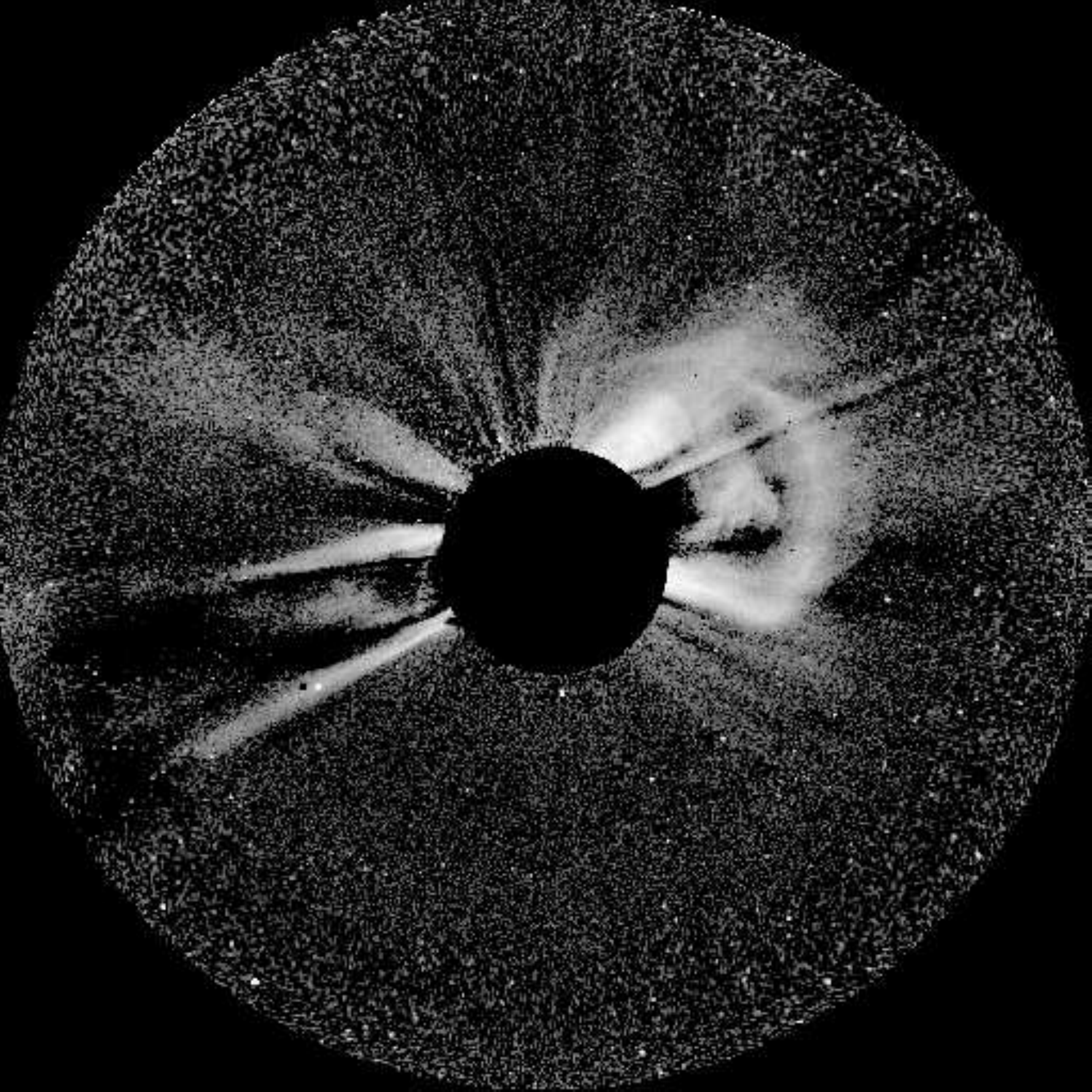}
               }
                 \centerline{\hspace*{0.0\textwidth}
              \includegraphics[width=0.4\textwidth,clip=]{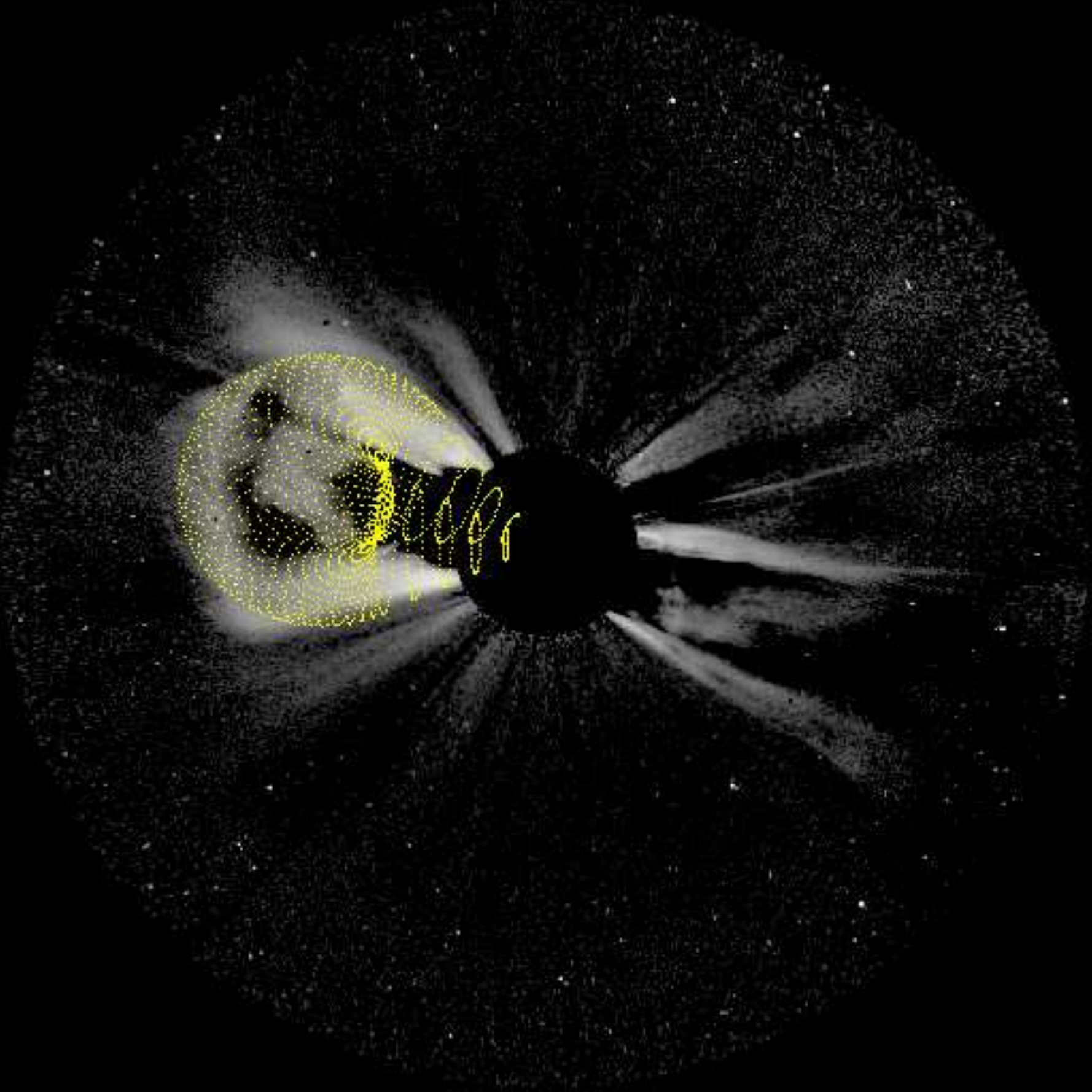}
               \hspace*{-0.02\textwidth}
               \includegraphics[width=0.4\textwidth,clip=]{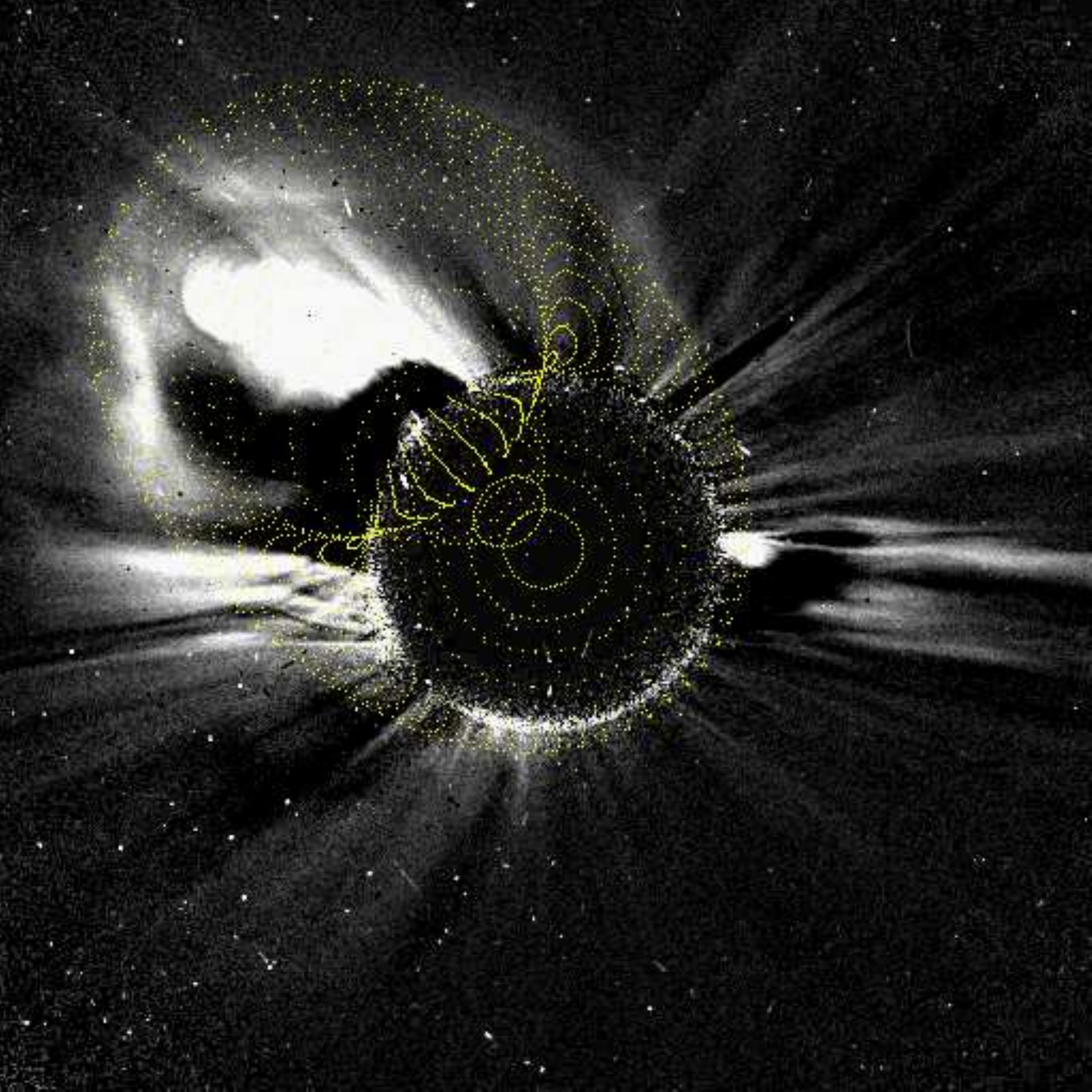}
              \hspace*{-0.02\textwidth}
               \includegraphics[width=0.4\textwidth,clip=]{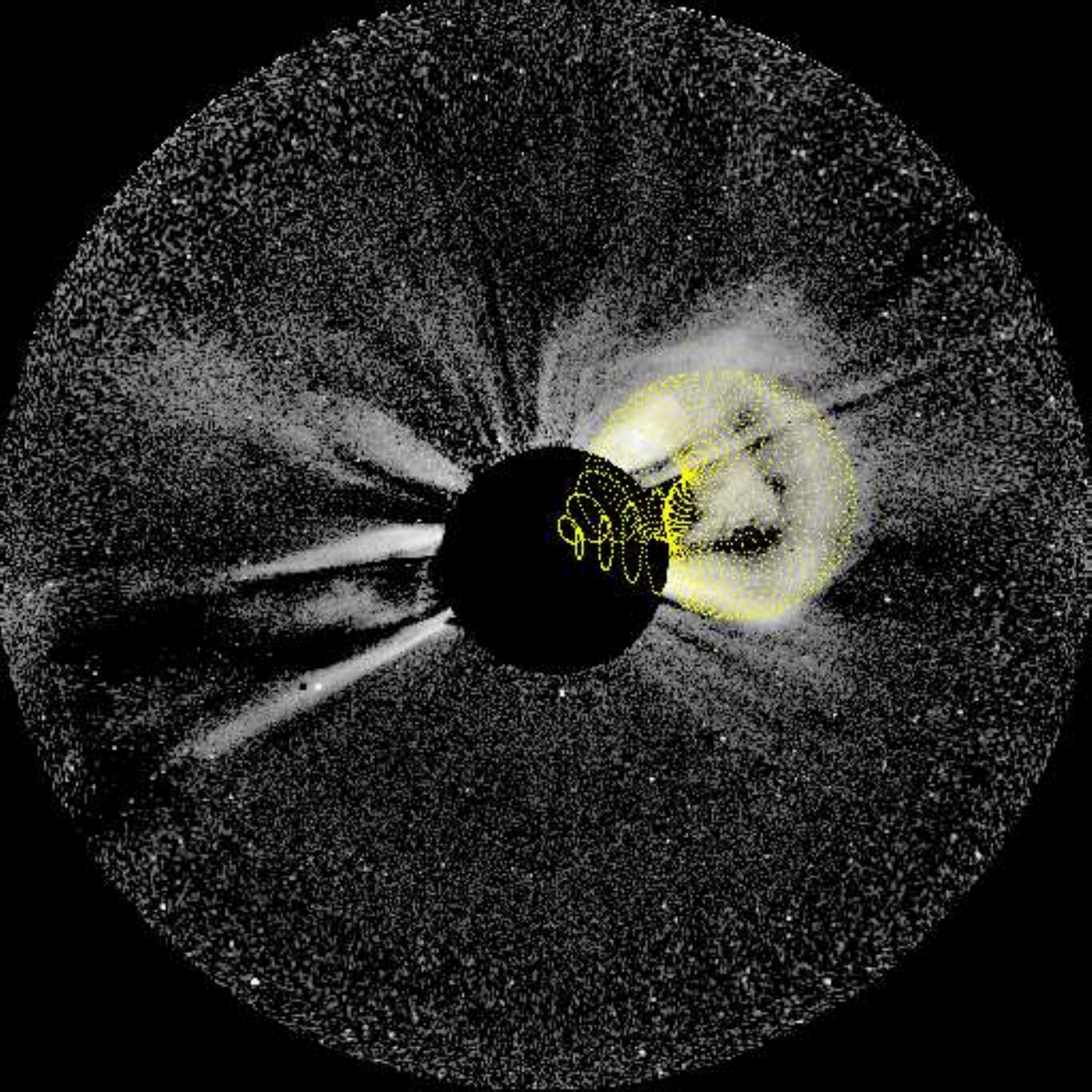}
                }
\vspace{0.0261\textwidth}  
\caption[GCS fit for CME 5 at 05:39]{GCS fit for CME 5 on September 11, 2010 at 05:39 UT at height $H=10.3$ \Rs. Table \ref{tblapp} 
lists the GCS parameters for this event.}
\label{figa5}
\end{figure}

\begin{figure}[h]    
  \centering                              
   \centerline{\hspace*{0.04\textwidth}
               \includegraphics[width=1.18\textwidth,clip=]{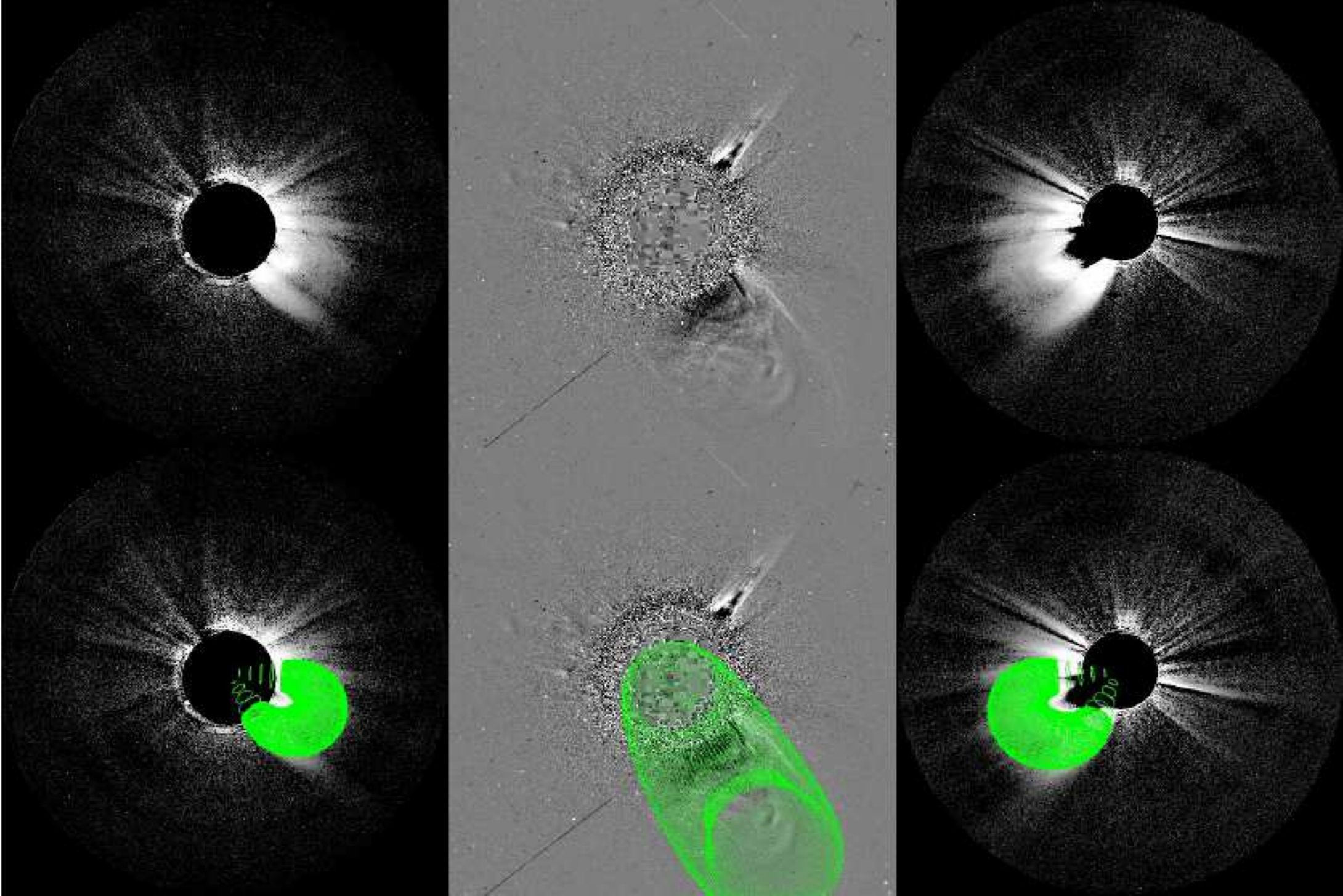}
                \hspace*{-0.02\textwidth}
                }
\vspace{0.0261\textwidth}  
\caption[GCS fit for CME 6 at 11:54]{GCS fit for CME 6 on October 26, 2010  at 11:54 at height $H=10.0$ \Rs. Table \ref{tblapp} 
lists the GCS parameters for this event. This figure is adapted from \citet{Col12} (Appendix).}
\label{figa6}
\end{figure}

\begin{figure}[h]    
  \centering                              
   \centerline{\hspace*{0.00\textwidth}
               \includegraphics[width=0.4\textwidth,clip=]{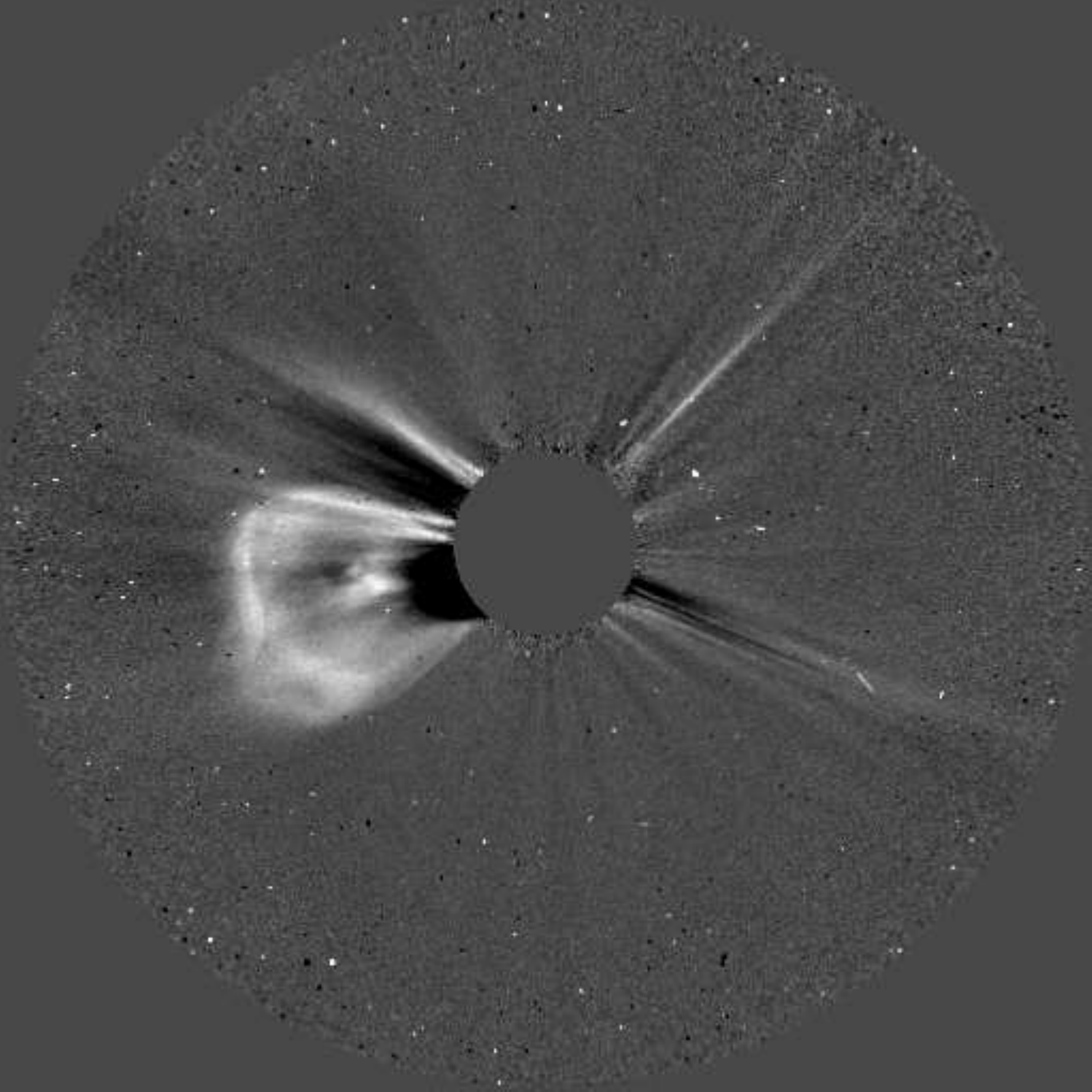}
                \hspace*{-0.02\textwidth}
               \includegraphics[width=0.4\textwidth,clip=]{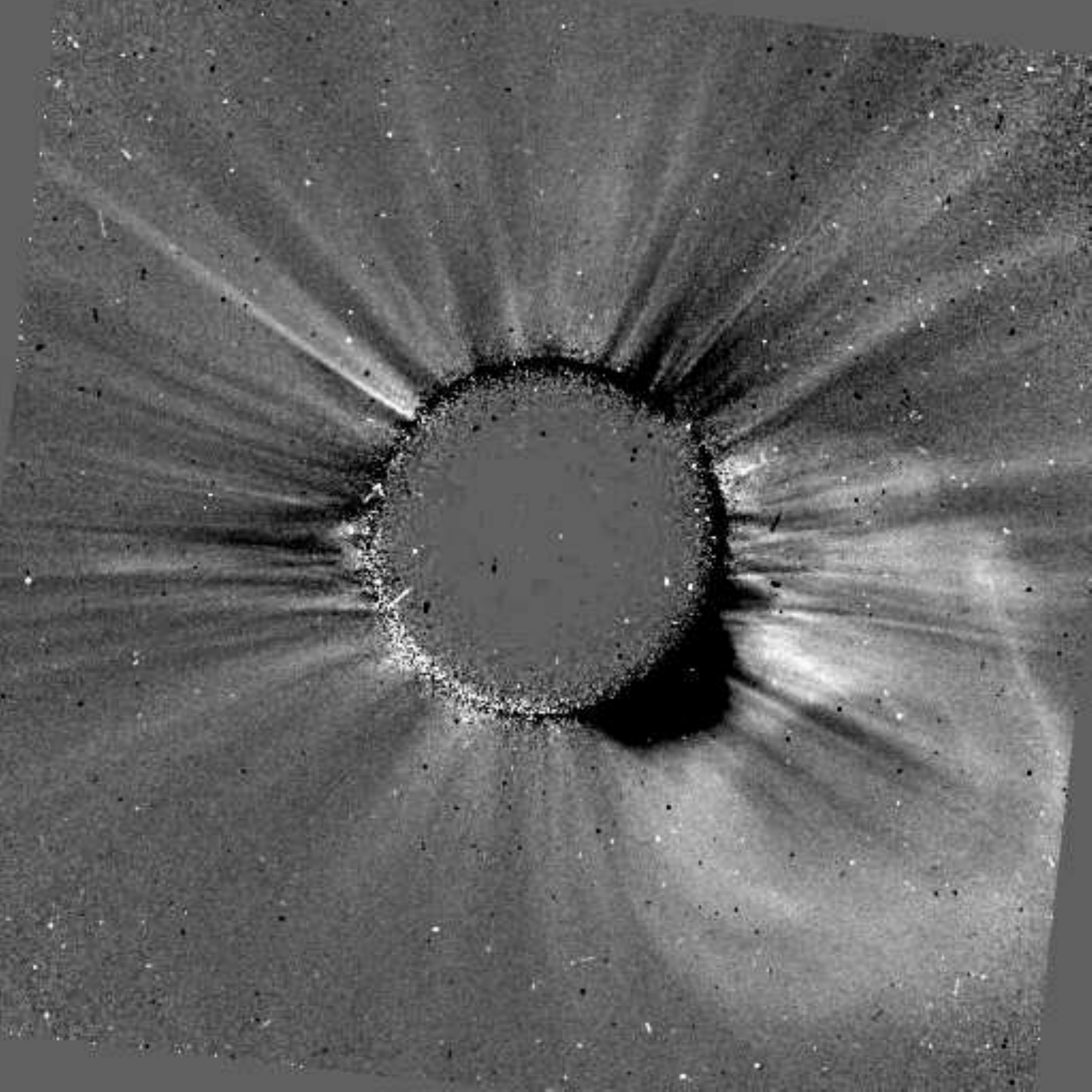}
             \hspace*{-0.02\textwidth}
               \includegraphics[width=0.4\textwidth,clip=]{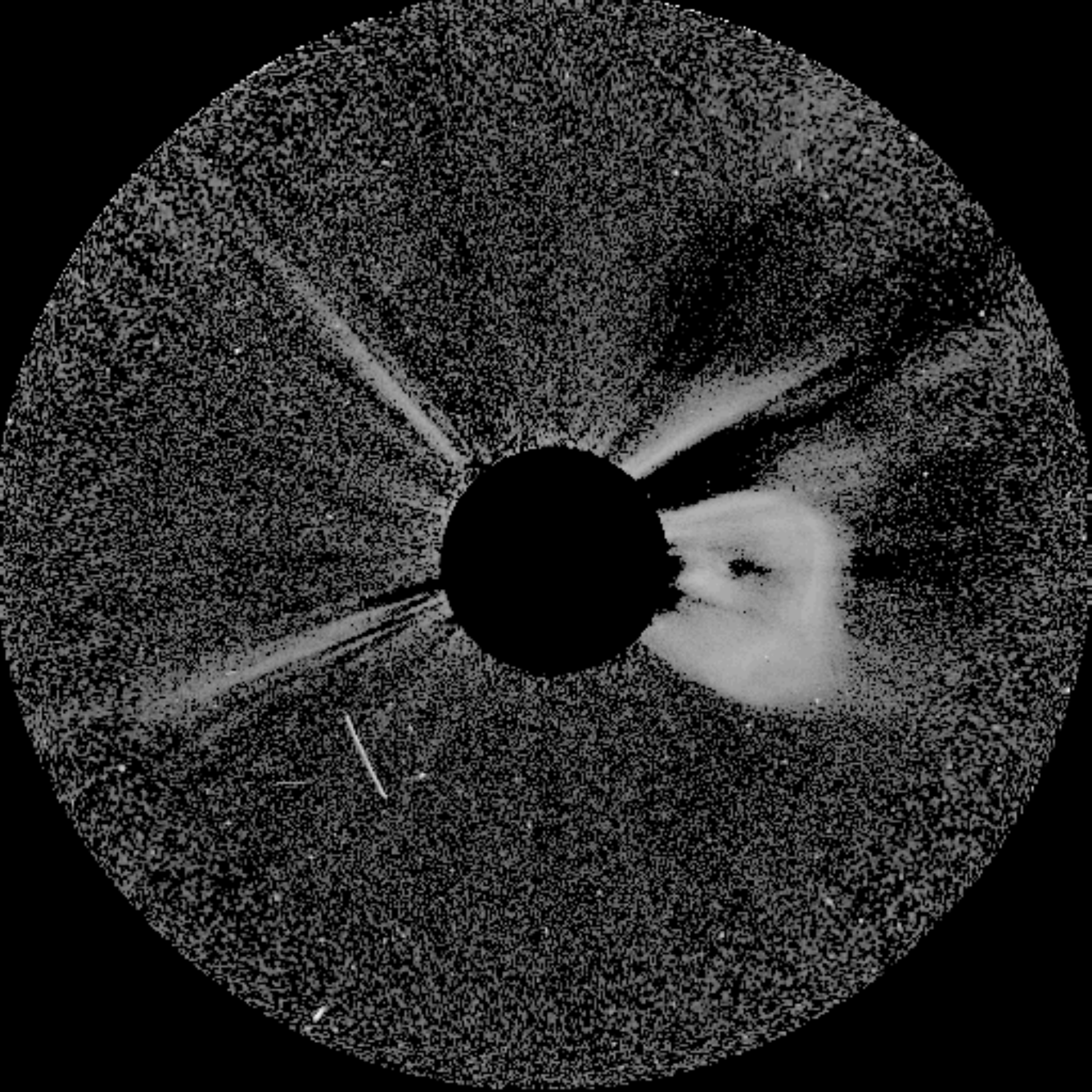}
               }
                 \centerline{\hspace*{0.0\textwidth}
              \includegraphics[width=0.4\textwidth,clip=]{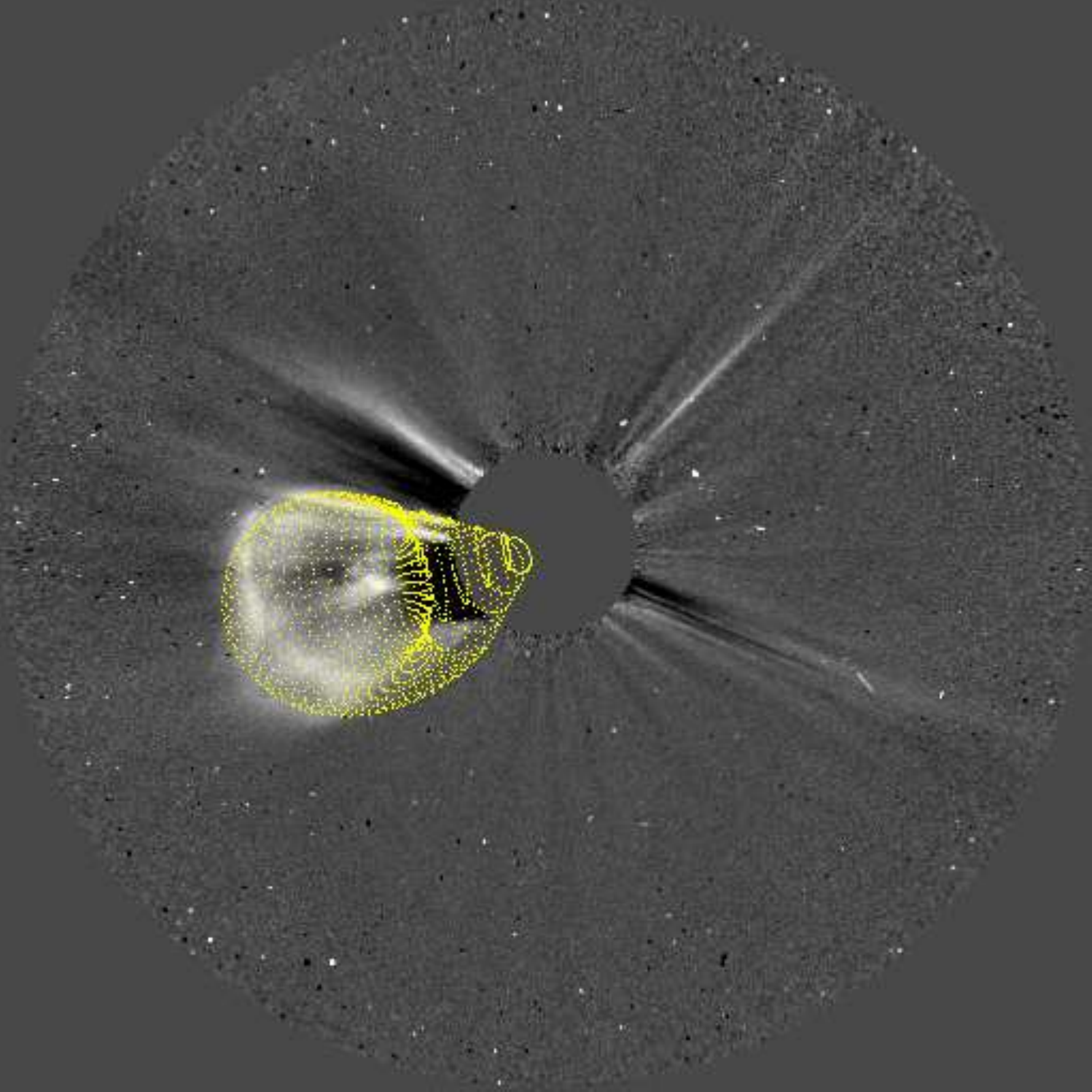}
               \hspace*{-0.02\textwidth}
               \includegraphics[width=0.4\textwidth,clip=]{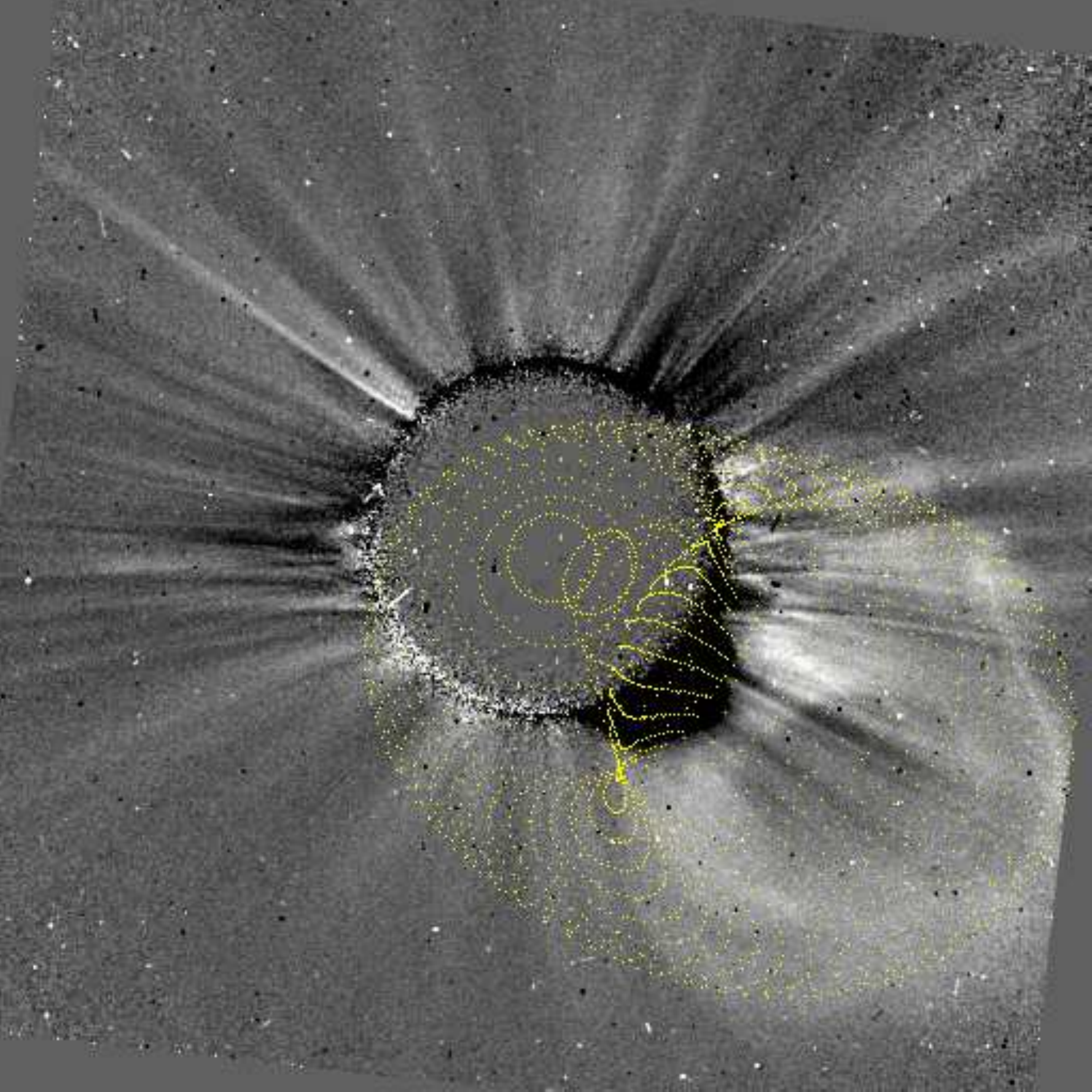}
              \hspace*{-0.02\textwidth}
               \includegraphics[width=0.4\textwidth,clip=]{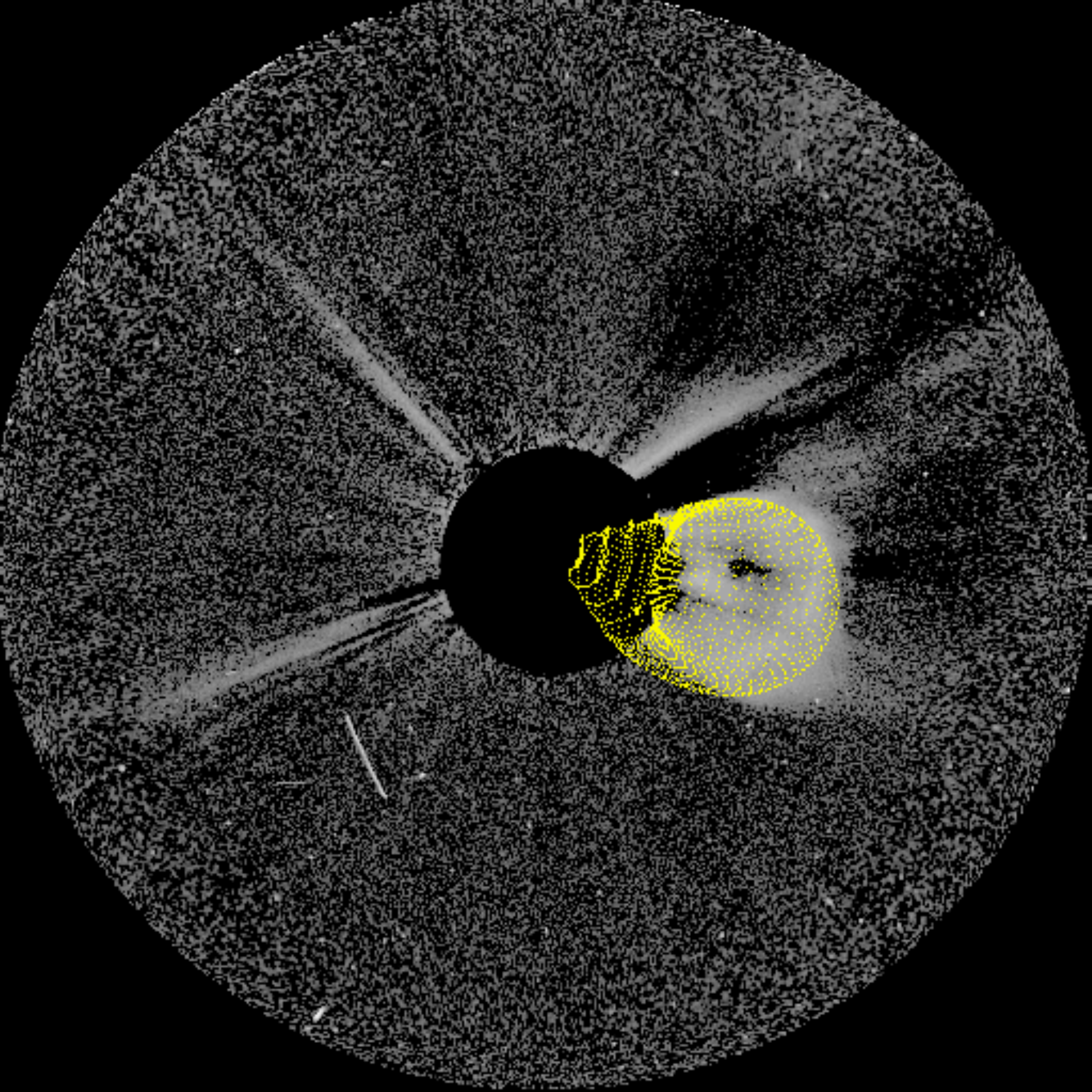}
                }
\vspace{0.0261\textwidth}  
\caption[GCS fit for CME 7 at 10:24]{GCS fit for CME 7 on December 23, 2010 at 10:24 UT at height $H=9.5$ \Rs. Table \ref{tblapp} 
lists the GCS parameters for this event.}
\label{figa7}
\end{figure}

\begin{figure}[h]    
  \centering                              
   \centerline{\hspace*{0.04\textwidth}
               \includegraphics[width=0.4\textwidth,clip=]{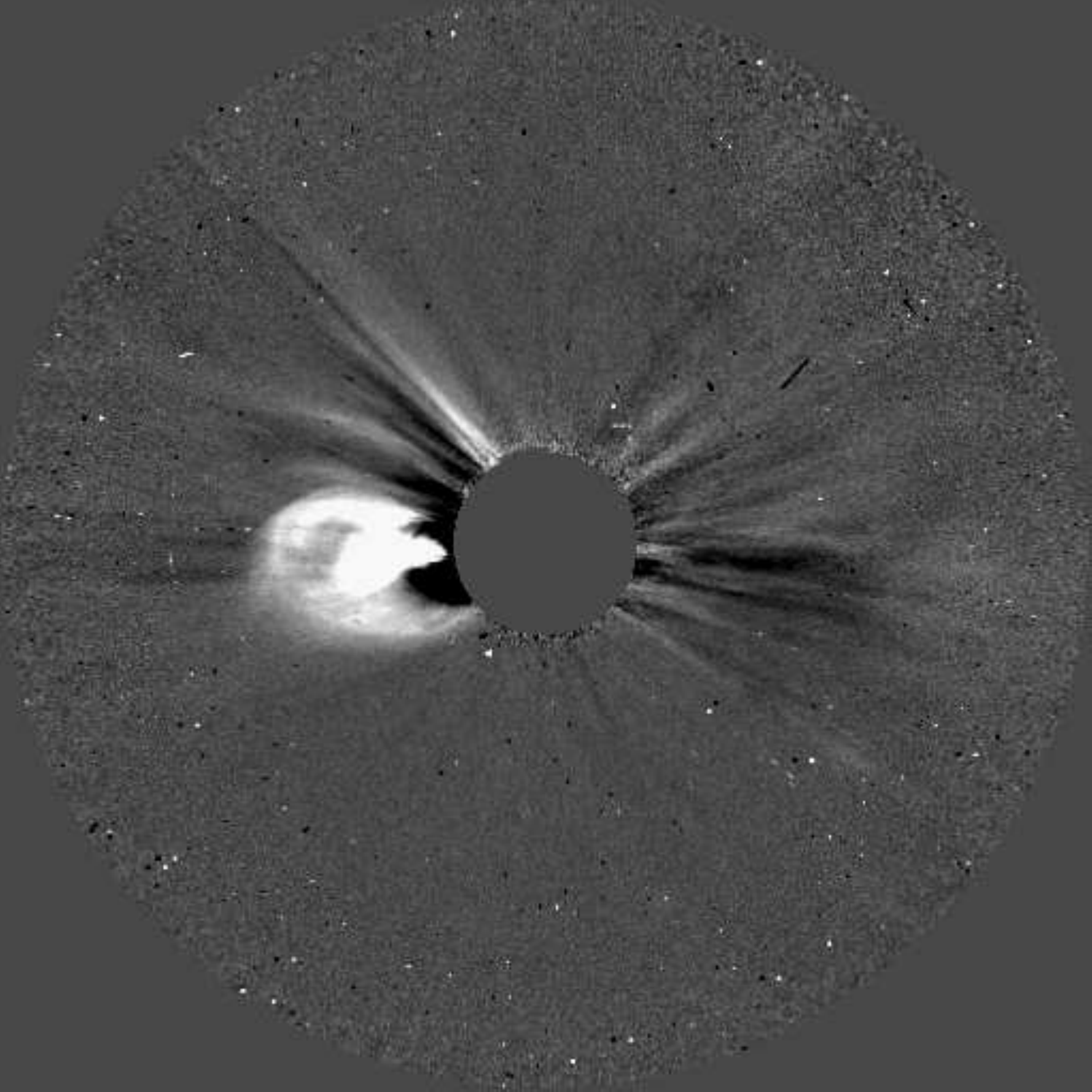}
                \hspace*{-0.02\textwidth}
               \includegraphics[width=0.4\textwidth,clip=]{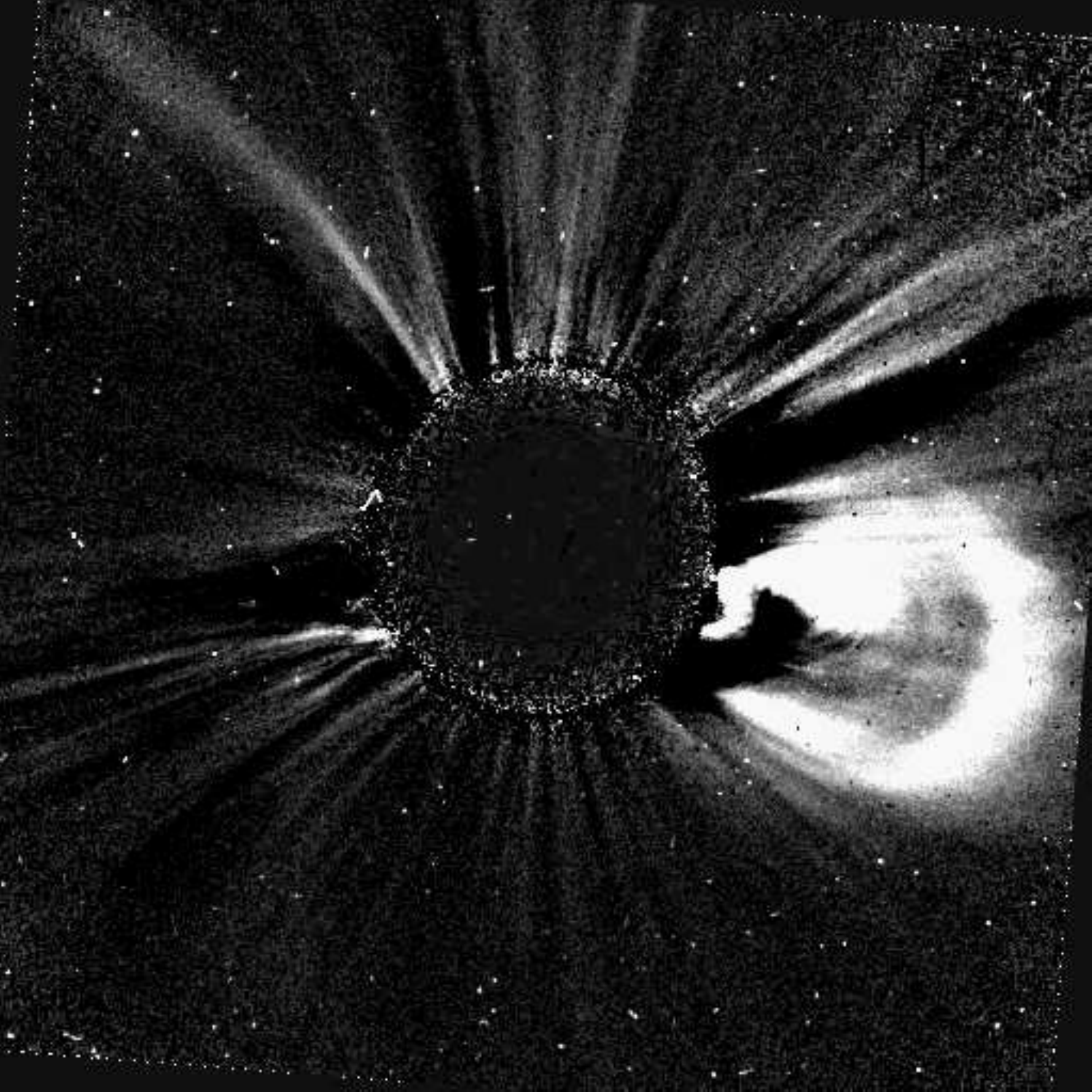}
             \hspace*{-0.02\textwidth}
               \includegraphics[width=0.4\textwidth,clip=]{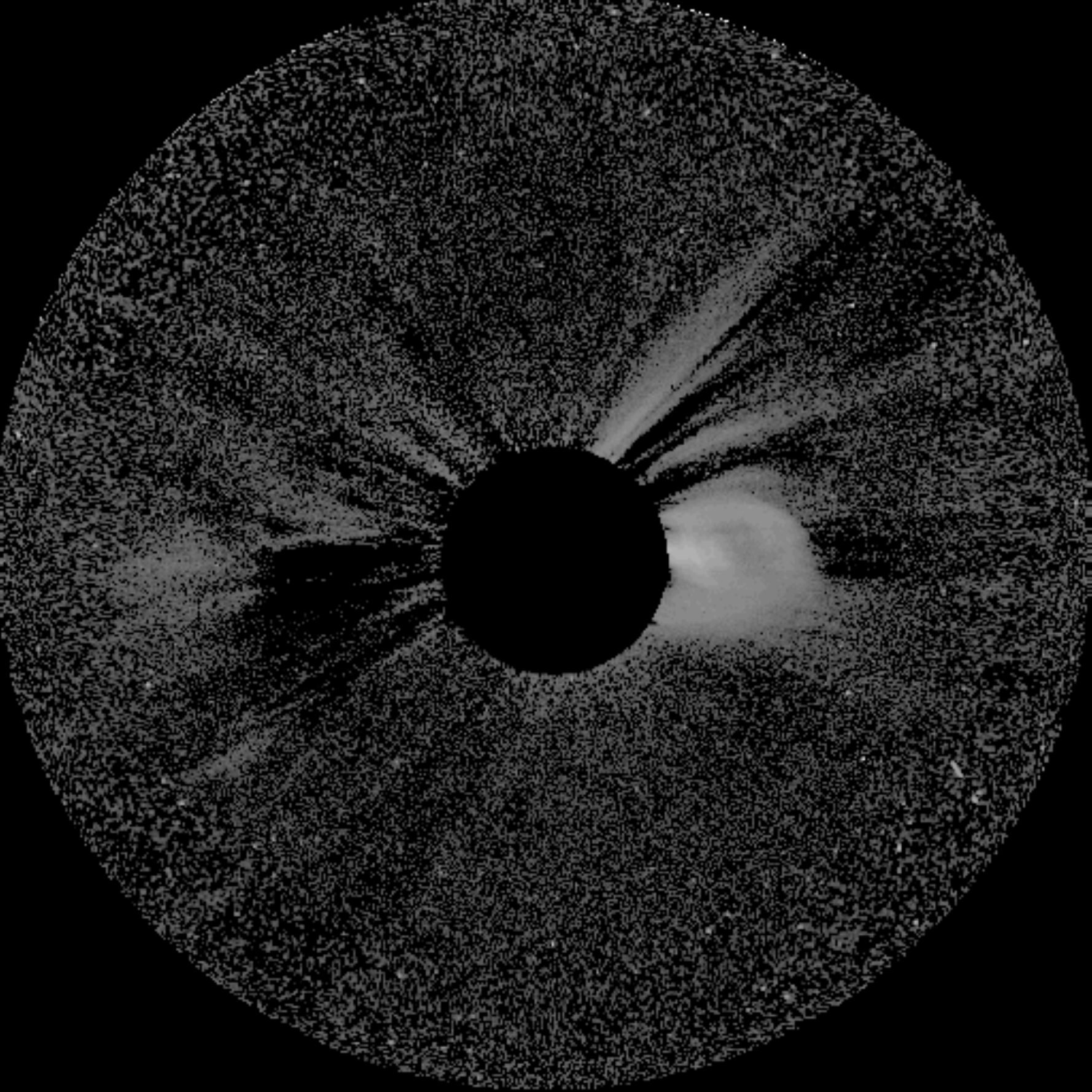}
               }
                 \centerline{\hspace*{0.04\textwidth}
              \includegraphics[width=0.4\textwidth,clip=]{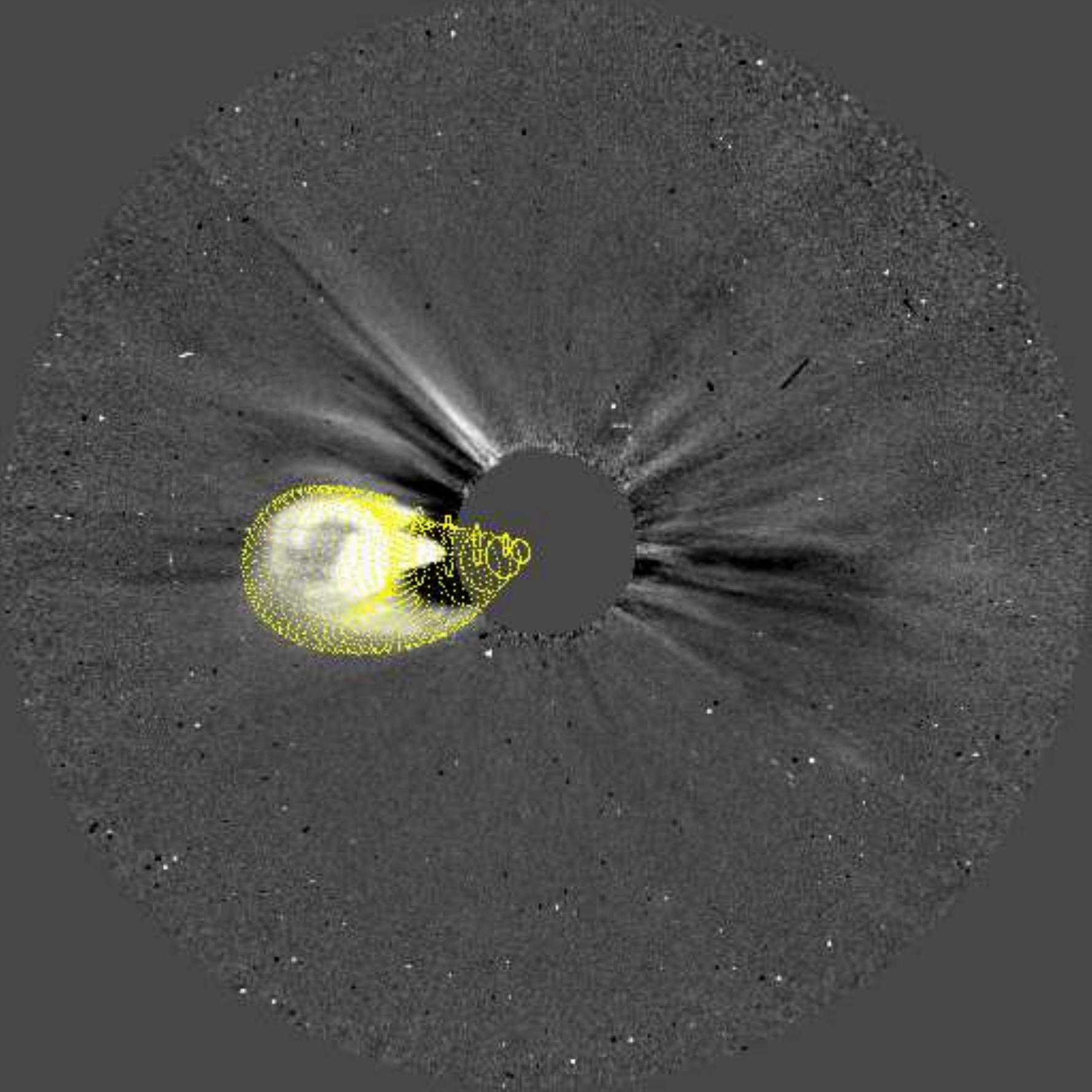}
               \hspace*{-0.02\textwidth}
               \includegraphics[width=0.4\textwidth,clip=]{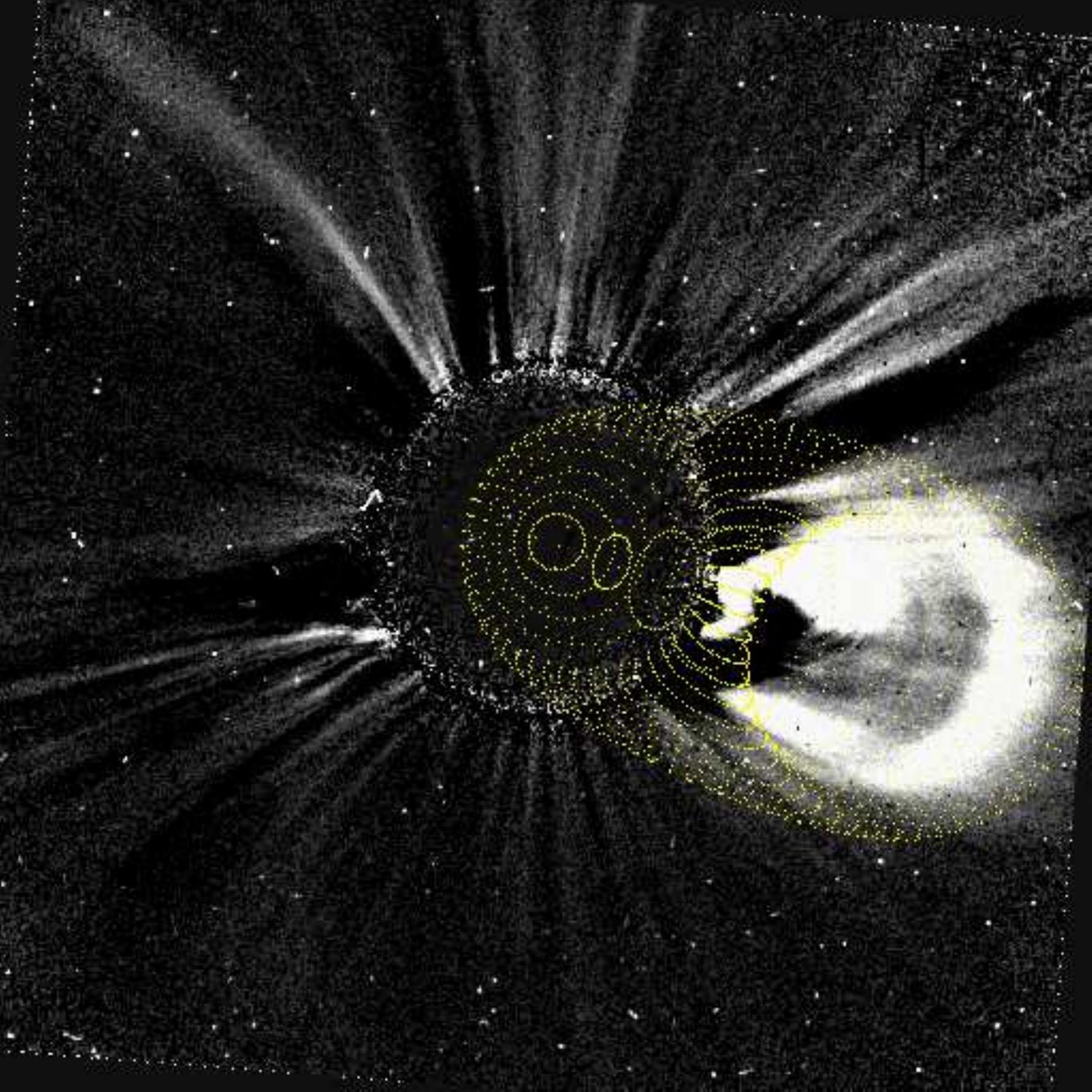}
              \hspace*{-0.02\textwidth}
               \includegraphics[width=0.4\textwidth,clip=]{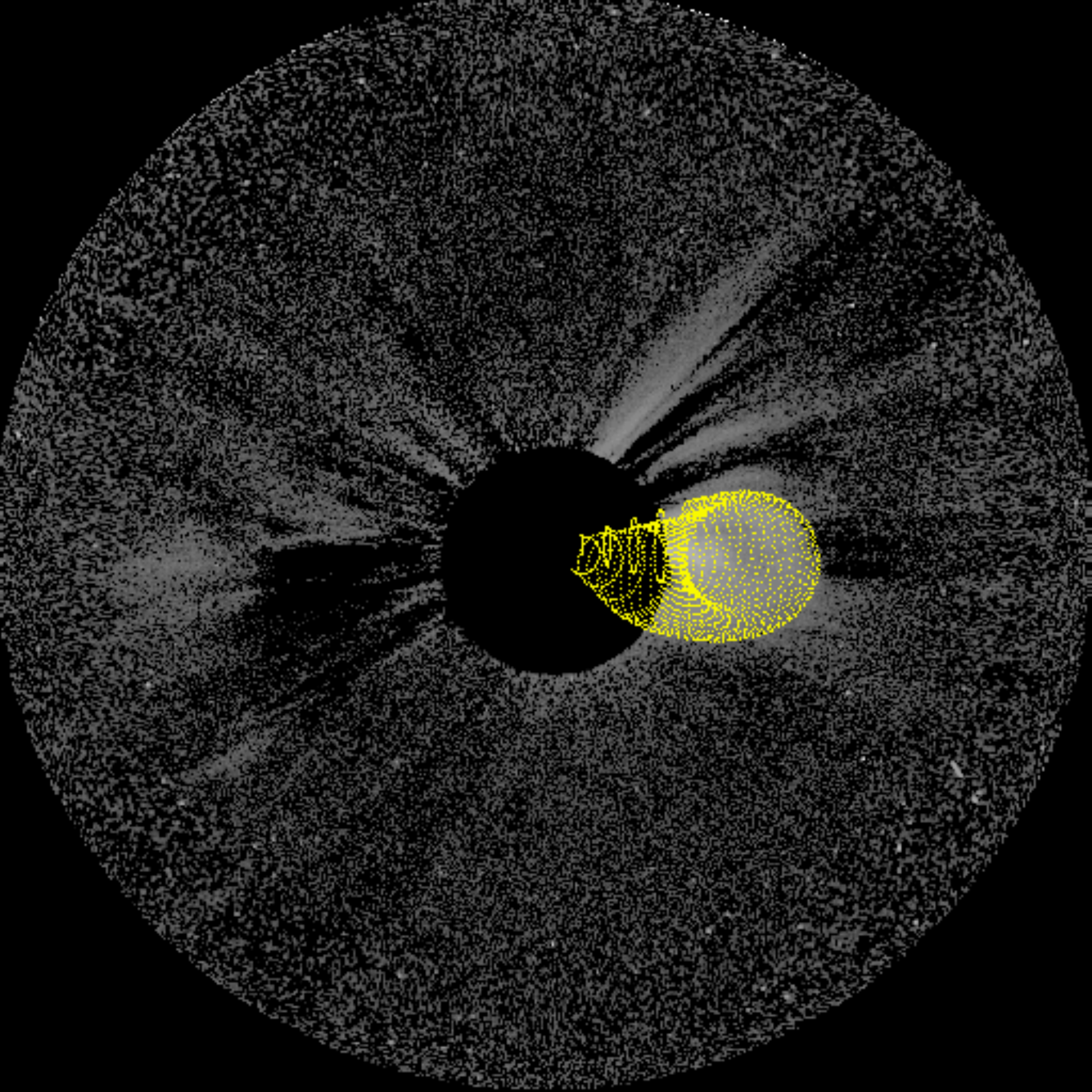}
                }
\vspace{0.0261\textwidth}  
\caption[GCS fit for CME 8 at 06:39]{GCS fit for CME 8 on January 24, 2011 at 06:39 UT at height $H=8.9$ \Rs. Table \ref{tblapp} 
lists the GCS parameters for this event.}
\label{figa8}
\end{figure}

\begin{figure}[h]    
  \centering                              
   \centerline{\hspace*{0.00\textwidth}
               \includegraphics[width=0.4\textwidth,clip=]{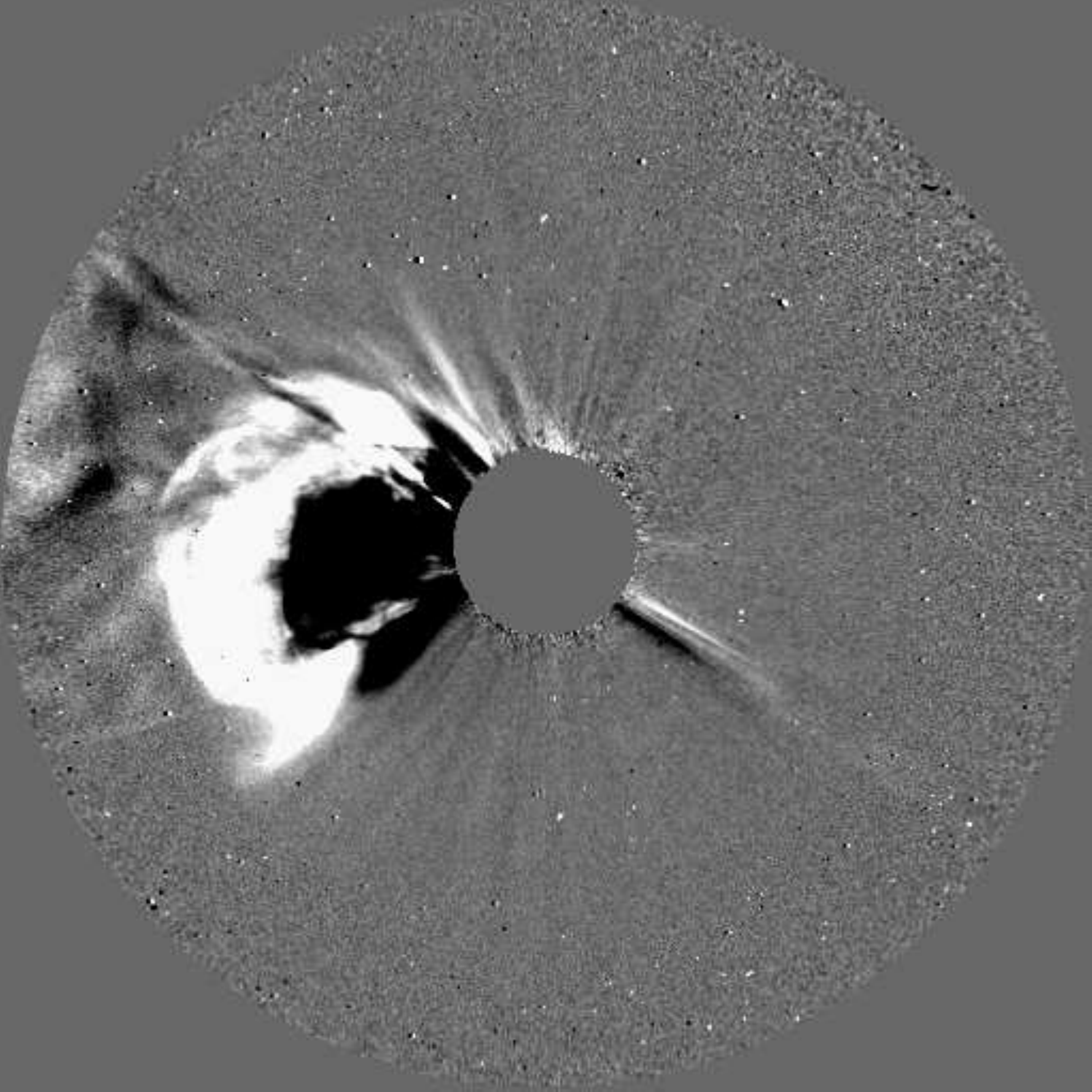}
                \hspace*{-0.02\textwidth}
               \includegraphics[width=0.4\textwidth,clip=]{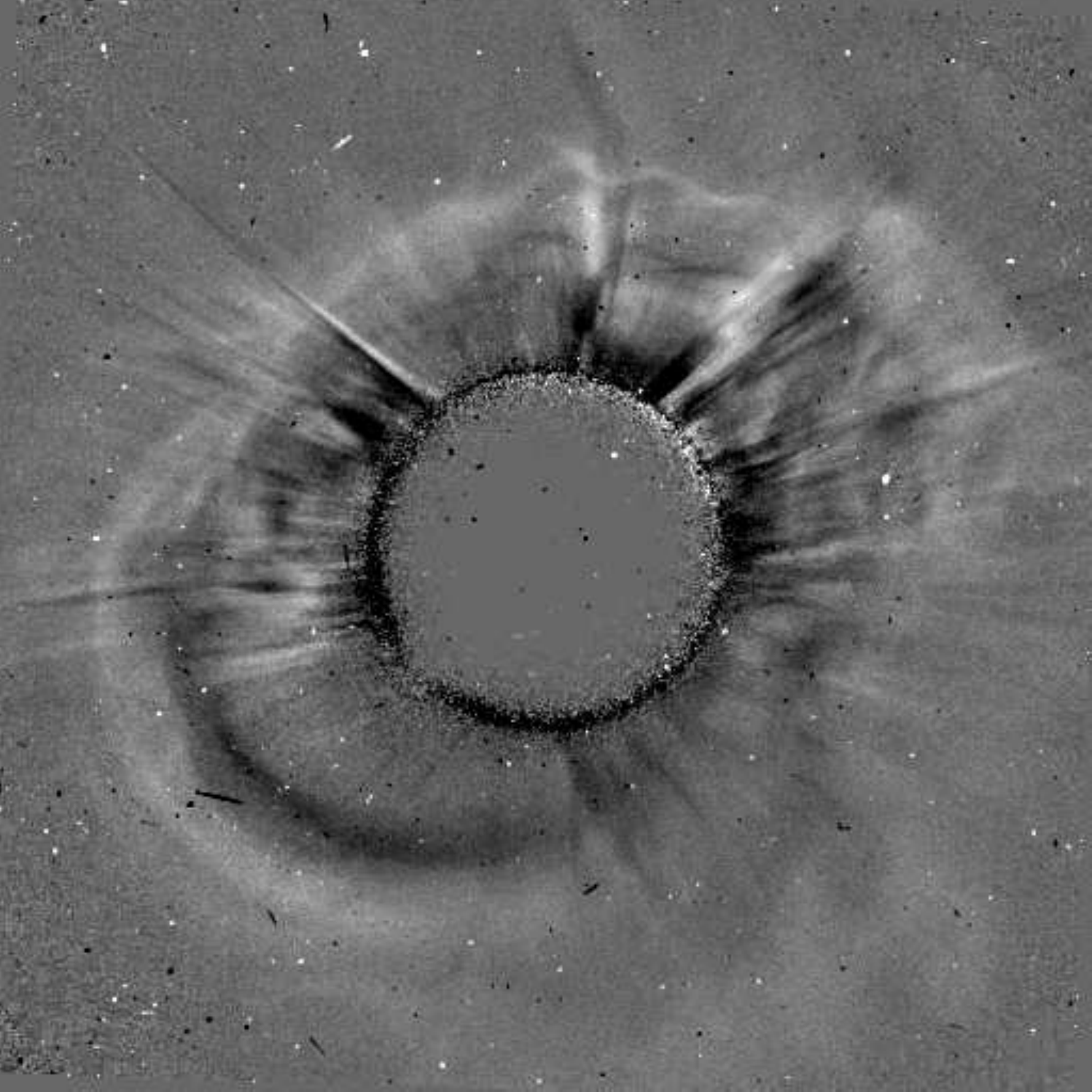}
             \hspace*{-0.02\textwidth}
               \includegraphics[width=0.4\textwidth,clip=]{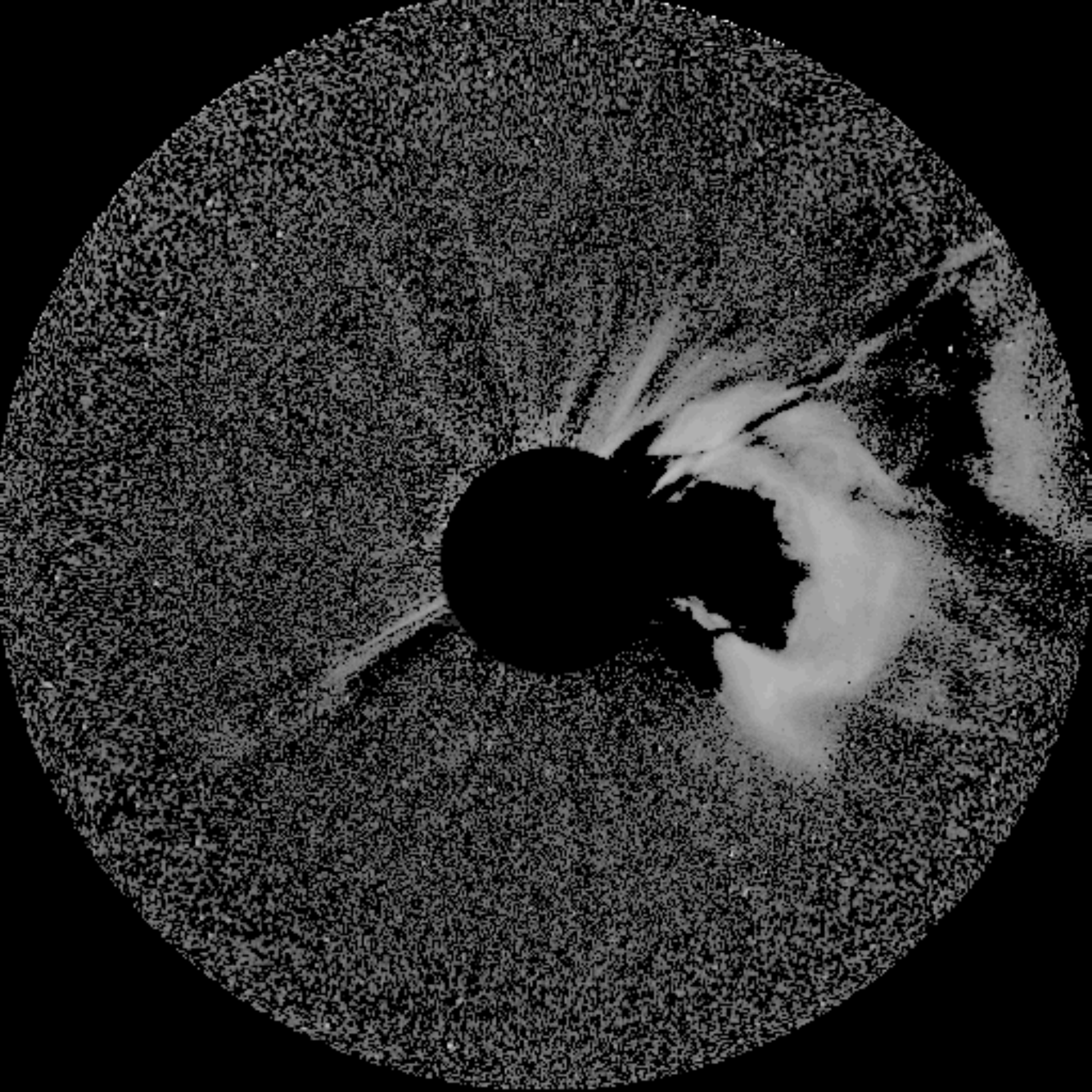}
               }
                 \centerline{\hspace*{0.0\textwidth}
              \includegraphics[width=0.4\textwidth,clip=]{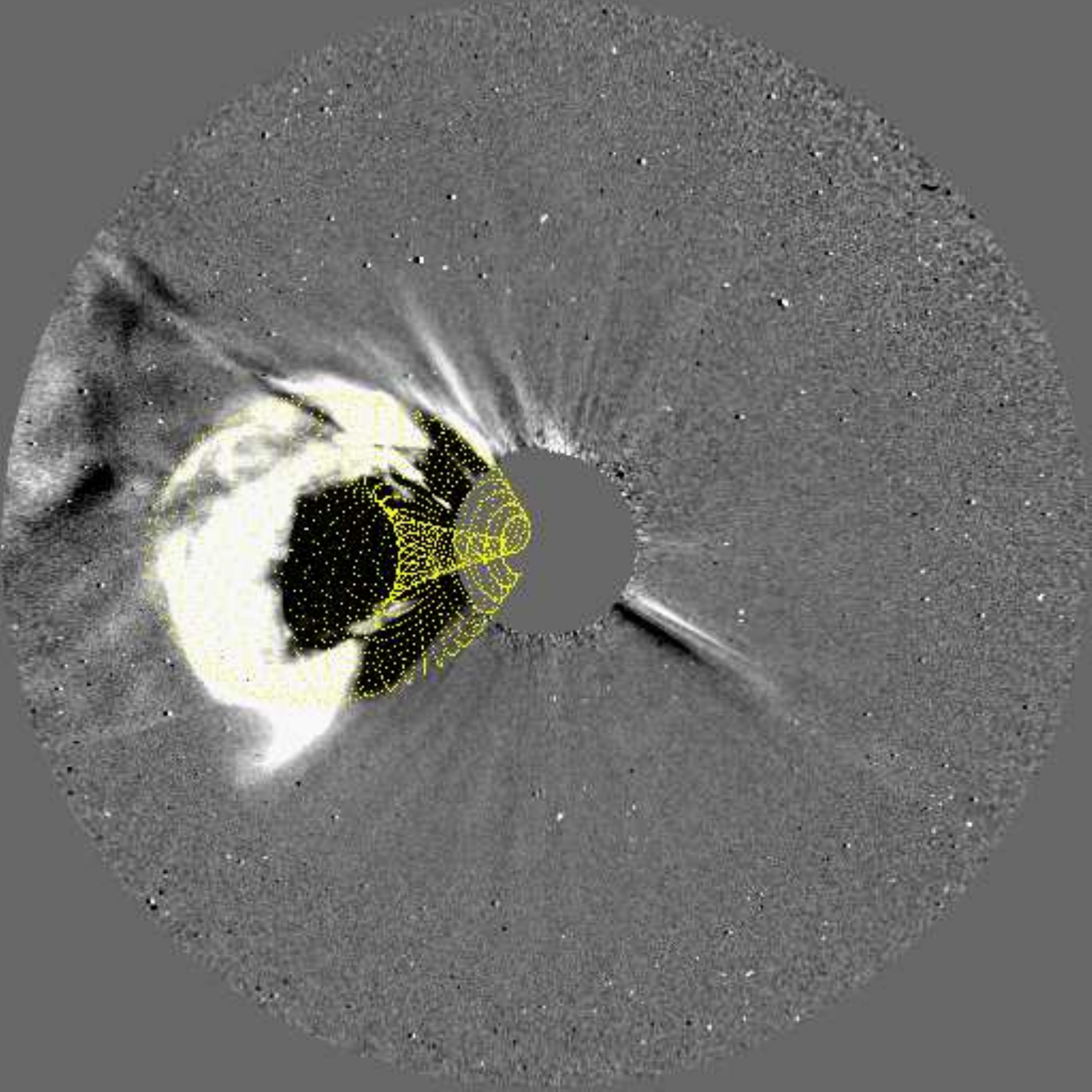}
               \hspace*{-0.02\textwidth}
               \includegraphics[width=0.4\textwidth,clip=]{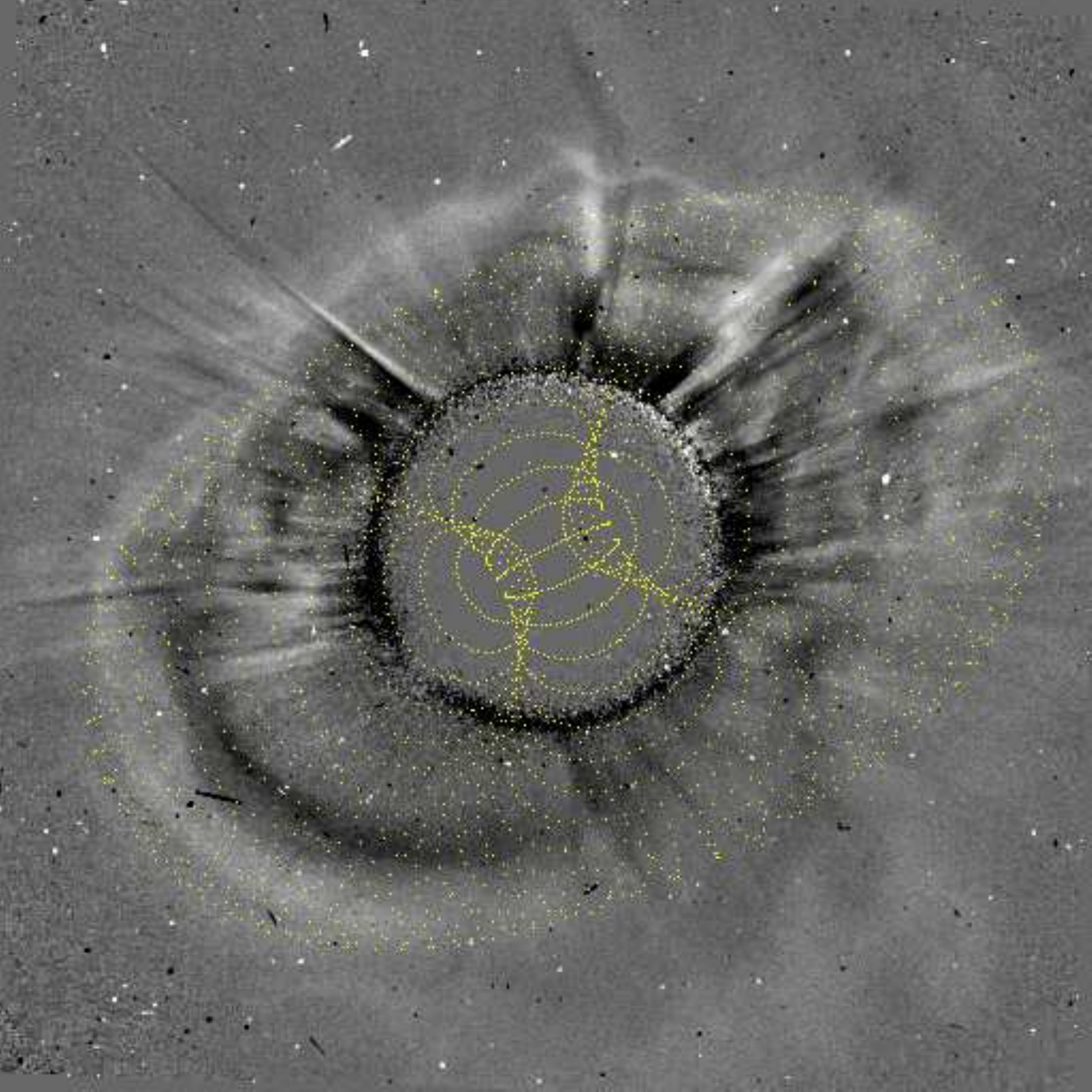}
              \hspace*{-0.02\textwidth}
               \includegraphics[width=0.4\textwidth,clip=]{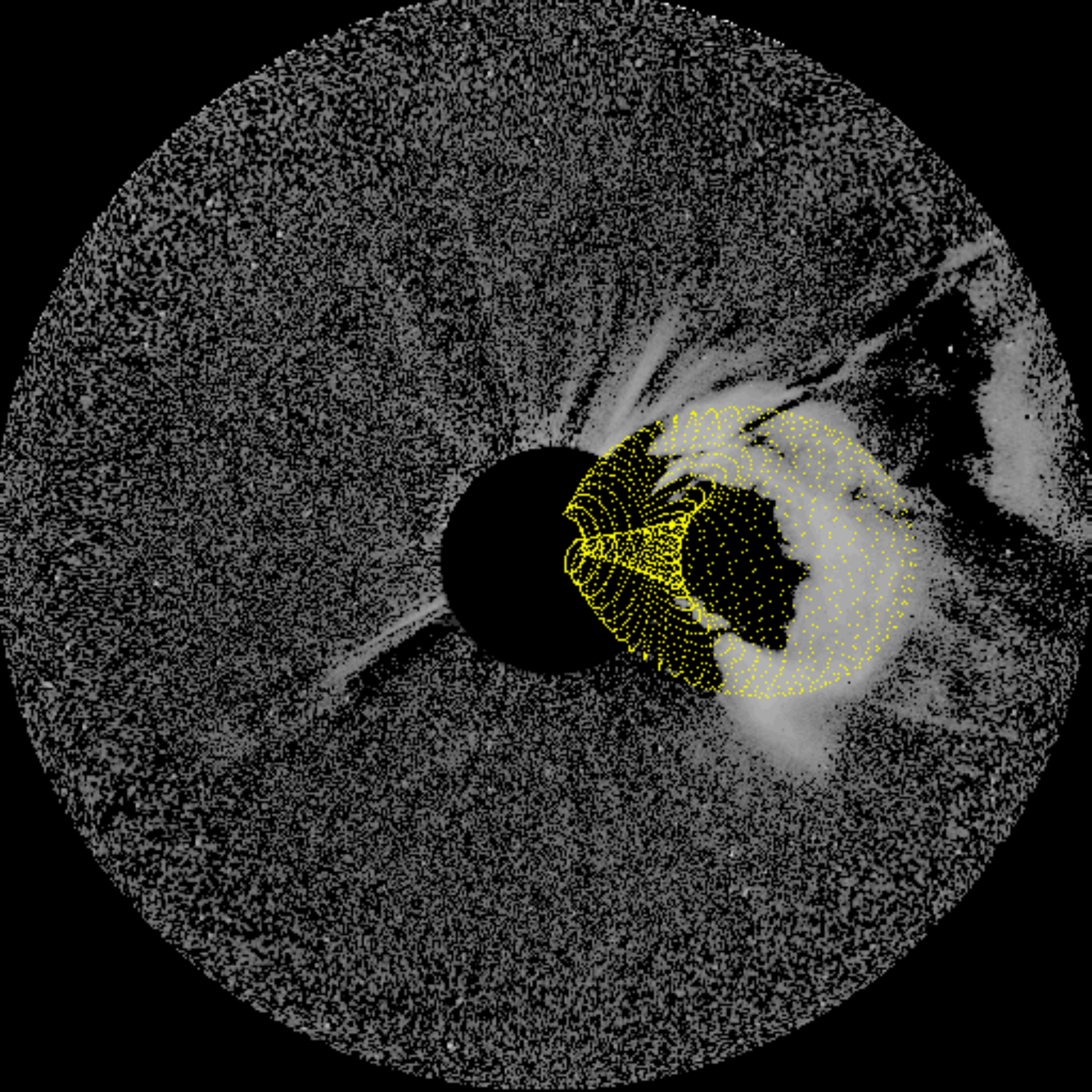}
                }
\vspace{0.0261\textwidth}  
\caption[GCS fit for CME 9 at 03:39]{GCS fit for CME 9 on February 15, 2011 at 03:39 UT at height $H=10.9$ \Rs. Table \ref{tblapp} 
lists the GCS parameters for this event.}
\label{figa9}
\end{figure}

\begin{figure}[h]    
  \centering                              
   \centerline{\hspace*{0.04\textwidth}
               \includegraphics[width=0.4\textwidth,clip=]{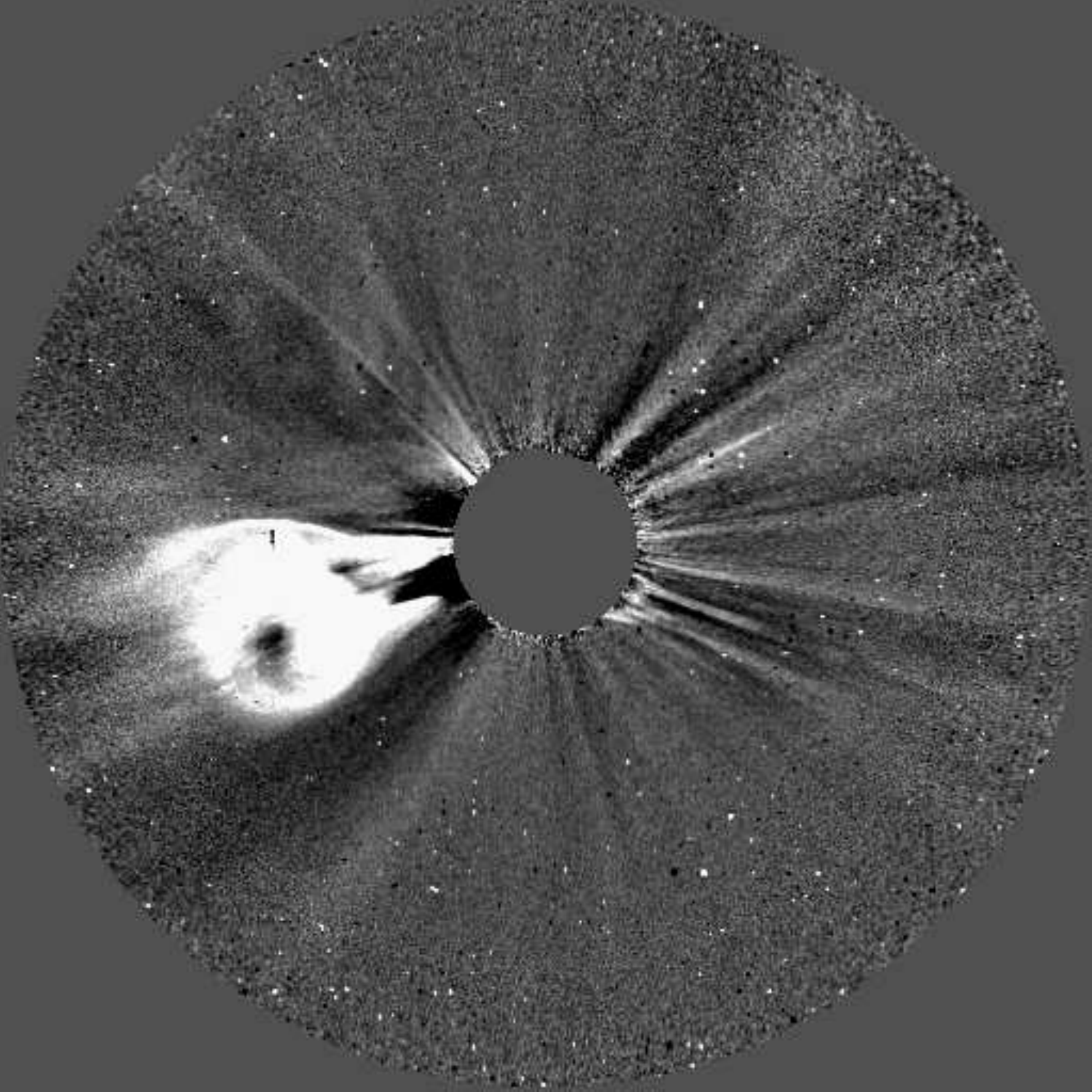}
                \hspace*{-0.02\textwidth}
               \includegraphics[width=0.4\textwidth,clip=]{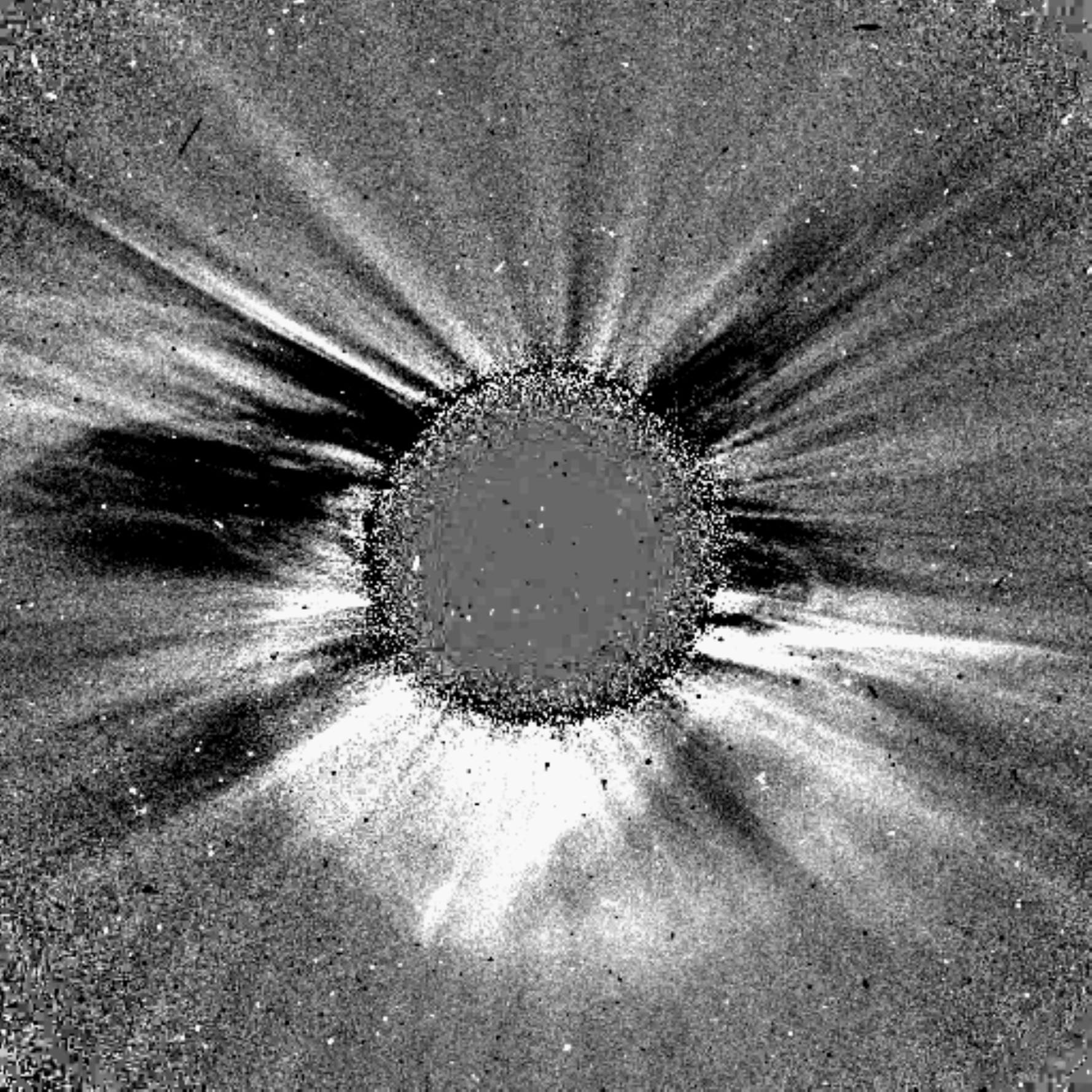}
             \hspace*{-0.02\textwidth}
               \includegraphics[width=0.4\textwidth,clip=]{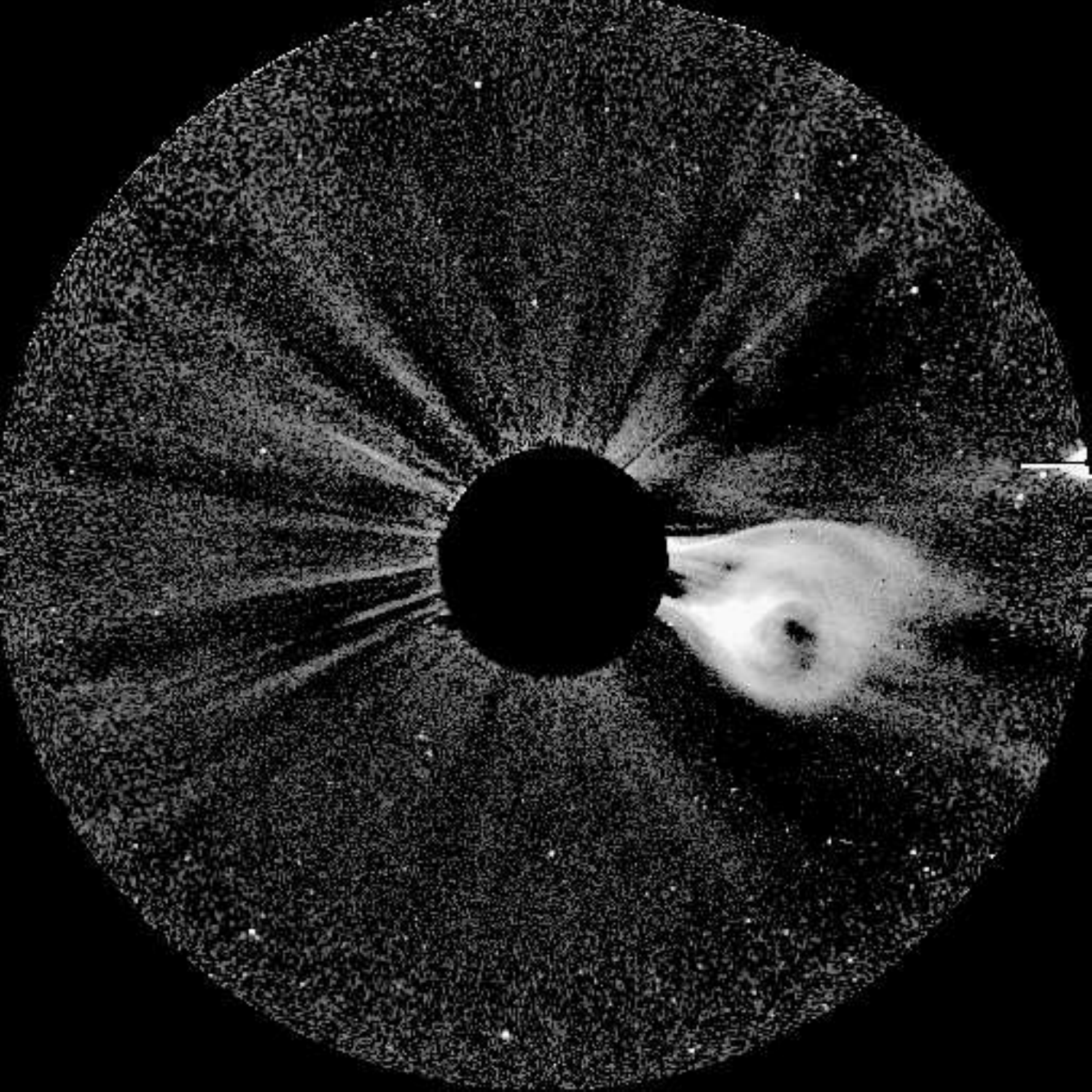}
               }
                 \centerline{\hspace*{0.04\textwidth}
              \includegraphics[width=0.4\textwidth,clip=]{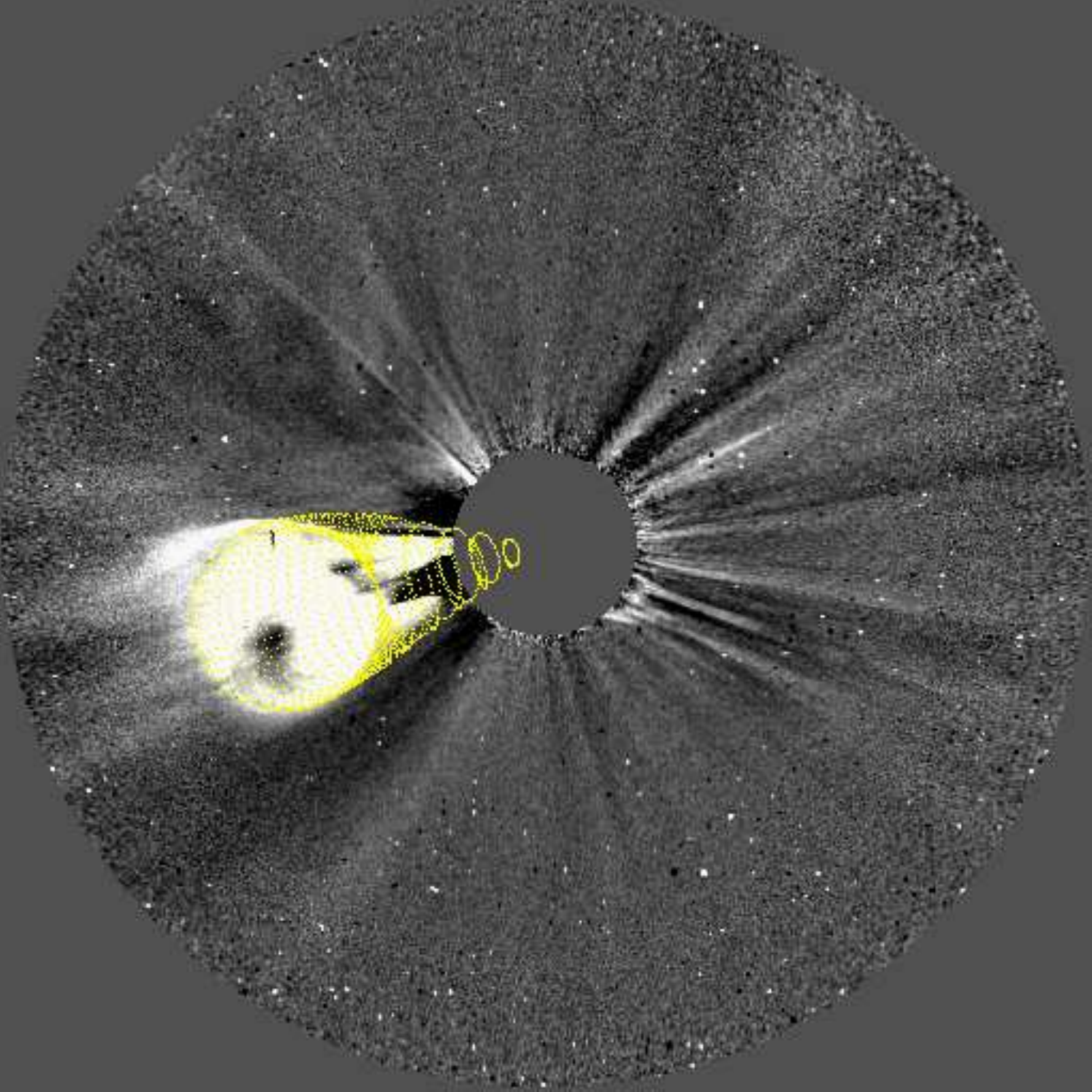}
               \hspace*{-0.02\textwidth}
               \includegraphics[width=0.4\textwidth,clip=]{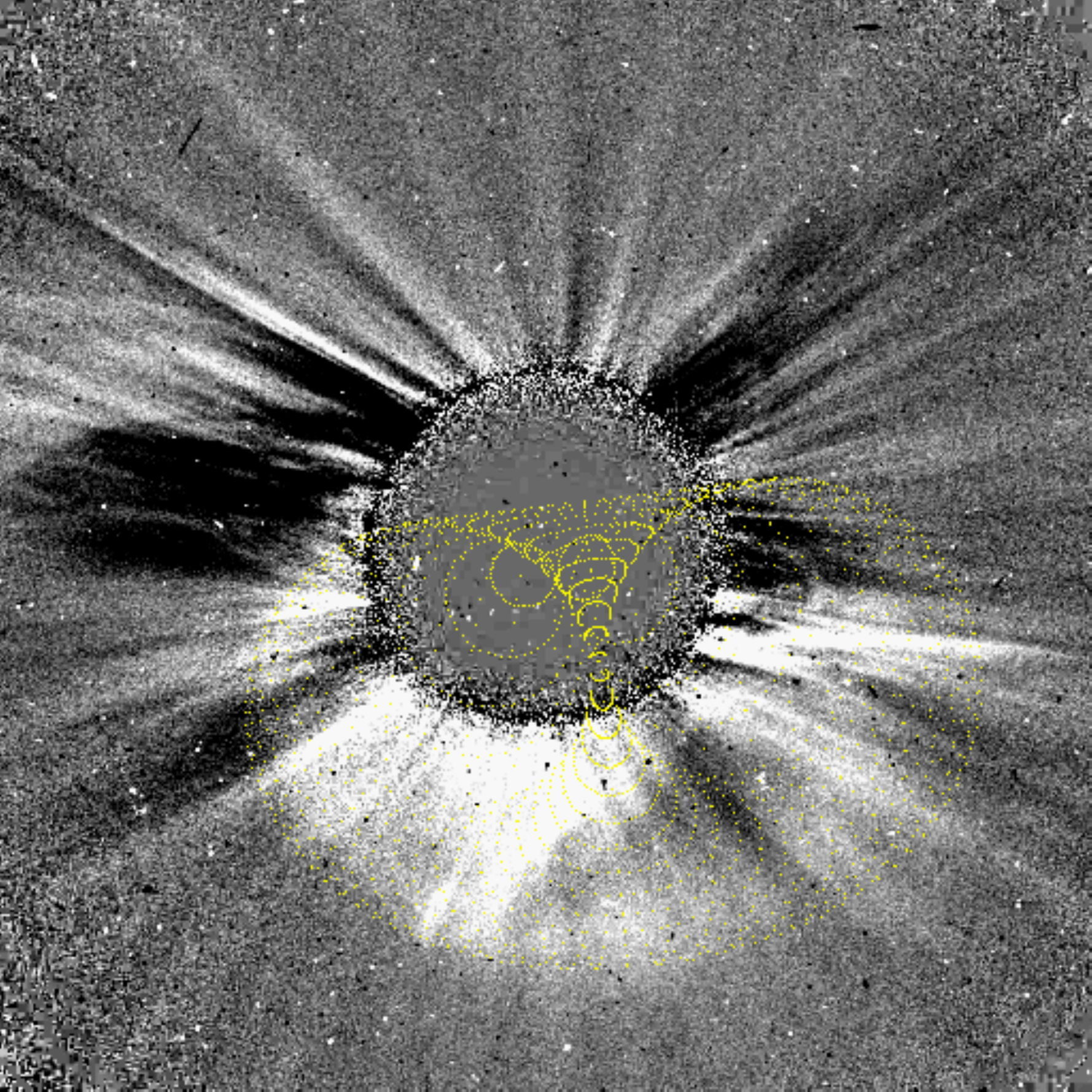}
              \hspace*{-0.02\textwidth}
               \includegraphics[width=0.4\textwidth,clip=]{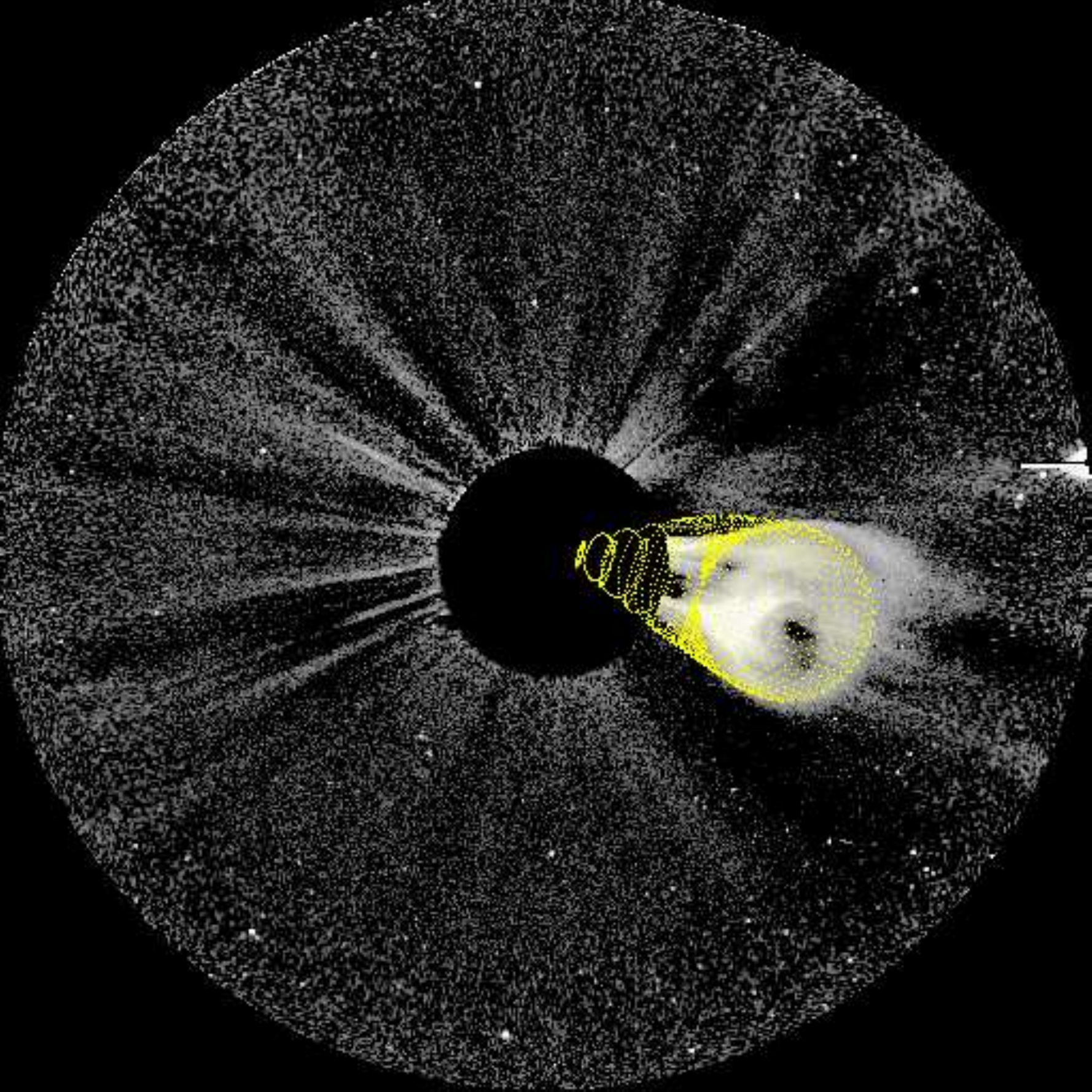}
                }
\vspace{0.0261\textwidth}  
\caption[GCS fit for CME 10 at 08:53]{GCS fit for CME 10 on March 03, 2011 at 08:53 UT at height $H=10.0$ \Rs. Table \ref{tblapp} 
lists the GCS parameters for this event.}
\label{figa10}
\end{figure}

\begin{figure}[h]    
  \centering                              
   \centerline{\hspace*{0.00\textwidth}
               \includegraphics[width=0.4\textwidth,clip=]{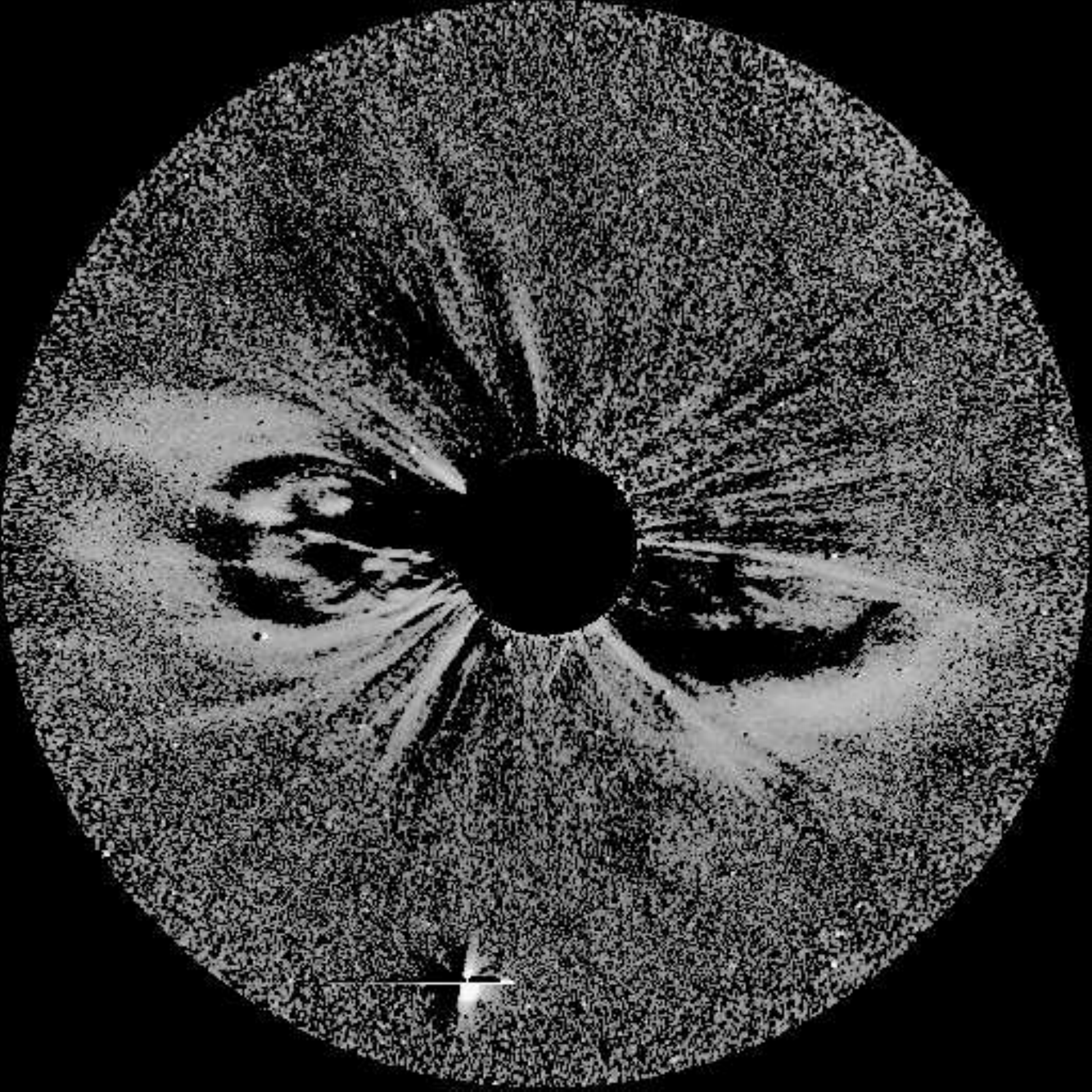}
                \hspace*{-0.02\textwidth}
               \includegraphics[width=0.4\textwidth,clip=]{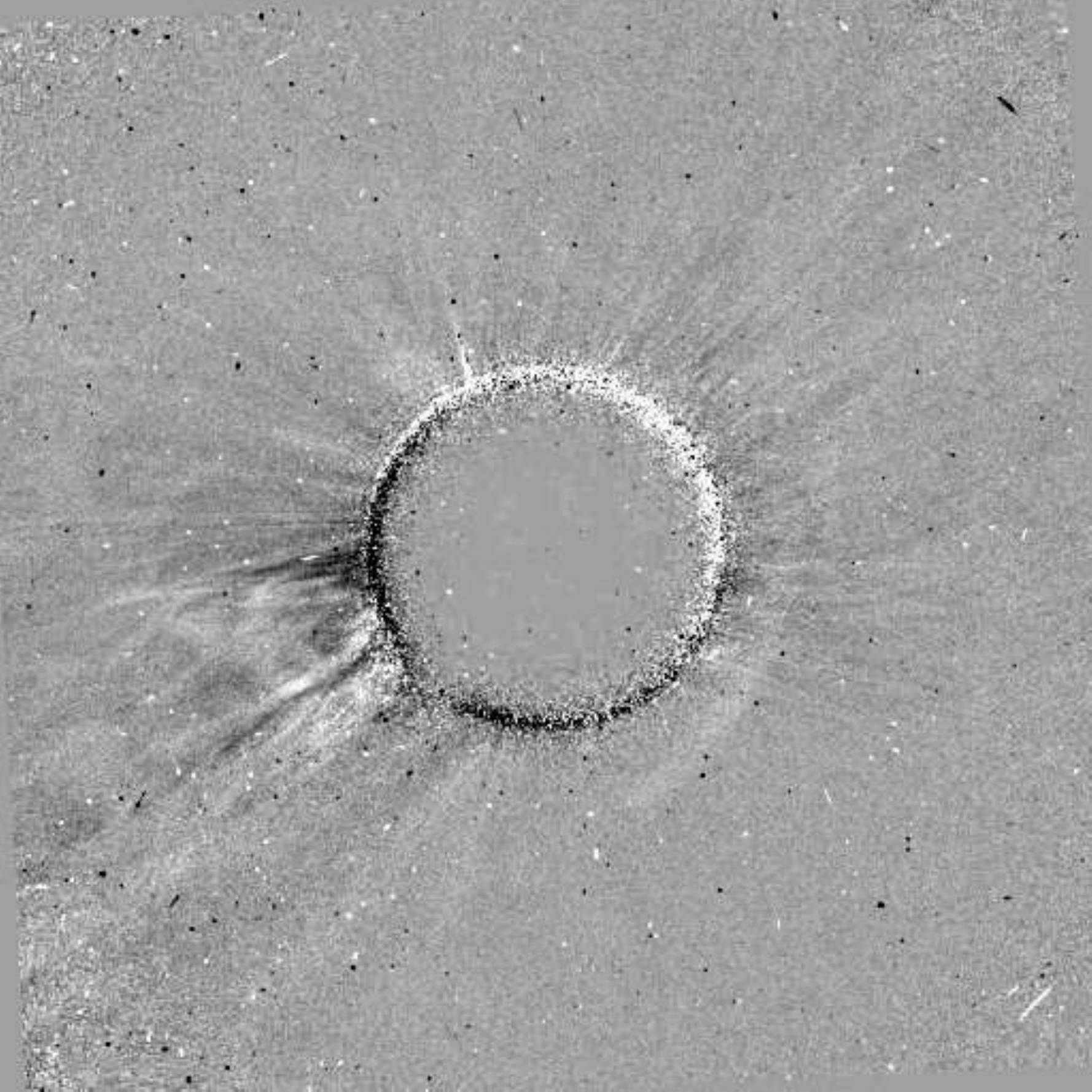}
             \hspace*{-0.02\textwidth}
               \includegraphics[width=0.4\textwidth,clip=]{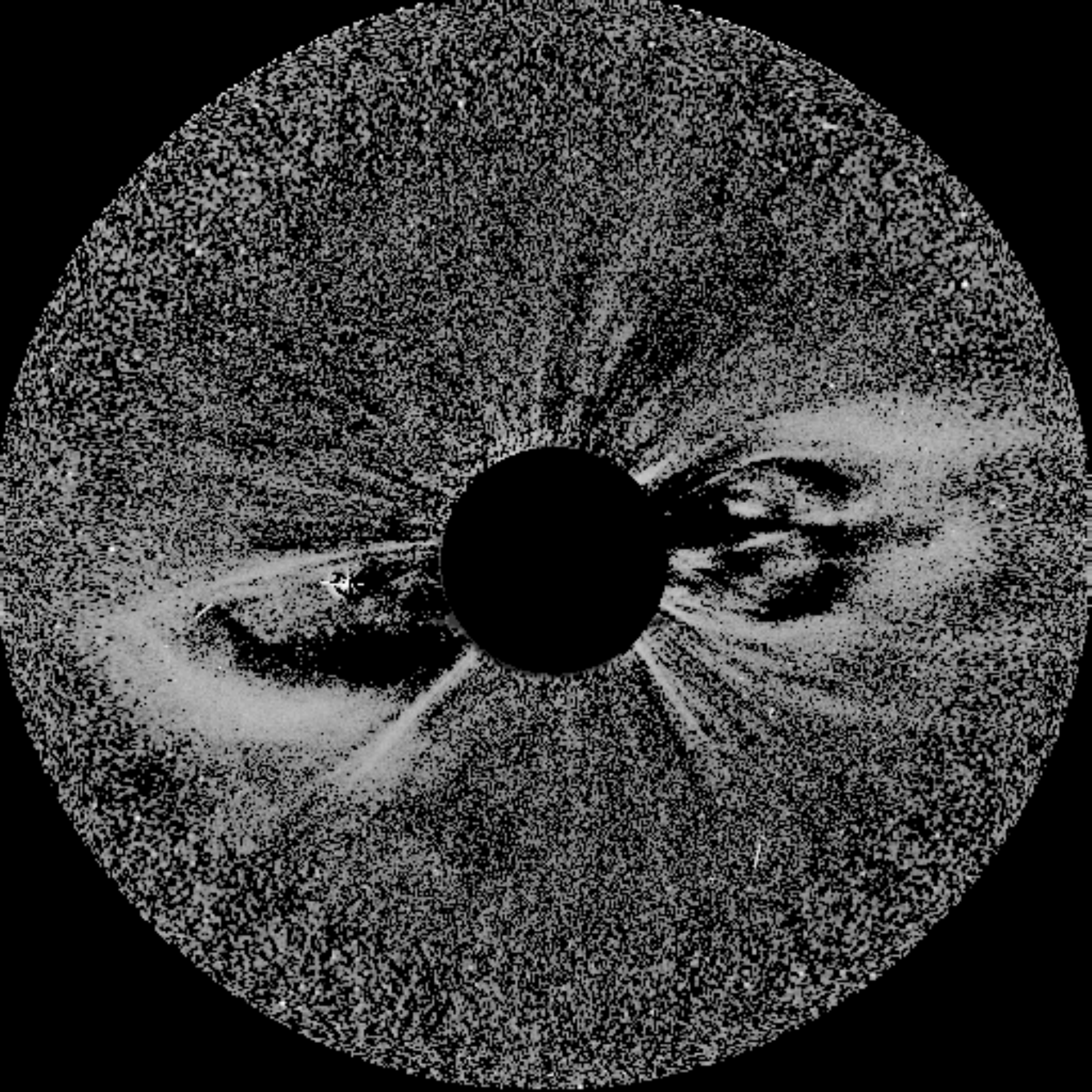}
               }
                 \centerline{\hspace*{0.0\textwidth}
              \includegraphics[width=0.4\textwidth,clip=]{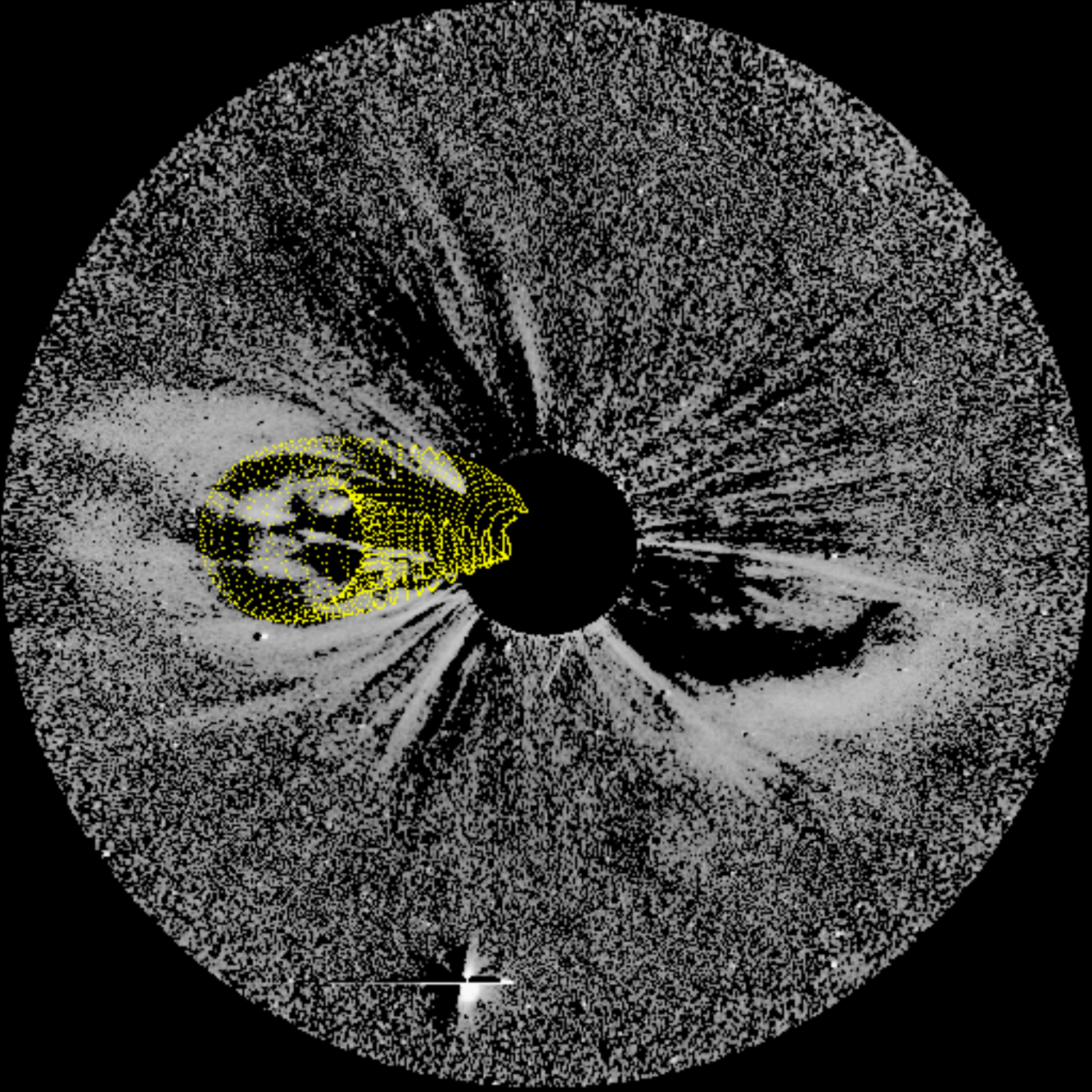}
               \hspace*{-0.02\textwidth}
               \includegraphics[width=0.4\textwidth,clip=]{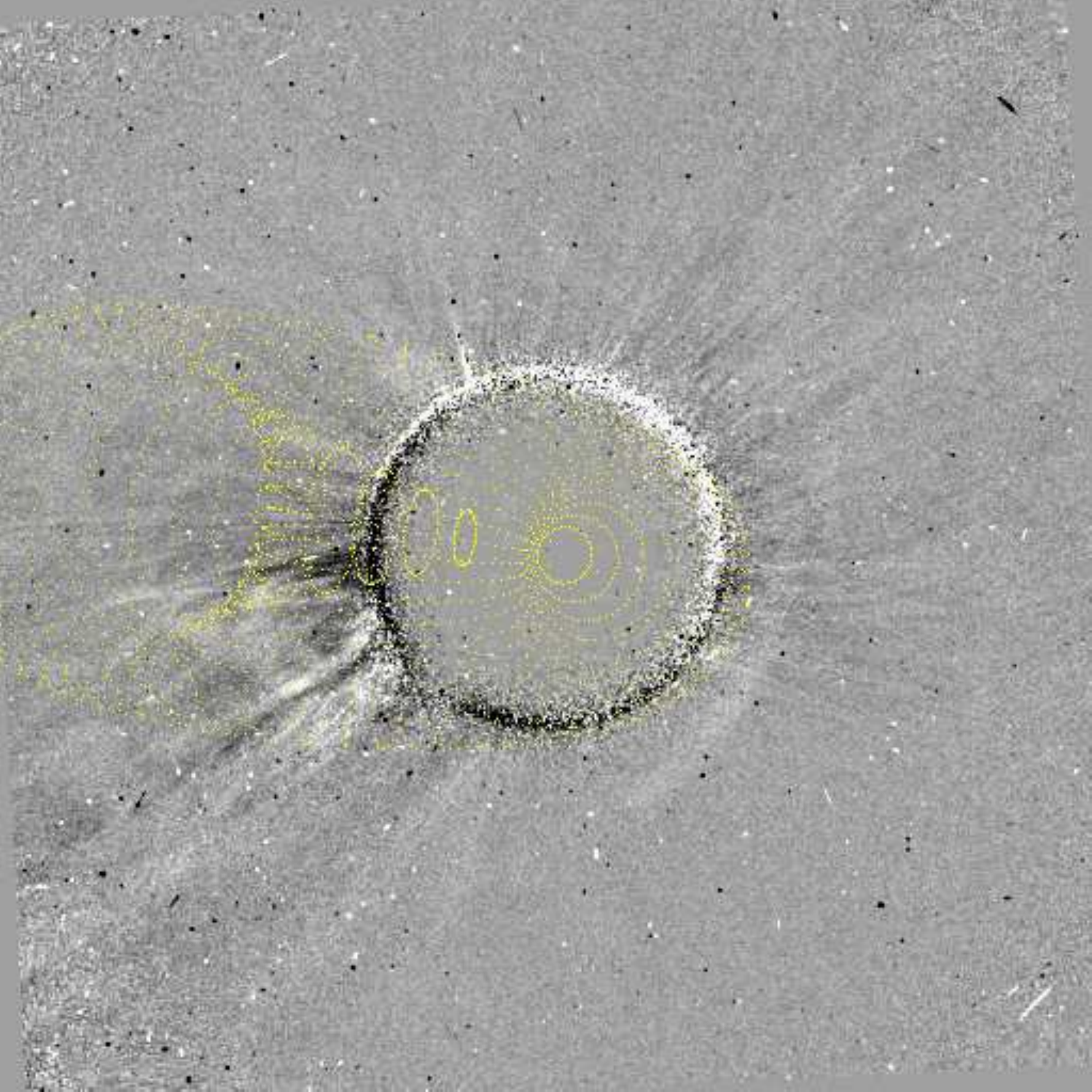}
              \hspace*{-0.02\textwidth}
               \includegraphics[width=0.4\textwidth,clip=]{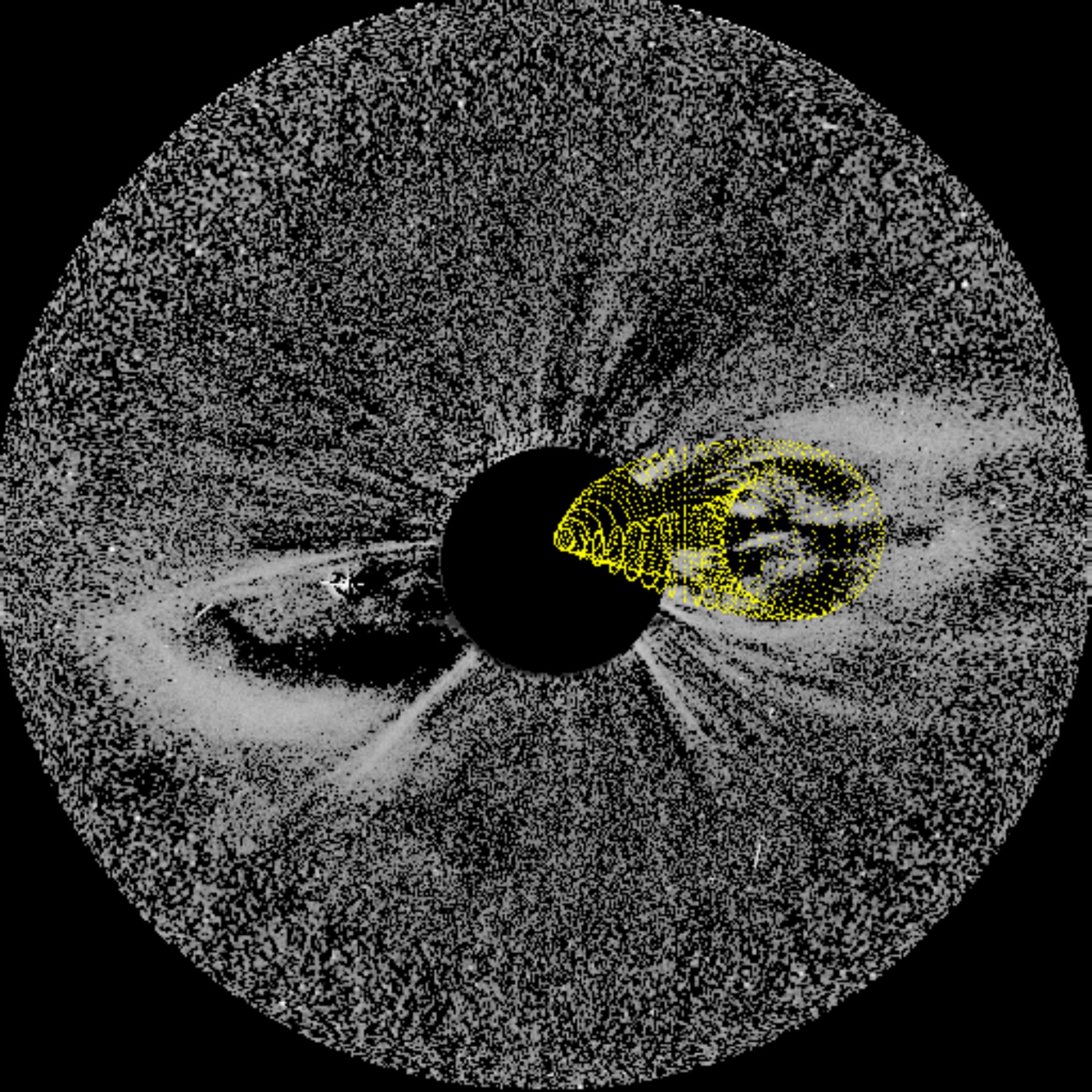}
                }
\vspace{0.0261\textwidth}  
\caption[GCS fit for CME 11 at 15:39]{GCS fit for CME 11 on March 25, 2011 at 15:39 UT at height $H=10.0$ \Rs. Table \ref{tblapp} 
lists the GCS parameters for this event.}
\label{figa11}
\end{figure}

\begin{figure}[h]    
  \centering                              
   \centerline{\hspace*{0.04\textwidth}
               \includegraphics[width=0.4\textwidth,clip=]{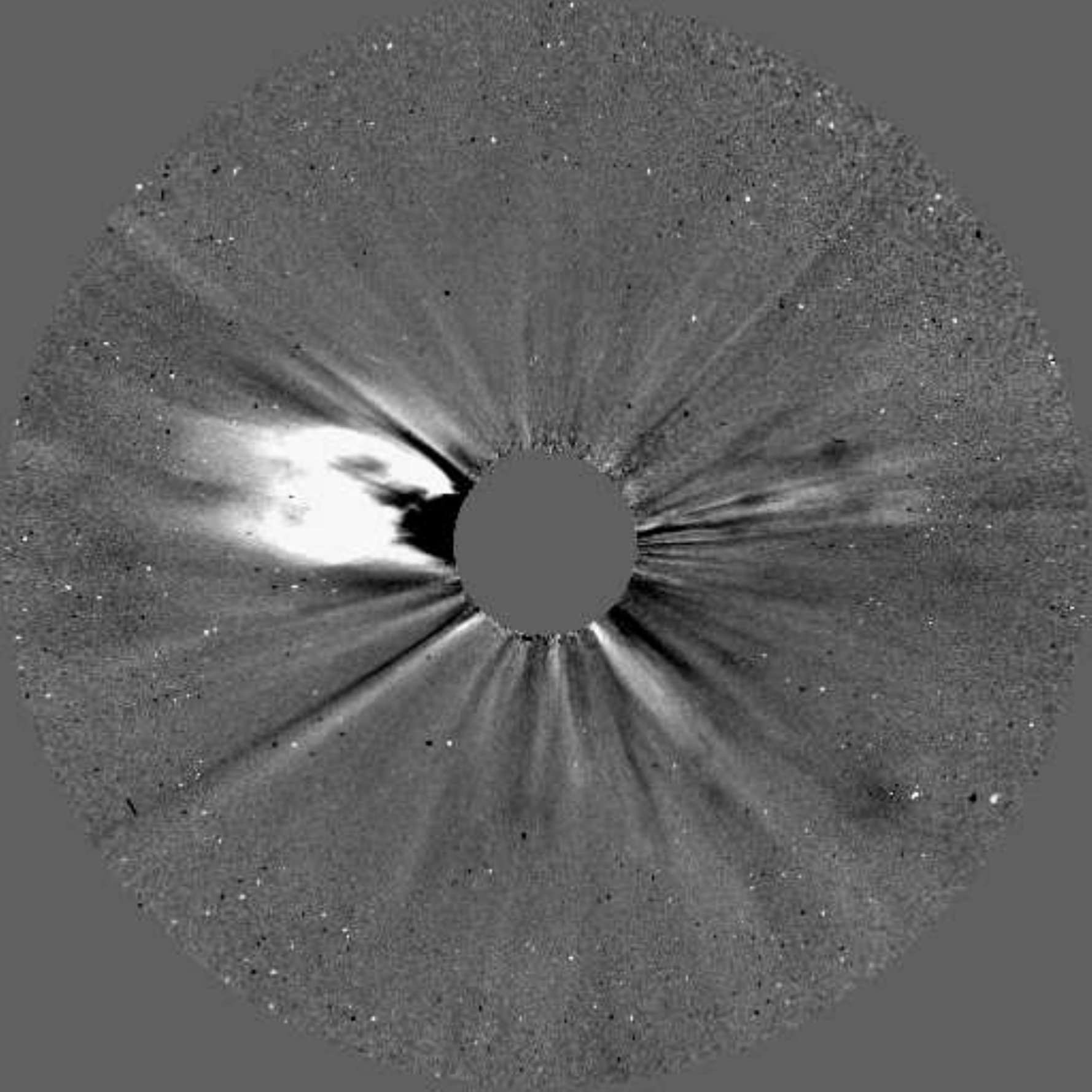}
                \hspace*{-0.02\textwidth}
               \includegraphics[width=0.4\textwidth,clip=]{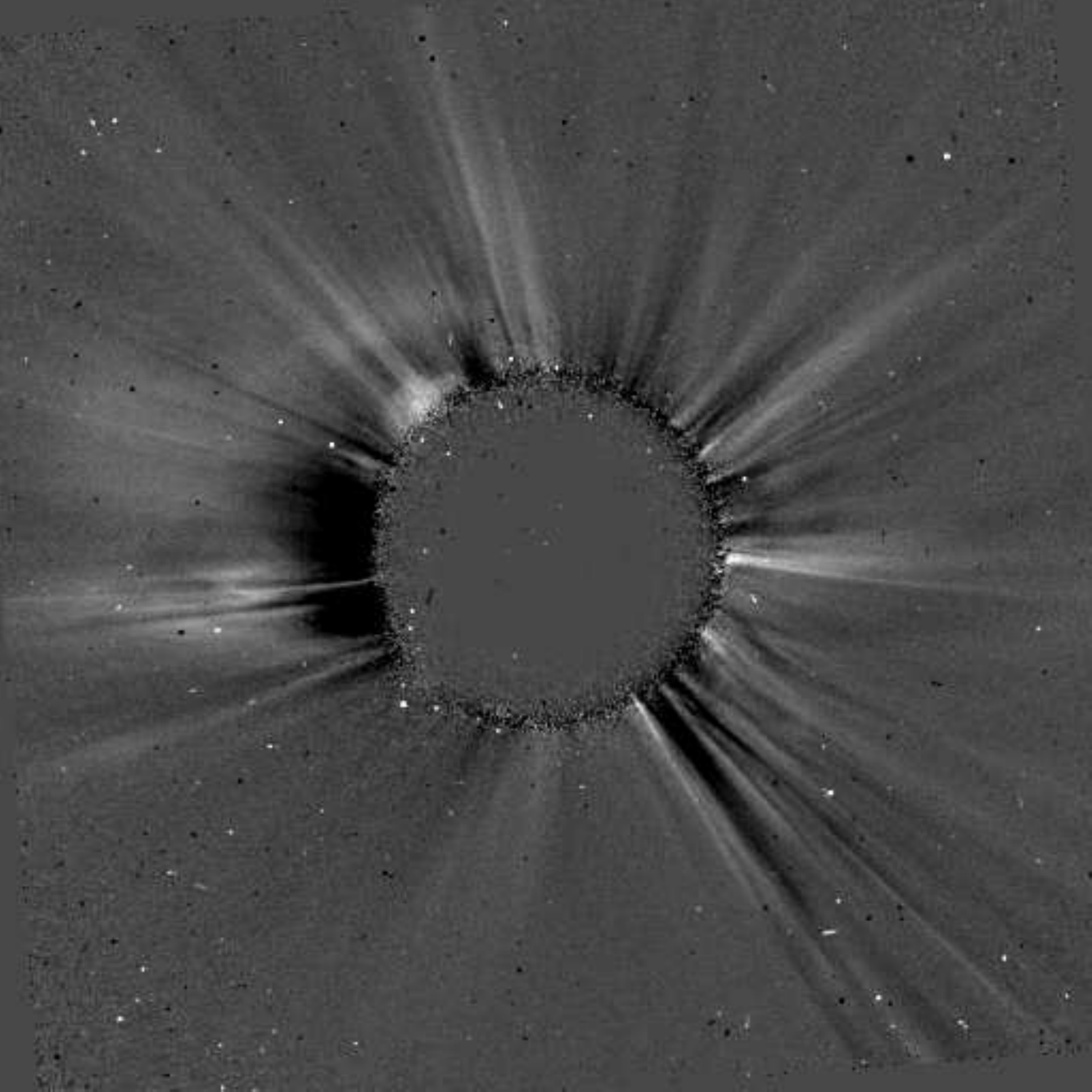}
             \hspace*{-0.02\textwidth}
               \includegraphics[width=0.4\textwidth,clip=]{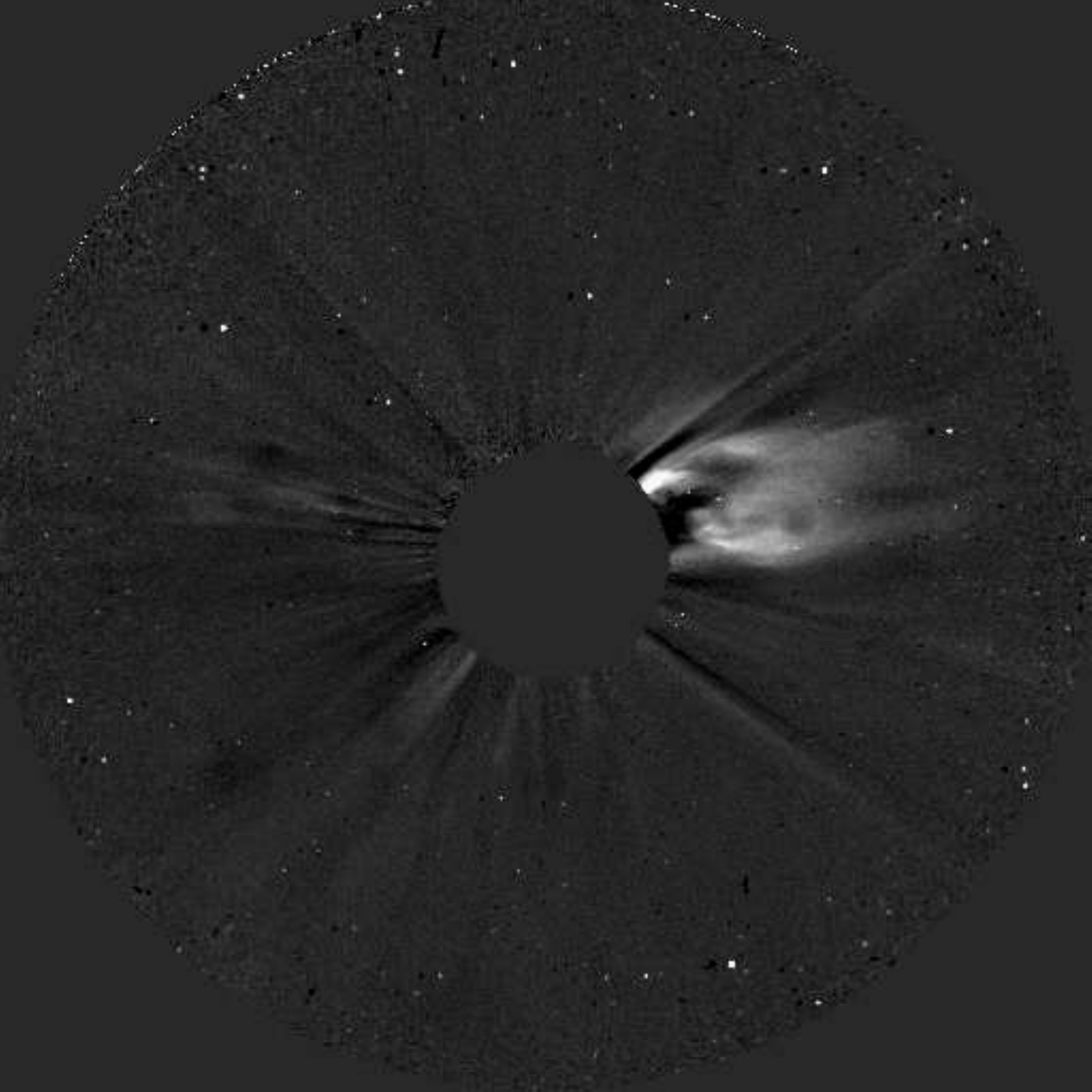}
               }
                 \centerline{\hspace*{0.04\textwidth}
              \includegraphics[width=0.4\textwidth,clip=]{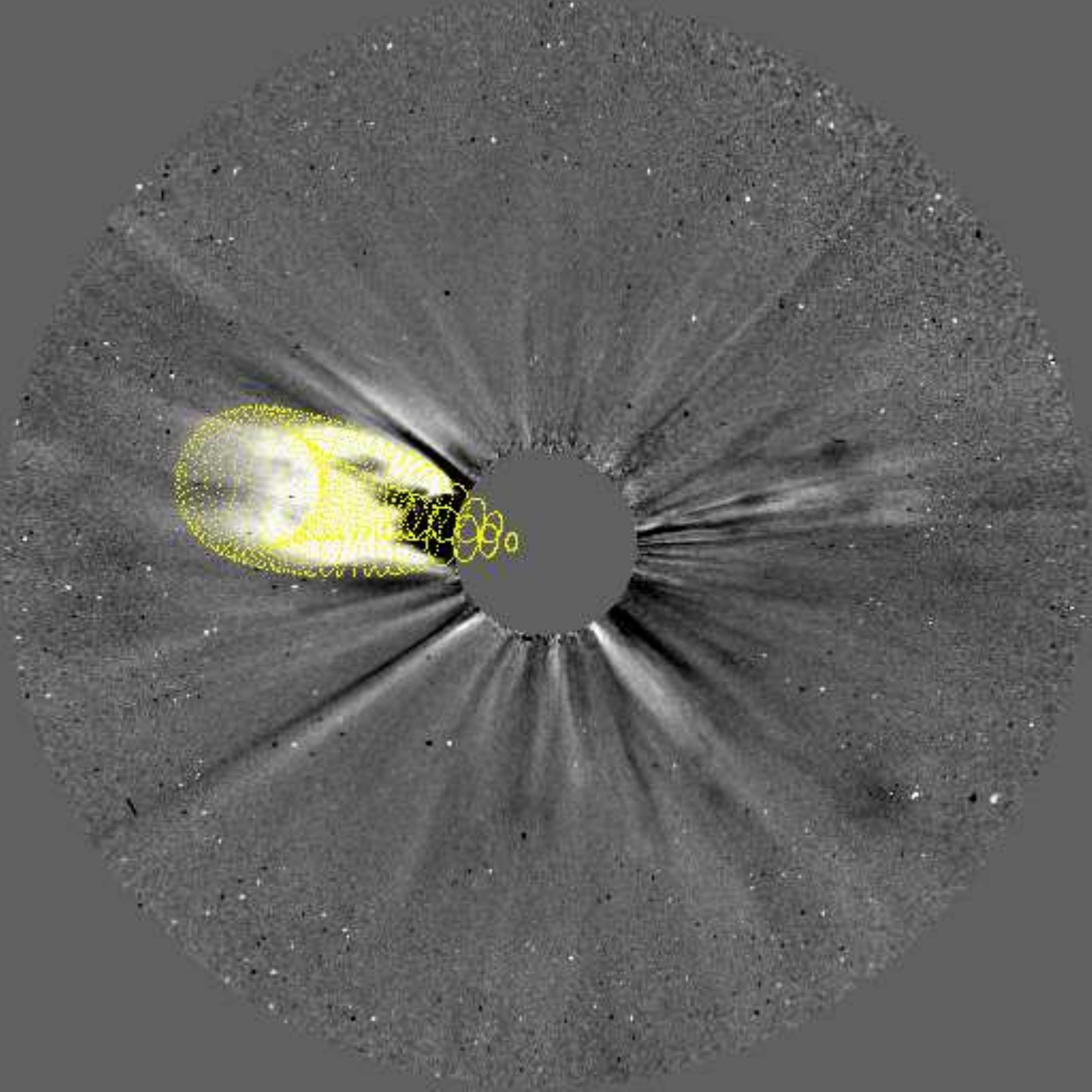}
               \hspace*{-0.02\textwidth}
               \includegraphics[width=0.4\textwidth,clip=]{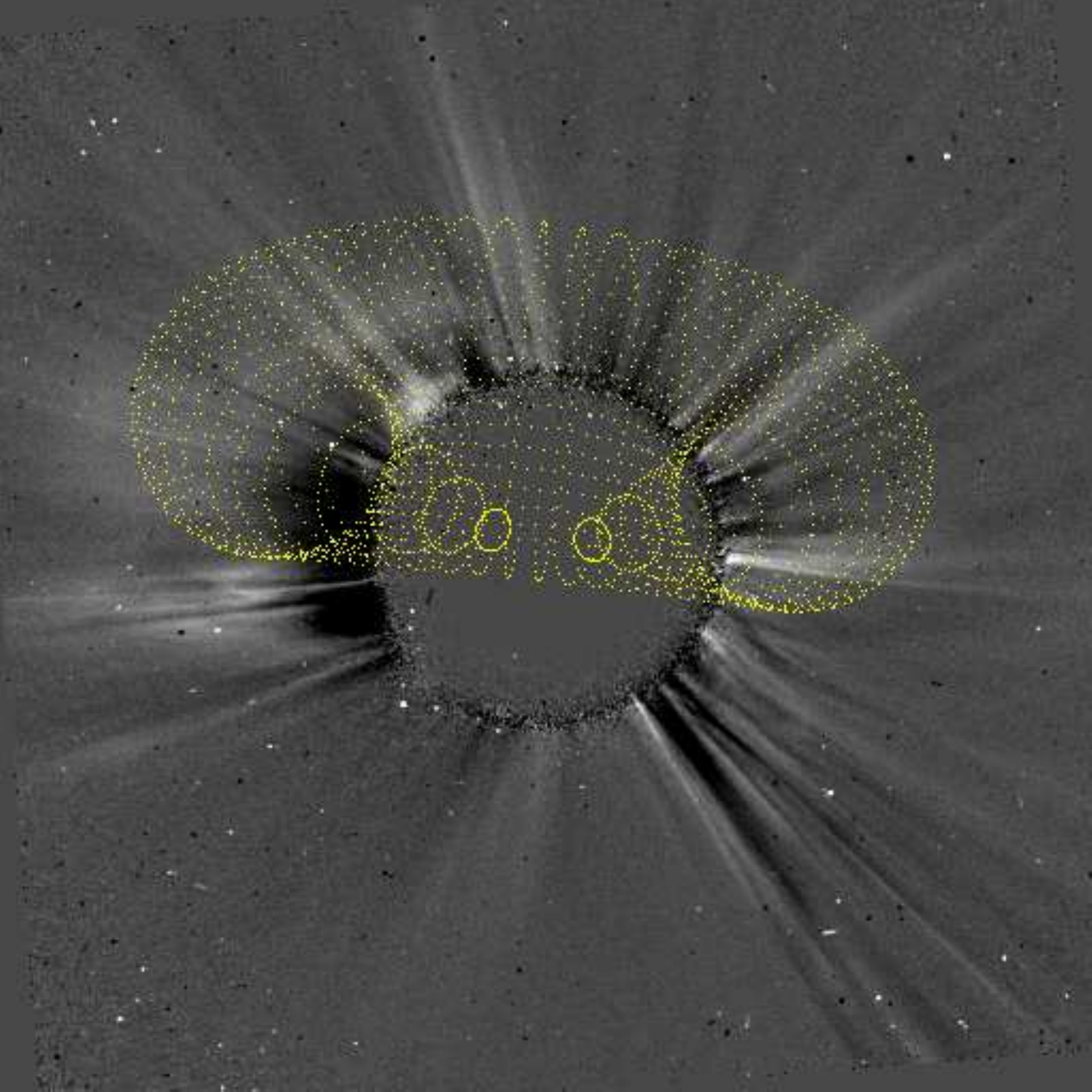}
              \hspace*{-0.02\textwidth}
               \includegraphics[width=0.4\textwidth,clip=]{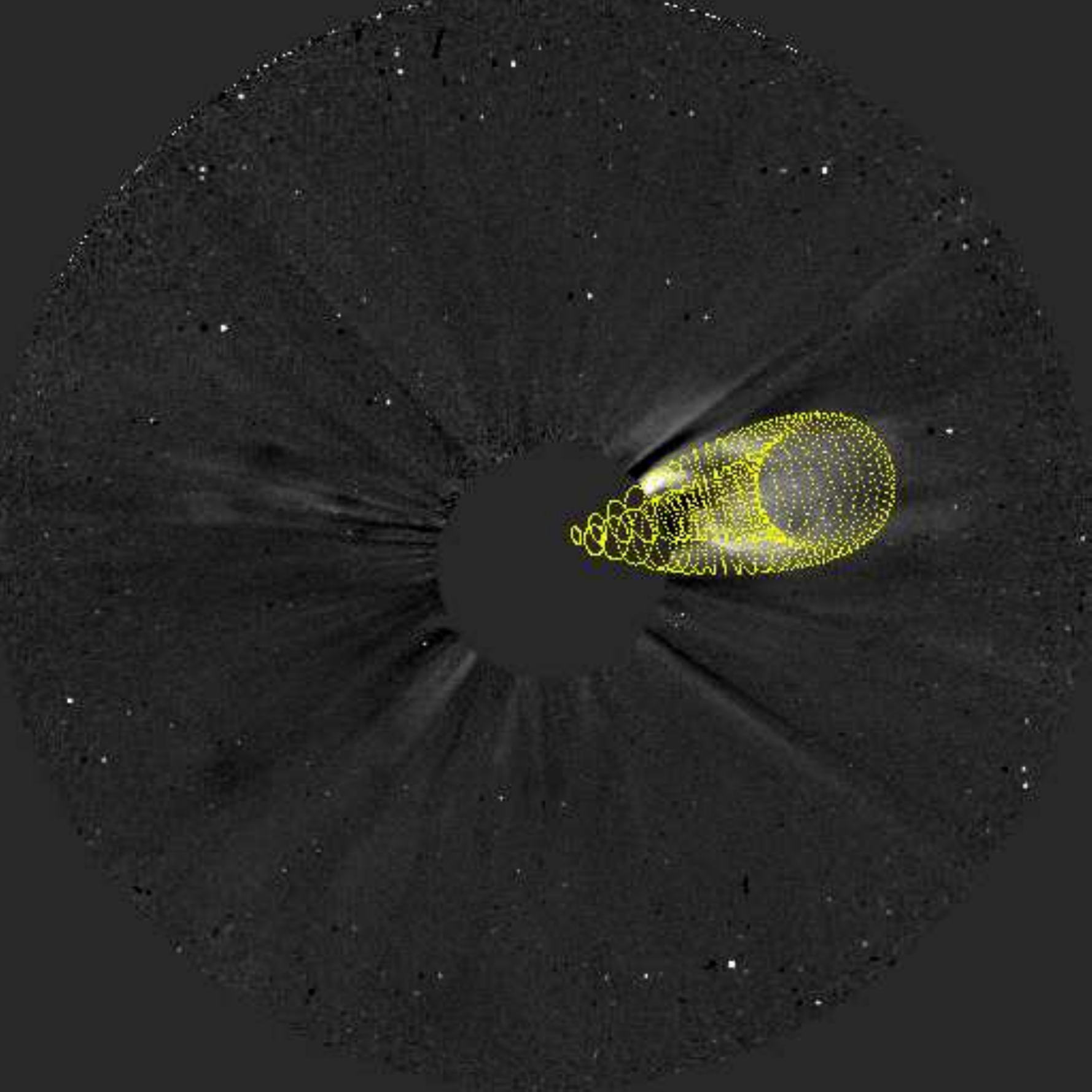}
                }
\vspace{0.0261\textwidth}  
\caption[GCS fit for CME 12 at 02:54]{GCS fit for CME 12 on April 09, 2011 at 02:54 UT at height $H=10.2$ \Rs. Table \ref{tblapp} 
lists the GCS parameters for this event.}
\label{figa12}
\end{figure}

\begin{figure}[h]    
  \centering                              
   \centerline{\hspace*{0.00\textwidth}
               \includegraphics[width=0.4\textwidth,clip=]{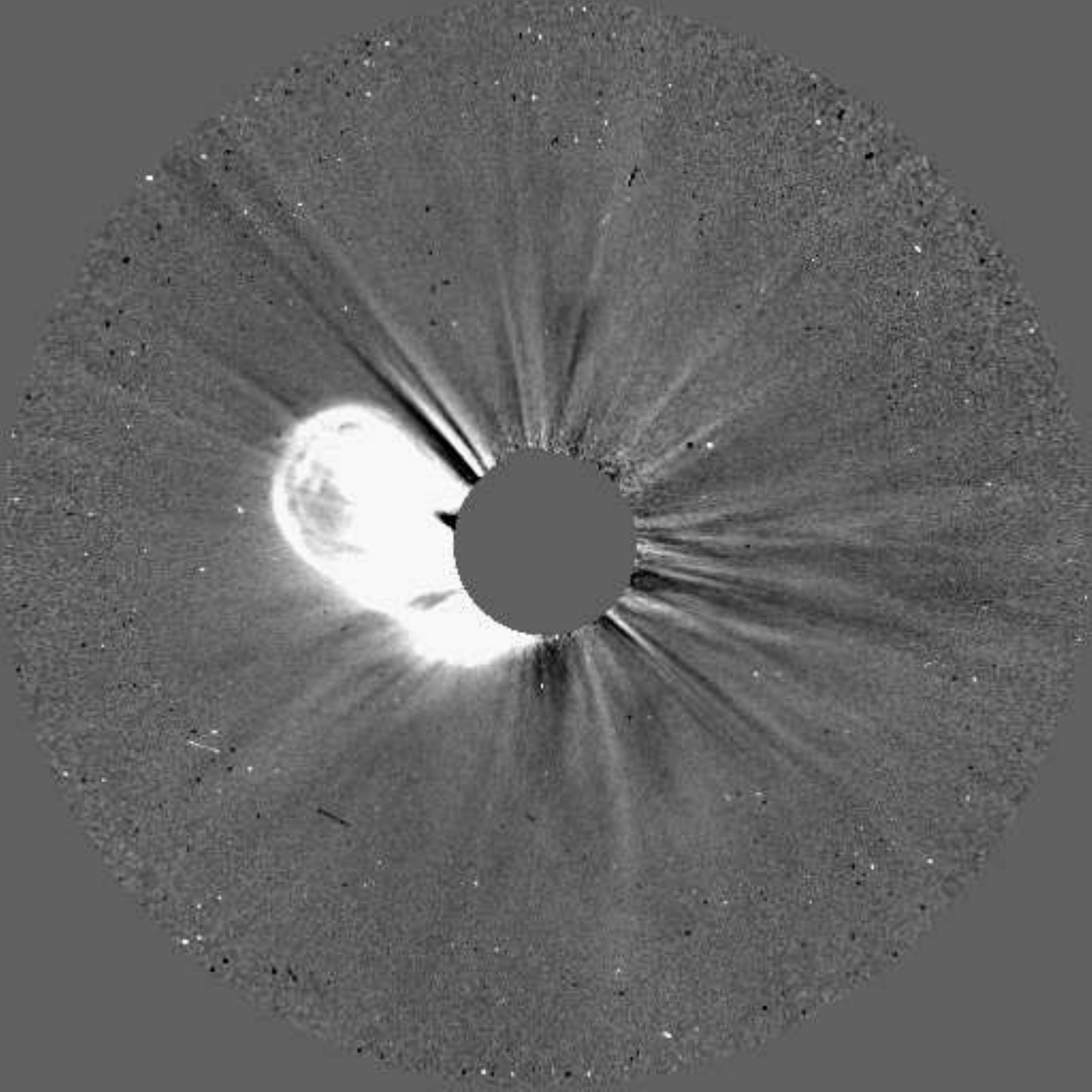}
                \hspace*{-0.02\textwidth}
               \includegraphics[width=0.4\textwidth,clip=]{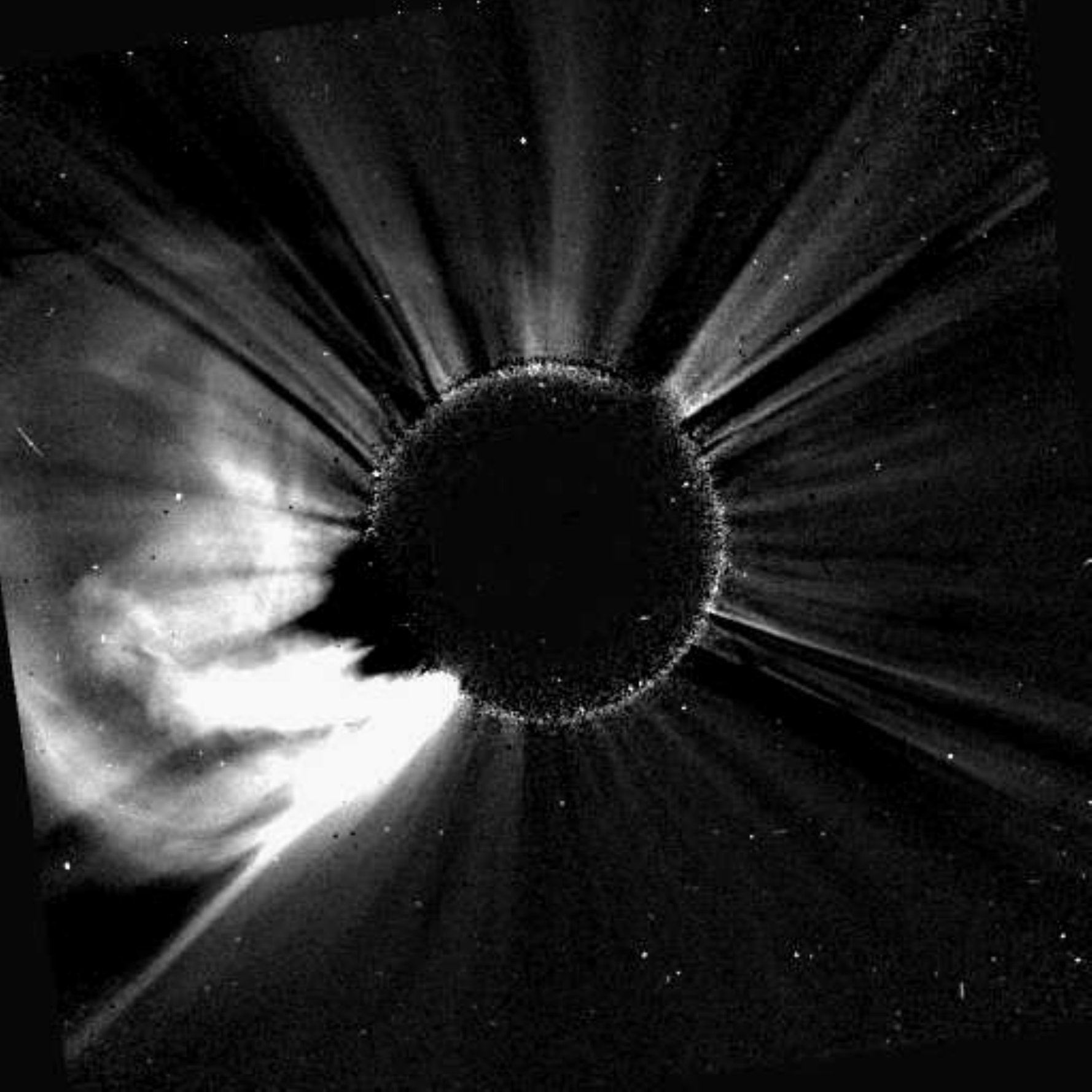}
             \hspace*{-0.02\textwidth}
               \includegraphics[width=0.4\textwidth,clip=]{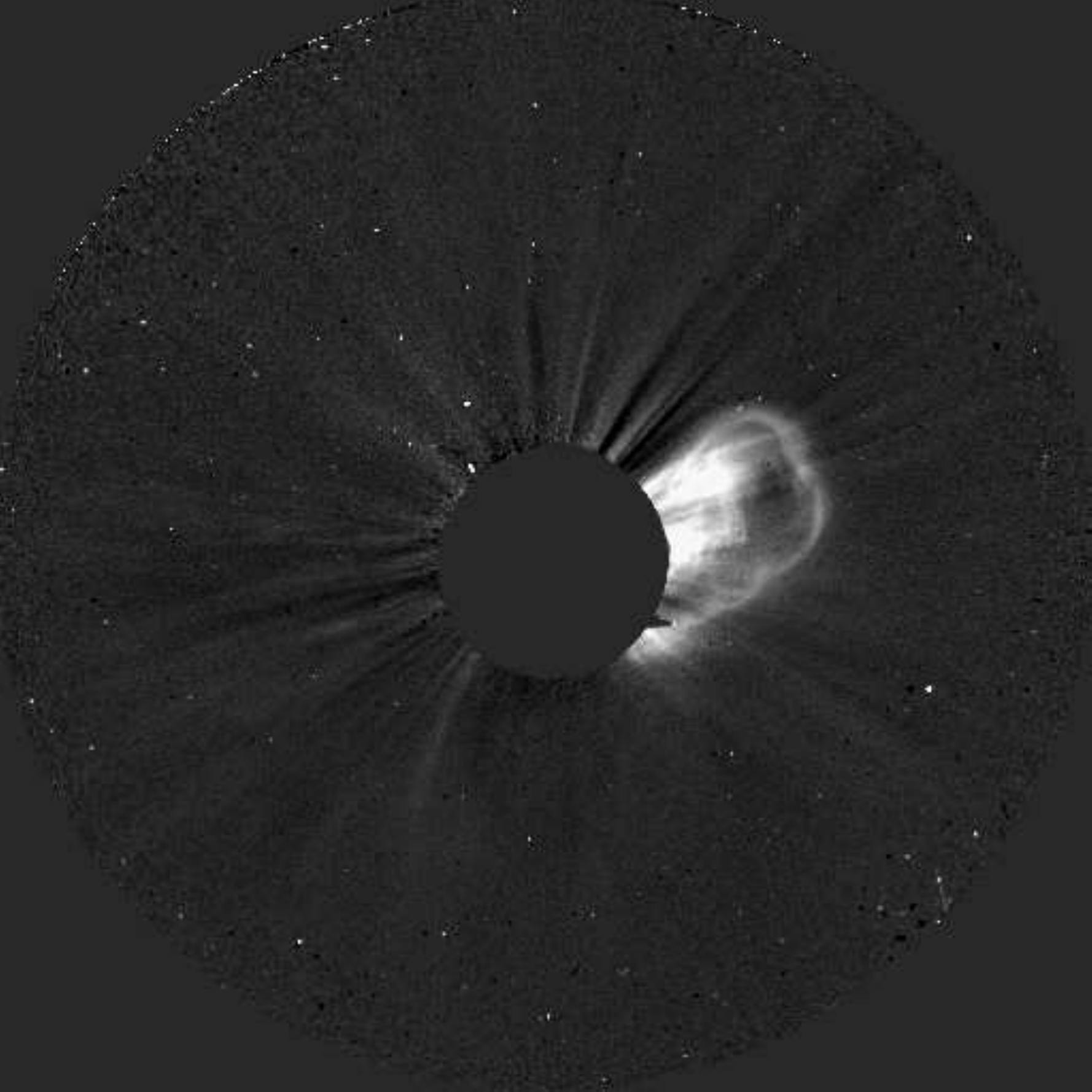}
               }
                 \centerline{\hspace*{0.0\textwidth}
              \includegraphics[width=0.4\textwidth,clip=]{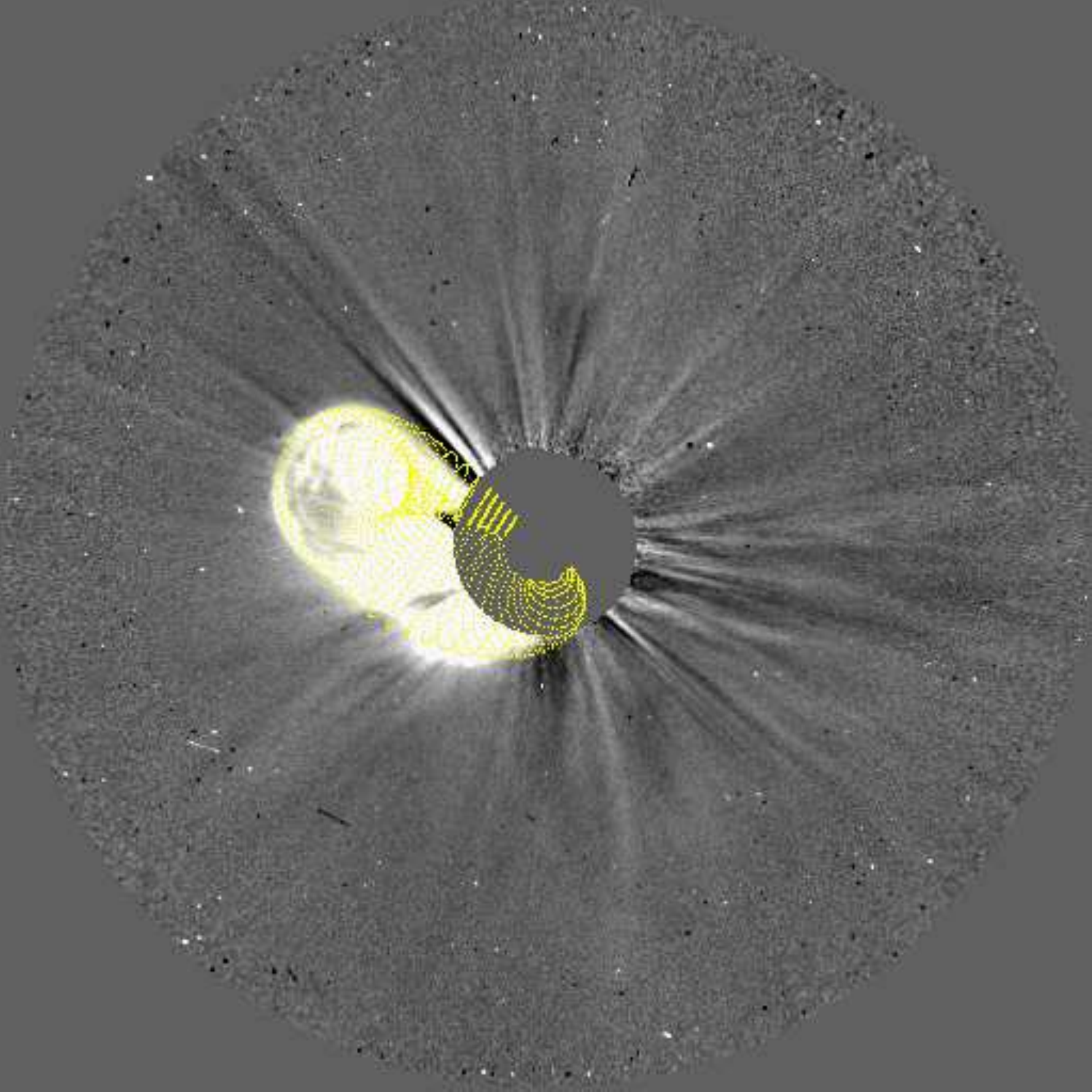}
               \hspace*{-0.02\textwidth}
               \includegraphics[width=0.4\textwidth,clip=]{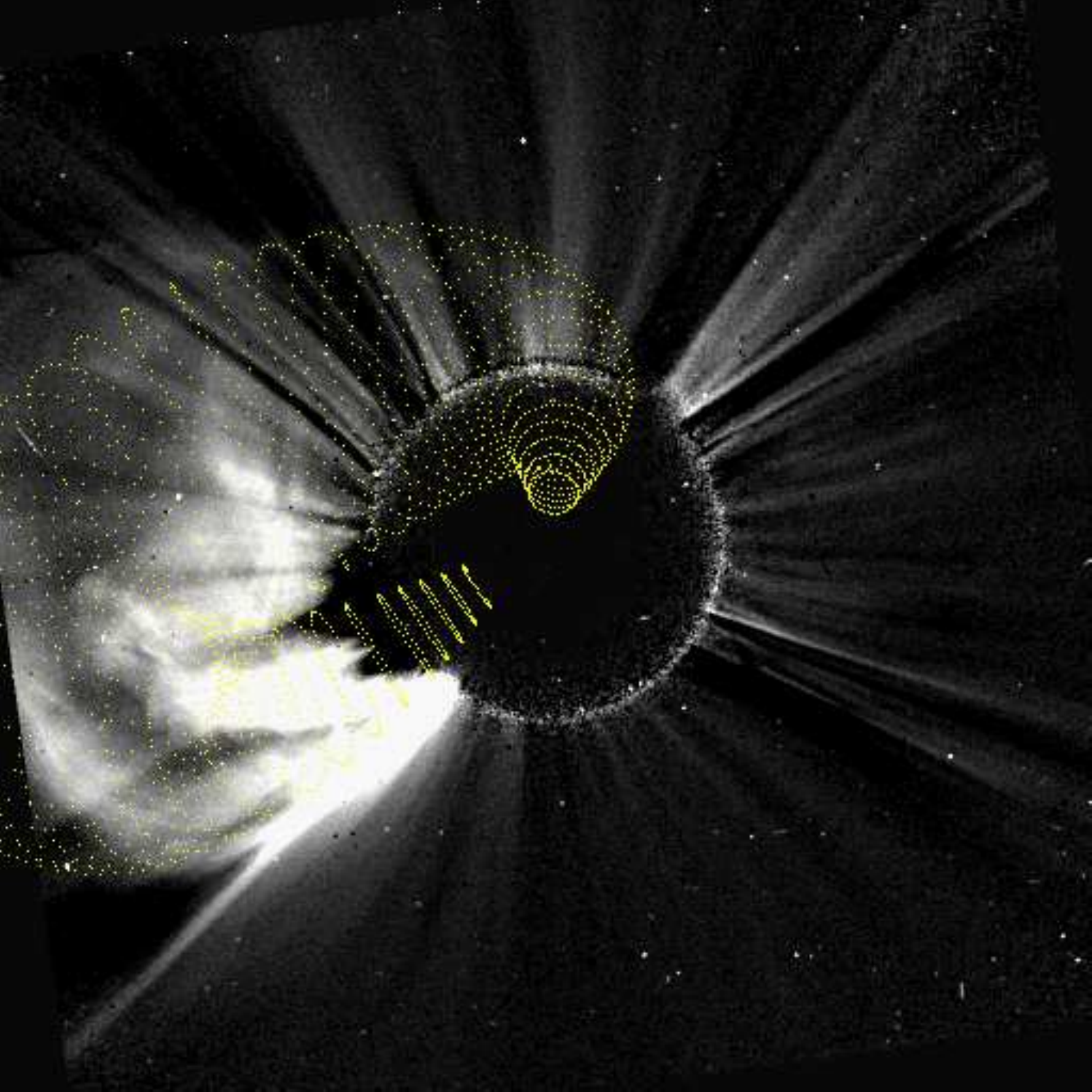}
              \hspace*{-0.02\textwidth}
               \includegraphics[width=0.4\textwidth,clip=]{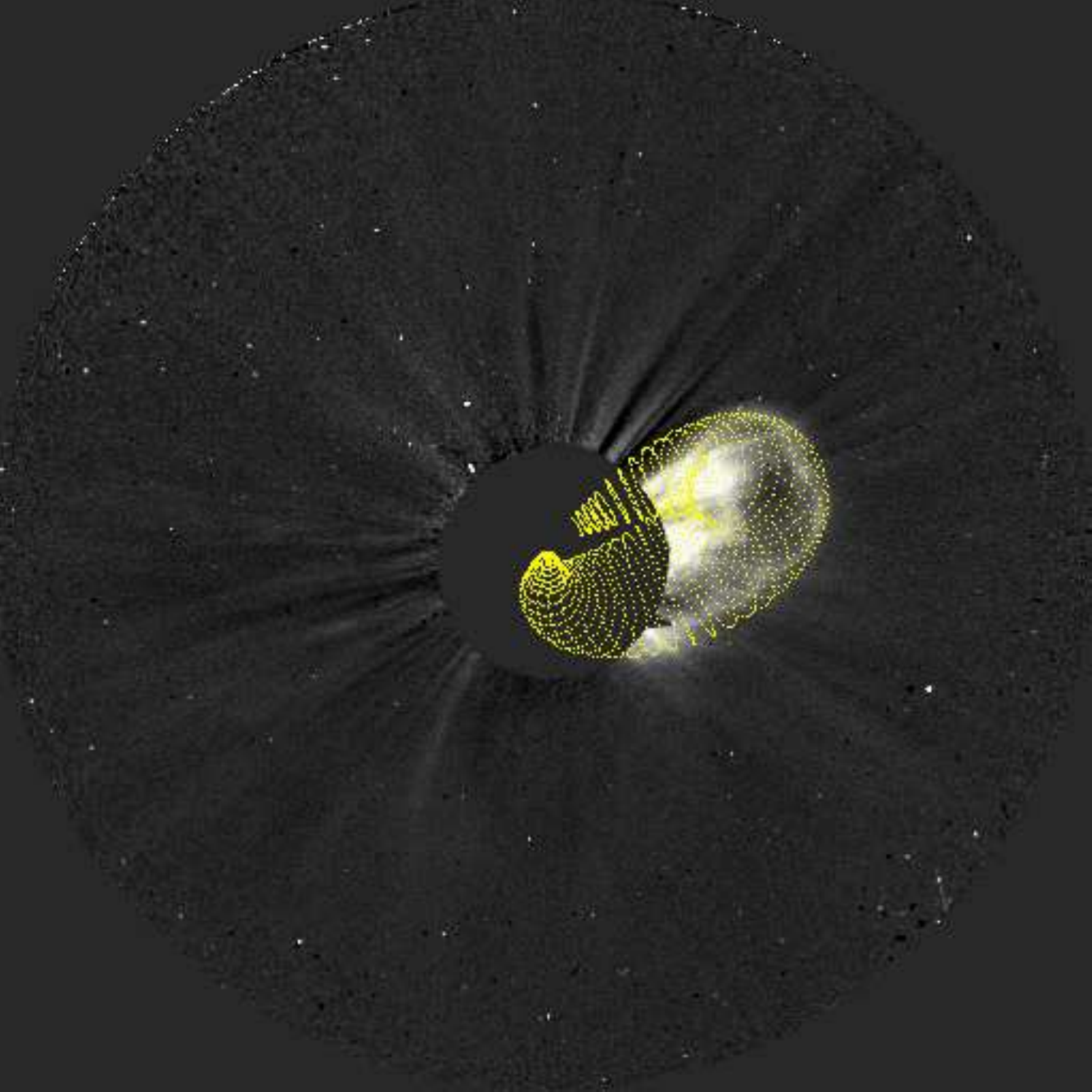}
                }
\vspace{0.0261\textwidth}  
\caption[GCS fit for CME 13 at 09:54]{GCS fit for CME 13 on June 14, 2011 at 09:54 UT at height $H=9.4$ \Rs. Table \ref{tblapp} 
lists the GCS parameters for this event.}
\label{figa13}
\end{figure}

\begin{figure}[h]    
  \centering                              
   \centerline{\hspace*{0.04\textwidth}
               \includegraphics[width=0.4\textwidth,clip=]{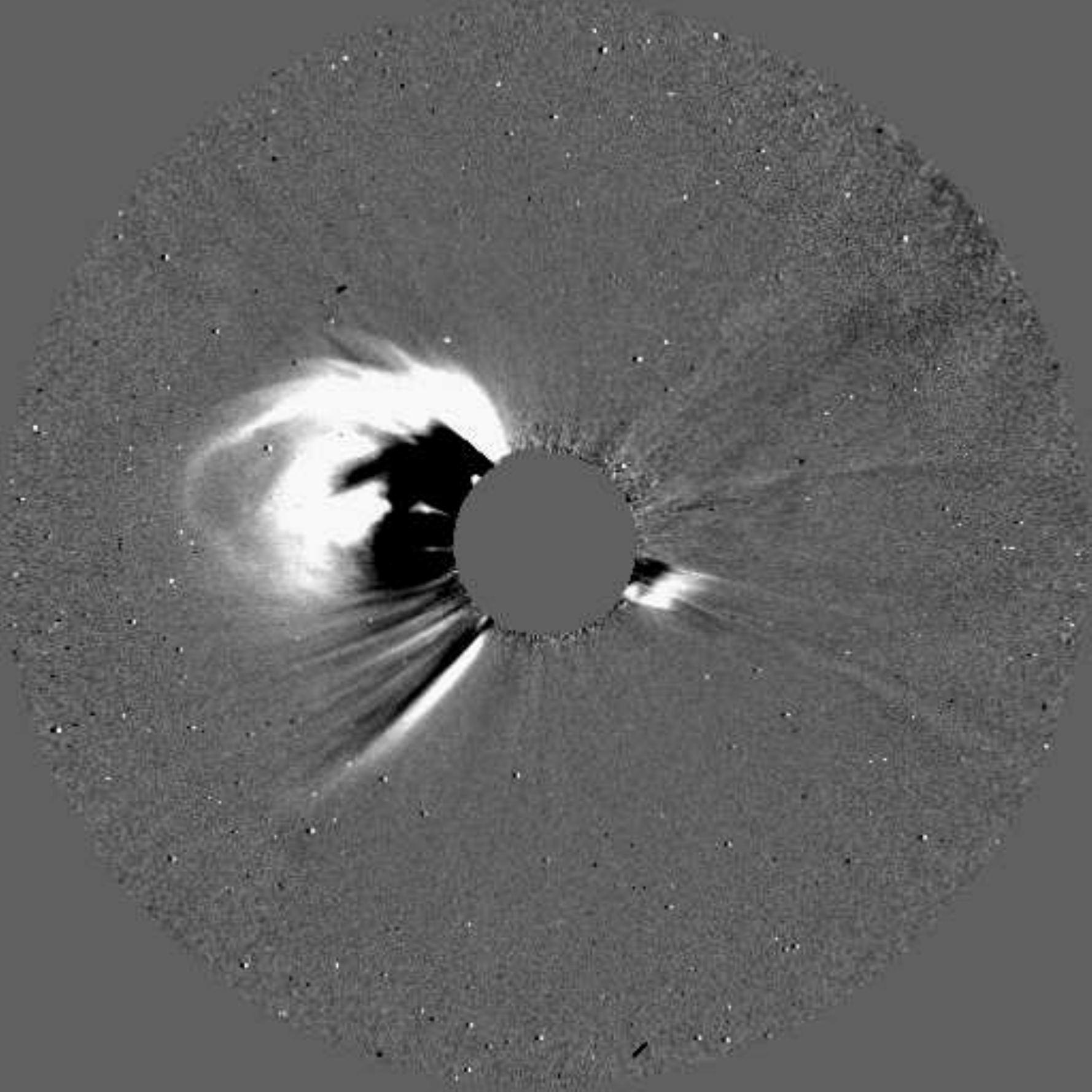}
                \hspace*{-0.02\textwidth}
               \includegraphics[width=0.4\textwidth,clip=]{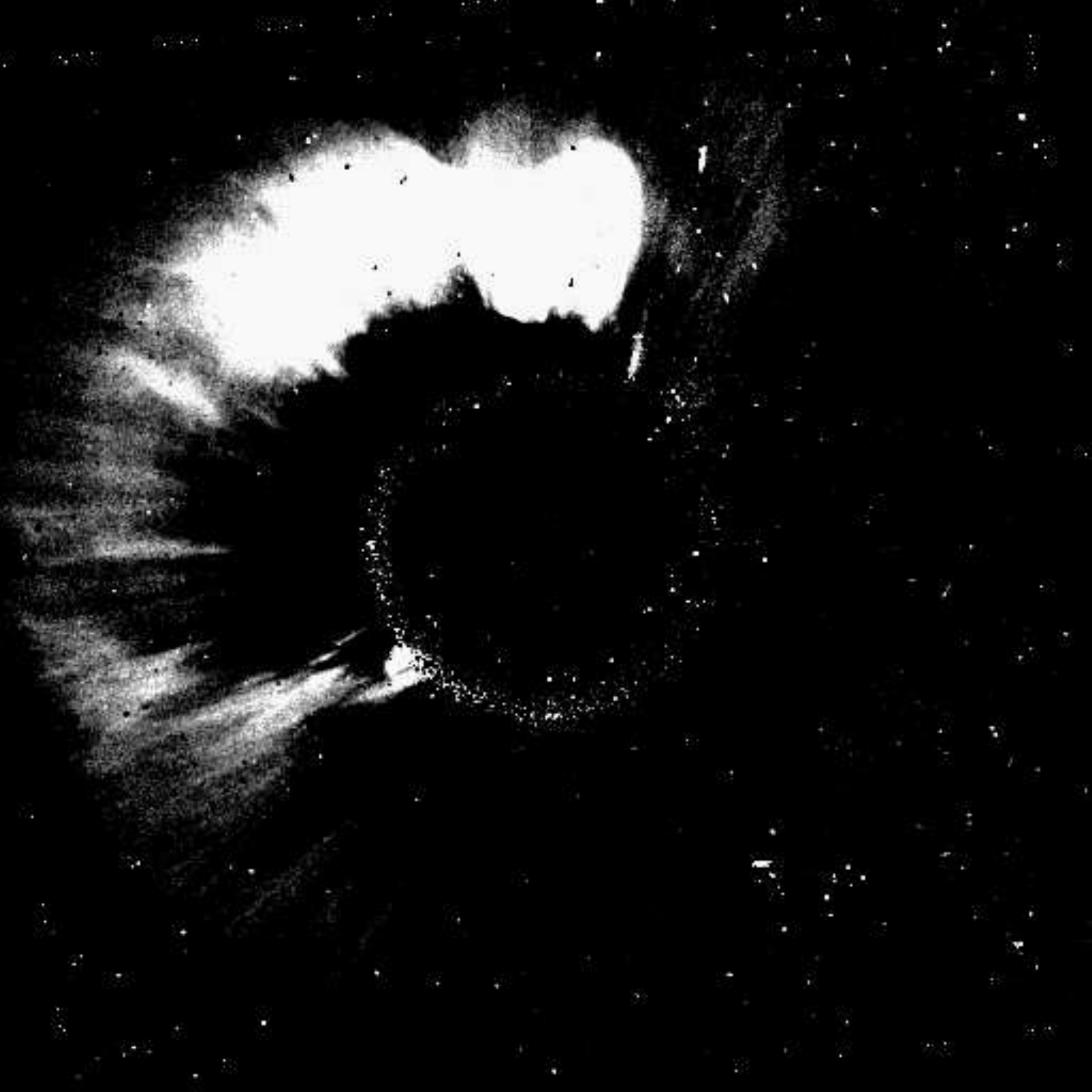}
             \hspace*{-0.02\textwidth}
               \includegraphics[width=0.4\textwidth,clip=]{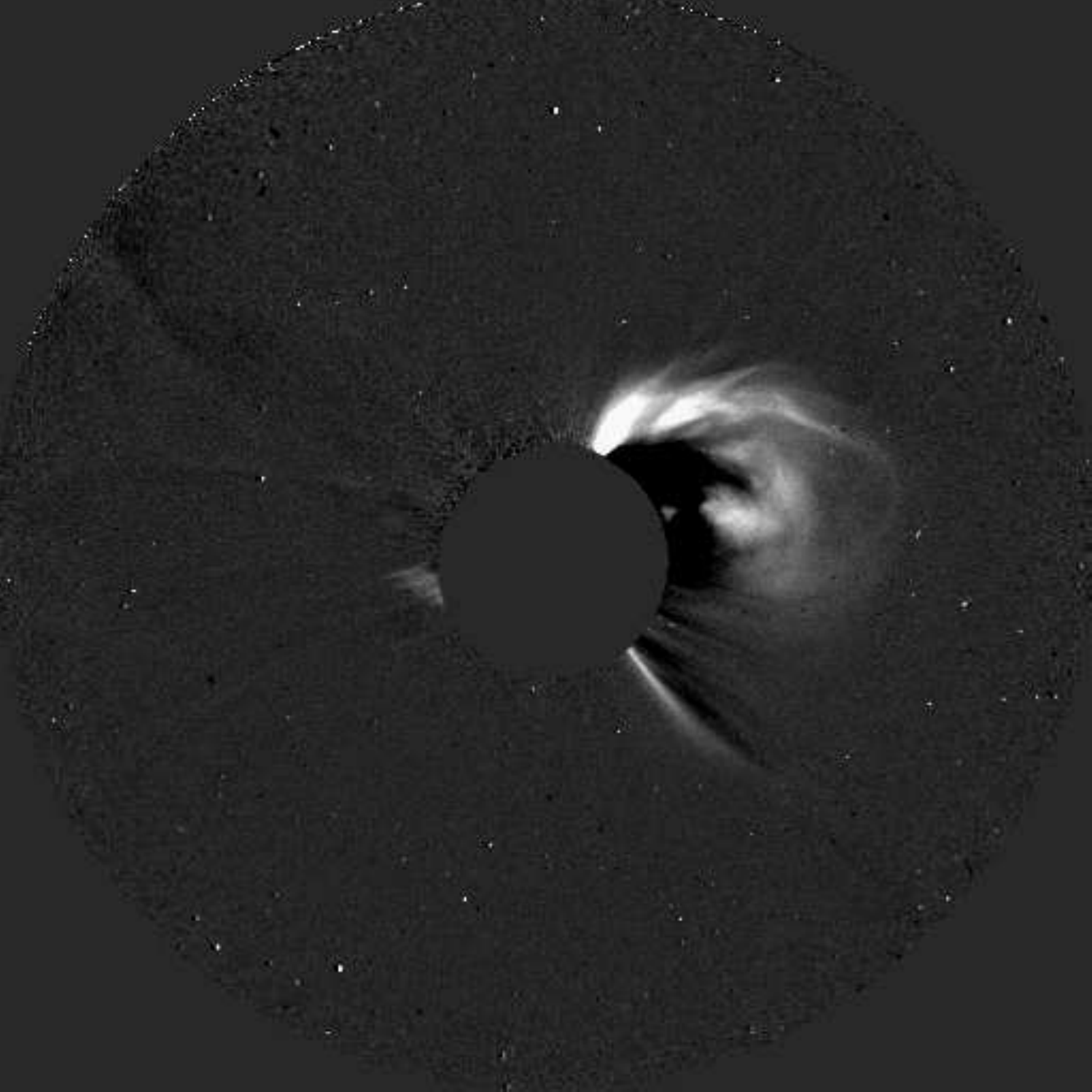}
               }
                 \centerline{\hspace*{0.04\textwidth}
              \includegraphics[width=0.4\textwidth,clip=]{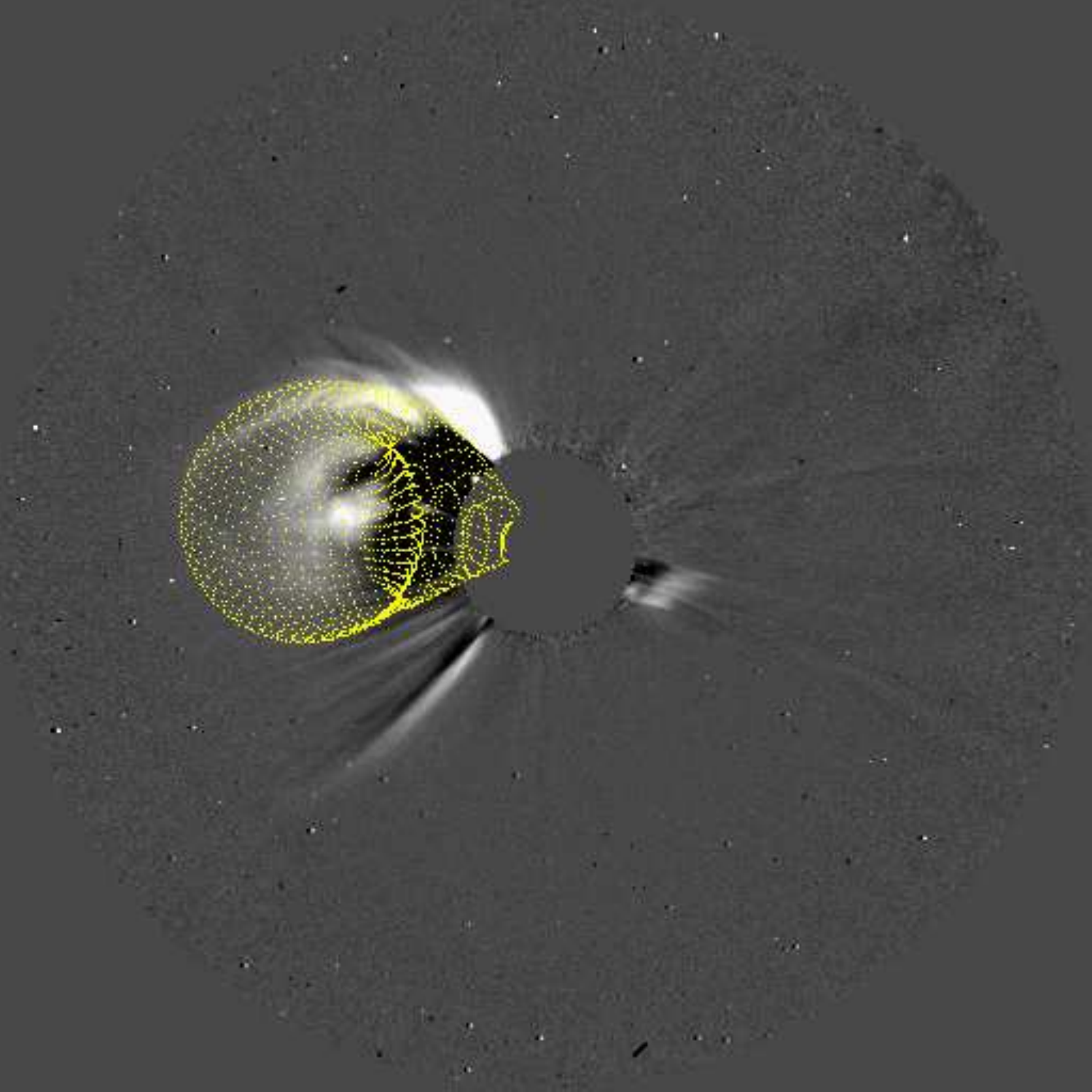}
               \hspace*{-0.02\textwidth}
               \includegraphics[width=0.4\textwidth,clip=]{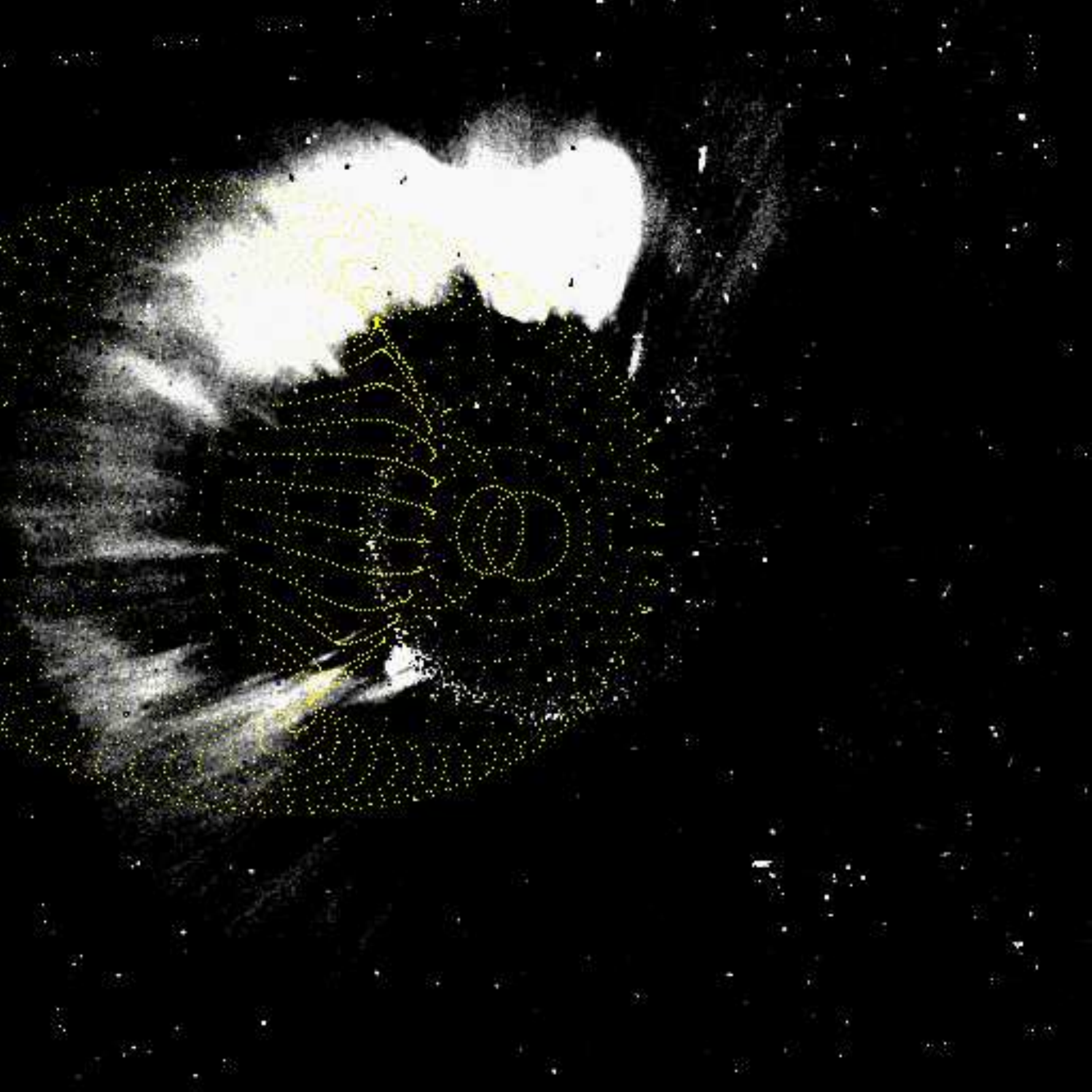}
              \hspace*{-0.02\textwidth}
               \includegraphics[width=0.4\textwidth,clip=]{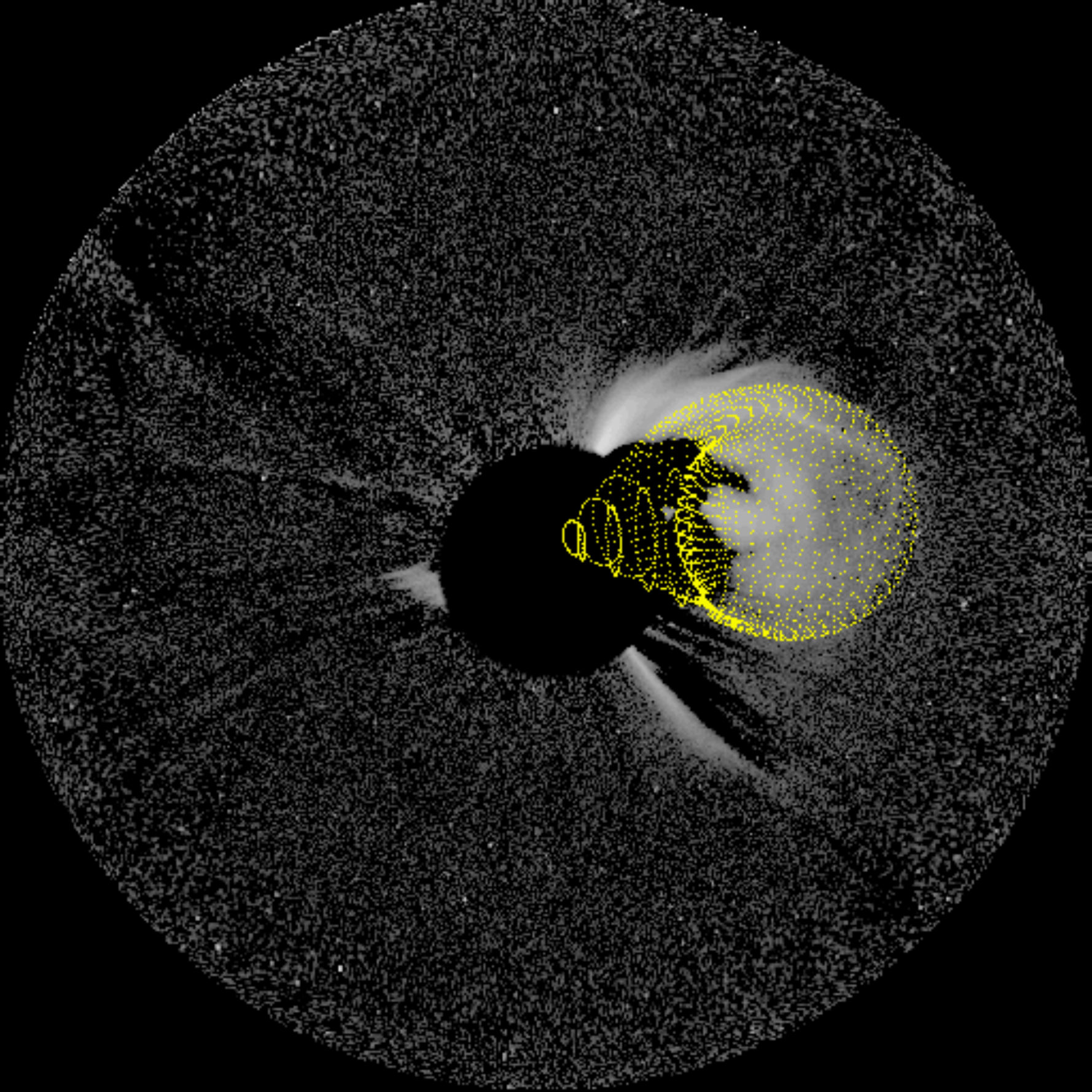}
                }
\vspace{0.0261\textwidth}  
\caption[GCS fit for CME 14 at 04:24]{GCS fit for June 21, 2011 (CME 14) at 04:24 at height 11.3 \Rs. Table \ref{tblapp} 
lists the GCS parameters for this event.}
\label{figa14}
\end{figure}

\begin{figure}[h]    
  \centering                              
   \centerline{\hspace*{0.00\textwidth}
               \includegraphics[width=0.4\textwidth,clip=]{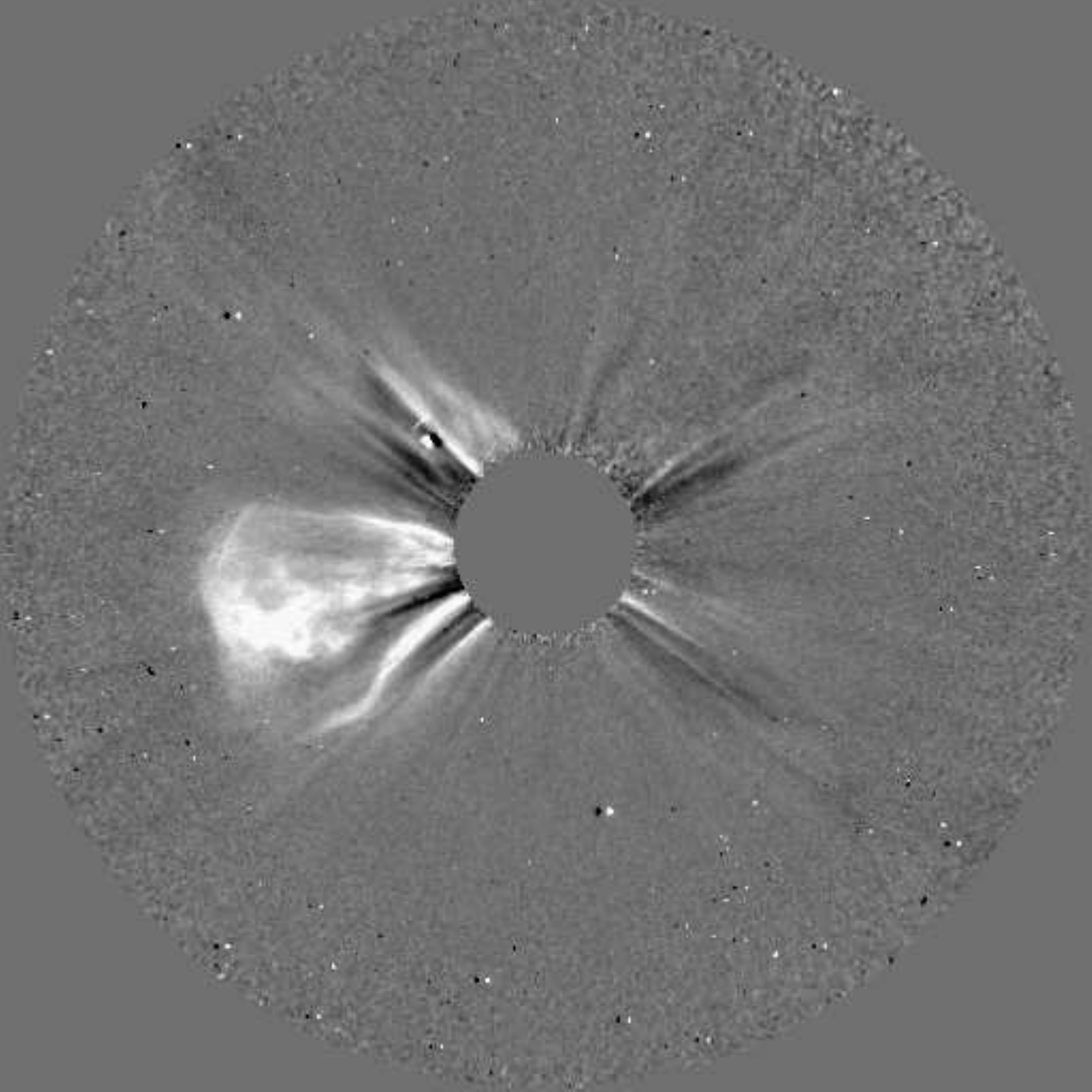}
                \hspace*{-0.02\textwidth}
               \includegraphics[width=0.4\textwidth,clip=]{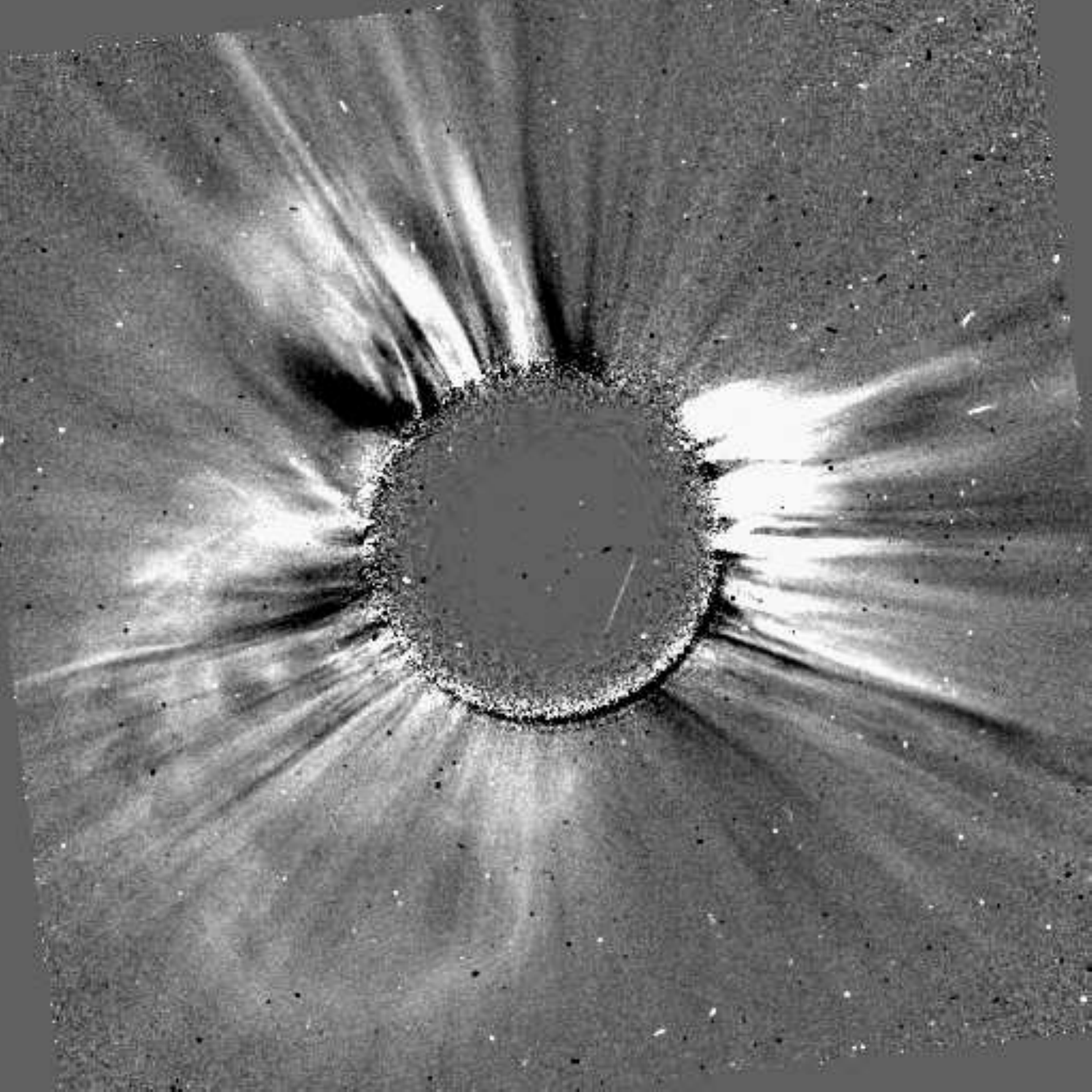}
             \hspace*{-0.02\textwidth}
               \includegraphics[width=0.4\textwidth,clip=]{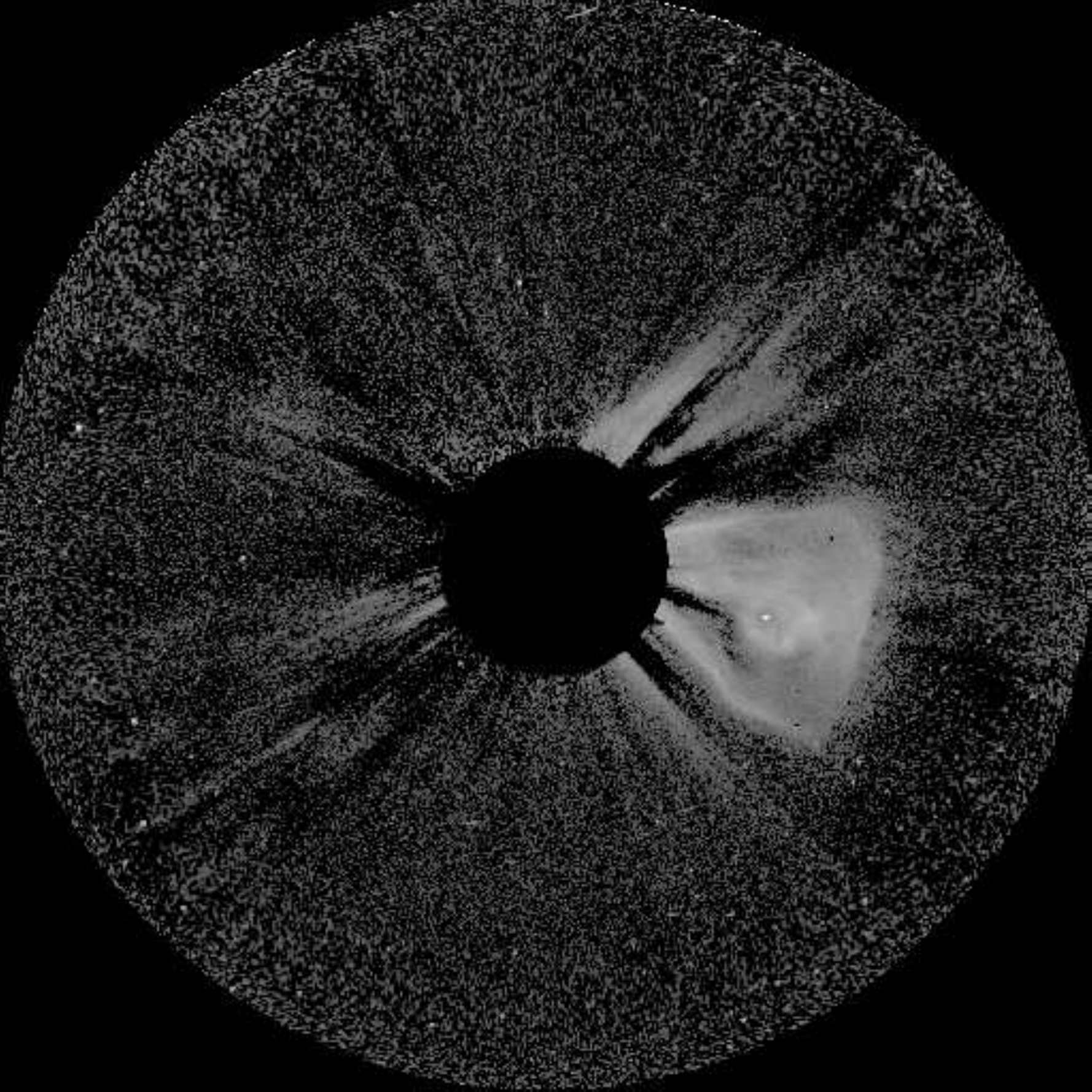}
               }
                 \centerline{\hspace*{0.0\textwidth}
              \includegraphics[width=0.4\textwidth,clip=]{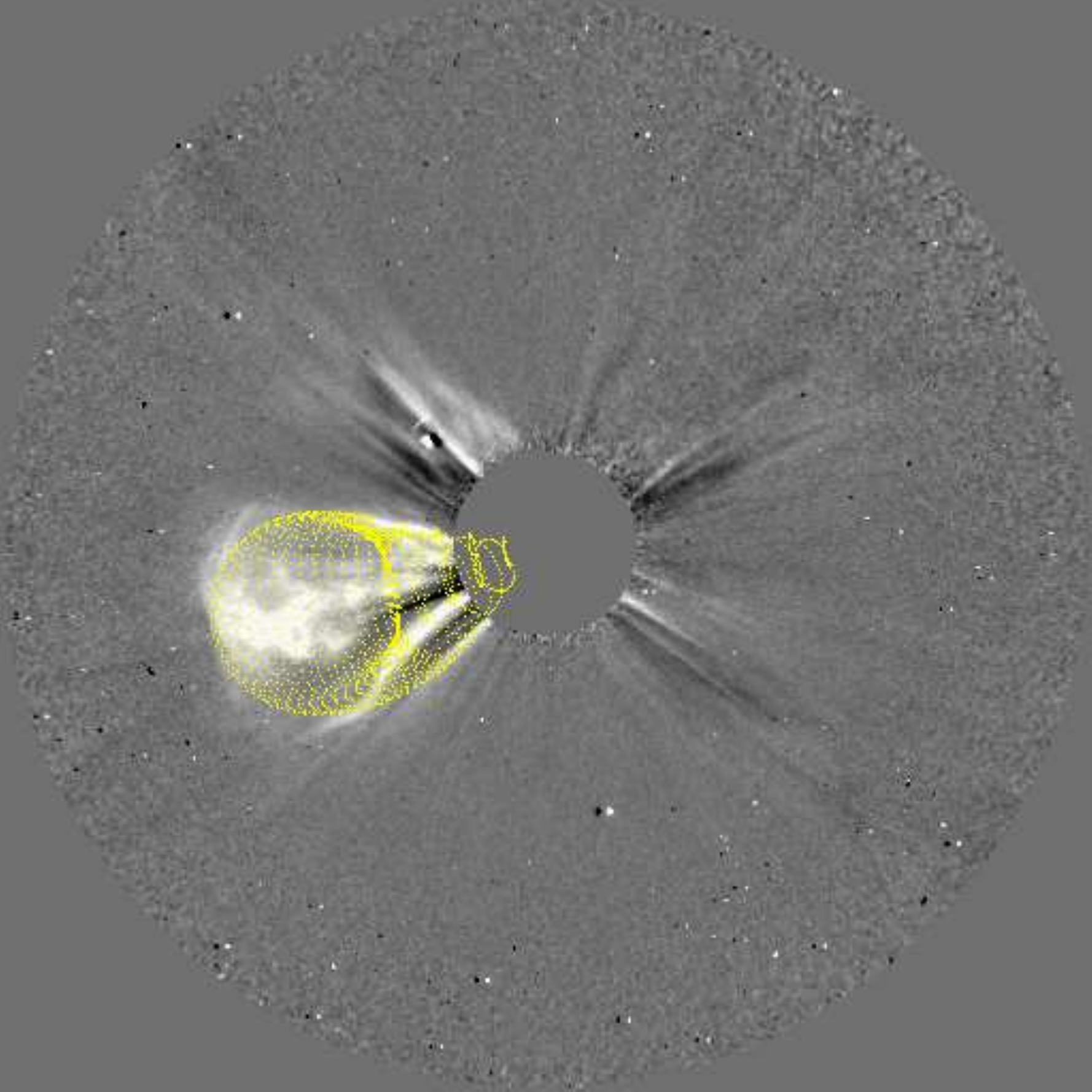}
               \hspace*{-0.02\textwidth}
               \includegraphics[width=0.4\textwidth,clip=]{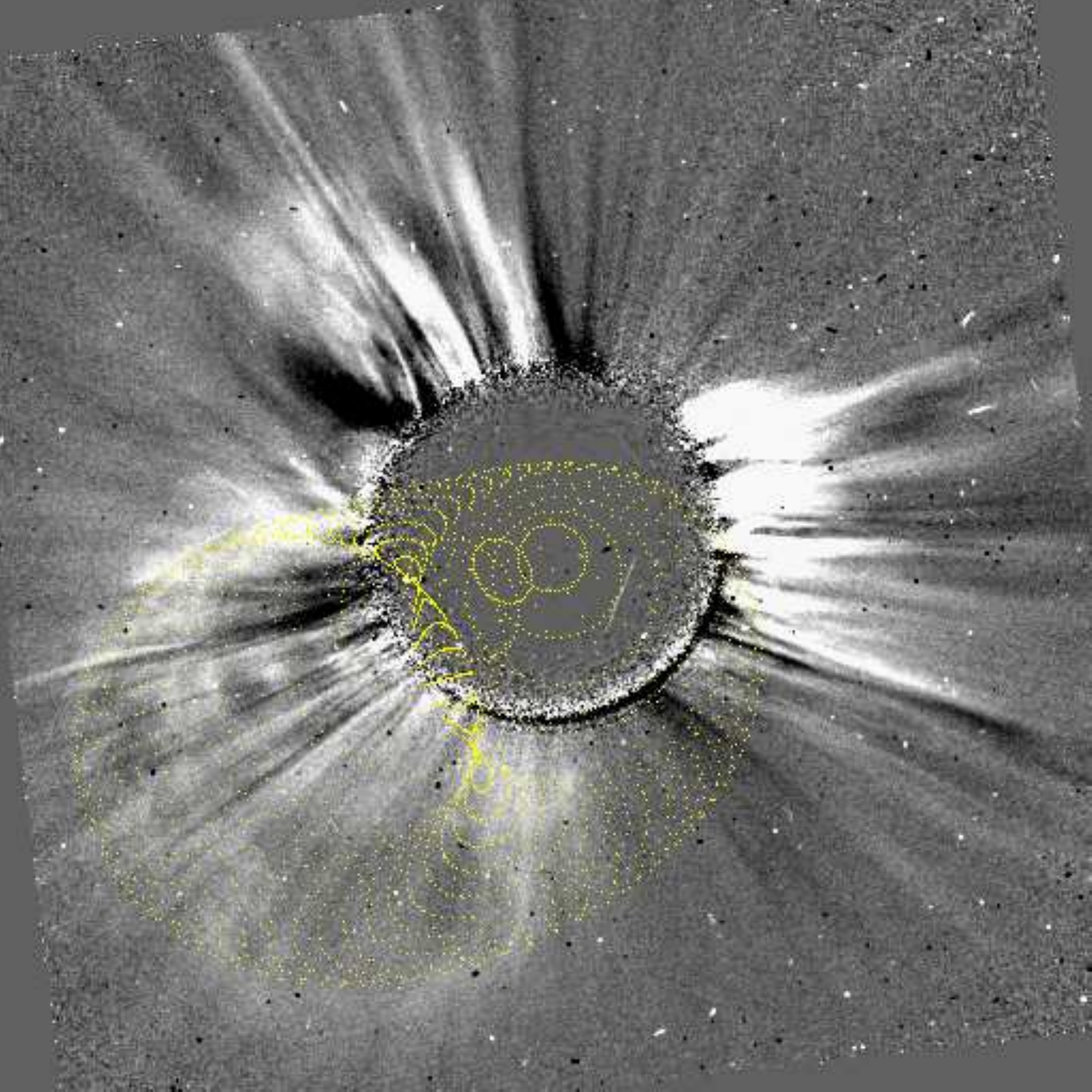}
              \hspace*{-0.02\textwidth}
               \includegraphics[width=0.4\textwidth,clip=]{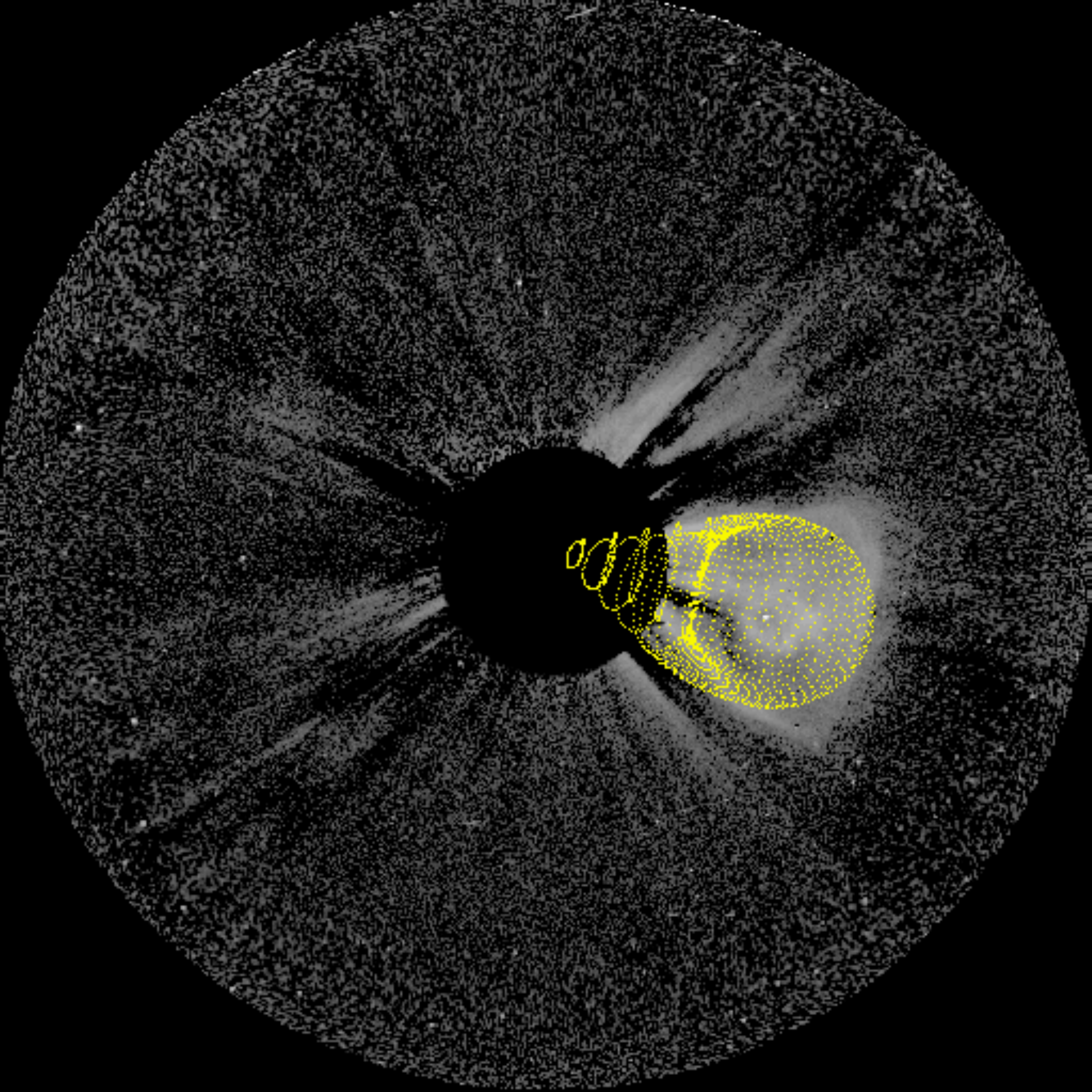}
                }
\vspace{0.0261\textwidth}  
\caption[GCS fit for CME 15 at 02:24]{GCS fit for CME 15 on July 09, 2011 at 02:24 UT at height $H=10.0$ \Rs. Table \ref{tblapp} 
lists the GCS parameters for this event.}
\label{figa15}
\end{figure}

\begin{figure}[h]    
  \centering                              
   \centerline{\hspace*{0.04\textwidth}
               \includegraphics[width=0.4\textwidth,clip=]{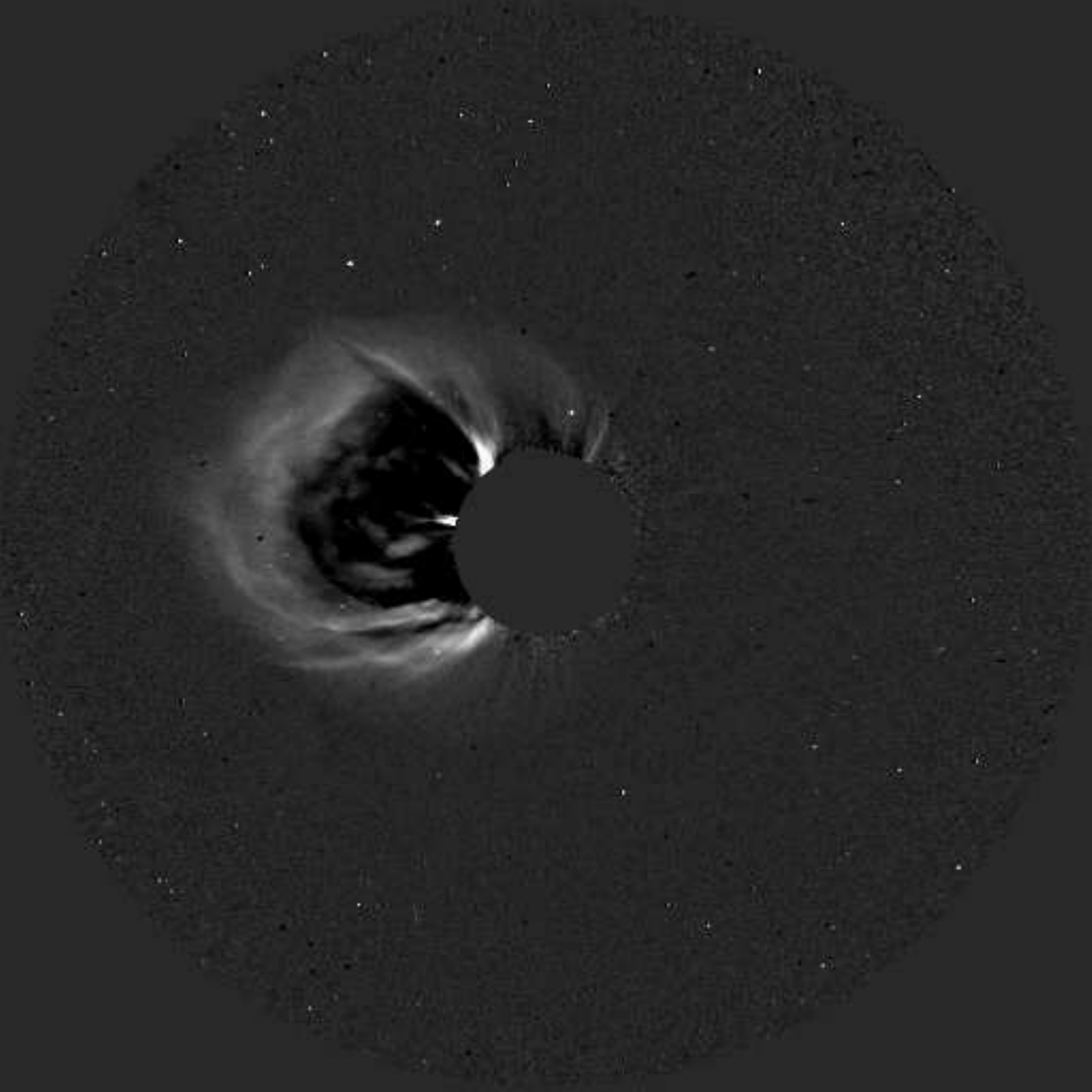}
                \hspace*{-0.02\textwidth}
               \includegraphics[width=0.4\textwidth,clip=]{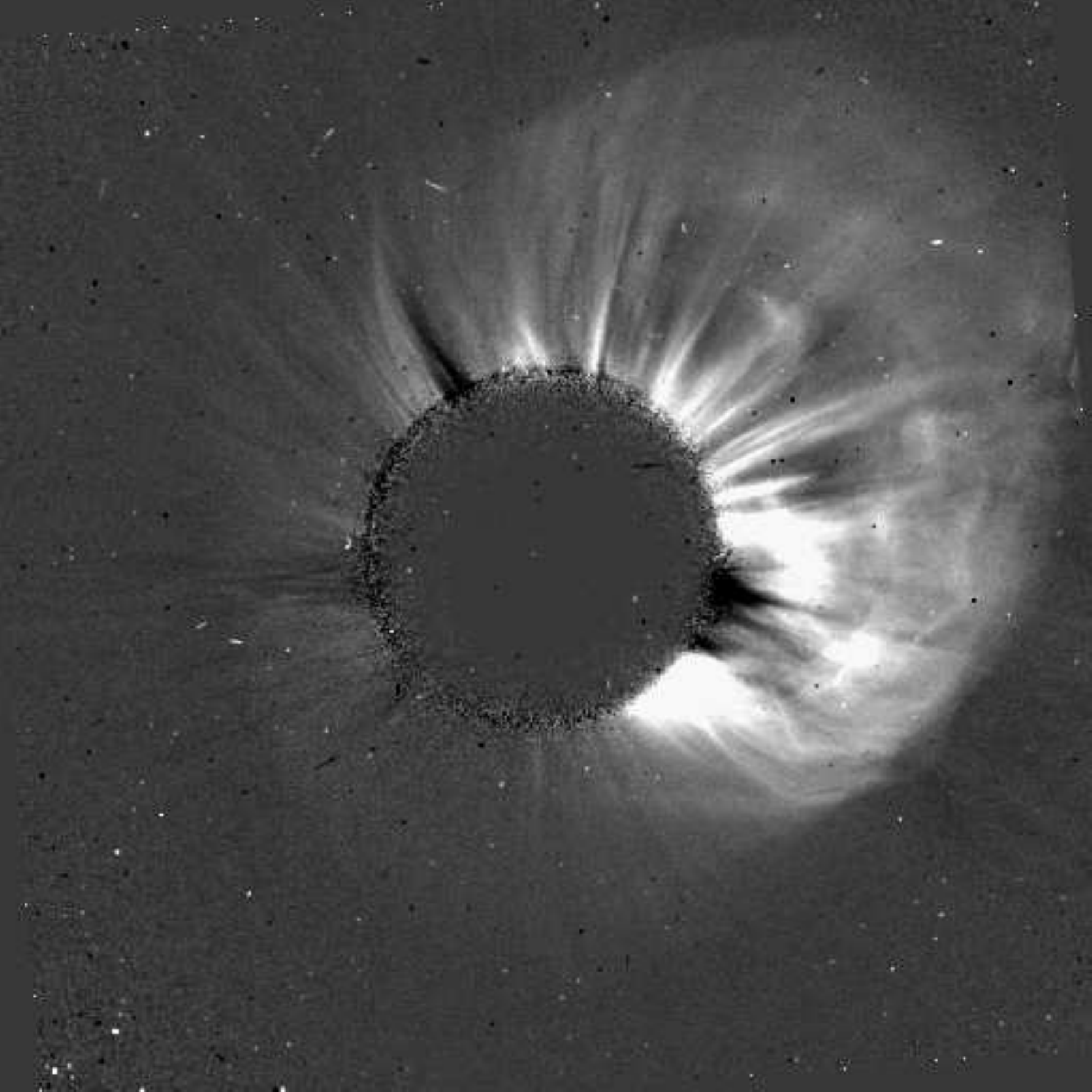}
             \hspace*{-0.02\textwidth}
               \includegraphics[width=0.4\textwidth,clip=]{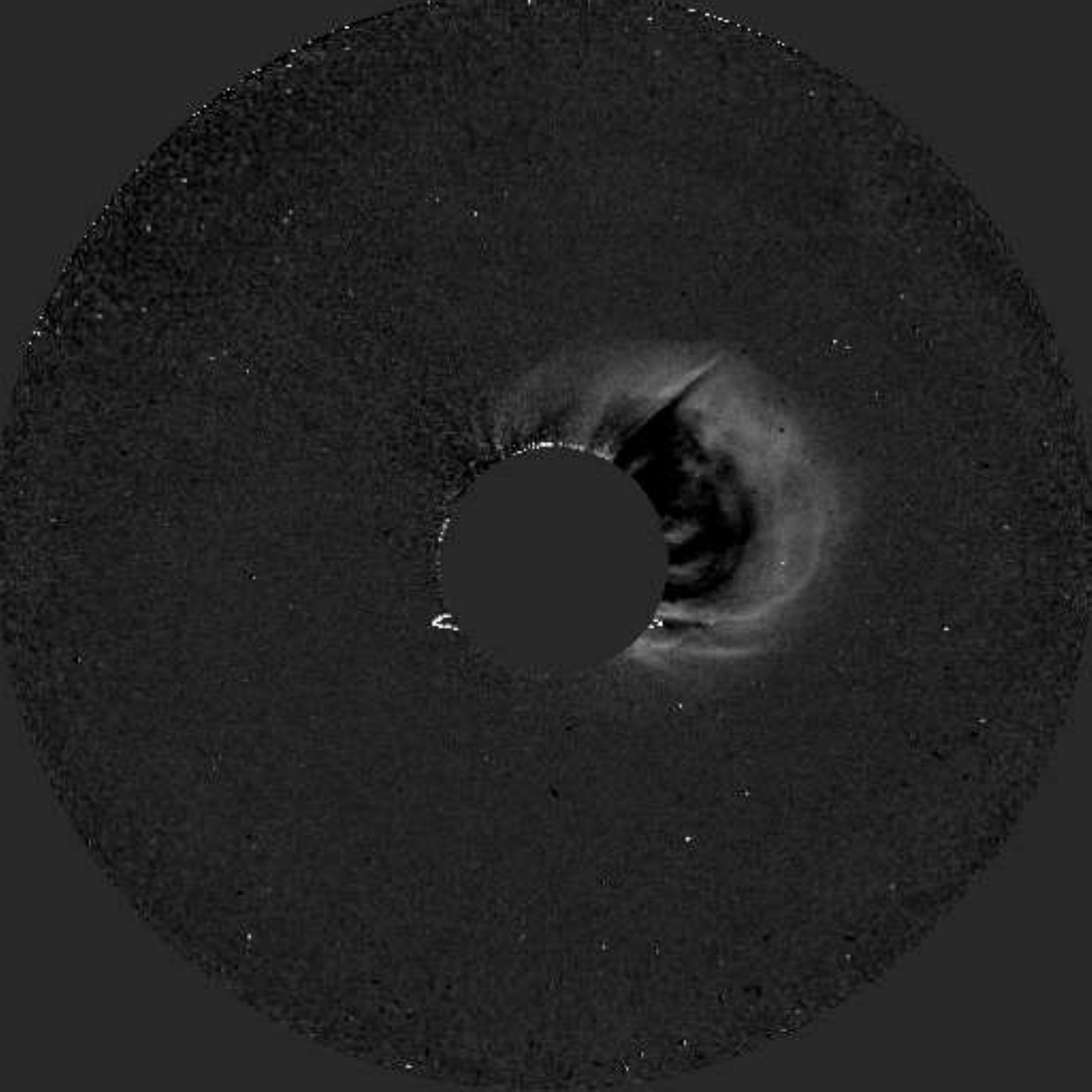}
               }
                 \centerline{\hspace*{0.04\textwidth}
              \includegraphics[width=0.4\textwidth,clip=]{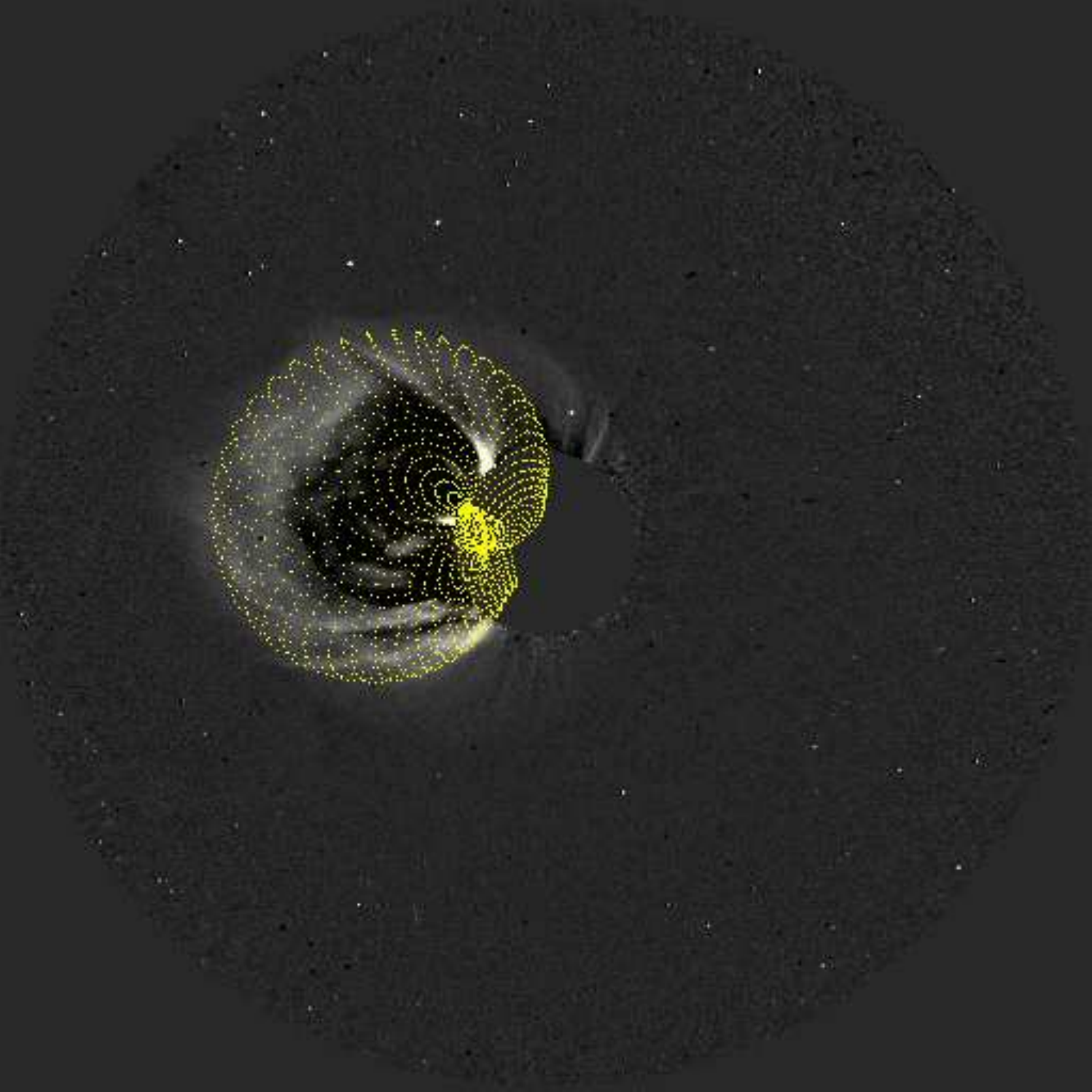}
               \hspace*{-0.02\textwidth}
               \includegraphics[width=0.4\textwidth,clip=]{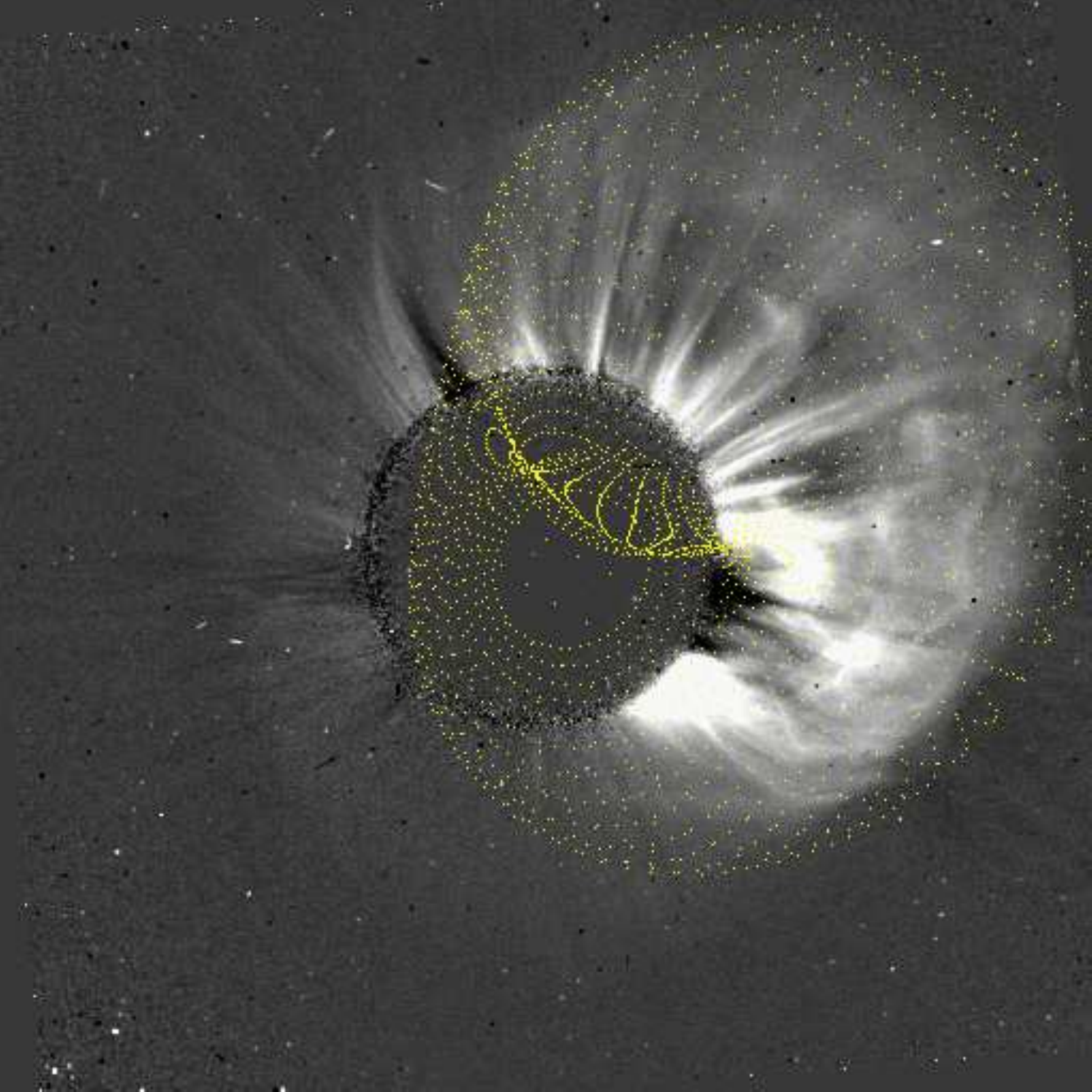}
              \hspace*{-0.02\textwidth}
               \includegraphics[width=0.4\textwidth,clip=]{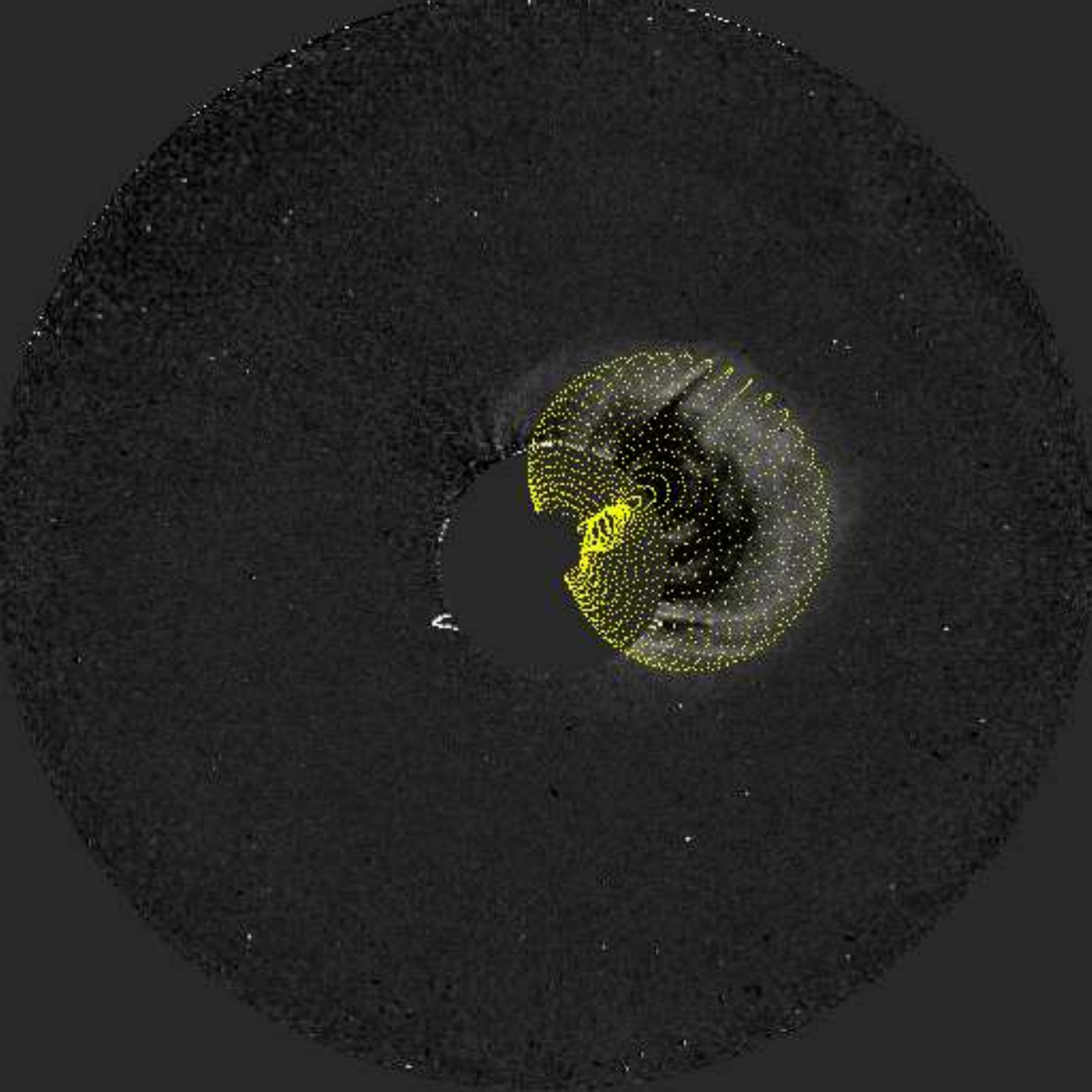}
                }
\vspace{0.0261\textwidth}  
\caption[GCS fit for CME 16 at 04:39]{GCS fit for CME 16 on August 04, 2011 at 04:39 UT at height $H=9.4$ \Rs. Table \ref{tblapp} 
lists the GCS parameters for this event.}
\label{figa16}
\end{figure}

\begin{figure}[h]    
  \centering                              
   \centerline{\hspace*{0.00\textwidth}
               \includegraphics[width=0.4\textwidth,clip=]{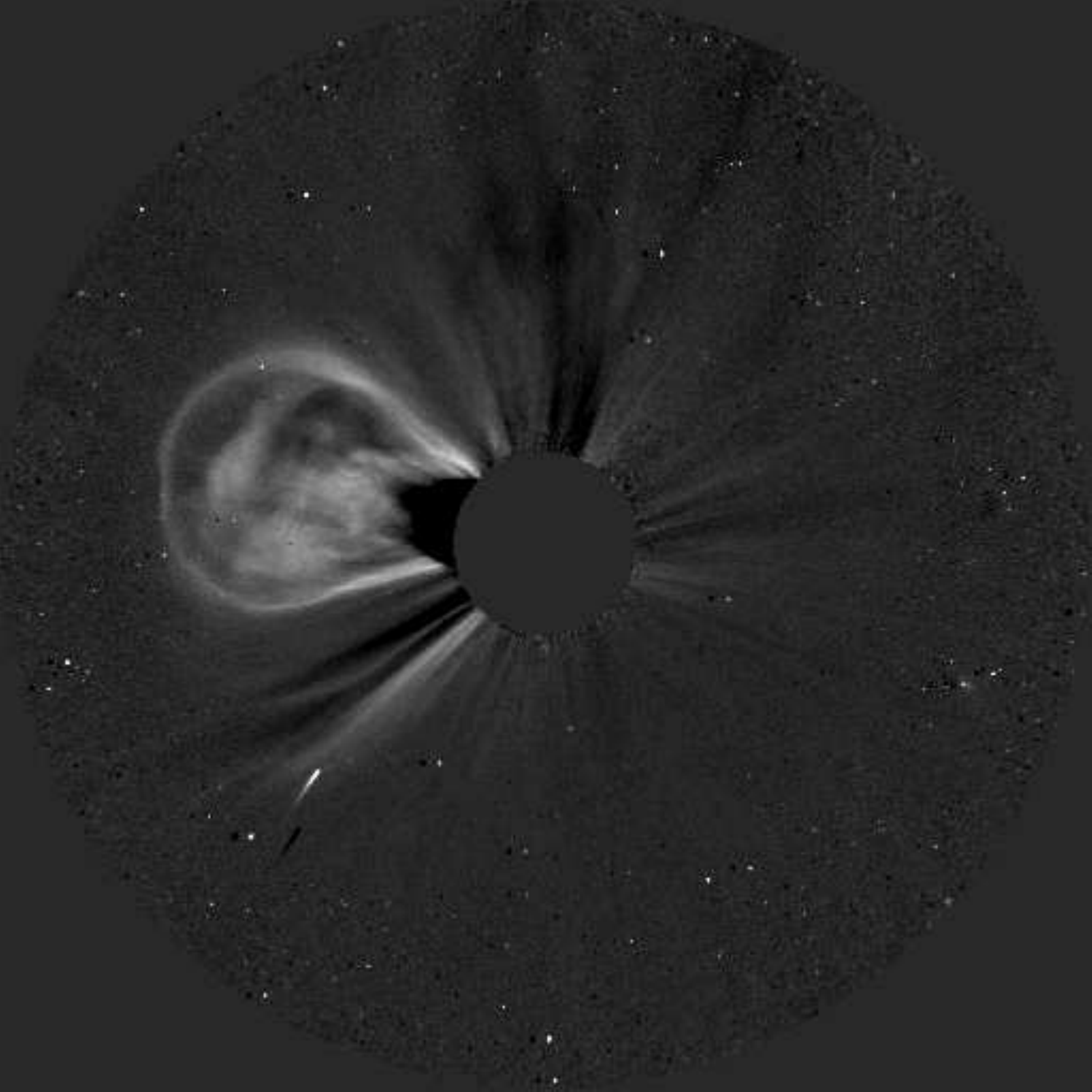}
                \hspace*{-0.02\textwidth}
               \includegraphics[width=0.4\textwidth,clip=]{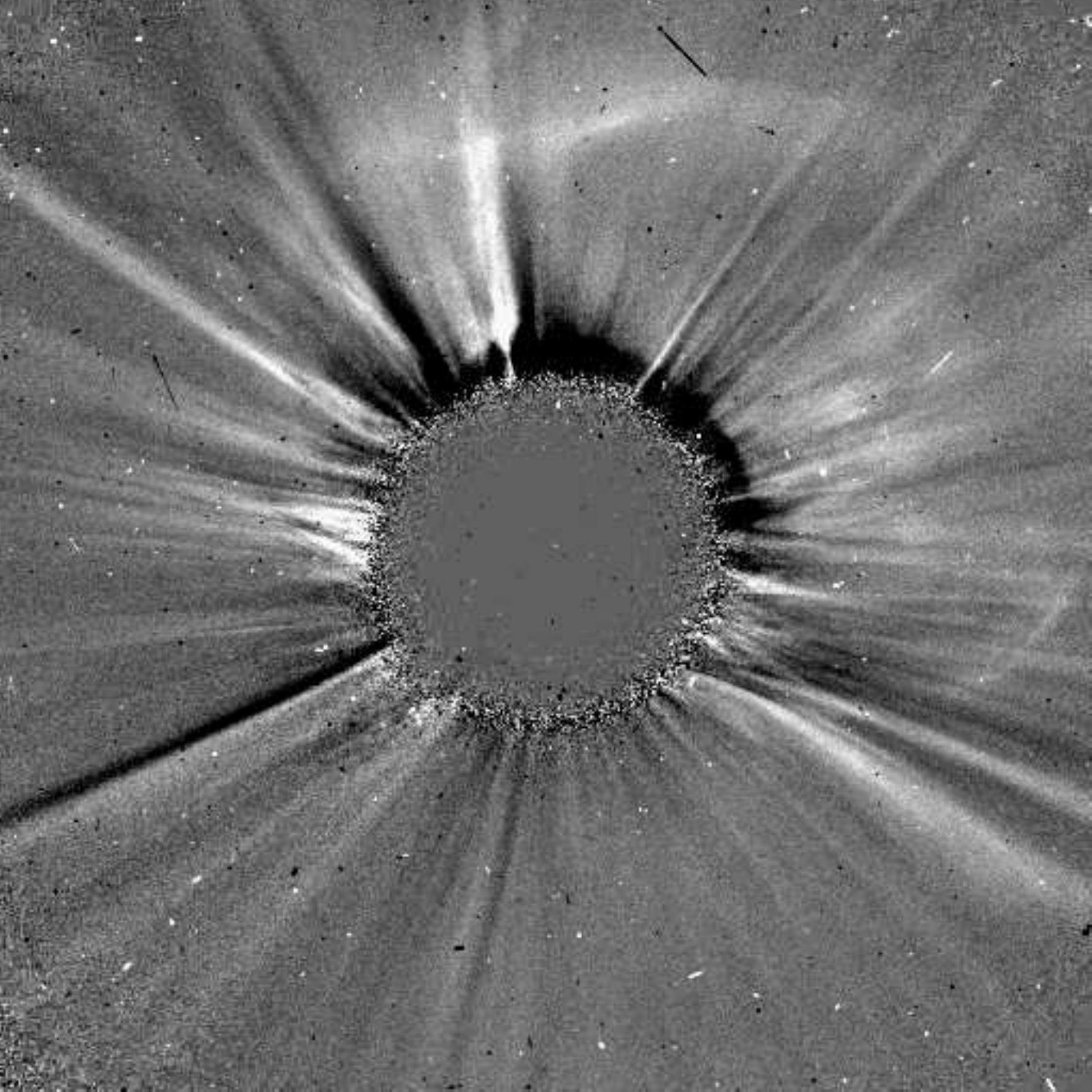}
             \hspace*{-0.02\textwidth}
               \includegraphics[width=0.4\textwidth,clip=]{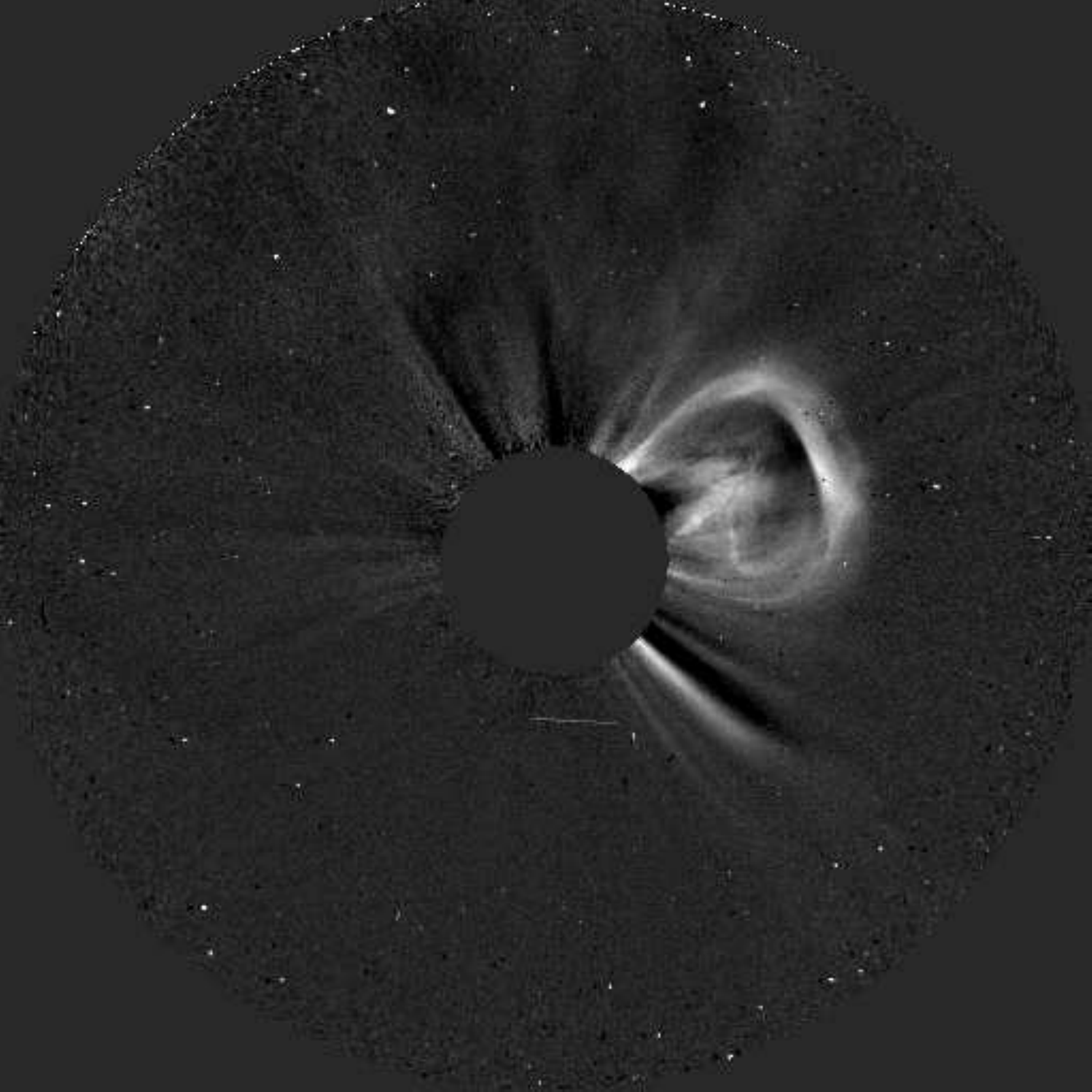}
               }
                 \centerline{\hspace*{0.0\textwidth}
              \includegraphics[width=0.4\textwidth,clip=]{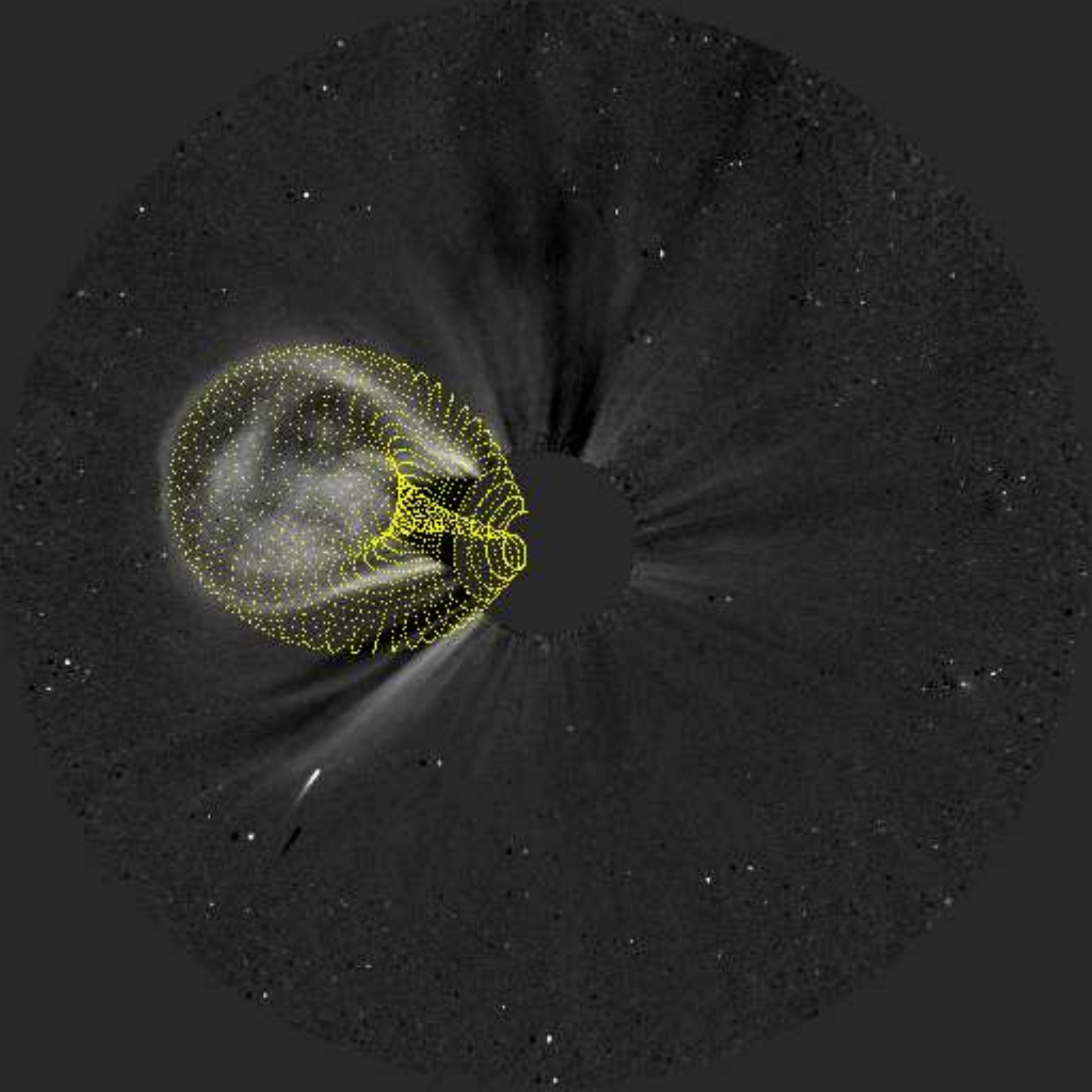}
               \hspace*{-0.02\textwidth}
               \includegraphics[width=0.4\textwidth,clip=]{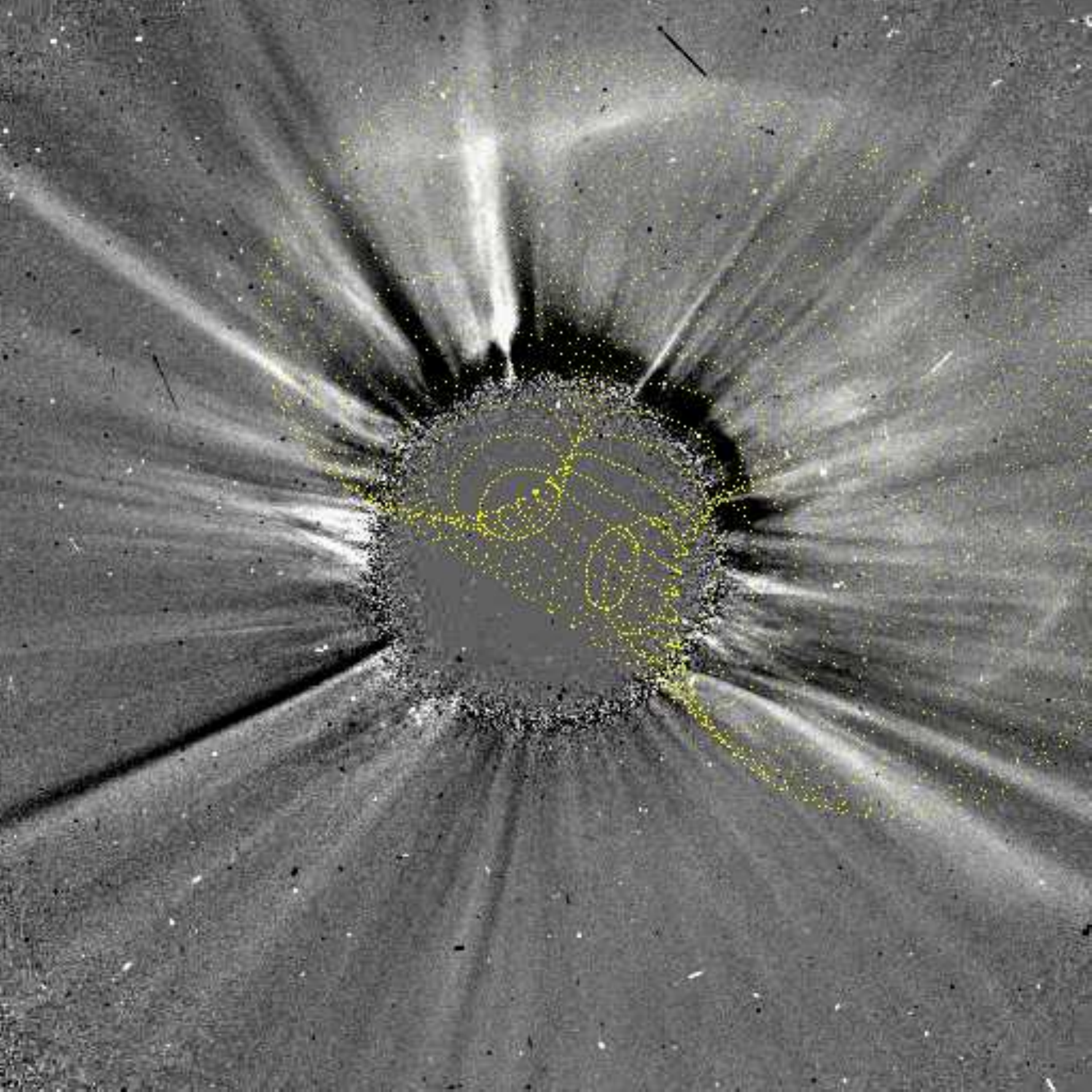}
              \hspace*{-0.02\textwidth}
               \includegraphics[width=0.4\textwidth,clip=]{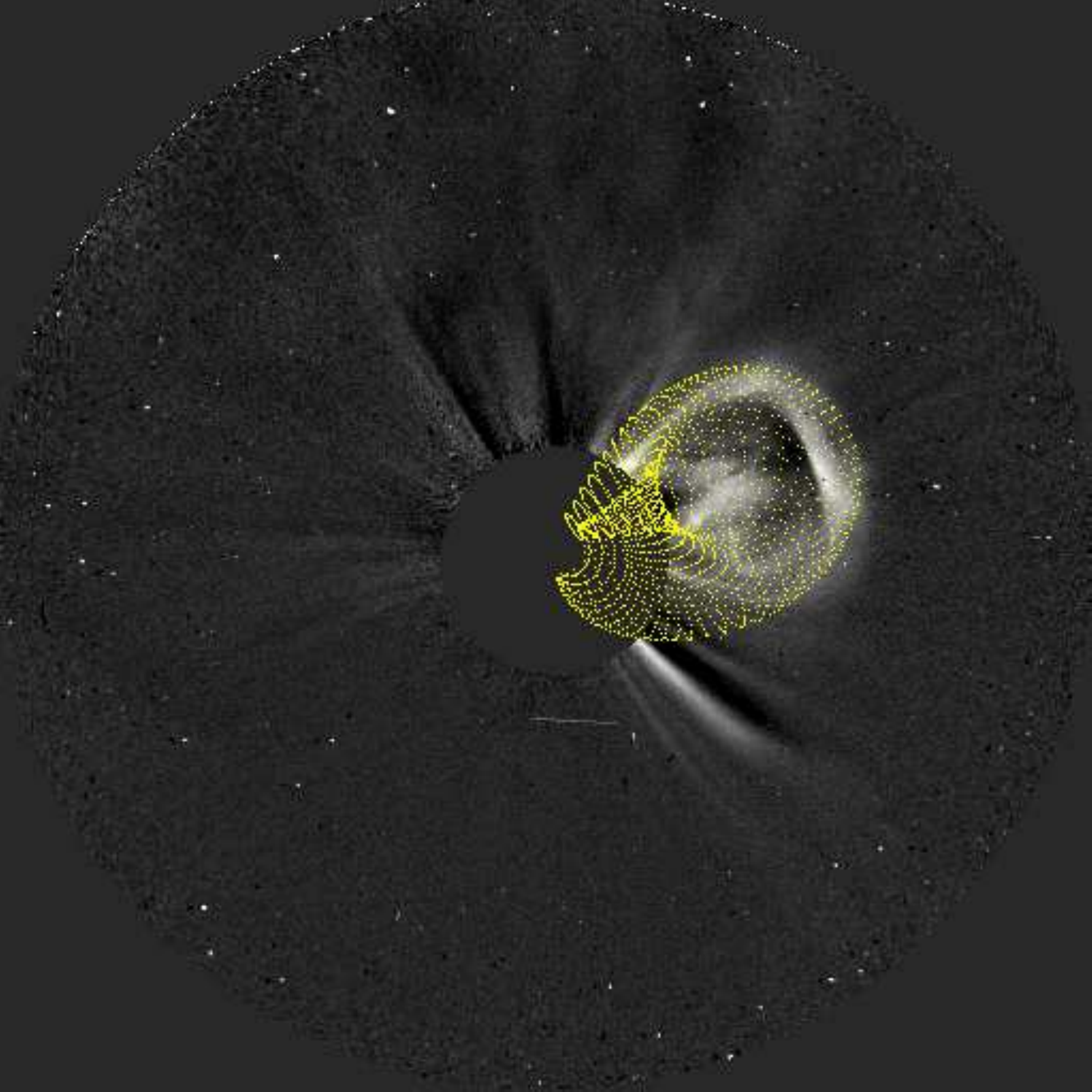}
                }
\vspace{0.0261\textwidth}  
\caption[GCS fit for CME 17 at 02:24]{GCS fit for CME 17 on September 14, 2011 at 02:24 UT at height $H=10.5$ \Rs. Table \ref{tblapp} 
lists the GCS parameters for this event.}
\label{figa17}
\end{figure}

\begin{figure}[h]    
  \centering                              
   \centerline{\hspace*{0.04\textwidth}
               \includegraphics[width=0.4\textwidth,clip=]{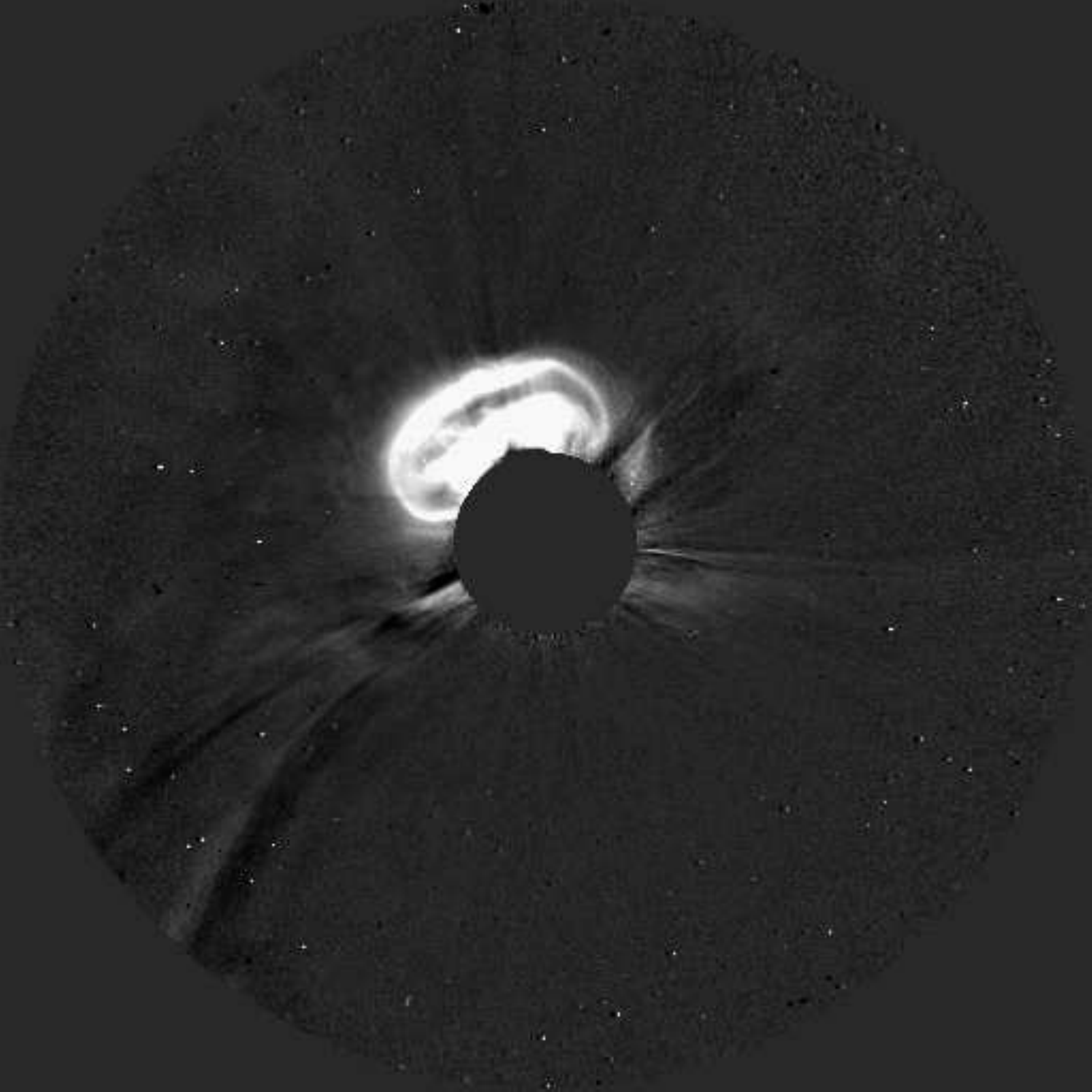}
                \hspace*{-0.02\textwidth}
               \includegraphics[width=0.4\textwidth,clip=]{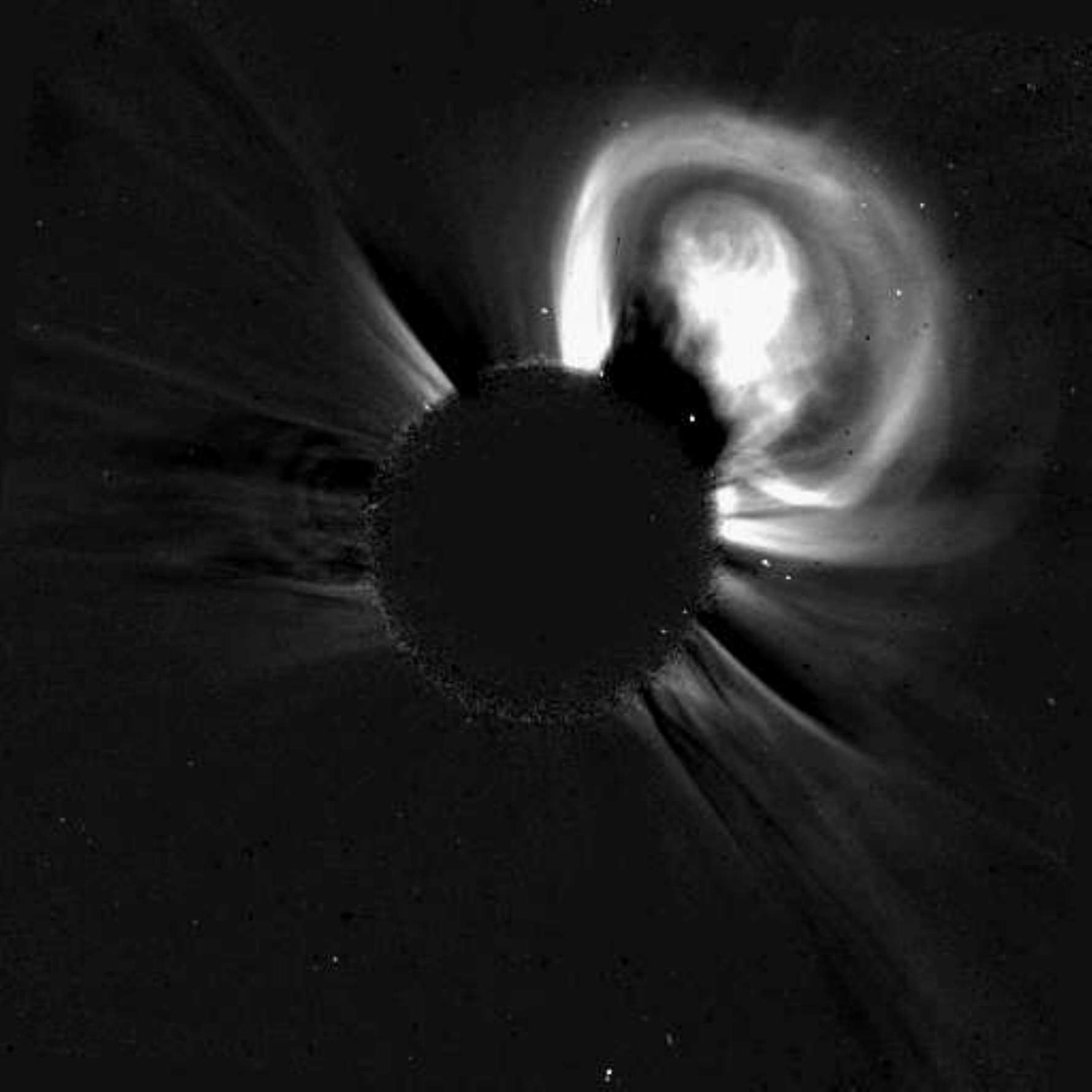}
             \hspace*{-0.02\textwidth}
               \includegraphics[width=0.4\textwidth,clip=]{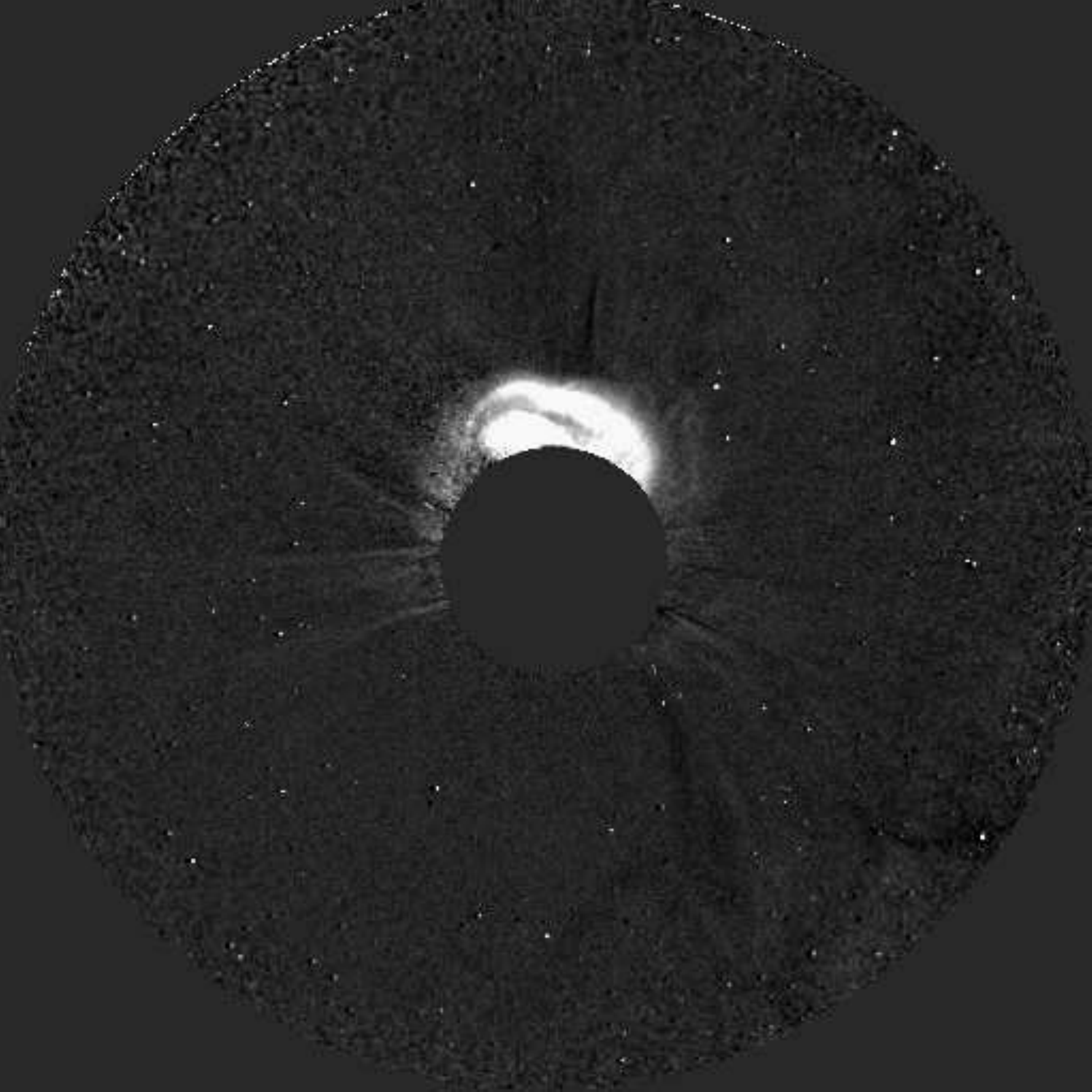}
               }
                 \centerline{\hspace*{0.04\textwidth}
              \includegraphics[width=0.4\textwidth,clip=]{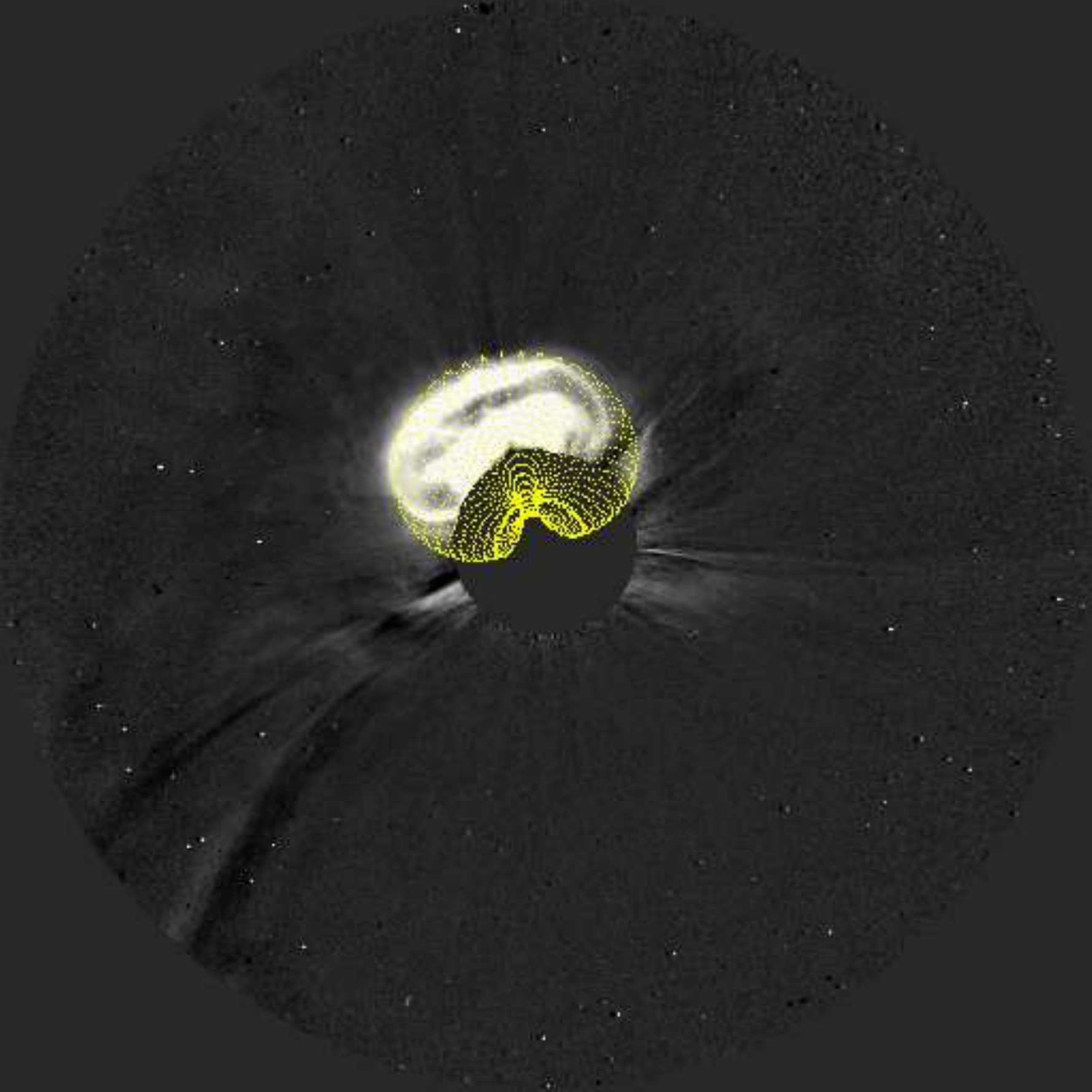}
               \hspace*{-0.02\textwidth}
               \includegraphics[width=0.4\textwidth,clip=]{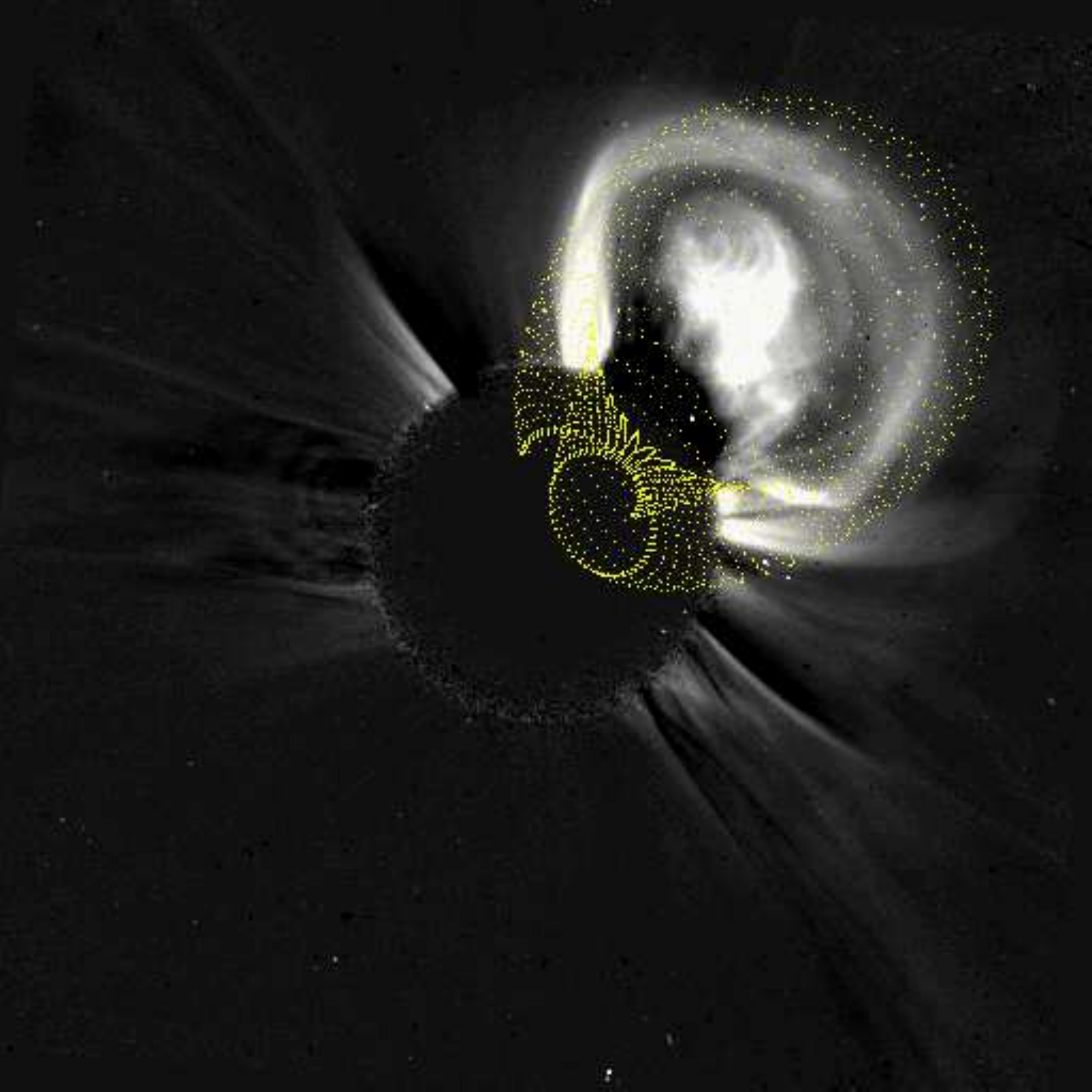}
              \hspace*{-0.02\textwidth}
               \includegraphics[width=0.4\textwidth,clip=]{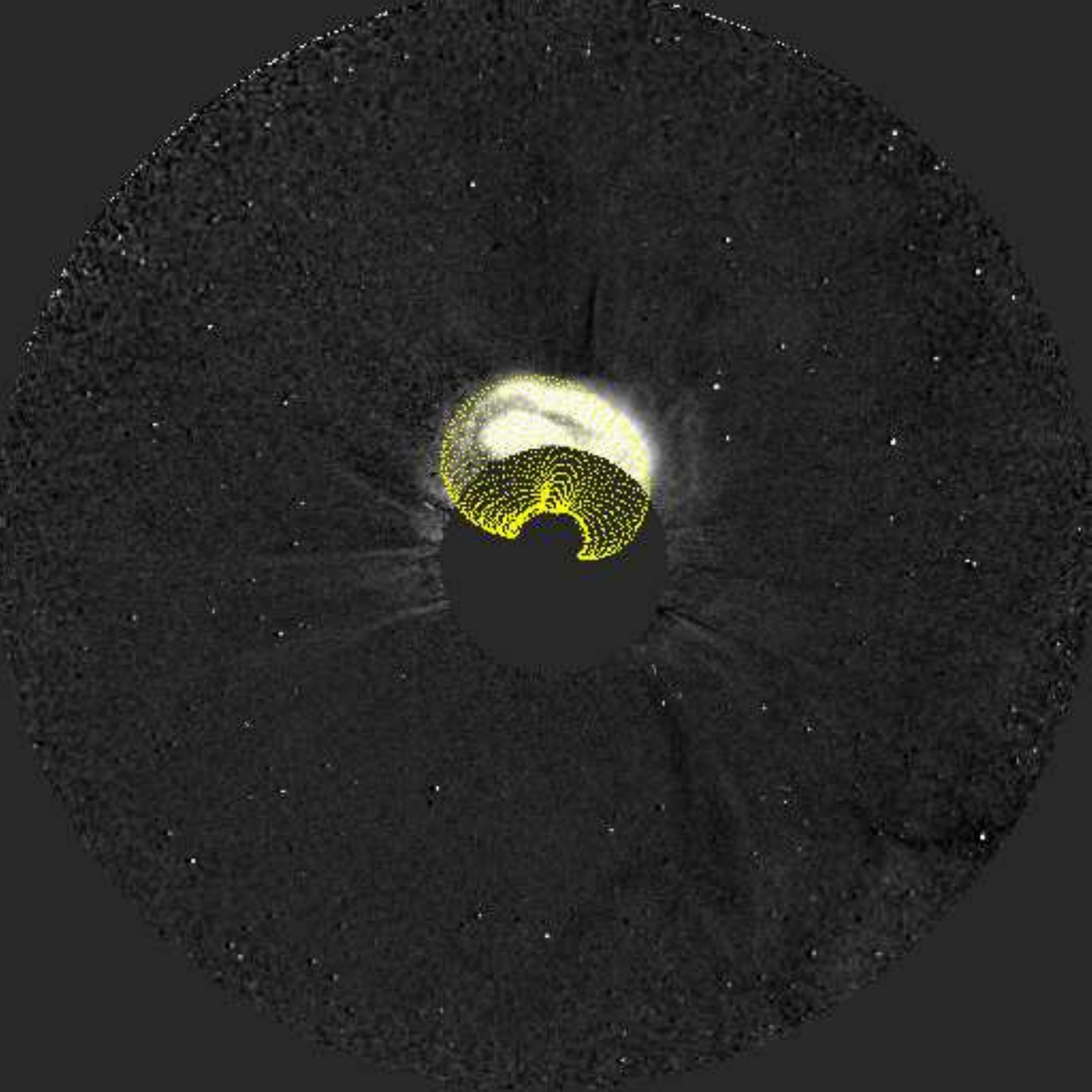}
                }
\vspace{0.0261\textwidth}  
\caption[GCS fit for CME 18 at 11:24]{GCS fit for CME 18 on October 22, 2011 at 11:24 at height $H=6.2$ \Rs. Table \ref{tblapp} 
lists the GCS parameters for this event.}
\label{figa18}
\end{figure}

\begin{figure}[h]    
  \centering                              
   \centerline{\hspace*{0.00\textwidth}
               \includegraphics[width=0.4\textwidth,clip=]{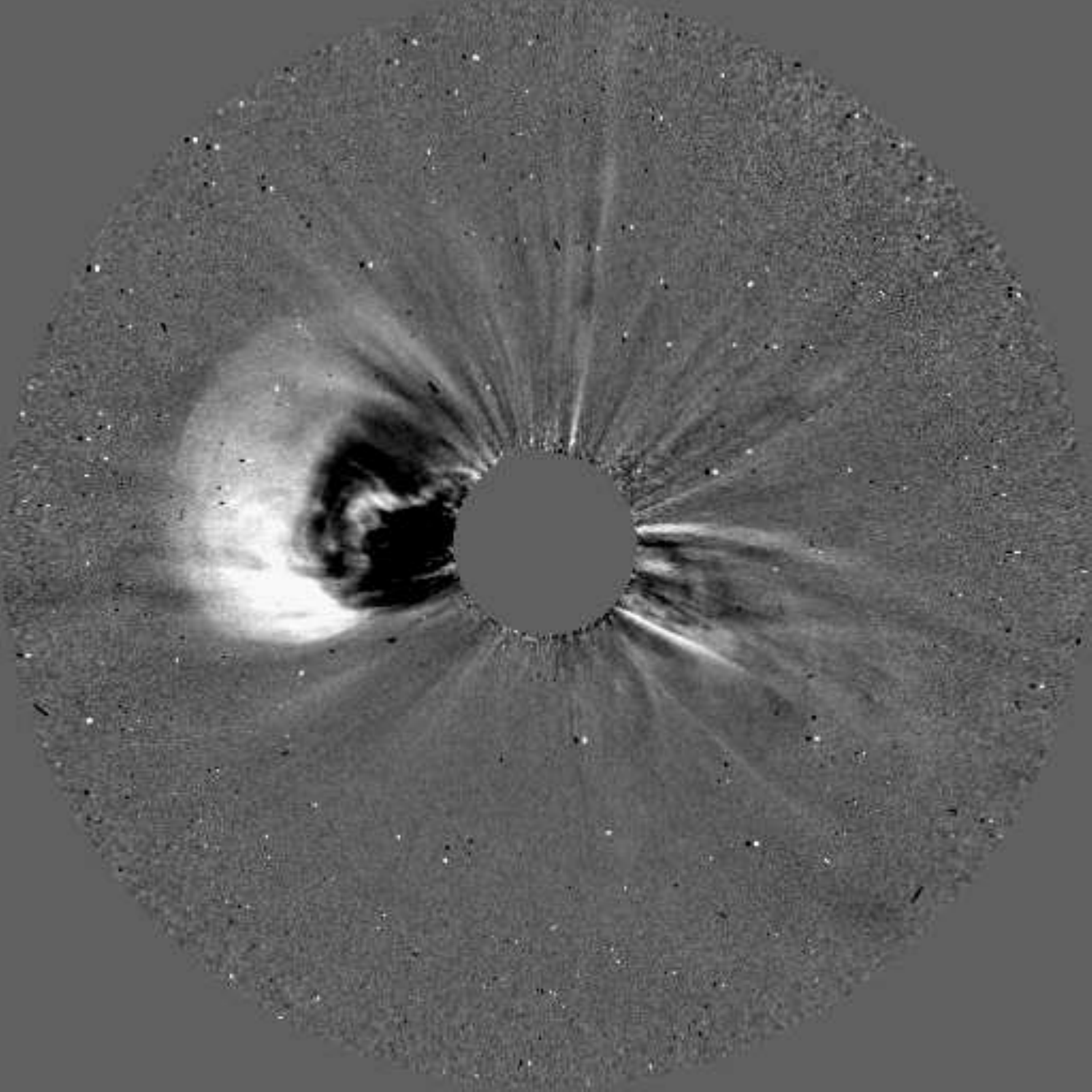}
                \hspace*{-0.02\textwidth}
               \includegraphics[width=0.4\textwidth,clip=]{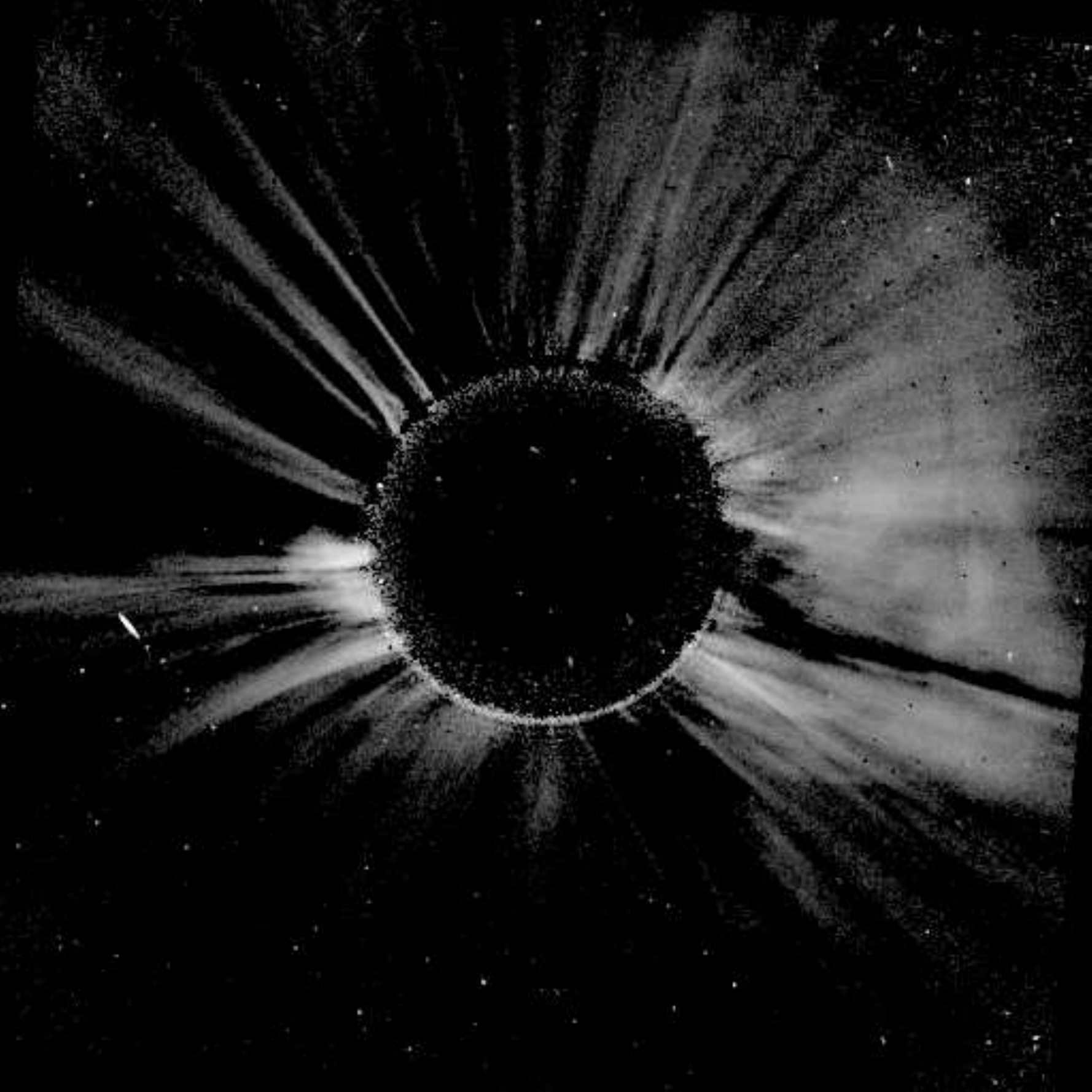}
             \hspace*{-0.02\textwidth}
               \includegraphics[width=0.4\textwidth,clip=]{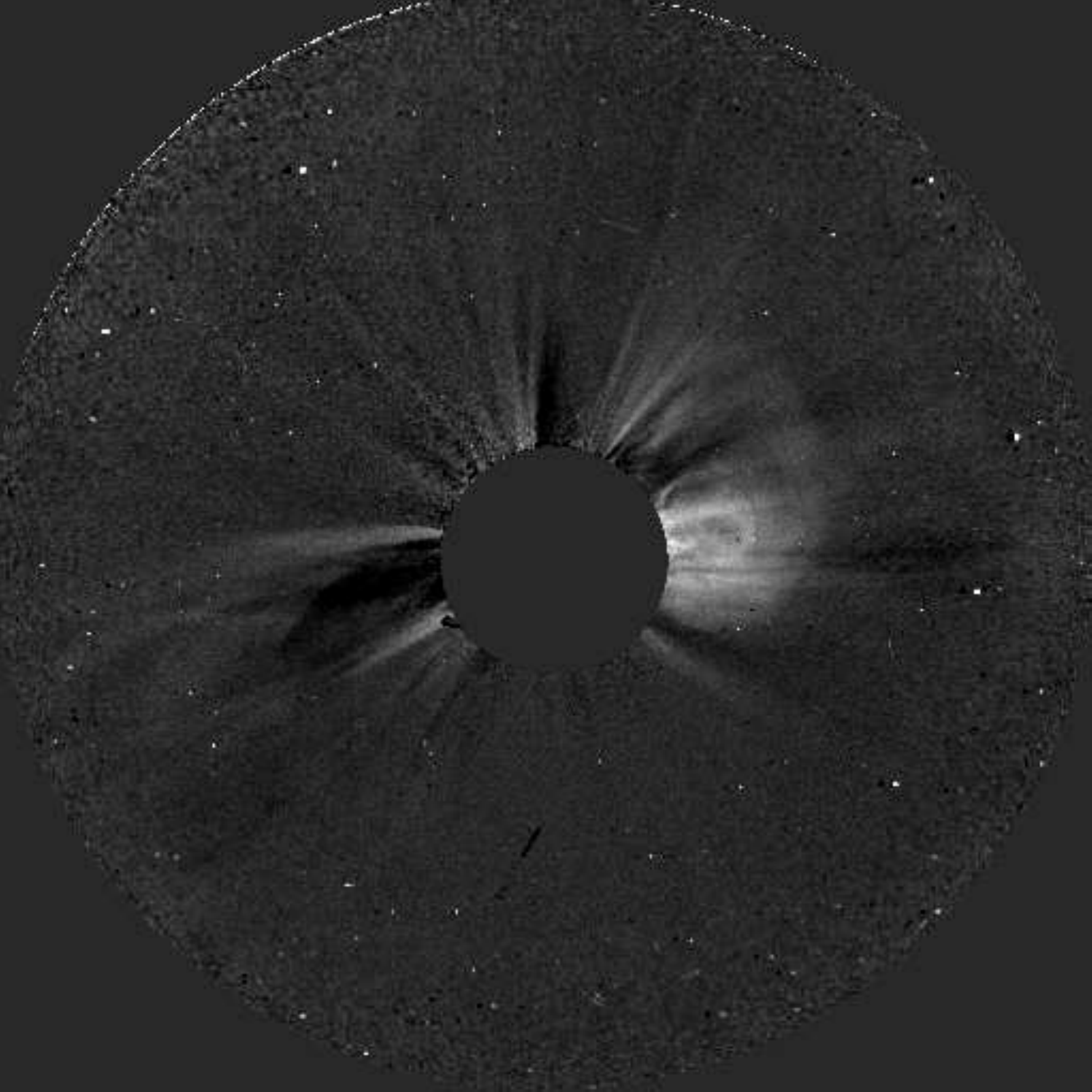}
               }
                 \centerline{\hspace*{0.0\textwidth}
              \includegraphics[width=0.4\textwidth,clip=]{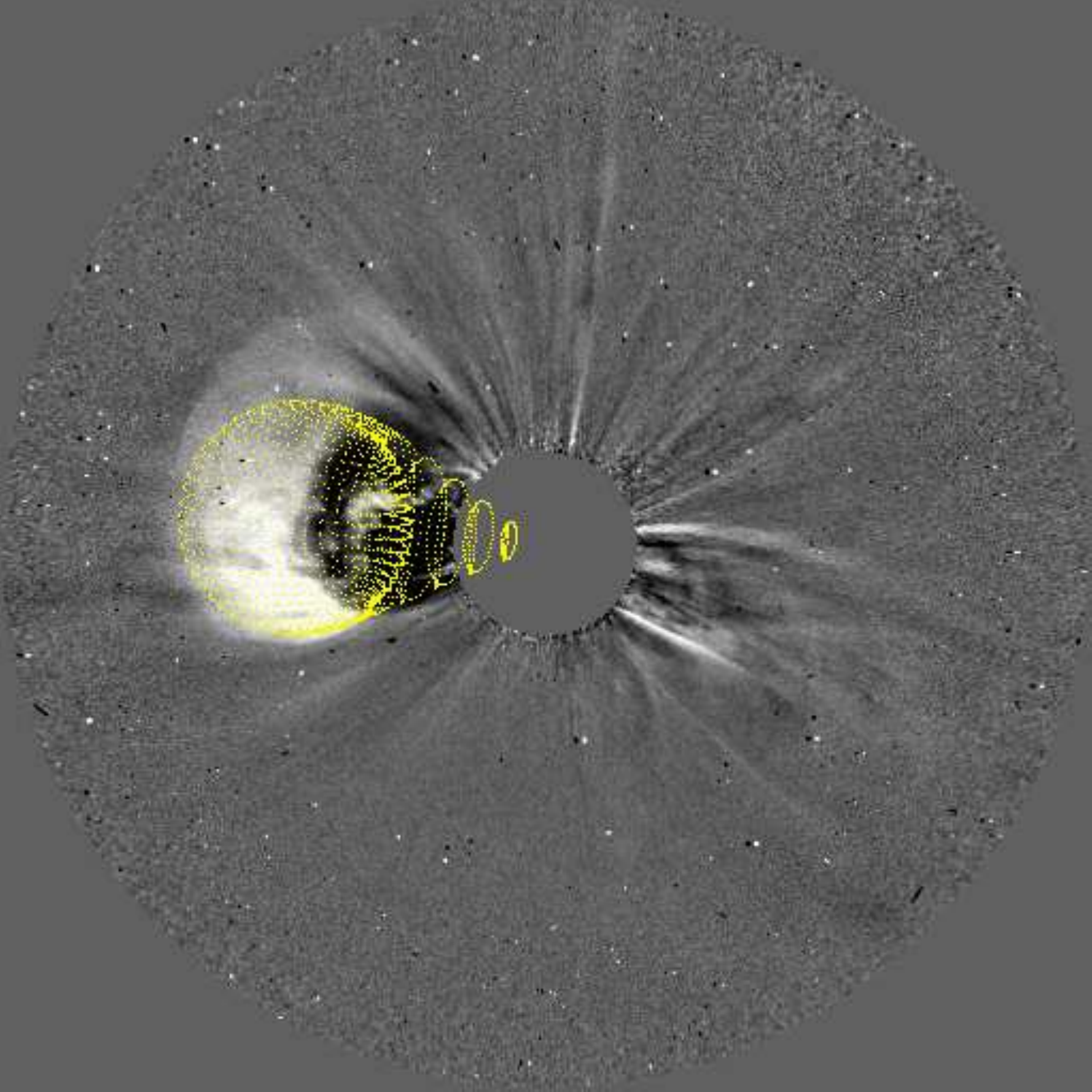}
               \hspace*{-0.02\textwidth}
               \includegraphics[width=0.4\textwidth,clip=]{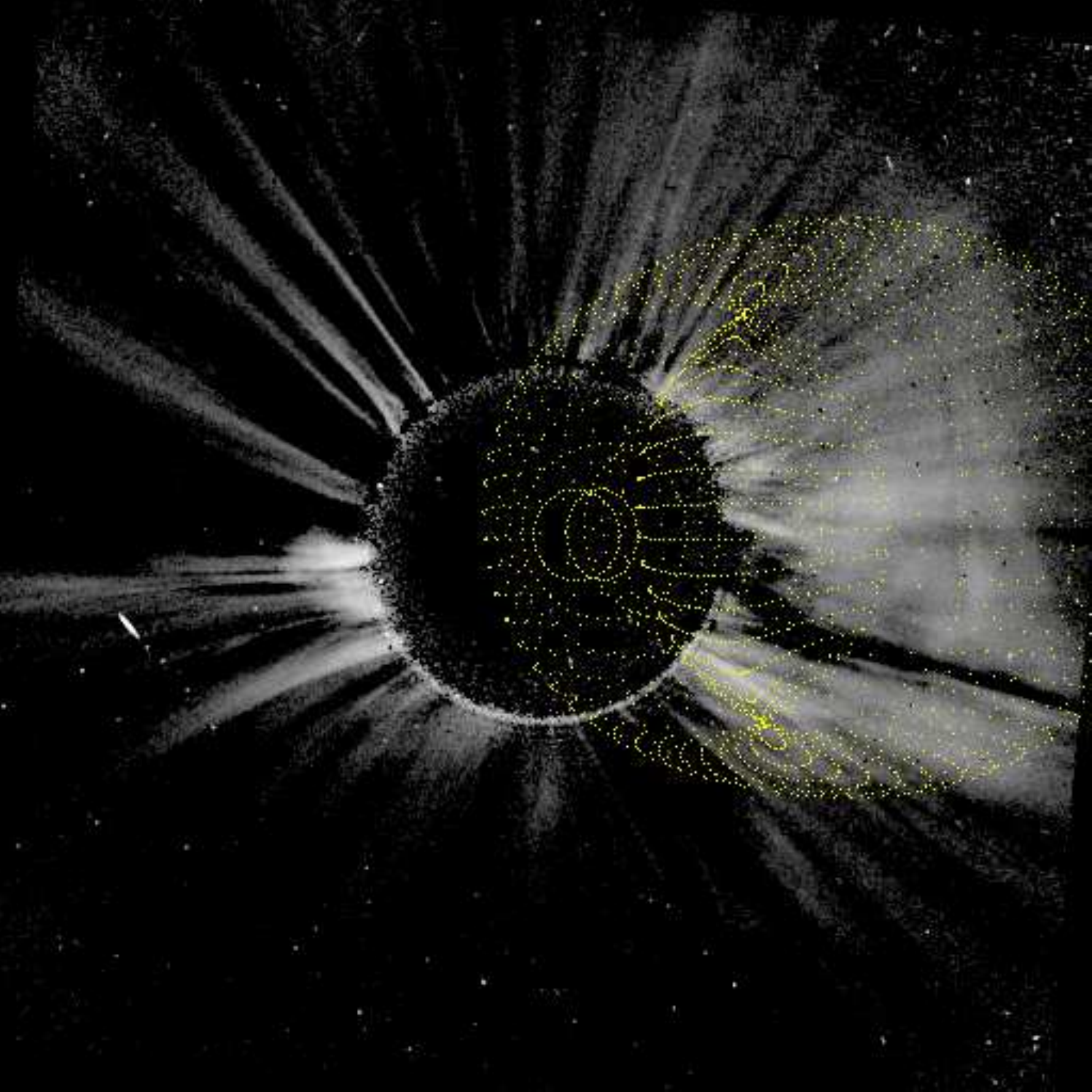}
              \hspace*{-0.02\textwidth}
               \includegraphics[width=0.4\textwidth,clip=]{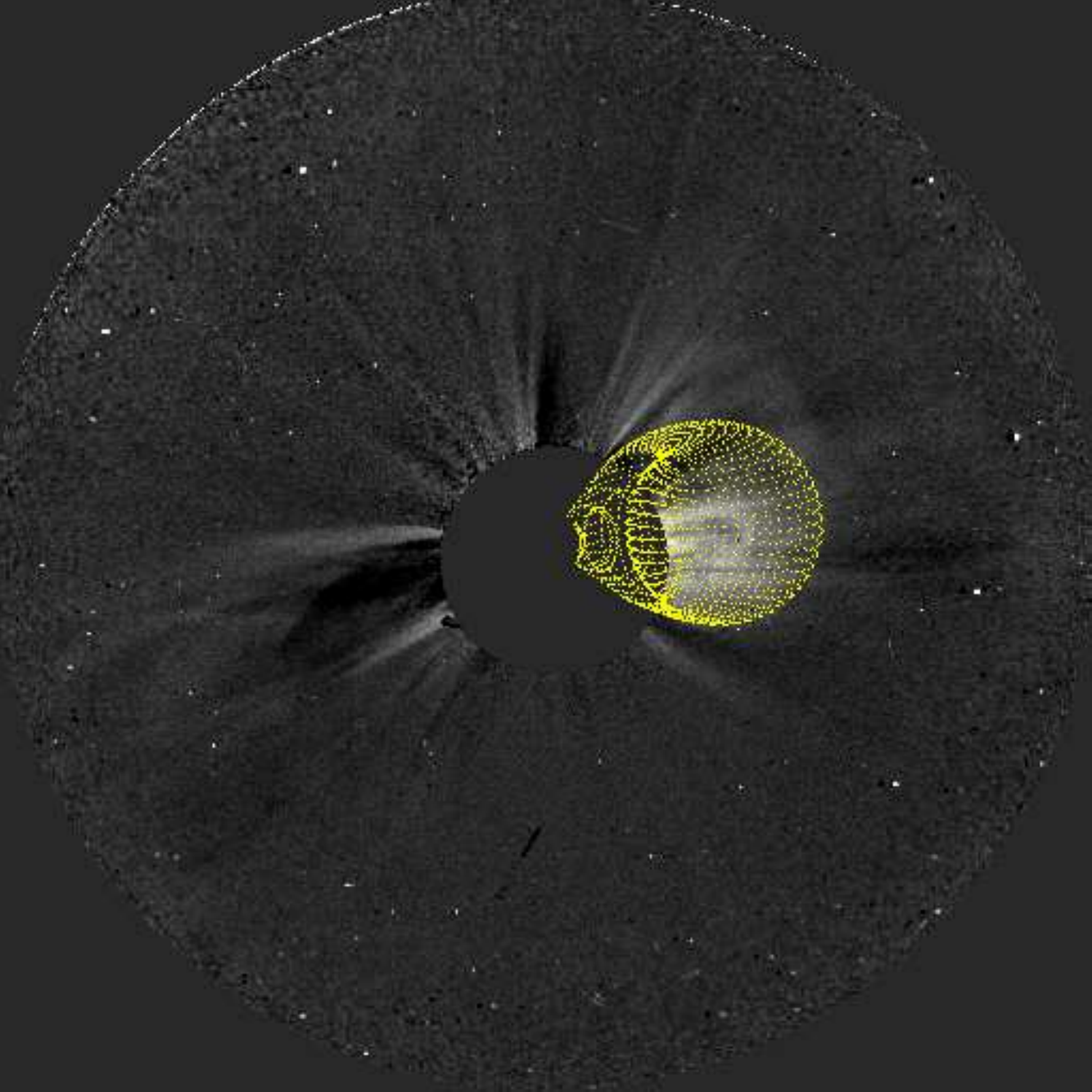}
                }
\vspace{0.0261\textwidth}  
\caption[GCS fit for CME 19 at 13:54]{GCS fit for CME 19 on October 26, 2011 at 13:54 UT at height $H=10.2$ \Rs. Table \ref{tblapp} 
lists the GCS parameters for this event.}
\label{figa19}
\end{figure}

\begin{figure}[h]    
  \centering                              
   \centerline{\hspace*{0.04\textwidth}
               \includegraphics[width=0.4\textwidth,clip=]{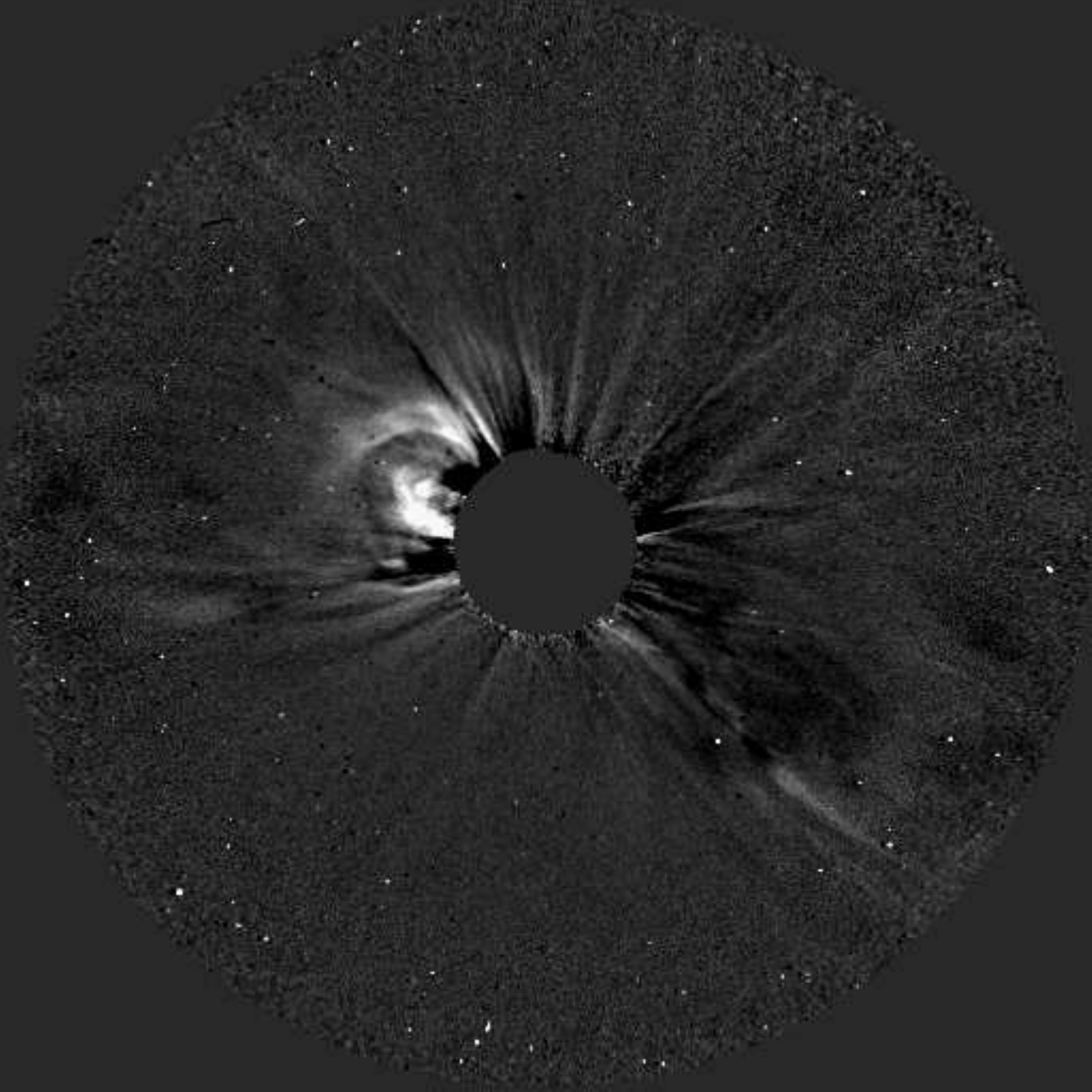}
                \hspace*{-0.02\textwidth}
               \includegraphics[width=0.4\textwidth,clip=]{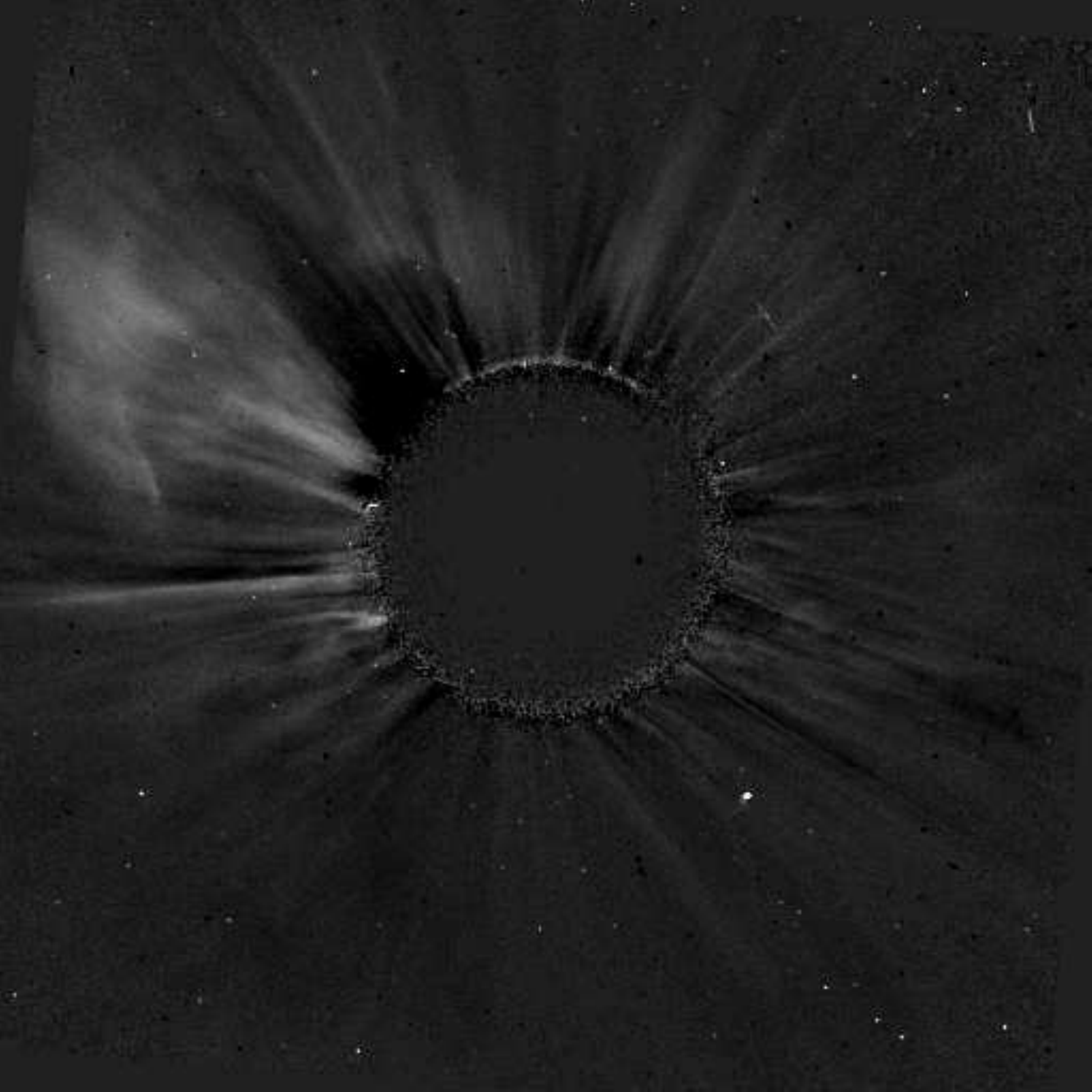}
             \hspace*{-0.02\textwidth}
               \includegraphics[width=0.4\textwidth,clip=]{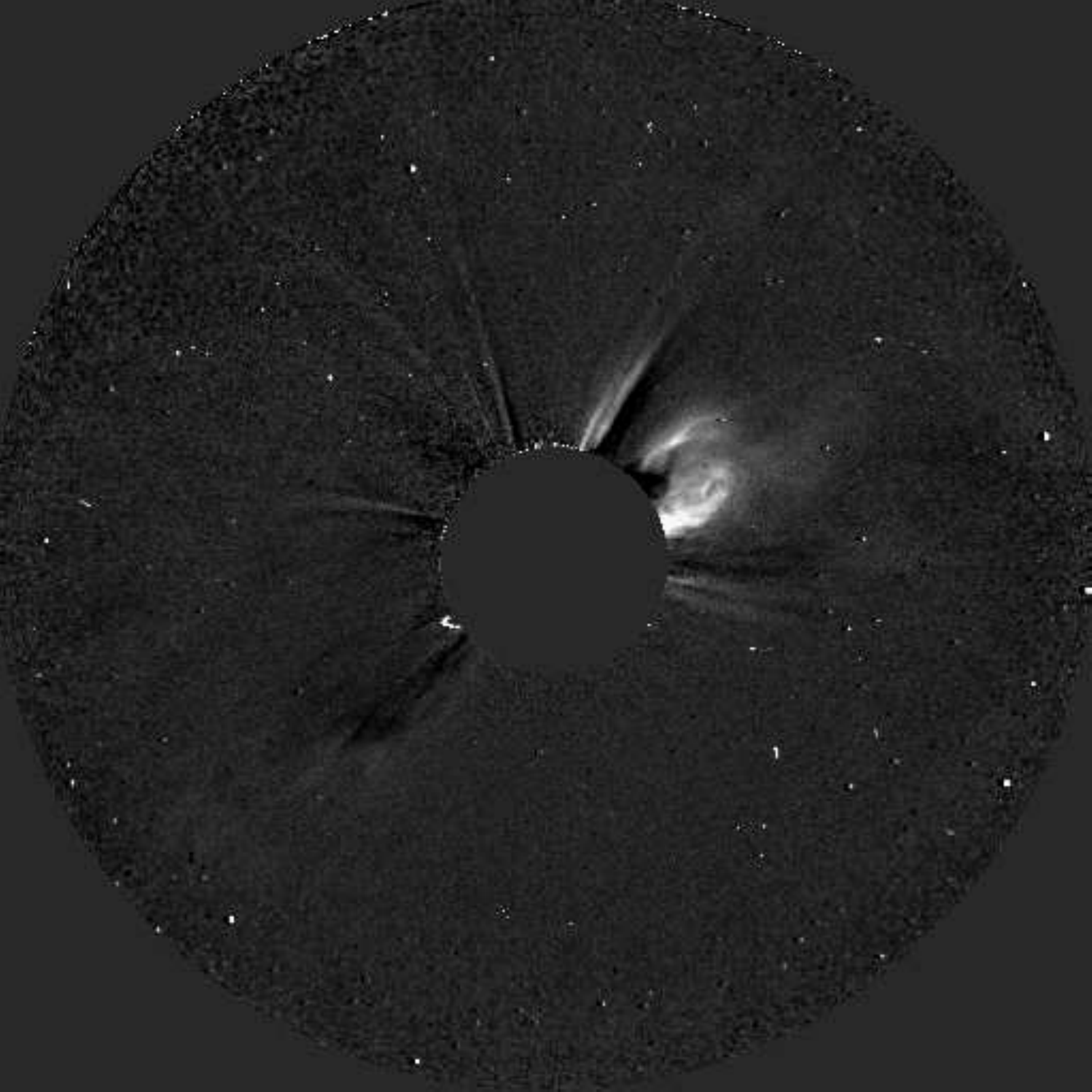}
               }
                 \centerline{\hspace*{0.04\textwidth}
              \includegraphics[width=0.4\textwidth,clip=]{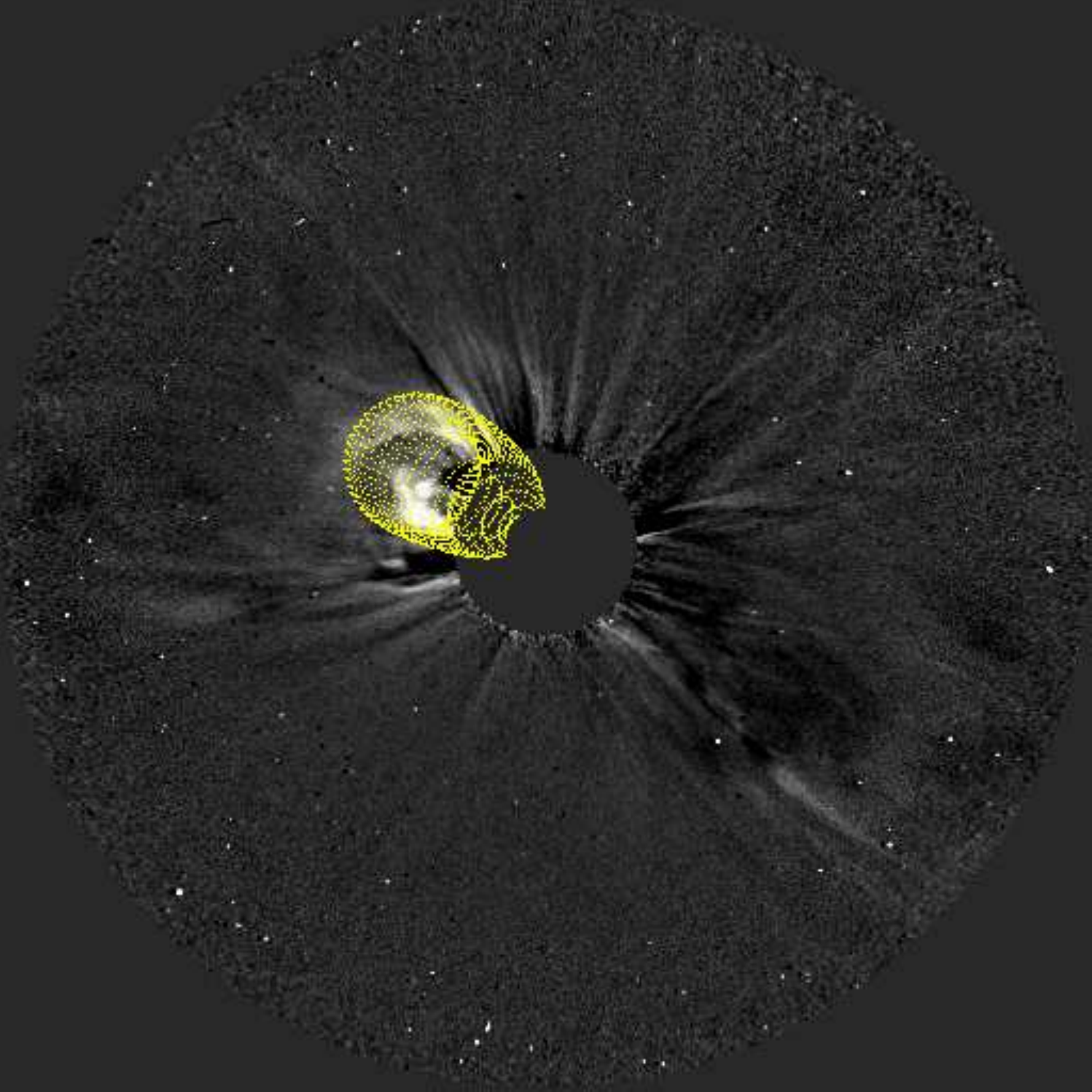}
               \hspace*{-0.02\textwidth}
               \includegraphics[width=0.4\textwidth,clip=]{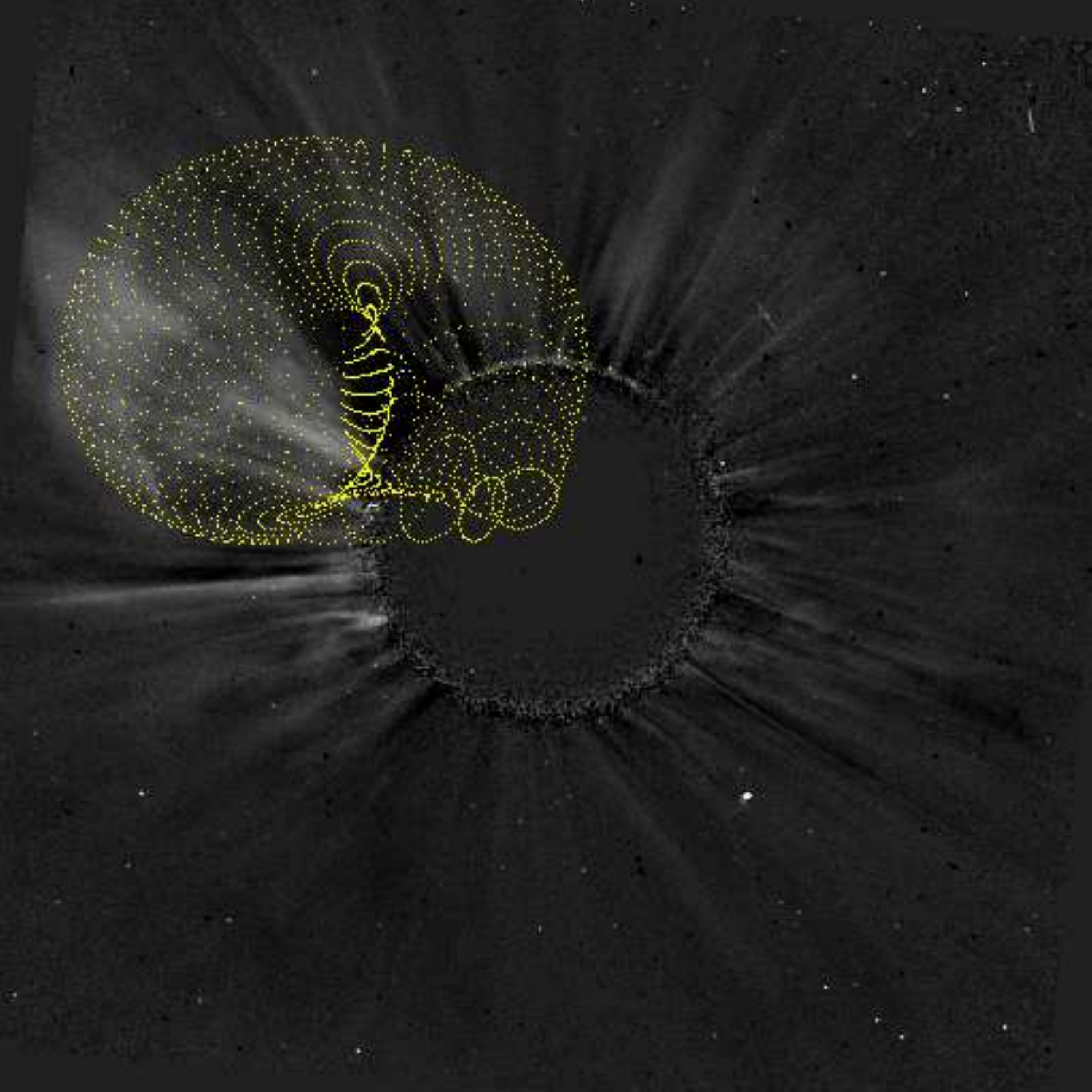}
              \hspace*{-0.02\textwidth}
               \includegraphics[width=0.4\textwidth,clip=]{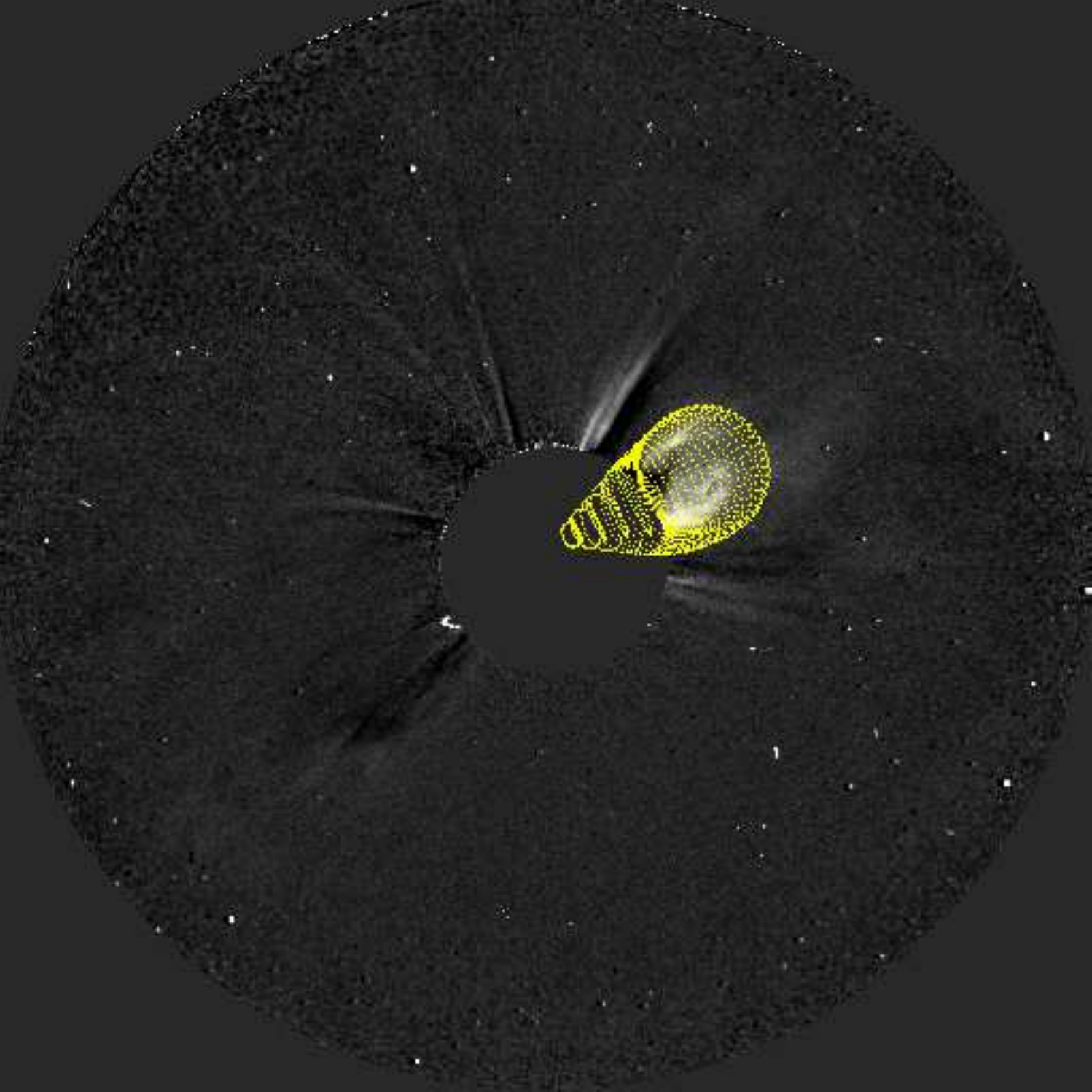}
                }
\vspace{0.0261\textwidth}  
\caption[GCS fit for CME 20 at 13:39]{GCS fit for CME 20 on October 27, 2011 at 13:39 UT at height $H=7.8$ \Rs. Table \ref{tblapp} 
lists the GCS parameters for this event.}
\label{figa20}
\end{figure}

\begin{figure}[h]    
  \centering                              
   \centerline{\hspace*{0.00\textwidth}
               \includegraphics[width=0.4\textwidth,clip=]{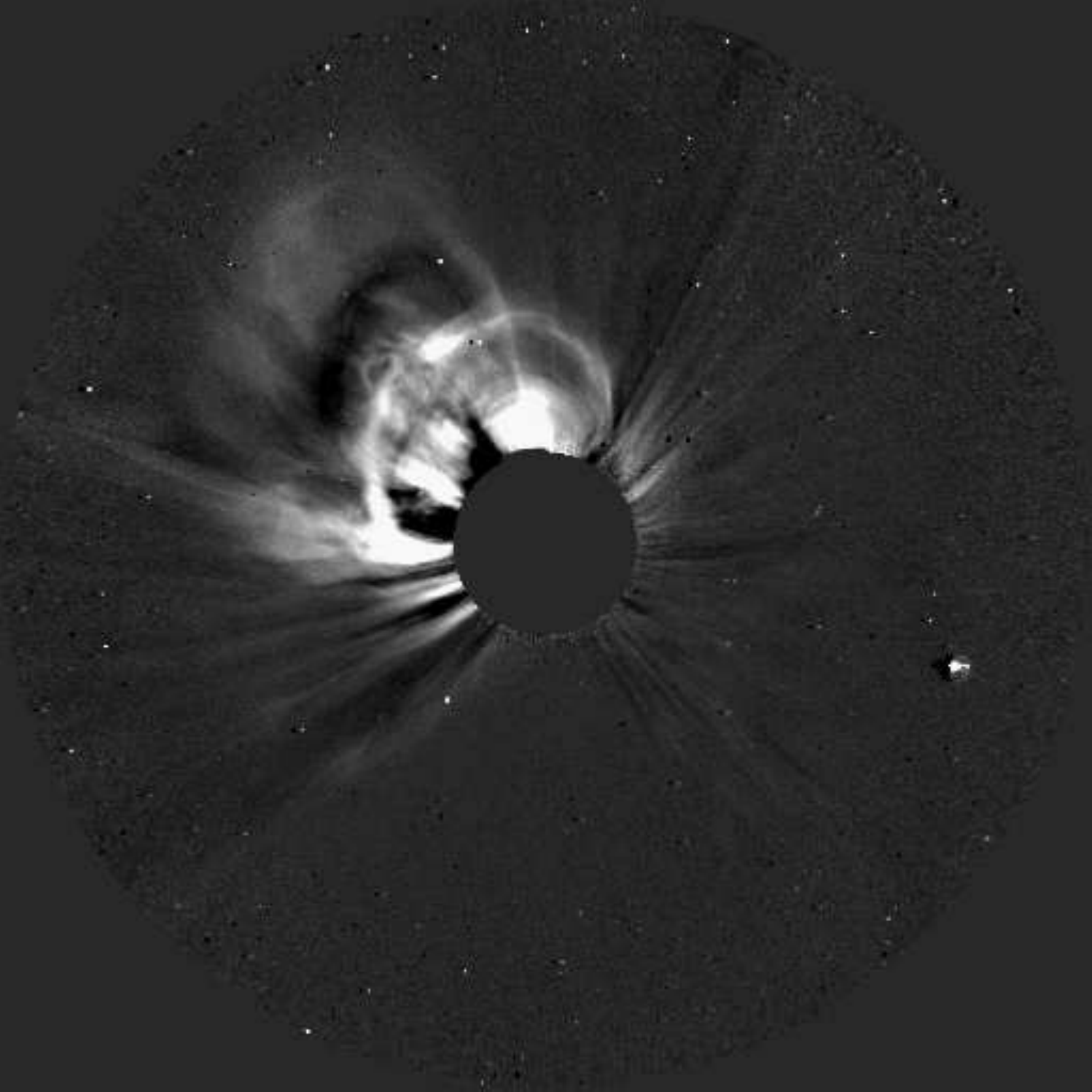}
                \hspace*{-0.02\textwidth}
               \includegraphics[width=0.4\textwidth,clip=]{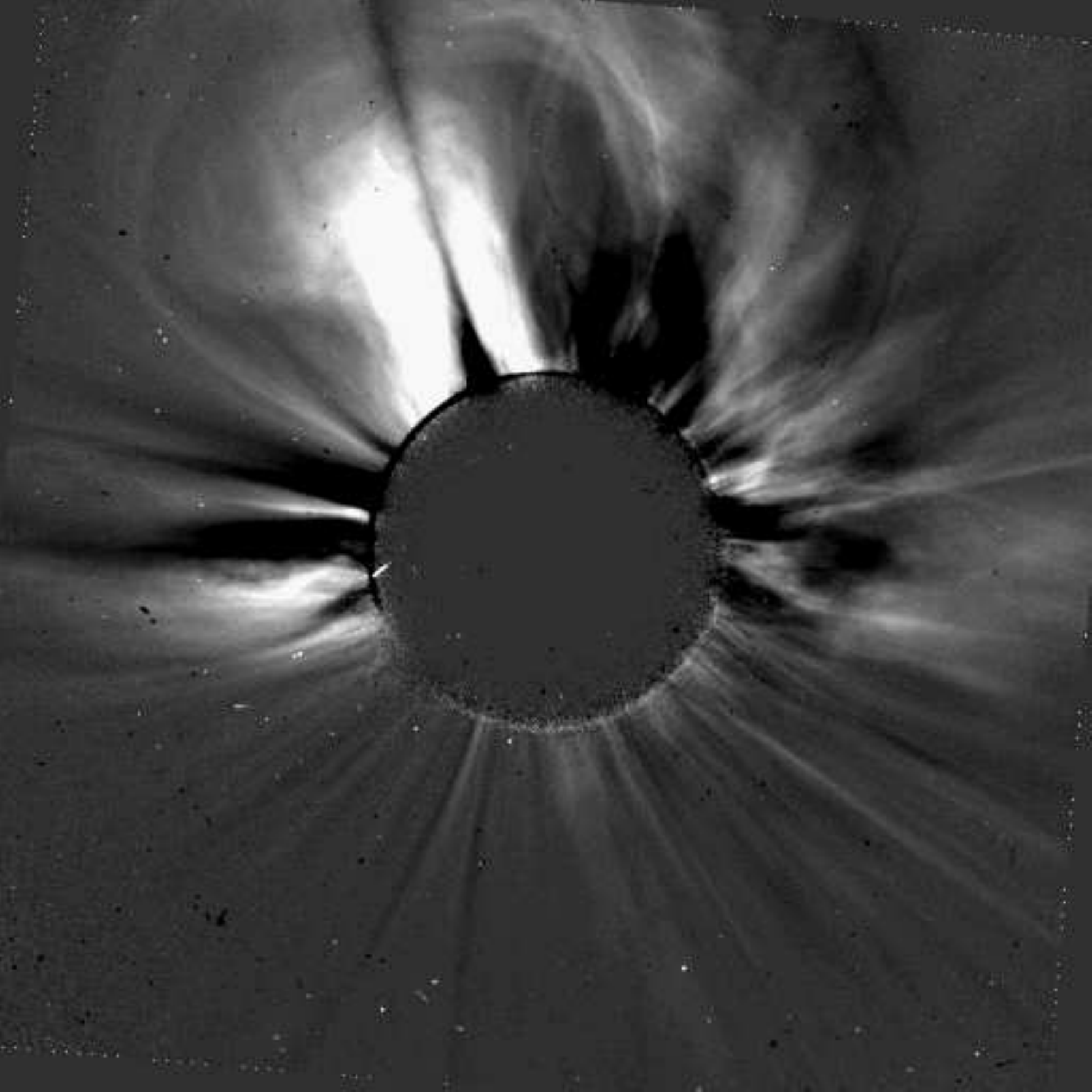}
             \hspace*{-0.02\textwidth}
               \includegraphics[width=0.4\textwidth,clip=]{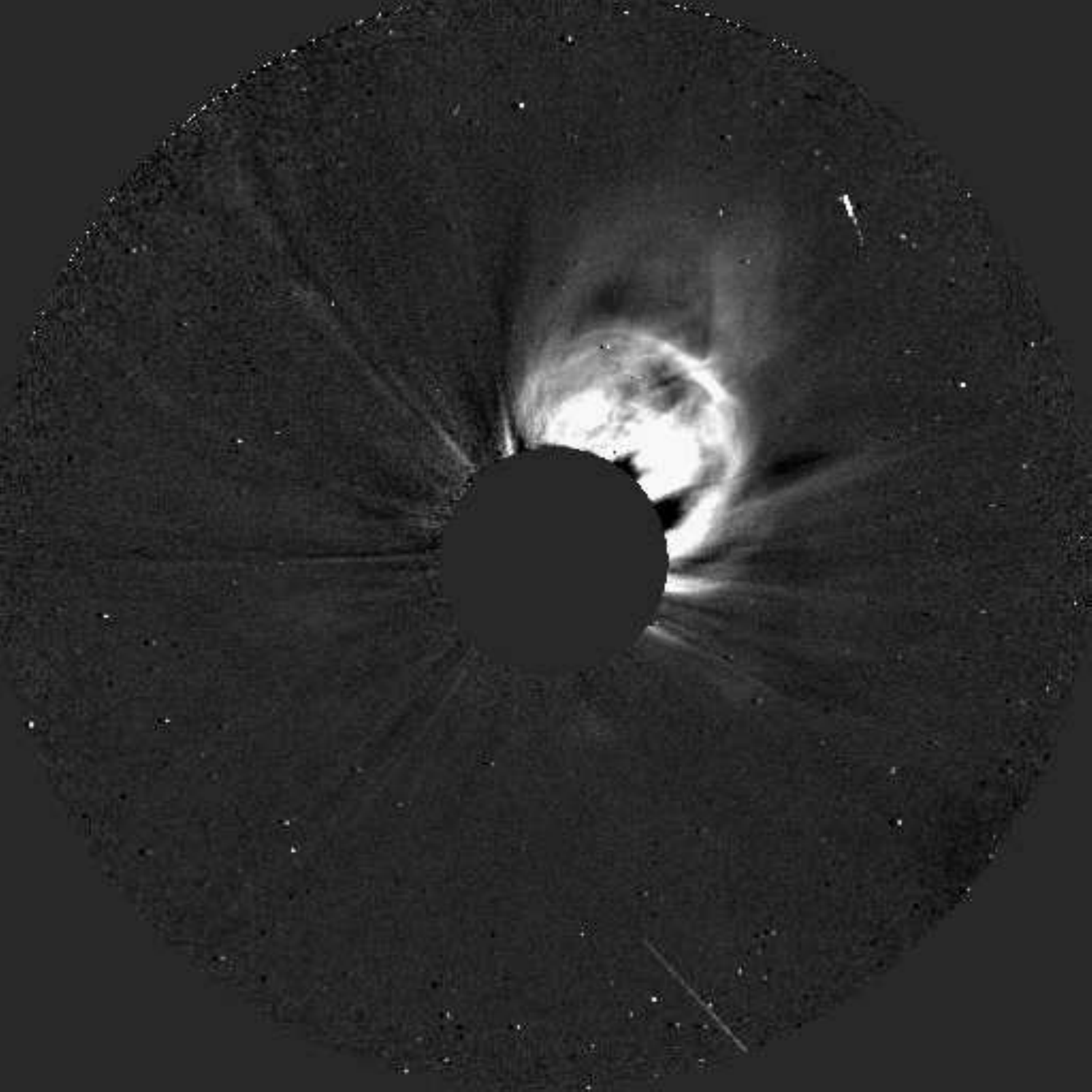}
               }
                 \centerline{\hspace*{0.0\textwidth}
              \includegraphics[width=0.4\textwidth,clip=]{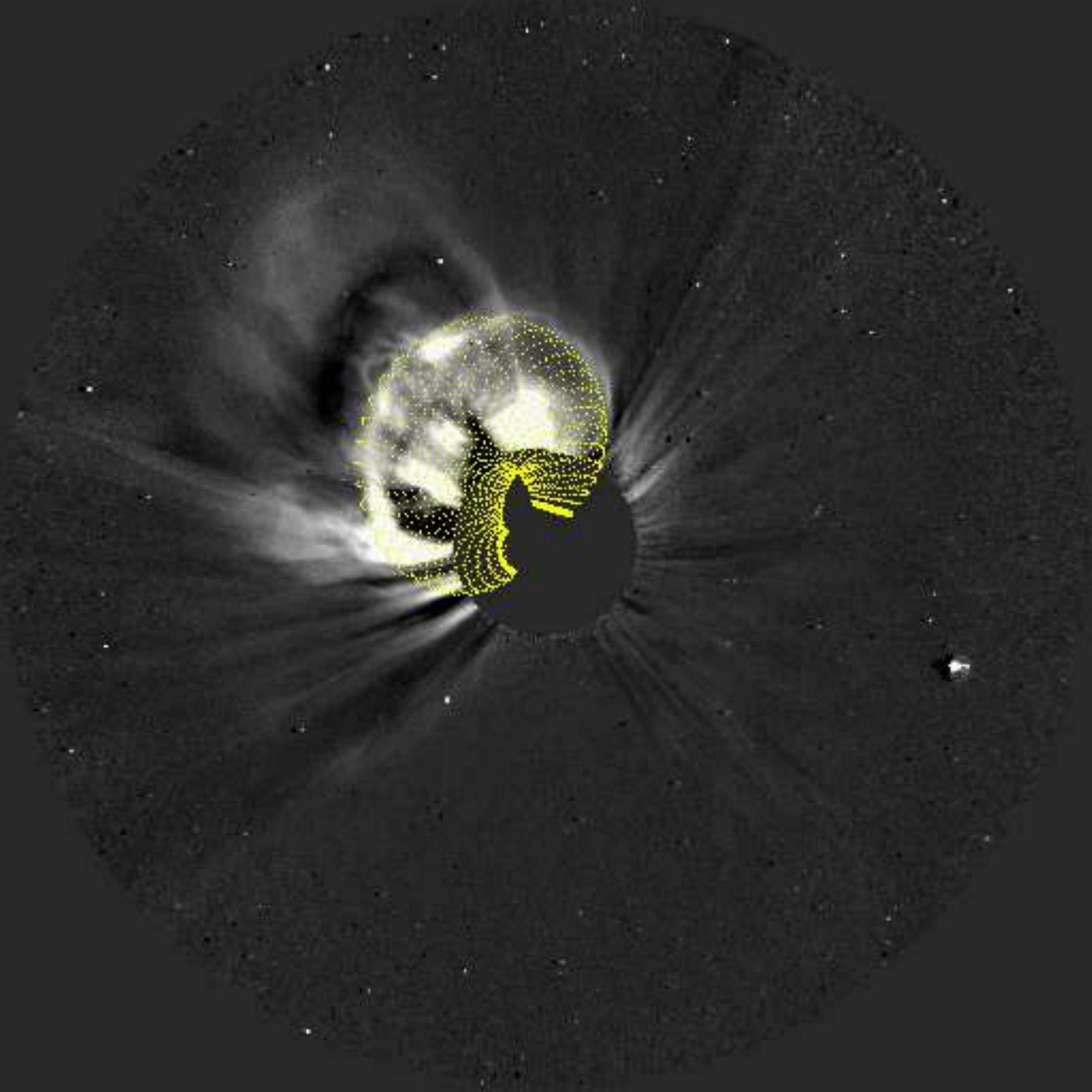}
               \hspace*{-0.02\textwidth}
               \includegraphics[width=0.4\textwidth,clip=]{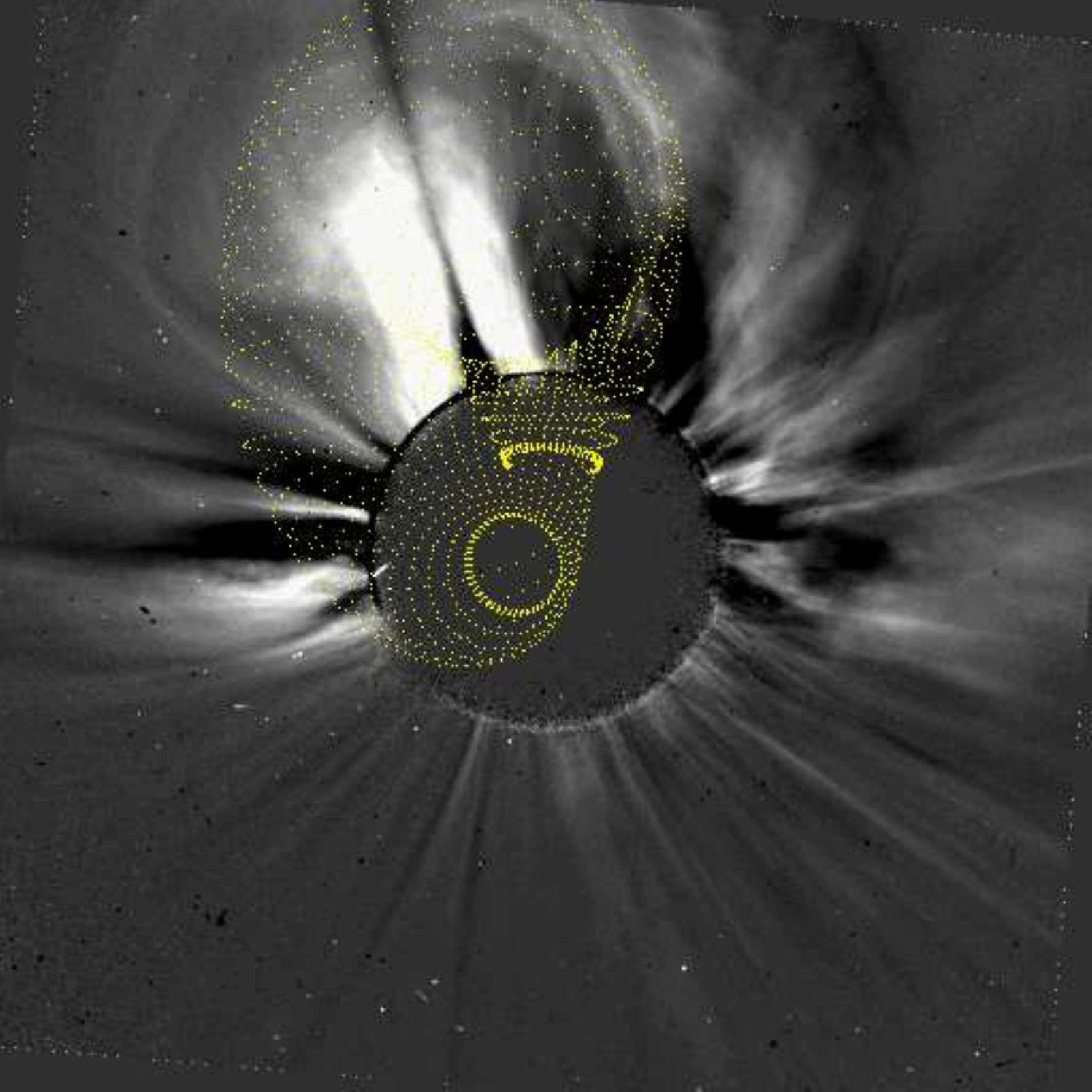}
              \hspace*{-0.02\textwidth}
               \includegraphics[width=0.4\textwidth,clip=]{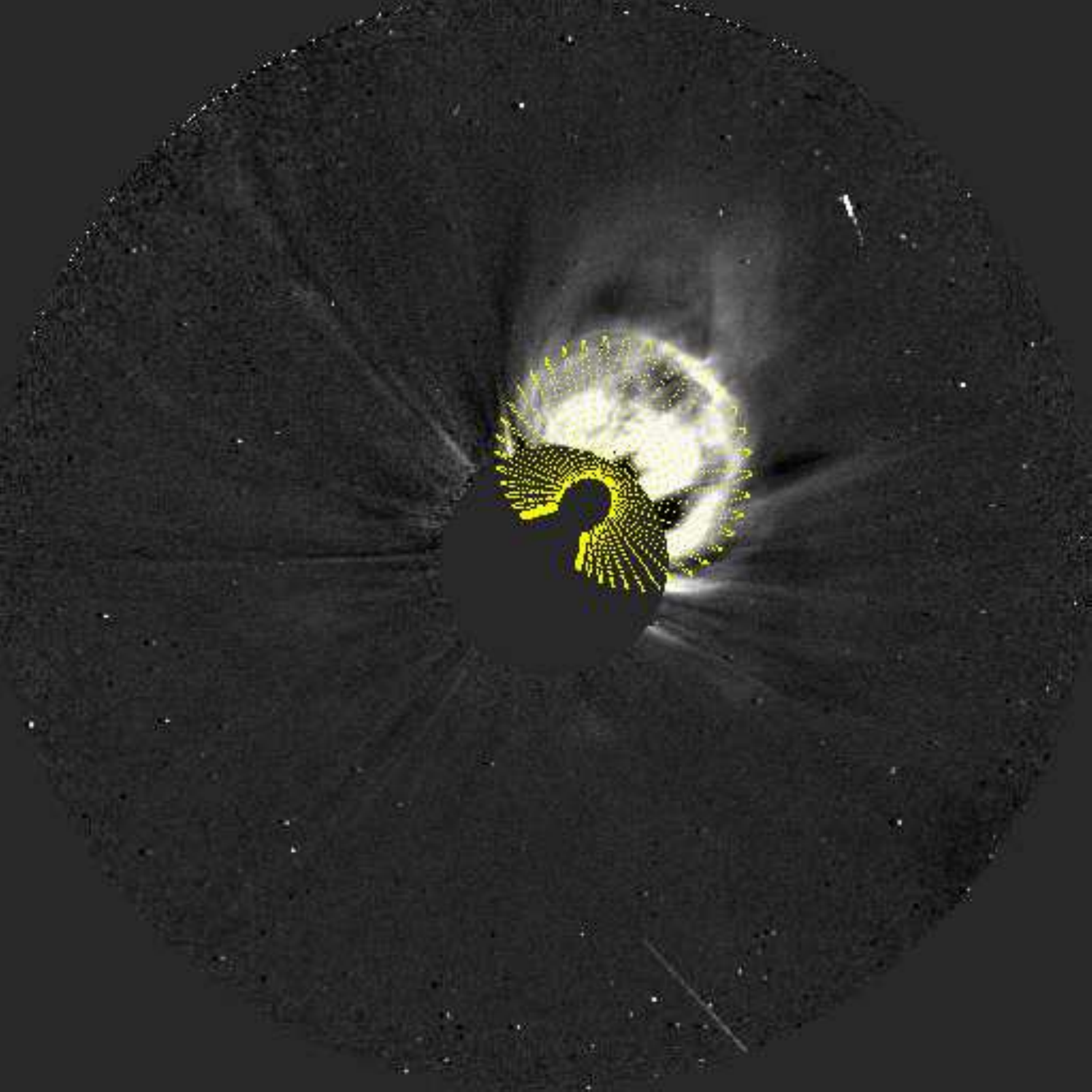}
                }
\vspace{0.0261\textwidth}  
\caption[GCS fit for CME 21 at 15:54]{GCS fit for CME 21 on November 19, 2011 at 15:54 UT at height $H=7.2$ \Rs. Table \ref{tblapp} 
lists the GCS parameters for this event.}
\label{figa21}
\end{figure}

\begin{figure}[h]    
  \centering                              
   \centerline{\hspace*{0.04\textwidth}
               \includegraphics[width=0.4\textwidth,clip=]{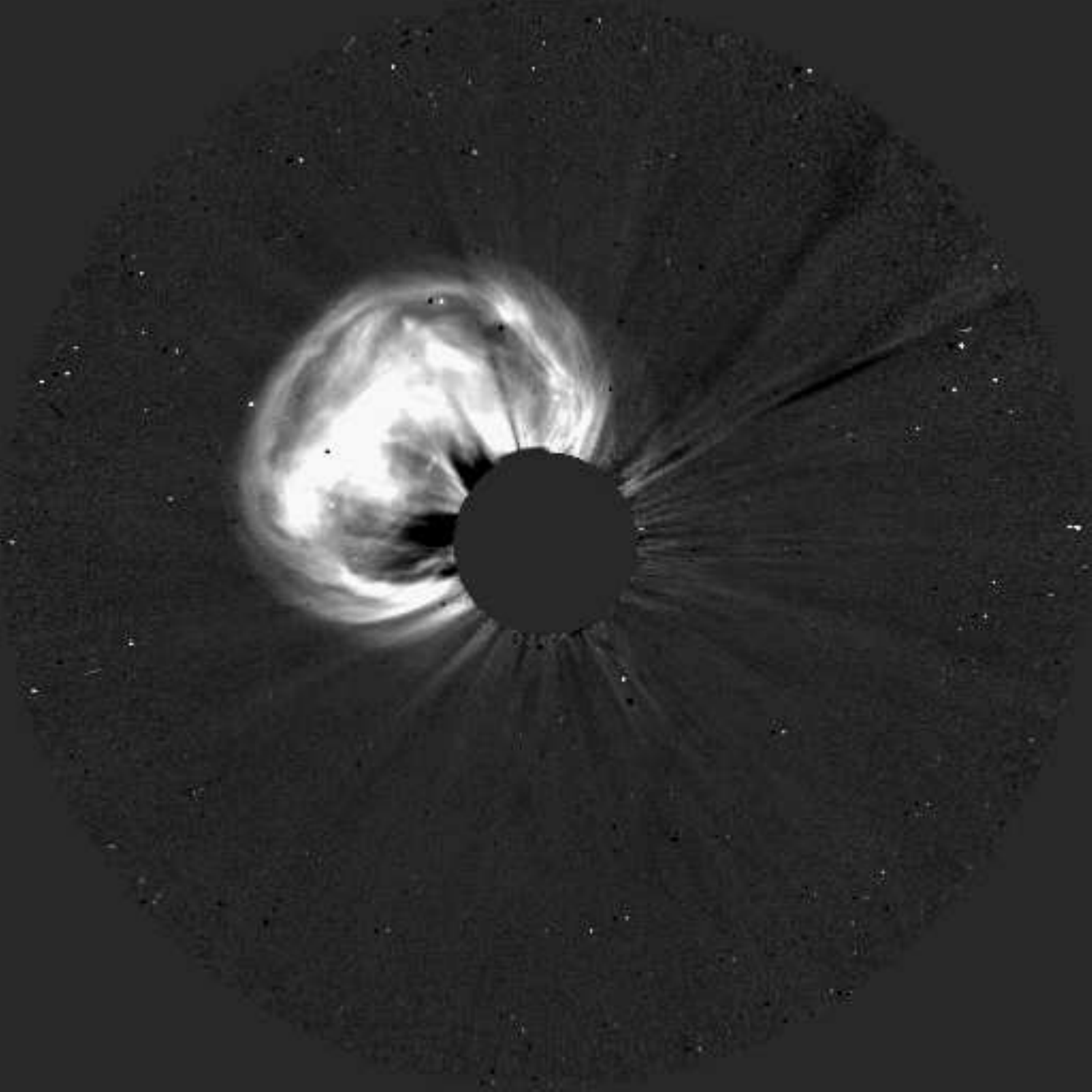}
                \hspace*{-0.02\textwidth}
               \includegraphics[width=0.4\textwidth,clip=]{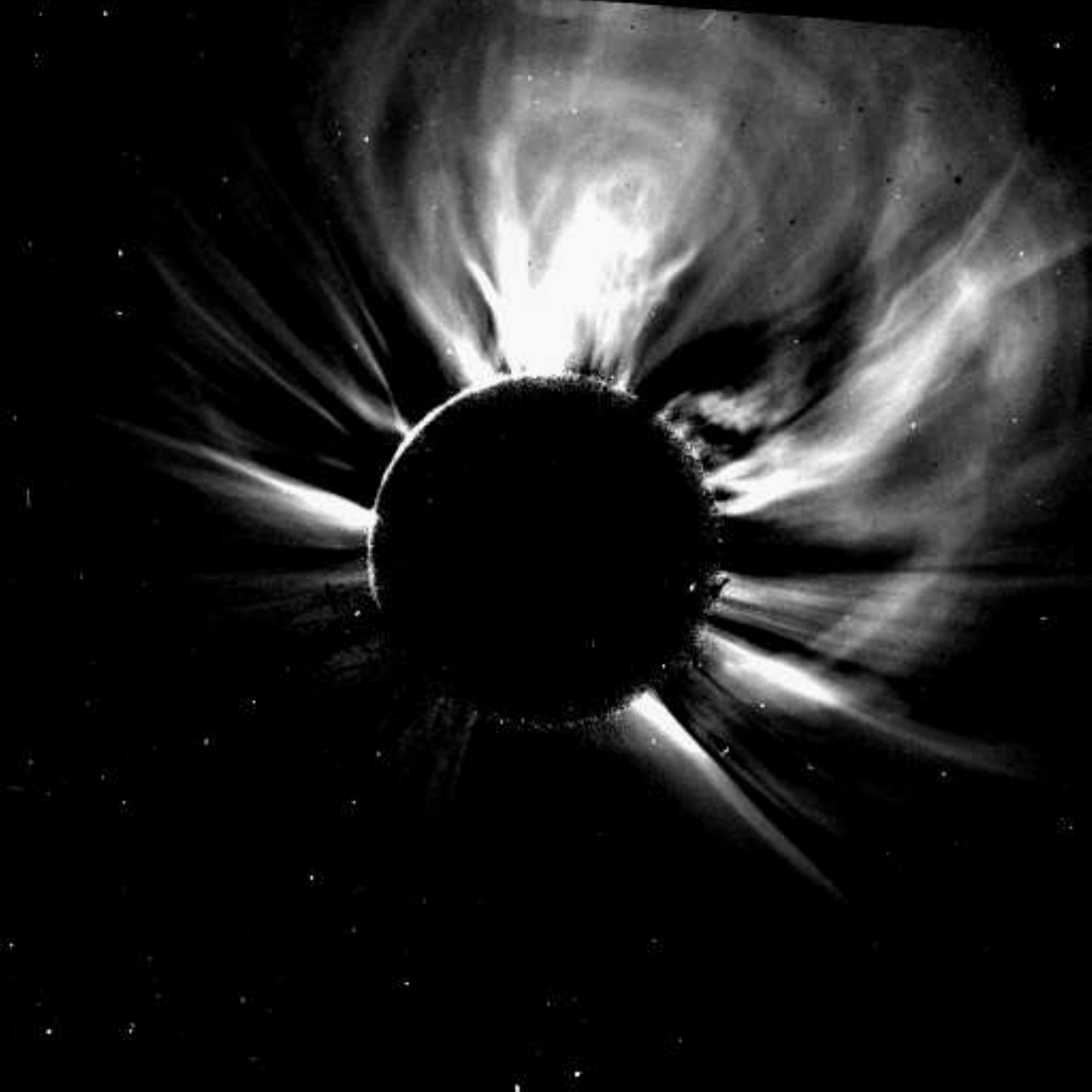}
             \hspace*{-0.02\textwidth}
               \includegraphics[width=0.4\textwidth,clip=]{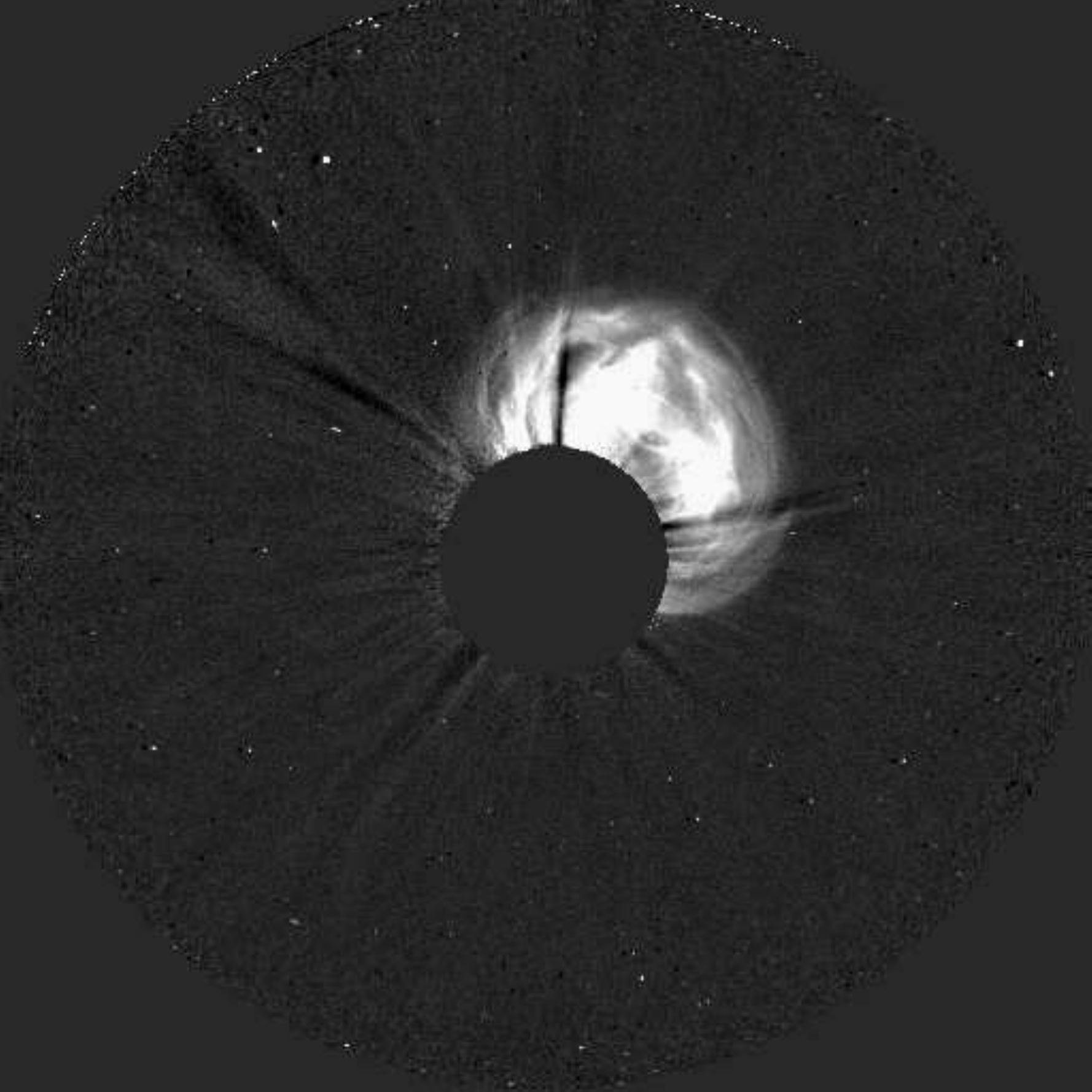}
               }
                 \centerline{\hspace*{0.04\textwidth}
              \includegraphics[width=0.4\textwidth,clip=]{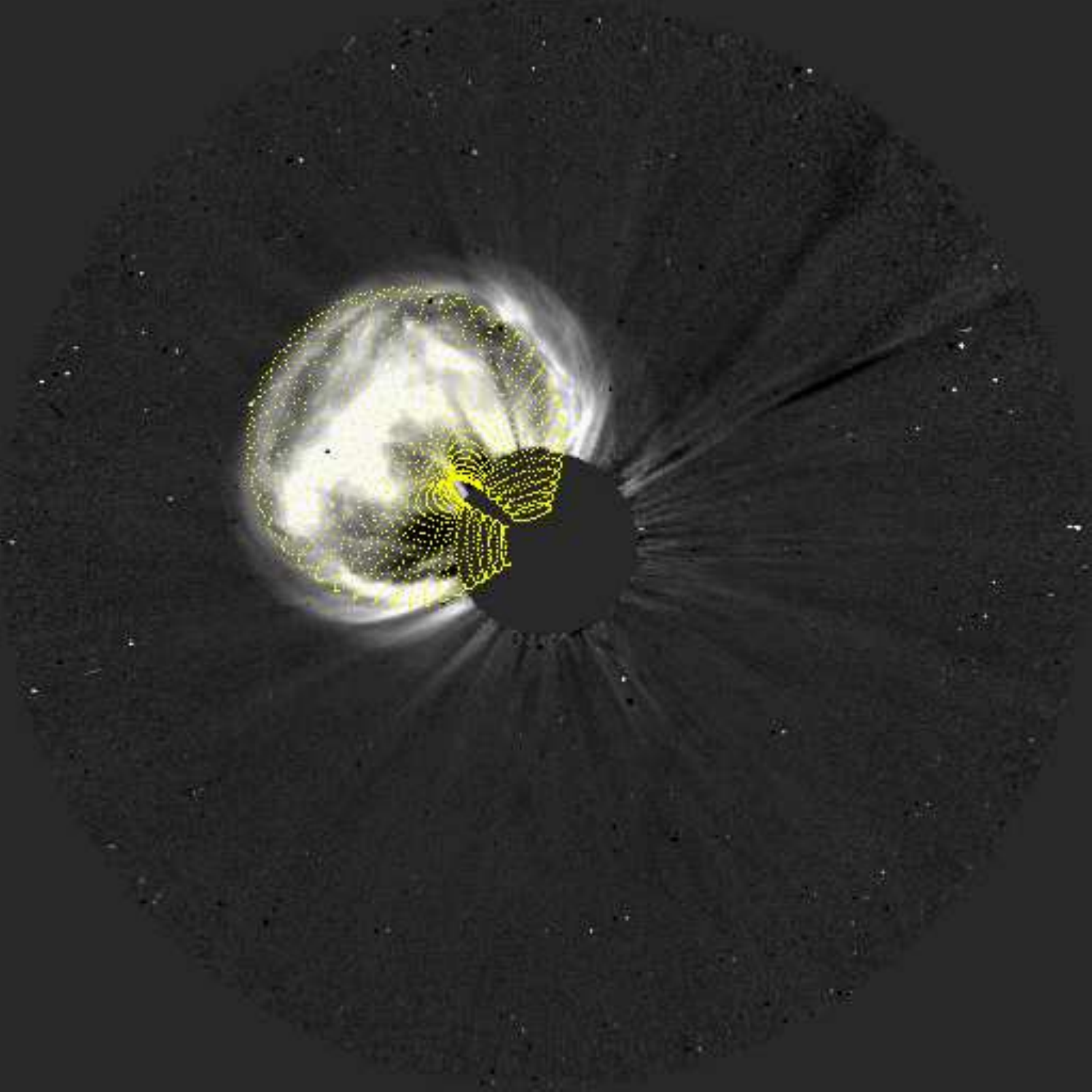}
               \hspace*{-0.02\textwidth}
               \includegraphics[width=0.4\textwidth,clip=]{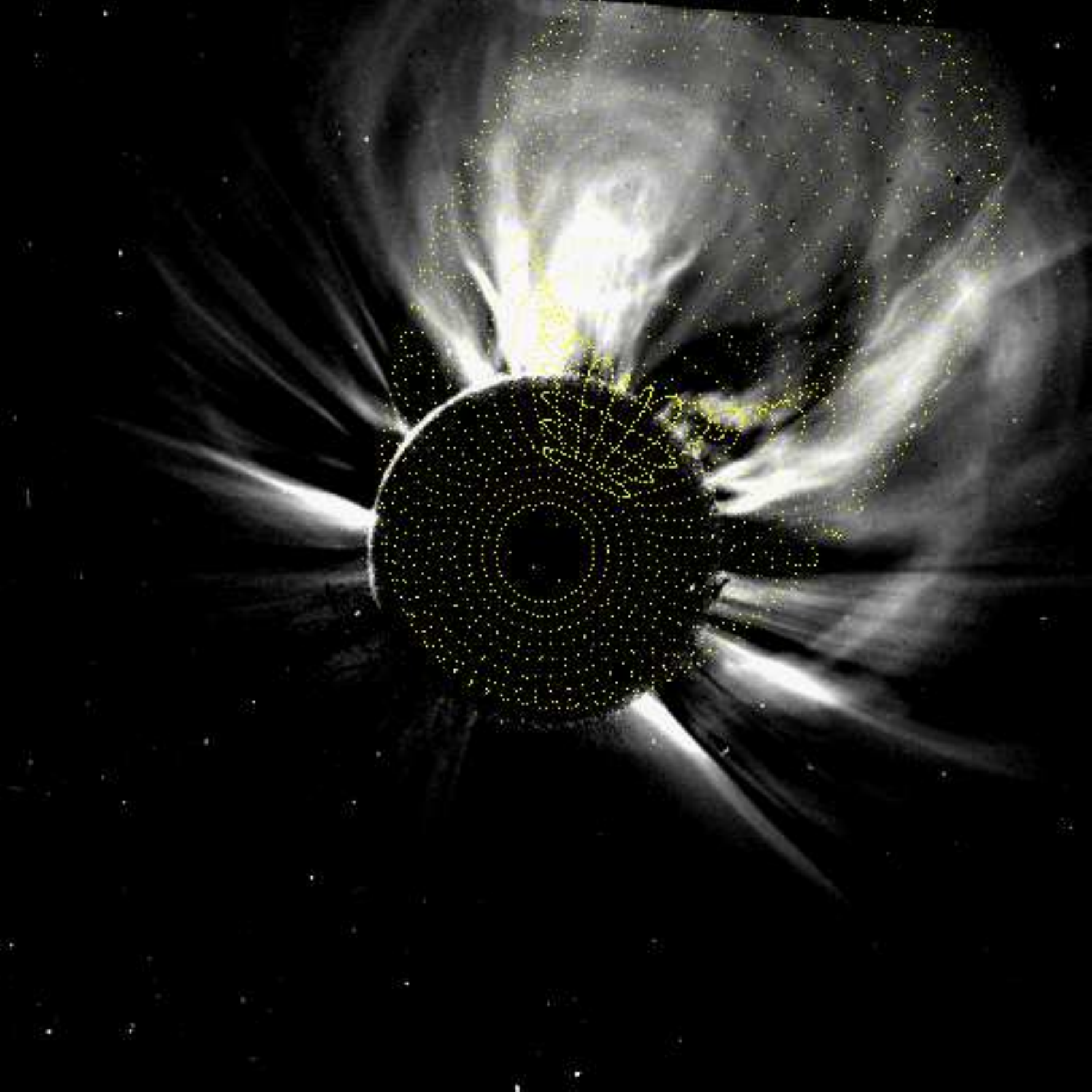}
              \hspace*{-0.02\textwidth}
               \includegraphics[width=0.4\textwidth,clip=]{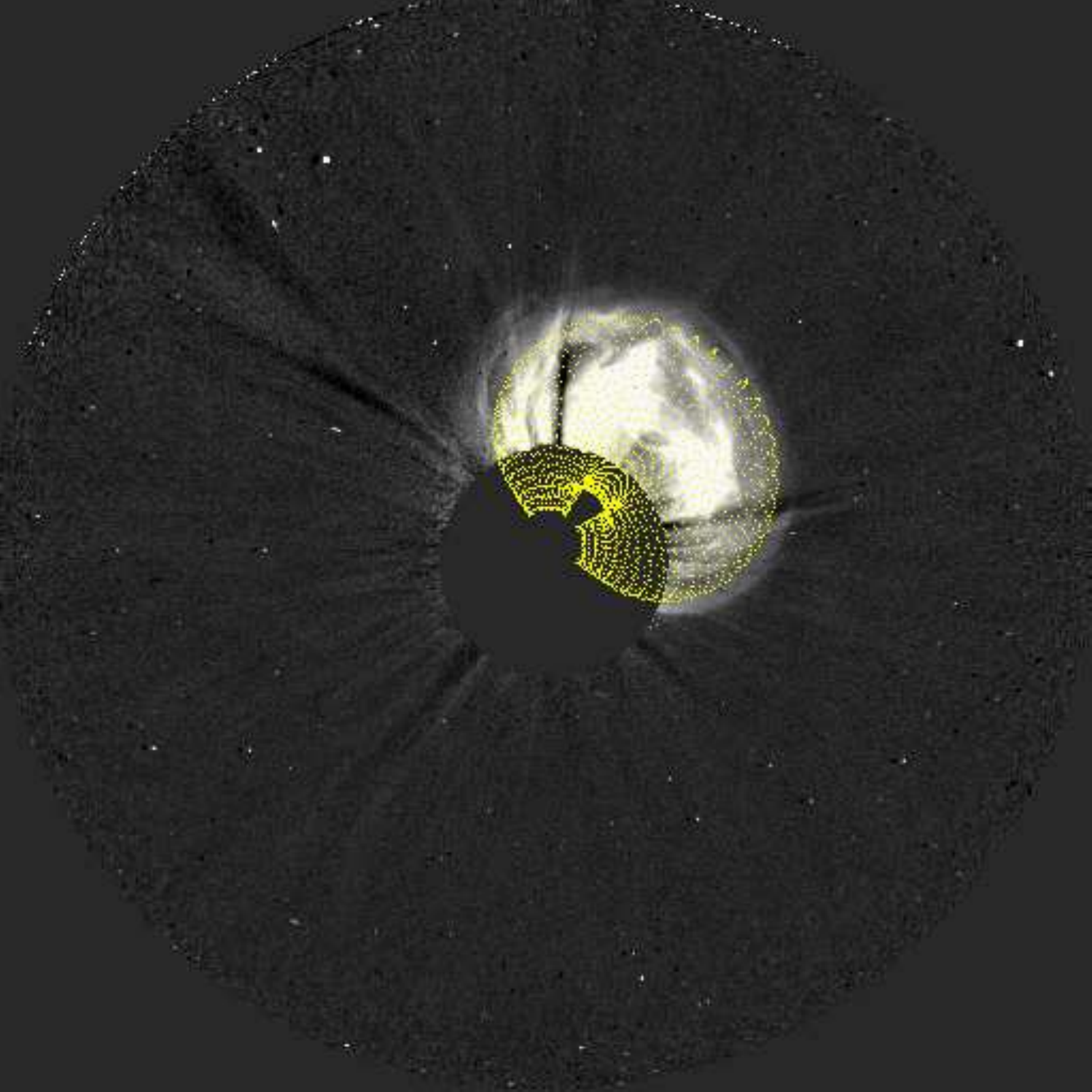}
                }
\vspace{0.0261\textwidth}  
\caption[GCS fit for CME 22 at 04:24]{GCS fit for CME 22 on January 23, 2012 at 04:24 UT at height $H=8.9$ \Rs. Table \ref{tblapp} 
lists the GCS parameters for this event.}
\label{figa22}
\end{figure}

\begin{figure}[h]    
  \centering                              
   \centerline{\hspace*{0.00\textwidth}
               \includegraphics[width=0.4\textwidth,clip=]{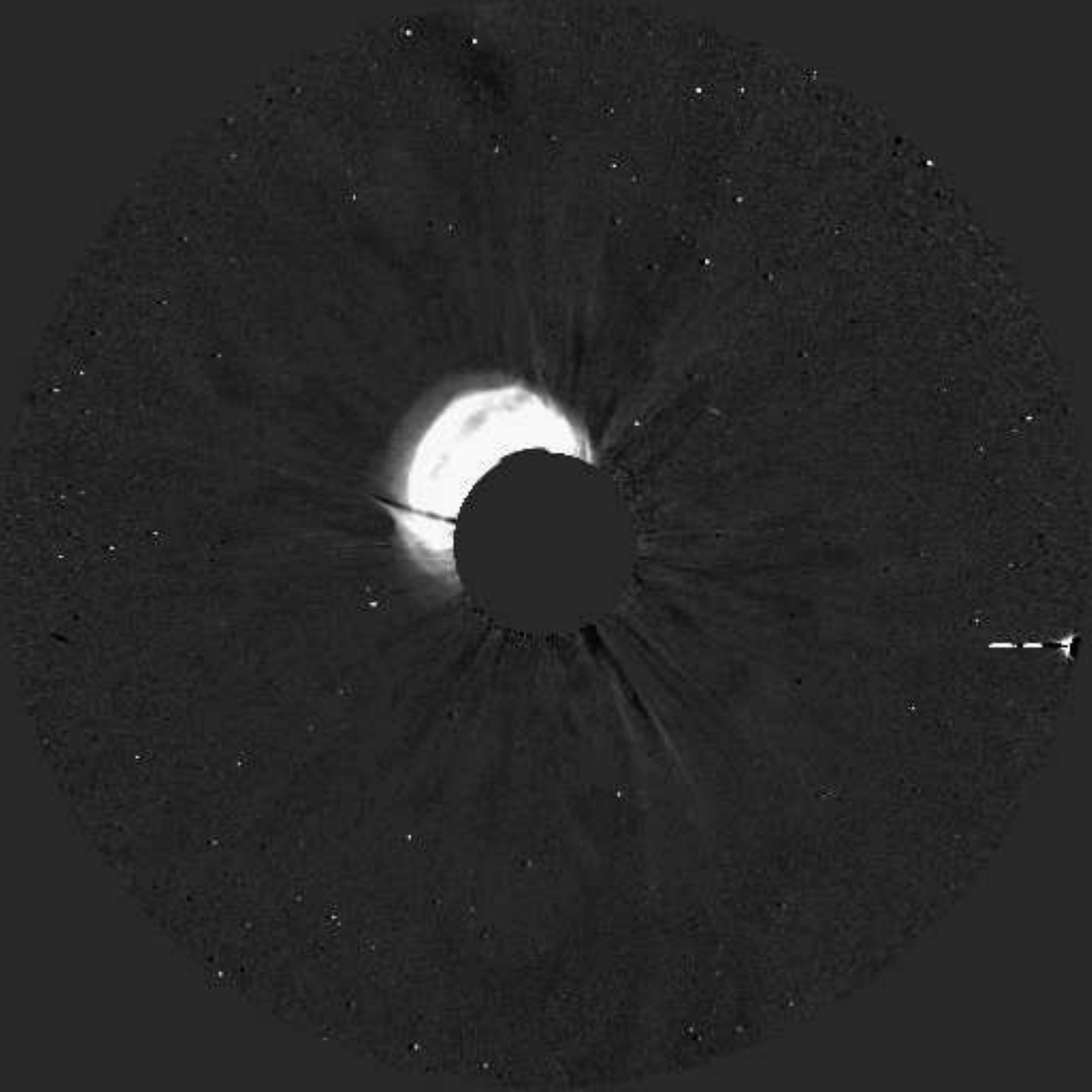}
                \hspace*{-0.02\textwidth}
               \includegraphics[width=0.4\textwidth,clip=]{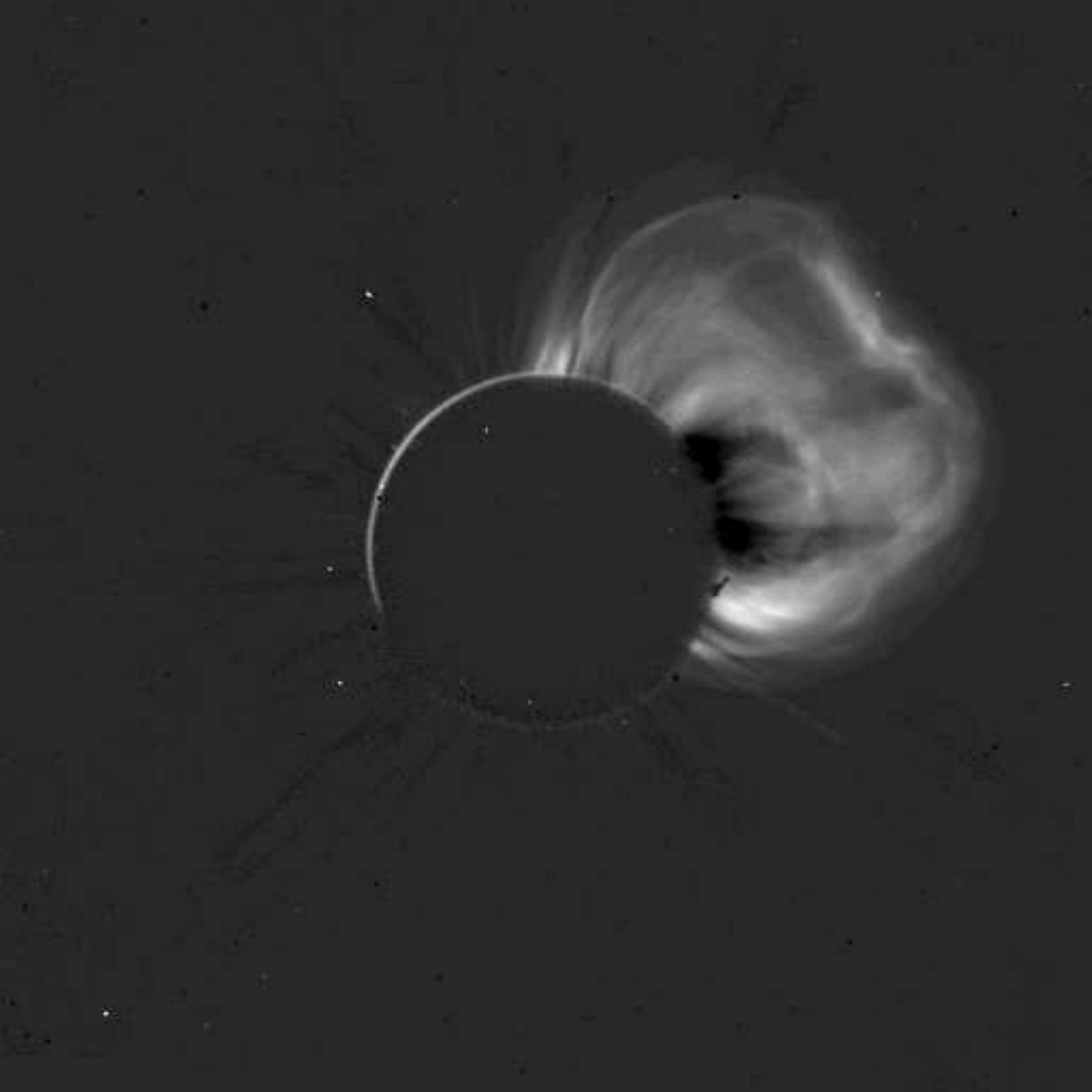}
             \hspace*{-0.02\textwidth}
               \includegraphics[width=0.4\textwidth,clip=]{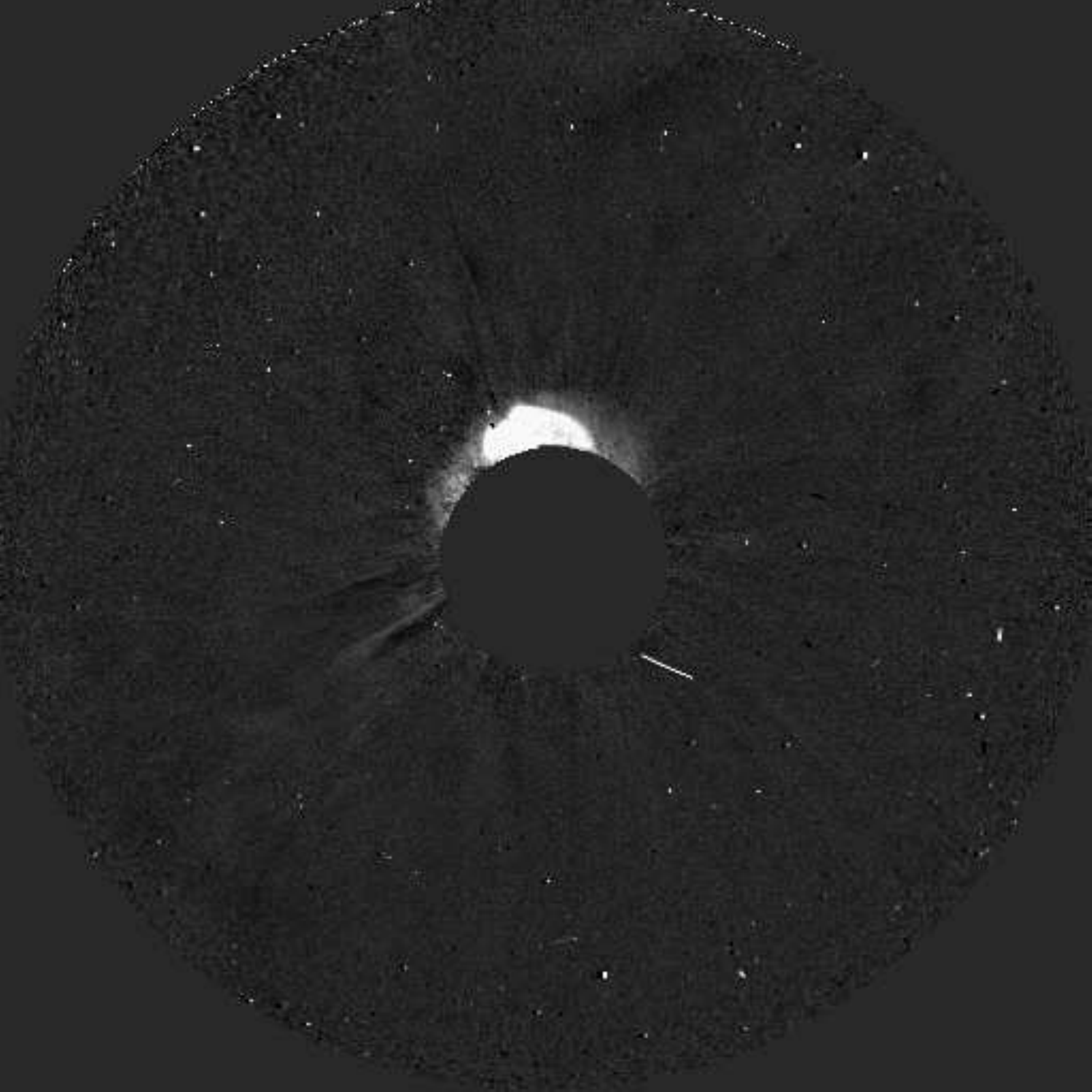}
               }
                 \centerline{\hspace*{0.0\textwidth}
              \includegraphics[width=0.4\textwidth,clip=]{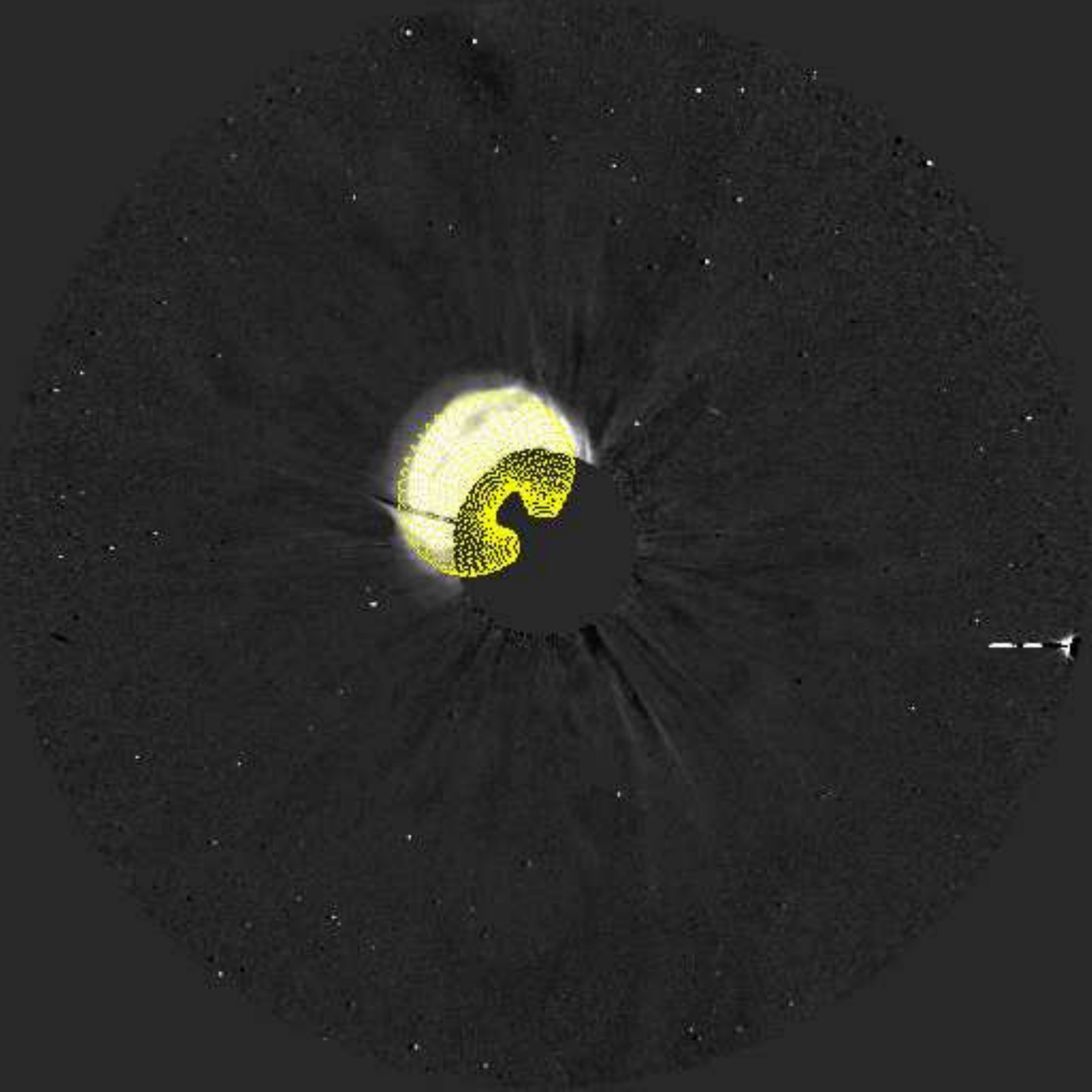}
               \hspace*{-0.02\textwidth}
               \includegraphics[width=0.4\textwidth,clip=]{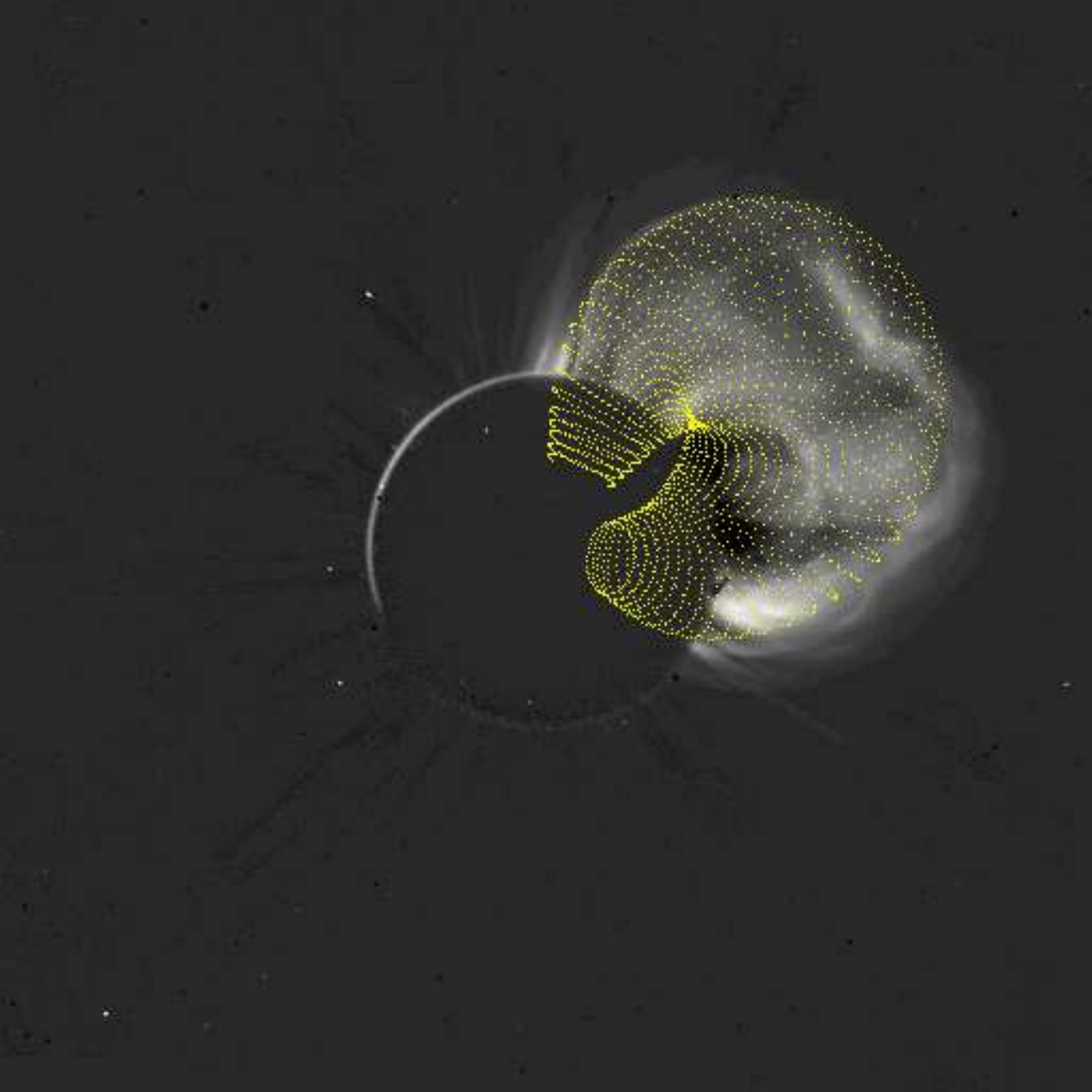}
              \hspace*{-0.02\textwidth}
               \includegraphics[width=0.4\textwidth,clip=]{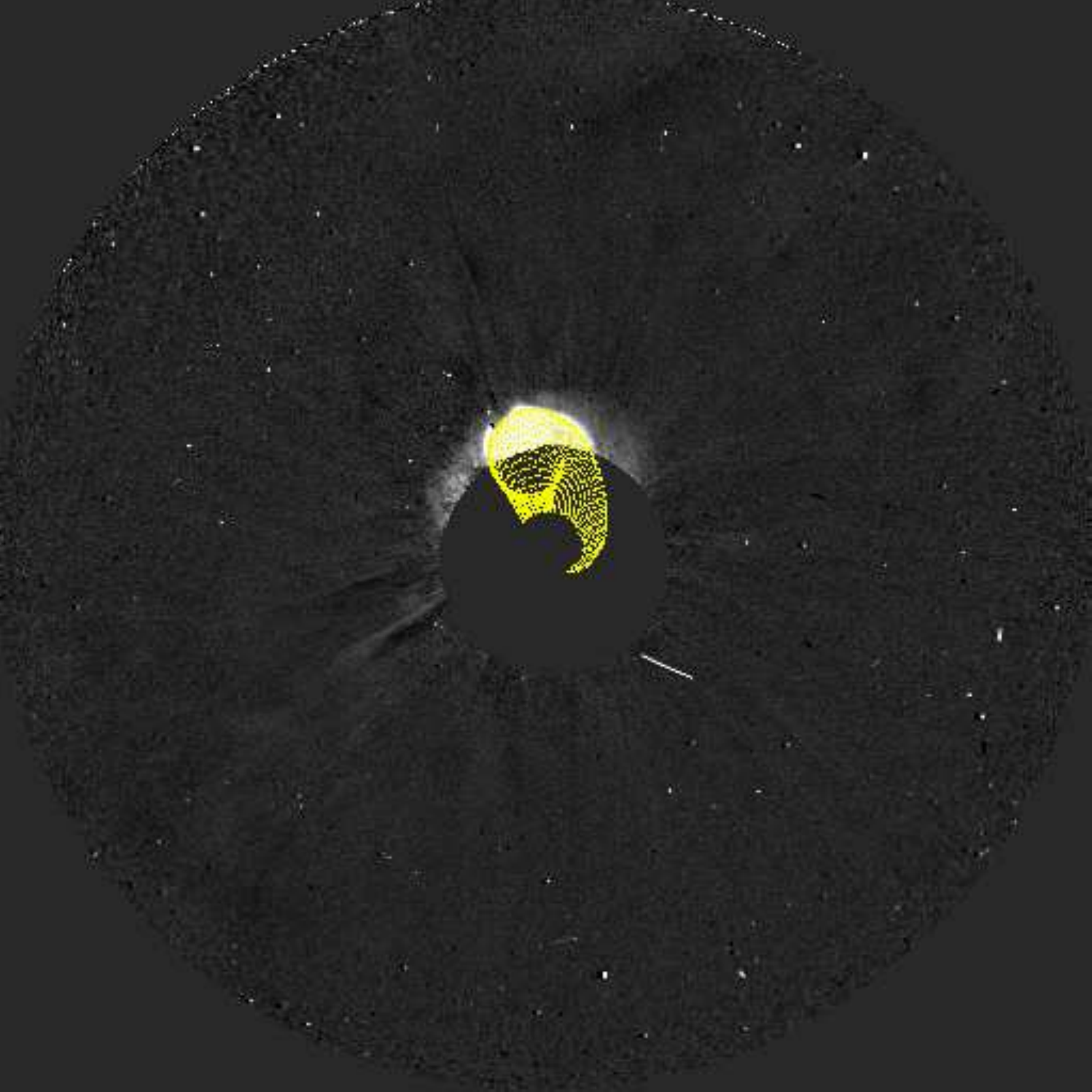}
                }
\vspace{0.0261\textwidth}  
\caption[GCS fit for CME 23 at 18:39]{GCS fit for CME 23 on January 27, 2012 at 18:39 UT at height $H=5.3$ \Rs. Table \ref{tblapp} 
lists the GCS parameters for this event.}
\label{figa23}
\end{figure}

\begin{figure}[h]    
  \centering                              
   \centerline{\hspace*{0.04\textwidth}
               \includegraphics[width=0.4\textwidth,clip=]{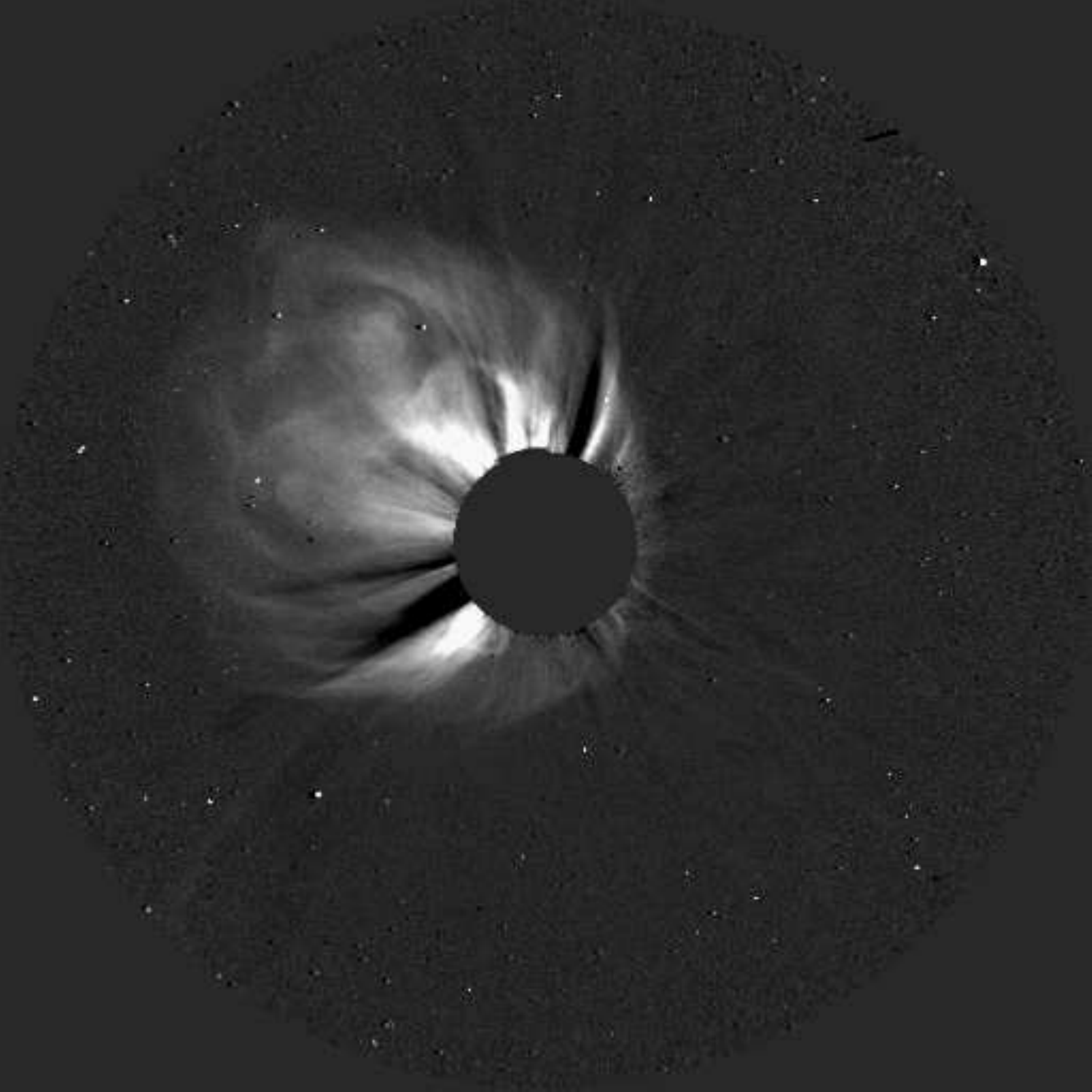}
                \hspace*{-0.02\textwidth}
               \includegraphics[width=0.4\textwidth,clip=]{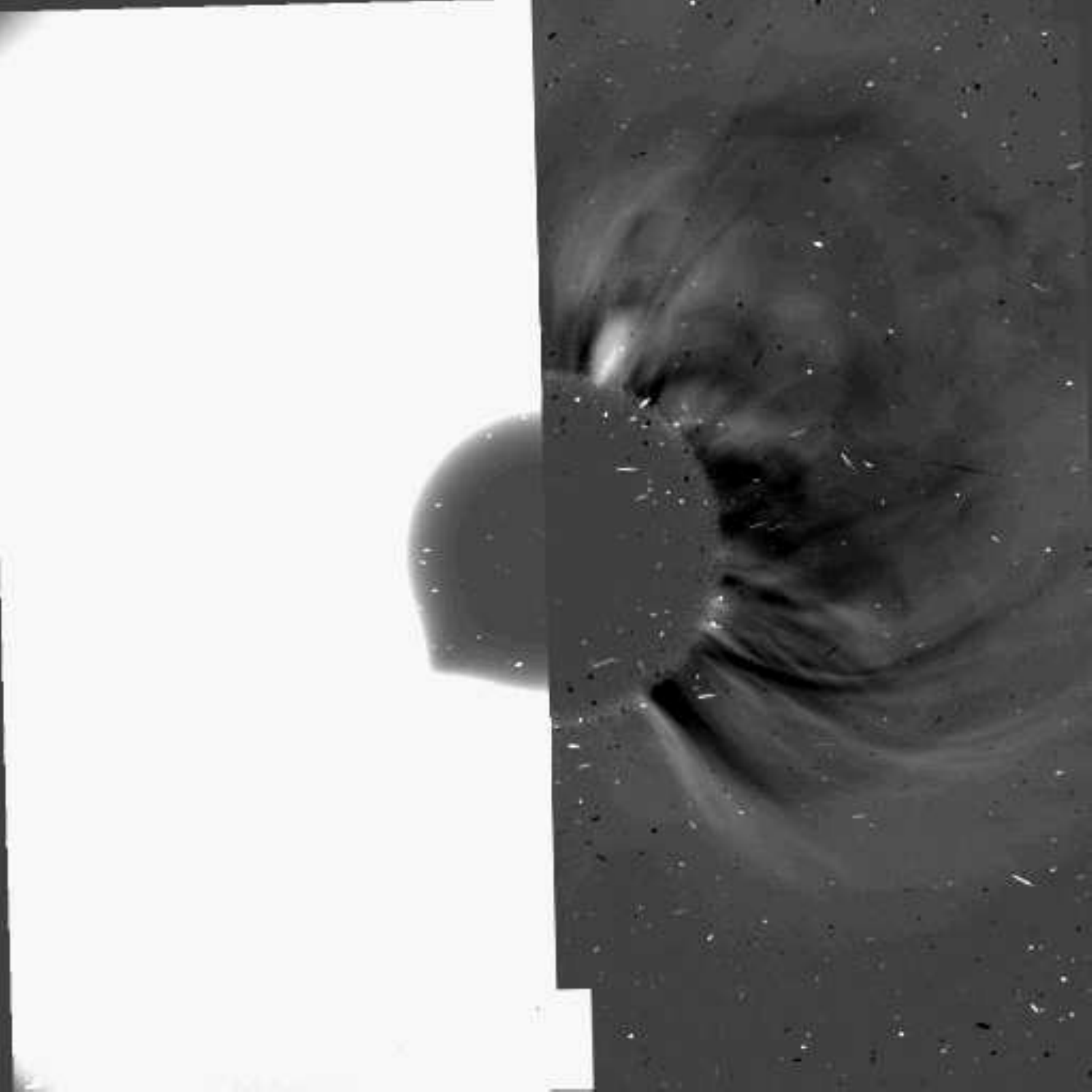}
             \hspace*{-0.02\textwidth}
               \includegraphics[width=0.4\textwidth,clip=]{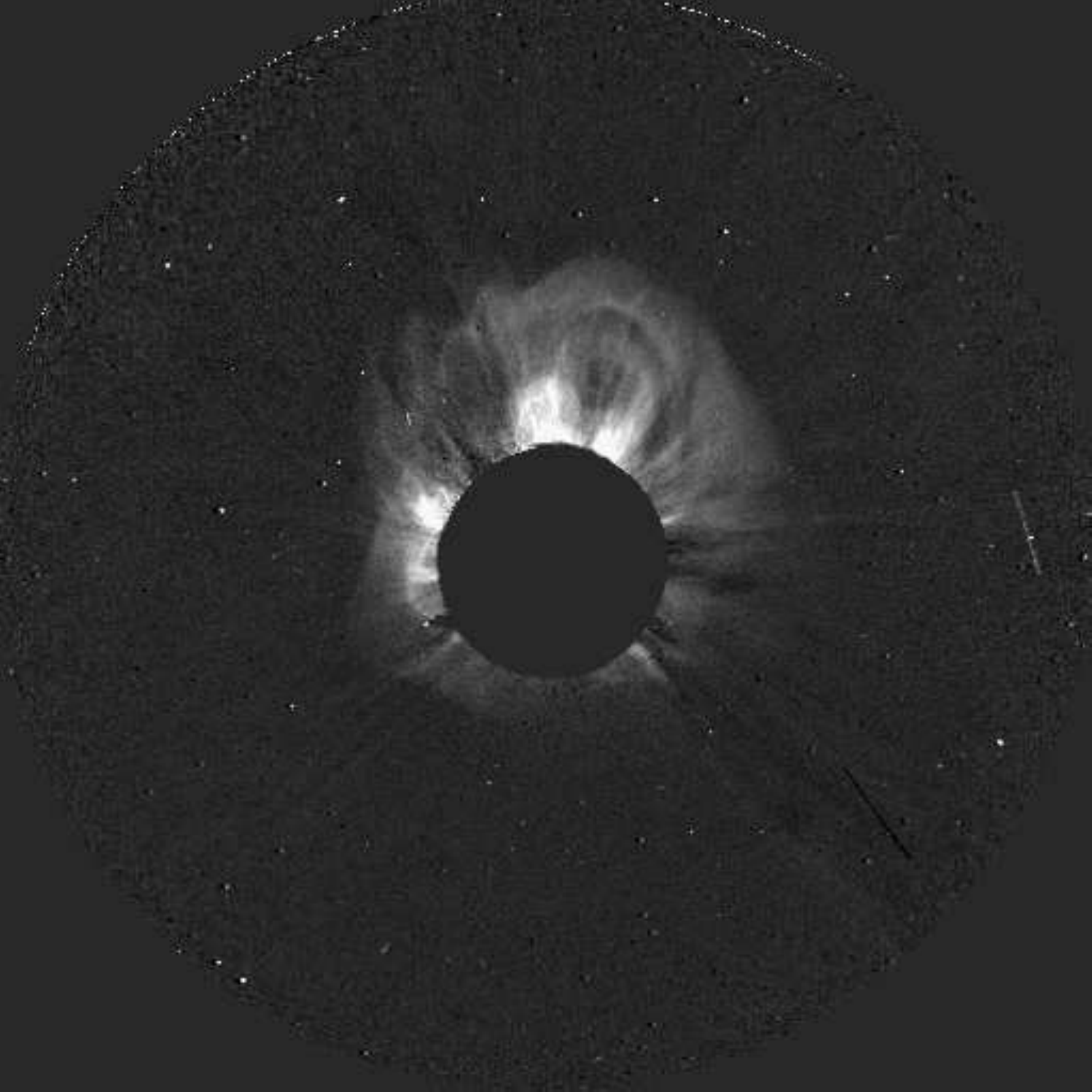}
               }
                 \centerline{\hspace*{0.04\textwidth}
              \includegraphics[width=0.4\textwidth,clip=]{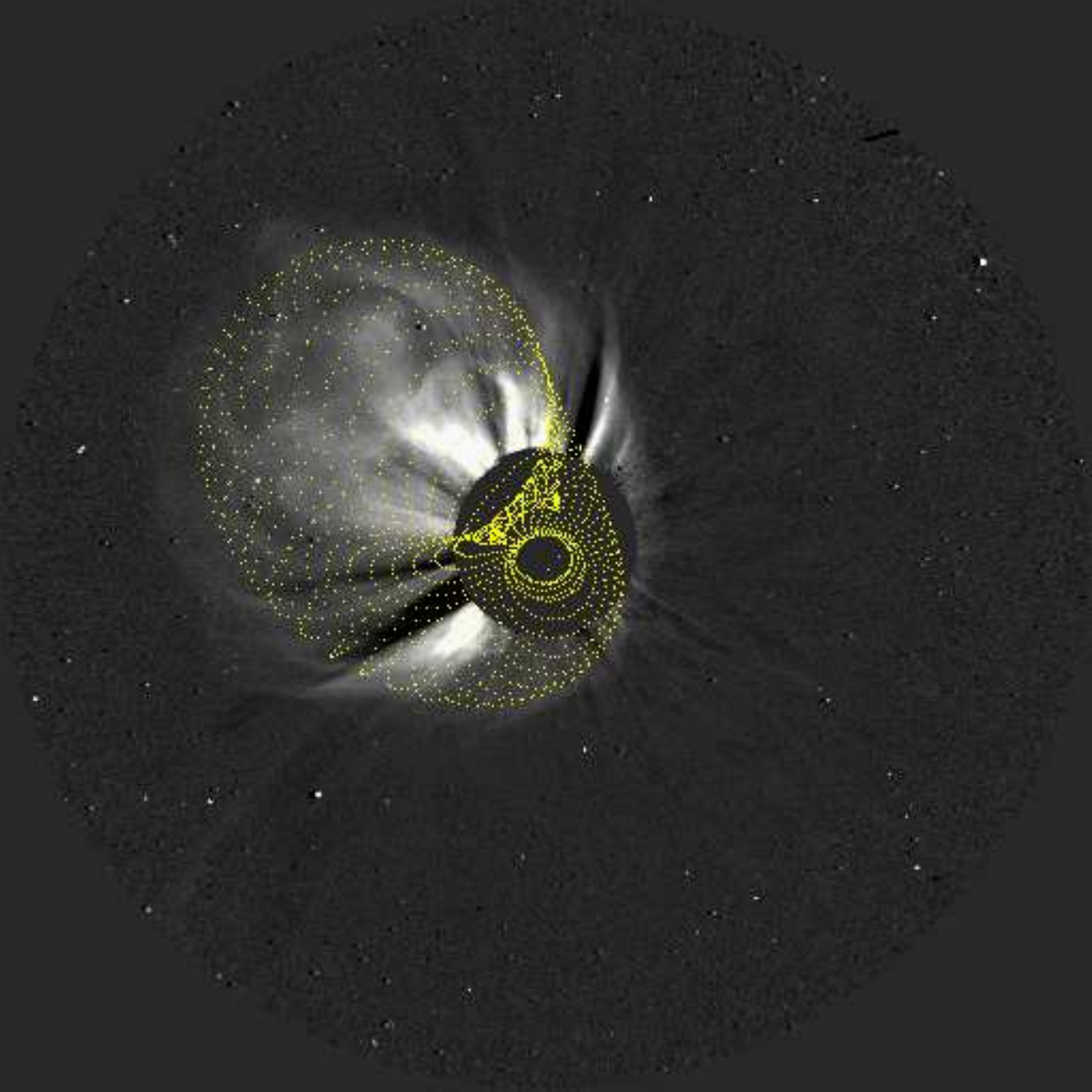}
               \hspace*{-0.02\textwidth}
               \includegraphics[width=0.4\textwidth,clip=]{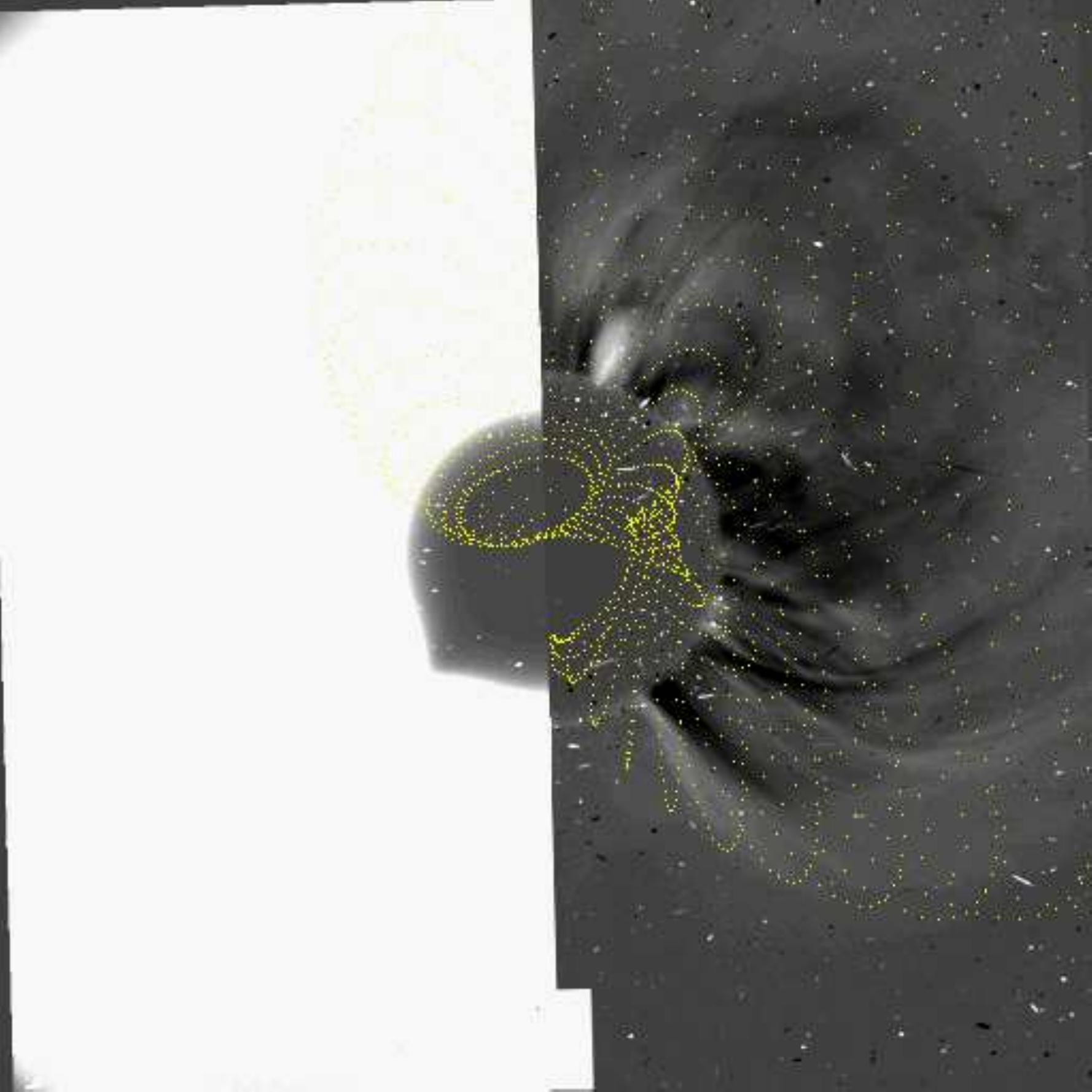}
              \hspace*{-0.02\textwidth}
               \includegraphics[width=0.4\textwidth,clip=]{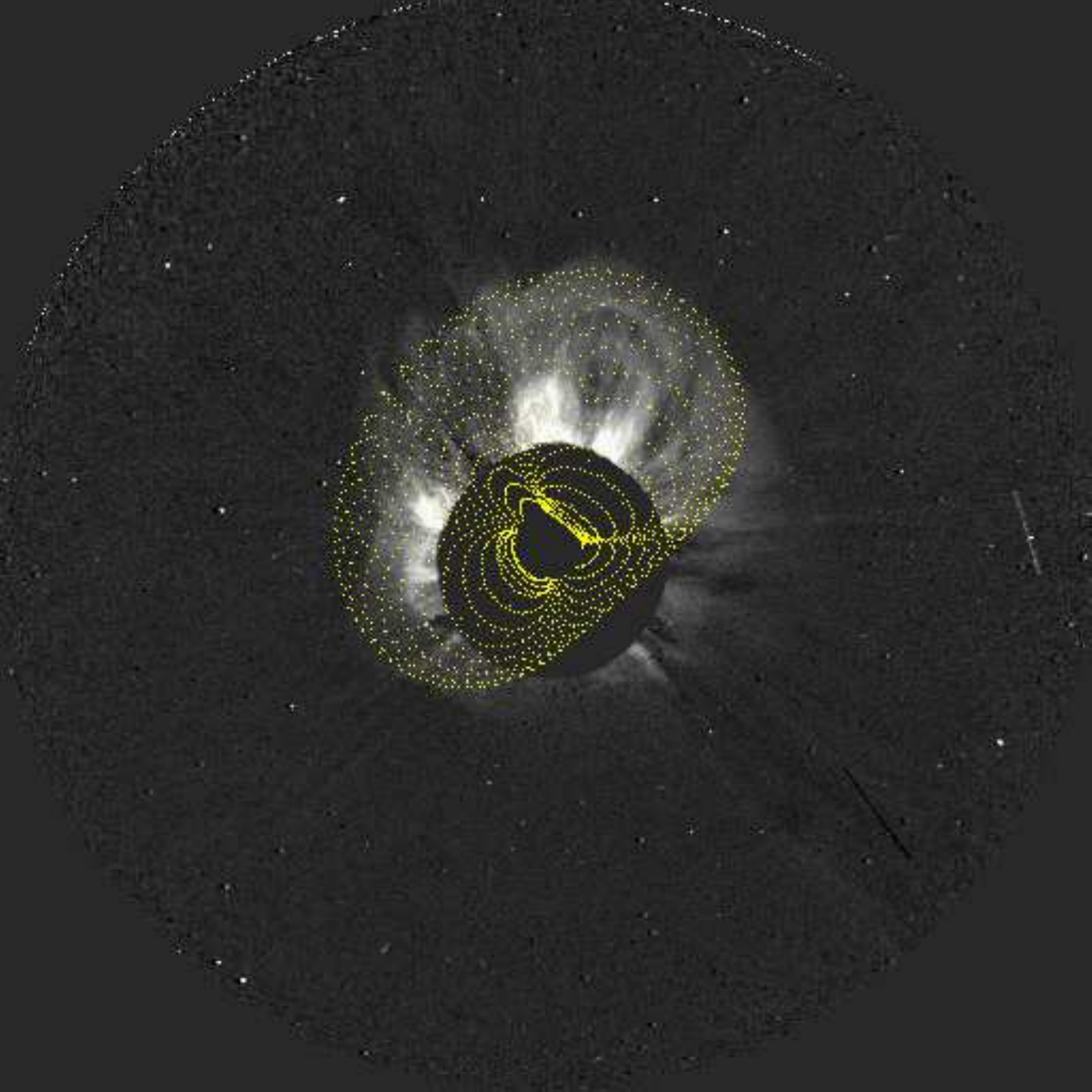}
                }
\vspace{0.0261\textwidth}  
\caption[GCS fit for CME 24 at 18:24]{GCS fit for CME 24 on March 13, 2012 at 18:24 UT at height $H=11.5$ \Rs. Table \ref{tblapp} 
lists the GCS parameters for this event.}
\label{figa24}
\end{figure}

\begin{figure}[h]    
  \centering                              
   \centerline{\hspace*{0.00\textwidth}
               \includegraphics[width=0.4\textwidth,clip=]{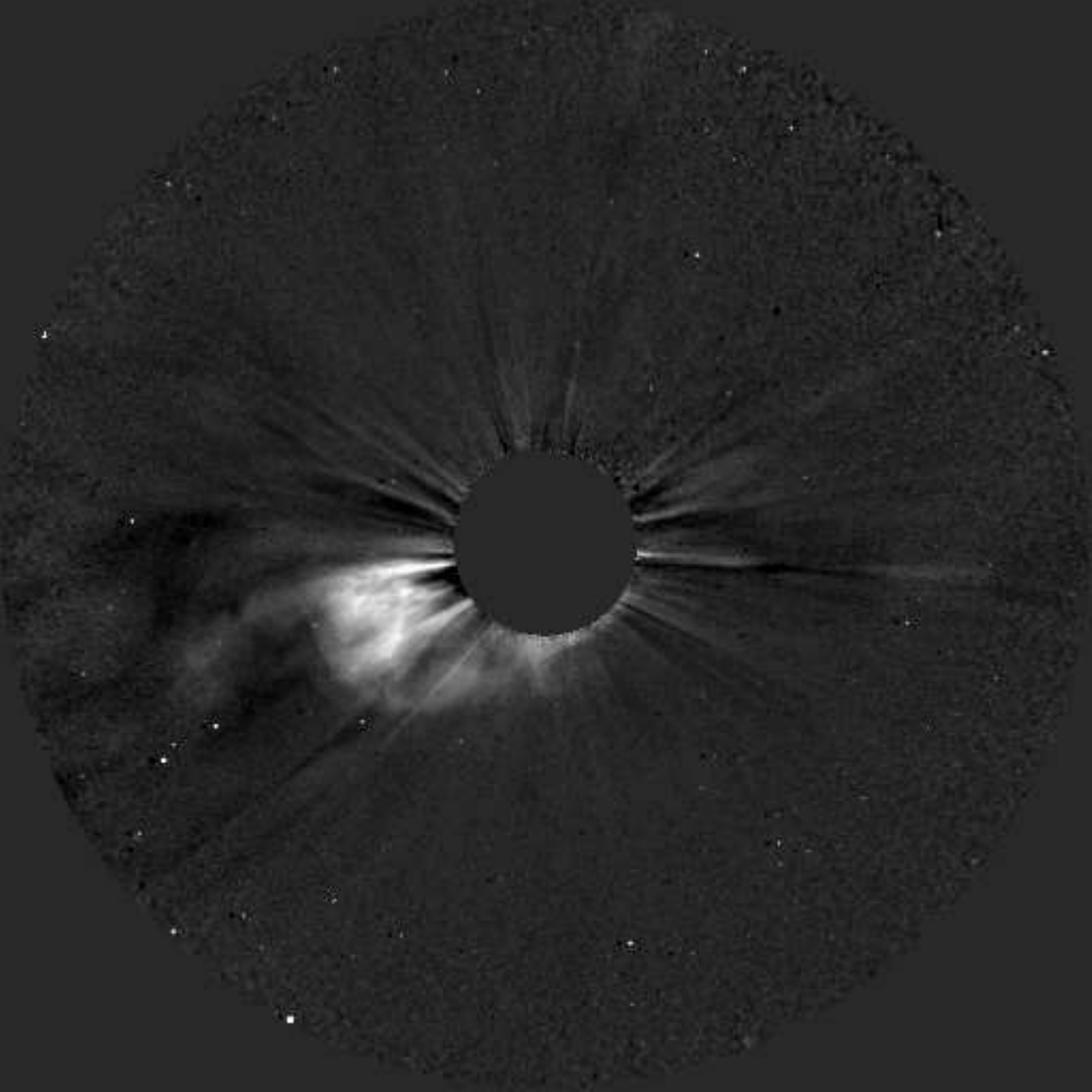}
                \hspace*{-0.02\textwidth}
               \includegraphics[width=0.4\textwidth,clip=]{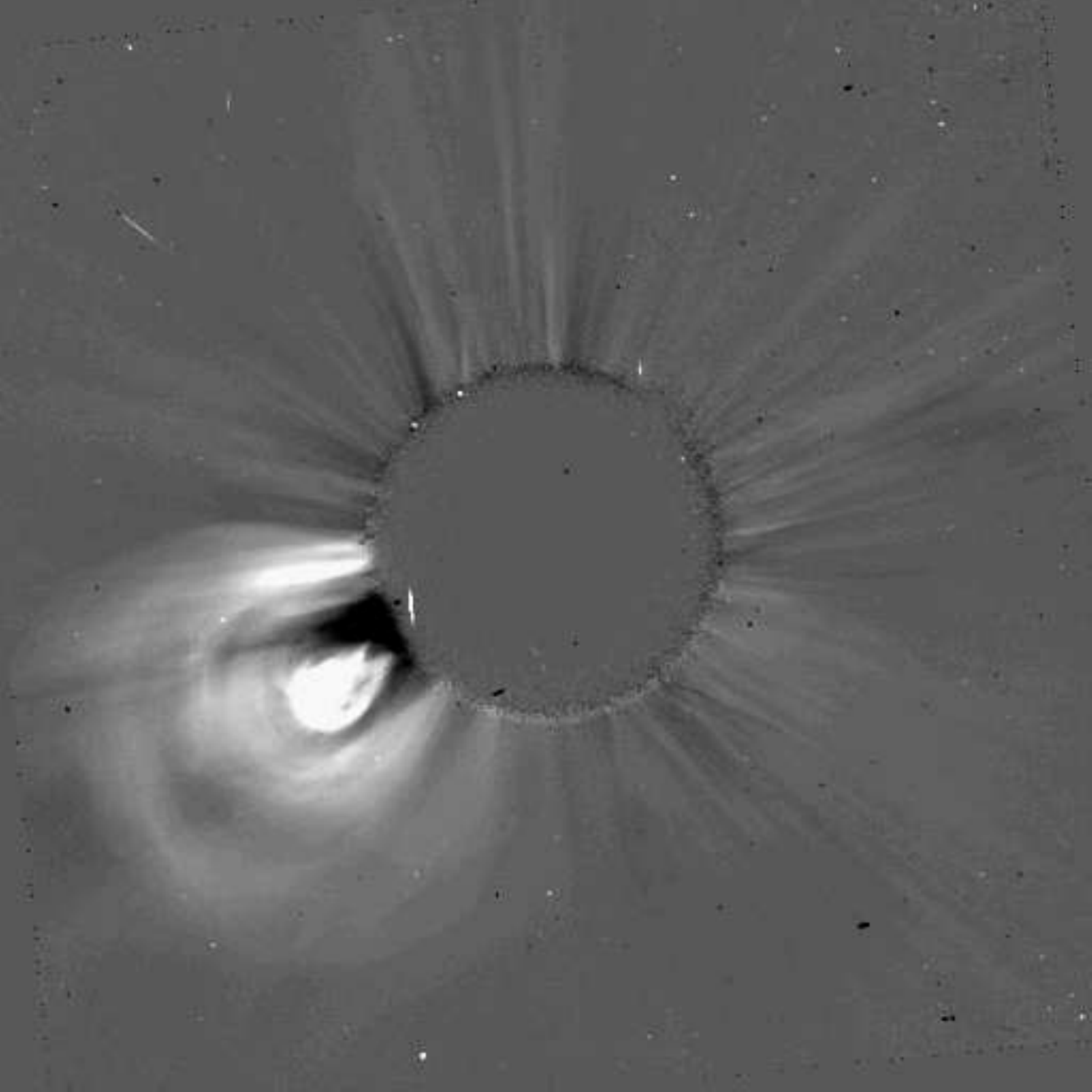}
             \hspace*{-0.02\textwidth}
               \includegraphics[width=0.4\textwidth,clip=]{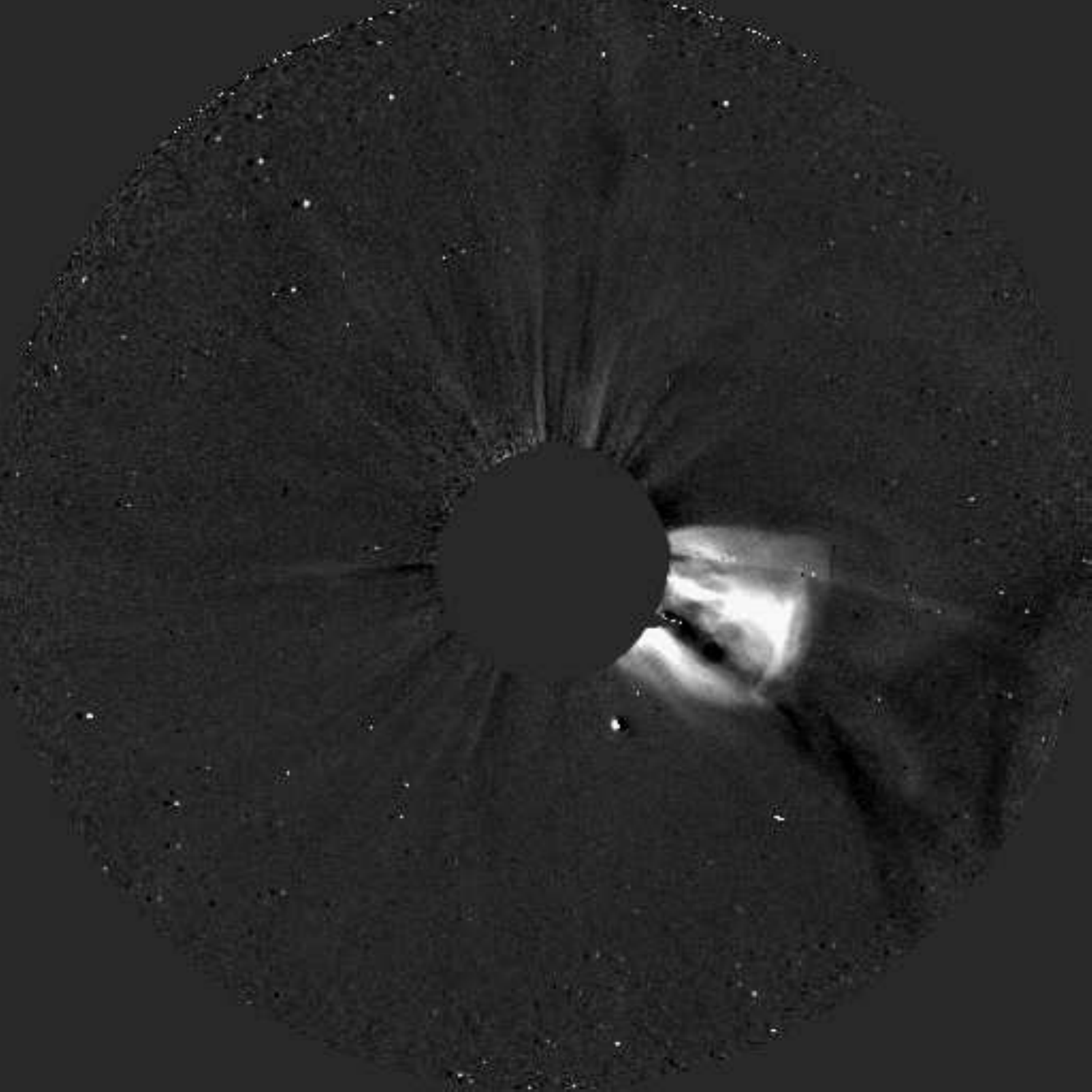}
               }
                 \centerline{\hspace*{0.0\textwidth}
              \includegraphics[width=0.4\textwidth,clip=]{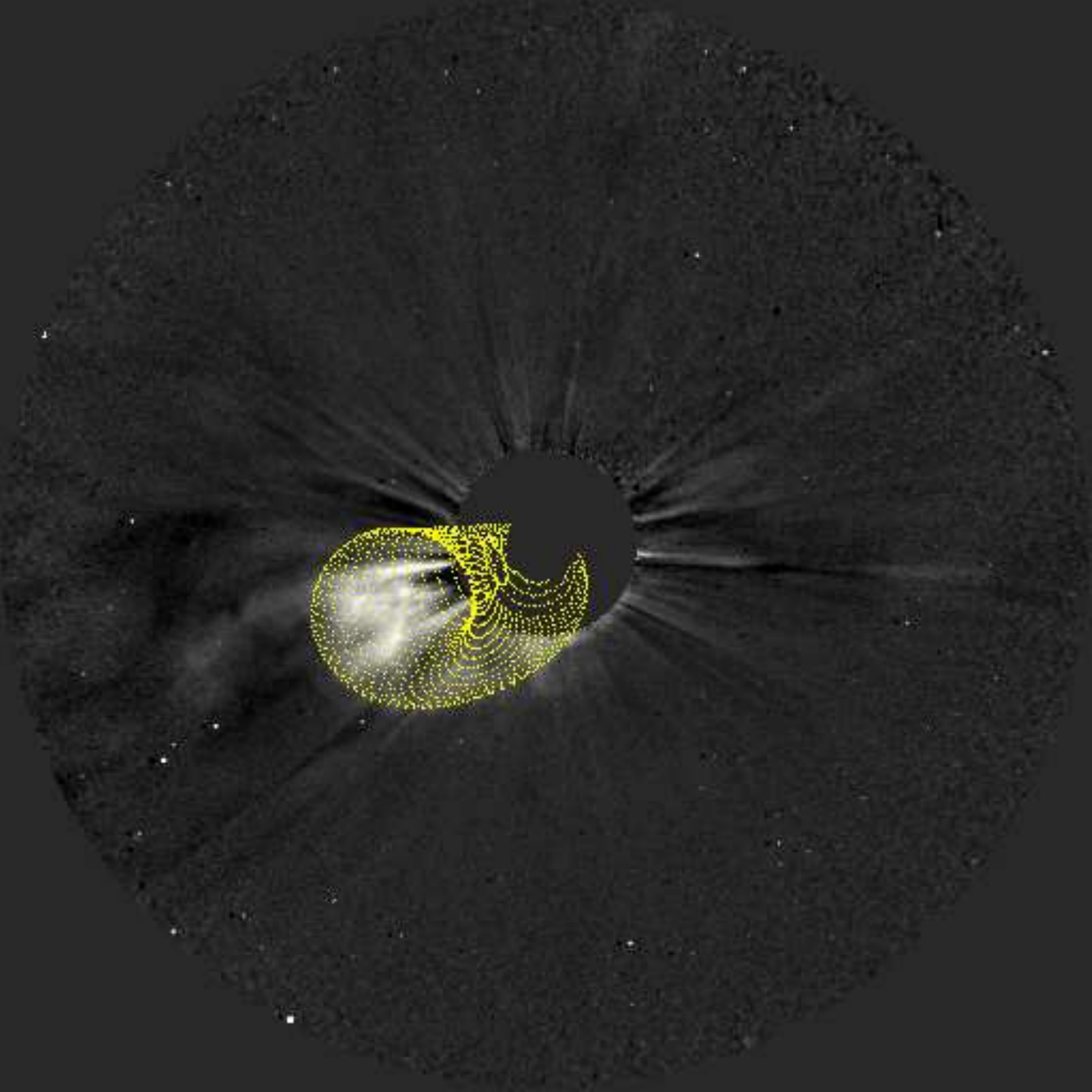}
               \hspace*{-0.02\textwidth}
               \includegraphics[width=0.4\textwidth,clip=]{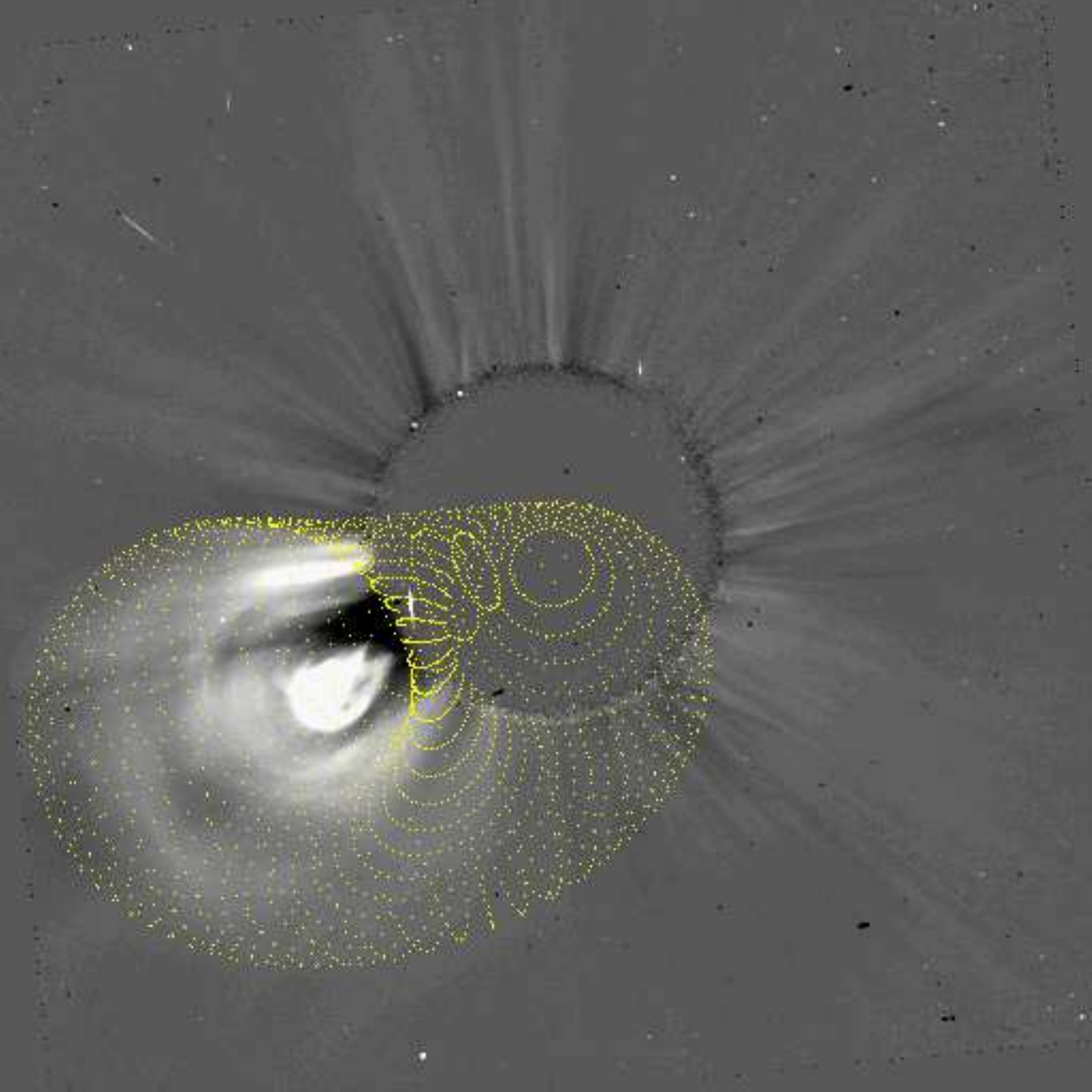}
              \hspace*{-0.02\textwidth}
               \includegraphics[width=0.4\textwidth,clip=]{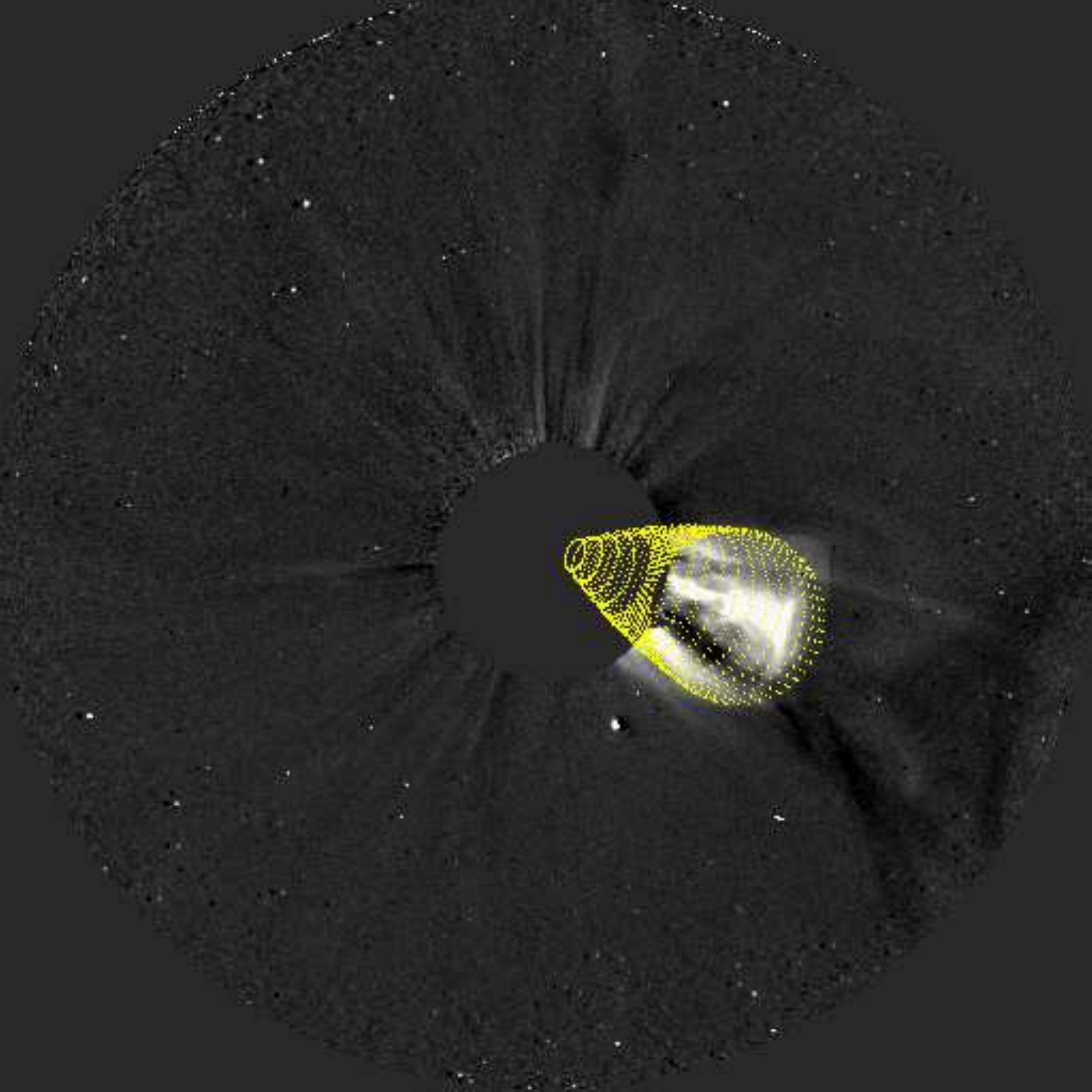}
                }
\vspace{0.0261\textwidth}  
\caption[GCS fit for CME 25 at 17:24]{GCS fit for CME 25 on April 19, 2012 at 17:24 UT at height $H=8.4$ \Rs. Table \ref{tblapp} 
lists the GCS parameters for this event.}
\label{figa25}
\end{figure}

\begin{figure}[h]    
  \centering                              
   \centerline{\hspace*{0.04\textwidth}
               \includegraphics[width=0.4\textwidth,clip=]{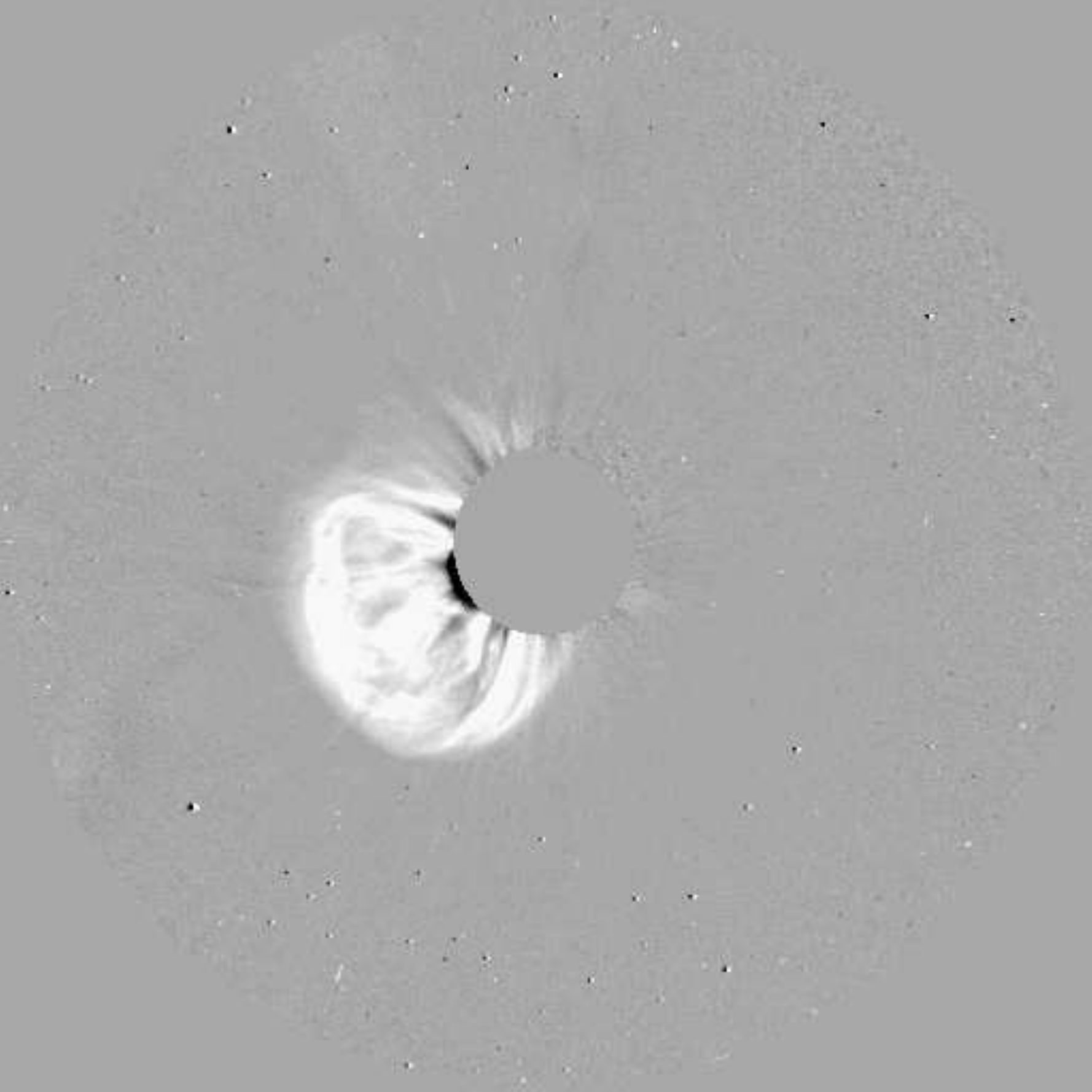}
                \hspace*{-0.02\textwidth}
               \includegraphics[width=0.4\textwidth,clip=]{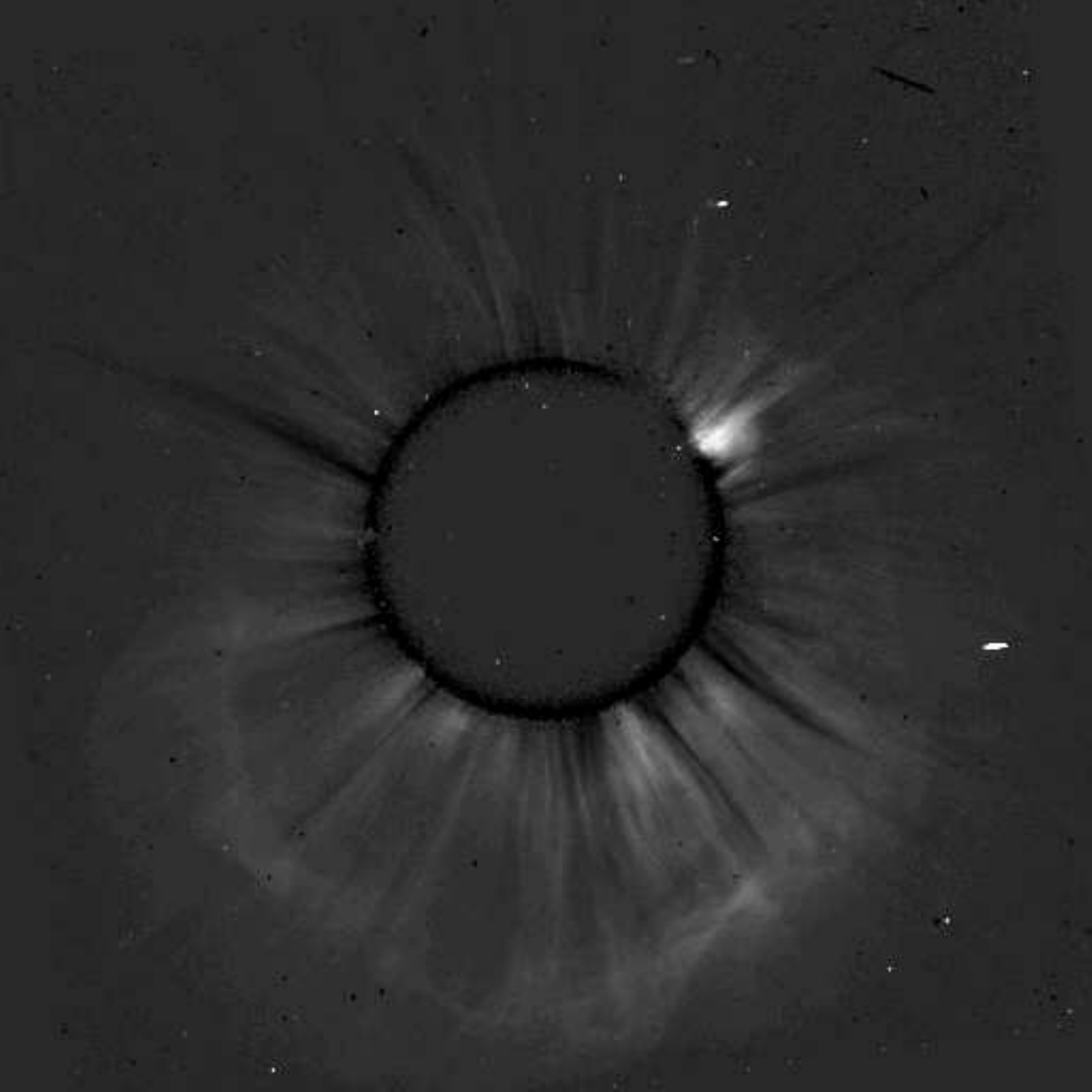}
             \hspace*{-0.02\textwidth}
               \includegraphics[width=0.4\textwidth,clip=]{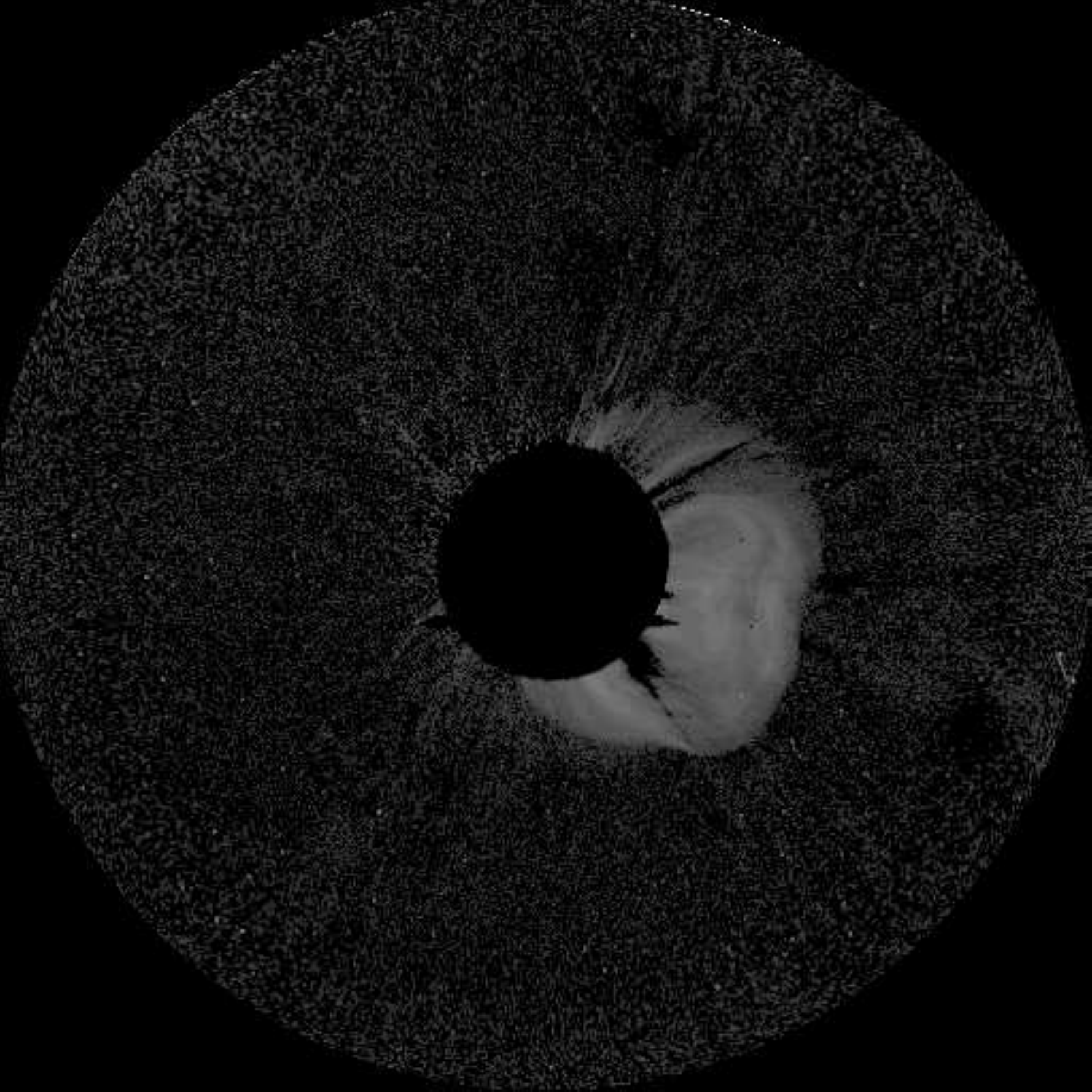}
               }
                 \centerline{\hspace*{0.04\textwidth}
              \includegraphics[width=0.4\textwidth,clip=]{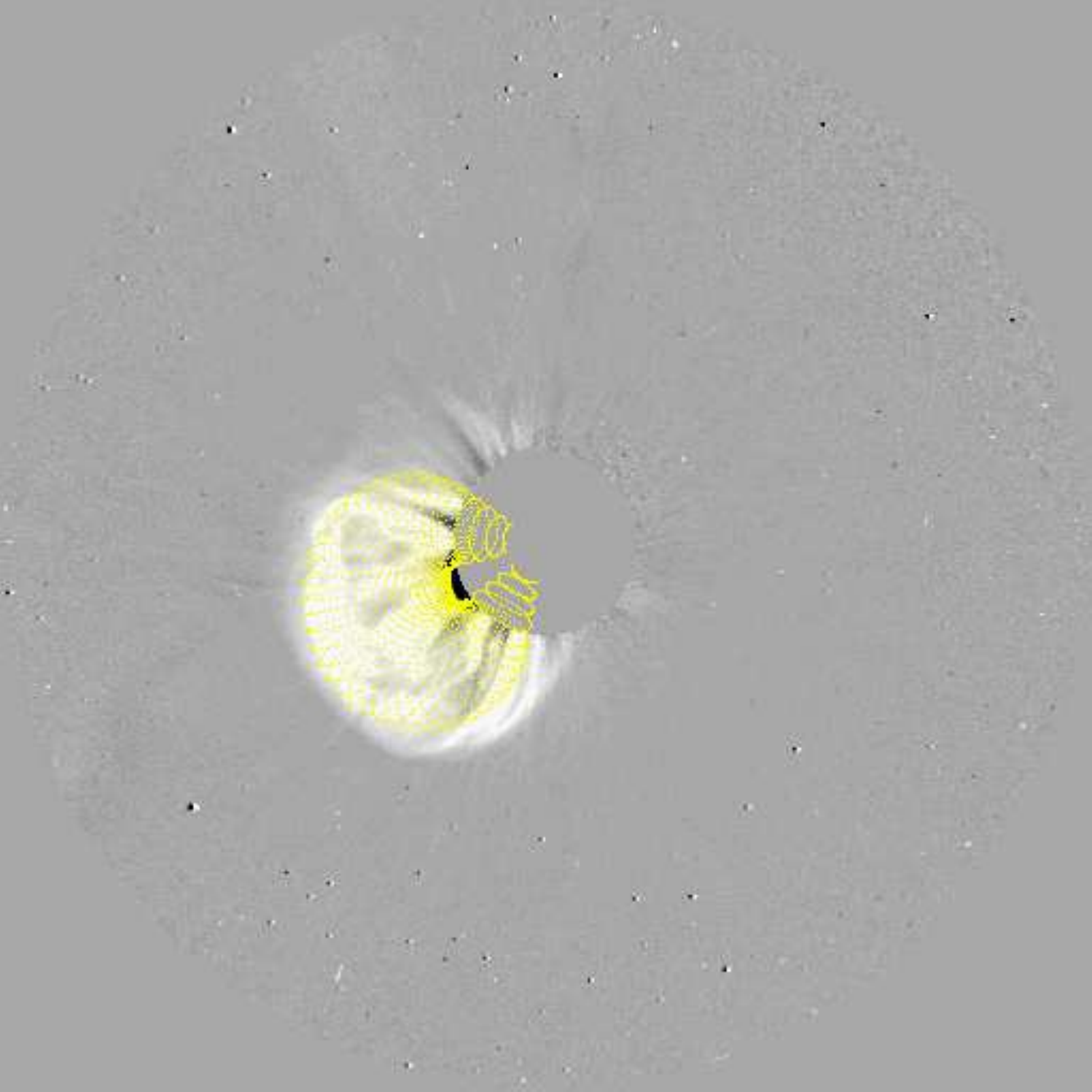}
               \hspace*{-0.02\textwidth}
               \includegraphics[width=0.4\textwidth,clip=]{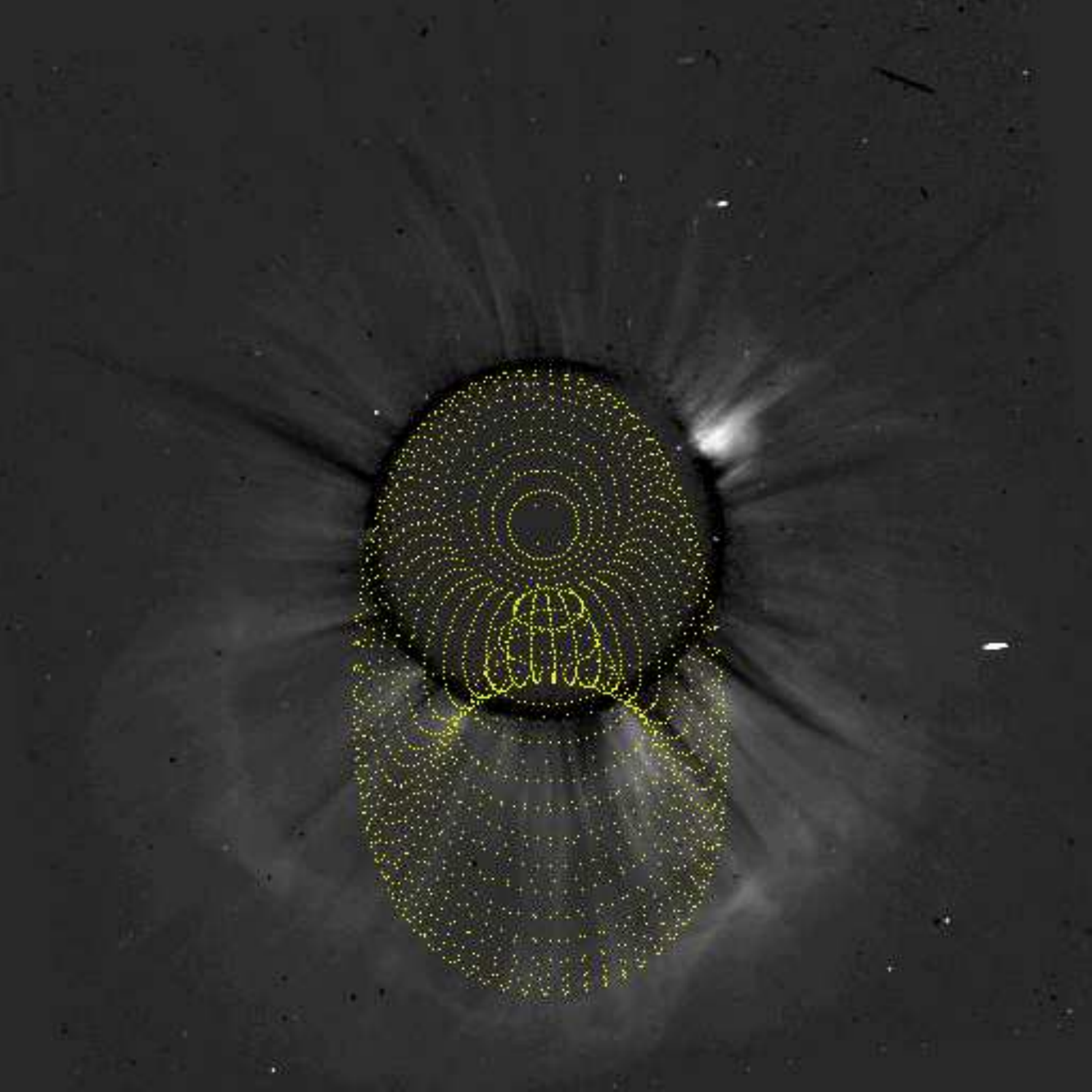}
              \hspace*{-0.02\textwidth}
               \includegraphics[width=0.4\textwidth,clip=]{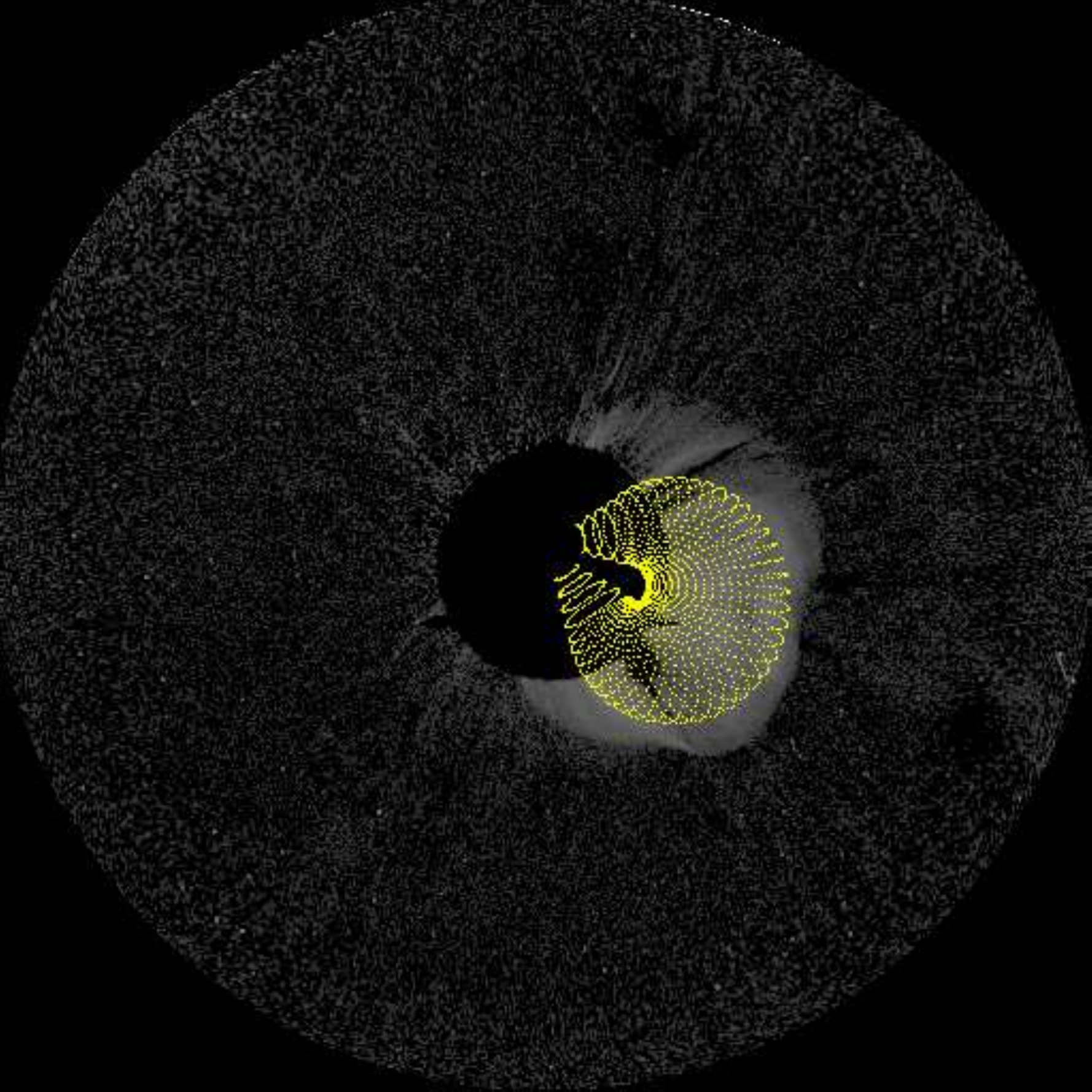}
                }
\vspace{0.0261\textwidth}  
\caption[GCS fit for CME 26 at 14:39]{GCS fit for CME 26 on June 14, 2012 at 14:39 UT at height $H=7.8$ \Rs. Table \ref{tblapp} 
lists the GCS parameters for this event.}
\label{figa26}
\end{figure}

\begin{figure}[h]    
  \centering                              
   \centerline{\hspace*{0.00\textwidth}
               \includegraphics[width=0.4\textwidth,clip=]{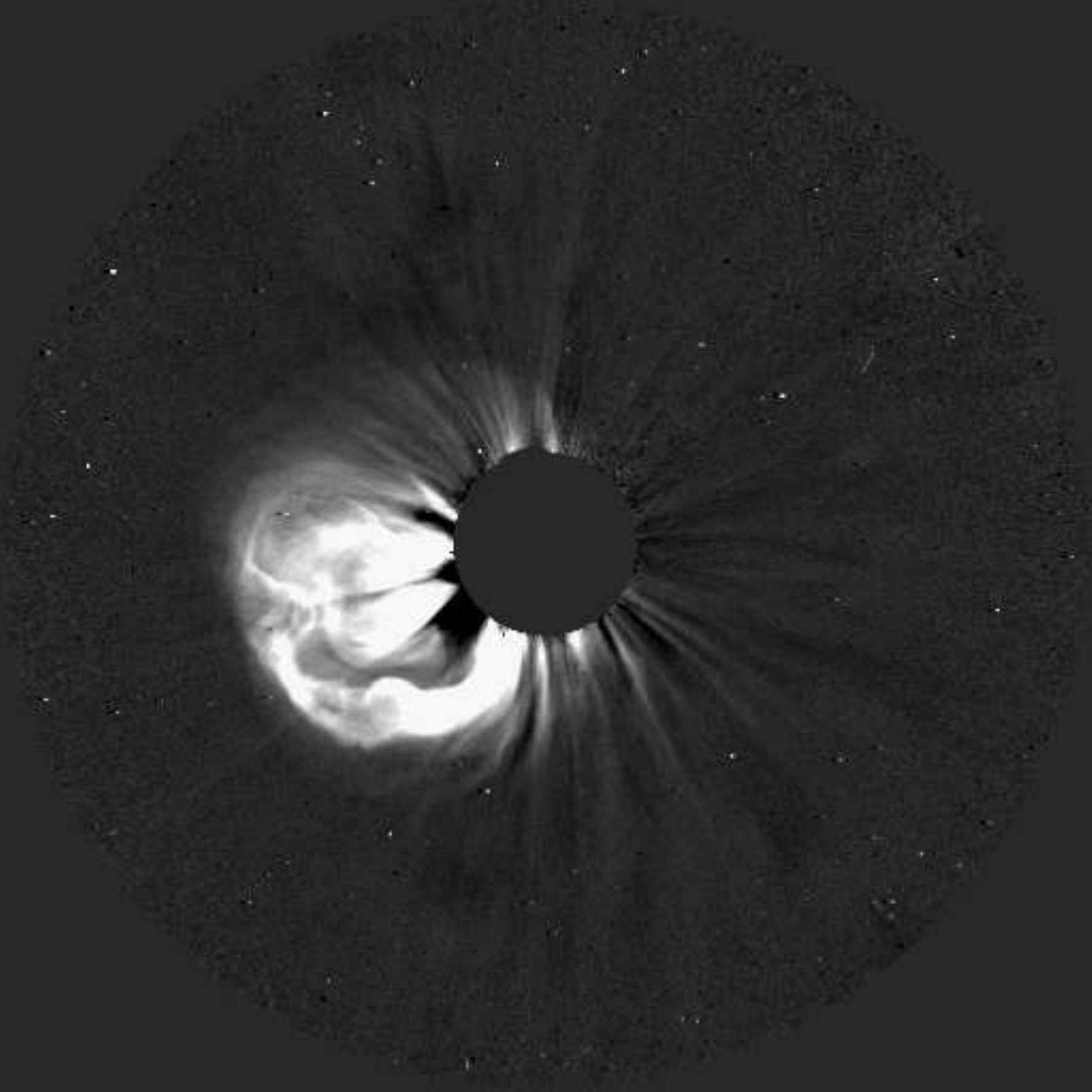}
                \hspace*{-0.02\textwidth}
               \includegraphics[width=0.4\textwidth,clip=]{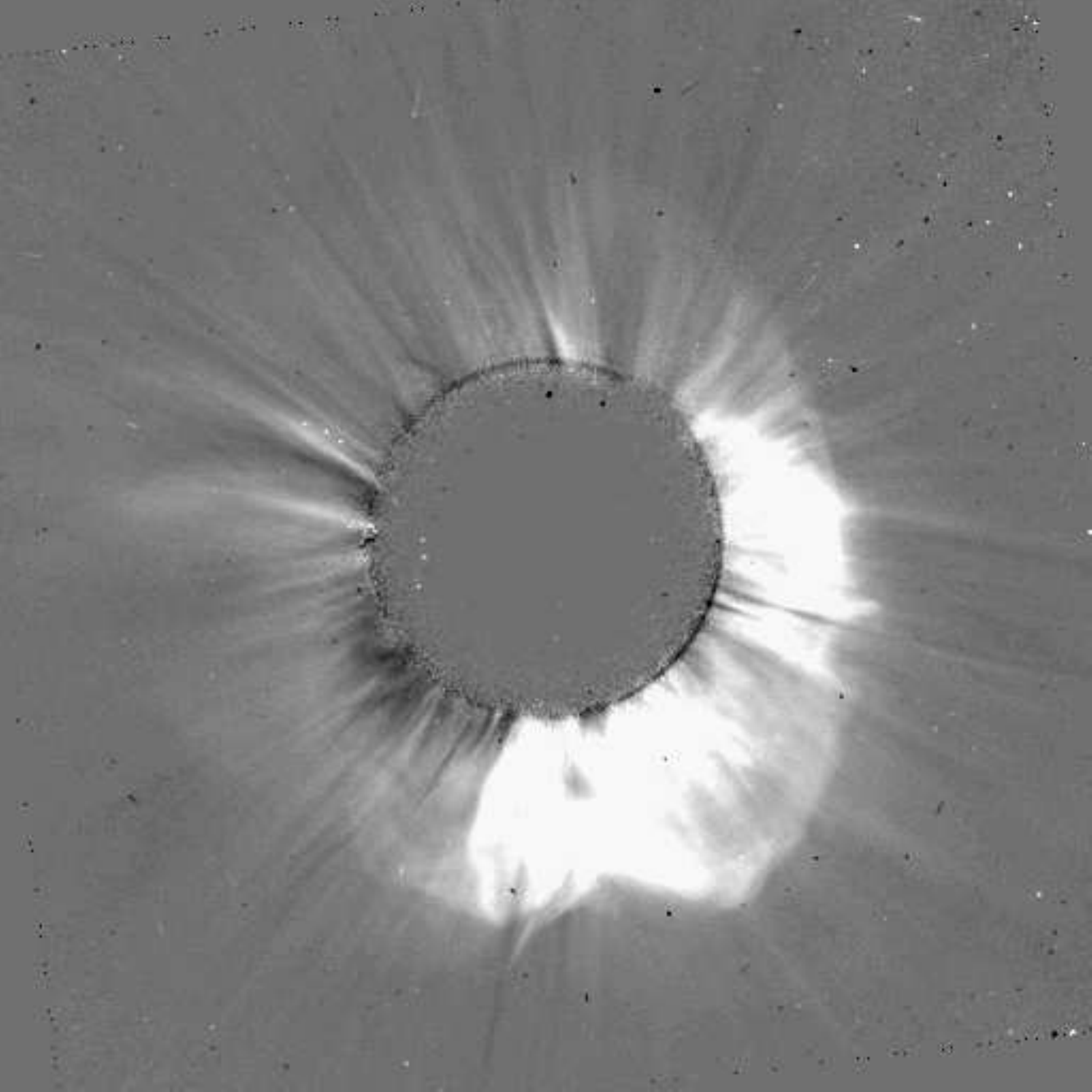}
             \hspace*{-0.02\textwidth}
               \includegraphics[width=0.4\textwidth,clip=]{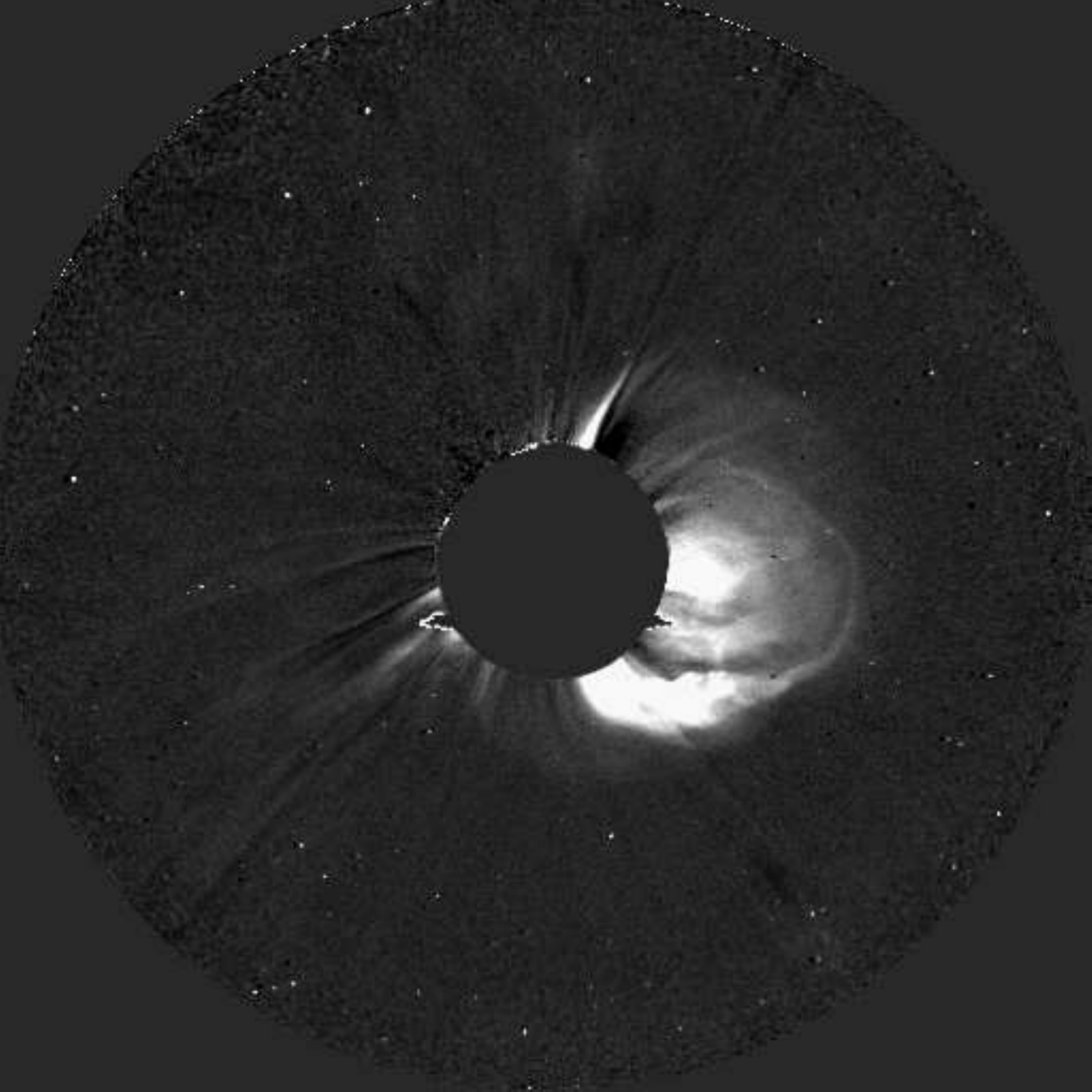}
               }
                 \centerline{\hspace*{0.0\textwidth}
              \includegraphics[width=0.4\textwidth,clip=]{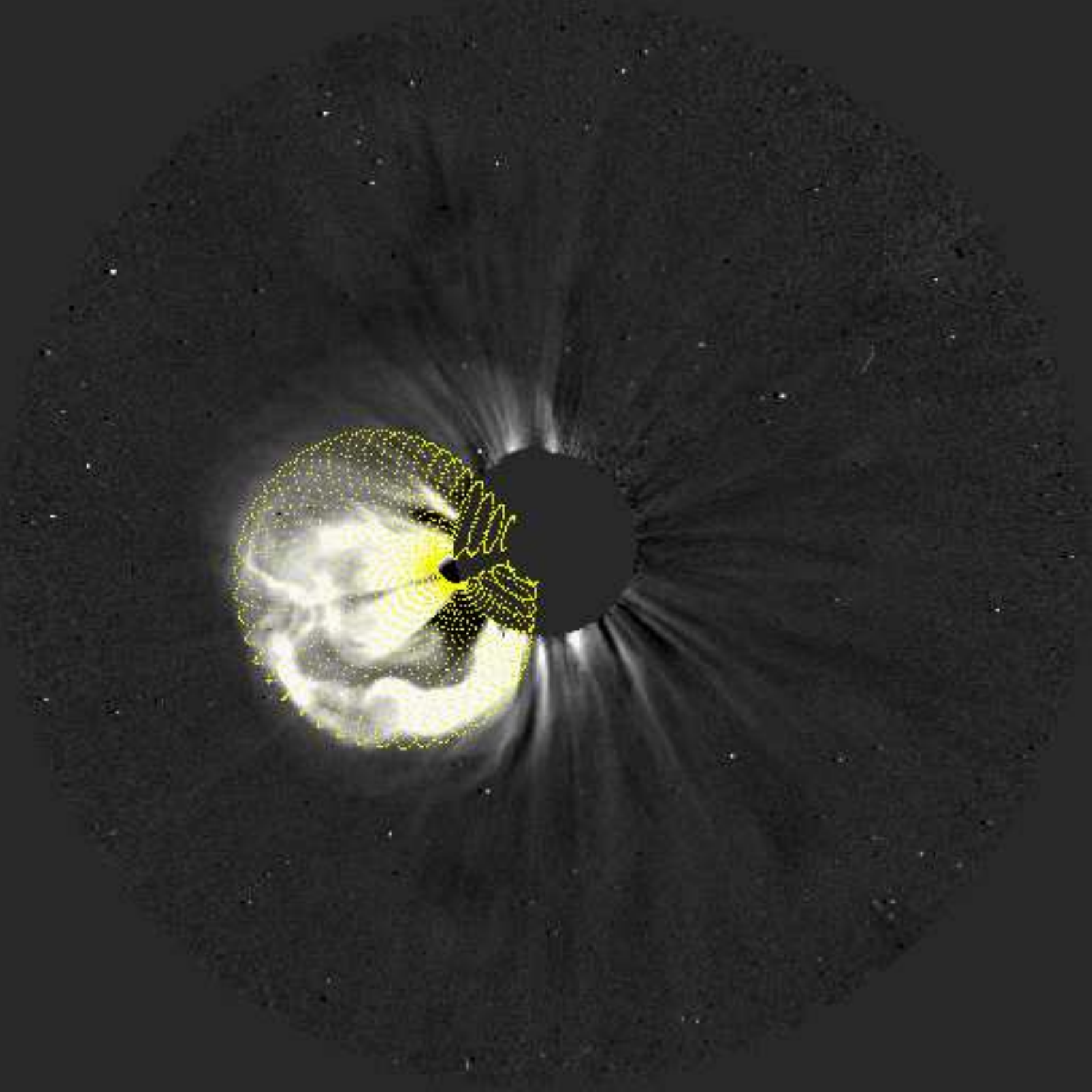}
               \hspace*{-0.02\textwidth}
               \includegraphics[width=0.4\textwidth,clip=]{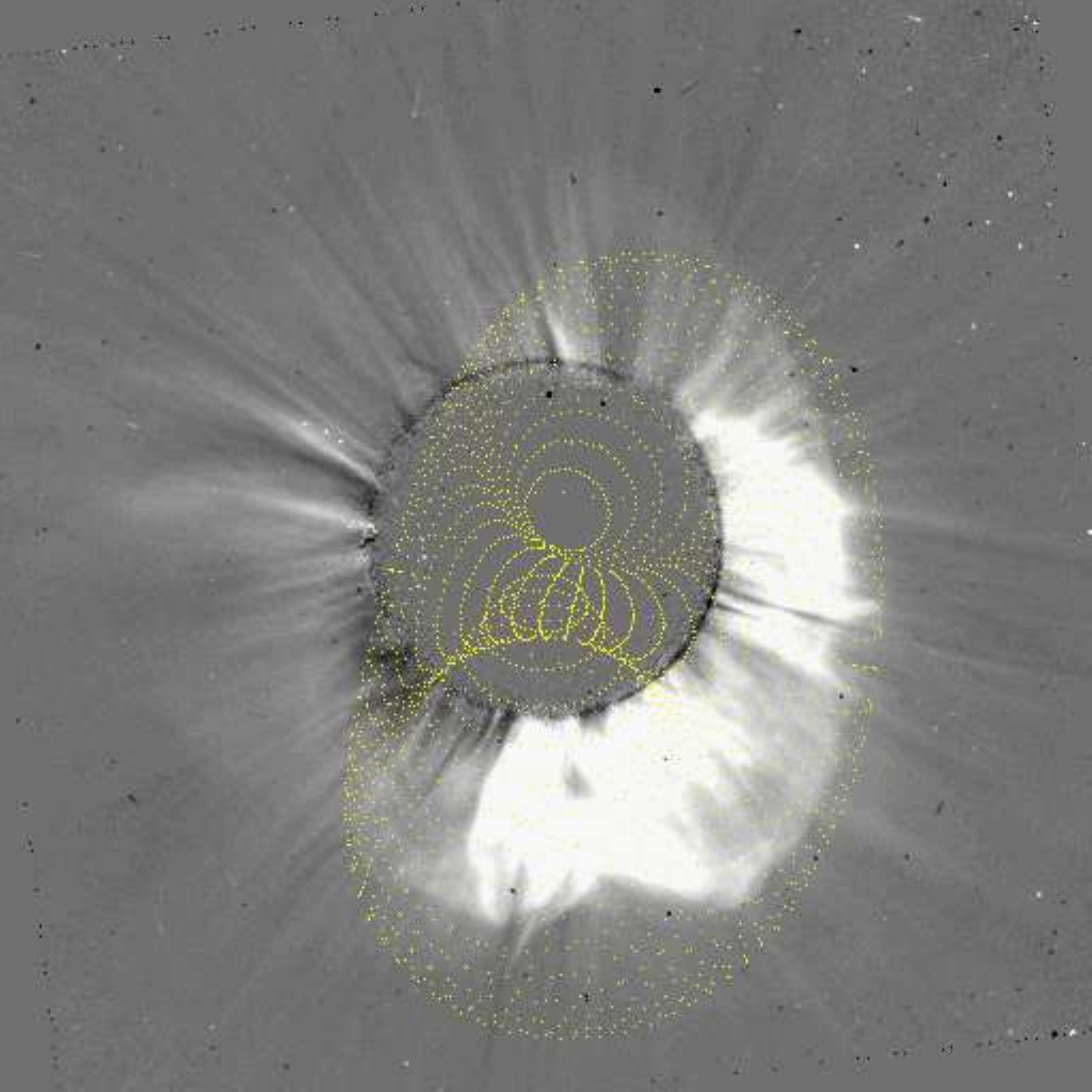}
              \hspace*{-0.02\textwidth}
               \includegraphics[width=0.4\textwidth,clip=]{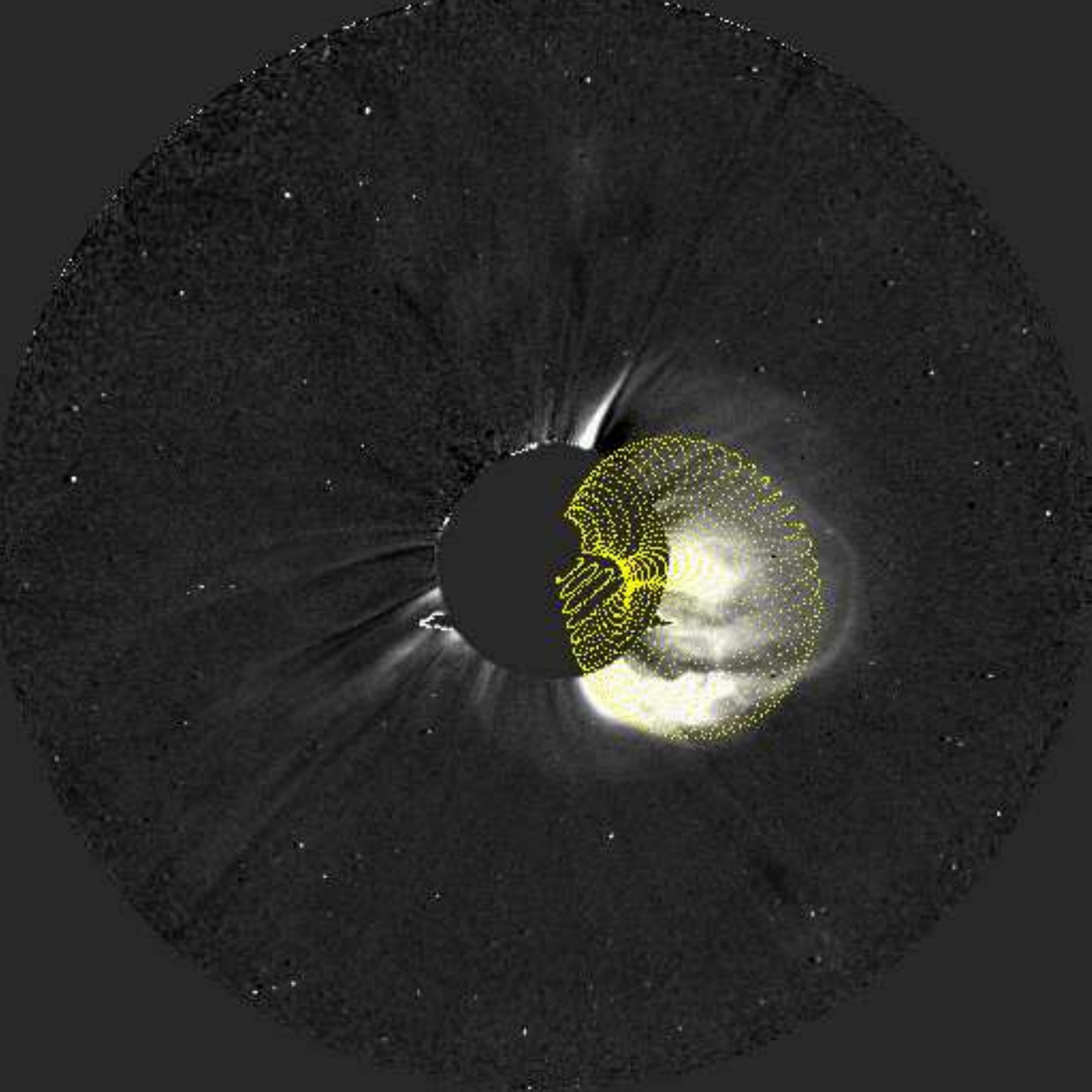}
                }
\vspace{0.0261\textwidth}  
\caption[GCS fit for CME 27 at 17:39]{GCS fit for CME 27 on July 12, 2012 at 17:39 UT at height $H=9.2$ \Rs. Table \ref{tblapp} 
lists the GCS parameters for this event.}
\label{figa27}
\end{figure}

\begin{figure}[h]    
  \centering                              
   \centerline{\hspace*{0.04\textwidth}
               \includegraphics[width=0.4\textwidth,clip=]{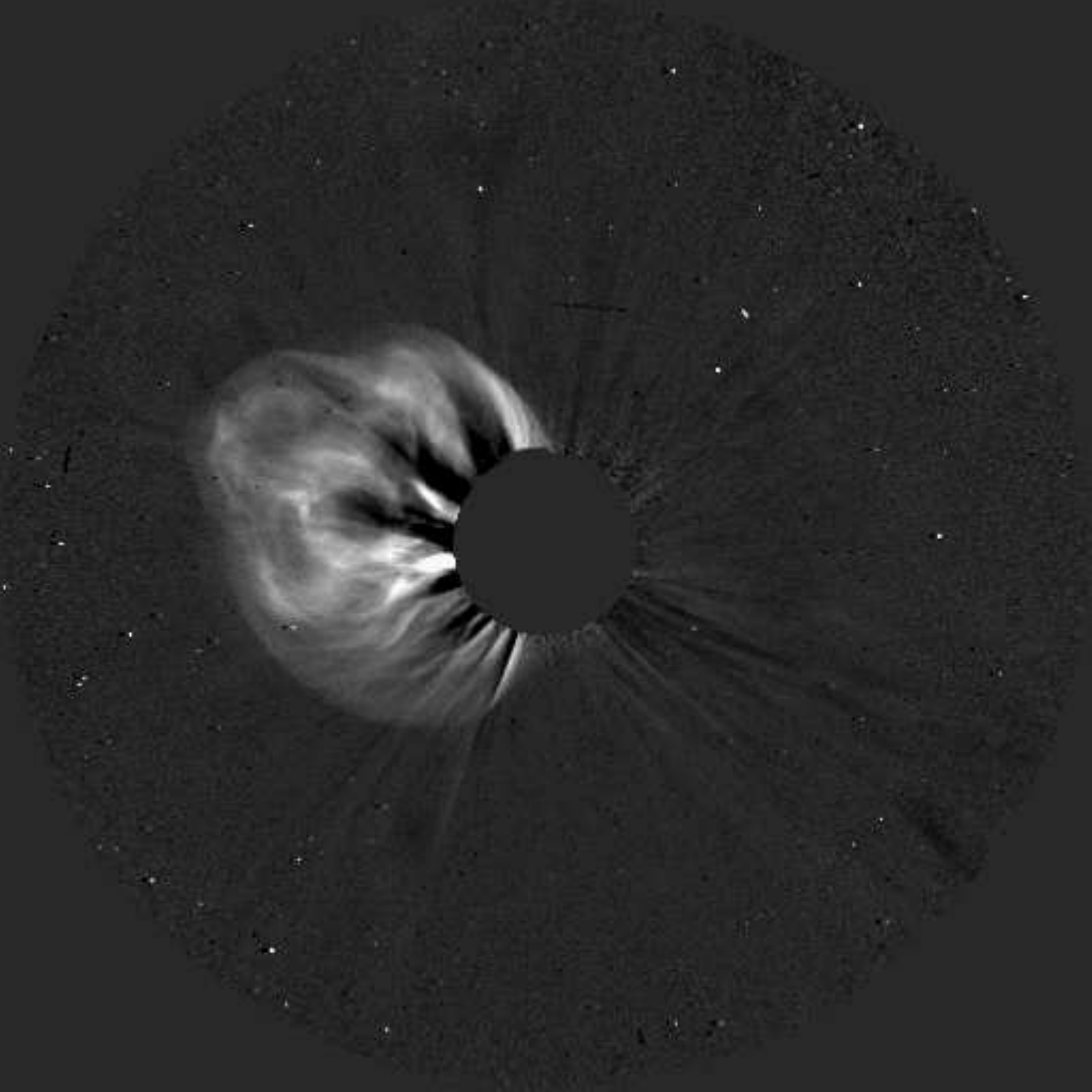}
                \hspace*{-0.02\textwidth}
               \includegraphics[width=0.4\textwidth,clip=]{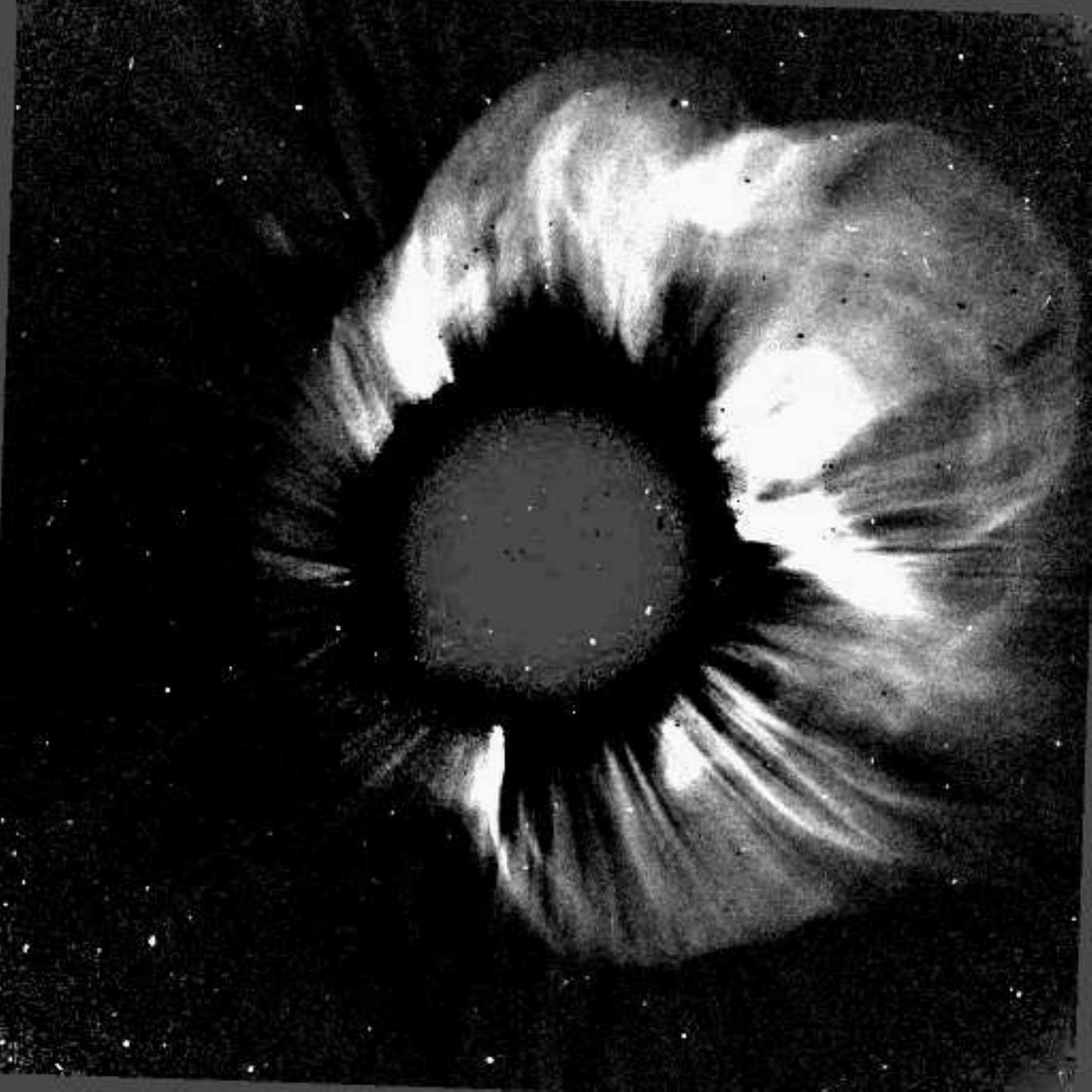}
             \hspace*{-0.02\textwidth}
               \includegraphics[width=0.4\textwidth,clip=]{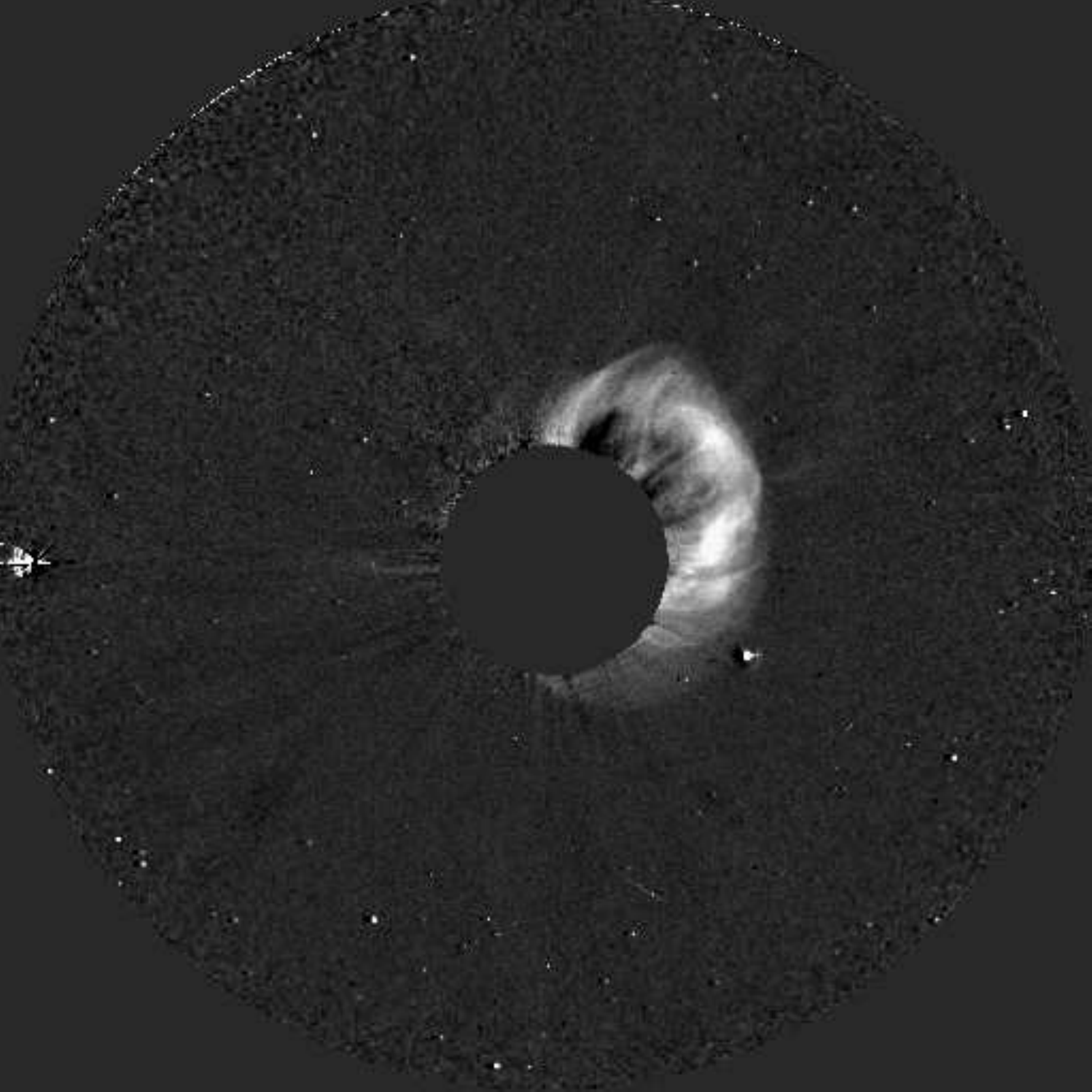}
               }
                 \centerline{\hspace*{0.04\textwidth}
              \includegraphics[width=0.4\textwidth,clip=]{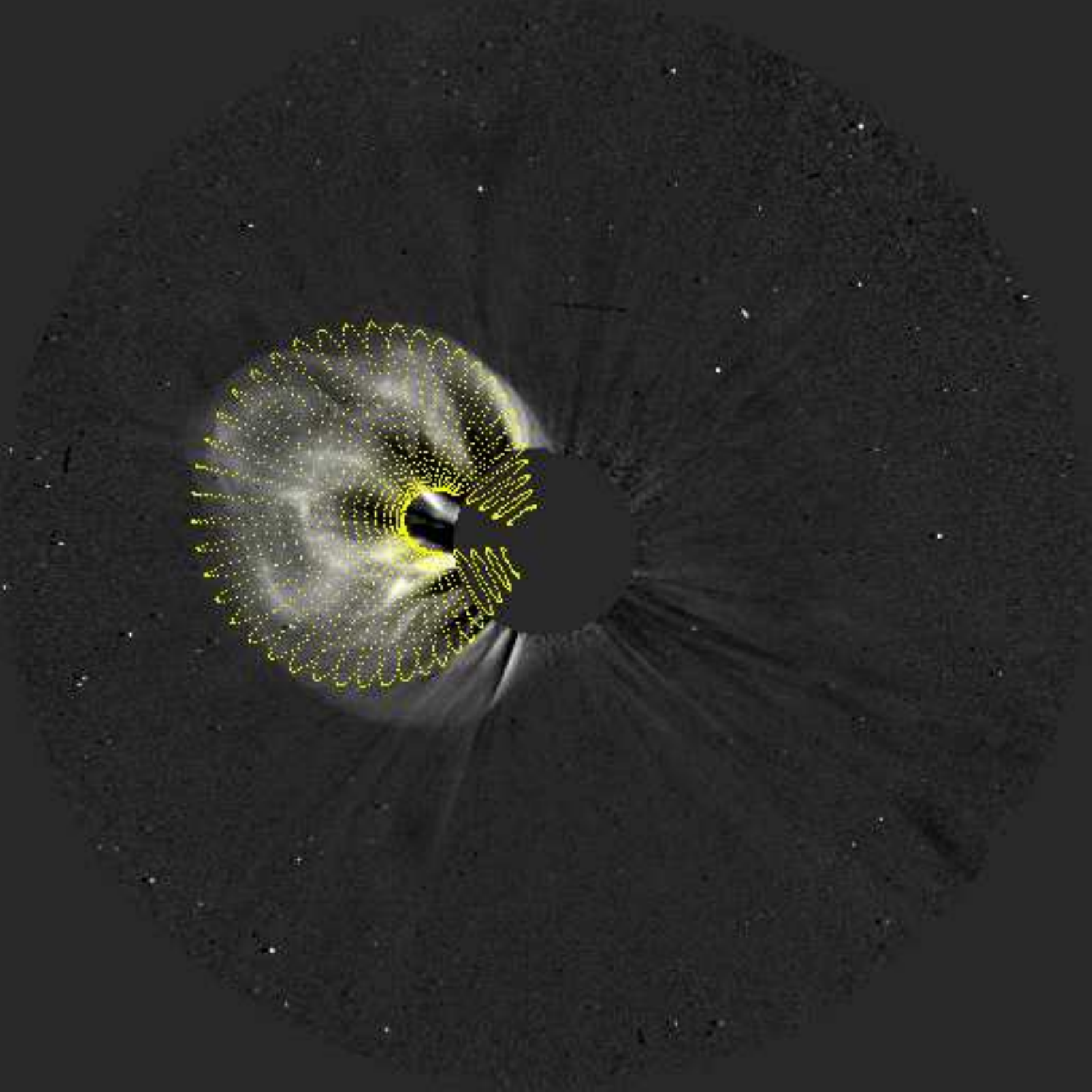}
               \hspace*{-0.02\textwidth}
               \includegraphics[width=0.4\textwidth,clip=]{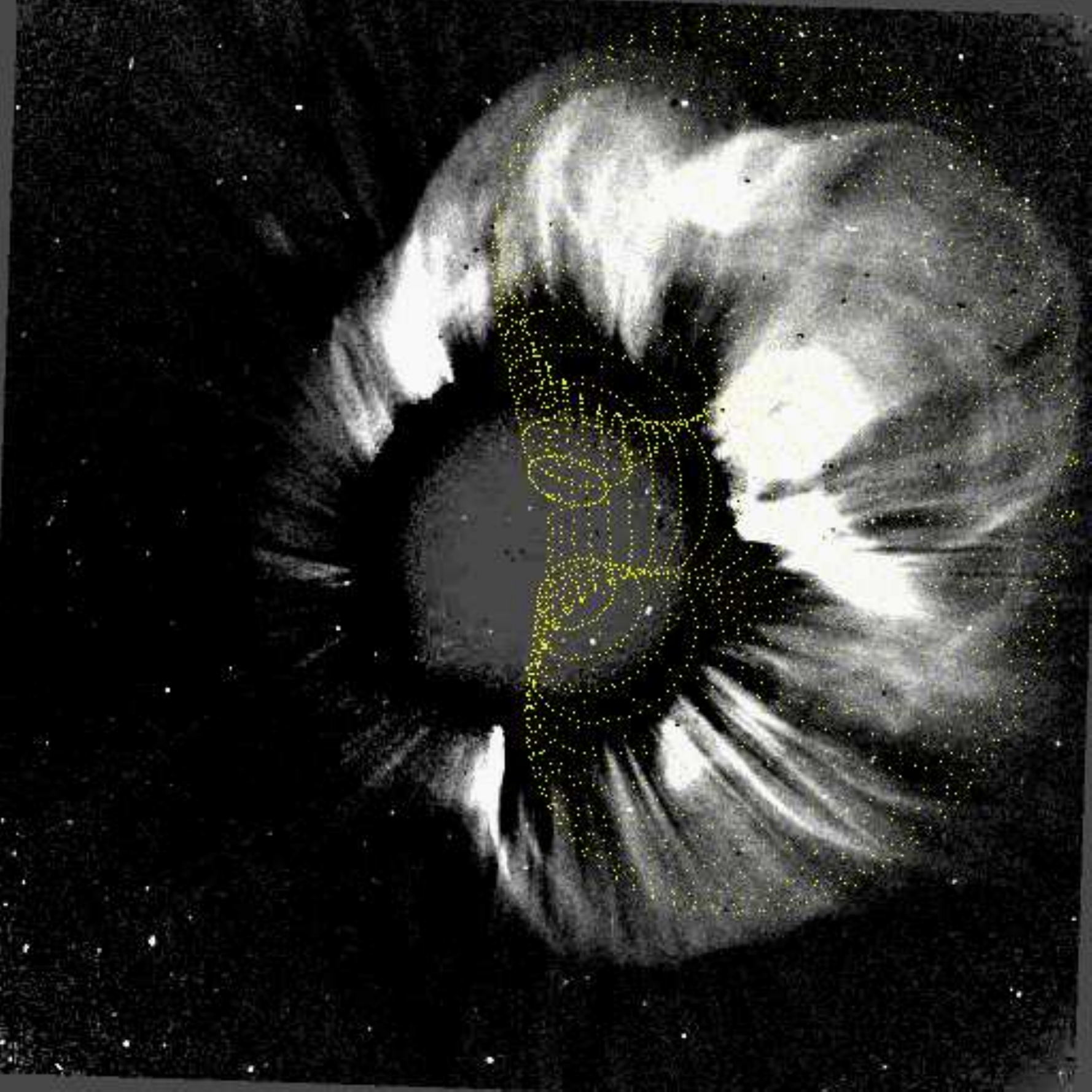}
              \hspace*{-0.02\textwidth}
               \includegraphics[width=0.4\textwidth,clip=]{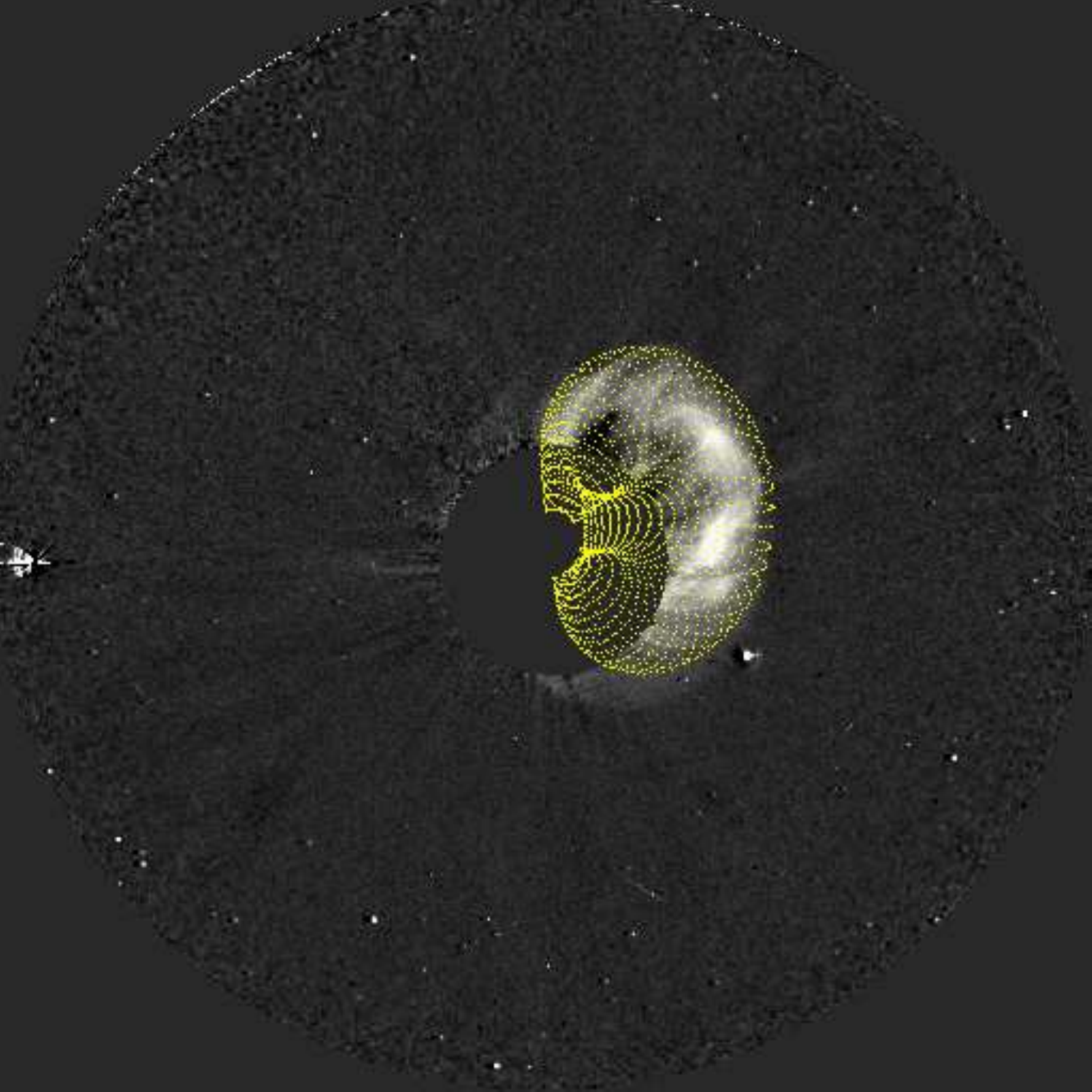}
                }
\vspace{0.0261\textwidth}  
\caption[GCS fit for CME 28 at 00:54]{GCS fit for CME 28 on September 28, 2012 at 00:54 UT at height $H=9.9$ \Rs. Table \ref{tblapp} 
lists the GCS parameters for this event.}
\label{figa28}
\end{figure}

\begin{figure}[h]    
  \centering                              
   \centerline{\hspace*{0.00\textwidth}
               \includegraphics[width=0.4\textwidth,clip=]{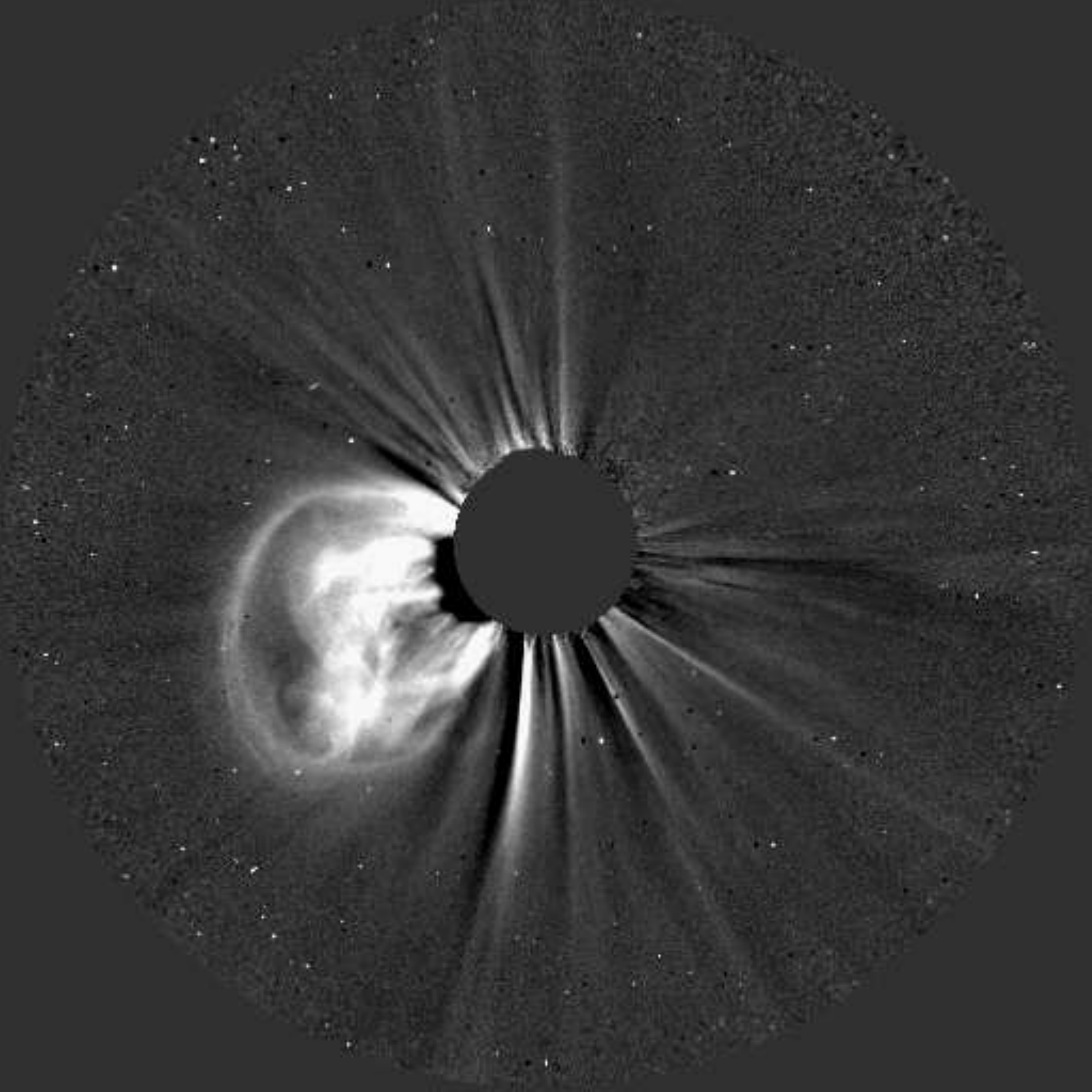}
                \hspace*{-0.02\textwidth}
               \includegraphics[width=0.4\textwidth,clip=]{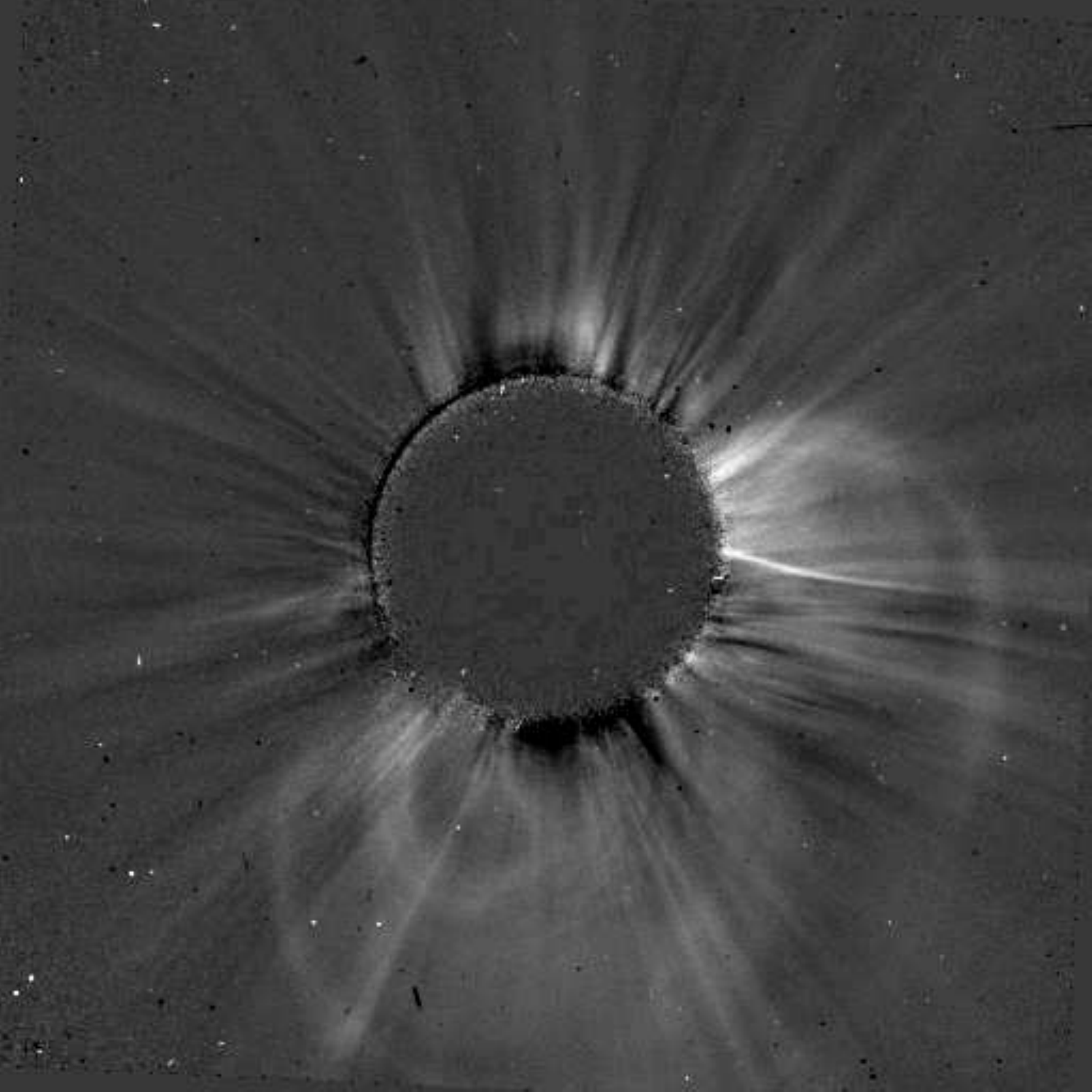}
             \hspace*{-0.02\textwidth}
               \includegraphics[width=0.4\textwidth,clip=]{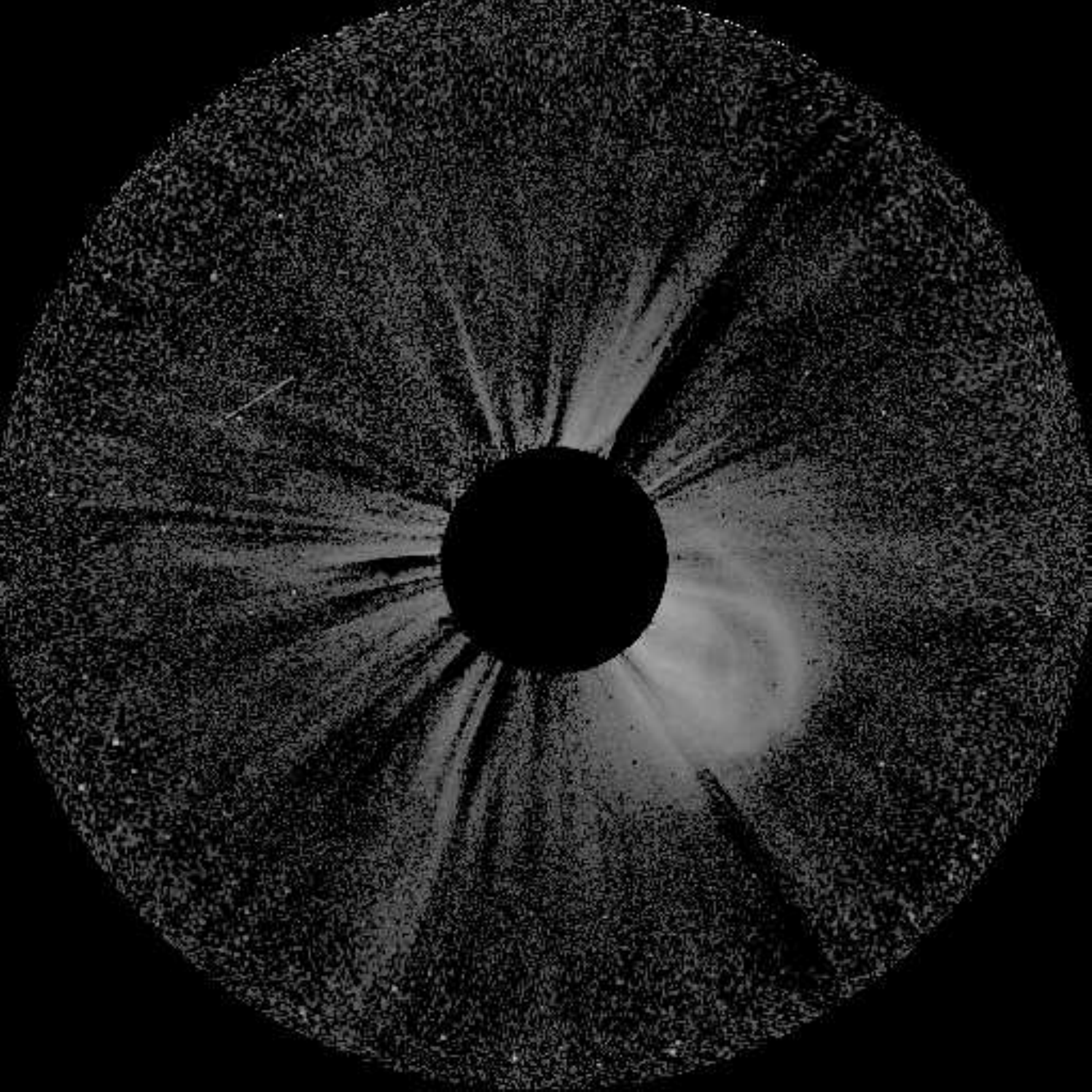}
               }
                 \centerline{\hspace*{0.0\textwidth}
              \includegraphics[width=0.4\textwidth,clip=]{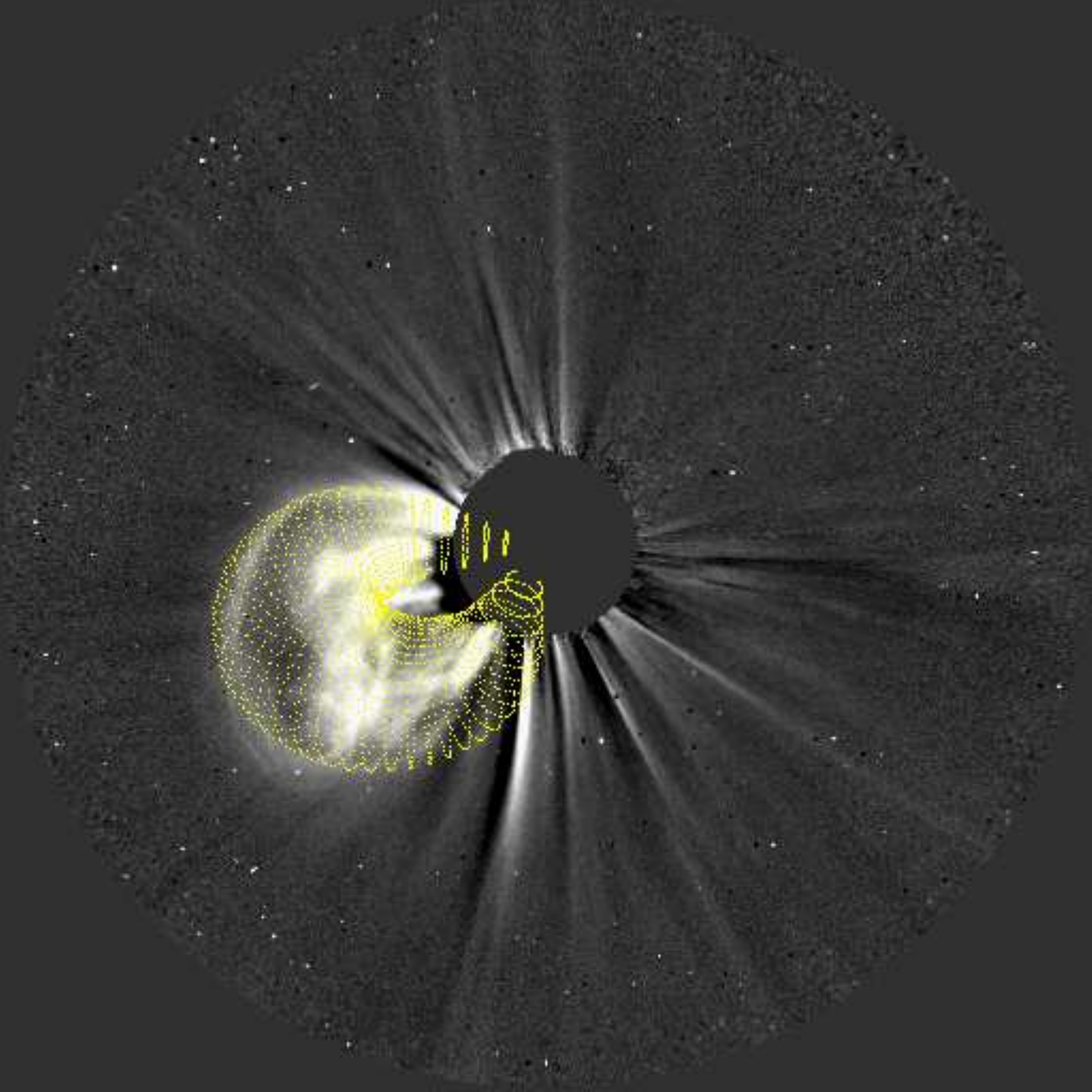}
               \hspace*{-0.02\textwidth}
               \includegraphics[width=0.4\textwidth,clip=]{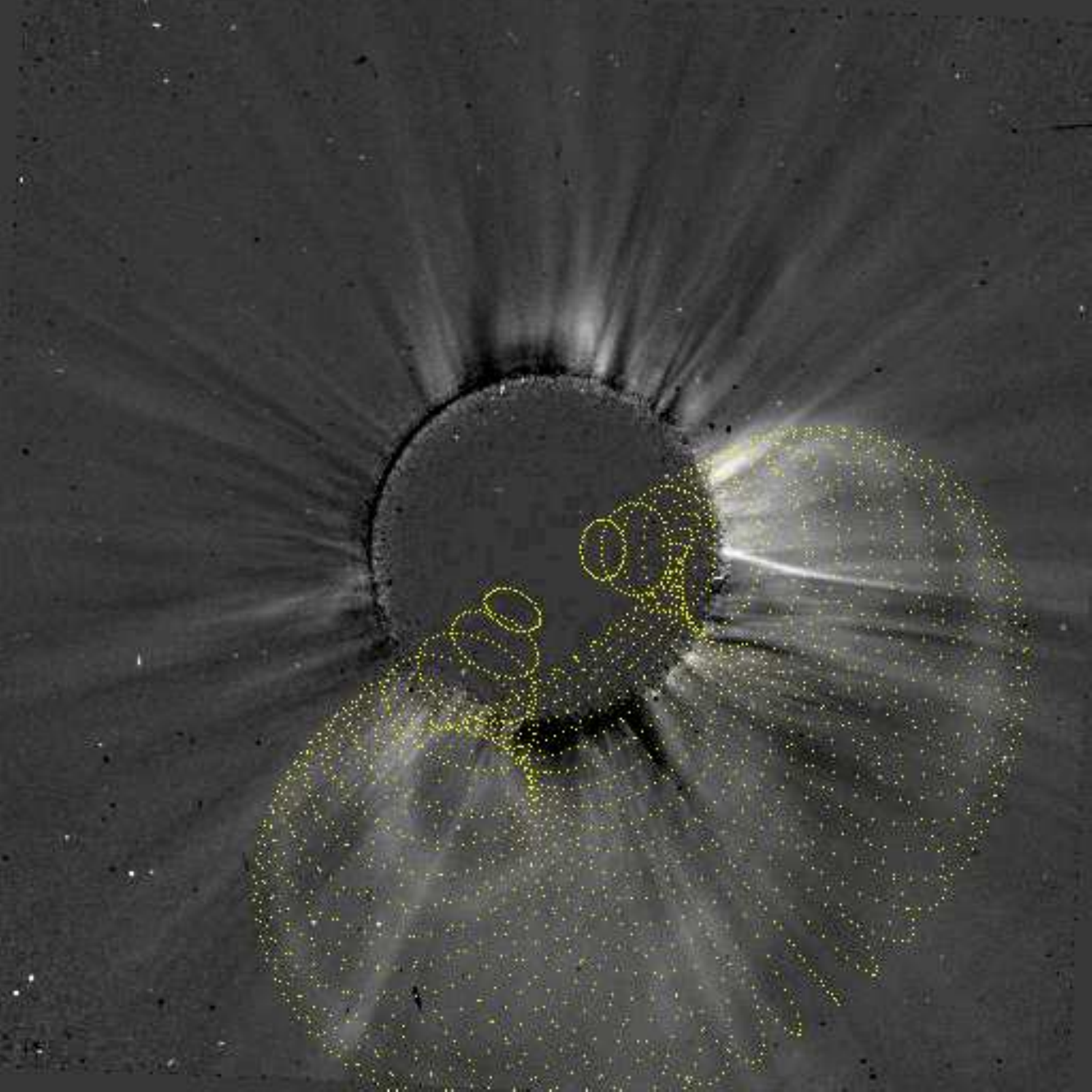}
              \hspace*{-0.02\textwidth}
               \includegraphics[width=0.4\textwidth,clip=]{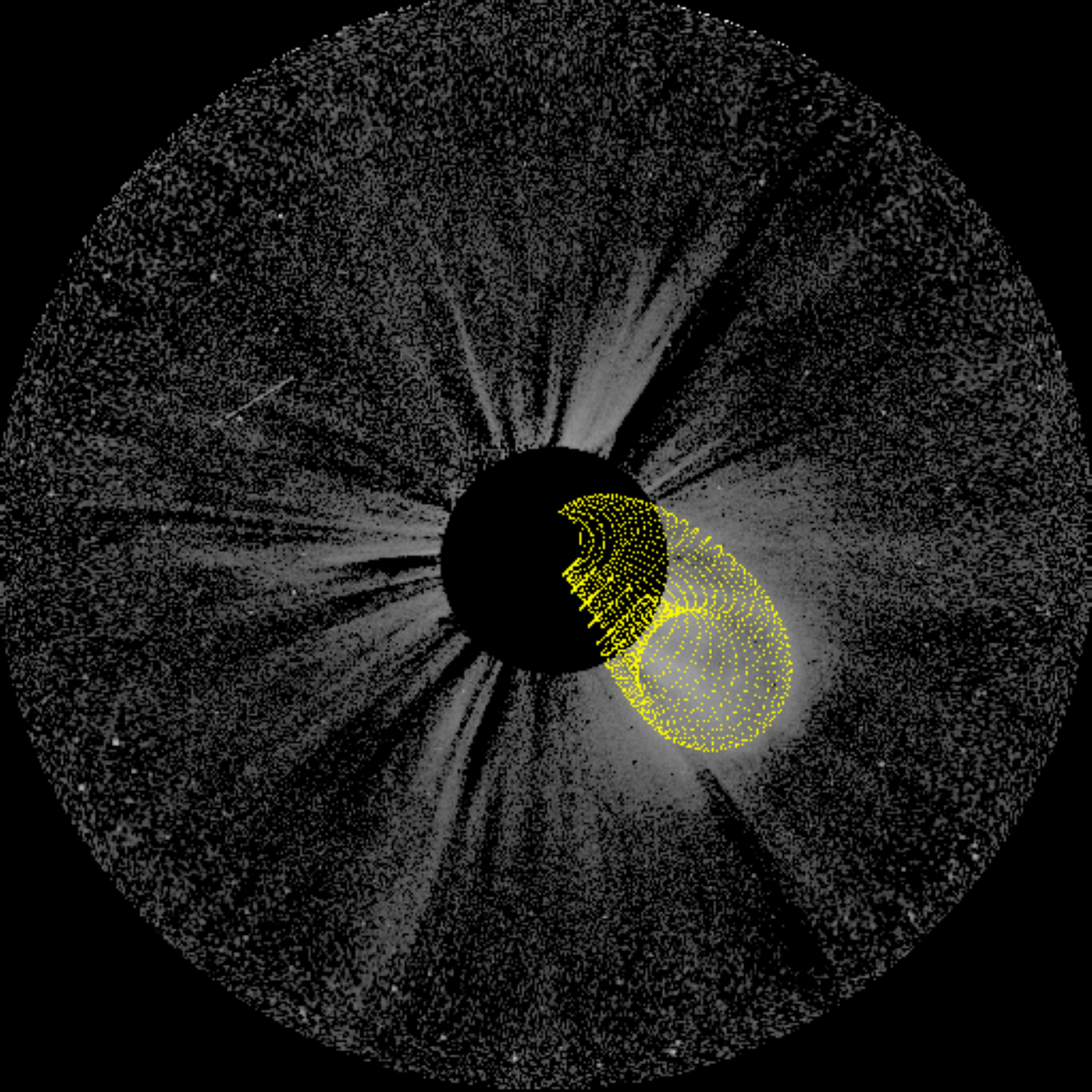}
                }
\vspace{0.0261\textwidth}  
\caption[GCS fit for CME 29 at 05:54]{GCS fit for CME 29 on October 05, 2012 at 05:54 UT at height $H=10.0$ \Rs. Table \ref{tblapp} 
lists the GCS parameters for this event.}
\label{figa29}
\end{figure}

\begin{figure}[h]    
  \centering                              
   \centerline{\hspace*{0.04\textwidth}
               \includegraphics[width=0.4\textwidth,clip=]{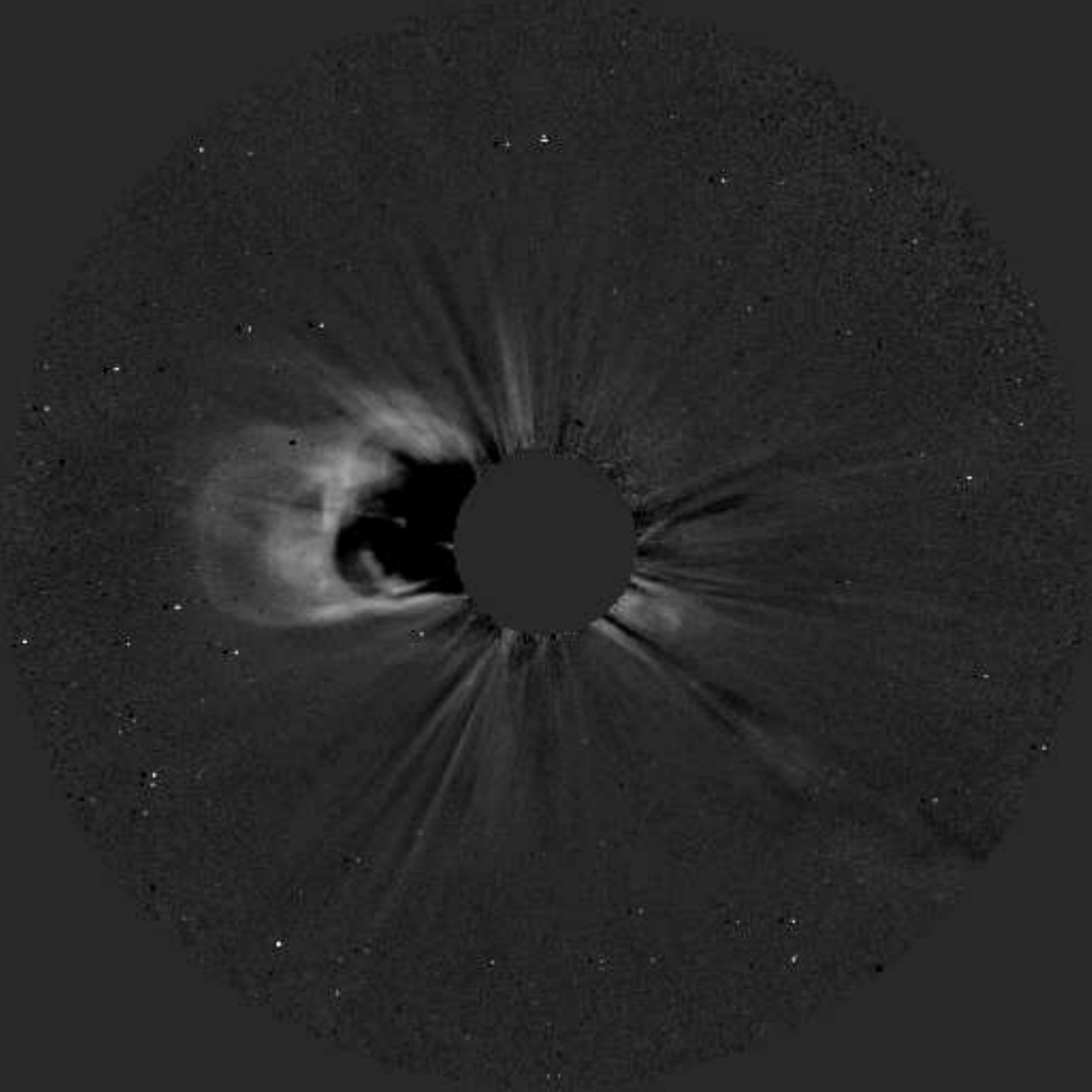}
                \hspace*{-0.02\textwidth}
               \includegraphics[width=0.4\textwidth,clip=]{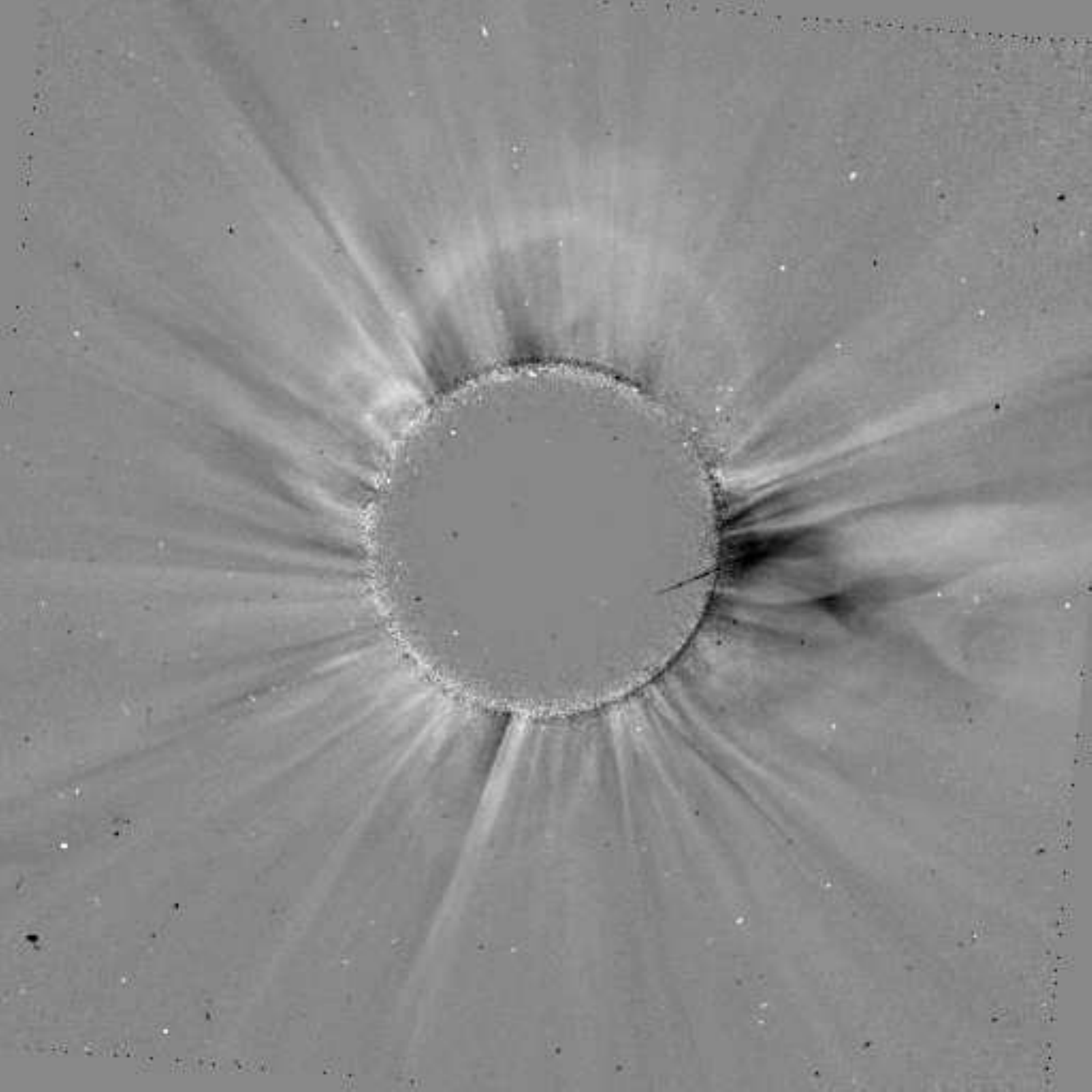}
             \hspace*{-0.02\textwidth}
               \includegraphics[width=0.4\textwidth,clip=]{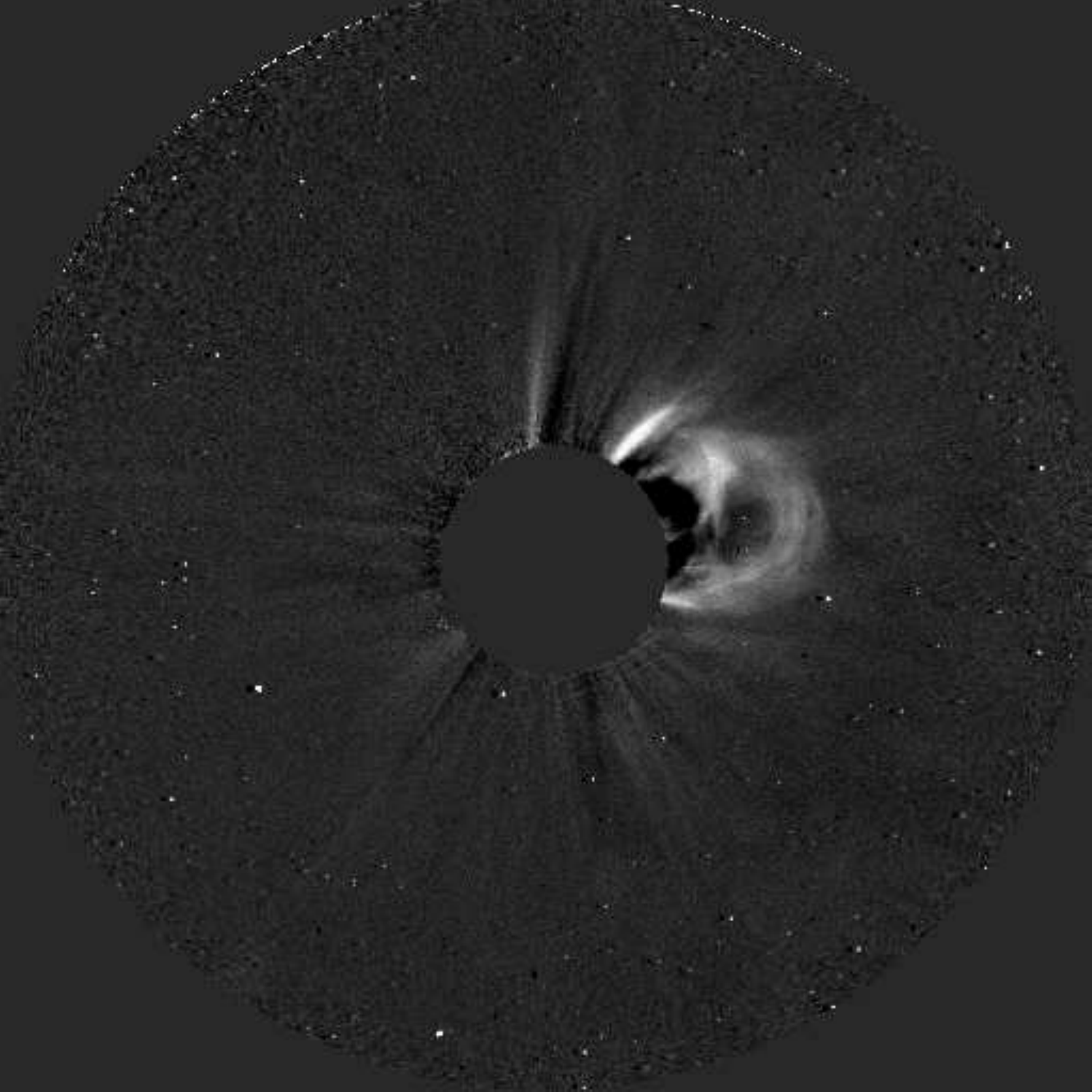}
               }
                 \centerline{\hspace*{0.04\textwidth}
              \includegraphics[width=0.4\textwidth,clip=]{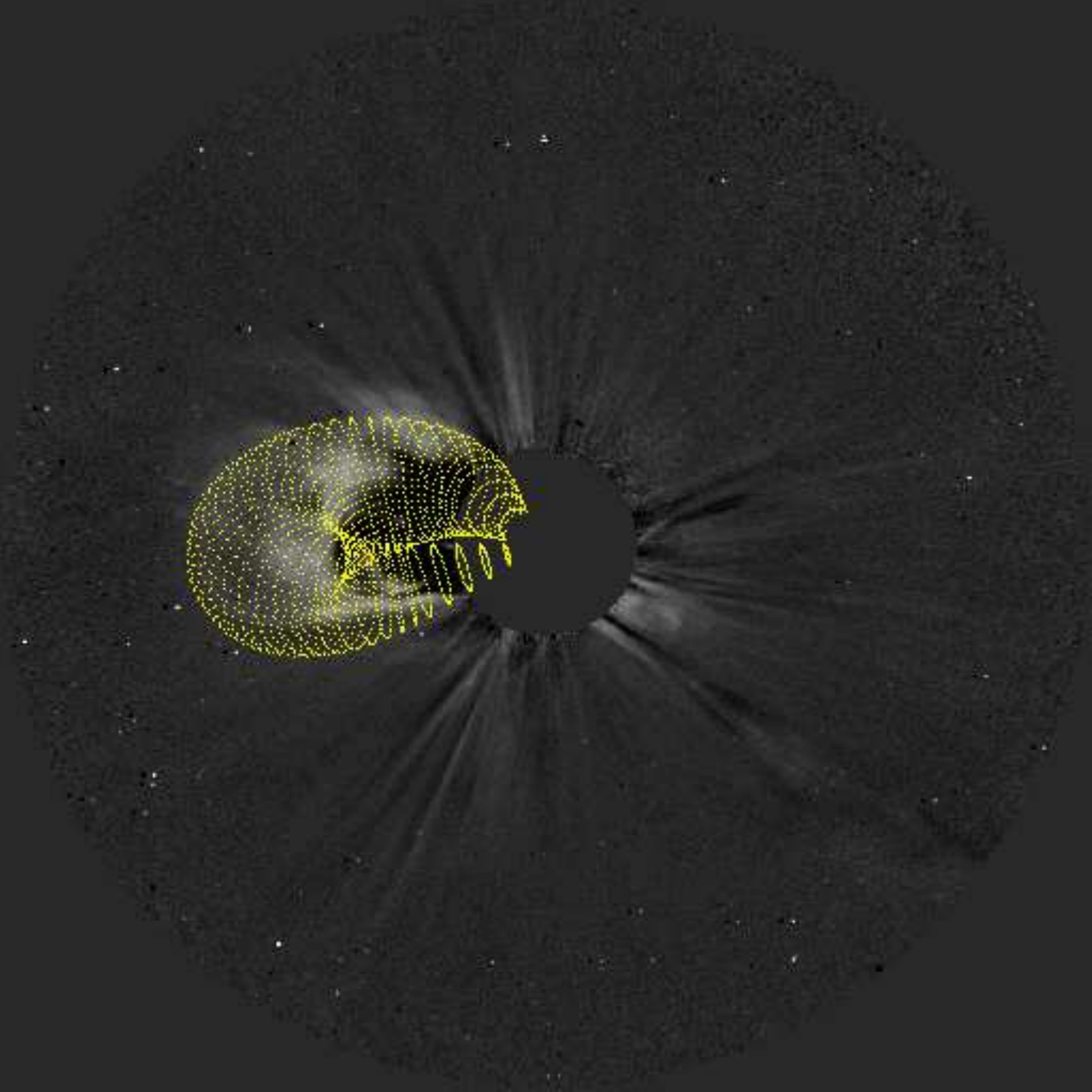}
               \hspace*{-0.02\textwidth}
               \includegraphics[width=0.4\textwidth,clip=]{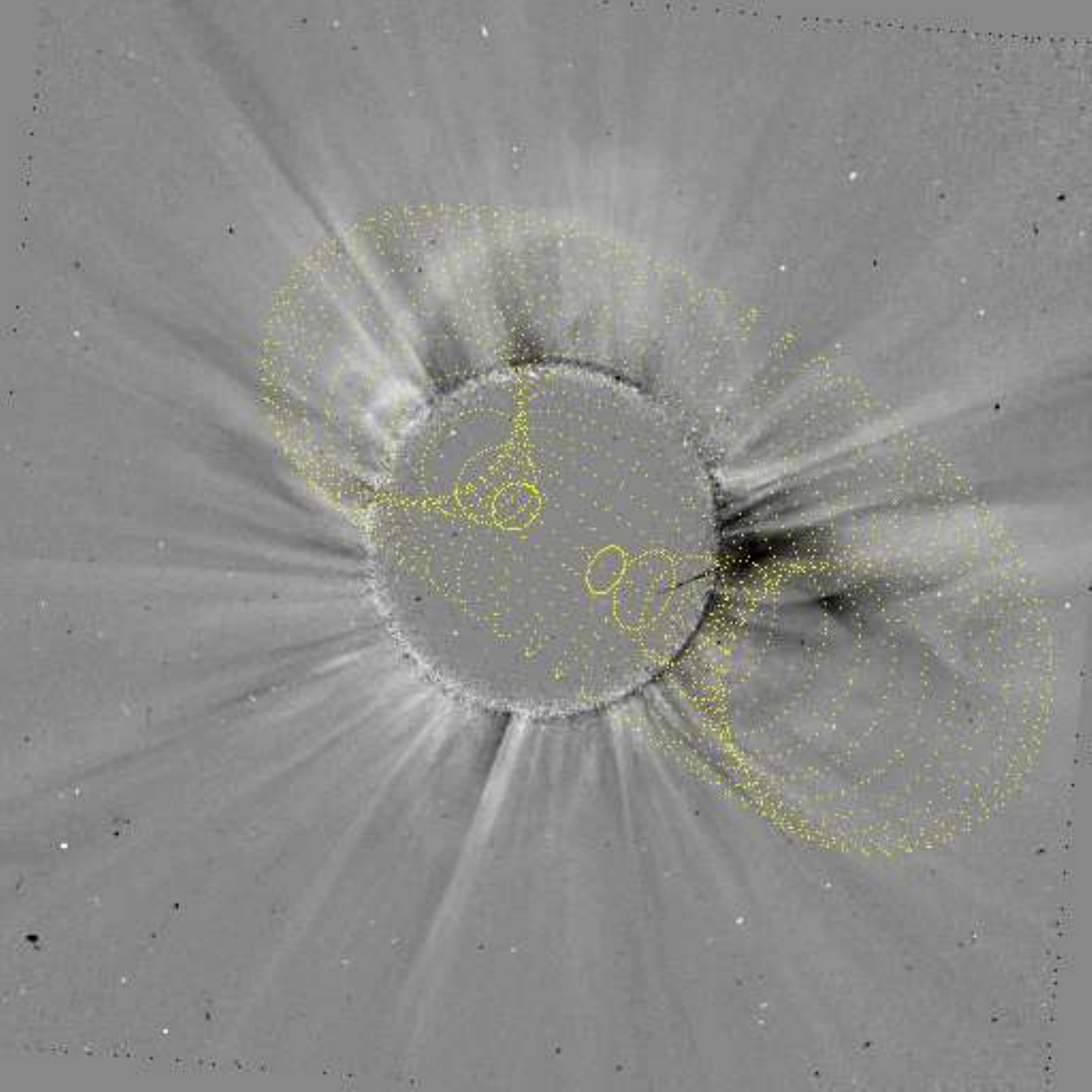}
              \hspace*{-0.02\textwidth}
               \includegraphics[width=0.4\textwidth,clip=]{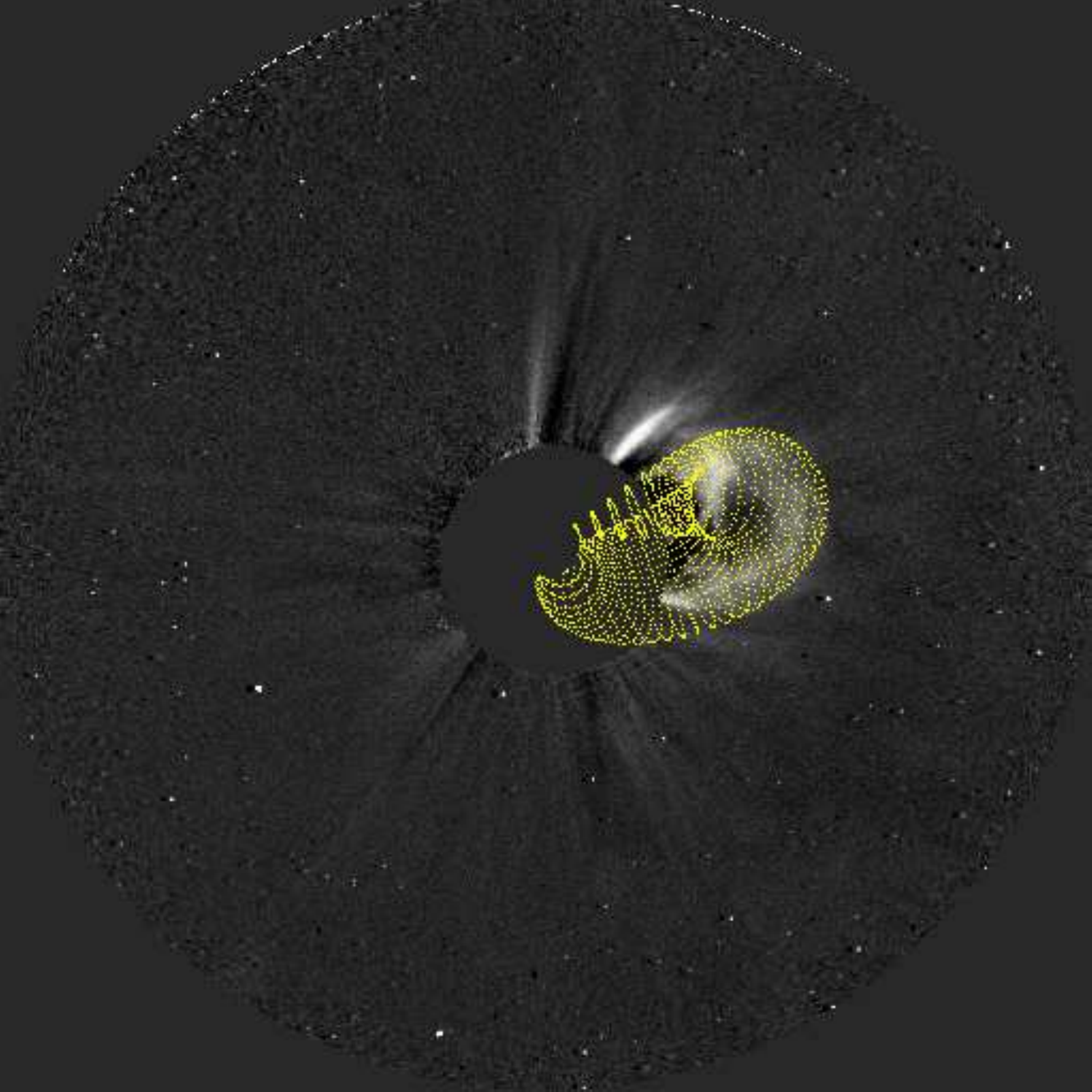}
                }
\vspace{0.0261\textwidth}  
\caption[GCS fit for CME 30 at 18:54]{GCS fit for CME 30 on October 27, 2012 at 18:54 UT at height $H=10.5$ \Rs. Table \ref{tblapp} 
lists the GCS parameters for this event.}
\label{figa30}
\end{figure}

\begin{figure}[h]    
  \centering                              
   \centerline{\hspace*{0.00\textwidth}
               \includegraphics[width=0.4\textwidth,clip=]{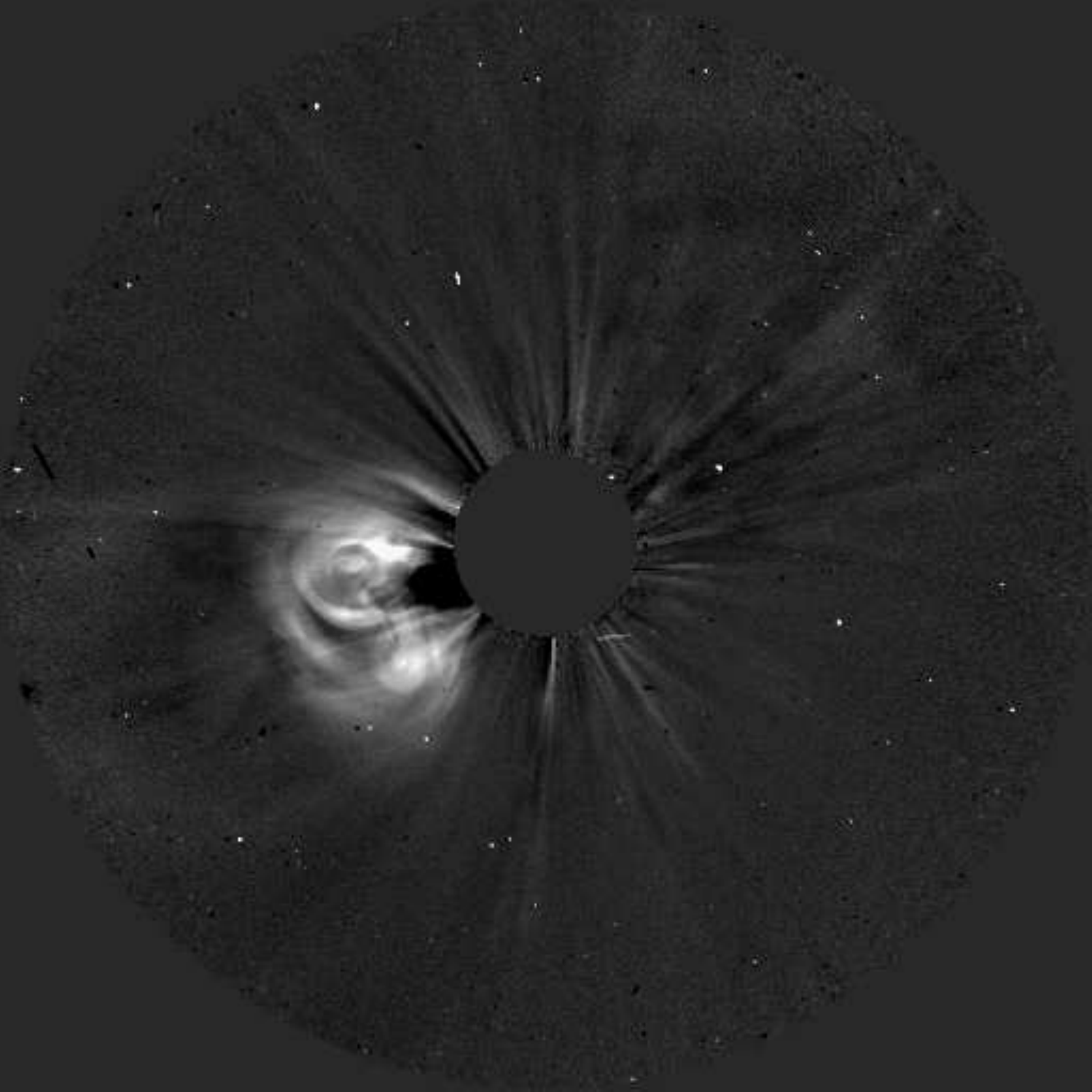}
                \hspace*{-0.02\textwidth}
               \includegraphics[width=0.4\textwidth,clip=]{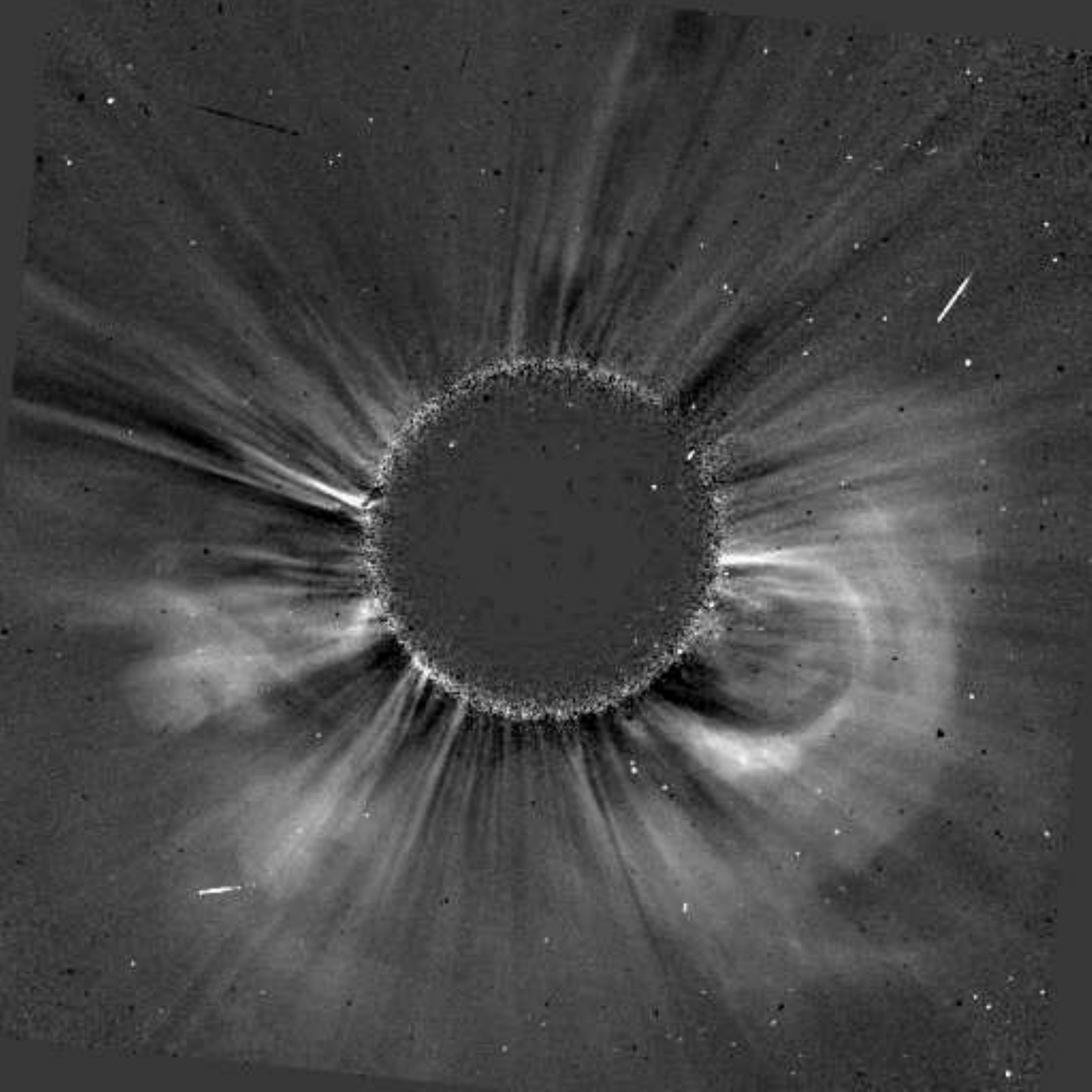}
             \hspace*{-0.02\textwidth}
               \includegraphics[width=0.4\textwidth,clip=]{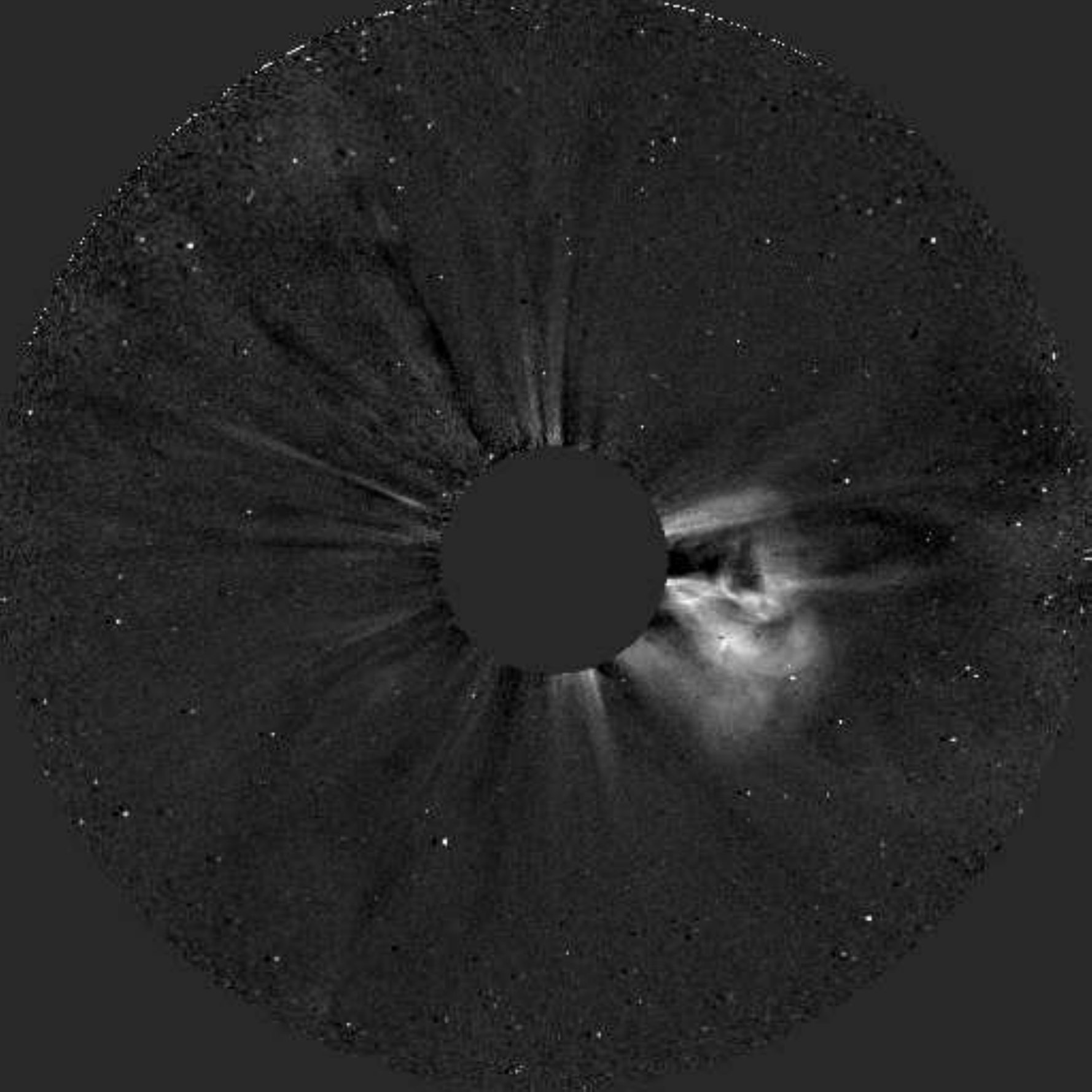}
               }
                 \centerline{\hspace*{0.0\textwidth}
              \includegraphics[width=0.4\textwidth,clip=]{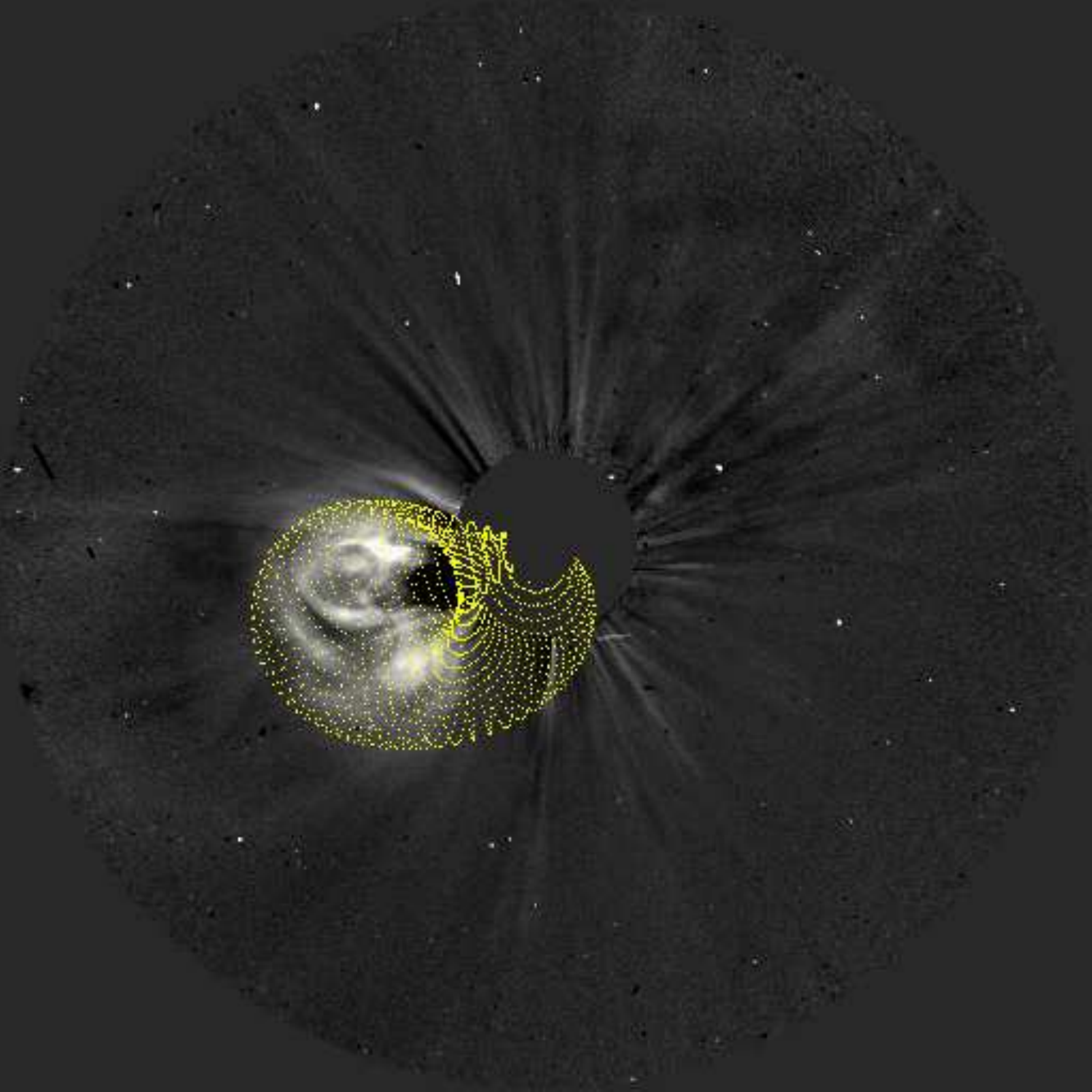}
               \hspace*{-0.02\textwidth}
               \includegraphics[width=0.4\textwidth,clip=]{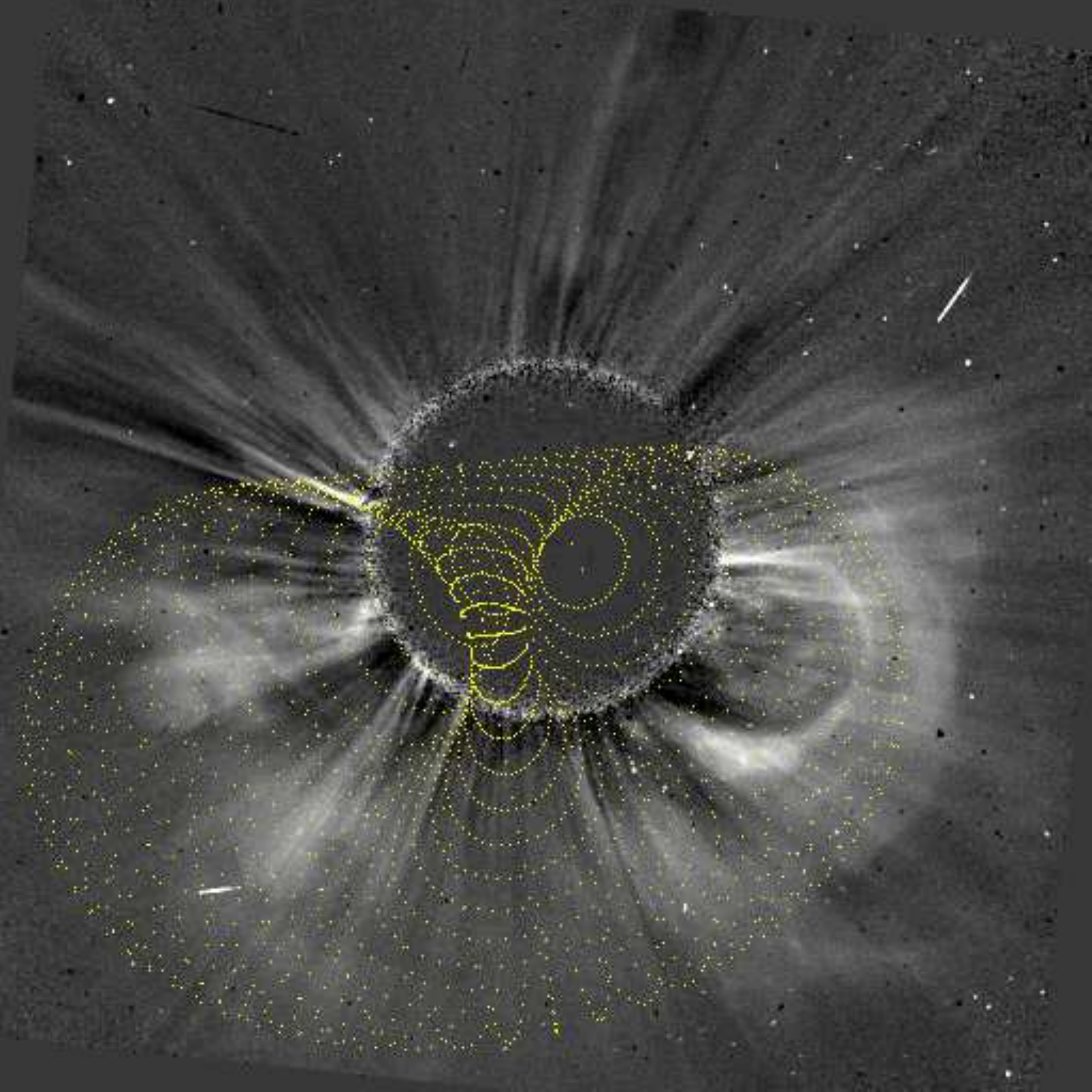}
              \hspace*{-0.02\textwidth}
               \includegraphics[width=0.4\textwidth,clip=]{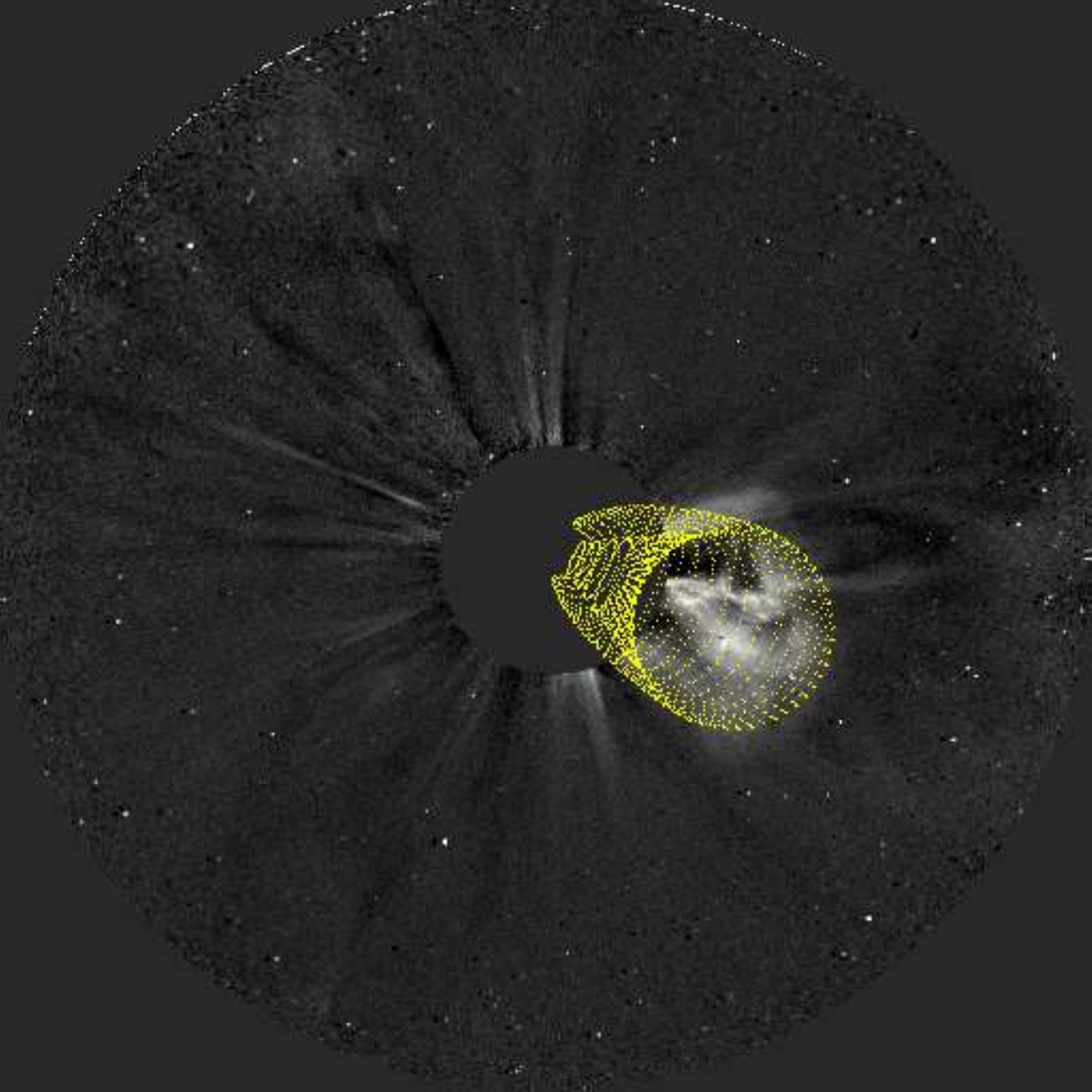}
                }
\vspace{0.0261\textwidth}  
\caption[GCS fit for CME 31 at 16:54]{GCS fit for CME 31 on November 09, 2012 at 16:54 UT at height $H=9.9$ \Rs. Table \ref{tblapp} 
lists the GCS parameters for this event.}
\label{figa31}
\end{figure}

\begin{figure}[h]    
  \centering                              
   \centerline{\hspace*{0.04\textwidth}
               \includegraphics[width=0.4\textwidth,clip=]{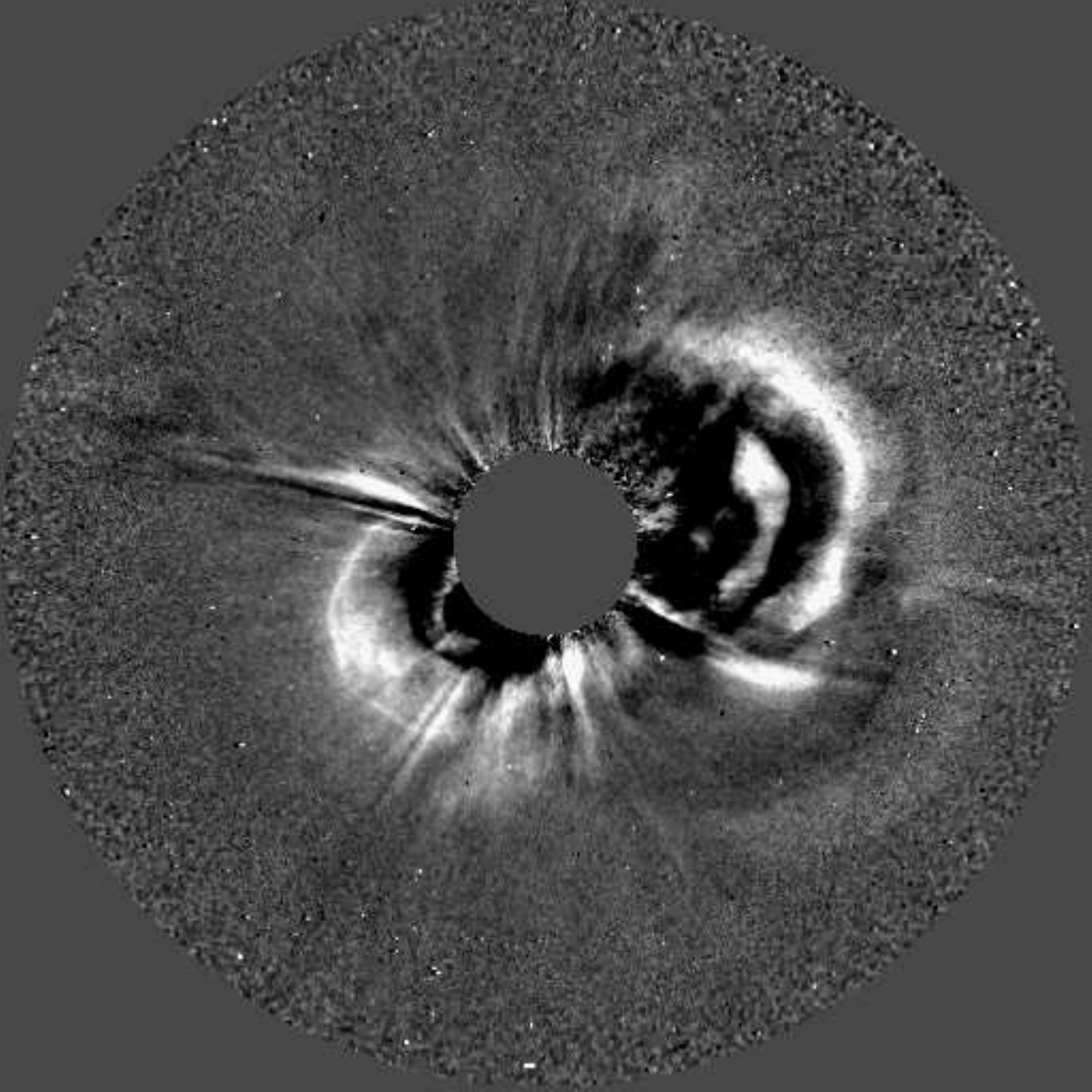}
                \hspace*{-0.02\textwidth}
               \includegraphics[width=0.4\textwidth,clip=]{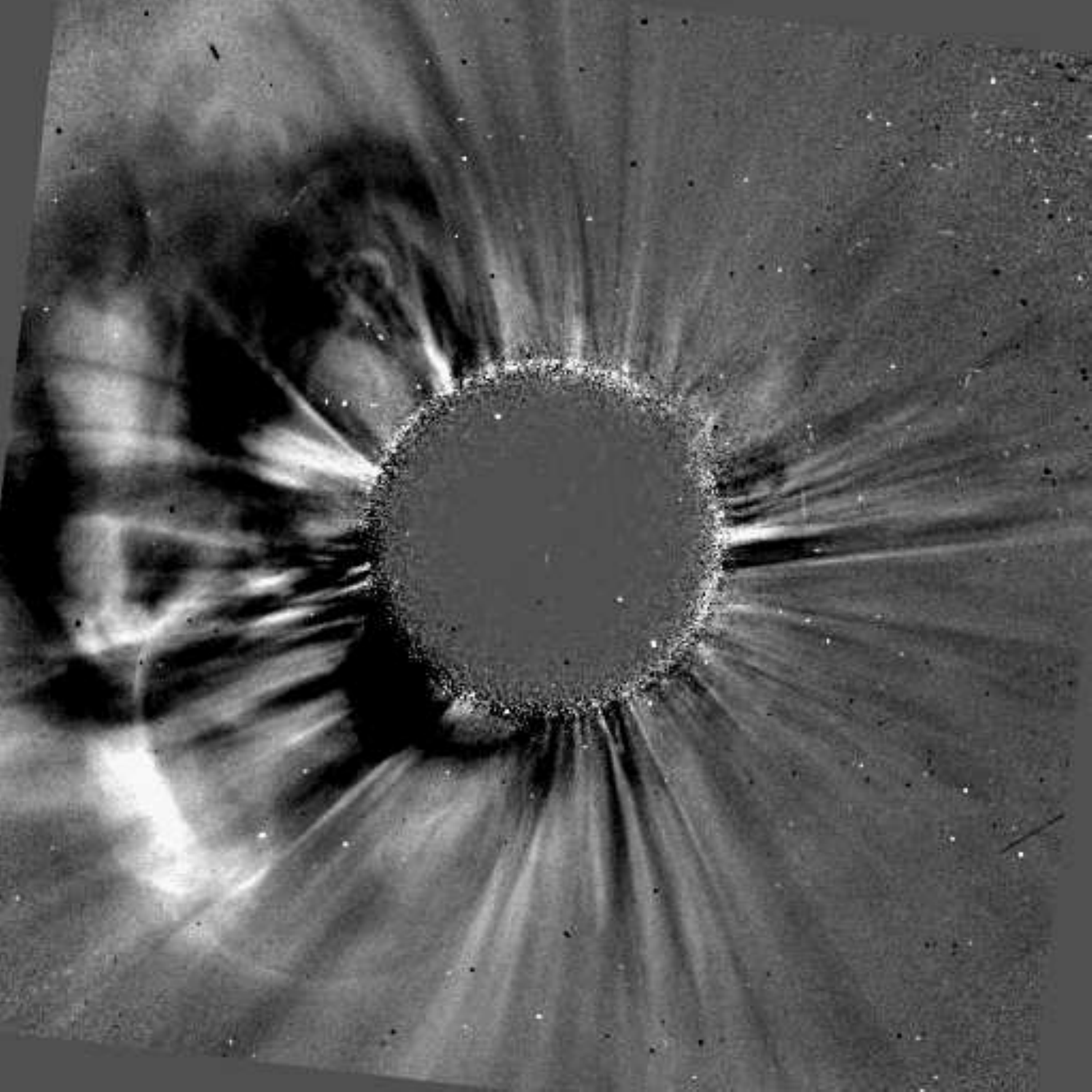}
             \hspace*{-0.02\textwidth}
               \includegraphics[width=0.4\textwidth,clip=]{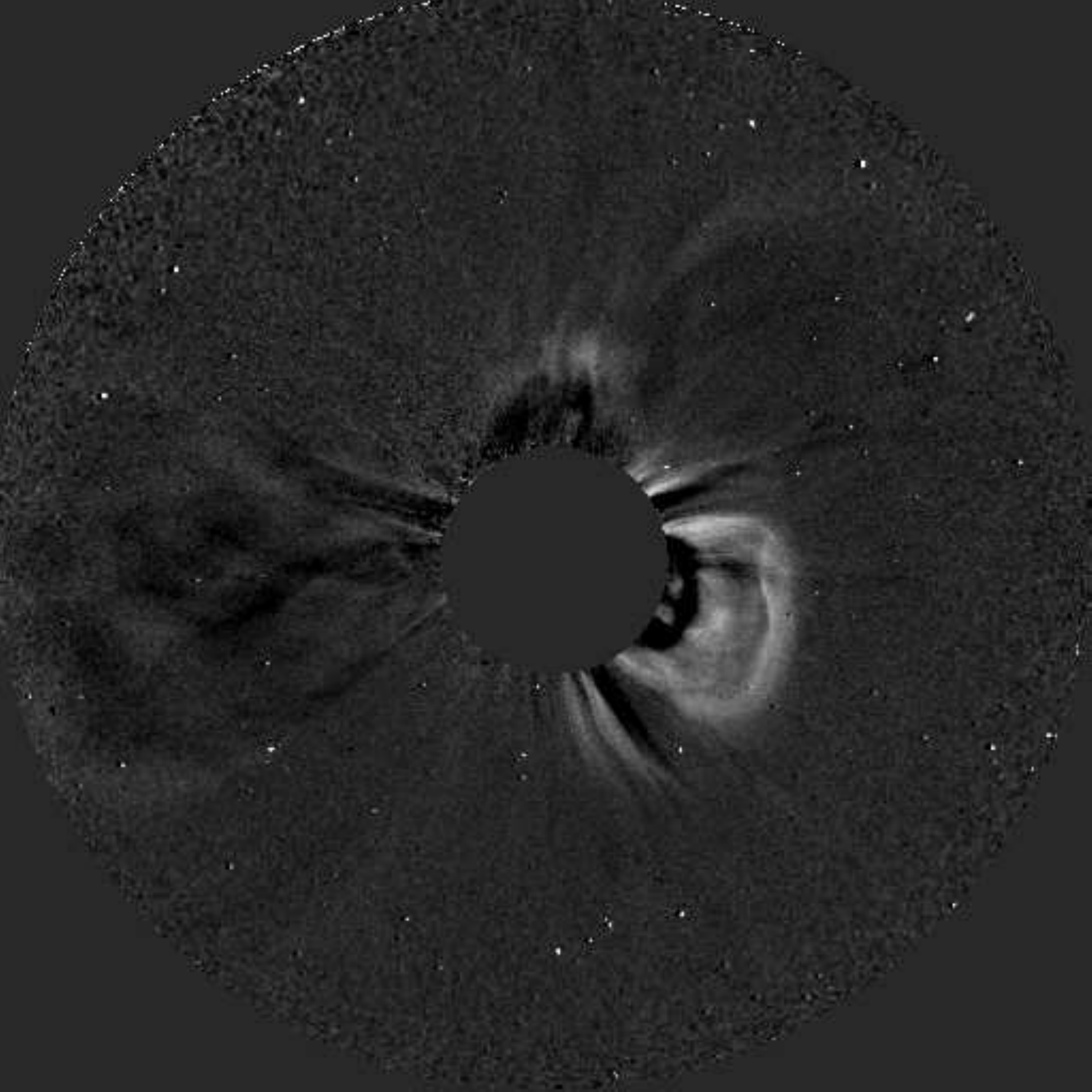}
               }
                 \centerline{\hspace*{0.04\textwidth}
              \includegraphics[width=0.4\textwidth,clip=]{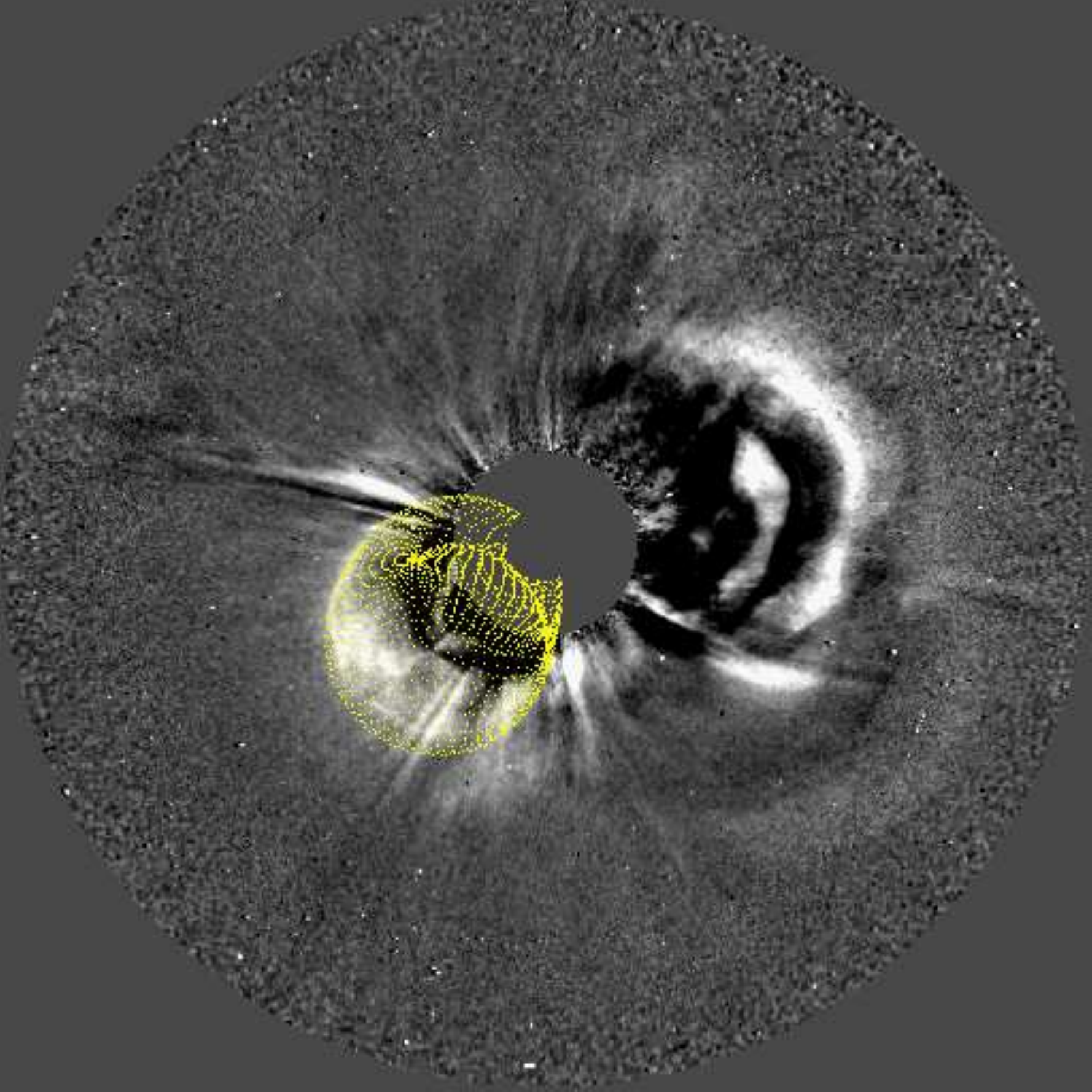}
               \hspace*{-0.02\textwidth}
               \includegraphics[width=0.4\textwidth,clip=]{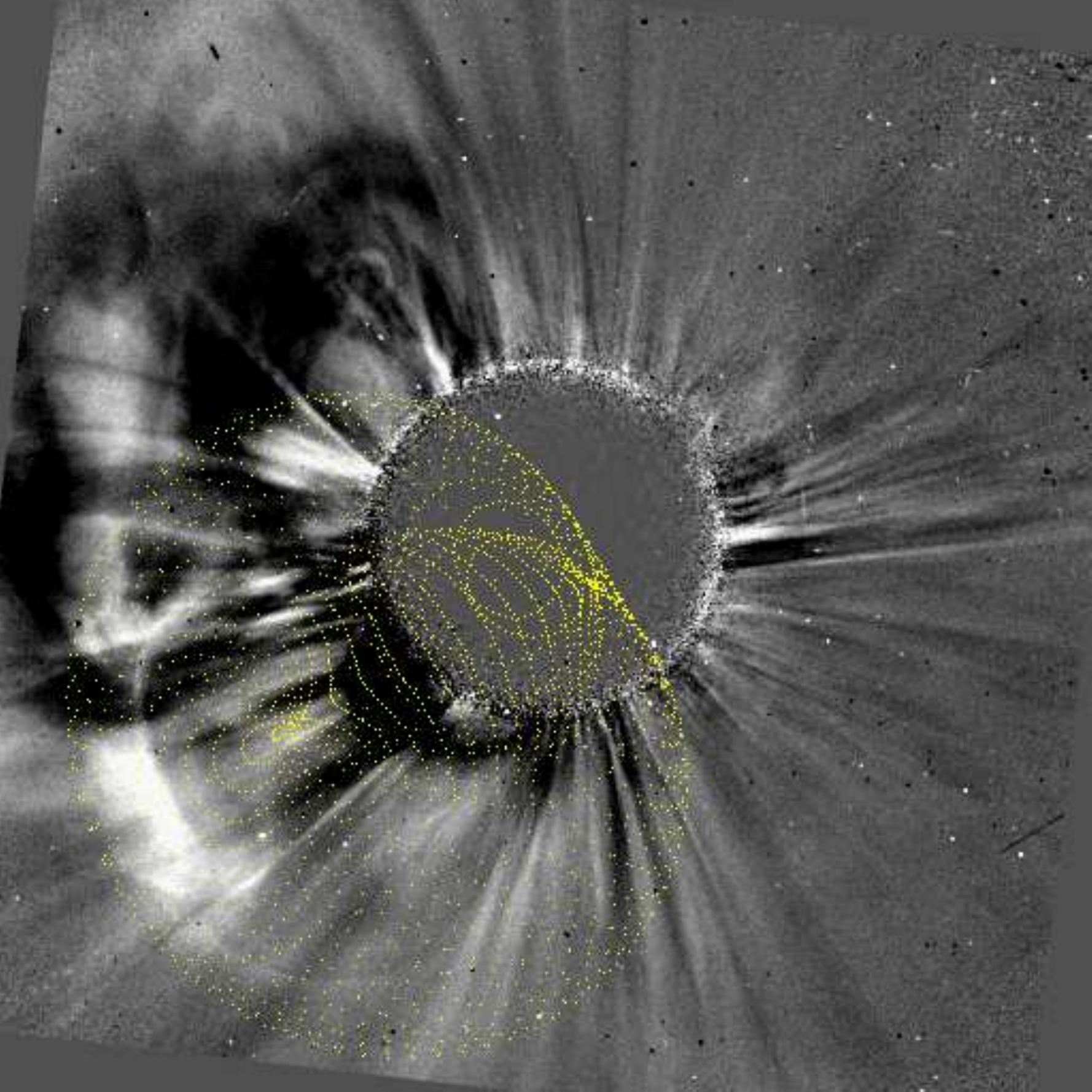}
              \hspace*{-0.02\textwidth}
               \includegraphics[width=0.4\textwidth,clip=]{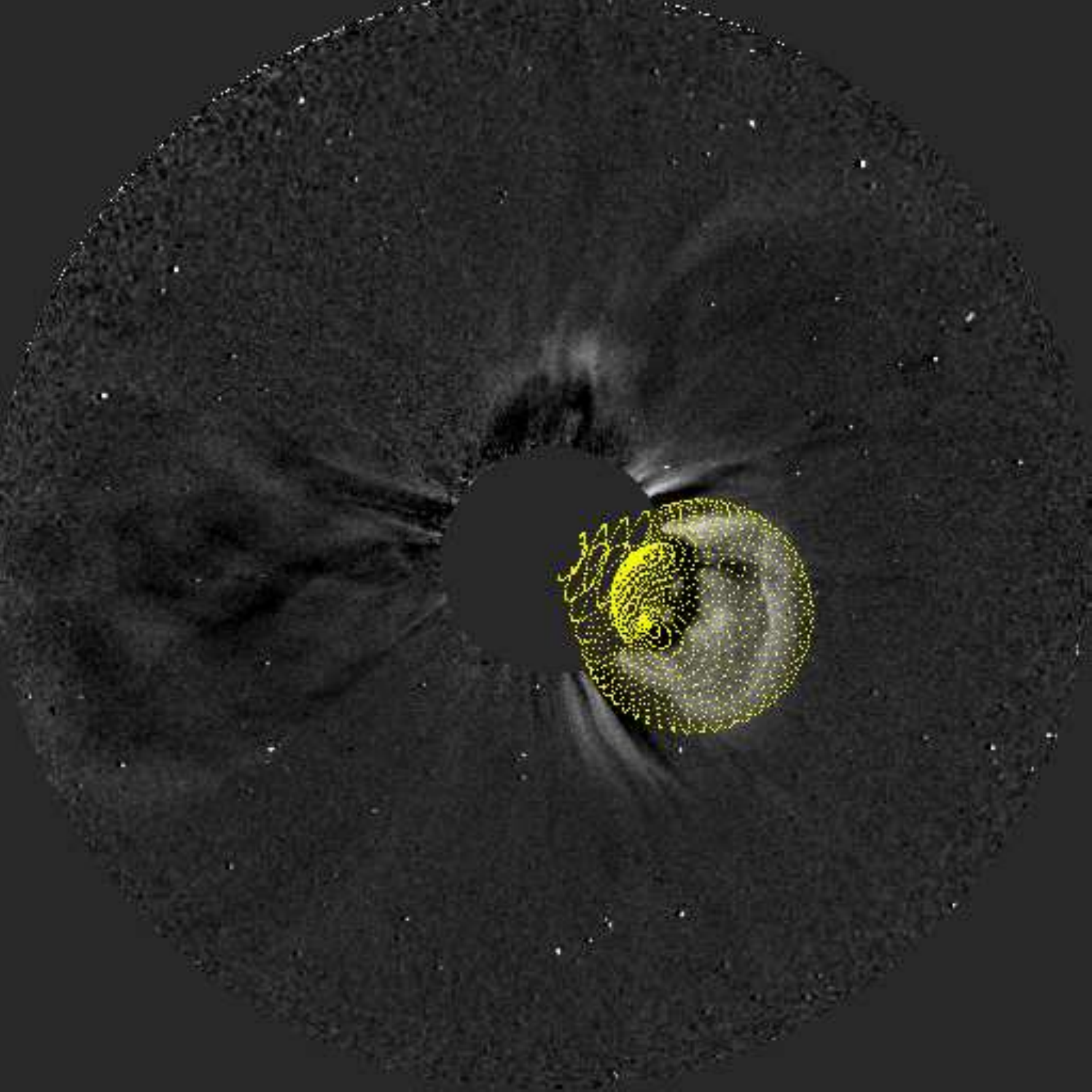}
                }
\vspace{0.0261\textwidth}  
\caption[GCS fit for CME 32 at 15:24]{GCS fit for CME 32 on November 23, 2012 at 15:24 UT at height $H=8.9$ \Rs. Table \ref{tblapp} 
lists the GCS parameters for this event.}
\label{figa32}
\end{figure}

\begin{figure}[h]    
  \centering                              
   \centerline{\hspace*{0.00\textwidth}
               \includegraphics[width=0.4\textwidth,clip=]{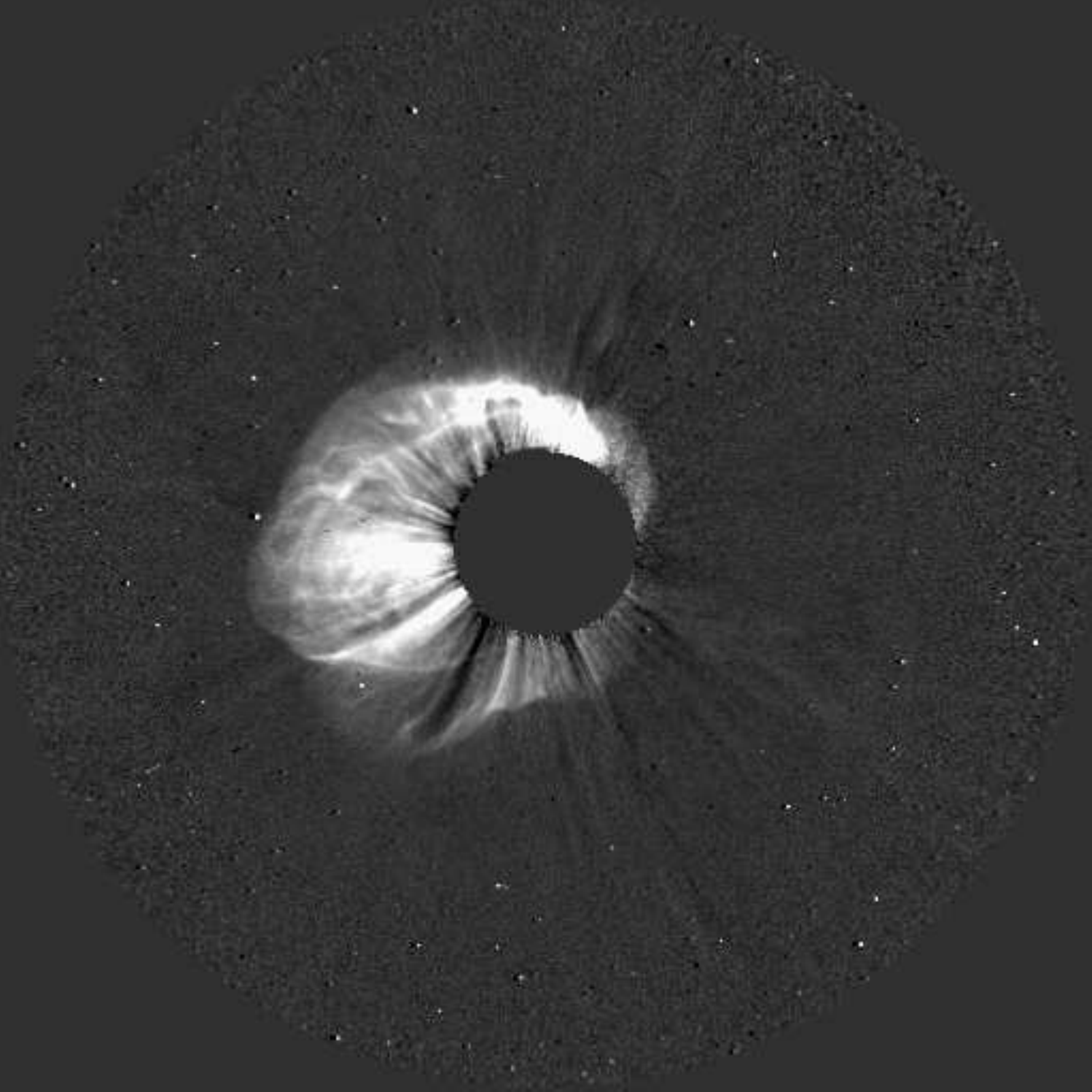}
                \hspace*{-0.02\textwidth}
               \includegraphics[width=0.4\textwidth,clip=]{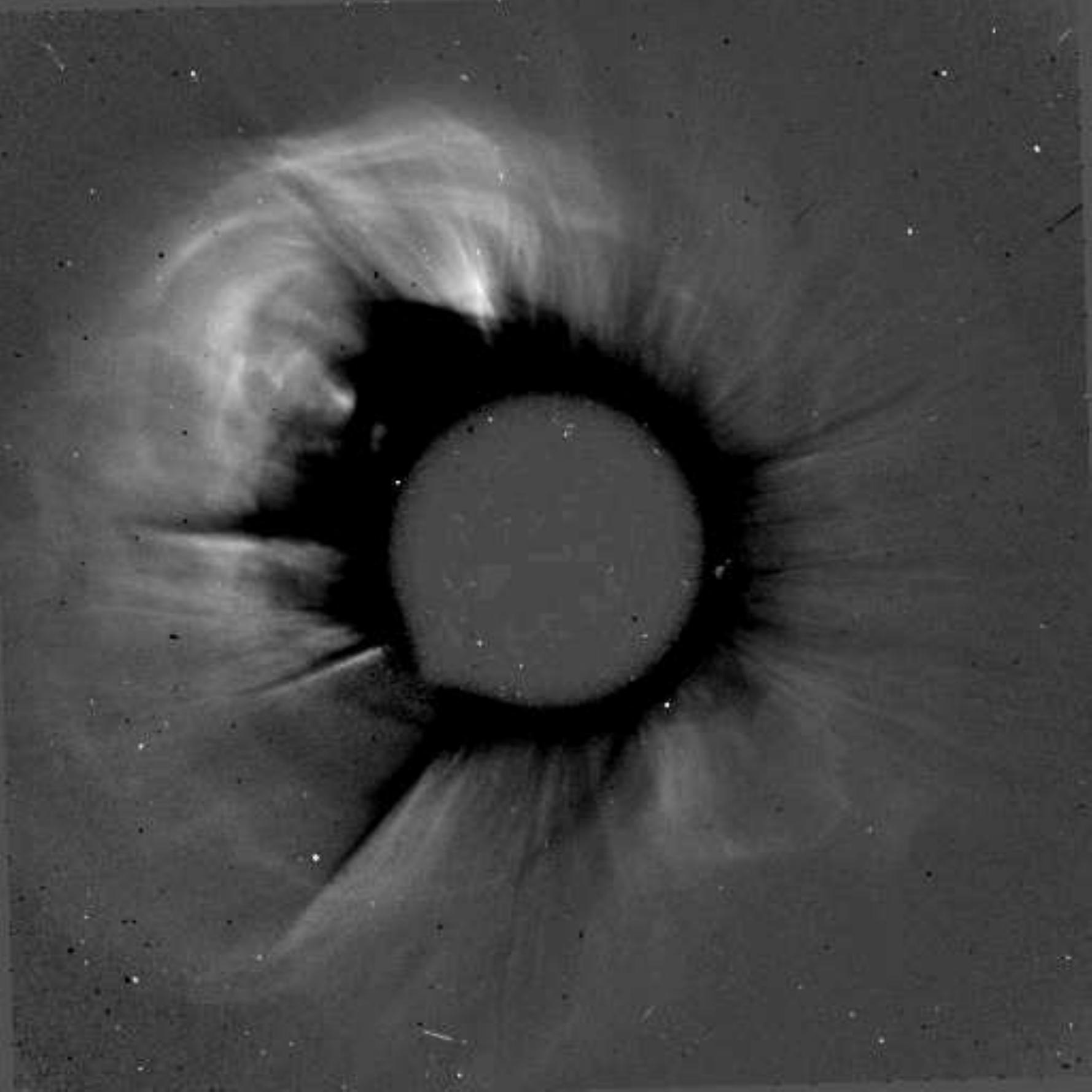}
             \hspace*{-0.02\textwidth}
               \includegraphics[width=0.4\textwidth,clip=]{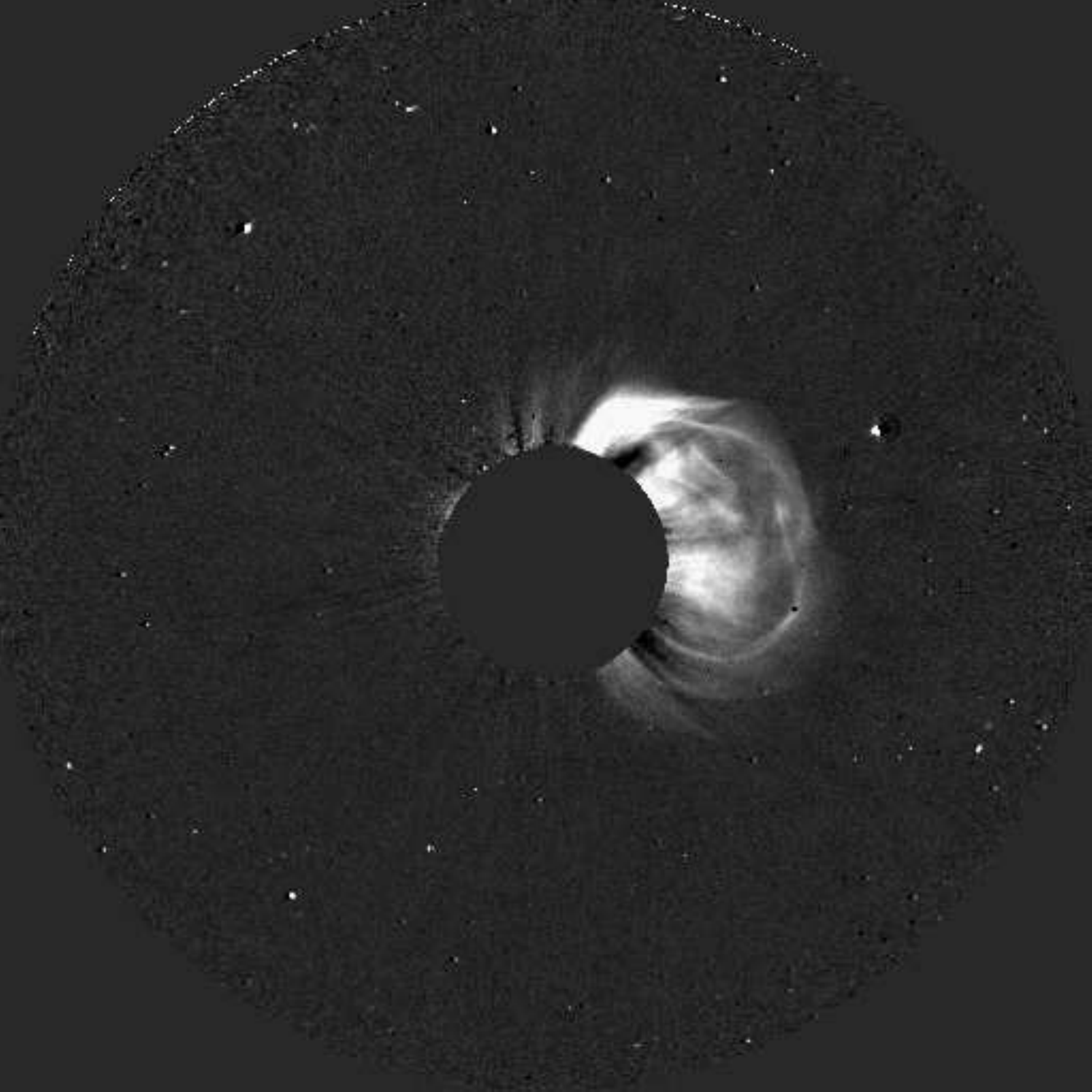}
               }
                 \centerline{\hspace*{0.0\textwidth}
              \includegraphics[width=0.4\textwidth,clip=]{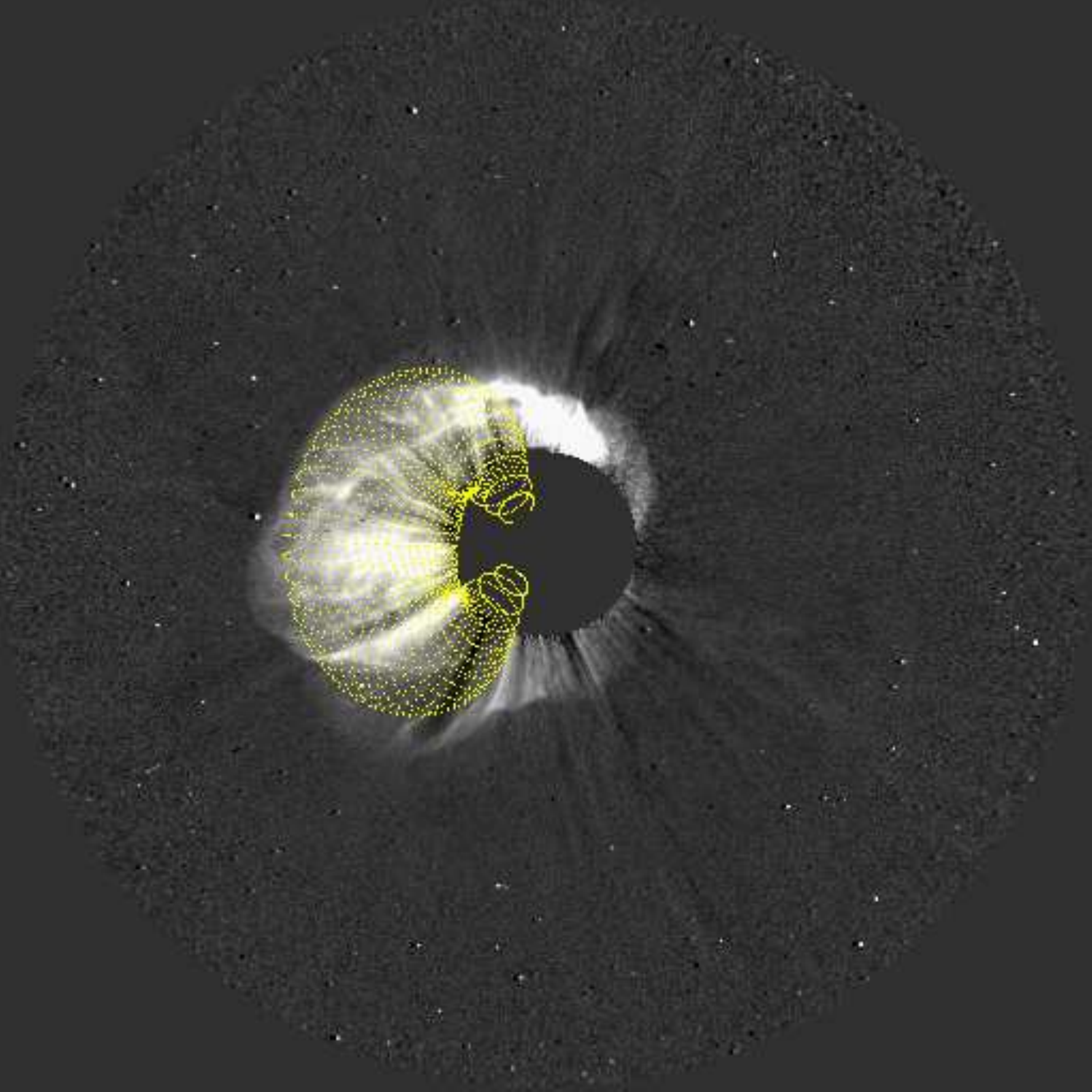}
               \hspace*{-0.02\textwidth}
               \includegraphics[width=0.4\textwidth,clip=]{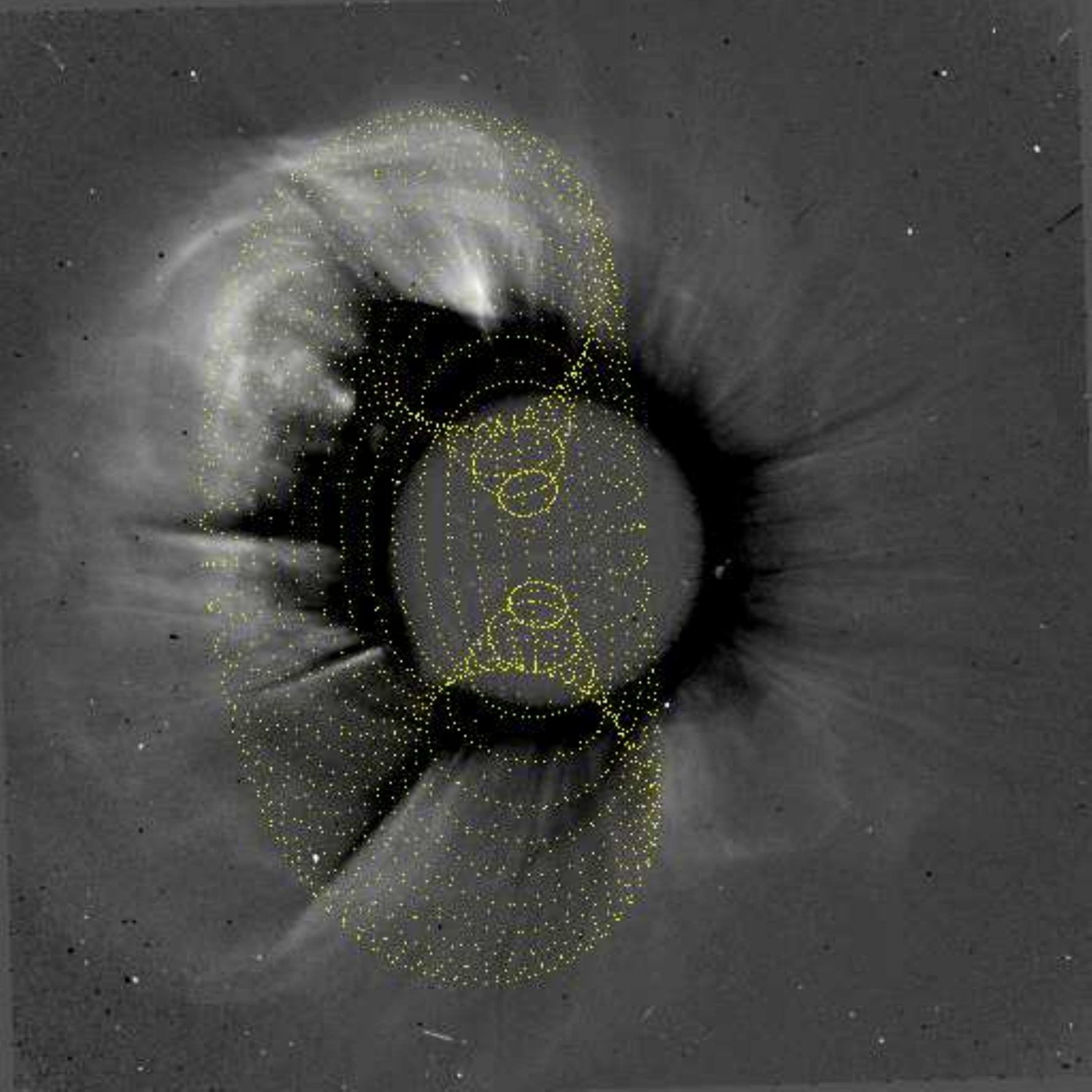}
              \hspace*{-0.02\textwidth}
               \includegraphics[width=0.4\textwidth,clip=]{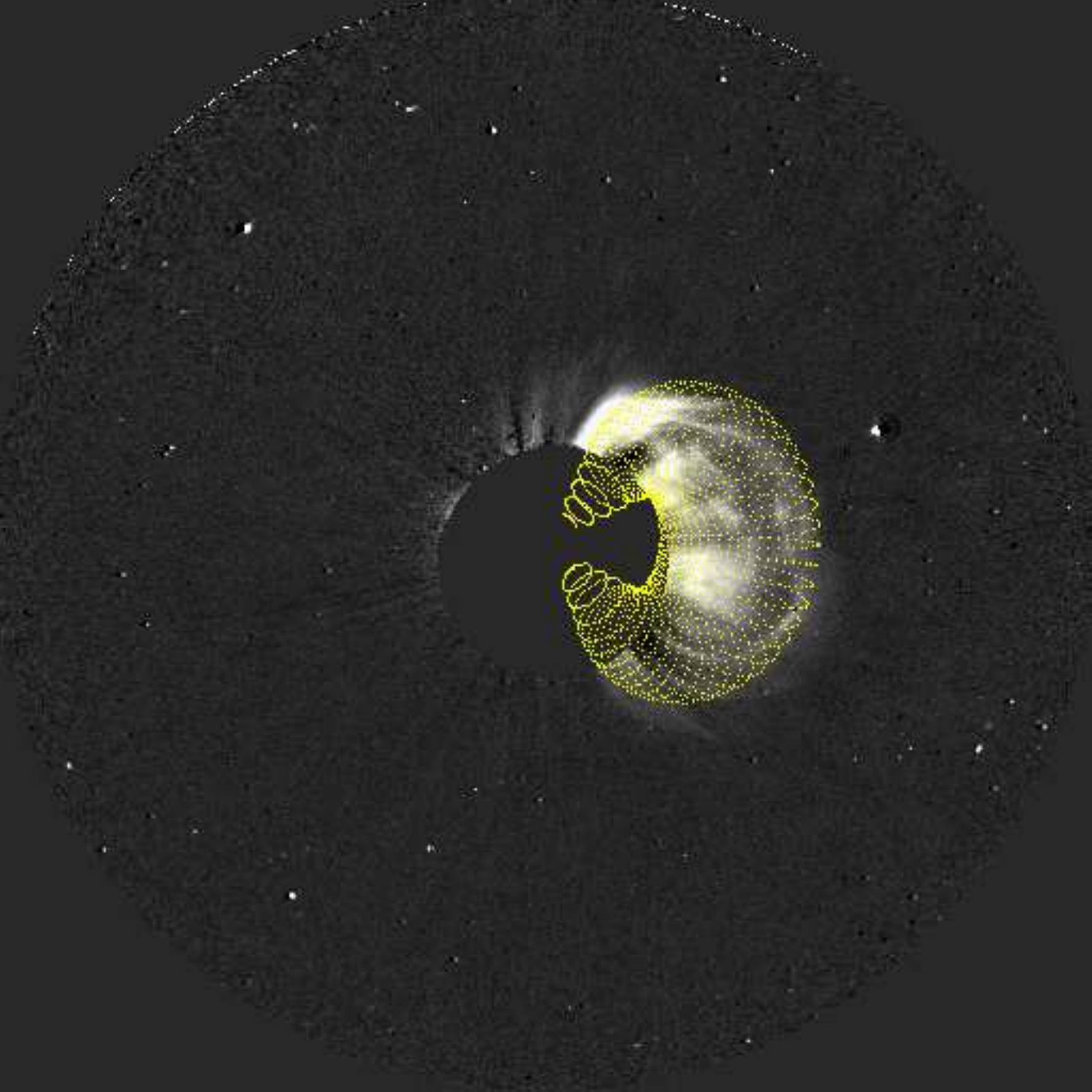}
                }
\vspace{0.0261\textwidth}  
\caption[GCS fit for CME 33 at 07:39]{GCS fit for CME 33 on March 15, 2013 at 07:39 UT at height $H=10.2$ \Rs. Table \ref{tblapp} 
lists the GCS parameters for this event.}
\label{figa33}
\end{figure}

\begin{figure}[h]    
  \centering                              
   \centerline{\hspace*{0.04\textwidth}
               \includegraphics[width=0.4\textwidth,clip=]{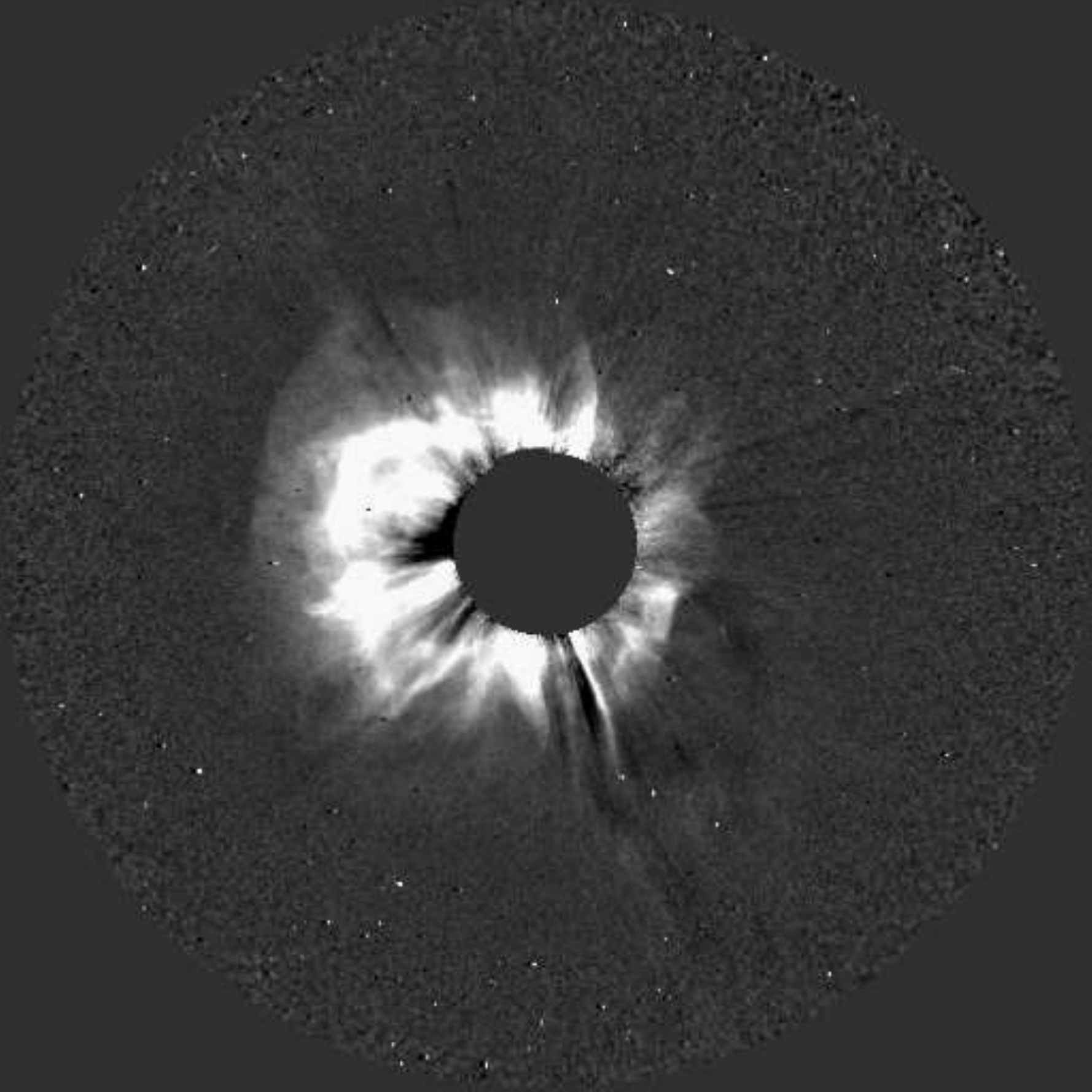}
                \hspace*{-0.02\textwidth}
               \includegraphics[width=0.4\textwidth,clip=]{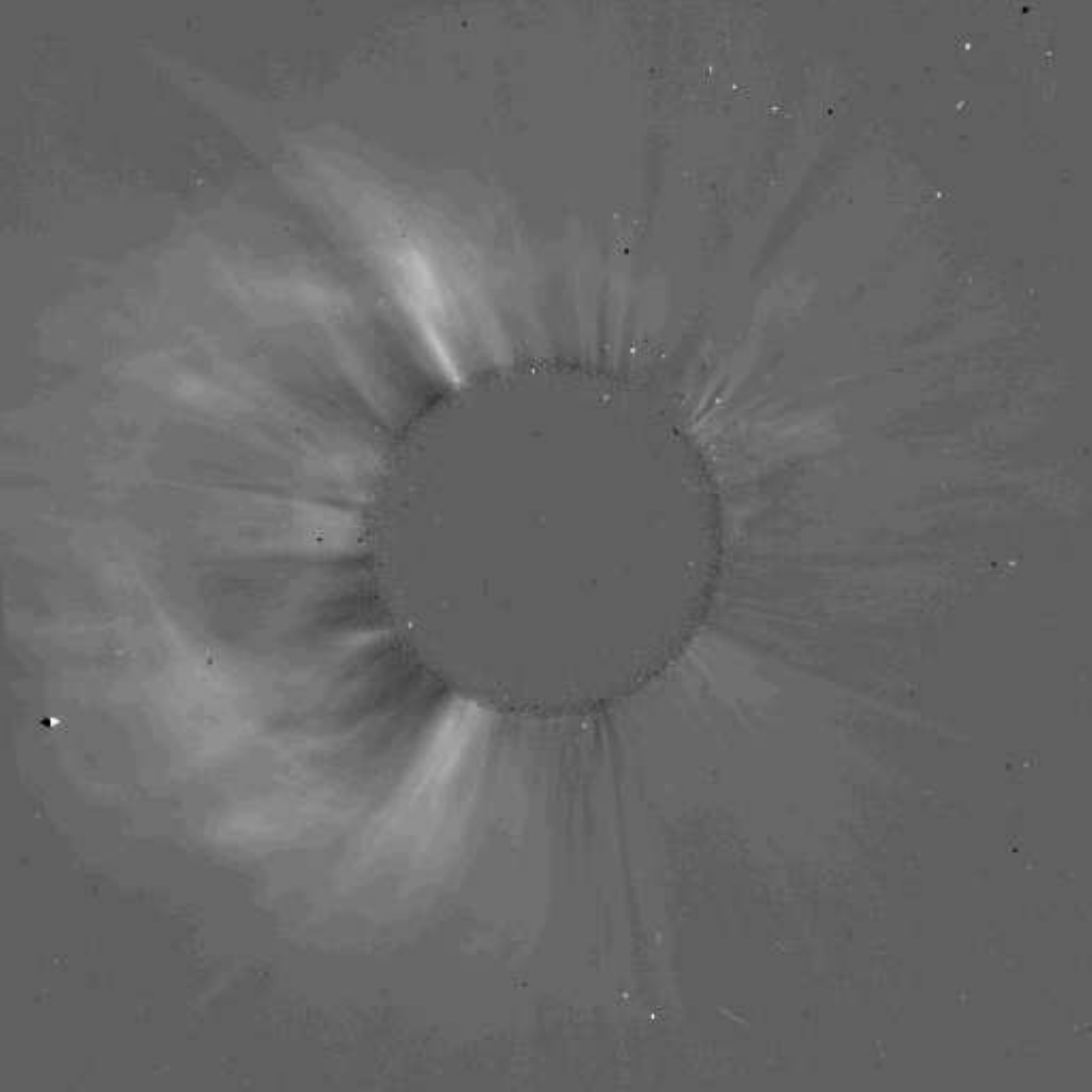}
             \hspace*{-0.02\textwidth}
               \includegraphics[width=0.4\textwidth,clip=]{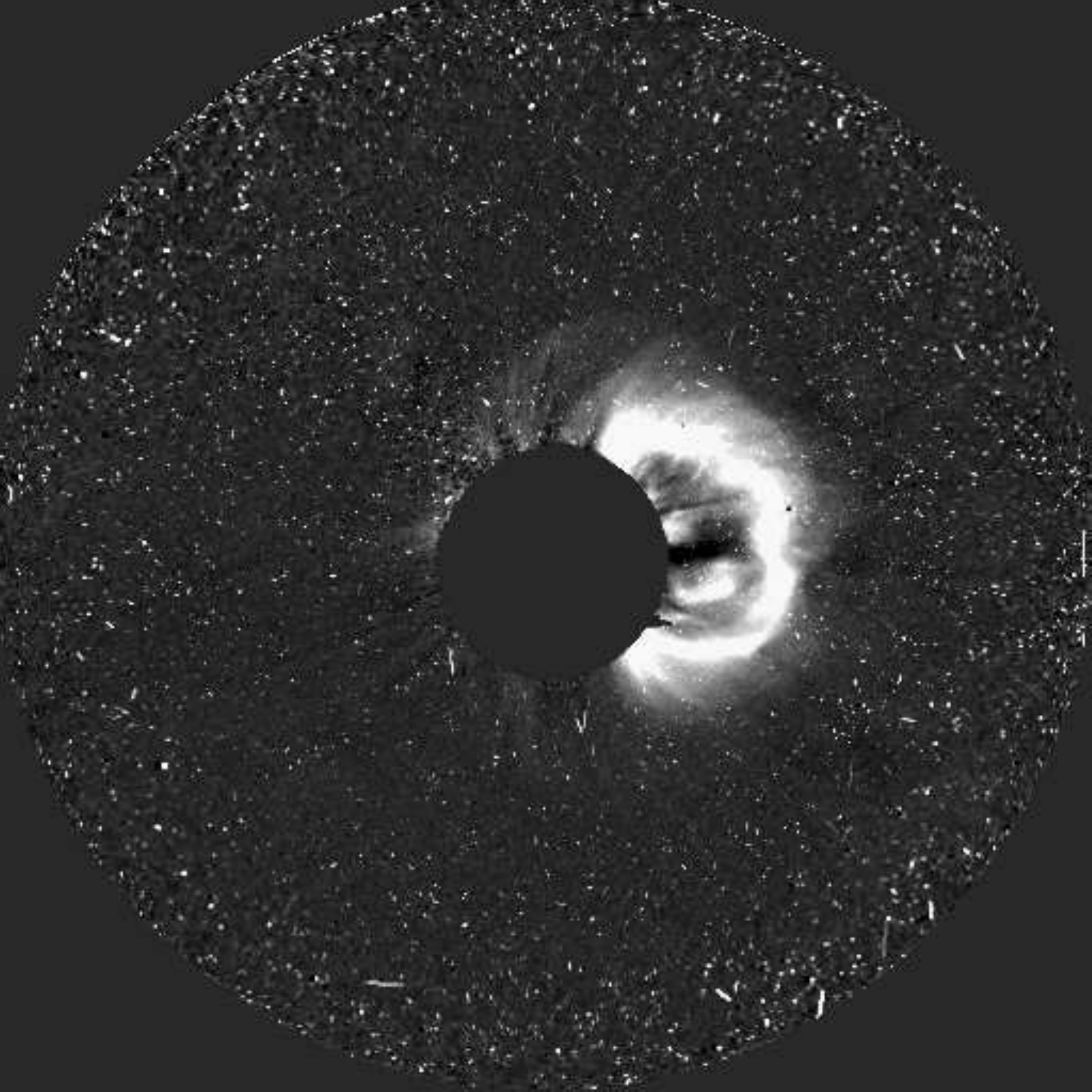}
               }
                 \centerline{\hspace*{0.04\textwidth}
              \includegraphics[width=0.4\textwidth,clip=]{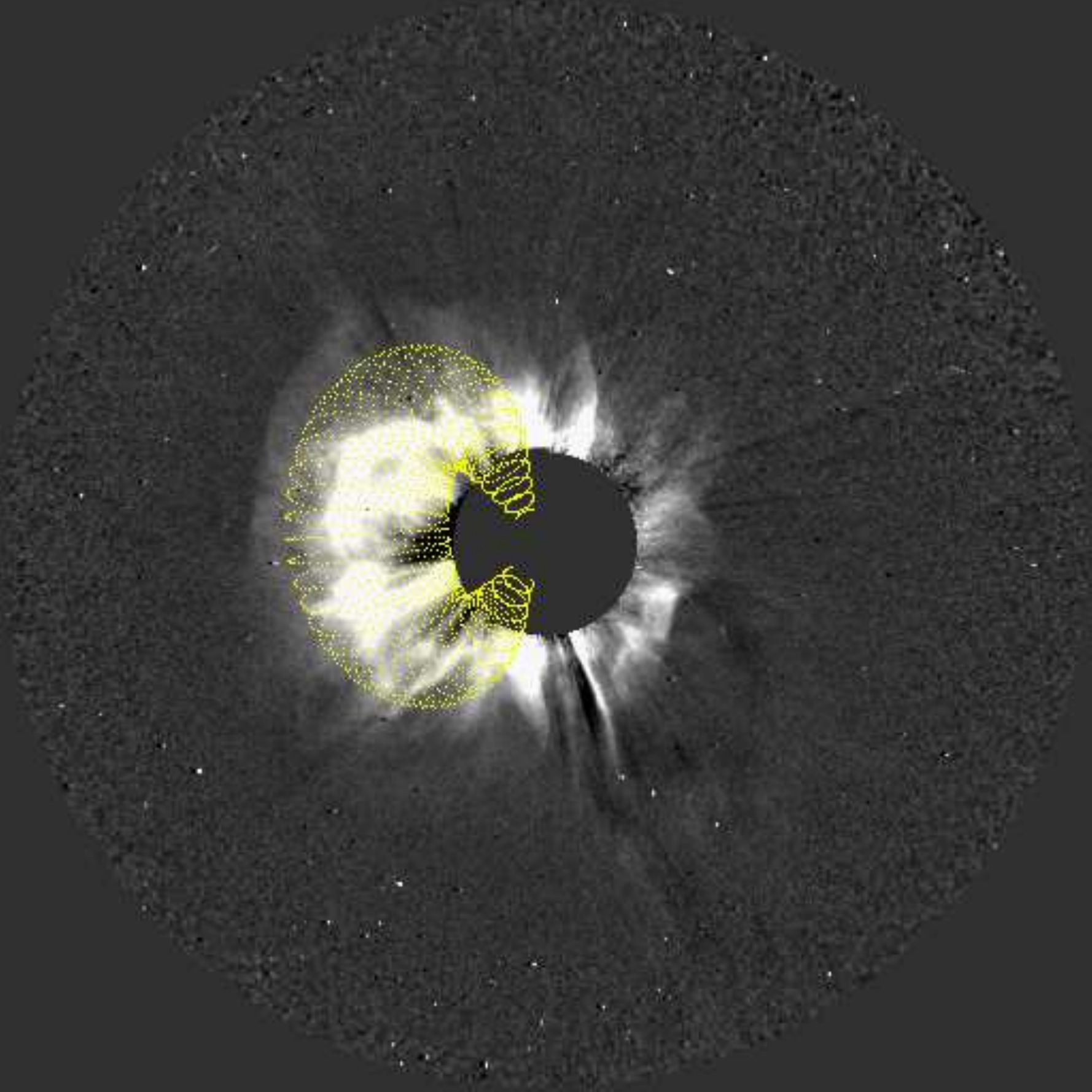}
               \hspace*{-0.02\textwidth}
               \includegraphics[width=0.4\textwidth,clip=]{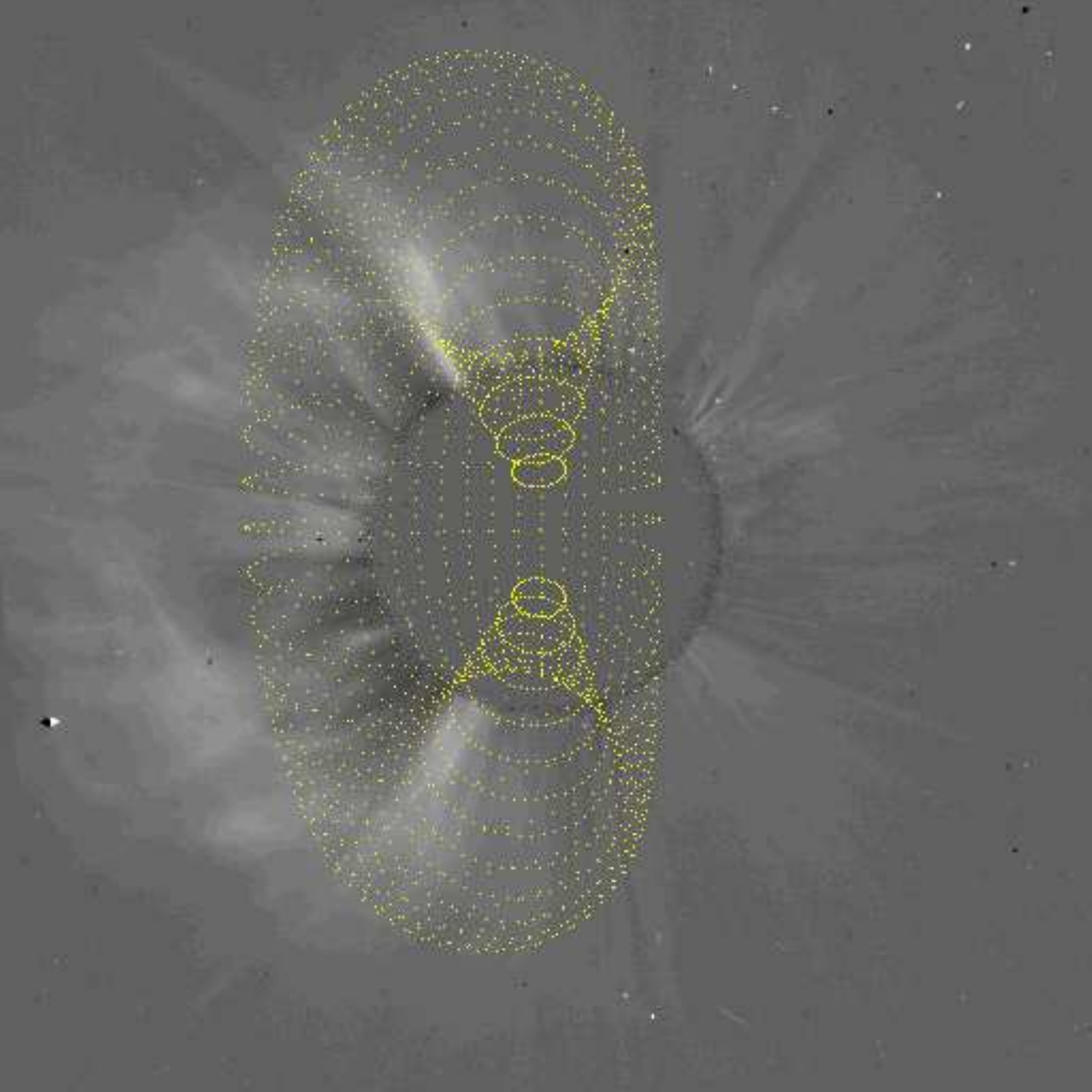}
              \hspace*{-0.02\textwidth}
               \includegraphics[width=0.4\textwidth,clip=]{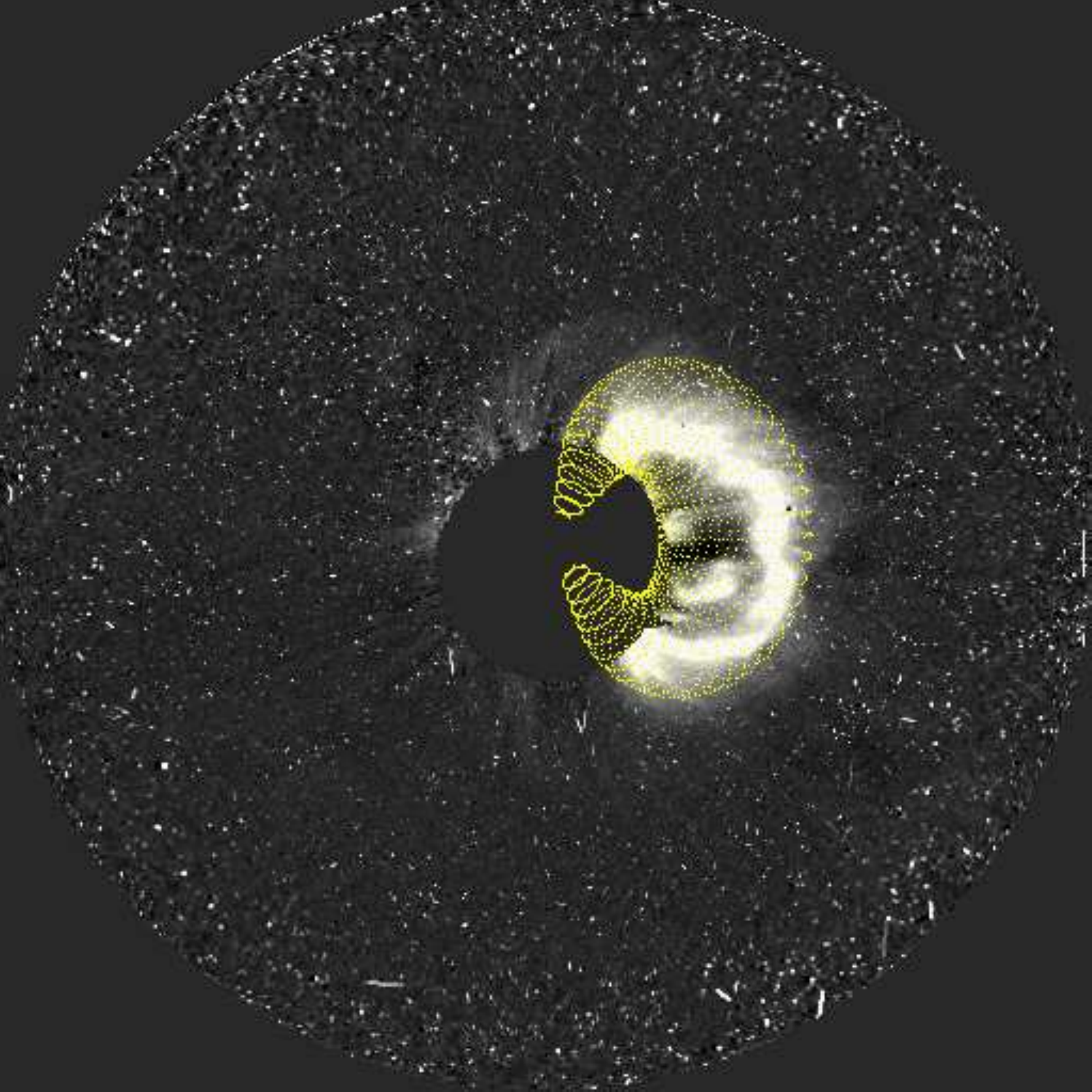}
                }
\vspace{0.0261\textwidth}  
\caption[GCS fit for CME 34 at 08:24]{GCS fit for CME 34 on April 11, 2013 at 08:24 UT at height $H=10.2$ \Rs. Table \ref{tblapp} 
lists the GCS parameters for this event.}
\label{figa34}
\end{figure}

\begin{figure}[h]    
  \centering                              
   \centerline{\hspace*{0.00\textwidth}
               \includegraphics[width=0.4\textwidth,clip=]{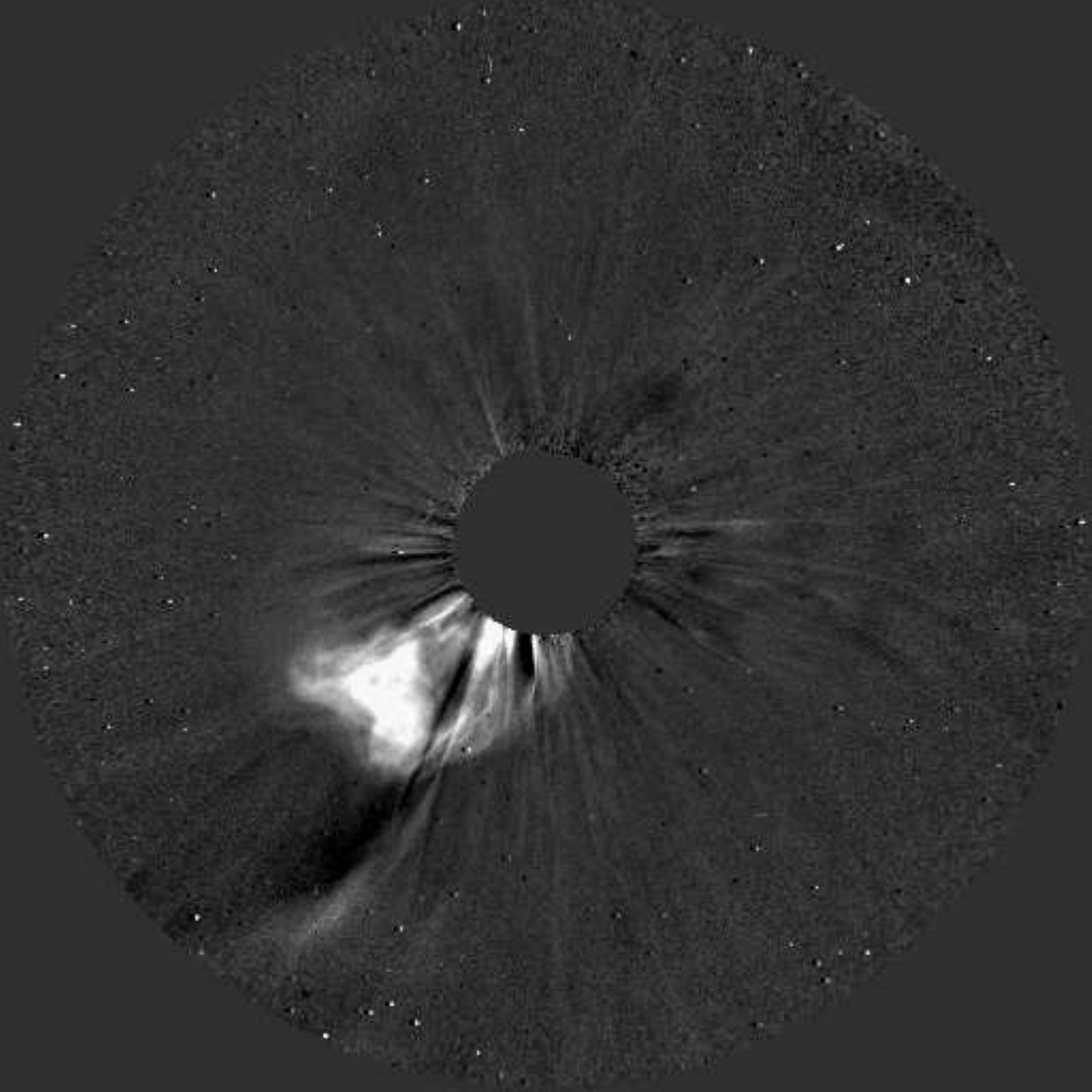}
                \hspace*{-0.02\textwidth}
               \includegraphics[width=0.4\textwidth,clip=]{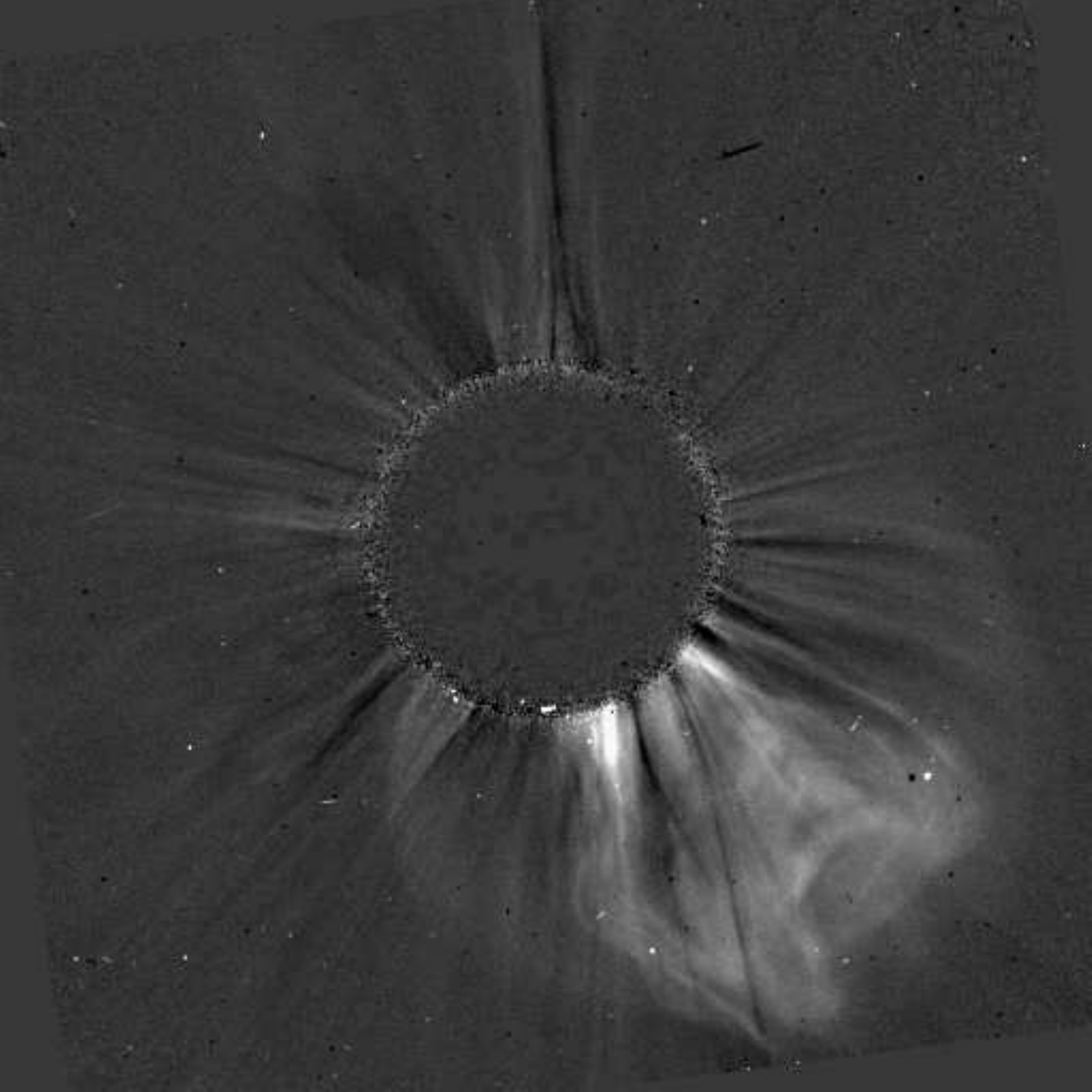}
             \hspace*{-0.02\textwidth}
               \includegraphics[width=0.4\textwidth,clip=]{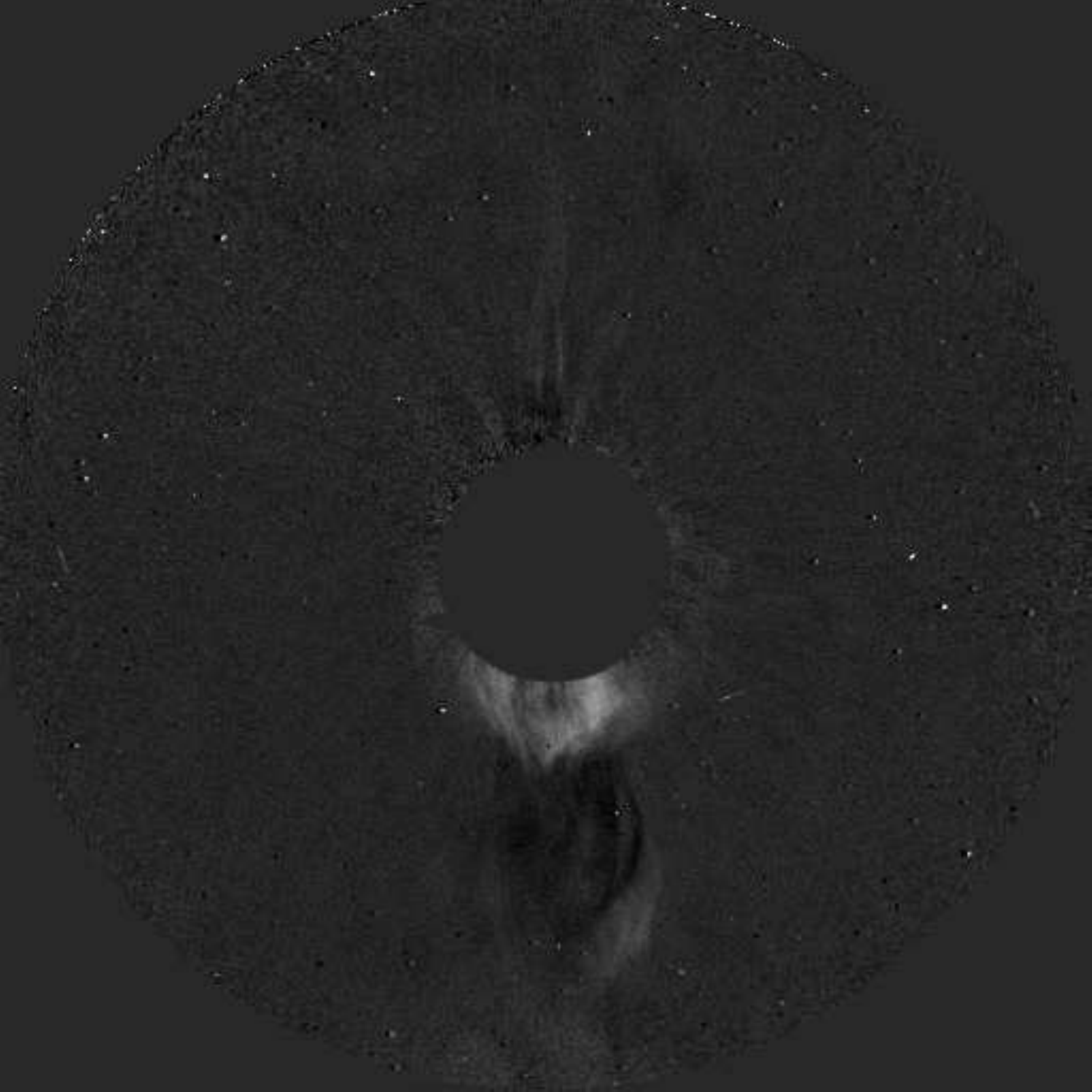}
               }
                 \centerline{\hspace*{0.0\textwidth}
              \includegraphics[width=0.4\textwidth,clip=]{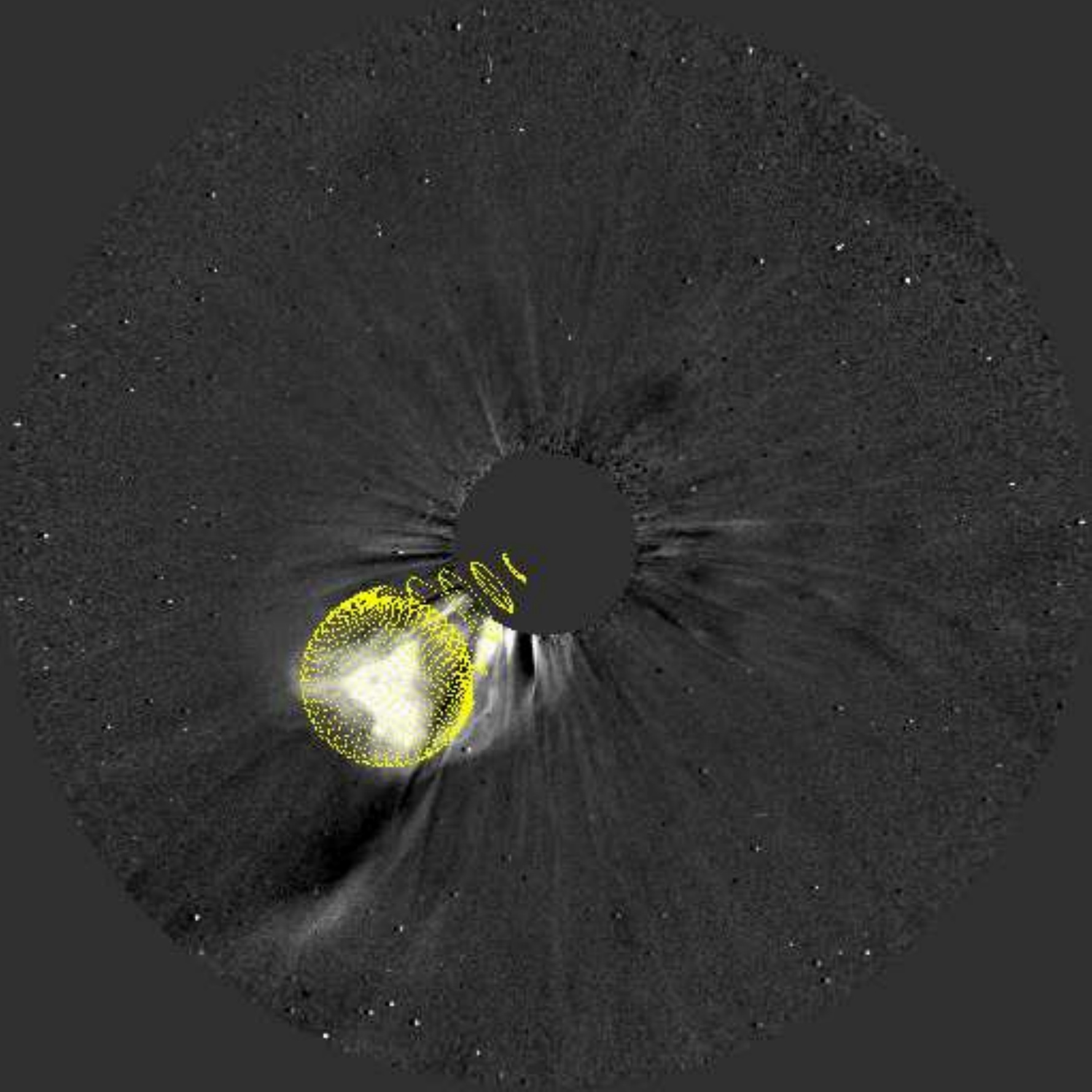}
               \hspace*{-0.02\textwidth}
               \includegraphics[width=0.4\textwidth,clip=]{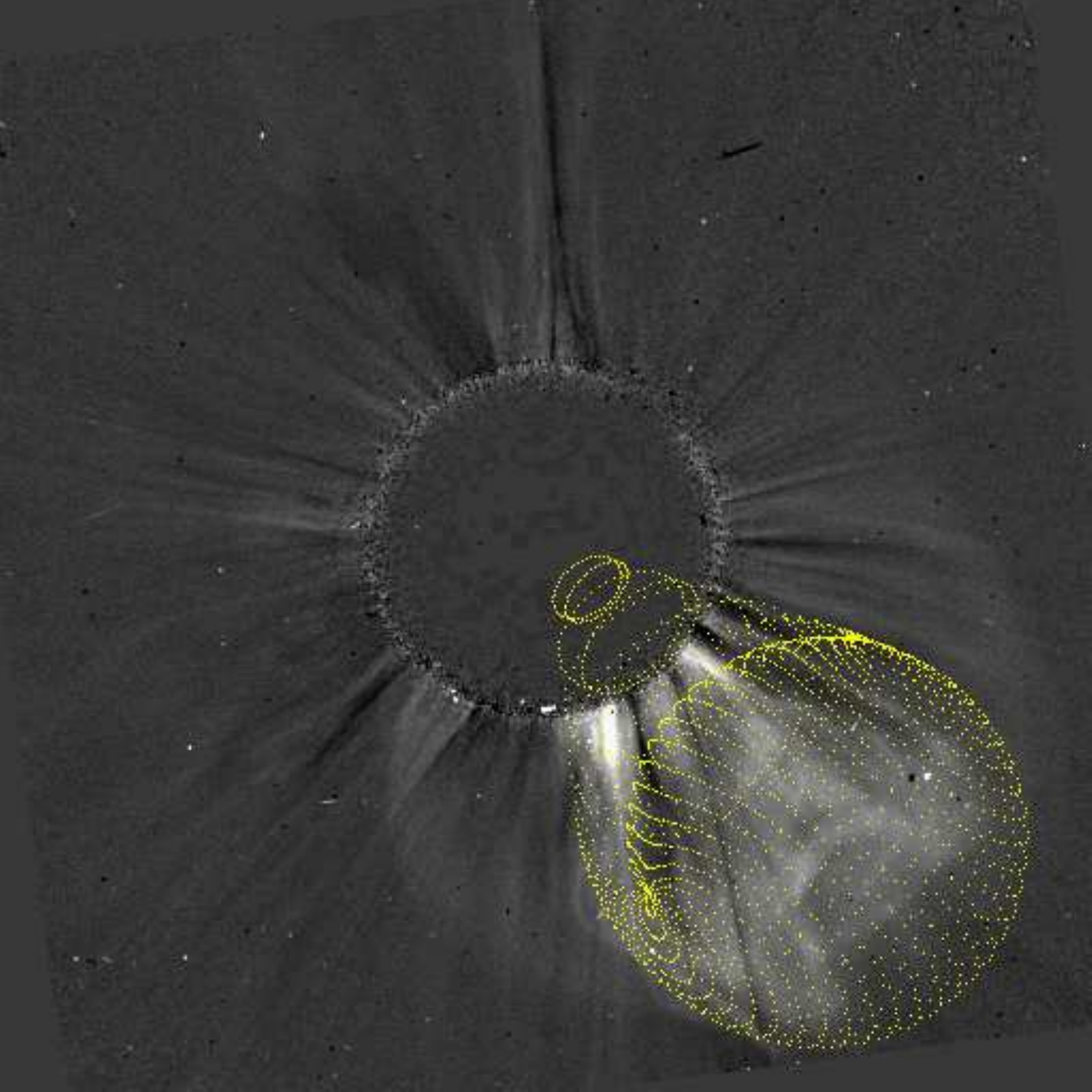}
              \hspace*{-0.02\textwidth}
               \includegraphics[width=0.4\textwidth,clip=]{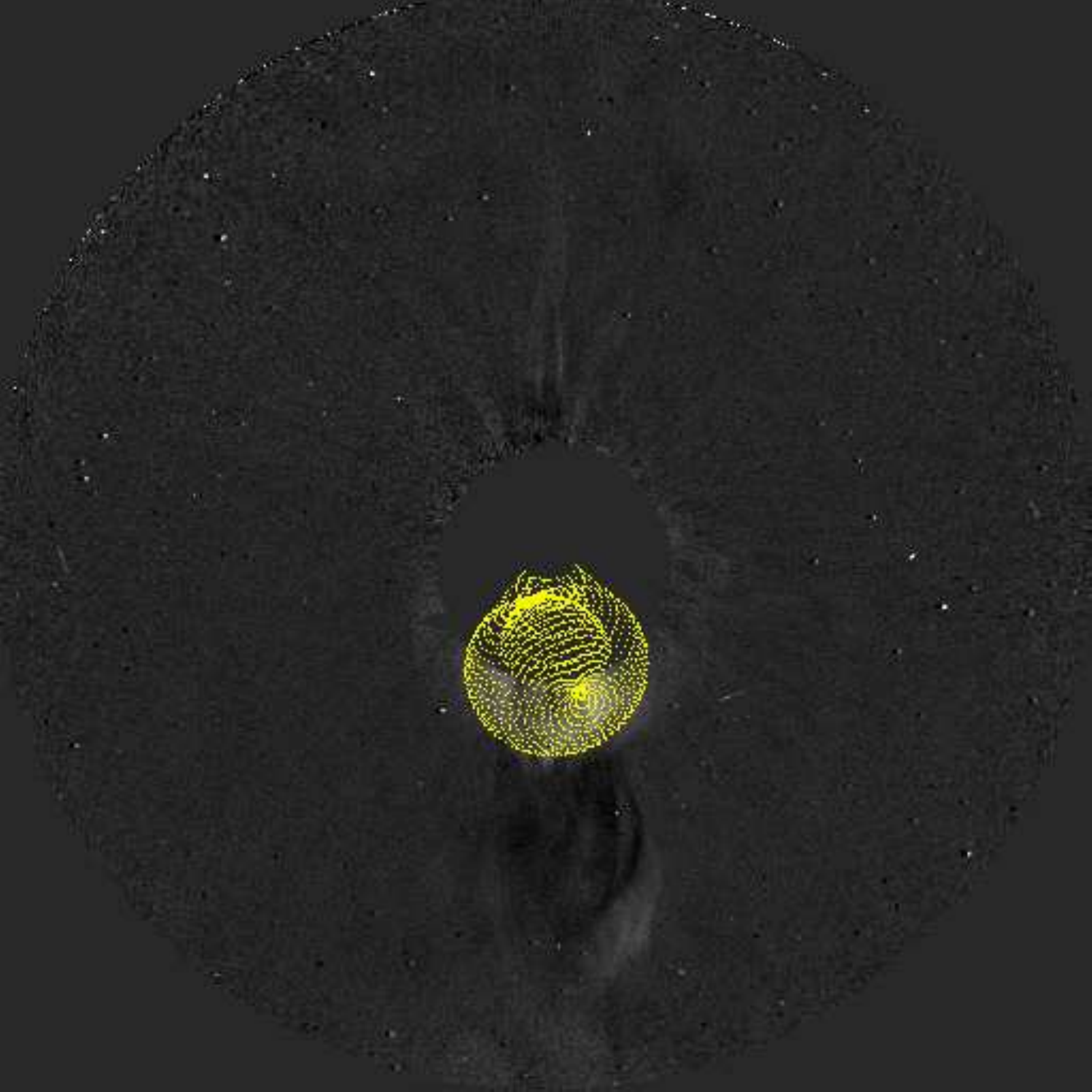}
                }
\vspace{0.0261\textwidth}  
\caption[GCS fit for CME 35 at 02:39]{GCS fit for CME 35 on June 28, 2013 at 02:39 UT at height $H=8.2$ \Rs. Table \ref{tblapp} 
lists the GCS parameters for this event.}
\label{figa35}
\end{figure}

\begin{figure}[h]    
  \centering                              
   \centerline{\hspace*{0.04\textwidth}
               \includegraphics[width=0.4\textwidth,clip=]{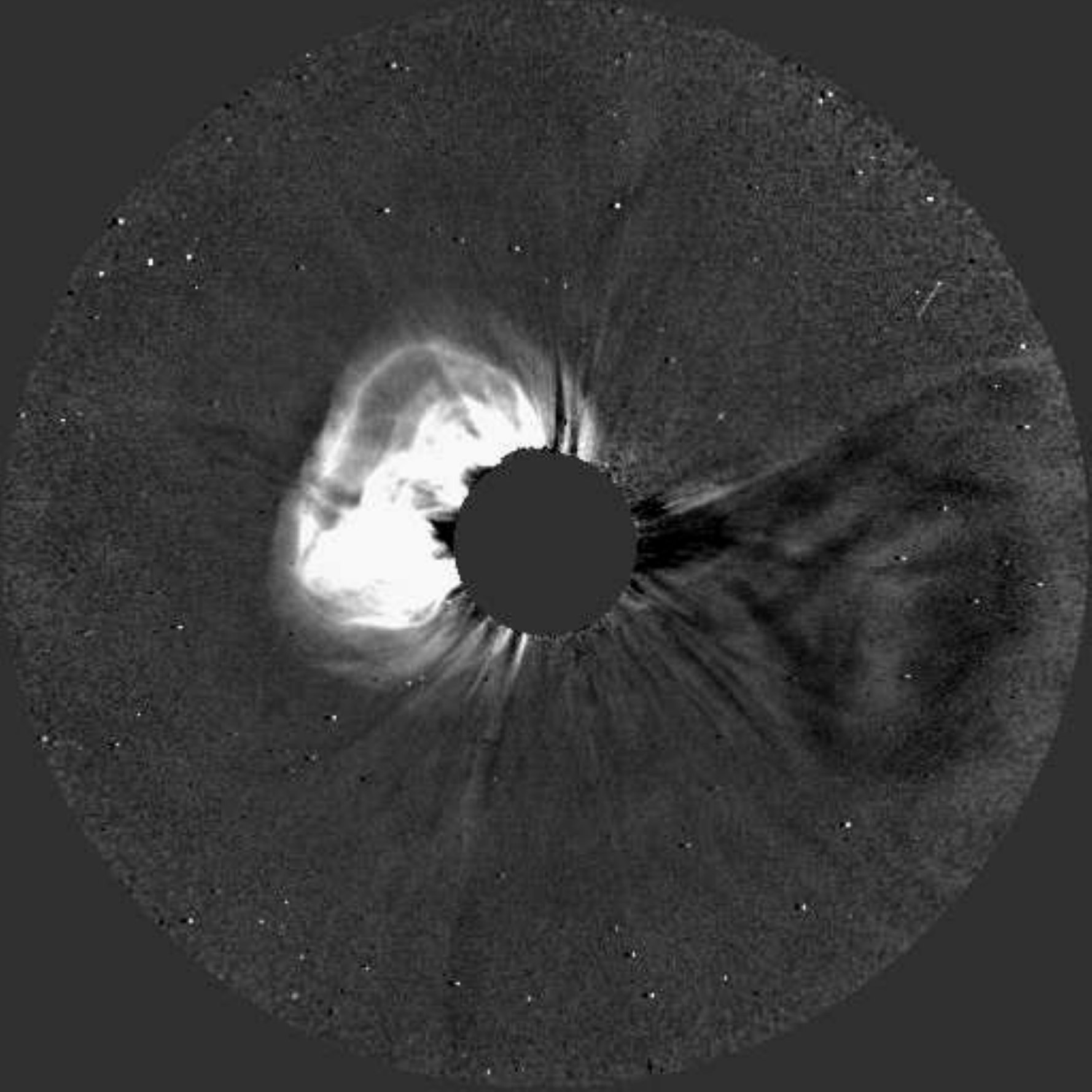}
                \hspace*{-0.02\textwidth}
               \includegraphics[width=0.4\textwidth,clip=]{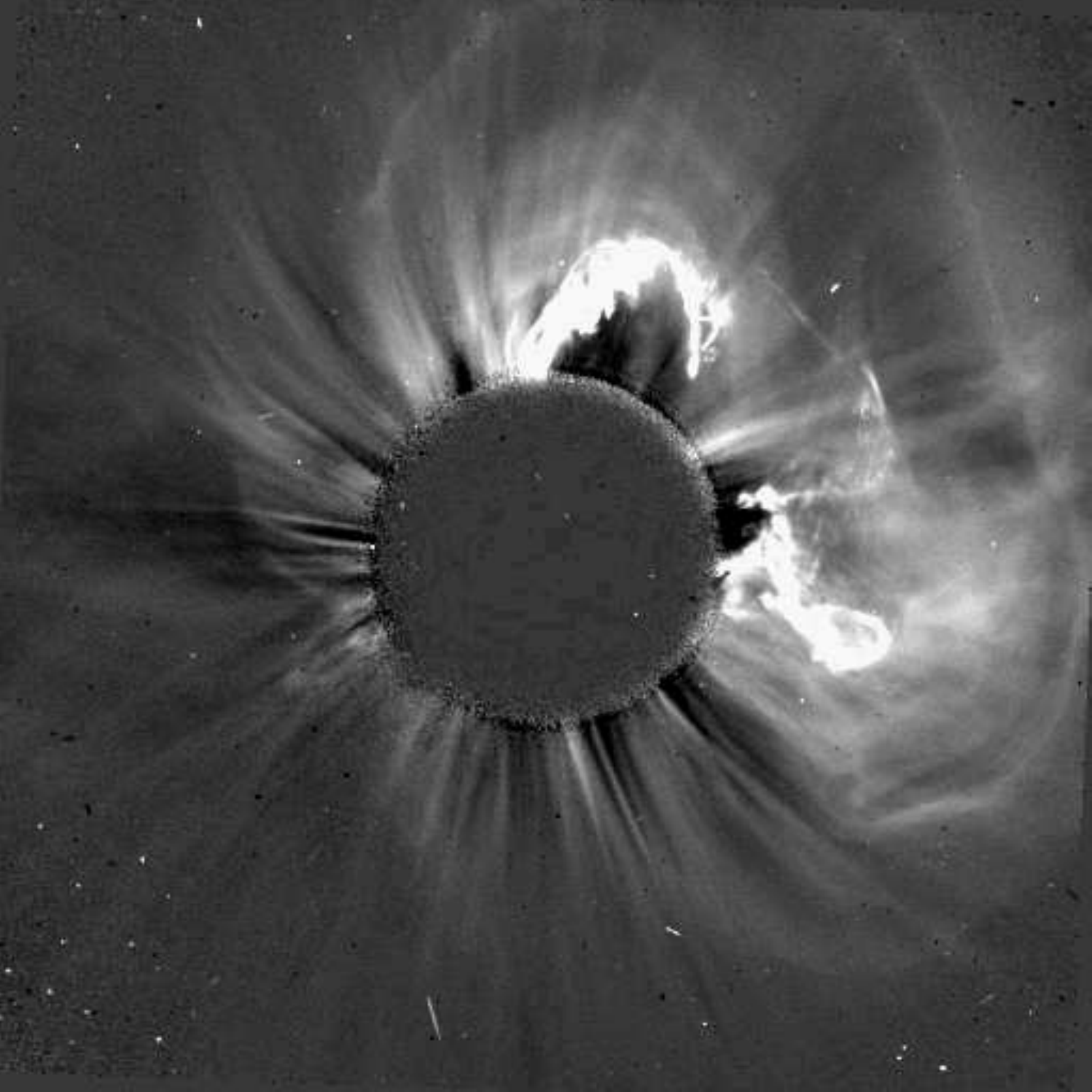}
             \hspace*{-0.02\textwidth}
               \includegraphics[width=0.4\textwidth,clip=]{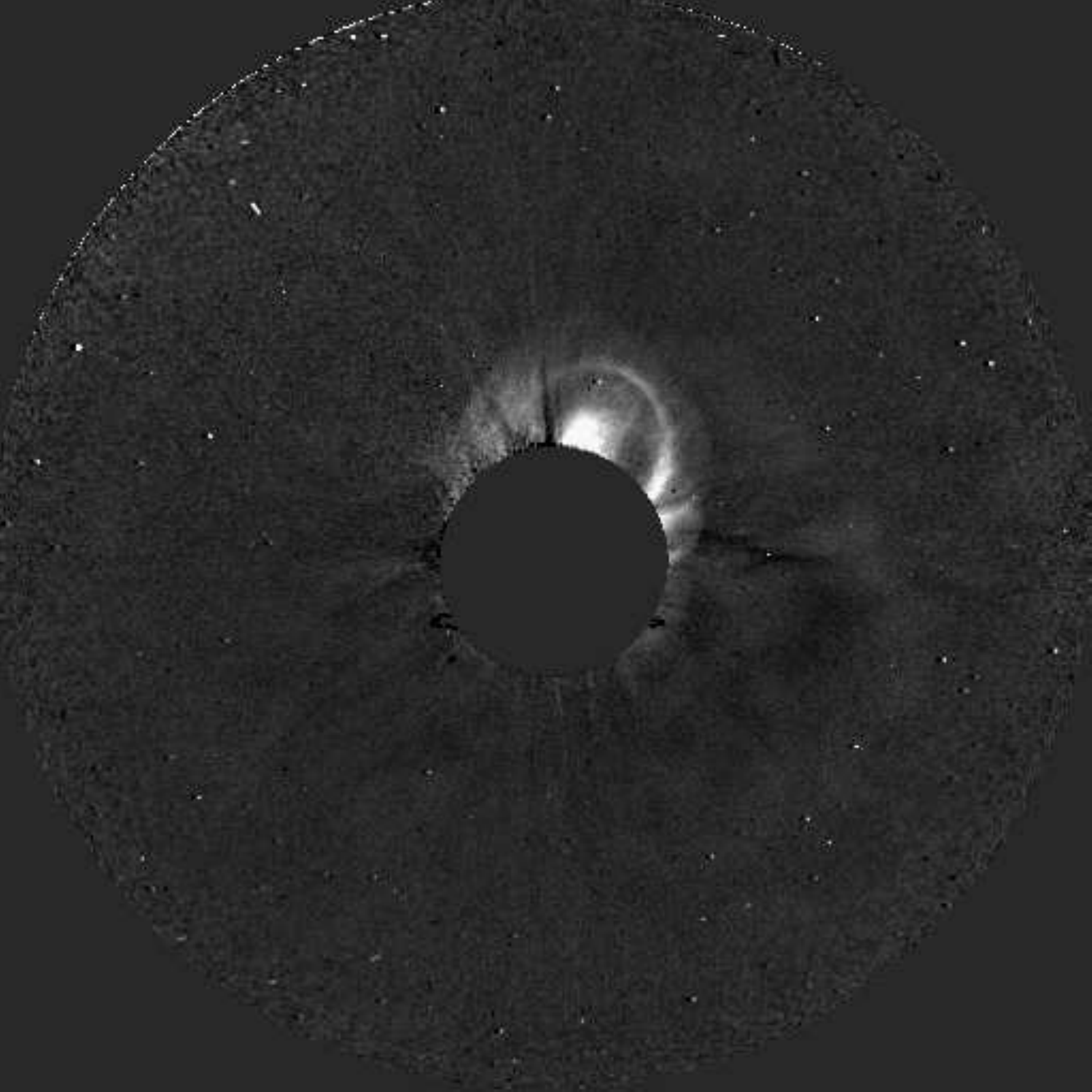}
               }
                 \centerline{\hspace*{0.04\textwidth}
              \includegraphics[width=0.4\textwidth,clip=]{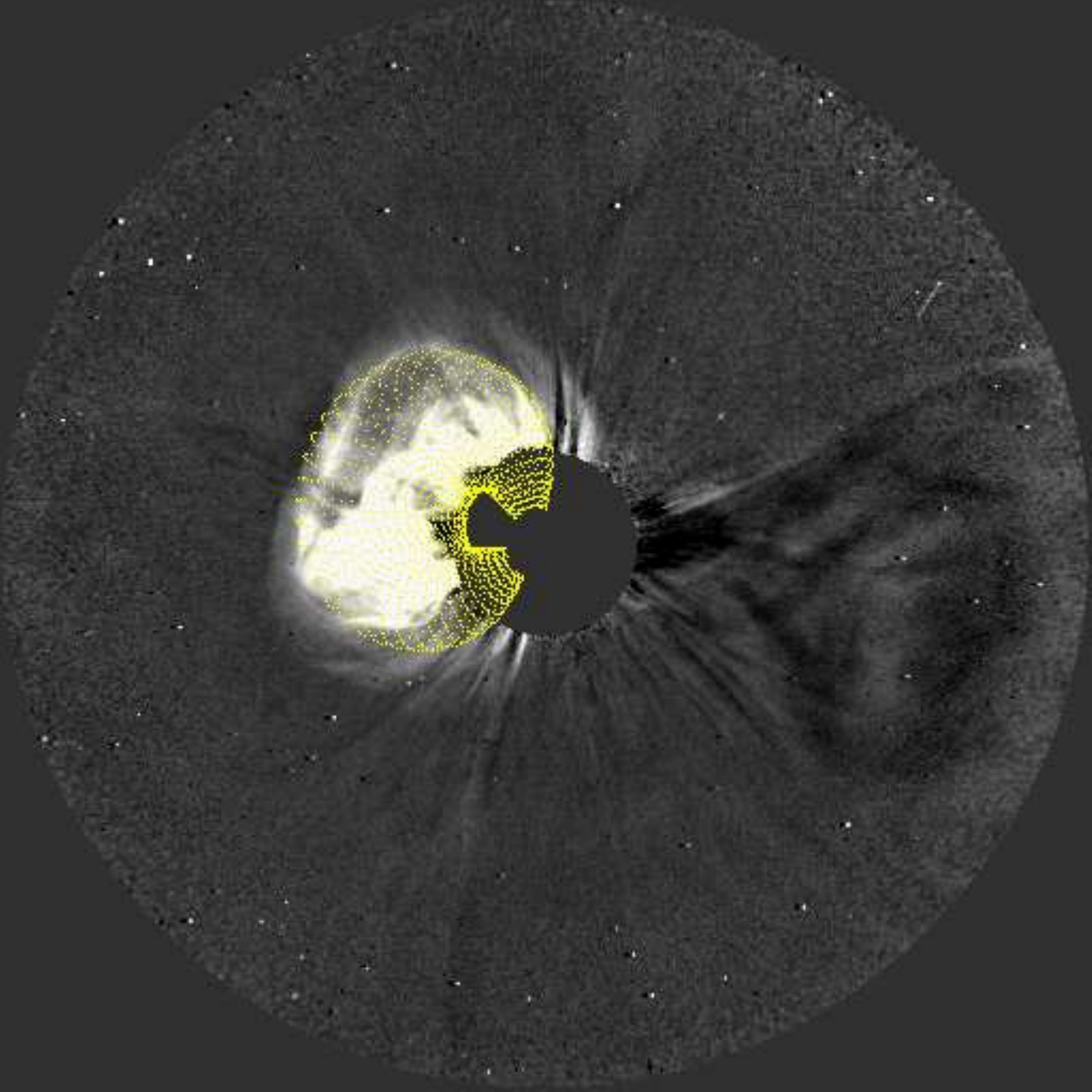}
               \hspace*{-0.02\textwidth}
               \includegraphics[width=0.4\textwidth,clip=]{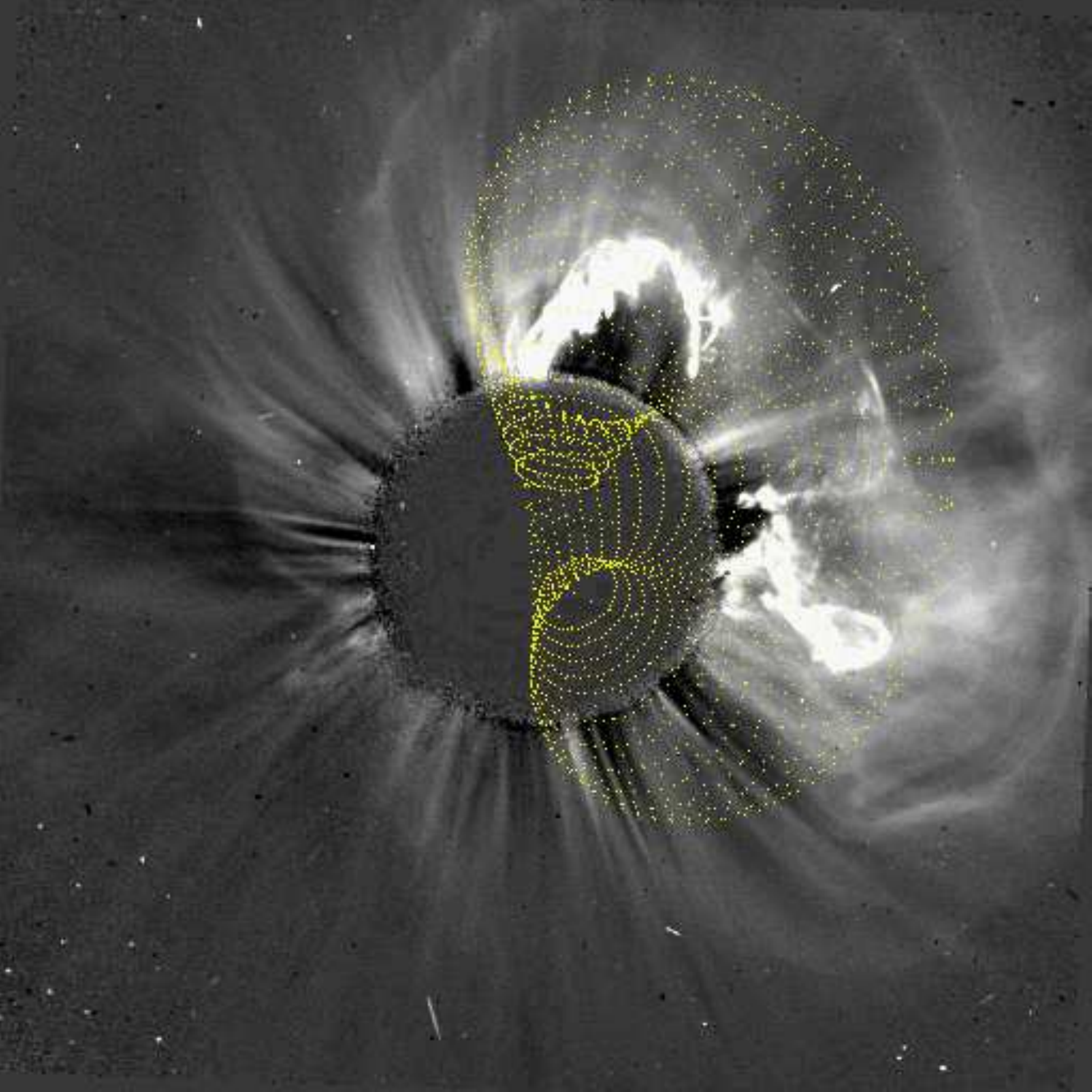}
              \hspace*{-0.02\textwidth}
               \includegraphics[width=0.4\textwidth,clip=]{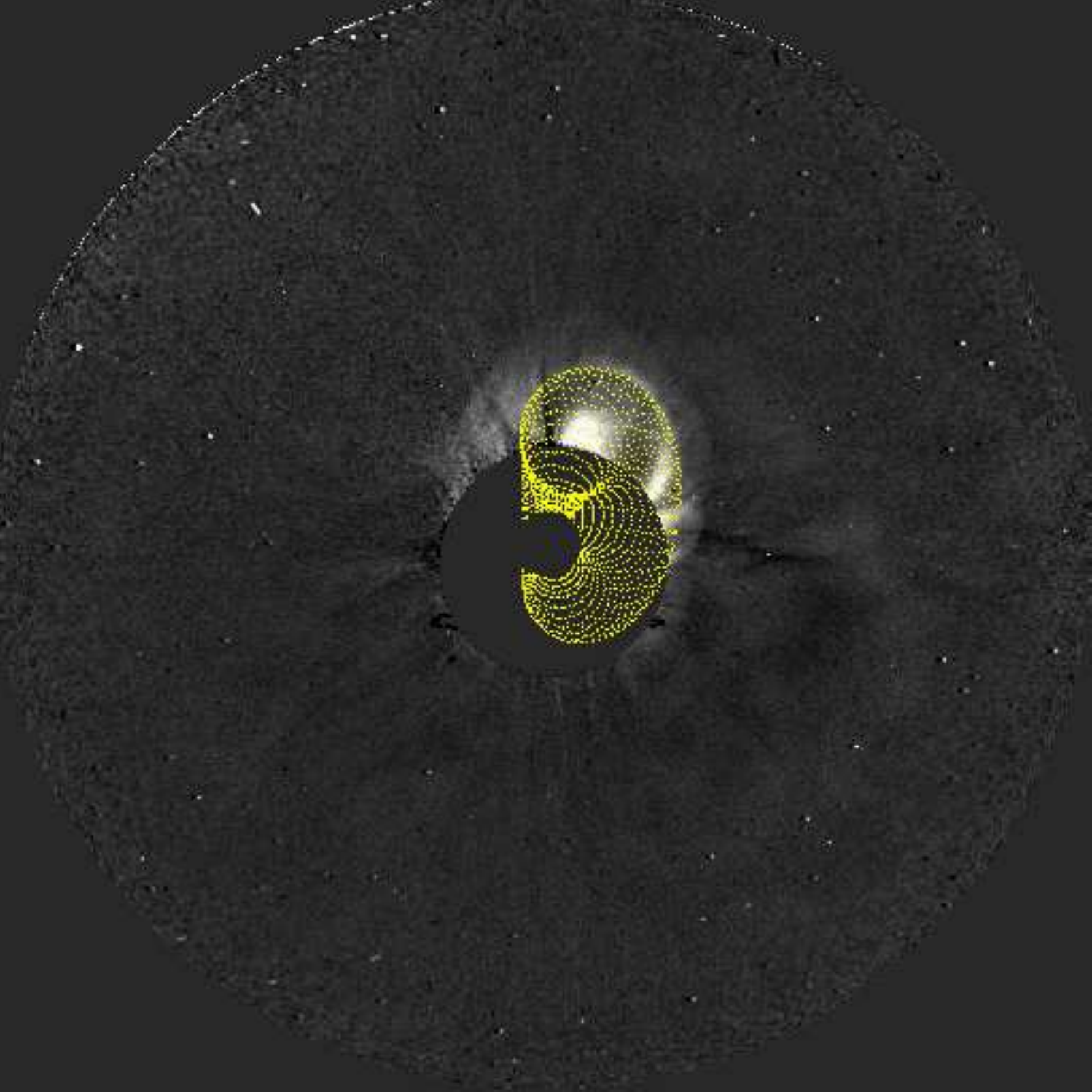}
                }
\vspace{0.0261\textwidth}  
\caption[GCS fit for CME 36 at 22:54]{GCS fit for CME 36 on September 29, 2013 at 22:54 UT at height $H=8.1$ \Rs. Table \ref{tblapp} 
lists the GCS parameters for this event.}
\label{figa36}
\end{figure}

\begin{figure}[h]    
  \centering                              
   \centerline{\hspace*{0.00\textwidth}
               \includegraphics[width=0.4\textwidth,clip=]{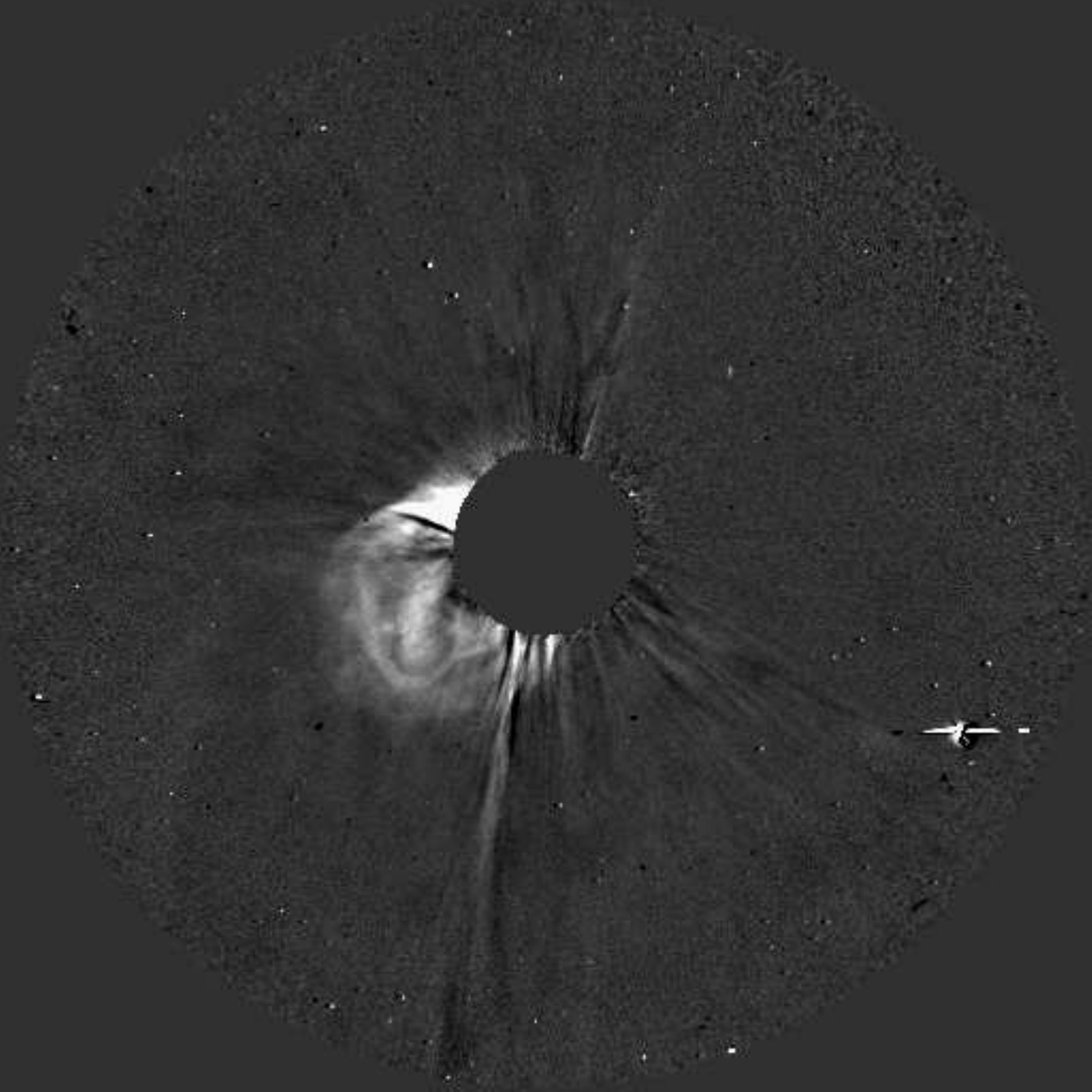}
                \hspace*{-0.02\textwidth}
               \includegraphics[width=0.4\textwidth,clip=]{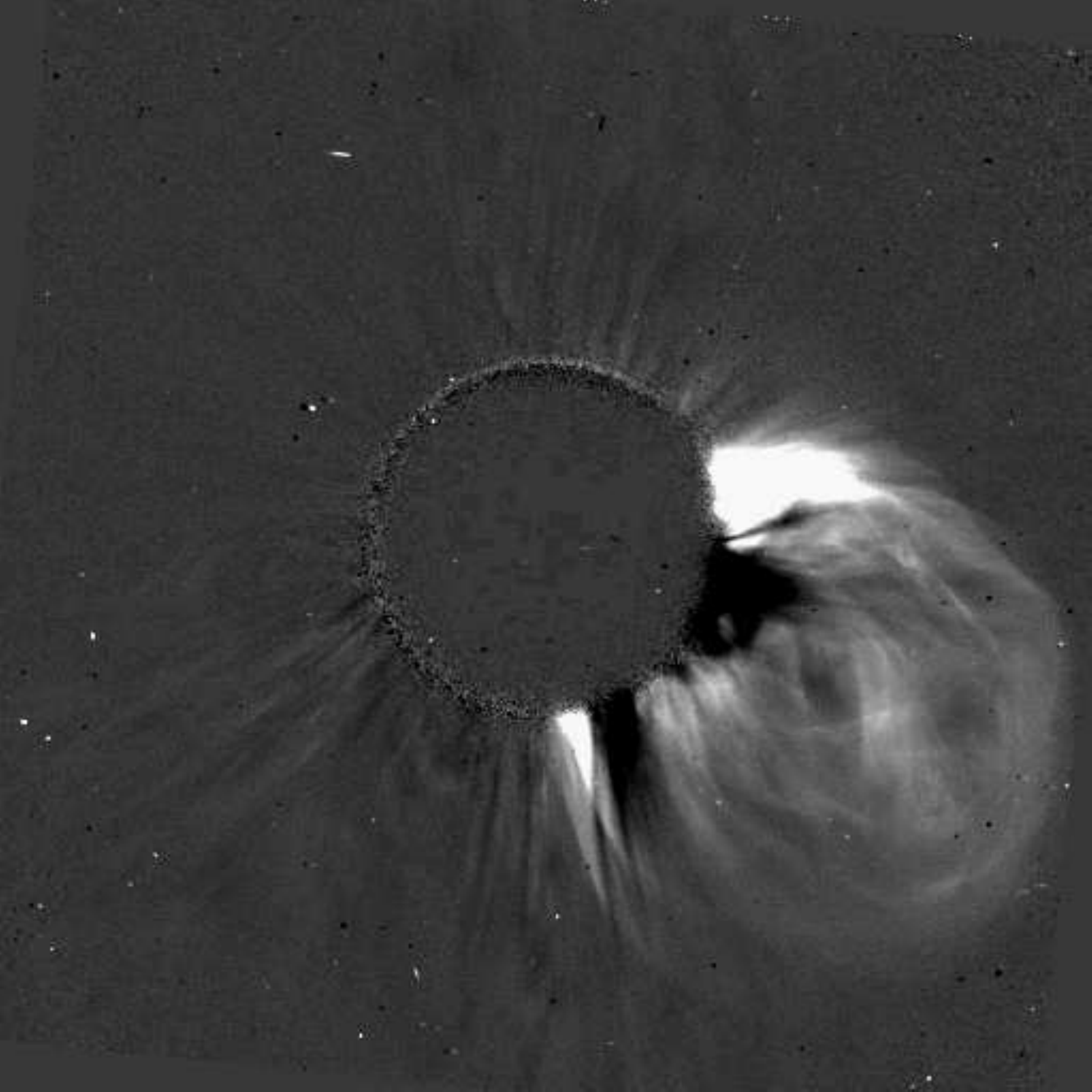}
             \hspace*{-0.02\textwidth}
               \includegraphics[width=0.4\textwidth,clip=]{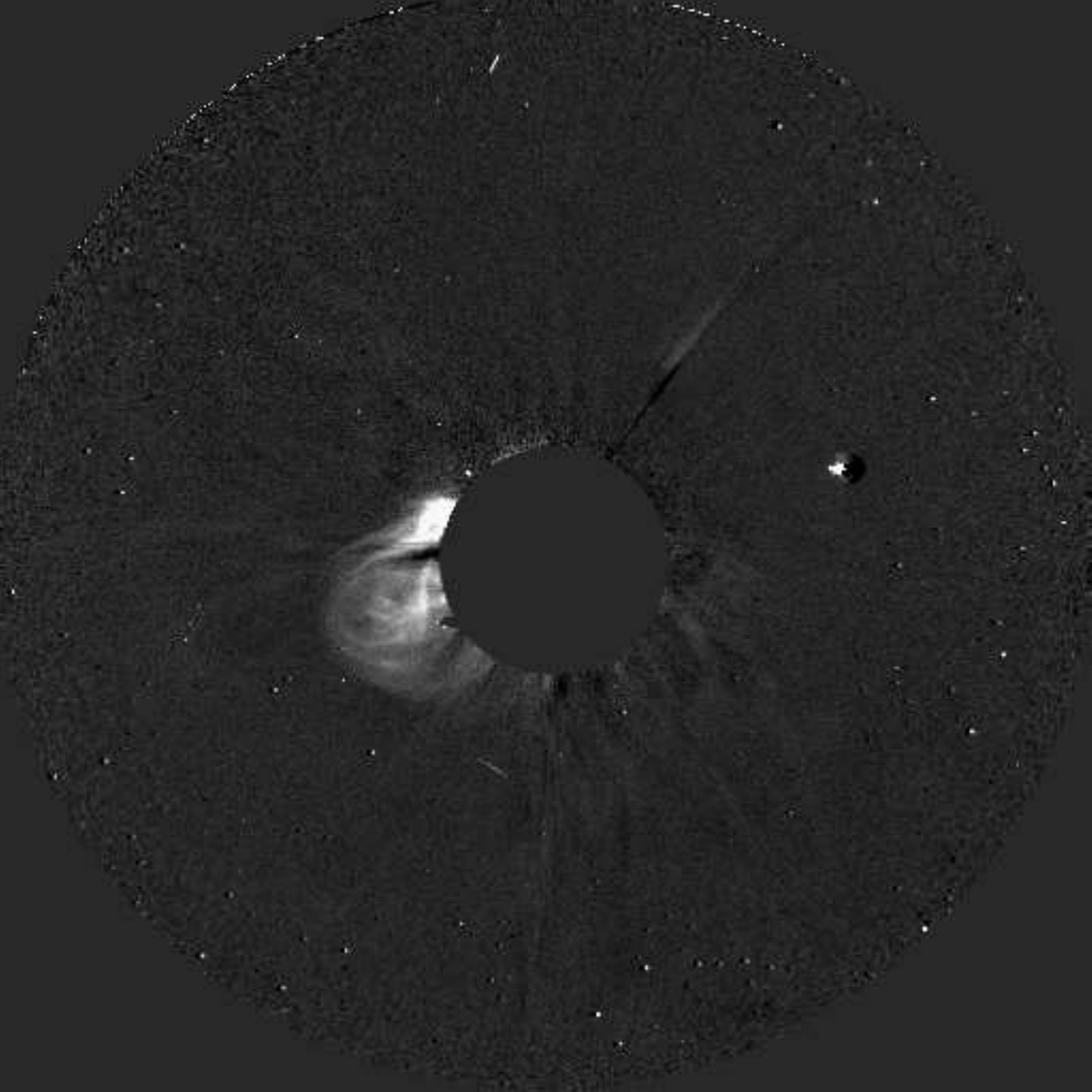}
               }
                 \centerline{\hspace*{0.0\textwidth}
              \includegraphics[width=0.4\textwidth,clip=]{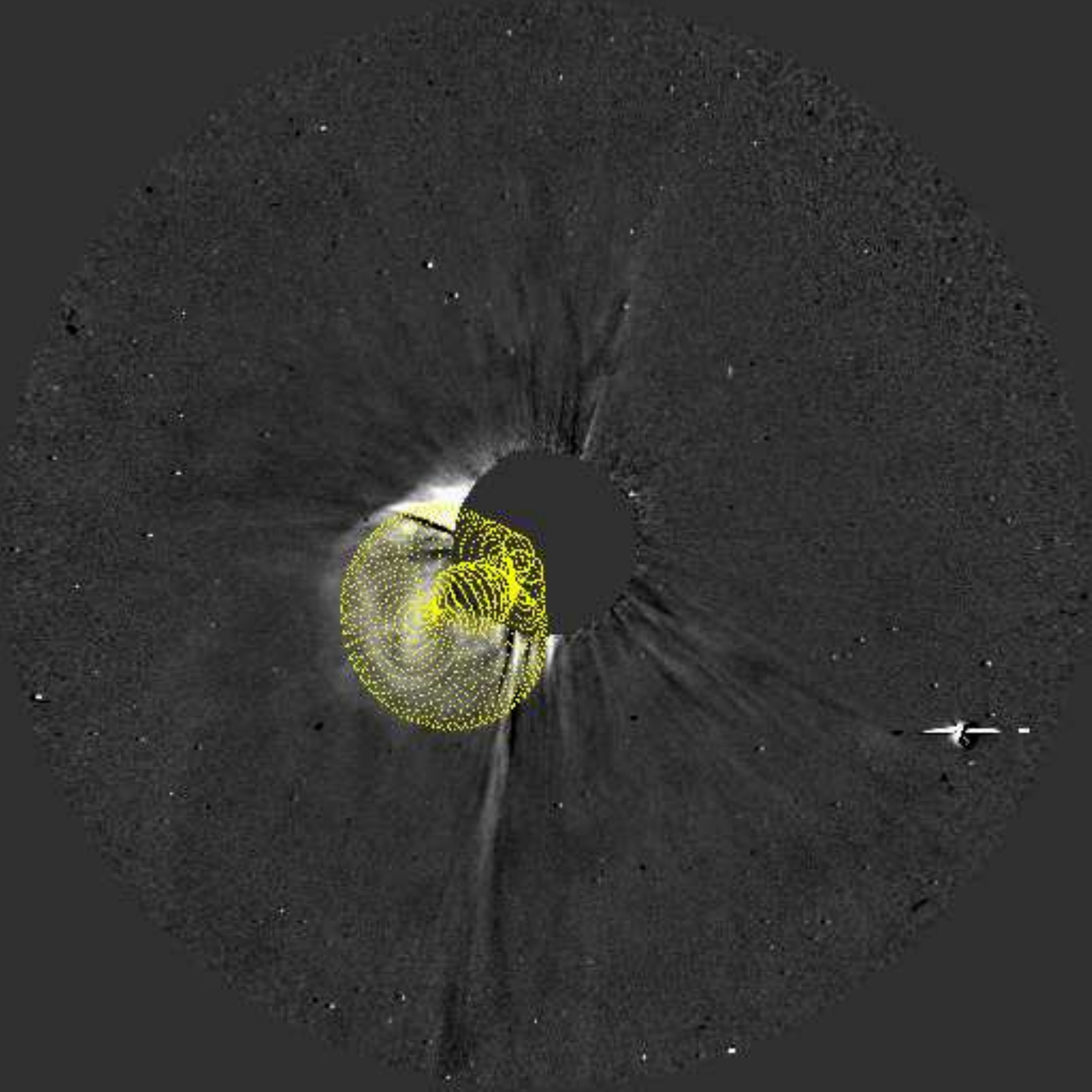}
               \hspace*{-0.02\textwidth}
               \includegraphics[width=0.4\textwidth,clip=]{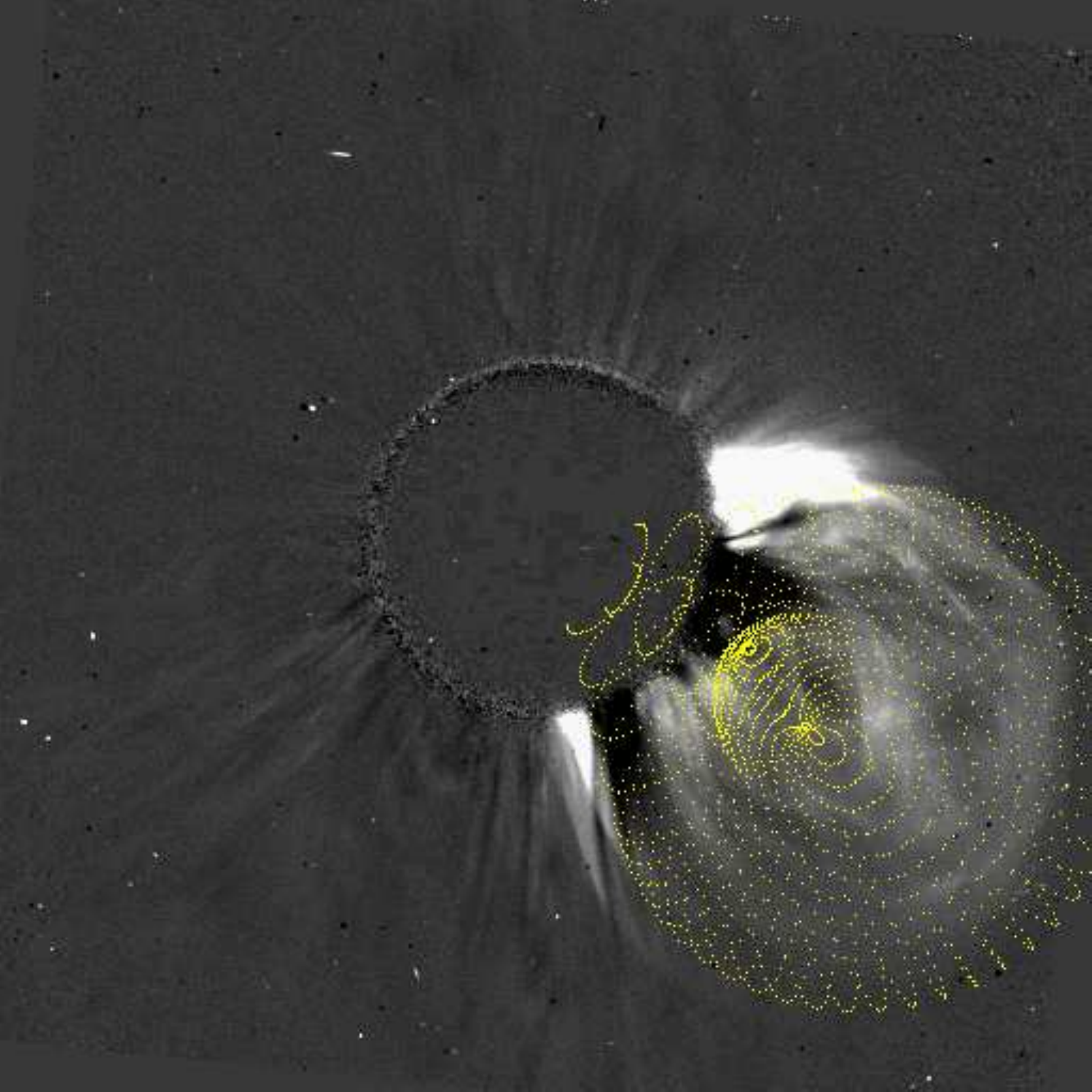}
              \hspace*{-0.02\textwidth}
               \includegraphics[width=0.4\textwidth,clip=]{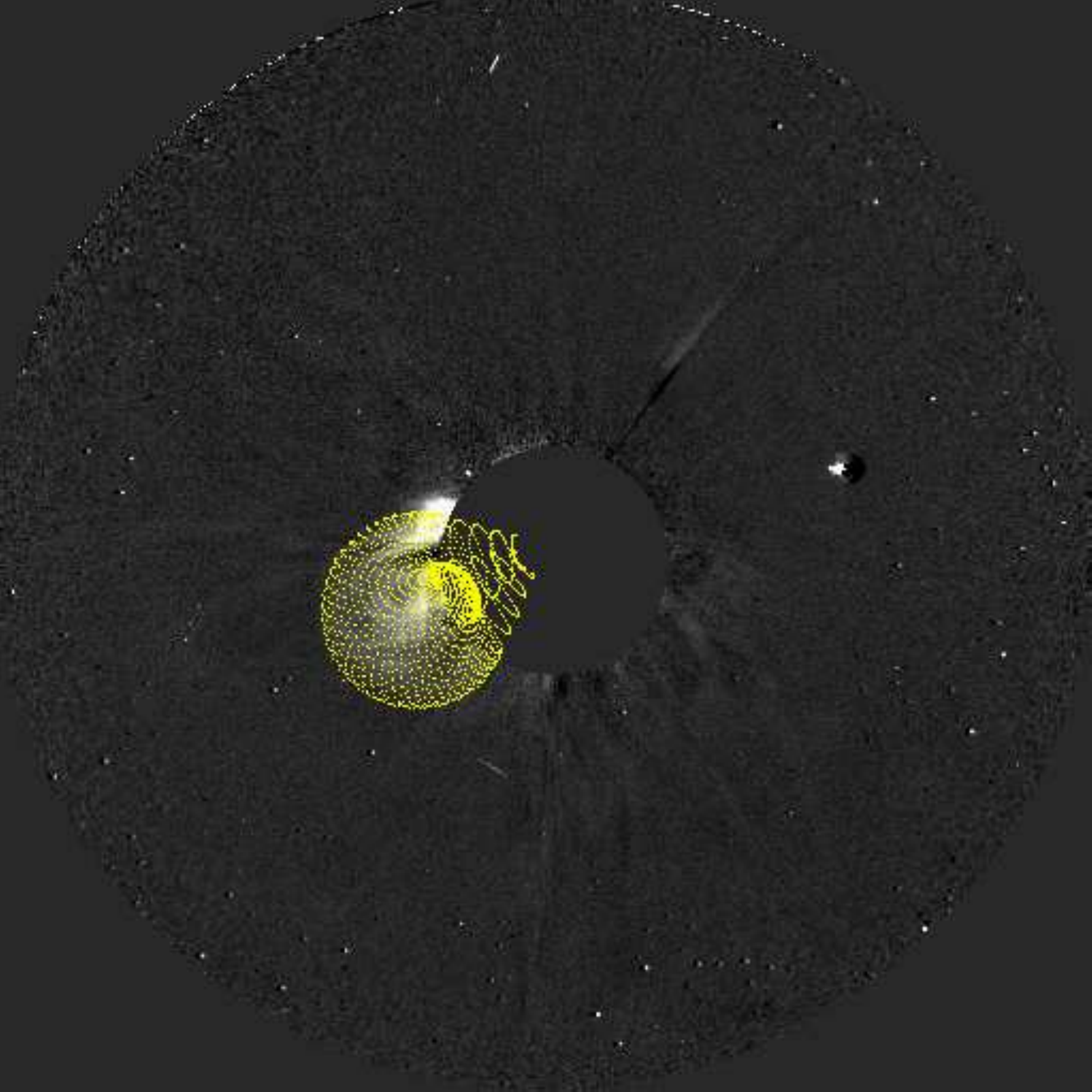}
                }
\vspace{0.0261\textwidth}  
\caption[GCS fit for CME 37 at 00:39]{GCS fit for CME 37 on November 07, 2013 at 00:39 UT at height $H=7.7$ \Rs. Table \ref{tblapp} 
lists the GCS parameters for this event.}
\label{figa37}
\end{figure}

\begin{figure}[h]    
  \centering                              
   \centerline{\hspace*{0.04\textwidth}
               \includegraphics[width=0.4\textwidth,clip=]{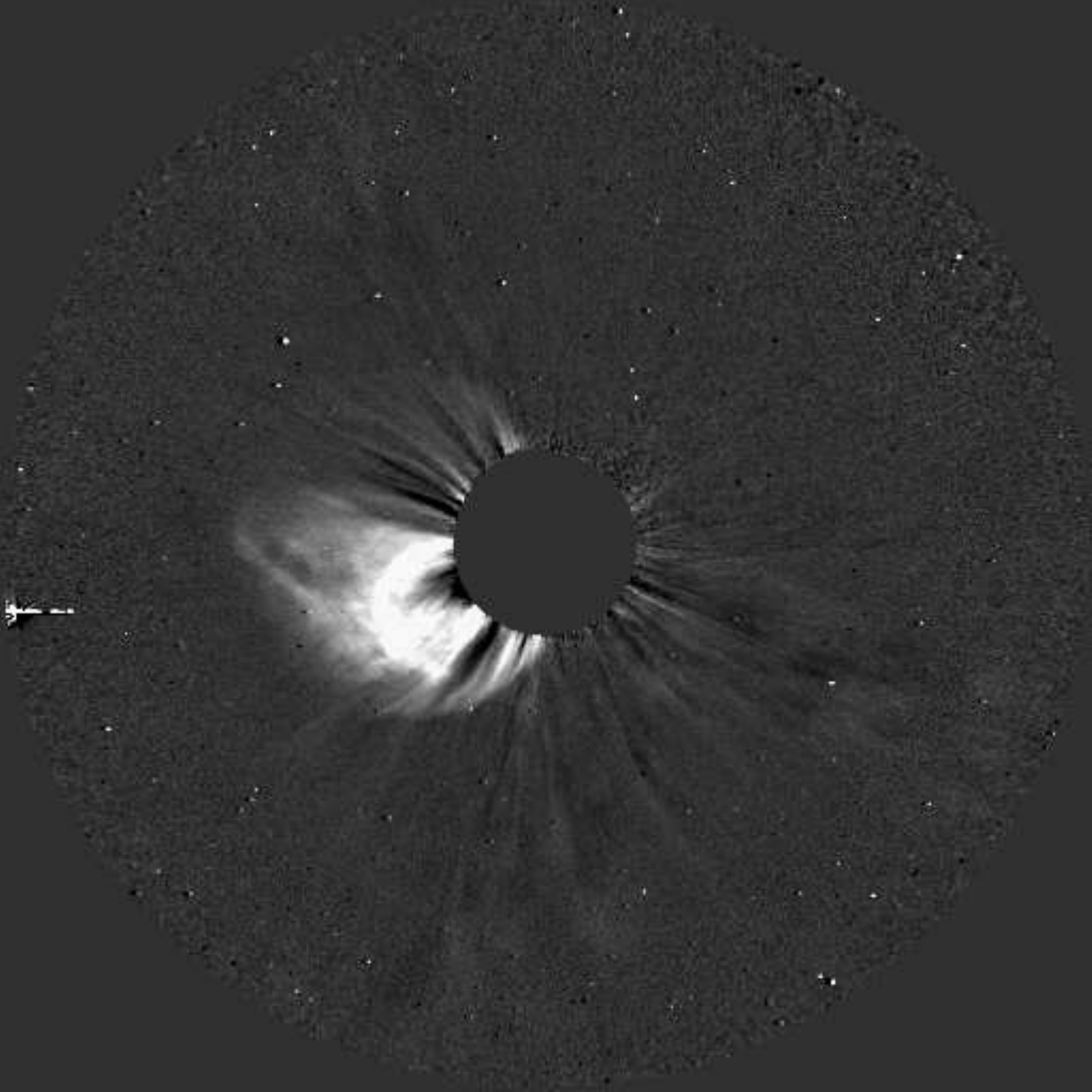}
                \hspace*{-0.02\textwidth}
               \includegraphics[width=0.4\textwidth,clip=]{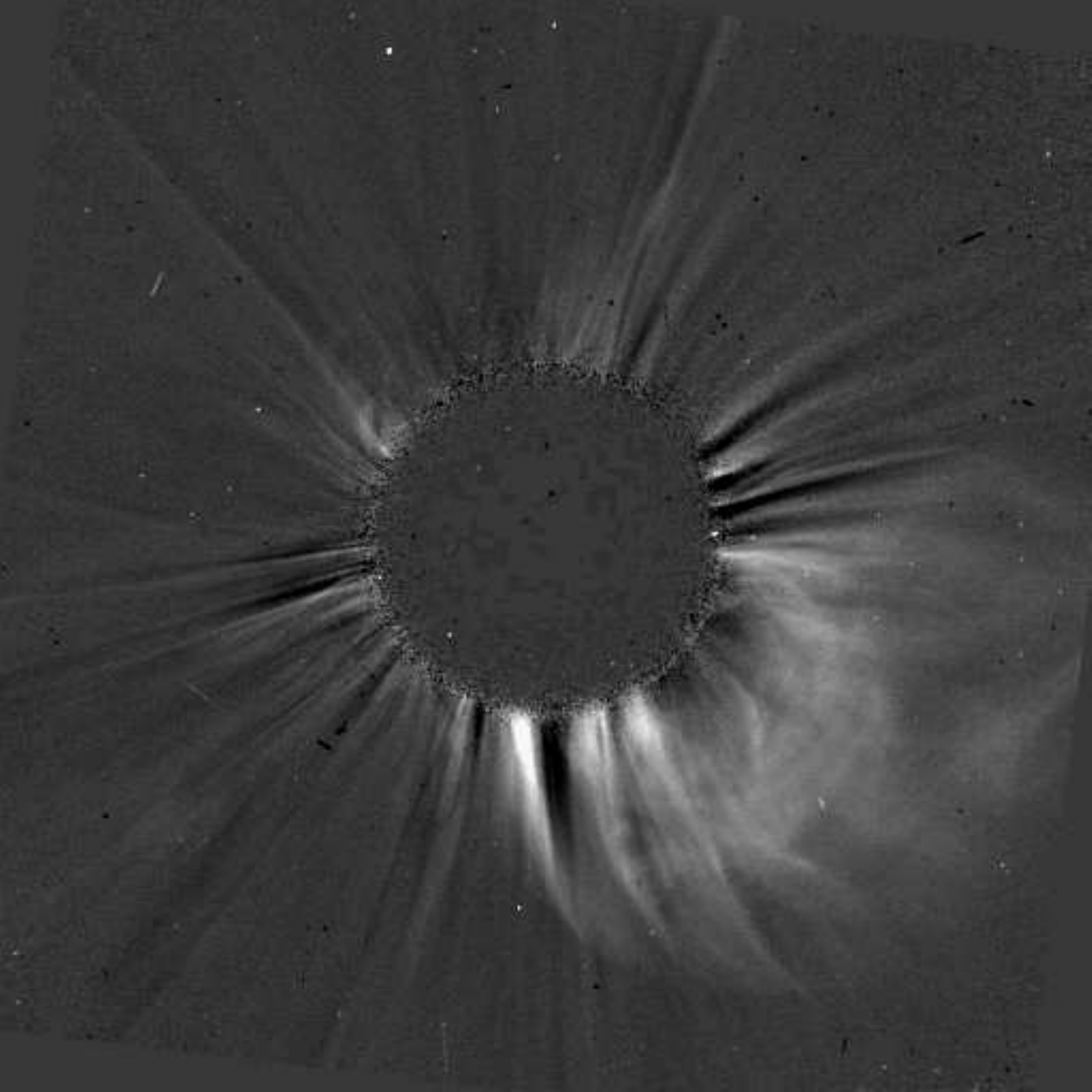}
             \hspace*{-0.02\textwidth}
               \includegraphics[width=0.4\textwidth,clip=]{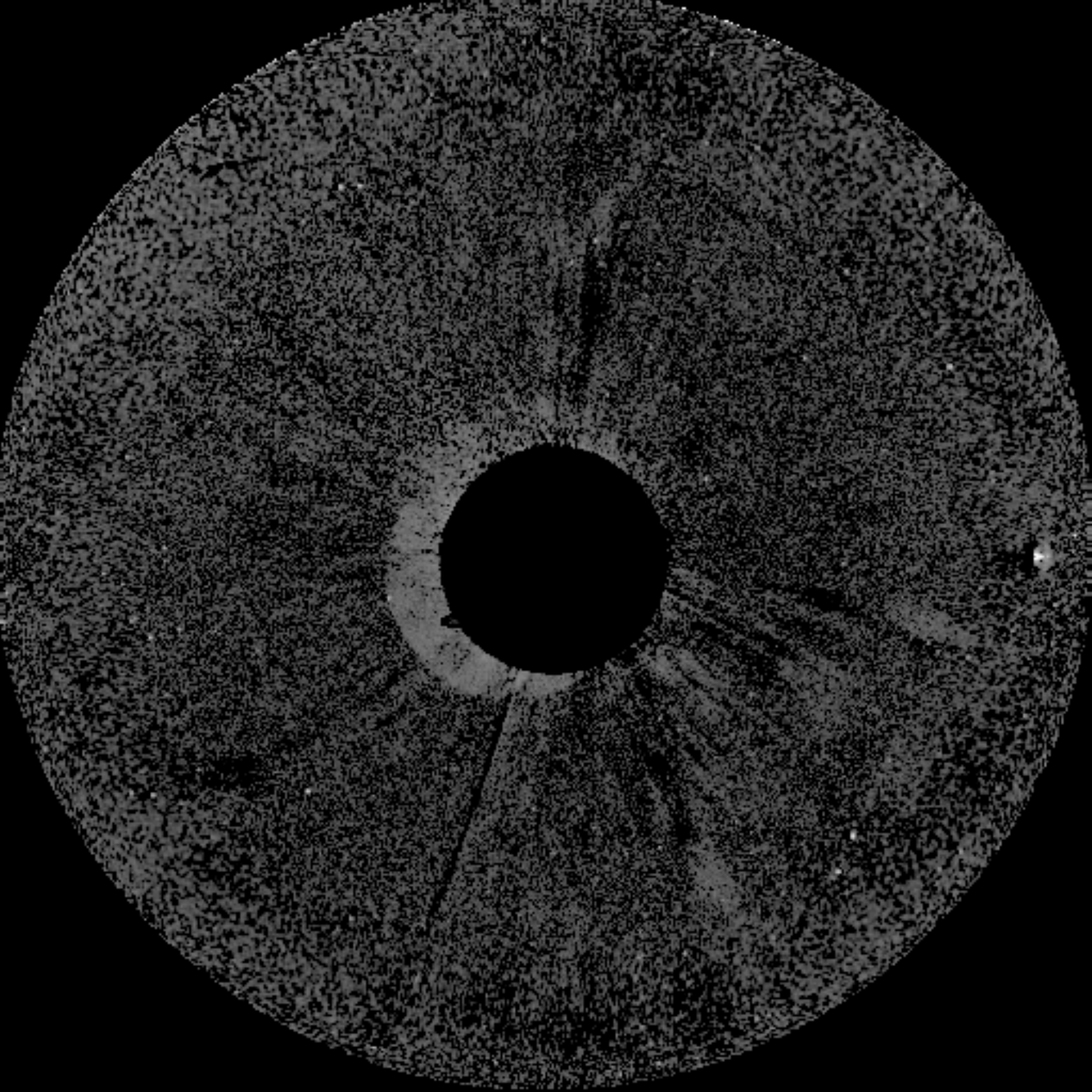}
               }
                 \centerline{\hspace*{0.04\textwidth}
              \includegraphics[width=0.4\textwidth,clip=]{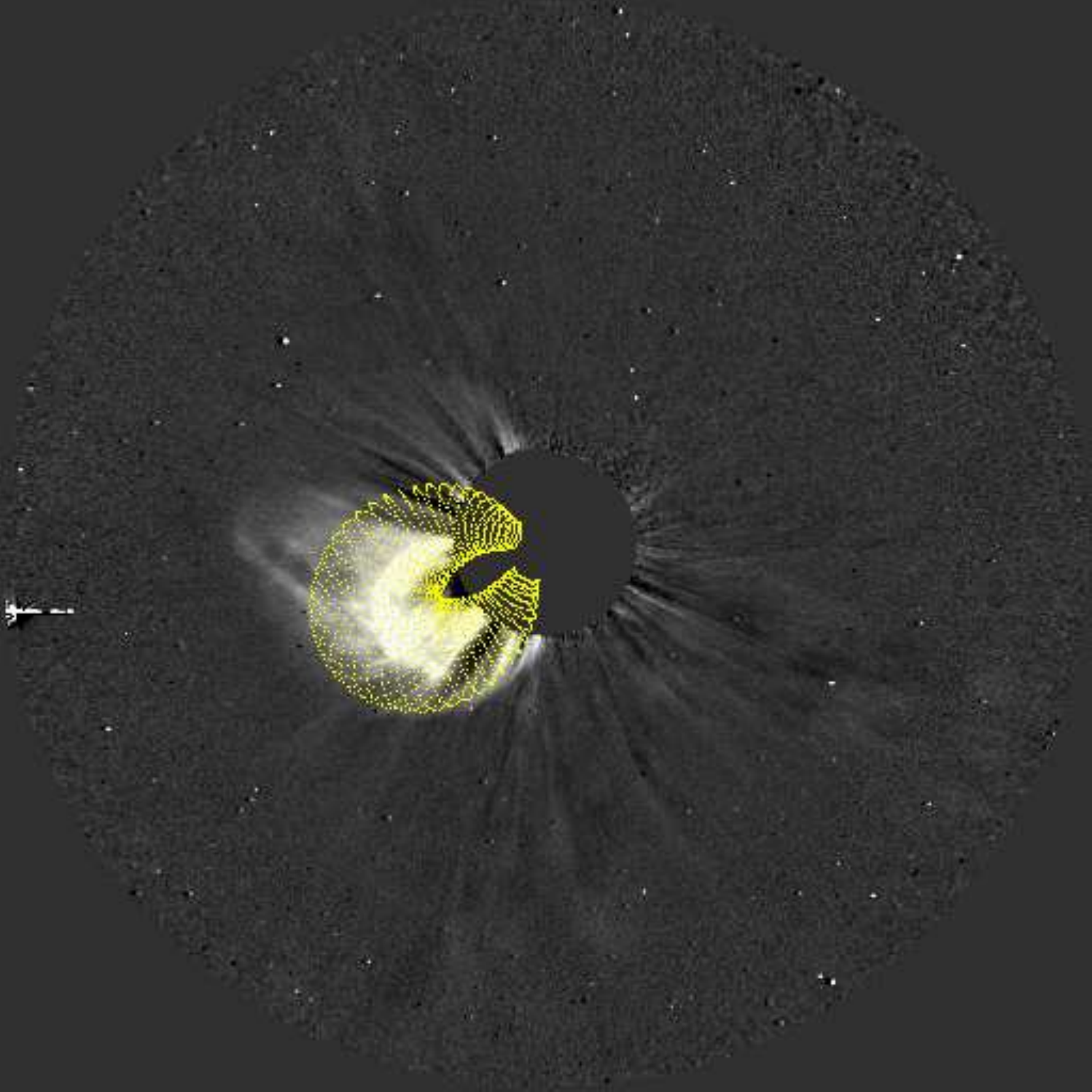}
               \hspace*{-0.02\textwidth}
               \includegraphics[width=0.4\textwidth,clip=]{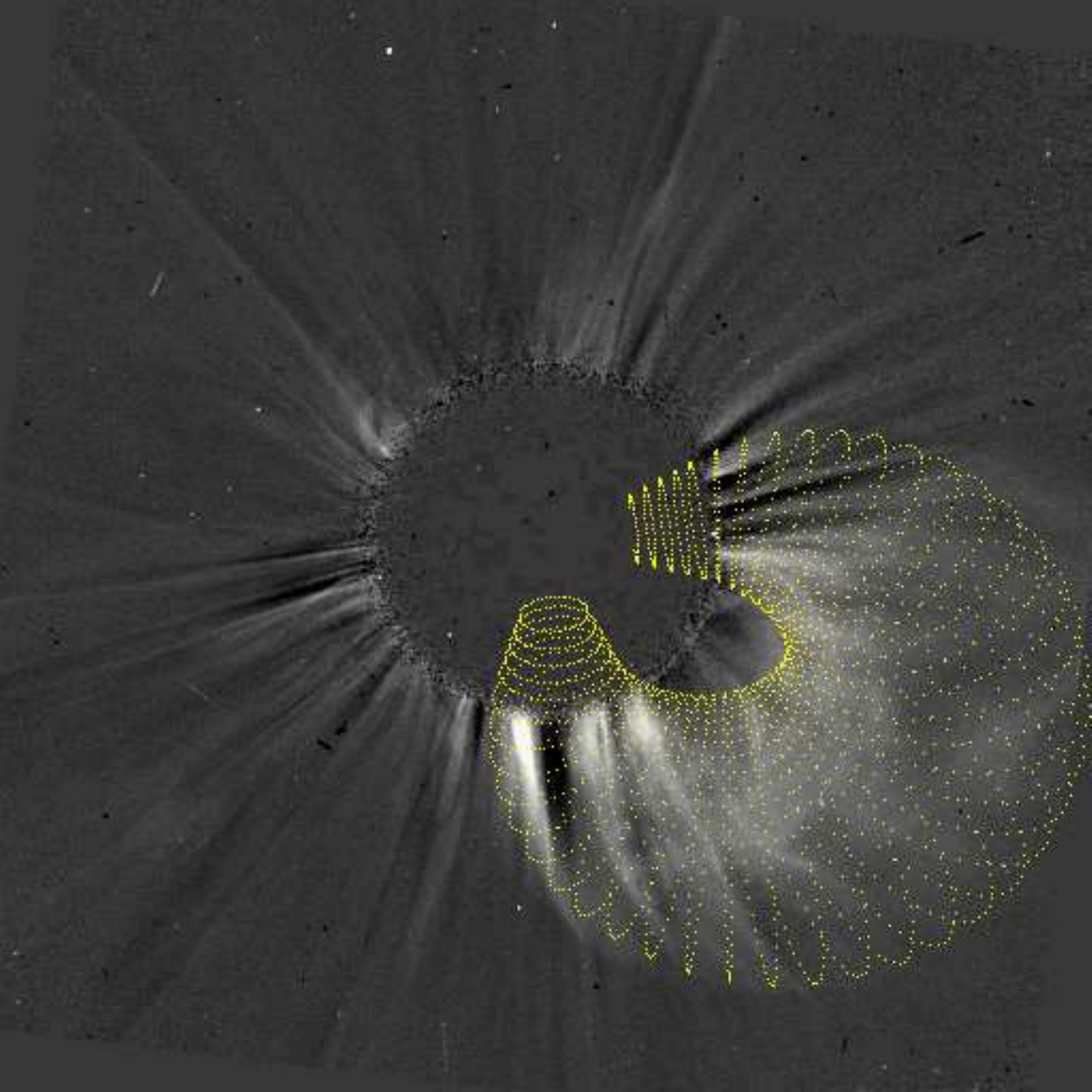}
              \hspace*{-0.02\textwidth}
               \includegraphics[width=0.4\textwidth,clip=]{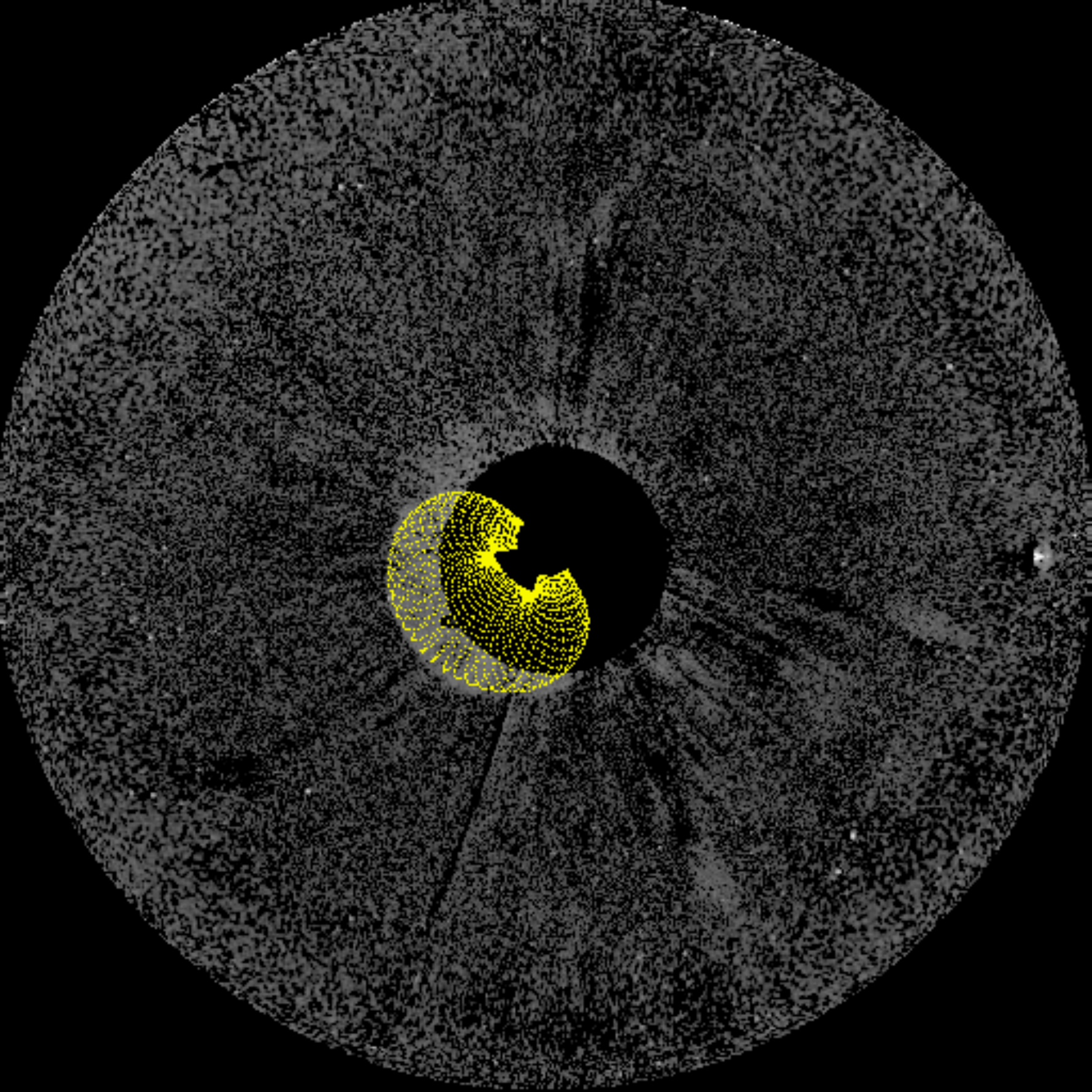}
                }
\vspace{0.0261\textwidth}  
\caption[GCS fit for CME 38 at 08:24]{GCS fit for CME 38 on December 07, 2013 at 08:24 UT at height $H=6.8$ \Rs. Table \ref{tblapp} 
lists the GCS parameters for this event.}
\label{figa38}
\end{figure}

We show the a GCS model fit to the LASCO C2, COR2 A and COR2 B data for all CMEs at one timestamp. 
In Figures \ref{figa1} to \ref{figa38}, the first row show the remote sensing observation data at a particular timestamp. The second row shows the 
GCS model fit to this data. The left panel is coronagraph image from STEREO COR 2A, middle 
panel is data from LASCO C2 and right panel is data from STEREO COR2 B. The flux-rope like wiremesh structure (yellow) represents the fitting
to each coronagraph image using GCS technique. Only one figure per event is shown to keep the file compact. 
The caption for each figure indicates the date of CME event, time of observation at which the fit is shown and the corresponding height at that time using GCS fitting.
Table \ref{tblapp} lists all the GCS parameters at the selected timestamp for each CME. 

\clearpage
\vspace*{3.cm}

\clearpage
\vspace*{3.cm}

\clearpage
\vspace*{3.cm}

\clearpage
\vspace*{3.cm}

\clearpage
\vspace*{3.cm}

\clearpage
\vspace*{3.cm}

\clearpage
\vspace*{3.cm}

\clearpage
\vspace*{3.cm}

\clearpage
\vspace*{3.cm}

\clearpage
\vspace*{3.cm}

\clearpage
\vspace*{3.cm}

\clearpage
\vspace*{3.cm}

\clearpage
\vspace*{3.cm}

\clearpage
\vspace*{3.cm}

\clearpage
\vspace*{3.cm}

\clearpage
\vspace*{3.cm}

\clearpage
\vspace*{3.cm}

\clearpage
\vspace*{3.cm}

\clearpage
\vspace*{3.cm}

\clearpage
\vspace*{3.cm}

\clearpage
\vspace*{3.cm}

\clearpage
\vspace*{3.cm}

\clearpage
\vspace*{3.cm}

\clearpage
\vspace*{3.cm}

\clearpage
\vspace*{3.cm}

\clearpage
\vspace*{3.cm}

\clearpage
\vspace*{3.cm}

\clearpage
\vspace*{3.cm}

\clearpage
\vspace*{3.cm}

\clearpage
\vspace*{3.cm}

\clearpage
\vspace*{3.cm}

\clearpage
\vspace*{3.cm}

\clearpage
\vspace*{3.cm}

\clearpage
\vspace*{3.cm}

\clearpage
\vspace*{3.cm}

\clearpage
\vspace*{3.cm}

\clearpage
\vspace*{3.cm}

\clearpage
\vspace*{3.cm}

\newpage
\begin{table}
\caption[GCS fitting parameters for each CME as corresponding to Figures \ref{figa1}-\ref{figa38}]{GCS parameters corresponding to the fittings in 
Figures \ref{figa1}-\ref{figa38} are given. CME no. indicates the serial number of the CME as referenced in Table 
\ref{tbl31}. Date and Time are the date of CME event and the timestamp at which the GCS fit for the CME at height $H$ is shown in the corresponding figure. 
GCS parameters at $H$ are given by Carrington longitude ($\phi$), latitude ($\theta$), 
tilt ($\gamma$), aspect ratio ($\kappa$) and half angle ($\alpha$). Fast CMEs are indicated by a 
superscript (f) and events from \citet{Sac15} are indicated by a superscript($\ast$) 
in their corresponding serial number.} 
\label{tblapp}
\centering
 \begin{tabular}{lcccccccc} 
  \hline
 No.& Date & Time  & $H$ & $\phi$& $\theta$ &  $\gamma$ & $\kappa$ &  $\alpha$ \\
    &  &(U.T.)&(\Rs) & ($^\circ$) &($^{\circ}$)  & ($^{\circ}$)& & ($^{\circ}$)  \\
  \hline
  1$^{*}$ & 2010 Mar. 19  & 17:54  & 10.07  & 116.5  & -11.7  & -35.2  &  0.28  & 10.9 \\
  2$^{* f}$ & 2010 Apr. 03 & 11:24 & 9.64 & 261.6 & -24.6 & 12.3 & 0.37 & 25.1\\
   3$^{*}$  & 2010 Apr. 08 & 06:54 & 10.2 & 189.3 & -8.9 & -29.1 & 0.19 & 32.7 \\
   4$^{*}$  & 2010 Jun. 16 & 18:54 & 9.9 & 334.8 & 3.3 & -32.9 & 0.26 & 9.5 \\
   5$^{*}$   & 2010 Sep. 11 & 05:39 & 10.3 & 276.1 & 20.7 & -43.0 & 0.41  &  18.1  \\
  6$^{*}$  & 2010 Oct. 26 & 11:54 & 10.0 & 70.3 & -25.2 & -55.3 &0.26 &  29.6\\
   7      & 2010 Dec. 23 & 10:24 & 9.5 & 29.1  &-19.6 & -15.6 &  0.45 & 18.2\\
   8      & 2011 Jan. 24 & 06:39 & 8.9 & 336.5 & -13.4 & -15.1 & 0.30 & 22.1 \\
   9$^{*}$  & 2011 Feb. 15 &  03:39 & 10.9 & 23.7 &  -8.9 & 26.8 &  0.47 & 40.8 \\
   10       & 2011 Mar. 03 & 08:53 & 10.0& 175.5 & -22.9  & 8.4  &  0.35 & 21.5\\
   11$^{*}$  & 2011 Mar. 25 & 15:39 & 10.0& 206.8 & -3.9 & -11.7&   0.31 & 40.8 \\
   12        & 2011 Apr. 09 & 02:54 &  10.2 & 41.4 &  5.6 &  -6.2 &   0.24 &  35.2 \\
   13        & 2011 Jun. 14 & 09:54 & 9.4 &202.4 &  -1.1 & 41.4 &  0.28 &  57.0 \\
   14$^{f}$  & 2011 Jun. 21 & 04:24 & 11.3 & 128.6 & 5.0 &  -8.4 & 0.46 & 13.7 \\
   15$^{f}$  & 2011 Jul. 09 & 02:24  &10.0 &  264.9 & -14.5 &  15.6 & 0.37 &  18.5\\
   16$^{f}$  & 2011 Aug. 04 & 04:39 & 9.4 & 324.2 & 19.6 &  65.4 & 0.68 &  29.6 \\
   17        & 2011 Sep. 14 & 02:24 & 10.5 & 134.2 &  19.0 & -38.0&  0.43 &  41.4 \\
   18$^{f}$  & 2011 Oct. 22 &11:24 & 6.2& 54.8 &  44.7 &  16.2 &  0.59 &  45.0 \\
   19        & 2011 Oct. 26 &13:54 & 10.2 &302.9 &  7.3 & -1.1& 0.46 &   9.5\\
   20        & 2011 Oct. 27 & 13:39 & 7.8 & 223.6 & 29.1 &  16.8 &  0.36 & 16.5 \\
   21$^{f}$  & 2012 Jan. 19 & 15:54 & 7.2 &212.4 &  44.2 &  90.0 &0.47 &  58.1 \\
   22$^{f}$  & 2012 Jan. 23 & 04:24 & 8.9 &206.8&  28.5 &  57.6 & 0.48 &  41.1\\
   23$^{f}$  & 2012 Jan. 27 &18:39 & 5.3 &188.9 &  29.6 &  68.7 & 0.38 & 41.4\\
   24$^{f}$  & 2012 Mar. 13 & 18:24 & 11.5 &301.9 &  20.7 &-39.7 & 0.74 &  73.2   \\
   25        & 2012 Apr. 19 &17:24 & 8.4 &  81.6& -27.9 &  0.0 & 0.43 & 30.5 \\
   26$^{f}$  & 2012 Jun. 14 & 14:39&  7.8 & 91.7 &-21.8 & -87.2& 0.37&  31.9 \\
   27$^{f}$  & 2012 Jul. 12 & 17:39 & 9.2 & 88.3 & -10.6&  77.7 & 0.45 &  34.9 \\
   28$^{f}$  & 2012 Sep. 28 & 00:54 & 9.9 & 165.5 &  17.3 &  86.1 &0.42 &  41.6\\
   29        & 2012 Oct. 05 & 05:54 &   10.0 & 55.9& -20.1 &  36.9 &  0.32 &  38.8 \\
   30        & 2012 Oct. 27 &18:54 & 10.5 &118.5 &  5.6 & -35.8 &0.28 &  39.9\\
   31        & 2012 Nov. 09 &16:54 & 9.9 & 285.1 & -17.9 &   6.1 & 0.48 &  34.6\\
   32        & 2012 Nov. 23 &15:24 & 8.9 &  90.6 & -21.2 & -65.9 &0.56 &  10.3\\
   33$^{f}$  & 2013 Mar. 15 &07:39 & 10.2 & 71.5 & -6.7 & -86.1 &  0.31 &  39.7 \\
   34$^{f}$  & 2013 Apr. 11 &08:24 & 10.2 & 77.1 &  -1.1 &  90.0 & 0.29 &  47.2\\
   35$^{f}$  & 2013 Jun. 28 & 02:39 & 8.2 & 176.6 & -34.6 & -19.6 &0.41 &   5.3 \\
   36$^{f}$  & 2013 Sep. 29 & 22:54 & 8.1 &360.0 &  22.9 &  90.0 &     0.43 &  47.2 \\
   37$^{f}$  & 2013 Nov. 07 & 00:39 & 7.7 &304.1 & -32.4 &-74.9 & 0.48 &  12.3\\
    38$^{f}$ & 2013 Dec. 07 & 08:24 & 6.8 & 221.4 & -31.8 &  51.4 & 0.36 &  46.9\\
\hline
\end{tabular}
\end{table}

%

\bibliographystyle{apalike}
\bibliography{references} 
\addcontentsline{toc}{section}{\textbf{Bibliography}}

\section*{Publications in international refereed journals}\addcontentsline{toc}{section}{\textbf{Publications in international refereed journals}}

\begin{enumerate}

\item \label{Sachdeva15} CME propagation – where does aerodynamic drag ``take over''?\ \\
\textit{\textbf {Sachdeva, N.}}, Subramanian, P., Colaninno, R.,  Vourlidas, A.  2015, The Astropysical Journal, 809, 158

\item \label{Sachdeva17} CME dynamics using STEREO and LASCO observations: The relative importance of Lorentz Forces and Solar wind drag. \\
\textit{ \textbf {Sachdeva, N.}}, 
Subramanian, P., Vourlidas, A., Bothmer, V.  2017, Solar Physics, 292, 118
\end{enumerate}
 \end{document}